%% file: phd.tex
\documentclass[12pt]{scrbook}

\usepackage[a4paper, margin=2.3cm,top=2.5cm,bottom=2.5cm,left=2.0cm]{geometry}
\usepackage[latin1]{inputenc}
\usepackage[T1]{fontenc}
\usepackage{graphicx}
\usepackage{placeins} 
\usepackage[intlimits]{amsmath}
\usepackage{mathtools}
\usepackage{cite}
\usepackage{amssymb}
\usepackage{marvosym}
\usepackage[seperr,repeatunits=false,obeybold]{siunitx}
\usepackage{subfigure}
\usepackage{float}
\usepackage{comment}
\usepackage{fancyhdr}
\usepackage{enumitem}
\usepackage{todonotes}
\usepackage{textcomp} 
\usepackage{url}
\usepackage{feynmp-auto}
\usepackage{scalerel}
\usepackage{aas_macros}
\usepackage[toc,title,header]{appendix}

\makeatletter
    \let\saved@bibitem\@bibitem
\makeatother
\usepackage{bibentry}
\usepackage{hyperref}

\newcommand{\emptypage}[0]{
  \newpage
  \thispagestyle{empty}
  \mbox{}
  \thispagestyle{empty}
}

\newcommand{\Hone}[0]{H\,\scriptsize I\normalsize}
\newcommand{\Htwo}[0]{H\,\scriptsize II\normalsize}
\newcommand{\Hmoc}[0]{$\mathrm{H_{2}}$}

\newcommand{\Rsun}[0]{R\textsubscript{\(\odot\)}}
\newcommand{\Msun}[0]{M\textsubscript{\(\odot\)}}
\newcommand\numberthis{\addtocounter{equation}{1}\tag{\theequation}}

\newcommand*{\hms}[2][]{{
    \def\SIUnitSymbolDegree{\textsuperscript{h}}%
    \def\SIUnitSymbolArcminute{\textsuperscript{m}}%
    \def\SIUnitSymbolArcsecond{\textsuperscript{s}}%
    \ang[#1]{#2}}%
}

\newcommand{\mylbrace}[2]{\vspace{#2pt}\hspace{2pt}\scaleleftright[\dimexpr6pt+#1\dimexpr0.11pt]{\lbrace}{\rule[\dimexpr2pt-#1\dimexpr0.5pt]{-4pt}{#1pt}}{.}}

\DeclareSIUnit\permille{\text{\textperthousand}}
\DeclareSIUnit\nucleon{n}
\DeclareSIUnit\year{yr}
\DeclareSIUnit\erg{erg}
\DeclareSIUnit\arcsecondtext{as}

\begin{document}

\include{titlepage}

\include{abstract}
\include{eid}

\include{publications}

\include{toc}

\include{introduction}
\include{cosmic_rays}
\include{detector}
\include{analysis}
\include{fermi}
\include{results}
\include{summary}


\include{bibliography}

\include{appendix}

\include{acknowledgments}

\end{document}

%% file: titlepage.tex

\begin{titlepage}
  \begin{center}
    \vspace*{1.0cm} \LARGE{\textbf{Measurement of High Energy Gamma Rays from
        \SI{200}{\mega\electronvolt} to
        \SI{1}{\tera\electronvolt} with the Alpha Magnetic Spectrometer on the International Space Station}} \\
    \bigskip
    \bigskip
    \bigskip
    \bigskip
    \normalsize{Von der Fakult\"at f\"ur Mathematik, Informatik und Naturwissenschaften der RWTH
      Aachen University zur Erlangung des akademischen Grades eines Doktors der
      Naturwissenschaften genehmigte Dissertation} \\
    \bigskip
    \bigskip
    \bigskip
    \bigskip
    \normalsize{vorgelegt von}\\
    \bigskip
    \bigskip
    \bigskip
    \bigskip
    \large{Dipl. Phys.} \\
    \large{\textbf{Bastian Beischer}} \\
    \bigskip
    \bigskip
    \bigskip
    \normalsize{aus Norden} \\
    \bigskip
    \bigskip
    \bigskip
    \bigskip
    \bigskip
    \bigskip
    \bigskip
    \bigskip
    \normalsize{
      \begin{tabular}{ll}
        Berichter:&Universit\"atsprofessor Prof. Dr. Stefan Schael \\
                  &Universit\"atsprofessor Prof. Dr. Christopher Wiebusch \\
      \end{tabular}
    }\\
    \bigskip
    \bigskip
    \bigskip
    \bigskip
    \normalsize{Tag der m\"undlichen Pr\"ufung: 17.06.2020} \\
    \bigskip
    \bigskip
    \bigskip
    \bigskip
    \small{Diese Dissertation ist auf den Internetseiten der
      Universit\"atsbibliothek online verf\"ugbar.}

  \end{center}
\end{titlepage}

\emptypage


%% file: abstract.tex

\pagestyle{empty}

\begin{center}
  \large
  \textbf{Abstract} \\
  \vspace*{1cm}
  Measurement of High Energy Gamma Rays from \SI{200}{\mega\electronvolt} to
  \SI{1}{\tera\electronvolt} with the Alpha Magnetic Spectrometer on the International Space Station
\end{center}
\vspace*{1cm}

In this thesis a measurement of the high energy $\gamma$-ray flux between
\SI{200}{\mega\electronvolt} and \SI{1}{\tera\electronvolt} with the Alpha Magnetic Spectrometer is
presented. The Alpha Magnetic Spectrometer (\mbox{AMS-02}) is a multi-purpose particle detector
mounted externally on the International Space Station. \mbox{AMS-02} is continuously collecting
scientific data since its installation in May 2011.

Although primarily designed for the measurement of charged cosmic rays \mbox{AMS-02} is capable of
measuring high energy $\gamma$-rays in two complementary modes. The large background of charged
particles is overcome with the help of the excellent particle detection efficiency of the detector.

In the first mode the electron and positron pair from a photon conversion in the upper part of the
detector is reconstructed with the help of the silicon tracker. In this mode the photon direction is
estimated from the two trajectories and its energy is inferred from the curvature of the two tracks
in the AMS magnetic field.

In the second mode the photon passes through almost the entire detector and produces an
electromagnetic shower in the calorimeter at the bottom of the experiment. In this case photon
direction and energy are estimated from the properties of the shower.

Two independent analyses are presented in this thesis, one for each of the two modes. The event
selection criteria and the associated resolution functions are presented in detail. The effective
area is estimated from a full detector Monte-Carlo simulation and corrected for the most important
differences between data and simulation. A full sky model for $\gamma$-rays is constructed from
diffuse emission predictions and recent $\gamma$-ray source catalogs. A dedicated analysis of
\mbox{Fermi-LAT} data is performed to fully enable a detailed comparison with the AMS result.

The measured flux of $\gamma$-rays is presented for various parts of the sky, including comparisons
with \mbox{Fermi-LAT} data and with the constructed model. The inner galaxy is studied in detail, as
an example of a region in which the photon flux is dominated by diffuse emission. The fluxes of
several $\gamma$-ray producing sources, including Vela, Geminga and the Crab pulsar are shown. The
Geminga pulsar is studied in detail, revealing its pulsed emission of $\gamma$-rays in the
\mbox{AMS-02} data, which allows to measure its frequency of rotation and to estimate its magnetic
field strength and age. Finally, \mbox{AMS-02} observed an outburst of the flaring blazar CTA-102 at
the end of 2016.

These important \mbox{AMS-02} results represent the first independent test of the \mbox{Fermi-LAT}
data in the energy range from \SI{200}{\mega\electronvolt} to \SI{1}{\tera\electronvolt}.

\clearpage

\begin{center}
  \large
  \textbf{Zusammenfassung} \\
  \vspace*{1cm}
  Messung von hoch-energetischer Gammastrahlung von \SI{200}{\mega\electronvolt} bis
  \SI{1}{\tera\electronvolt} mit dem Alpha Magnet Spektrometer auf der Internationalen Raumstation
\end{center}
\vspace*{1cm}

In dieser Arbeit wird eine Messung des hoch-energetischen $\gamma$-ray Flusses zwischen
\SI{200}{\mega\electronvolt} und \SI{1}{\tera\electronvolt} mit dem Alpha Magnet Spektrometer
vorgestellt. Das Alpha Magnet Spektrometer (\mbox{AMS-02}) ist ein Mehrzweck-Teilchendetektor,
welcher extern auf der Internationalen Raumstation angebracht ist. Seit seiner Installation im Mai
2011 zeichnet \mbox{AMS-02} kontinuierlich wissenschaftliche Daten auf.

Obwohl \mbox{AMS-02} prim{\"a}r f{\"u}r die Messung von geladener kosmischer Strahlung konzipiert
wurde, ist es in der Lage hoch-energertische $\gamma$-Strahlung auf zwei komplement{\"a}re Arten zu
messen. Der gro{\ss}e Untergrund an geladenen Teilchen wird mit Hilfe der exzellenten
Teilchennachweiseffizienz des Detektors unterdr{\"u}ckt.

Im ersten Modus werden die Spuren je eines Elektrons und eines Positrons aus einer Photonkonversion
im oberen Detektor mit dem Siliziumspurdetektor rekonstruiert. Dabei wird die Photonrichtung aus den
beiden Trajektorien bestimmt und die Energie des Photons {\"u}ber die Kr{\"u}mmung der beiden Spuren
im AMS Magnetfeld gemessen.

Im zweiten Modus passieren Photonen fast den gesamten Detektor und produzieren dann im Kalorimeter
einen elektromagnetischen Schauer am unteren Ende des Experiments. In diesem Fall werden die
Photonrichtung und Energie aus den Eigenschaften des Schauers bestimmt.

Zwei unabh{\"a}ngige Analysen werden in dieser Arbeit vorgestellt, eine f\"ur jeden der beiden
Modi. Die Ereignisselektionskriterien werden dargelegt und die dazugeh\"origen
Aufl\"osungsfunktionen im Detail bestimmt. Die effektive Fl\"ache wird aus einer Monte-Carlo
Simulation des gesamten Detektors berechnet und die gr\"o{\ss}ten Unterschiede zwischen Daten und
Simulation werden korrigiert. Ein Modell der $\gamma$-Strahlung, welches f\"ur den gesamten Himmel
g\"ultig ist, wird aus Vorhersagen f\"ur die diffuse Emission und aktuellen Katalogen von
$\gamma$-Strahlungsquellen konstruiert. Eine dedizierte Analyse von \mbox{Fermi-LAT} Daten wird
durchgef\"uhrt, um einen detaillierten Vergleich mit dem AMS Ergebnis zu erm\"oglichen.

Die gemessenen $\gamma$-ray Fl{\"u}sse werden f\"ur verschiedene Regionen am Himmel vorgestellt und
mit den \mbox{Fermi-LAT} Daten und dem konstruierten Modell verglichen. Die innere Galaxie, als
Beispiel f\"ur eine Region in der die diffuse Emission dominiert, wird im Detail studiert. Die
Fl\"usse von mehreren $\gamma$-Strahlung produzierenden Quellen (z.B. Vela, Geminga und der Pulsar
im Krebsnebel) werden gezeigt. Im Besonderen wird der Geminga Pulsar untersucht, wodurch die
gepulste Emission von $\gamma$-Strahlung in dem AMS Daten sichtbar wird. Daraus wird die
Rotationsfrequenz, die St\"arke des Magnetfeldes und das Alter des Pulsars ermittelt. Desweiteren
hat \mbox{AMS-02} einen Ausbruch des Blasaren CTA-102 Ende 2016 beobachtet.

Diese wichtigen \mbox{AMS-02} Ergebnisse stellen den ersten unabh\"angigen Test der \mbox{Fermi-LAT}
Daten im Energiebereich zwischen \SI{200}{\mega\electronvolt} und \SI{1}{\tera\electronvolt} dar.



%% file: eid.tex

\thispagestyle{empty}
\pagestyle{empty}
\begin{center}
  \large
  \textbf{Eidesstattliche Erkl\"arung}
\end{center}

\bigskip
\bigskip

Bastian Beischer erkl\"art hiermit, dass diese Dissertation und die darin dargelegten Inhalte die
eigenen sind und selbstst\"andig, als Ergebnis der eigenen origin\"aren Forschung, generiert wurden.

\bigskip
\bigskip

Hiermit erkl\"are ich an Eides statt

\begin{enumerate}
\bigskip
\item{Diese Arbeit wurde vollst\"andig oder gr\"o{\ss}tenteils in der Phase als Doktorand dieser
    Fakult\"at und Universit\"at angefertigt;}

\bigskip
\item{Sofern irgendein Bestandteil dieser Dissertation zuvor für einen akademischen Abschluss oder
    eine andere Qualifikation an dieser oder einer anderen Institution verwendet wurde, wurde dies
    klar angezeigt;}

\bigskip
\item{Wenn immer andere eigene- oder Ver\"offentlichungen Dritter herangezogen wurden, wurden diese
    klar benannt;}

\bigskip
\item{Wenn aus anderen eigenen- oder Ver\"offentlichungen Dritter zitiert wurde, wurde stets die
    Quelle hierfür angegeben. Diese Dissertation ist vollst\"andig meine eigene Arbeit, mit der
    Ausnahme solcher Zitate;}

\bigskip
\item{Alle wesentlichen Quellen von Unterst\"utzung wurden benannt;}

\bigskip
\item{Wenn immer ein Teil dieser Dissertation auf der Zusammenarbeit mit anderen basiert, wurde von
    mir klar gekennzeichnet, was von anderen und was von mir selbst erarbeitet wurde;}

\bigskip
\item{Kein Teil dieser Arbeit wurde vor deren Einreichung ver\"offentlicht.}
\end{enumerate}


%% file: publications.tex

\thispagestyle{empty}
\pagestyle{empty}
\begin{center}
  \large
  \textbf{List of Publications}
\end{center}


\begingroup
  \makeatletter
    \let\@bibitem\saved@bibitem
    \nobibliography*
  \makeatother
\endgroup

\makeatletter
\renewcommand\bibentry[1]{\nocite{#1}{\frenchspacing
     \@nameuse{BR@r@#1\@extra@b@citeb}}}
\makeatother

The ACsoft software package, of which I am one of the principal authors, was used extensively for
these publications (which were selected as Editor's Suggestions):

\begin{itemize}
\item \bibentry{AMS02_ElecPos_2014}
\item \bibentry{AMS02_ElecPosTime_2018}
\end{itemize}

My work on the Transition Radiation Detector of AMS has contributed to the following publication:

\begin{itemize}
\item \bibentry{AMS02_Detector_TRD_MC_2017}
\end{itemize}

My work for the successful operation of the TRD and of AMS as a whole was relevant for these
publications:

\begin{itemize}
\item \bibentry{AMS02_PosFrac1_2013}
\item \bibentry{AMS02_PosFrac2_2014}
\item \bibentry{AMS02_AllElec_2014}
\item \bibentry{AMS02_Proton_2015}
\item \bibentry{AMS02_Helium_2015}
\item \bibentry{AMS02_AntiProton_2016}
\item \bibentry{AMS02_BoverC_2016}
\item \bibentry{AMS02_HeCO_2017}
\item \bibentry{AMS02_LiBeB_2018}
\item \bibentry{AMS02_Nitrogen_2018}
\item \bibentry{AMS02_ProtonHeliumTime_2018}
\item \bibentry{AMS02_Positrons_2019}
\item \bibentry{AMS02_Electrons_2019}
\item \bibentry{AMS02_HeliumIsotopes_2019}
\end{itemize}

\thispagestyle{empty}
\pagestyle{empty}

\emptypage
\KOMAoptions{open=right}


%% file: toc.tex

\thispagestyle{empty}
\pagestyle{empty}
\begingroup
\renewcommand*{\chapterpagestyle}{empty}
\tableofcontents
\endgroup
\thispagestyle{empty}
\pagestyle{empty}
\emptypage
\emptypage


%% file: introduction.tex

\pagestyle{fancy}
\setcounter{page}{1}
\chapter{Introduction}
\label{sec:introduction}

The physics of high energy $\gamma$-rays is a gold mine for scientific discovery and full of unique
possibilities. Since $\gamma$-rays form the high energy limit of electromagnetic radiation, they are
associated with the most violent phenomena in the cosmos. It takes spectacular objects, such as
pulsars or blazars to produce photons at \si{\giga\electronvolt} and \si{\tera\electronvolt}
energies. In addition, new physics such as the ominous dark matter, is predicted to manifest itself
in an excess of $\gamma$-rays in many models~\cite{Dodelson2009,Serpico_DM_2009,Bringmann_2012}.

At the same time, because photons can pass through the universe almost undisturbed, they can be
directly associated with their sources, making them the perfect messenger.

As an example, measurements of dwarf spheroidal galaxies provide some of the most stringent limits
on the dark matter annihilation cross
section\cite{Fermi_DM_DwarfSphG_2015,Fermi_DM_Limits_2011}. Because the photon energy does not
change (which is in stark contrast to cosmic ray energies), $\gamma$-rays allow to search for line
signatures of dark matter decays, which if detected, would allow to directly reconstruct the mass of
the dark matter particle.

Gamma ray bursts (GRBs) are among the most violent and least understood phenomena in the
universe. The enormous energy released within the course of a few seconds, manifests itself in
massive $\gamma$-ray flares.

Within our own galaxy, the study of diffuse emission of $\gamma$-rays opens a new window to unveil
the mysteries of cosmic rays\cite{Fermi_Diffuse_2012,Fermi_Diffuse_IEM_2016}, which can otherwise
only be studied in the vicinity of the solar system

Excess diffuse emission produced by the annihilation of dark matter particles, for example in the
galactic center, is another topic that has sparked enormous
interest~\cite{DM_Fermi_GC_2011,DM_Fermi_GC_2016,Fermi_GC_2017}. Large scale structures of unknown
origin, the Fermi bubbles~\cite{Fermi_Bubbles_2010_ADS} have been identified in the residuals and
continue to puzzle astronomers.

Measurements of $\gamma$-rays have contributed to the discovery of gravitational
waves~\cite{GW170817_2017}, and to the association of a cosmic neutrinos with flaring
blazars~\cite{IceCube_170922_2017}.

These are only some of the reasons why $\gamma$-ray astronomy is such a vital field.

On the other hand, experiments capable of studying $\gamma$-rays are relatively scarce. Because the
Earth's atmosphere is opaque to $\gamma$-radiation, experiments can be divided into two groups:
Satellites in space, which directly observe the radiation, but are expensive to launch and operate,
and telescopes on the Earth's surface which indirectly measure the electromagnetic showers produced
when the $\gamma$-ray hits the atmosphere. These telescopes are limited to the high energy end of
the $\gamma$-ray spectrum, and suffer from a limited field of view.

\begin{table}[t]
  \centering
  \caption{List of satellite based $\gamma$-ray experiments.}
  \label{tab:gamma-ray-experiments}
  \vspace*{5mm}
  \begin{tabular}[h]{l|l|l}
    Experiment & Energy Range & Start of Operations \\
    \hline
    OSO-3      &  \SI{50}{\mega\electronvolt} - \SI{300}{\mega\electronvolt}   & 1967 \\
    SAS-2      &  \SI{20}{\mega\electronvolt} - \SI{300}{\mega\electronvolt}   & 1972 \\
    COS-B      &  \SI{50}{\mega\electronvolt} - \SI{5}{\giga\electronvolt}     & 1975 \\
    EGRET      &  \SI{30}{\mega\electronvolt} - \SI{30}{\giga\electronvolt}    & 1991 \\
    AGILE      &  \SI{30}{\mega\electronvolt} - \SI{50}{\giga\electronvolt}    & 2007 \\
    Fermi-LAT  &  \SI{20}{\mega\electronvolt} - > \SI{300}{\giga\electronvolt} & 2008 \\
    AMS-02     & \SI{200}{\mega\electronvolt} - \SI{1}{\tera\electronvolt}     & 2011
  \end{tabular}
\end{table}

Table~\ref{tab:gamma-ray-experiments} provides a historic overview of $\gamma$-ray satellites. In
the 1960s the OSO-3 satellite discovered the existence of cosmic $\gamma$-rays~\cite{OSO_3_1972} and
reported early measurements. In the 1970s, the satellites \mbox{SAS-2}~\cite{SAS_2_Results_1975} and
\mbox{COS-B}~\cite{COS_B_Detector_1975} were able to coarsely map the $\gamma$-ray sky and the first
sources were identified and studied. This included the discovery of $\gamma$-ray pulsars, such as
Geminga~\cite{SAS_2_Results_1975}.

In the 1990s the EGRET\cite{EGRET_Detector_1992} instrument on the Compton Gamma Ray Observatory
(CGRO), part of NASA's Great Observatories Program, was able to extend the list of
sources~\cite{EGRET_Catalog3_1999} and to study diffuse emission~\cite{EGRET_Diffuse_1997} in some
detail. The CGRO also contained the BATSE~\cite{BATSE_1992} and COMPTEL~\cite{COMPTEL_1993}
instruments, which were specifically designed to study GRBs and to extend the lower energy reach of
EGRET down to \SI{1}{\mega\electronvolt}, respectively.

Nowadays, the most sensitive experiment by far is the Large Area Telescope (LAT)~\cite{Atwood2009}
on the Fermi satellite. The satellite is also equipped with a Gamma Ray Burst monitor
(GBM)~\cite{Fermi_GBM_2009} for the detection of GRBs.

\begin{figure}[t]
  \centering
  \includegraphics[width=1.0\linewidth]{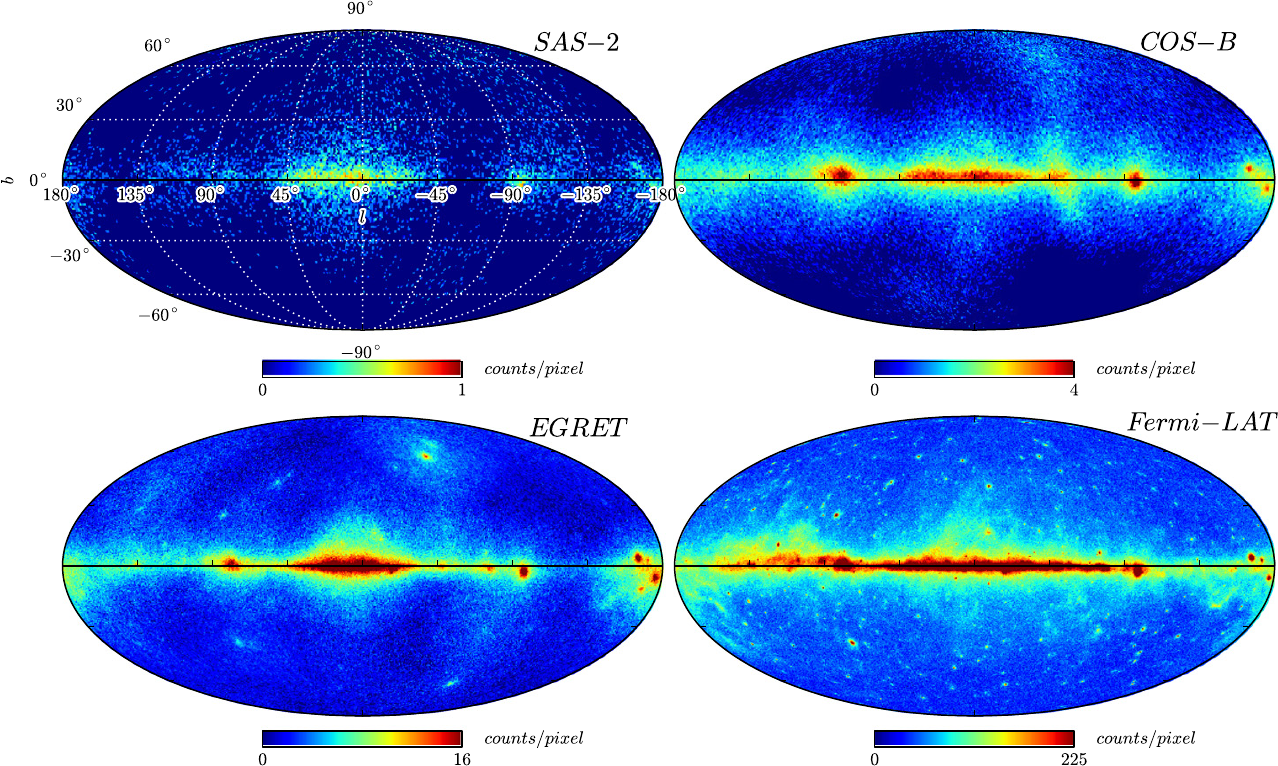}
  \caption{The $\gamma$-ray sky as seen by four different satellites~\cite{Fermi_Diffuse_IEM_2016}:
    \mbox{SAS-2} (1972, top left), \mbox{COS-B} (1975, top right), EGRET (1991, bottom left) and
    \mbox{Fermi-LAT} (2008, bottom right). The maps are Mollweide projections of galactic
    coordinates.}
  \label{fig:gamma-ray-experiments}
\end{figure}

Figure~\ref{fig:gamma-ray-experiments} shows the improvement of the instrumental technique, starting
with the \mbox{SAS-2} satellite, all the way to the present day \mbox{Fermi-LAT} experiment. Both
resolution and statistics improve as time progresses and more and more sources, structures and
phenomena can be identified.

The experimental results from the \mbox{Fermi-LAT} instrument results have revolutionized
$\gamma$-ray astronomy, with their unprecedented statistical accuracy and outstanding instrumental
performance.

Even though cosmic photons at \si{\giga\electronvolt} and \si{\tera\electronvolt} energies can not
be detected directly in ground based telescopes, there is a second class of experiments in which the
Cherenkov light produced by relativistic particles in the atmospheric showers initiated by
$\gamma$-rays is measured. Observatories which follow this approach are referred to as Imaging
Atmospheric Cherenkov Telescopes (IACT).

These experiments generally observe photons at very high energies (VHE), with sensitivities which
extend from approximately \SI{50}{\giga\electronvolt} all the way to
\SI{50}{\tera\electronvolt}. The major Cherenkov telescopes currently in operation are
MAGIC~\cite{MAGIC_Detector_2005}, H.E.S.S.~\cite{HESS_Project_2004} and
VERITAS~\cite{Veritas_Detector_2002}.

IACTs have excellent angular resolution and energy reach, with acceptable energy resolution. The
major difference with respect to satellite based $\gamma$-ray experiments is that these telescopes
can not be operated continuously and have a limited field of view. This means that they generally
study specific point sources, and are not well suited for studies of large scale diffuse
emission. This is also a disadvantage when trying to catch transient phenomena such as GRBs, since
alert notifications from other experiments are required and time is needed to reorient the
telescope.

It is interesting to note that bigger telescopes are required to extend the energy reach to lower
energies. The Cherenkov Telescope Array (CTA)~\cite{CTA_Design_2011} is aiming to improve the lower
energy limit down to \SI{20}{\giga\electronvolt}, and will generally improve the sensitivity. It is
currently under construction.

At the present time, there are only very few experiments which can measure photons in the energy
range between \SI{200}{\mega\electronvolt} and \SI{1}{\tera\electronvolt}. In fact, there is only
one experiment which covers the entirety of this energy range: The \mbox{Fermi-LAT} experiment.

Results obtained with the Fermi satellite have excellent statistical accuracy. But on the other hand
the experiment, like any other, suffers from systematic uncertainties related to calibrations and
imperfect understanding of the detector. Therefore, it is vital to have independent measurements as
cross checks, in particular given the scientific relevance of the Fermi results.

The \mbox{AMS-02} detector is a device which was built for the measurement of charged cosmic
rays. It was installed as an external payload on the International Space Station (ISS) on May 19th
2011 and is operational ever since. It is designed as a multi-\si{\tera\electronvolt} spectrometer
in space and is supported by the efforts of more than 500 international scientists.

Although AMS was designed for charged cosmic ray measurements, the tracker and calorimeter of the
experiment are also able to measure the properties of photons with outstanding precision. In
addition, due to its excellent detection efficiency of charged cosmic rays, the AMS detector allows
for a reliable reduction of charged particle backgrounds in $\gamma$-ray measurements.

The single photon pointing accuracy of the \mbox{AMS-02} tracker is comparable to, and at high
energies even better than, the \mbox{Fermi-LAT} pointing resolution. This is a result of the
excellent single point resolution of the AMS tracker.

In addition, the AMS calorimeter is easily capable of measuring photons with \si{\tera\electronvolt}
energies, due to its 17 radiation length thickness. At these energies, calorimeter shower lateral
leakage is a major problem in the \mbox{Fermi-LAT} calorimeter, and part of the reason why the
energy reach was originally limited to \SI{300}{\giga\electronvolt}~\cite{Fermi_LAT_Instrument_2009}
and only gradually increased later.

The resolution of the reconstructed energy in AMS calorimeter showers is
outstanding~\cite{AMS02_Detector_Ecal3D}. The fine calorimeter granularity allows to reconstruct the
photon direction with good accuracy~\cite{AMS02_Detector_Ecal3D}. In contrast to the
\mbox{Fermi-LAT} calorimeter, the AMS flight model ECAL energy scale was calibrated in a dedicated
test beam at CERN~\cite{AMS02_Detector_ECAL,AMS02_Detector_Ecal3D}. The in flight absolute energy
scale of the LAT has only been calibrated indirectly using $\sim$ \SI{10}{\giga\electronvolt}
electrons~\cite{Fermi_EnergyScale_2012,Fermi_AllElec_2017}.

Finally, the AMS detector was built with redundancy in mind. Because of this very important aspect,
the photon analysis is possible in two complementary modes: With the tracker, using photons which
converted in the upper detector, and with the calorimeter. These two modes are entirely
complementary, which allows to reduce systematic uncertainties.

All these aspects make the \mbox{AMS-02} detector very well suited for the measurement of high
energy $\gamma$-rays.

Still, AMS measurements will not be able to compete with the \mbox{Fermi-LAT} satellite in terms of
pure statistics, because of the limited acceptance of the detector in the two photon modes. But on
the other hand, there are many regions of the sky in which the Fermi measurement is dominated by
systematic uncertainties.

A good example is the inner galaxy, in which diffuse emission is the dominant process of
$\gamma$-ray production. The study of these photons allows to infer enormous amounts of information
about the galaxy and about cosmic rays, which is otherwise unavailable. It will be shown in this
thesis that AMS can contribute significantly to the measurement of diffuse emission.

Spectra from strong $\gamma$-ray sources, such as Vela, Geminga and the Crab pulsar are other
examples, in which AMS is able to add valuable information. AMS also surveys a sizable portion of
the sky at any given time. For this reason it is well suited for the study of transient phenomena,
in particular for the measurement of flaring sources.

In chapter~\ref{sec:understanding} the ingredients to describe the high energy $\gamma$-ray sky will
be assembled. This includes a short discussion of elementary processes relating to high energy
photon physics. The charged cosmic ray fluxes, as well as the interstellar structure of gas and
radiation fields in the Milky Way will be discussed. A short summary of a few important types of
$\gamma$-ray sources will also be given. These ingredients will then be put together to form a
predictive model of the $\gamma$-ray sky.

Chapter~\ref{sec:experimental-setup} will introduce the \mbox{AMS-02} detector as the experimental
apparatus whose measured data are the foundation for the analyses in this thesis. The detector is
located on the International Space Station, which provides the operational support and the
scientific environment for AMS. Both are described in detail, together with a review of astronomical
coordinate systems and transformations.

The data and its analysis is explained in chapter~\ref{sec:data-analysis}. The chapter contains a
description of the selection techniques and the response functions of the experiment, which include
the angular and energy resolution functions and the effective area. The response functions are used
to construct the exposure maps, which are in turn applied to the model of diffuse emission and
$\gamma$-ray sources in order to construct photon count predictions maps for the entire sky. This
chapter concludes with a discussion of a few necessary corrections to the photon Monte-Carlo
simulation and an overview of the systematic uncertainties relevant to the analysis.

Chapter~\ref{sec:fermi-lat-analysis} contains a short description of a complementary
\mbox{Fermi-LAT} analysis, which includes all necessary steps to construct photon fluxes from the
publicly available \mbox{Fermi-LAT} data. It also includes an extensive discussion of
\mbox{Fermi-LAT} systematic uncertainties.

Finally, the results for the photon fluxes in several regions of interest are presented and
discussed in chapter~\ref{sec:results}, followed by a short summary in chapter~\ref{sec:summary}.

\emptypage


%% file: cosmic_rays.tex

\chapter{Understanding the $\gamma$-ray Sky}
\label{sec:understanding}

The observable $\gamma$-ray sky is a complex superposition of many different processes. In order to
understand it, a complete picture of the structure and contents of the Milky Way must be combined
with elementary particle physics, which describes the fundamental interactions of particles. In
addition, some of the strongest $\gamma$-ray sources are violent extra galactic objects, such as
blazars, which are extremely compact and are able to accelerate particles to the highest
energies. These sources must also be incorporated in a realistic model.

In this chapter the fundamental processes relating to photons at the highest energies are described
and combined with recent measurements of galactic gas and radiation field distributions in order to
construct a model which can be compared to the \mbox{AMS-02} data.

An overview of the physical processes for the production and detection of $\gamma$-rays will be
given in section~\ref{sec:physics}. These processes are generally well understood, because they can
be studied in laboratories on Earth.

In order to predict the diffuse component of $\gamma$-rays it is important to understand the
distribution of gas and radiation fields in the Milky Way. A short review of recent measurements and
their results will be given in section~\ref{sec:galaxy}.

Charged cosmic rays (CRs), such as a protons, electrons and $\alpha$ particles are the projectiles
which in turn interact with the interstellar matter and produce diffuse $\gamma$-rays. Therefore the
flux and density of the most relevant cosmic ray species are needed for the calculation. In
addition, cosmic rays are a major background in the detection of photons in the \mbox{AMS-02}
measurement. For these reasons a summary of recent cosmic ray measurement is given in
section~\ref{sec:cosmic-rays}.

In addition to diffuse emission $\gamma$-rays are also produced in the vicinity of sources, such as
pulsars and Active Galactic Nuclei (AGN). A brief summary of the most relevant types of sources and
the physical phenomena related to the production of $\gamma$-rays is given in
section~\ref{sec:sources}.

Finally, all of these results be used to construct a model for the full $\gamma$-ray sky in
section~\ref{sec:modeling}. This model will be used for comparisons with \mbox{AMS-02} and
\mbox{Fermi-LAT} data in chapters~\ref{sec:data-analysis} and~\ref{sec:results}.

\section{Elementary Physical Processes}
\label{sec:physics}

\subsection{Processes for Gamma Ray Production}
\label{sec:gamma-ray-production}

Most of the photons detected by \mbox{AMS-02} are produced in galactic diffuse emission
processes. Three types of interactions are important in particular: Pion Decays, bremsstrahlung and
the inverse Compton scattering.

\subsubsection*{Pion Decay}

When cosmic ray protons collide with protons at rest in the galactic gas, hadronic interactions can
lead to the production of new particles. In this fixed target collision it is possible to produce
neutral pions:

\begin{equation}
  \label{eq:pion-production}
  p_{\mathrm{CR}} + p_{\mathrm{gas}} \rightarrow p + p + \pi^0 \,.
\end{equation}

This requires the kinetic energy $T_p$ of the incoming proton to be greater than the pion production
threshold:

\begin{displaymath}
T_p > T_{p,\mathrm{thr}} = E_{p,\mathrm{thr}} - m_p = \frac{m_\pi^2 + 4 m_\pi m_p}{2 m_p} \approx
\SI{280}{\mega\electronvolt} \,.
\end{displaymath}

Production of $\pi^0$ mesons can also occur in collisions with other forms of gas, such as molecular
hydrogen or neutral helium gas and with other projectiles, such as cosmic ray
$\alpha$-particles. The discussion here will focus on proton-proton collisions which is the most
important effect.

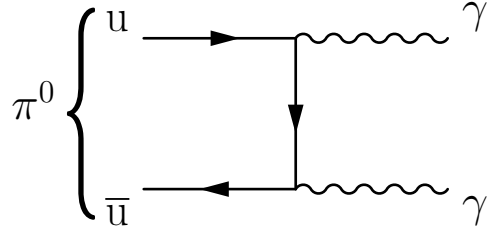
\begin{figure}[t]
  \centering
  \vspace*{5mm}
  \begin{fmffile}{feynmp/pion_decay}
      \unitlength = 1cm
      \begin{Large}
        \begin{fmfgraph*}(4,2)
          \fmfstraight
          \fmfleft{i1,P,i2}
          \fmfright{o1,o2}
          \fmf{fermion}{i2,t2}
          \fmf{fermion}{t1,i1}
          \fmf{fermion, tension=0}{t2,t1}
          \fmflabel{u}{i2}
          \fmflabel{$\overline{\text{u}}$}{i1}
          \fmfv{l.d=20,l.a=180,l={$\pi^0$\mylbrace{80}{0}}}{P}
          \fmf{photon}{t1,o1}
          \fmf{photon}{t2,o2}
          \fmflabel{$\gamma$}{o1}
          \fmflabel{$\gamma$}{o2}
        \end{fmfgraph*}
      \end{Large}
  \end{fmffile}
  \vspace*{5mm}
  \caption{Leading order Feynman diagram for the decay of $\pi^0$ mesons into two photons.}
  \label{fig:feyn-pion-decay}
\end{figure}

Once produced the $\pi^0$ meson immediately decays electromagnetically into two photons as shown in
figure~\ref{fig:feyn-pion-decay}.

Because the pion is a scalar particle, the emission of $\gamma$-rays is isotropic in the rest frame
of the pion and the energy spectrum of each of the photons is flat, centered around
$E_\pi / 2 = m_\pi / 2$. The energy of the produced photons is limited by:

\begin{align}
  \label{eq:pion-decay-spectrum-min}
  E_{\gamma}^{\mathrm{min}} &= \frac{E_\pi}{2}(1 - \beta_\pi) \,,\\
  \label{eq:pion-decay-spectrum-max}
  E_{\gamma}^{\mathrm{max}} &= \frac{E_\pi}{2}(1 + \beta_\pi) \,,
\end{align}

where $\beta_\pi = v_\pi / c$ is the velocity. The differential photon emission spectrum is thus:

\begin{displaymath}
  \frac{\mathrm{d}N_{\pi \rightarrow \gamma}}{\mathrm{d}E_{\gamma}} \left(E_{\gamma}, E_{\pi}\right)
  = \frac{1}{E_{\gamma}^{\mathrm{max}} - E_{\gamma}^{\mathrm{min}}}
  = \frac{1}{E_{\pi} \beta_{\pi}} \,,
  \qquad E_\gamma \in [E_\gamma^{\mathrm{min}}, E_\gamma^{\mathrm{max}}] \,.
\end{displaymath}

The rate of emission of $\gamma$-rays of energy $E_\gamma$ is the product of the pion production
rate $\dot{N}_\pi$ with the photon emission spectrum, integrated over the pion energy $E_\pi$:

\begin{equation}
  \label{eq:gamma-ray-rate}
  \dot{N}_{\gamma} (E_\gamma) =
  \int_{E_\pi^{\mathrm{min}}(E_\gamma)}^{\infty}{\dot{N}_\pi \frac{1}{E_\pi \beta_\pi} \,
    \mathrm{d}E_\pi} \,.
\end{equation}

The pion production rate depends on the flux of cosmic ray protons, so the expression is complex in
general. However only the lower limit of the integral depends on the photon energy $E_\gamma$. This
lower limit is the minimum energy the pion must exceed in order to produce a photon of energy
$E_\gamma$. Using equations~(\ref{eq:pion-decay-spectrum-min})
and~(\ref{eq:pion-decay-spectrum-max}) this minimum energy can be calculated:

\begin{displaymath}
  E_\pi^{\mathrm{min}}(E_\gamma) = E_\gamma + \frac{m_\pi^2}{4 E_\gamma} =
  \frac{m_\pi}{2}\left(\frac{2 E_\gamma}{m_\pi} + \frac{m_\pi}{2 E_\gamma}\right) \,.
\end{displaymath}

This expression is symmetric in log space about half of the pion mass:

\begin{displaymath}
  E_\pi^{\mathrm{min}}(x \frac{m_\pi}{2}) = E_\pi^{\mathrm{min}}(\frac{m_\pi}{2 x})
\end{displaymath}

\begin{figure}[t!]
  \centering
  \includegraphics[width=0.7\linewidth]{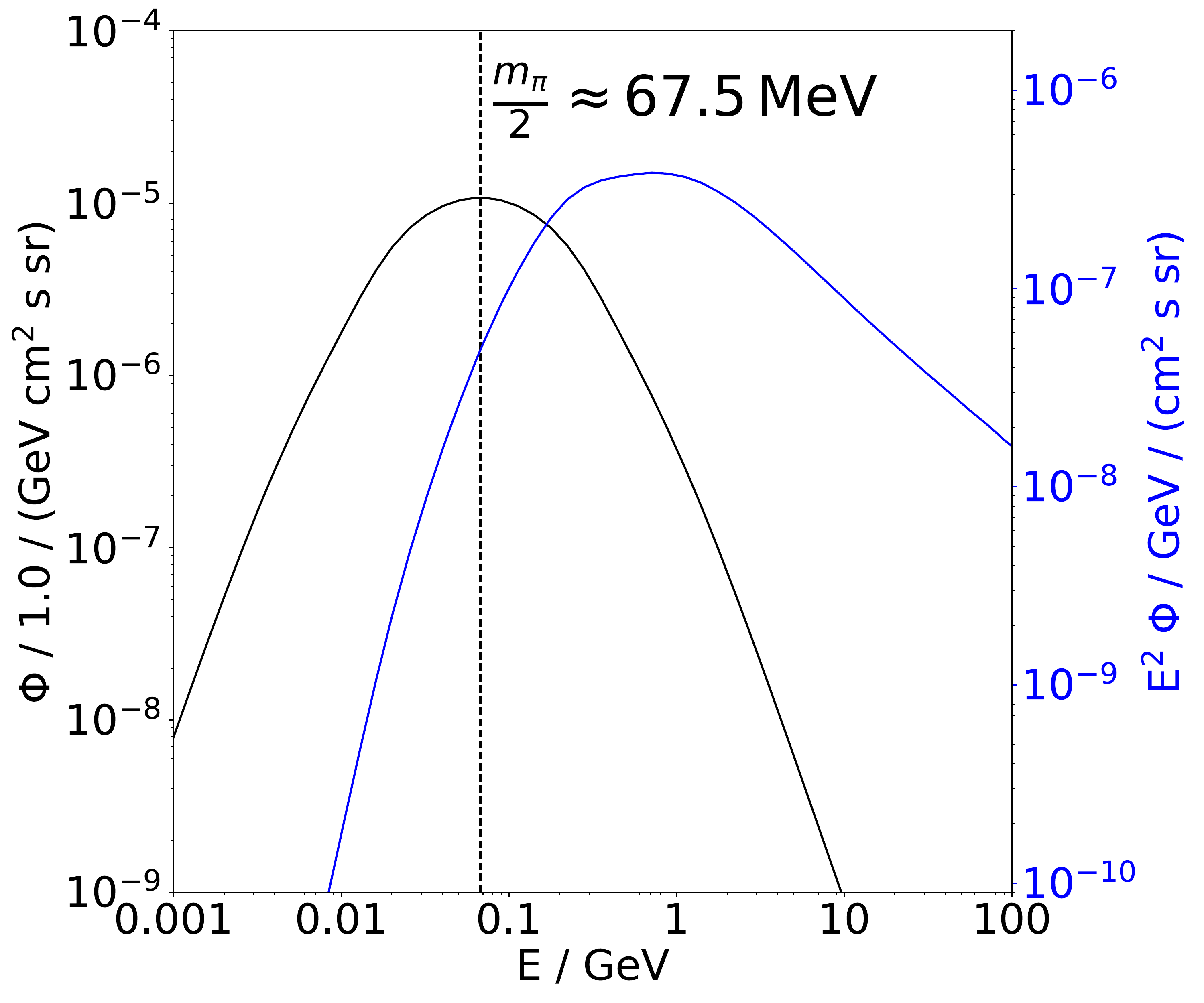}
  \caption{Black, left axis: Example for model calculation of $\gamma$-ray flux from pion
    decays. The peak position is located at $m_\pi / 2 \approx \SI{67.5}{\mega\electronvolt}$. Blue,
    right axis: The same spectrum, but scaled with $E^2$.}
  \label{fig:pion-decay-spectrum}
\end{figure}

where $x$ is an arbitrary factor~\footnote{This is because $\log{(xE)} = \log{E} + \log{x}$ and
  $\log{(\frac{E}{x})} = \log{E} - \log{x}$.}. As a result the $\gamma$-ray emission rate in
equation~\ref{eq:gamma-ray-rate} is symmetric about
$m_\pi / 2 \approx \SI{67.5}{\mega\electronvolt}$ when plotted as a function of $\log(E_\gamma)$ as
shown in figure~\ref{fig:pion-decay-spectrum} in black. However, in order to improve the visual
appearance, it is customary to scale the flux with the square of the photon energy ($y = E^2 \Phi$)
when it is plotted. In that case the symmetry about $m_\pi / 2$ is no longer apparent and the peak
is shifted towards the region of \SIrange[]{0.5}{1.0}{\giga\electronvolt}, depending on the
specifics of the spectrum, as can be seen from the blue curve in the same figure.

This feature of the pion decay spectrum is the ``pion bump'', which is a unique signature of this
process. It was used to identify pion decays in the spectra of the Supernova Remnants IC~443 and W44
by the Fermi-LAT collaboration~\cite{Fermi_PionDecay_2013}, which provided direct evidence that
cosmic ray protons are accelerated in Supernova Remnants.

To either side of the peak located at $m_\pi / 2$ the spectrum falls with energy as a power law,
whose spectral index is directly related to that of the cosmic ray protons. This connection provides
a unique way to indirectly infer properties of the cosmic ray proton spectrum in locations other
than the solar system.

As can be seen from equation~(\ref{eq:pion-production}) the spatial distribution of the pion decay
component of diffuse emission depends on the distribution of the gas in the Milky Way, which is
discussed in section~\ref{sec:galaxy-gas}. It also depends on three-dimensional variation of the
flux of cosmic ray protons, which can only be measured at the location of the solar system and must
be extrapolated to other regions.

\subsubsection*{Bremsstrahlung}

\begin{figure}[t]
  \centering
  \begin{fmffile}{feynmp/bremsstrahlung}
    \unitlength = 1cm
    \begin{fmfgraph*}(6,3)
      \fmfleft{i_nuc,i_e}
      \fmfright{o_nuc,o_e,o_gamma}
      \fmf{heavy, label=Nuc}{i_nuc,v1}
      \fmf{fermion, label=e}{i_e,v3}
      \fmf{fermion}{v3,v2}
      \fmf{photon, label=$\gamma$}{v1,v2}
      \fmf{heavy, label=Nuc}{v1,o_nuc}
      \fmf{fermion, label=e}{v2,o_e}
      \fmf{photon, label=$\gamma$}{v3,o_gamma}
    \end{fmfgraph*}
  \end{fmffile}
  \vspace*{5mm}
  \caption{Leading order Feynman diagram for the bremsstrahlung process.}
  \label{fig:feyn-bremsstrahlung}
\end{figure}
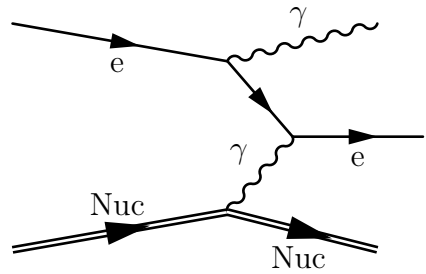

When passing through matter high energy electrons and positrons predominantly lose energy by
bremsstrahlung. This also occurs when cosmic ray electrons (and positrons) interact with the nuclei
of the interstellar gas. In bremsstrahlung there is a probability for emission of hard photons. The
leading order diagram for the process is shown in figure~\ref{fig:feyn-bremsstrahlung}.

In this process an energetic electron radiates away a portion of its energy, producing a
$\gamma$-ray. Because of momentum conservation this process does not occur in free space, but
requires exchange of a photon with a nucleus.

Compared to the pion decay process the bremsstrahlung component is governed by the population of
cosmic ray electrons and positrons, which are less abundant than protons. In addition the cosmic ray
electron spectrum is softer than the proton spectrum (see
section~\ref{sec:cosmic-ray-electrons-positrons}), which makes the bremsstrahlung spectrum fall more
steeply, too.

However in both processes cosmic rays are interacting with the interstellar gas, so the spatial
distribution is very similar.

When an electron passes through a section of matter, the typical length scale for the bremsstrahlung
process is the radiation length $X_0$, which is a property of the traversed material and usually
given in \si{\gram\per\centi\meter\squared}. The radiation length corresponds to the mean distance
over which an electron loses all but $1 / e$ of its energy by bremsstrahlung. The radiation length
for most materials can be reasonable well approximated by Tsai's formula~\cite{Tsai_1974}.

For an electron with energy $E$ in a material with radiation length $X_0$ the differential cross
section can be approximated in the ``complete screening case'' by~\cite{Tsai_1974}:

\begin{equation}
  \label{eq:cross-section-bremsstrahlung}
  \frac{\mathrm{d}\sigma}{\mathrm{d}k}
  = \frac{A}{X_0 N_A k} \left( \frac{4}{3} - \frac{4}{3} y + y^2 \right) \,,
\end{equation}

where $k$ is the energy of the radiated photon, $y = k / E$ is the fraction of the radiated energy,
$A$ is the molar mass of the traversed material in \si{\gram\per\mole} and
$N_A \approx \SI{6.022e23}{\per\mole}$ is the Avogadro constant. The approximation is valid except
near the two extremes of $y = 0$ and $y = 1$. It also becomes invalid for electron energies above
approximately $\SI{100}{\giga\electronvolt}$. The infrared divergence for $k \rightarrow 0$ is
canceled by the Landau-Pomeranchuk-Migdal (LPM) effect~\cite{Landau_1953,Migdal_1956}, which is a
result of quantum mechanical interference of interactions with different scattering centers.

\begin{figure}[t]
  \centering
  \includegraphics[width=0.75\linewidth]{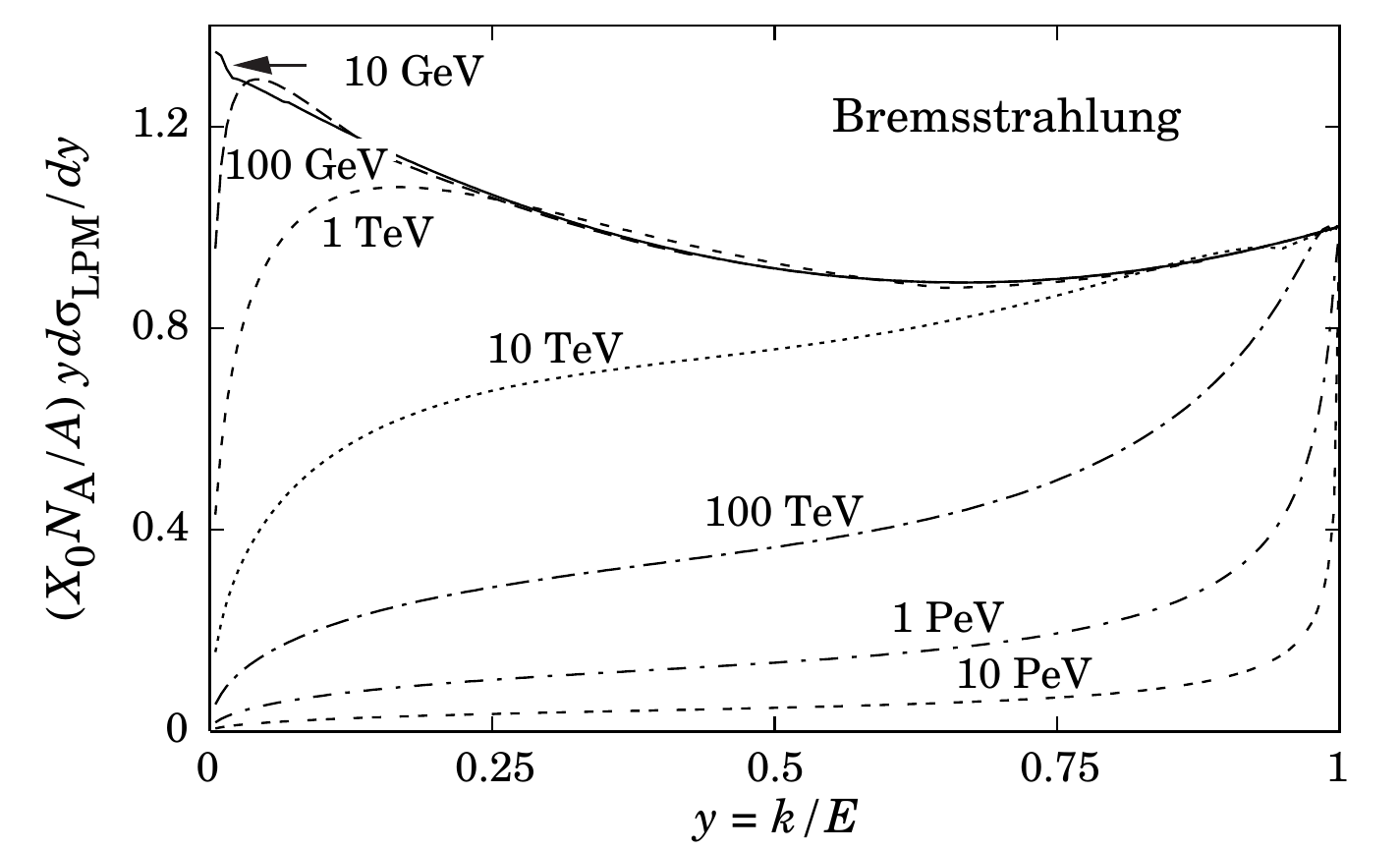}
  \caption{Normalized differential cross section $k \mathrm{d}\sigma / \mathrm{d}k$ for electrons of
    various energies in lead. Figure taken from the Particle Data Group's Review of Particle
    Physics~\cite{PDG_2018}.}
  \label{fig:bremsstrahlung-cross-section}
\end{figure}

Figure~\ref{fig:bremsstrahlung-cross-section} shows the differential cross section for various
electron energies. The solid line corresponds to the cross section in the complete screening
approximation, as given by equation~(\ref{eq:cross-section-bremsstrahlung}). For high energy
energies above approximately $\SI{100}{\giga\electronvolt}$ the LPM effect becomes important and
modifies the cross section as shown in the dashed and dotted curves. For energies below
\SI{1}{\tera\electronvolt} the probability distribution for the energy fraction transferred to the
photon is rather flat, so that emission of photons with any energy $0 < E_\gamma < E_{e^-}$ is
approximately equally likely.

The bremsstrahlung process is also important for the interaction of electrons and positrons with the
\mbox{AMS-02} detector material. Emission of a hard photon significantly changes the measured
properties of the primary particle in the detector. This will become important in
section~\ref{sec:corrections-ecal-trigger}.

\subsubsection*{Inverse Compton Emission}

The Compton effect is the scattering of a photon on an electron and is one of the three important
energy loss mechanisms for photons. In the Compton effect the incoming photon transfers some of its
energy to the electron and escapes with reduced energy.

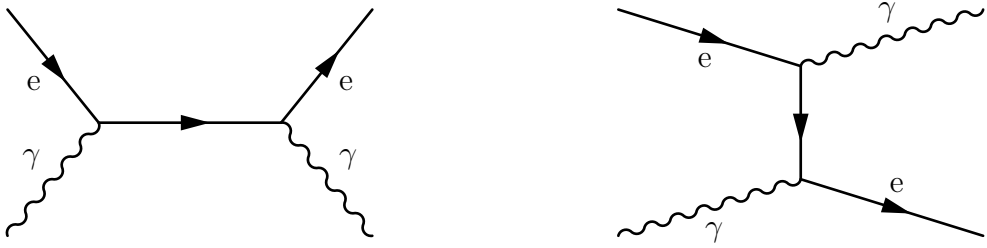
\begin{figure}[t]
  \begin{minipage}{0.48\linewidth}
    \centering
    \begin{fmffile}{feynmp/inverse_compton_1}
      \unitlength = 1cm
      \begin{fmfgraph*}(6,3)
        \fmfleft{i1,i2}
        \fmfright{o1,o2}
        \fmf{fermion, label=e}{i2,v1}
        \fmf{fermion}{v1,v2}
        \fmf{fermion, label=e}{v2,o2}
        \fmf{photon, label=$\gamma$}{i1,v1}
        \fmf{photon, label=$\gamma$}{v2,o1}
      \end{fmfgraph*}
    \end{fmffile}
    \end{minipage}
  \hspace{0.01\linewidth}
  \begin{minipage}{0.48\linewidth}
    \centering
    \begin{fmffile}{feynmp/inverse_compton_2}
      \begin{center}
        \unitlength = 1cm
        \begin{fmfgraph*}(6,3)
          \fmfleft{i1,i2}
          \fmfright{o1,o2}
          \fmf{fermion, label=e}{i2,v2}
          \fmf{fermion}{v2,v1}
          \fmf{photon, label=$\gamma$}{i1,v1}
          \fmf{fermion, label=e}{v1,o1}
          \fmf{photon, label=$\gamma$}{v2,o2}
        \end{fmfgraph*}
      \end{center}
    \end{fmffile}
  \end{minipage}
  \vspace*{5mm}
  \caption{Leading order Feynman diagrams for (inverse) Compton scattering. Left: s-channel, Right:
    t-channel.}
  \label{fig:feyn-inverse-compton}
\end{figure}

In the inverse Compton (IC) effect the incoming particle is the electron which interacts with a
low-energy photon and transfers enough energy to it to turn it into an energetic $\gamma$-ray. The
underlying physical process is the same as the regular Compton effect, whose leading order Feynman
diagrams are shown in figure~\ref{fig:feyn-inverse-compton}.

In the Milky Way the Cosmic Microwave Background (CMB) is an isotropic and homogeneous source of
low-energy photons. Energetic cosmic ray electrons and positrons can up-scatter these photons to
$\gamma$-ray energies. The Interstellar Radiation Field (ISRF) also includes other sources for
low-energy photons, such as starlight and thermal emission from heated dust.

In the rest frame of the electron the energy of the photon after scattering $E'_\gamma$ is given by
the Klein Nishina formula~\cite{Klein_1929}:

\begin{equation}
  \label{eq:compton-scattering}
  E'_\gamma(\phi) = \frac{E_\gamma}{1 + \frac{E_\gamma}{m_e c^2} \left(1 - \cos{\phi}\right)}
\end{equation}

where $E_\gamma$ is the energy of the photon before scattering, $m_e$ is the electron mass and
$\phi$ is the scattering angle.

The inverse Compton emission traces the spatial population of cosmic ray electrons. Because of the
isotropy of the CMB the spatial distribution of the diffuse IC emission is far less structured than
the pion decay and bremsstrahlung components.

In the vicinity of astrophysical sources such as Active Galactic Nuclei (AGN), Supernova Remnants
(SNRs) and Pulsar Wind Nebulae (PWNe) IC emission is one of the most important mechanisms of high
energy $\gamma$-ray production. In the Synchrotron Self Compton (SSC) model~\cite{SSC_Pulsars_1995}
the initial photons for the IC interaction are the result of synchrotron radiation of electrons in
the compact object's magnetic field. The same electrons later up-scatter the photon to the highest
energies in IC processes.

The energy spectra of photons produced in IC processes are often harder than those produced in pion
decays or emitted by bremsstrahlung. This makes the IC process particularly important for the study
of very high energy (VHE) photons with ground based Cherenkov telescopes, such as H.E.S.S., MAGIC
and Veritas.



\subsection{Processes for Gamma Ray Detection}
\label{sec:gamma-ray-detection}

The three most important processes for the interaction of photons with matter are the photoelectric
effect, Compton scattering and the production of an $e^-$ / $e^+$ pair. At low energies
($E \ll \SI{1}{\mega\electronvolt}$) the photoelectric effect is the most important process, in
which a photon transfers parts of its energy and excites and liberates an electron from the
material. At intermediate energies ($E \approx \SI{1}{\mega\electronvolt}$) Compton scattering
dominates the interactions of photons with matter. For energies above
$E = 2 m_e \approx \SI{1.02}{\mega\electronvolt}$, which covers the $\gamma$-ray energy regime, pair
production dominates the total photon interaction cross section. In this section we will discuss
pair production as the most important mechanism for the detection of $\gamma$-rays.

\subsubsection*{Pair Production}

At energies above a few tenths of \si{\mega\electronvolt} the most important physical process for
the detection of $\gamma$-rays in the detector is the production of an $e^-$ / $e^+$ pair. This
process is strongly linked to bremsstrahlung, since the Feynman diagrams are variants of each other.

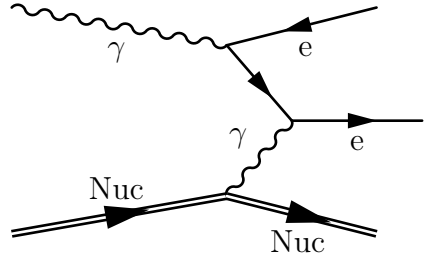
\begin{figure}[t]
  \centering
  \begin{fmffile}{feynmp/pair_production}
    \unitlength = 1cm
    \begin{fmfgraph*}(6,3)
      \fmfleft{i_nuc,i_gamma}
      \fmfright{o_nuc,o_e1,o_e2}
      \fmf{heavy, label=Nuc}{i_nuc,v1}
      \fmf{photon, label=$\gamma$}{i_gamma,v3}

      \fmf{fermion}{v3,v2}
      \fmf{photon, label=$\gamma$}{v1,v2}

      \fmf{heavy, label=Nuc}{v1,o_nuc}
      \fmf{fermion, label=e}{v2,o_e1}
      \fmf{fermion, label=e}{o_e2,v3}
    \end{fmfgraph*}
  \end{fmffile}
  \vspace*{5mm}
  \caption{Leading order Feynman diagram for production of an electron and positron pair by a
    $\gamma$-ray photon.}
  \label{fig:feyn-pair-production}
\end{figure}

The Feynman diagram for $e^-$ / $e^+$ pair production is shown in
figure~\ref{fig:feyn-pair-production}. It differs from that of bremsstrahlung only by an interchange
of the incoming electron with the outgoing photon. For this reason the physical properties of the
processes are also tightly linked.

Similarly to equation~(\ref{eq:cross-section-bremsstrahlung}) the differential cross section for
pair production can be expressed as:

\begin{equation}
  \label{eq:cross-section-pair-production}
  \frac{\mathrm{d}\sigma}{\mathrm{d}x}
  = \frac{A}{X_0 N_A} \left(1 - \frac{4}{3} x \left( 1 - x \right) \right) \,,
\end{equation}

where $x = E_{e^-} / E_\gamma$ is the fractional energy transferred to the electron in the
production.

\begin{figure}[t]
  \centering
  \includegraphics[width=0.75\linewidth]{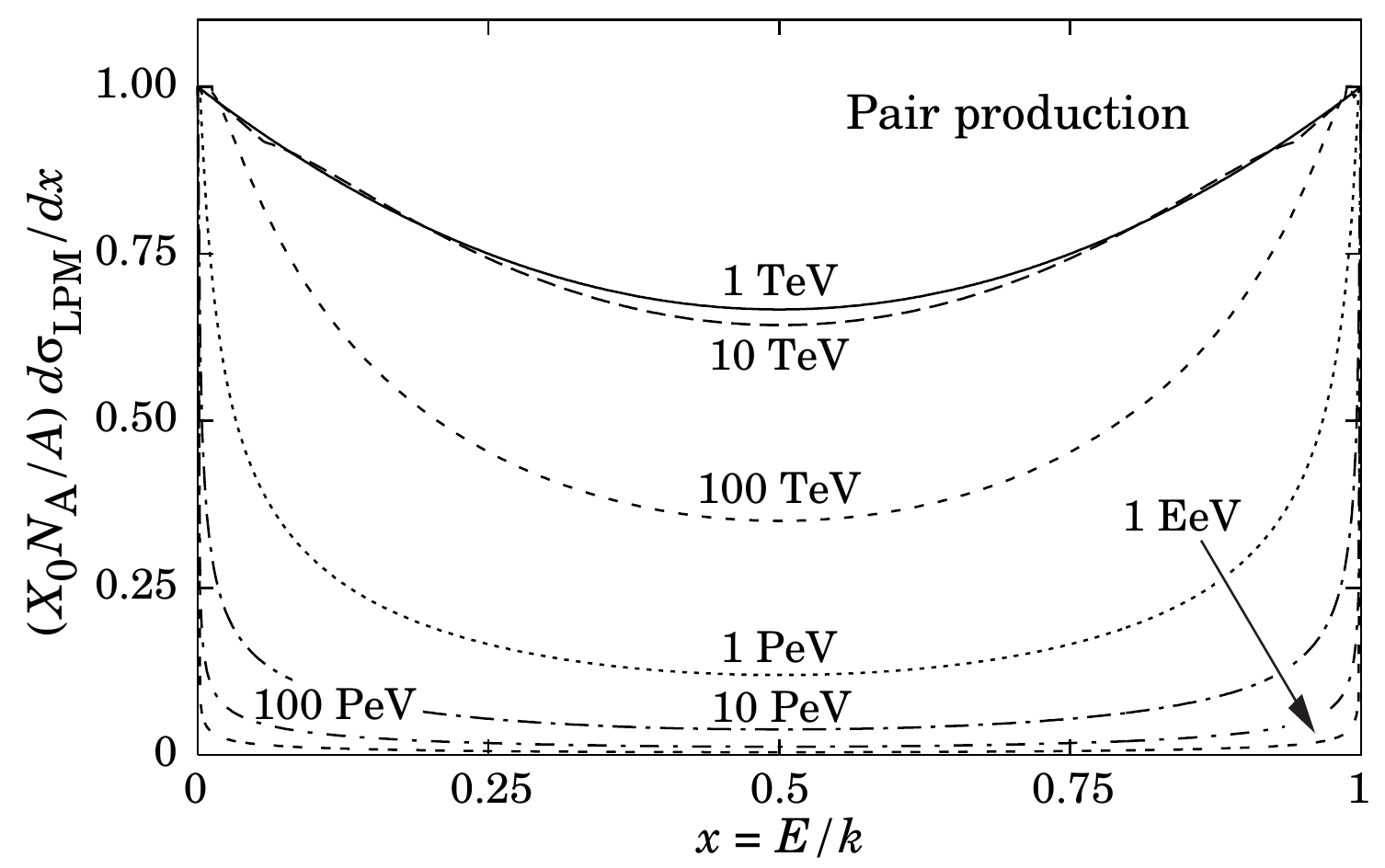}
  \caption{Normalized differential cross section $\mathrm{d}\sigma / \mathrm{d}x$ for pair
    production as a function of the fraction electron energy $x = E_{e^-} / E_\gamma$. Figure taken
    from the Particle Data Group's Review of Particle Physics~\cite{PDG_2018}.}
  \label{fig:pair-production-cross-section}
\end{figure}

The differential cross section is shown in the solid line in
figure~\ref{fig:pair-production-cross-section}. For energies below \SI{1}{\tera\electronvolt} the
distribution is flat, but slightly rises for $x \rightarrow 0$ and $x \rightarrow 1$. For photon
energies higher than \SI{1}{\tera\electronvolt} equation~(\ref{eq:cross-section-pair-production})
becomes inadequate and must be corrected for the LPM effect, which leads
to the dashed curves shown in figure~\ref{fig:pair-production-cross-section}.

As a result most partitions of the incoming photon's energy on the electron and positron are equally
likely, with a slight preference for asymmetric partitions, in which one of the two particles
carries most of the energy of the incoming photon. This preference becomes more pronounced as the
photon energy increases.

The total cross section for pair production can be found by integration of
equation~(\ref{eq:cross-section-pair-production}):

\begin{equation}
  \label{eq:cross-section-pair-production-total}
  \sigma = \frac{7}{9} \frac{A}{X_0 N_A} \,.
\end{equation}

After passing through a material of thickness $d$ the intensity of a beam of photons drops
exponentially:

\begin{equation}
  \label{eq:pair-production-intensity}
  I = I_0 e^{-\mu d} = I_0 e^{-\frac{\mu}{\rho} \rho d} = I_0 e^{-\frac{\mu}{\rho} x} \,,
\end{equation}

where $\mu$ is the attenuation coefficient, $\rho$ is the material's density and $x = \rho d$ is the
mass thickness. The relation between the pair production cross section $\sigma$ and the attenuation
coefficient $\mu$ is:

\begin{equation}
  \label{eq:pair-production-attenuation-cross-section}
  \frac{\mu}{\rho} = \sigma \frac{N_A}{A} = \frac{7}{9} \frac{1}{X_0} \,.
\end{equation}

The probability for a photon to convert after passing through material with mass thickness $x$ can
therefore be expressed as:

\begin{equation}
  \label{eq:pair-production-probability}
  P(x) = 1 - e^{-\frac{7}{9} \frac{x}{X_0}} \,.
\end{equation}

Pair production and bremsstrahlung are also the processes which govern the development of
electromagnetic cascades in dense materials, such as lead. The characteristic length scale in the
cascade is the radiation length $X_0$. Such cascades are important in electromagnetic calorimeters
where photons and electrons develop showers. The measurement of these showers enables the
identification of electrons and positrons, and provides a good way to estimate their energy and
incoming direction.


\section{Cosmic Rays}
\label{sec:cosmic-rays}

Cosmic rays such as protons, helium nuclei, electrons and positrons are important in $\gamma$-ray
physics both as a projectile for diffuse $\gamma$-ray production and as a background for the
measurement of photons in the detector.

Protons as well as helium, carbon and oxygen nuclei are among the primary cosmic ray species, which
are directly accelerated at the cosmic ray sources. Supernova Remnants (SNRs) were shown to
accelerate protons by measurements of their $\gamma$-ray spectra with the Fermi
satellite~\cite{Fermi_PionDecay_2013}. Fermi acceleration of the first order was put forward as the
mechanism for acceleration, which generates a particle flux at injection with the form of a power
law with spectral index of -2. The exact spatial distribution of the cosmic ray sources in the
galaxy is unknown and different models are currently under study~\cite{Johannesson_APJ_856,
  Johannesson_APJ_856}.

After production in the CR sources the primary cosmic rays propagate through the galaxy. Because of
the random orientation of the magnetic field and its turbulences the process is similar to
diffusion. More complicated phenomena such as re-acceleration and convection are likely also
important in the process. The effect of propagation changes the spectral index of the CR flux,
because particles can escape the Milky Way. In addition particles lose energy when they collide with
the gas in the ISM.

Secondary CRs such as lithium, beryllium and boron are produced in these collisions. These CR
species exhibit a distinctly different spectrum than the primaries~\cite{AMS02_LiBeB_2018}. Ratios
of secondaries to primaries, such as the boron over carbon ratio~\cite{AMS02_BoverC_2016}, can be
used to study propagation in detail.

Electrons are also thought to be primary cosmic rays. However, their energy spectrum is softer than
that of protons, because different physical processes govern the interactions of leptons. In
particular, bremsstrahlung and inverse Compton scattering cause energy losses, which scale with the
particle's energy squared ($\dot{E} \sim E^2$). For this reason the sources for energetic electrons
and positrons must be ``local'', i.e. at distances less than about \SI{1}{\kilo\parsec}.

After propagation CRs must enter the heliosphere before they can be observed at Earth. In the
magnetic field and the solar wind produced by the Sun the particle fluxes change: This process is
solar modulation. The effect is time dependent, because the activity of the Sun changes with
time. Solar modulation affects the spectra mostly at low rigidities ($R < \SI{30}{\giga\volt}$),
which means that high energy fluxes measured at Earth are representative of the Local Interstellar
Spectrum (LIS).

Positrons were originally believed to be secondary CRs. However measurements from
\mbox{AMS-01}~\cite{Aguilar2007} and the PAMELA satellite~\cite{Adriani2009} have revealed that the
positron spectrum is incompatible with the expectation for pure secondary production. A primary
component is likely present. Dark matter~\cite{Hooper2009d} and a nearby positron source, such as a
pulsar~\cite{Hooper2009b}, have been put forward as possible explanations for the excess of
positrons.

\subsection{Protons and Helium nuclei}
\label{sec:cosmic-ray-protons-helium}

Protons make up the majority of the cosmic rays, at least for energies below the knee
($E < \SI{e15}{\electronvolt}$). Above those energies the exact composition of cosmic rays is not
well known, and is the subject of measurements by ground based experiments such as the Pierre Auger
Observatory~\cite{Auger_Detector_2015}.

\begin{figure}[p]
  \centering
  \includegraphics[width=0.85\linewidth]{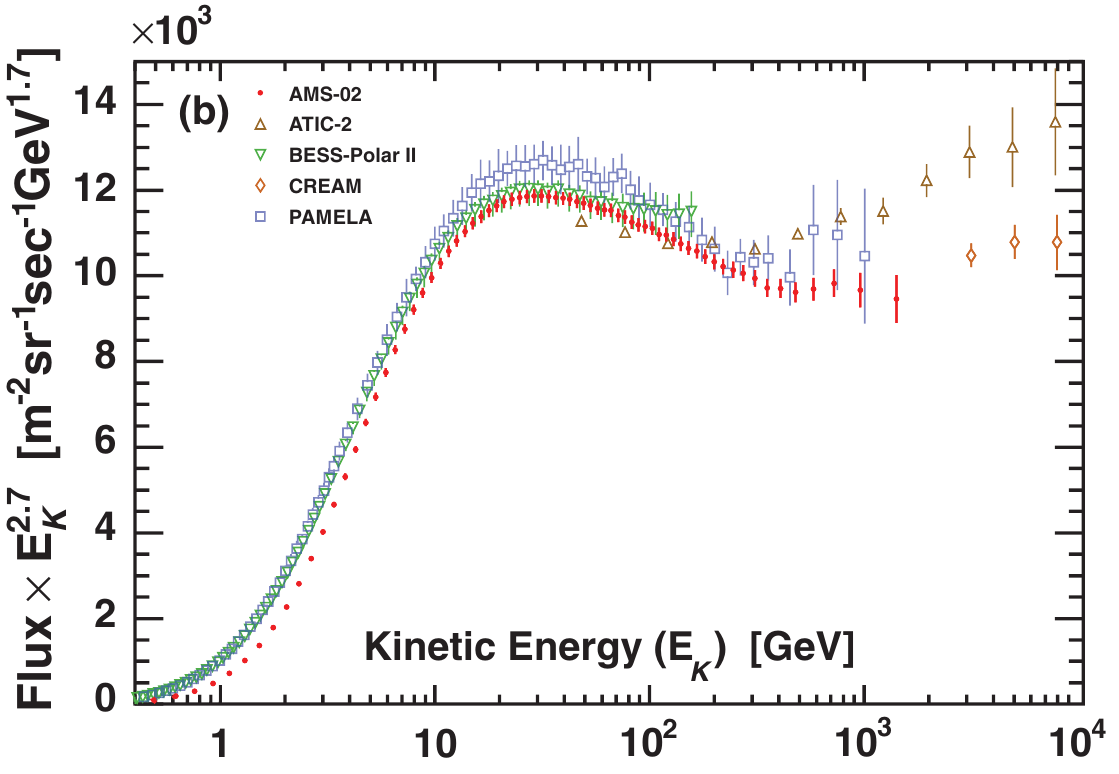}
  \caption{Flux of cosmic ray protons as measured by \mbox{AMS-02}, multiplied by $E_K^{2.7}$, as a
    function of proton kinetic energy~\cite{AMS02_Proton_2015}. Also shown are several prior
    experimental results.}
  \label{fig:ams02-proton-flux}

  \vspace*{\floatsep}
  \vspace*{\floatsep}

  \centering
  \includegraphics[width=0.85\linewidth]{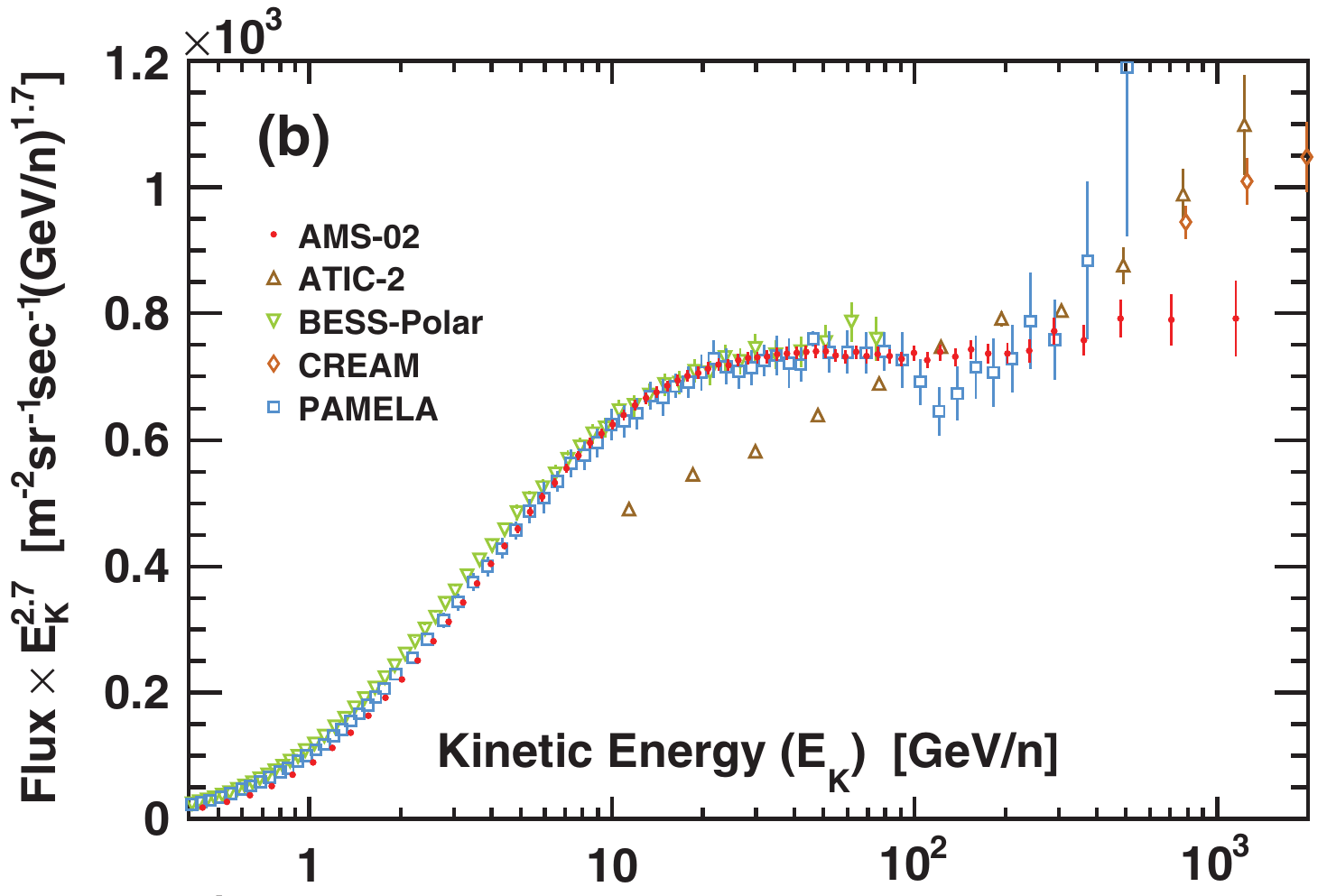}
  \caption{Flux of cosmic ray helium nuclei as measured by \mbox{AMS-02}, multiplied by $E_K^{2.7}$,
    as a function of kinetic energy per nucleon~\cite{AMS02_Helium_2015}. Also shown are several
    prior experimental results.}
  \label{fig:ams02-helium-flux}
\end{figure}

Figure~\ref{fig:ams02-proton-flux} shows the proton flux as measured by \mbox{AMS-02}, based on data
collected between May 2011 and November 2013, together with several earlier results. Between
approximately \SI{30}{\giga\electronvolt} and \SI{200}{\giga\electronvolt} the flux can be
reasonably well described by a power law with spectral index of approximately $-2.8$. At lower
energies solar modulation causes the spectrum to fall. This effect is time dependent and is studied
in detail in later publications~\cite{AMS02_ProtonHeliumTime_2018,AMS02_ElecPosTime_2018}. Because
of solar modulation it is expected that the results from the various experiments disagree at low
energies, since the data was collected in different time intervals.

Unexpectedly the proton flux begins to harden around a few hundred \si{\giga\electronvolt}. This was
first reported by the PAMELA group~\cite{Adriani2011} and is confirmed with good accuracy in the
\mbox{AMS-02} proton flux measurement.


Recently, the DAMPE collaboration~\cite{DAMPE_Detector_2017} has reported another break in the
cosmic ray proton spectrum around approximately \SI{13}{\tera\electronvolt} kinetic
energy~\cite{DAMPE_Protons_2019}, where the spectrum appears to soften.

Figure~\ref{fig:ams02-helium-flux} shows the \mbox{AMS-02} measurement of the cosmic ray helium
flux, also based on the data collected between May 2011 and November 2013. Based on the measured
fluxes the helium component in cosmic rays is between 4 and 7 times less abundant than the proton
component, for rigidities below \SI{2}{\tera\volt}. Like the CR proton flux, the helium flux hardens
around \SIrange{200}{300}{\giga\volt} rigidity. The same behavior was also observed in other primary
cosmic ray nuclei such as carbon and oxygen~\cite{AMS02_HeCO_2017}.

Protons and helium nuclei are responsible for the production of diffuse $\gamma$-rays through pion
decays when they interact with the gas in the ISM. The measured fluxes are therefore used as
ingredients when predicting the $\gamma$-ray flux from $\pi^0$ decays. However, the charged particle
fluxes can only be measured directly at the location of the solar system. For the calculation of
diffuse emission the flux of these charged particles must be known in the entire Milky Way. It is
customary to use numerical models of CR propagation and diffusion, such as
GALPROP~\cite{Strong1998,Galprop_WWW}, to calculate these fluxes. The measured fluxes at Earth can
then be used to constrain the models.

Because of the effect of diffusion the galactic proton and helium fluxes are almost perfectly
isotropic. Both of the species are a lot more abundant than $\gamma$-rays, even in regions of the
sky in which the $\gamma$-ray flux is at its highest, such as the galactic center, the ratio of
photons to protons is much smaller than \num{e-3}. Therefore the proton and helium fluxes form an
important background in the identification of photons in the detector.

\subsection{Electrons and Positrons}
\label{sec:cosmic-ray-electrons-positrons}

Electrons and positrons are of special interest in cosmic rays. These species behave differently
compared to other components such as nuclei which interact hadronically with the ISM. As a result
they probe a different, more local, region of the galaxy. Also exotic processes, such as those
predicted by extensions of the Standard Model, often produce observable signatures in the spectra of
electrons and positrons in particular.

They are also directly connected to some of the mechanisms for diffuse emission of
$\gamma$-rays. Electrons and positrons play a vital role in the physics of $\gamma$-ray producing
sources. Because of pair production and emission processes such as bremsstrahlung, inverse Compton
scattering and curvature radiation leptons and photons are directly linked.

Various techniques have been used to measure the spectra of electrons and positrons near
Earth. Space based experiments with a magnet include PAMELA, \mbox{AMS-01} and \mbox{AMS-02}. These
experiments are able to directly measure the individual fluxes of electrons and positrons.

Detectors without a magnet can not distinguish the two species. For that reason many experiments
measured the sum of the electron and positron fluxes (although this sum is often incorrectly
referred to as ``the electron flux''). Often a calorimeter is used to identify electrons or
positrons and to measure their energy. This technique was used in the Fermi-LAT~\cite{Atwood2009},
DAMPE~\cite{DAMPE_Detector_2017} and CALET~\cite{CALET_Detector_2011} experiments, for example.

Ground based Cherenkov telescopes, such as H.E.S.S.~\cite{HESS_Project_2004} and
MAGIC~\cite{MAGIC_Detector_2005}, are also unable to discriminate electrons from positrons, but
measure the summed flux instead. These experiments measure the air showers induced by CR electrons
or positrons in the Earth's atmosphere.

Instead of measuring the individual fluxes of electrons and positrons a simpler alternative is to
measure the ratio of positrons to the sum of electrons plus positrons ($e^+ / (e^- + e^+)$), because
some of the systematic uncertainties associated with the measurement of the individual fluxes cancel
in the ratio. The positron fraction is sensitive to the signals predicted by many extensions of the
Standard Model, such as models which predict annihilation of dark matter into electrons and
positrons. For these reasons the positron fraction serves as a good observable to study new physics.

Overall there are four types of measurements related to the fluxes of electrons and positrons: The
sum of electrons and positrons, the ratio of positrons to the sum of both, and the two individual
fluxes.

\begin{figure}[t]
  \centering
  \includegraphics[width=0.8\linewidth]{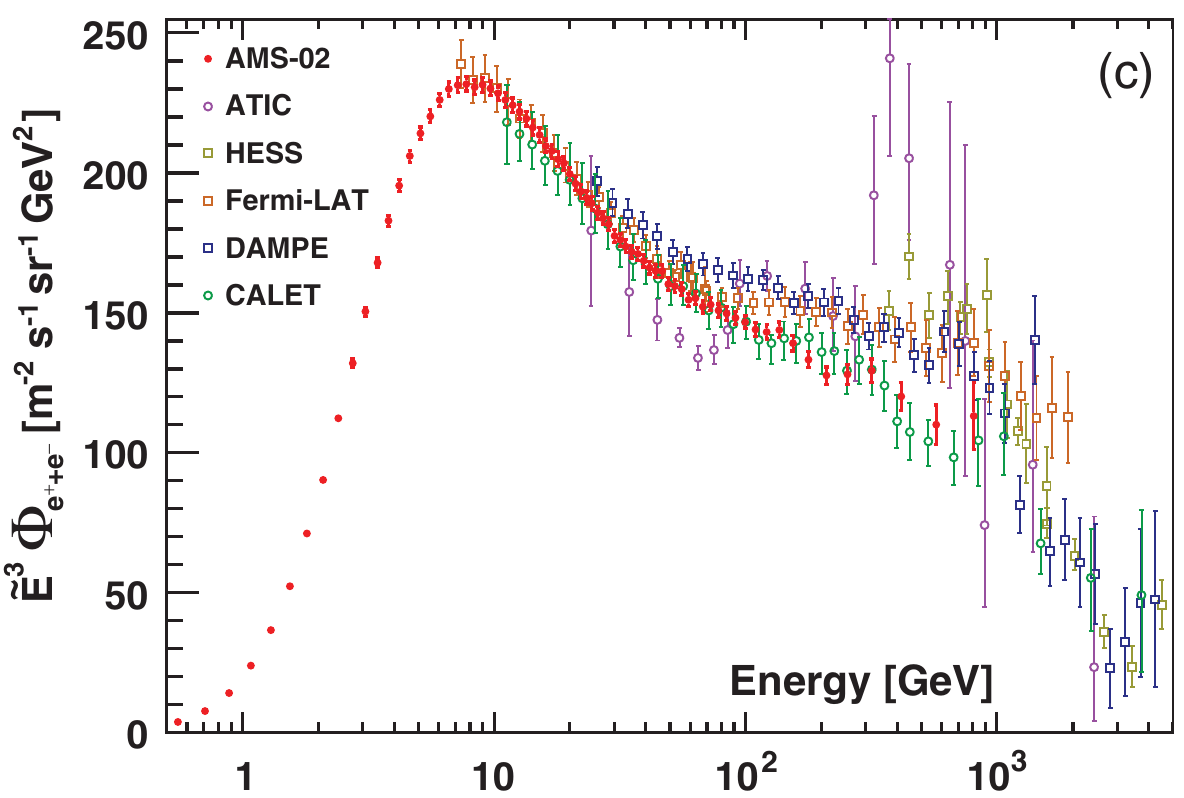}
  \caption{Sum of the fluxes of cosmic ray electrons and positrons, multiplied by $E^3$, as a
    function of the electron/positron energy~\cite{AMS02_Electrons_2019}.}
  \label{fig:ams02-all-elec-flux}
\end{figure}

Figure~\ref{fig:ams02-all-elec-flux} shows the latest \mbox{AMS-02} result for the summed flux of
electrons and positrons~\cite{AMS02_Electrons_2019}, together with earlier measurements. The data
was collected between May 2011 and November 2017. Below \SI{1}{\tera\electronvolt} the best
measurement is from \mbox{AMS-02}. The spectral index of the electron plus positron flux is about
$-3.2$ for energies above \SI{30}{\giga\electronvolt}. It is compatible with a single power law. At
lower energies the flux is modulated by solar modulation. The measurement by
CALET~\cite{CALET_AllElec_2017} agrees with the \mbox{AMS-02} data and extends the energy reach to
approximately \SI{4.8}{\tera\electronvolt}.

At energies below approximately \SI{100}{\giga\electronvolt} the measurement by the \mbox{Fermi-LAT}
satellite~\cite{Fermi_AllElec_2017} also agrees with the \mbox{AMS-02} data. However, above the
spectrum measured by Fermi hardens. The results by the H.E.S.S.~\cite{Aharonian2009,Aharonian2008}
and DAMPE~\cite{DAMPE_AllElec_2017} collaborations agree with the \mbox{Fermi-LAT} results. In
addition, they measured a break in the summed electron plus positron flux at approximately
\SI{0.9}{\tera\electronvolt}~\cite{DAMPE_AllElec_2017}.

The experimental results apparently split into two groups at high energies: The results by
\mbox{AMS-02} and CALET are compatible with each other, as are those by \mbox{Fermi-LAT} and
DAMPE. A possible explanation for the disagreement are systematic uncertainties associated with the
absolute energy scale of the experiments.

The main purpose of the \mbox{Fermi-LAT} satellite is the measurement of high energy
$\gamma$-rays. A measurement of photons at the same energies by \mbox{AMS-02} will therefore allow a
second comparison between the energy scales of the two experiments, which might help to understand
the differences in the measured electron plus positron fluxes.

\begin{figure}[t]
  \centering
  \includegraphics[width=0.8\linewidth]{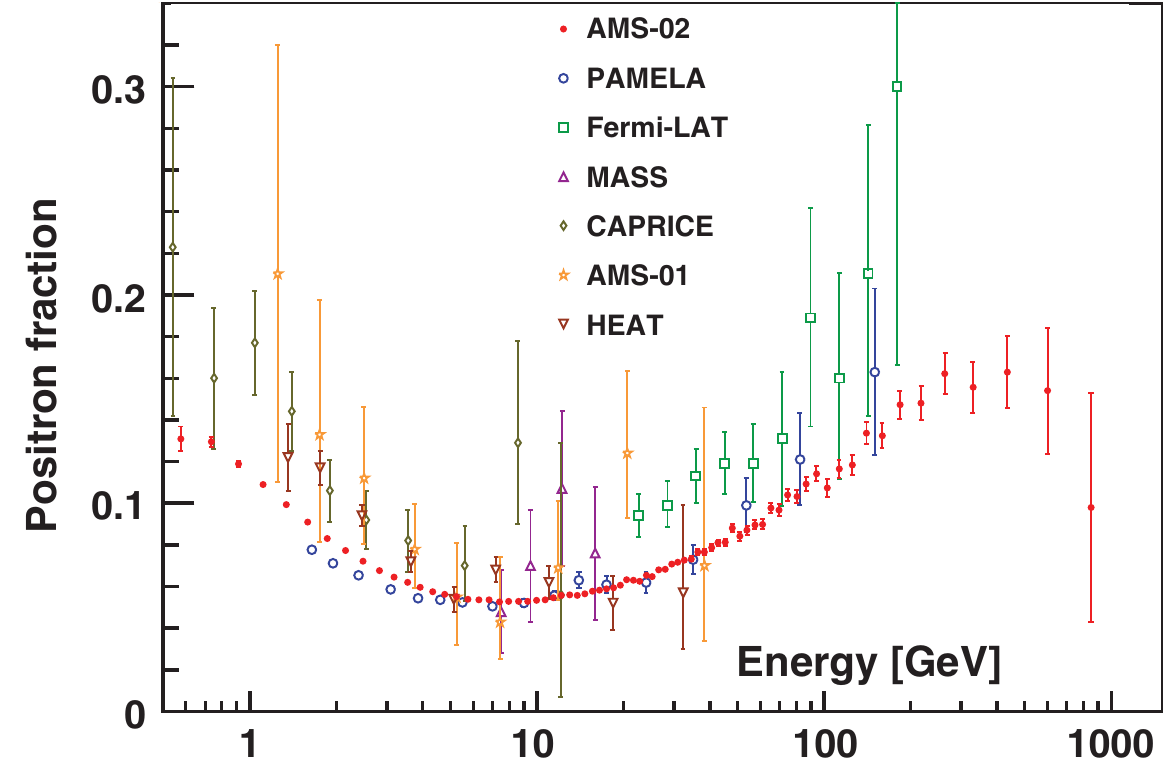}
  \caption{Ratio of cosmic ray positrons to the sum of electrons and positrons ($e^+ / (e^- + e^+)$)
    as a function of electron/positron energy~\cite{AMS02_Electrons_2019}.}
  \label{fig:ams02-positron-fraction}
\end{figure}

The result for the \mbox{AMS-02} measurement of the positron fraction is shown in
figure~\ref{fig:ams02-positron-fraction}. The data was collected between May 2011 and November
2017. Even though it is a ratio of CR species, the positron fraction is time dependent at low
energies, because solar modulation affects electrons and positrons
differently~\cite{AMS02_ElecPosTime_2018}. The standard theory of cosmic ray positrons as a
secondary species predicts a positron fraction which strictly falls with energy. The observed data
agrees with this hypothesis only below \SI{10}{\giga\electronvolt}, at which point the positron
fraction starts to rise.

This unexpected result was first observed by HEAT~\cite{Barwick1997} and then confirmed with better
precision by \mbox{AMS-01}~\cite{Aguilar2007} and PAMELA~\cite{Adriani2009}. Today, the precision of
the \mbox{AMS-02} data~\cite{AMS02_Electrons_2019} confirms the rise with unprecedented accuracy.

The \mbox{AMS-02} result extends the energy reach by almost one order of magnitude, up to
approximately \SI{1}{\tera\electronvolt}. In this region the positron fraction reaches a maximum
around \SI{350}{\giga\electronvolt} and begins to drop at even higher energies.

Many different models which explain the rise in the positron fraction by Dark Matter particle
annihilation and decay have been proposed~\cite{Hooper2009d,Cirelli2008a}. However, as of today,
other explanations, such as the presence of a nearby pulsar, remain viable
alternatives~\cite{Chowdhury2009}. The sharpness of the drop in the positron fraction at high
energies, as well as possible anisotropy in the flux of positrons (or in the positron fraction),
might help to differentiate between these alternatives~\cite{Ando2007}.

\begin{figure}[p]
  \centering
  \includegraphics[width=0.85\linewidth]{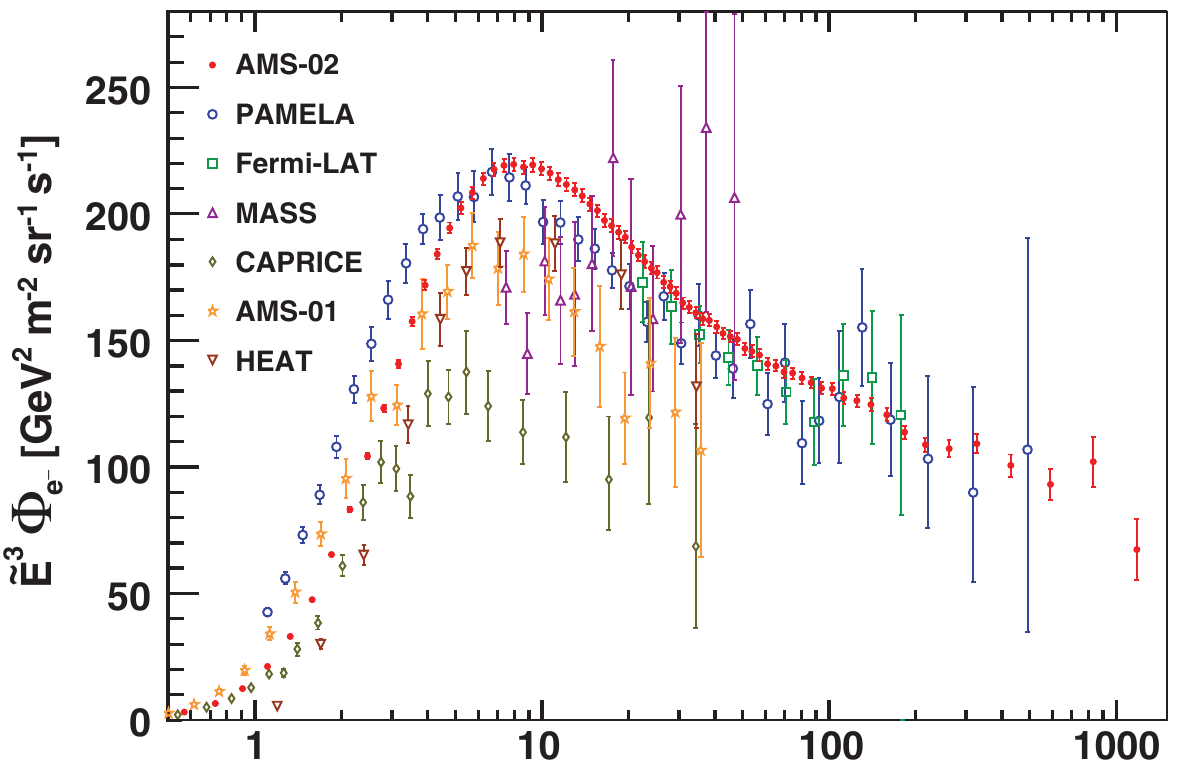}
  \caption{Flux of cosmic ray electrons, multiplied by $E^3$, as a function of electron
    energy~\cite{AMS02_Electrons_2019}.}
  \label{fig:ams02-electron-flux}

  \vspace*{\floatsep}
  \vspace*{\floatsep}

  \centering
  \includegraphics[width=0.85\linewidth]{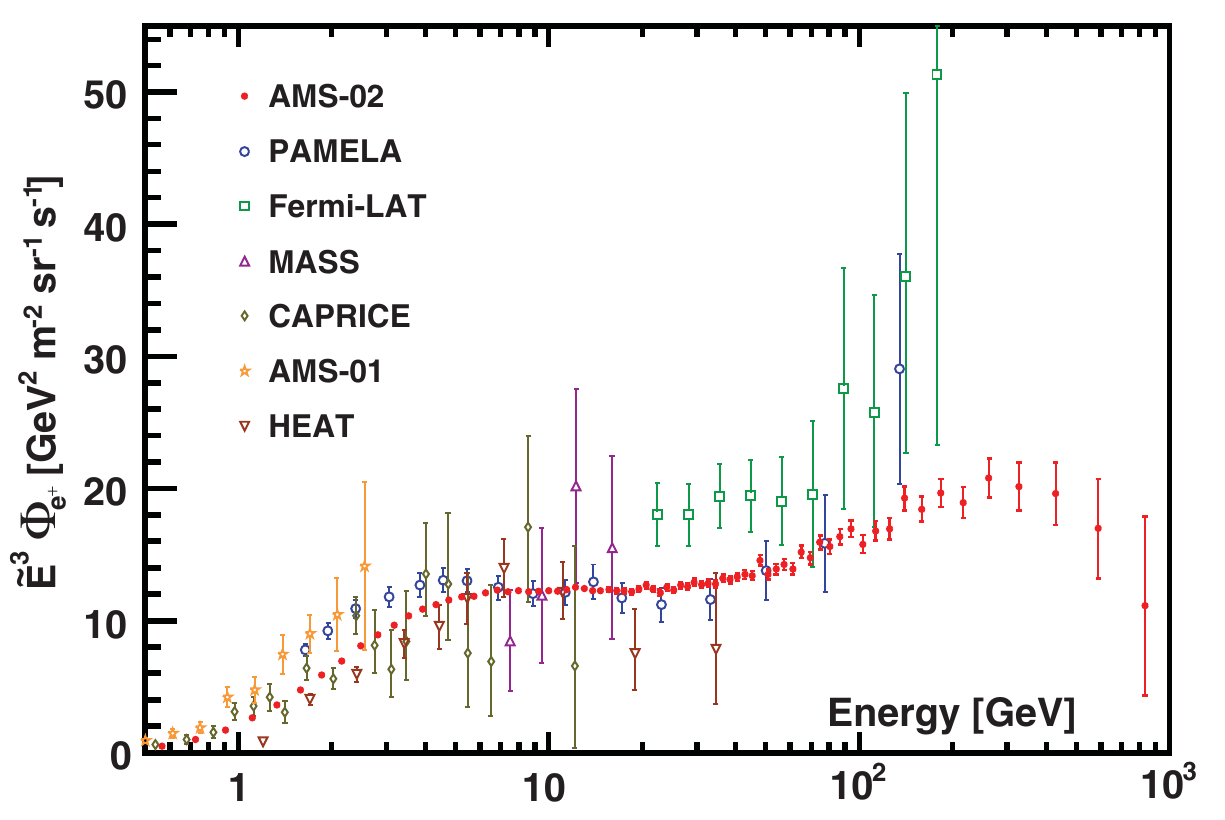}
  \caption{Flux of cosmic ray positrons, multiplied by $E^3$, as a function of positron
    energy~\cite{AMS02_Positrons_2019}.}
  \label{fig:ams02-positron-flux}
\end{figure}

The \mbox{AMS-02} measurements of the fluxes of cosmic ray electrons~\cite{AMS02_Electrons_2019} and
positrons~\cite{AMS02_Positrons_2019} are shown in figures~\ref{fig:ams02-electron-flux}
and~\ref{fig:ams02-positron-flux}, respectively. Both measurements are also based on data collected
between May 2011 and November 2017.

The \mbox{AMS-02} experiment is the only spectrometer in space, capable of measuring the individual
fluxes of electrons and positrons up to \si{\tera\electronvolt} energies, improving upon prior
results by almost one order of magnitude in energy reach. The results also show that the drop in the
positron fraction is related to a softening in the positron flux, and not to a hardening of the
electrons. In addition, the rise of the positron fraction at around \SI{10}{\giga\electronvolt} can
indeed be traced back to a hardening of the positron flux.

In addition, the different influence of solar modulation on the spectra of electrons and positrons
requires a separate measurement of the two species, in order to fully understand the behavior of the
positron fraction~\cite{AMS02_ElecPosTime_2018}.

Although cosmic ray electrons are less abundant than protons, they form an important background in
the measurement, in particular because the electromagnetic showers they induce are hard to
distinguish from those induced by $\gamma$-ray photons.

\section{Structure of the Milky Way}
\label{sec:galaxy}

The three dimensional structure of the Galaxy is vital in the understanding of diffuse emission of
$\gamma$-rays, since both $\pi^0$-decay and bremsstrahlung are directly correlated with the spatial
distribution of the gas. Instead, the IC emission is produced by interactions of energetic electrons
with the ISRF.

\subsection{Interstellar Gas}
\label{sec:galaxy-gas}

The interstellar matter (ISM) consists of more than $99\percent$ gas, more than $70\percent$ of
which is hydrogen. There are three different types of hydrogen which are important for the modeling
of gamma ray production: Atomic neutral hydrogen ({\Hone}), molecular neutral hydrogen ({\Hmoc}) and
ionized atomic hydrogen ({\Htwo}). In addition the gas can be either cold, warm or hot.

The distribution of the (warm) neutral atomic hydrogen can be traced by the well known
\SI{21}{\centi\meter} line. Photons with a wavelength of \SI{21}{\centi\meter} are emitted when a
transition between the two hyperfine levels of the hydrogen 1s state occurs. This corresponds to a
``spin-flip'' of the electron in the hydrogen atom. The radiation can pass through large parts of
the galaxy without being reabsorbed, because the interstellar dust is particularly transparent for
electromagnetic radiation at this wavelength.

A comprehensive {\Hone} survey of the entire sky was carried out in the Leiden/Argentine/Bonn
survey~\cite{LAB_HI_2005}, which combined data from two radio telescopes in order to map both the
northern and the southern hemispheres.

\begin{figure}[t]
  \centering
  \includegraphics[width=0.7\linewidth]{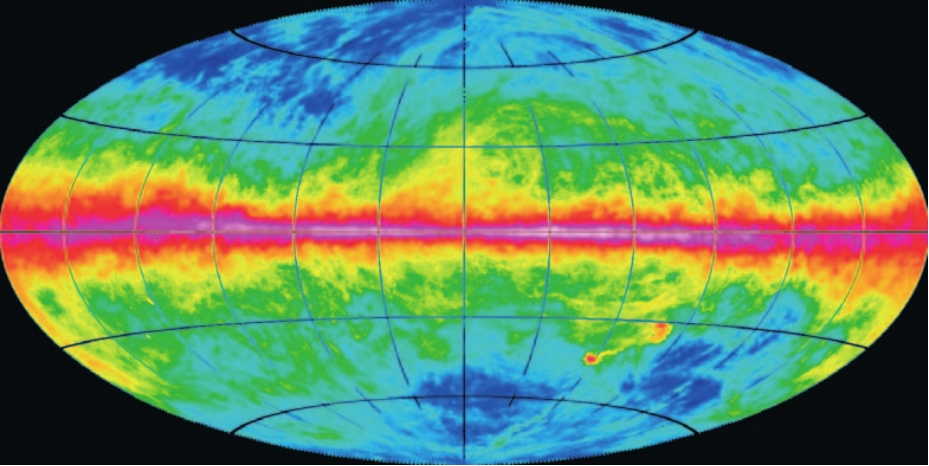}
  \caption{{\Hone} emissivity as measured in the LAB survey, shown in an Aitoff projection of
    galactic coordinates, integrated over the velocity range from
    $\SI{-400}{\kilo\meter\per\second}$ to $\SI{400}{\kilo\meter\per\second}$~\cite{LAB_HI_2005}.}
  \label{fig:lab-hone-emissivity}
\end{figure}

Figure \ref{fig:lab-hone-emissivity} shows the emissivity of the {\Hone} gas as a function of
galactic coordinates. As expected, the \SI{21}{\centi\meter} emission is strongest in the galactic
plane, but complex structures are observed.

The LAB survey was recently superseded by the HI4PI survey~\cite{HI4PI_Gas_2016}, which is based on
the Effelsberg-Bonn {\Hone} Survey (EBHIS) and the Galactic All-Sky Survey (GASS) and features
better angular resolution of approximately \SI{16.2}{\arcmin} and better sensitivity.

Because the {\Hone} gas is not entirely optically thin, it is required to know the spin temperature
$T_{S}$ of the hydrogen gas (related to its excitation), in order to convert the observed brightness
into a number density of hydrogen atoms. Measurements of the radial velocity of the gas clouds via
the Doppler shift of the \SI{21}{\centi\meter} line combined with a model for the rotation curve of
the Milky Way can be used to construct density maps in galactocentric rings. A recent model for the
rotation curve of the Milky Way is given in reference~\cite{Sofue_2015}, based on a solar system
distance of ${\Rsun} = \SI{8}{\kilo\parsec}$ and a local velocity of
$\Theta_\odot = \SI{238}{\kilo\meter\per\second}$.

The molecular hydrogen cannot be observed directly, one typically uses the $J = 1 \rightarrow 0$
transition line of the $\mathrm{^{12}CO}$ molecule as tracer. The carbon monoxide molecules cluster
in the same regions as the molecular hydrogen. In addition the collisions between {\Hmoc} and CO
molecules provide the excitation required for the line emission. A scaling factor (referred to as
$X_{CO}$) is commonly employed to convert the CO density into the {\Hmoc} density, which assumes a
constant ratio of CO to {\Hmoc} everywhere in the galaxy.

The ionized component in the form of {\Htwo} regions is the most difficult to locate. Studies of
dispersion measures of radio pulsars were used to compare the free electron column densities with
the integrated column density of {\Hone}~\cite{He2013}. This study puts the ratio of {\Htwo} to
{\Hone} to approximately \SIrange[]{7}{14}{\percent}, indicating that the collisions of protons with
ionized hydrogen play only a subdominant role. In addition, it is expected that the spatial
distribution of ionized hydrogen closely follows that of the atomic hydrogen, which means that it is
not necessarily required to construct independent gamma-ray templates for the two components.

Helium atoms are usually assumed to be uniformly mixed with the hydrogen gas, with a relative
abundance of approximately $11\percent$.

\begin{figure}[t]
  \begin{minipage}{0.48\linewidth}
    \centering
    \includegraphics[width=1.0\linewidth]{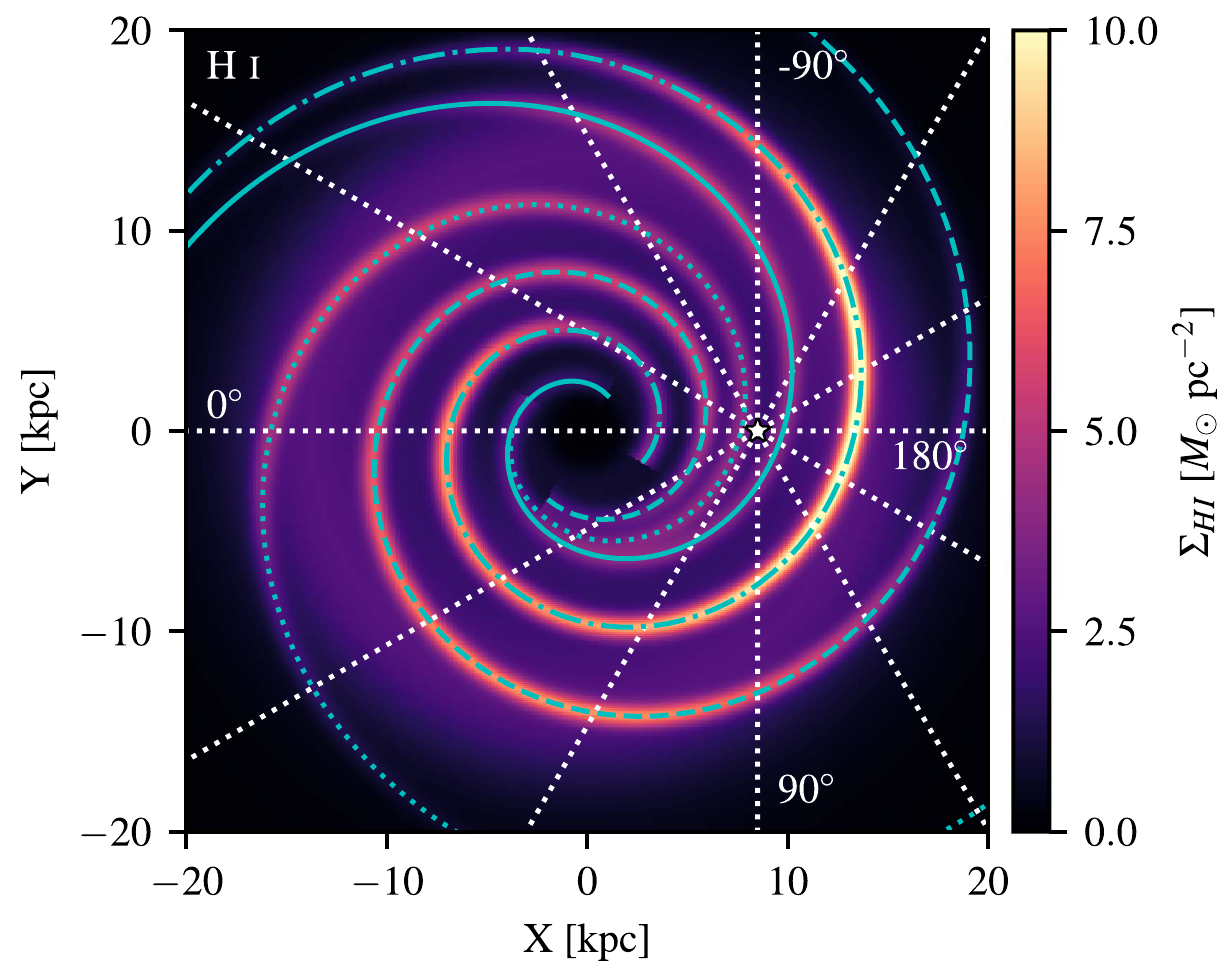}
  \end{minipage}
  \hspace{0.01\linewidth}
  \begin{minipage}{0.48\linewidth}
    \centering
    \includegraphics[width=1.0\linewidth]{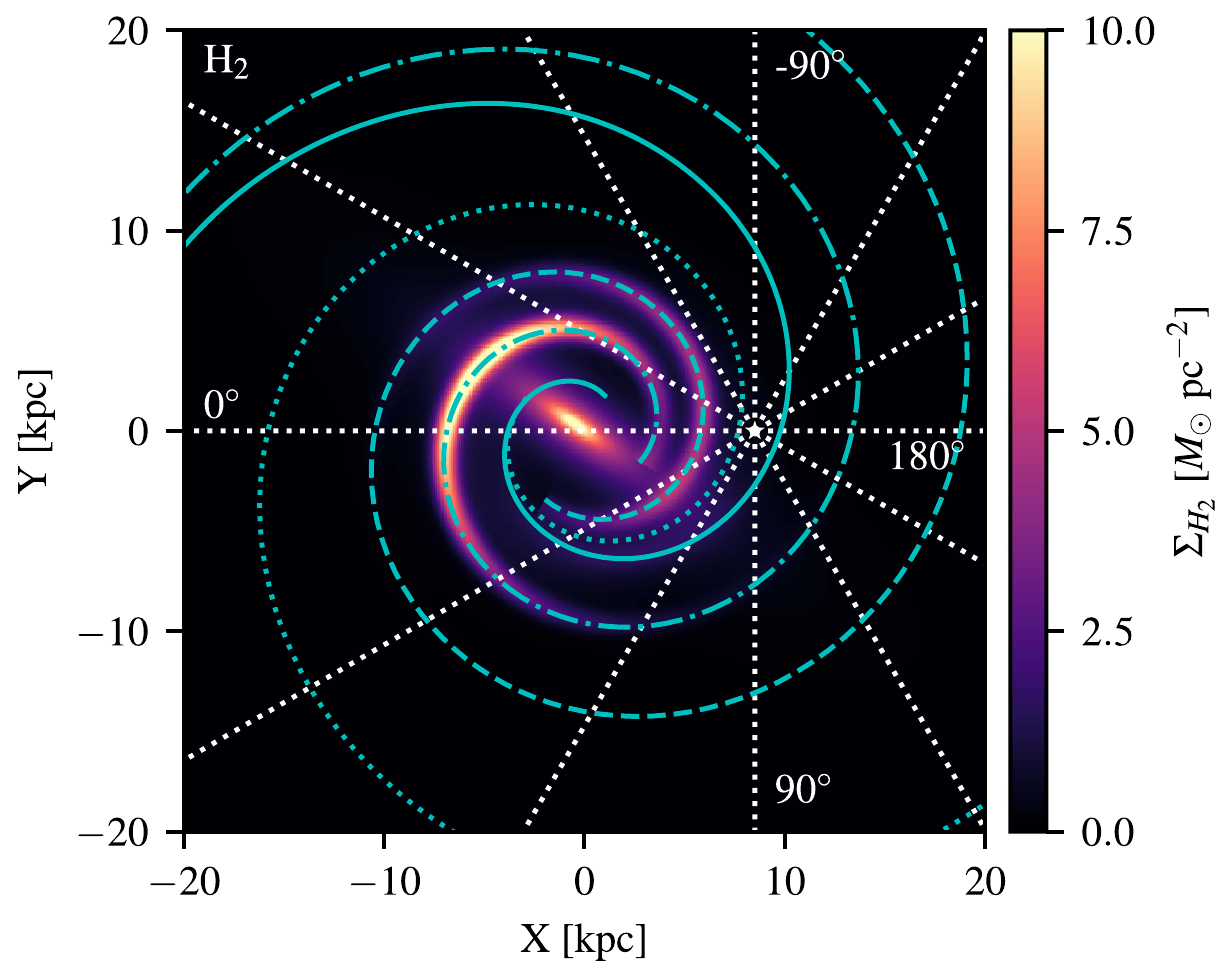}
  \end{minipage}
  \caption{Models for the three dimensional distribution of {\Hone} (left) and {\Hmoc} (right) gas
    in the Milky Way~\cite{Johannesson_APJ_856}. The galaxy is viewed along the North Galactic Pole,
    the Sun is located at the location of the white star. The white dashed lines represent lines of
    constant galactic longitude with a \SI{30}{\degree} spacing. The cyan curves on the density
    distribution mark the locations of the four spiral arms in the model.}
  \label{fig:galprop-3dgas-distribution}
\end{figure}

Figure~\ref{fig:galprop-3dgas-distribution} shows the three dimensional distribution of the gas
components {\Hone} and {\Hmoc} in a recent model~\cite{Johannesson_APJ_856}. The density of the gas
is correlated with the spiral arm structure of the Milky Way. {\Hmoc} dominates the central part of
the galaxy and forms the so called Central Molecular Zone. Models such as the one shown in
figure~\ref{fig:galprop-3dgas-distribution} are important ingredients in cosmic ray propagation
models.

\subsection{Radiation Fields}
\label{sec:galaxy-isrf}

Electrons and positrons can up-scatter photons to gamma-ray energies in the inverse Compton
scattering process. In order to calculate this contribution to the gamma-ray diffuse flux one needs
to know the energy density distribution of the radiation field as a function of the wavelength and
spatial coordinate.

Photons in the interstellar radiation field (ISRF) are emitted by stars and are subject to
absorption and re-emission in the interstellar dust. Although it is not possible to directly observe
the radiation field, elaborate models of the ISRF exist and are based on surveys of stellar
populations combined with measurements of the dust and its emissivity which are typically carried
out in the infrared band. Models for the stellar disk components were built by
Freudenreich~\cite{Freudenreich1998} based on COBE satellite data. The distribution of {\Hone} and
{\Hmoc} is also important to understand the dust emissivity.

Another important component of the ISRF is the almost completely isotropic cosmic microwave
background with its well-known black body spectrum which provides an abundant source of photons for
inverse Compton scattering.

A recent review of the structure of the ISRF is provided in~\cite{Porter_APJ_846}, where models by
Robitaille~\cite{Robitaille_2012} and Freudenreich~\cite{Freudenreich1998} are compared to
COBE/DIRBE, IRAS and Spitzer data and the implications for galactic gamma rays are studied.

\begin{figure}[t]
  \begin{minipage}{0.48\linewidth}
    \centering
    \includegraphics[width=1.0\linewidth]{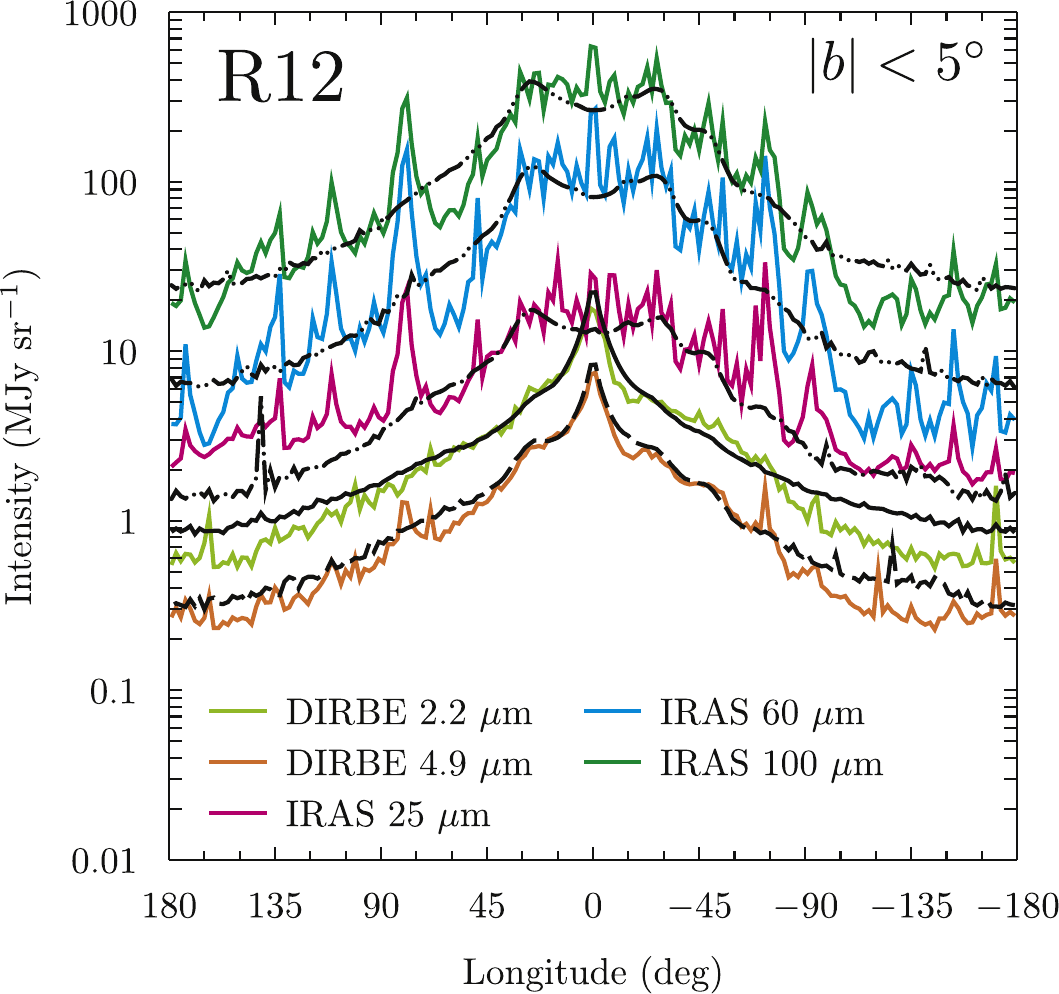}
  \end{minipage}
  \hspace{0.01\linewidth}
  \begin{minipage}{0.48\linewidth}
    \centering
    \includegraphics[width=1.0\linewidth]{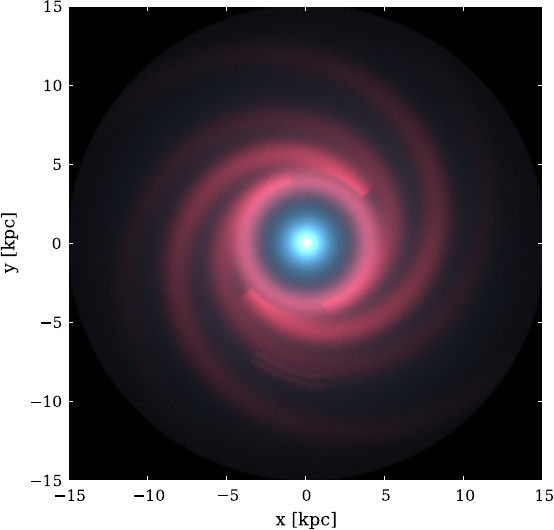}
  \end{minipage}
  \caption{Left: Comparison of the ISRF model of Robitaille~\cite{Robitaille_2012} with infrared
    data from the COBE/DIRBE~\cite{Sodroski_1997} and IRAS\cite{Miville_Deschenes_2005}
    experiments~\cite{Porter_APJ_846}. Black lines correspond to the model prediction and the
    colored lines represent various wavelengths as indicated in the legend. Right: Color composite
    view of the galactic ISRF model of Robitaille\cite{Robitaille_2012} as seen along the North
    Galactic Pole. The Sun is located at ($x$, $y$) = ($0$, $\SI{-8.5}{\kilo\parsec}$). The colors
    are IRAC \SI{8.0}{\micro\meter} (red), IRAC \SI{4.5}{\micro\meter} (green), and IRAC
    \SI{3.6}{\micro\meter} (blue) intensities. IRAC is the InfraRed Array Camera on the Spitzer
    telescope~\cite{IRAC_2004}.}
  \label{fig:isrf-r12}
\end{figure}

Figure~\ref{fig:isrf-r12} shows a comparison of the intensity of the ISRF in a spiral arm
model~\cite{Robitaille_2012} to infrared data. The left figure shows integrated intensity for
latitudes $|b| < \SI{5}{\degree}$. The model generally compares well with the data, although the
data is generally more structured. The model slightly over-predicts the data in the third and fourth
sector ($\SI{90}{\degree} < l < \SI{270}{\degree}$). Shown on the right hand side is a composite RGB
image, showing the intensity of the ISRF in three different infrared wavelengths. The spiral arm
structure is clearly seen, in particular in the \SI{8}{\micro\meter} channel. However, other models,
in which the spiral structure of the Milky Way is much less pronounced, are also viable
alternatives~\cite{Porter_APJ_846}.

\section{Gamma Ray Sources}
\label{sec:sources}

In addition to diffuse emission $\gamma$-rays can also be produced in sources, which often appear
point-like in the sky. These photons are particularly interesting, since they carry direct
information about the physical processes in the vicinity of the sources. These processes are
typically very energetic and the sources are often among the most compact objects in the Universe.

\subsection{Supernova Remnants}
\label{sec:sources-snr}

Supernova remnants are the results of supernova explosions. After the explosion, an expanding shock
wave transports ejected material out into the interstellar medium and creates a bubble with a
relatively sharp edge. The deceleration of the shock wave lasts for several 10000 years and finally
stops when the velocity of the ejected material has reached the speed of the surrounding material,
at which point the SNR slowly merges with its surrounding.

\begin{figure}[t]
  \centering
  \includegraphics[width=0.5\linewidth]{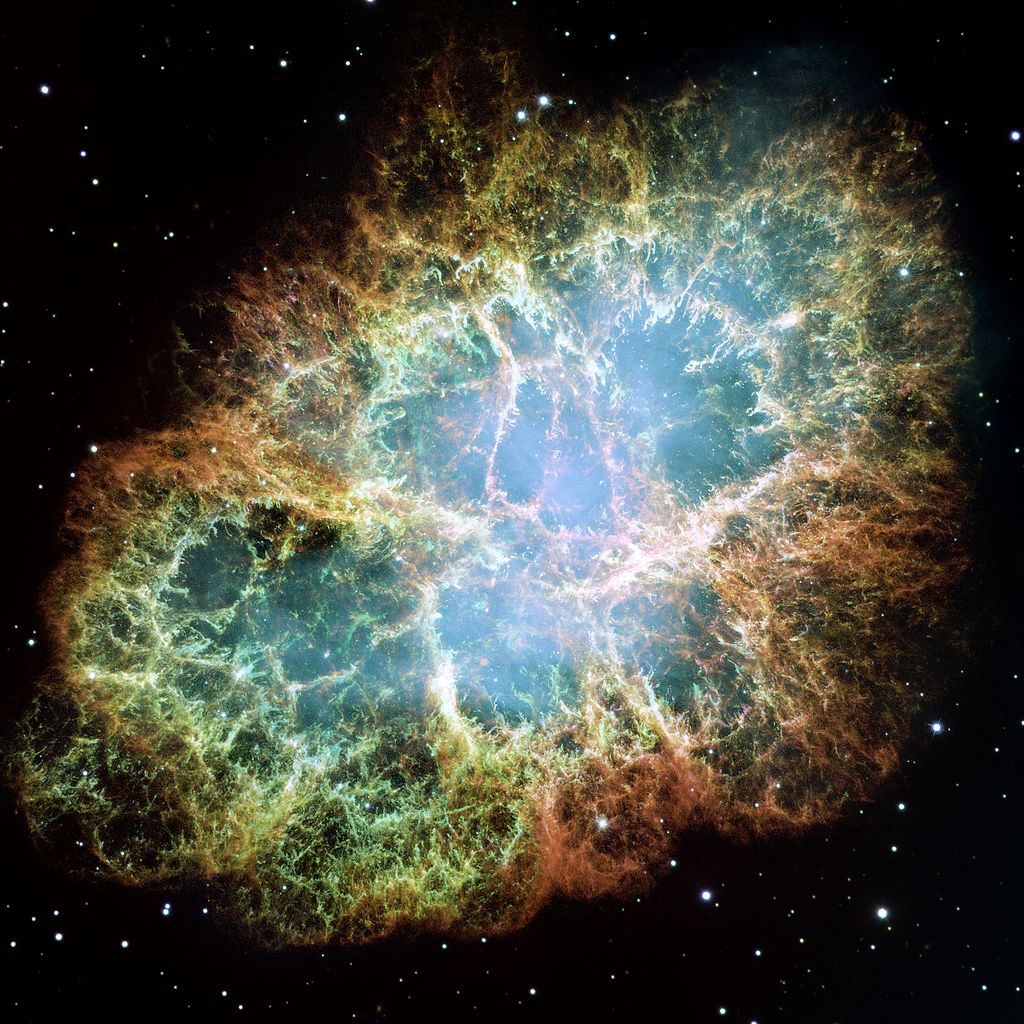}
  \caption{Hubble space telescope mosaic image of the Crab Nebula SNR~\cite{Hubble_Crab_SNR_2005}.}
  \label{fig:crab-nebula-snr-hubble}
\end{figure}

Figure~\ref{fig:crab-nebula-snr-hubble} shows a mosaic image of the Hubble Space Telescope of the
Crab Nebula, which is the remnant of the supernova explosion SN 1054, observed by Chinese
astronomers in 1054, approximately 965 years ago. The explosion lead to the formation of a rotating
neutron star, the Crab pulsar, in the center of the nebula. The filaments on the exterior are formed
by ejected material from the original star's atmosphere. Synchrotron emission from the curved
trajectories of electrons in the pulsar's magnetic field is believed to be responsible for the
diffuse blue light observed in the interior of the nebula~\cite{Crab_Optical_1957}.

SNRs are assumed to be the predominant sources of cosmic ray acceleration. Primary cosmic rays, such
as protons, electrons and helium nuclei are believed to be accelerated in first order Fermi
acceleration, in which particles gain energy when they are reflected by magnetic turbulences on both
sides of the shock front. In this way they become more and more energetic and the resulting spectrum
is a power law with spectral index of -2.

Collisions of cosmic ray protons in the SNR with other nuclei lead to $\pi^0$ production and
subsequent decay into $\gamma$-rays, as described in
section~\ref{sec:gamma-ray-production}. Bremsstrahlung from electrons also creates a $\gamma$-ray
signal from the SNR.

\subsection{Pulsars and Pulsar Wind Nebulae}
\label{sec:sources-pulsars}

When stars with masses between 10 and 29 solar masses collapse formation of a neutron star is
possible. Neutron stars are extremely compact objects, made almost exclusively of neutrons. They
withstand gravitational collapse by the neutron degeneracy pressure generated by the Pauli exclusion
principle of fermion quantum states, which acts because the matter density is on the scale of
nuclear matter ($\approx \SI{4e17}{\kilo\gram\per\cubic\meter}$).

The lower limit for the mass of a neutron star is the Chandrasekhar mass of approximately
$1.4 \, \Msun$~\cite{Chandrasekhar_1931}. Objects with lower masses are typically white dwarfs,
which are supported against collapse by electron degeneracy pressure. Conversely, the upper limit
for neutron star masses is the
Tolman-Oppenheimer-Volkoff~\cite{Tolman_1939,Oppenheimer_Volkoff_1939} limit of approximately
$2.17 \, \Msun$. Heavier stellar remnants collapse further and form a black hole. Thus, neutron star
masses are fairly confined between the two limits.

Typical neutron stars have radii of about \SI{10}{\kilo\meter}, which means that their matter is
about \num{e14} times more dense than the Sun.

Because of the non-zero magnetic moment of the neutron, a rotating neutron star often generates a
net magnetic dipole field along a magnetic axis which does not necessarily coincide with the
rotational axis. Similar to the effect of gravitational precession the magnetic axis of the pulsar
rotates around the rotational axis on a cone. This creates a dynamo which emits low frequency
($< \SI{1}{\kilo\hertz}$) electromagnetic radiation along the magnetic axis. This emission can not
directly be observed, as it would be absorbed in the ISM. Instead the radiated power heats up the
material surrounding the pulsar.

\begin{figure}[t]
  \begin{minipage}{0.48\linewidth}
    \centering
    \includegraphics[width=0.98\linewidth]{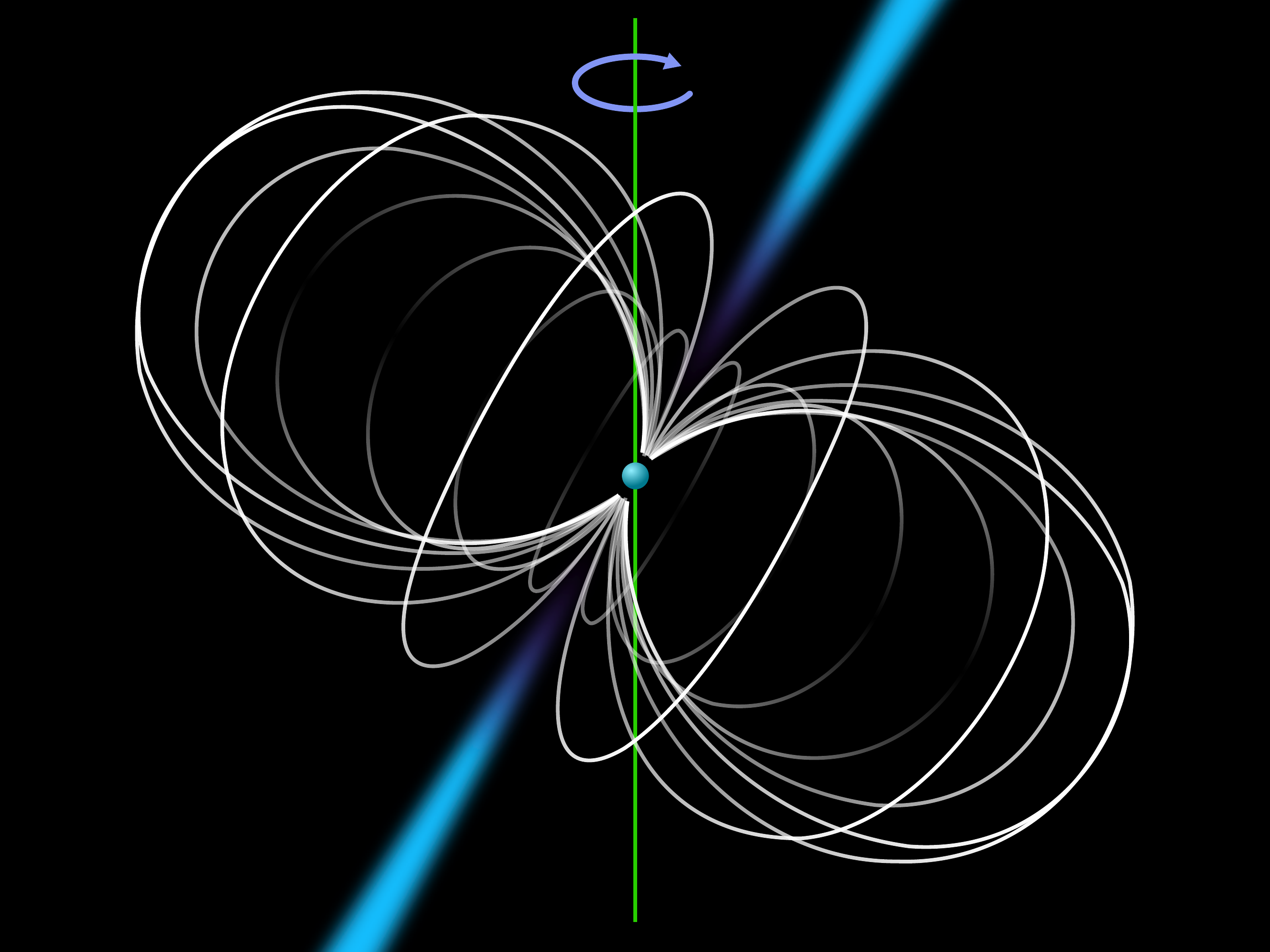}
  \end{minipage}
  \hspace{0.01\linewidth}
  \begin{minipage}{0.48\linewidth}
    \centering
    \includegraphics[width=1.0\linewidth]{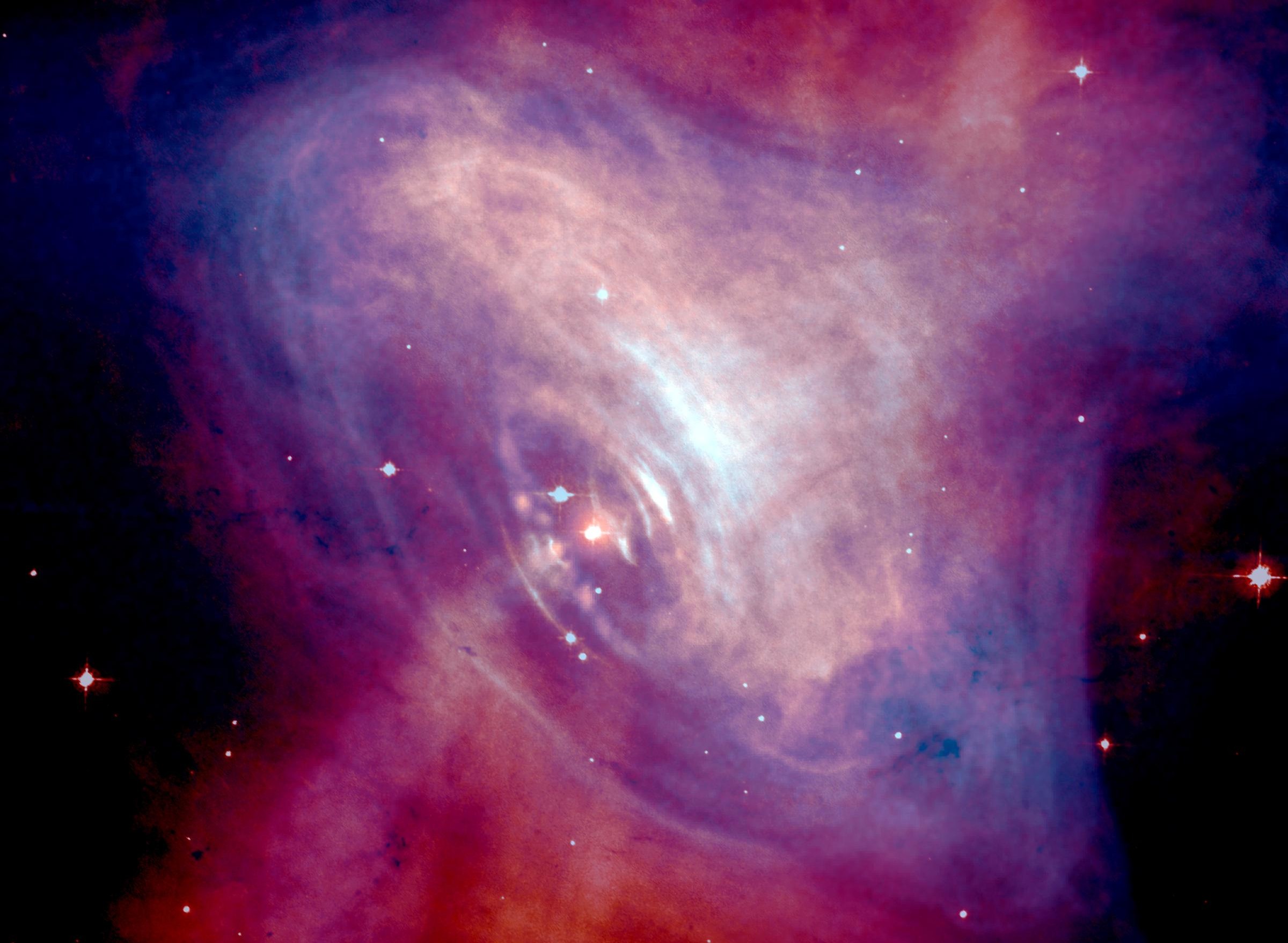}
  \end{minipage}
  \caption{Left: Schematic depiction of a pulsar, the rotational axis is shown in green, magnetic
    field lines in white~\cite{Pulsar_Schematic_2007}. The blue beams correspond to the jets emitted
    along the magnetic axis. Right: Composite overlay image of the Crab pulsar in X-rays (blue) and
    optical (red) wavelengths. The X-ray image is from the Chandra X-ray space telescope, and the
    optical image was recorded by the Hubble Space Telescope~\cite{Chandra_Hubble_Crab_2002}.}
  \label{fig:pulsar-schematic-crab-chandra-hst}
\end{figure}

Due to the strong magnetic field electrons and positrons from regions close to the pulsars surface
are pulled along the field lines and accelerated. The bending of electrons and positrons in the
extremely strong magnetic field of pulsars causes emission of $\gamma$-ray photons by curvature
radiation. In addition, energetic electrons can up-scatter photons from the environment or from the
CMB to $\gamma$-ray energies through the inverse Compton process. This results in a particle
cascade, which becomes beamed if the process occurs close to the magnetic axis of the
pulsar. Curvature radiation from electrons and positrons is most likely also responsible for the
emission of radio waves. The $\gamma$-ray spectra of most pulsars cut off at energies around
\SI{10}{\giga\electronvolt}. In the polar cap model, this is a result of pair production
attenuation, where, depending on the strength of the magnetic field, photons convert back into
electron positron pairs~\cite{Harding_2001}.

Figure~\ref{fig:pulsar-schematic-crab-chandra-hst} shows a schematic of the configuration of the
magnetic field of a pulsar. As the pulsar rotates the beam of photons and relativistic particles
sweeps across the sky like a lighthouse beam. The right hand side of the figure shows a composite
image of the Crab nebula in the X-ray and radio bands. The synchrotron emission of relativistic
electrons in the jet of the pulsar is clearly visible in the X-ray band.

In addition ring like structures in the equatorial plane of the pulsar are the result of
relativistic electrons which travel along the magnetic field lines and create a shock front when
colliding with the surrounding nebula. This is the Pulsar Wind Nebula (PWN) of the Crab
Pulsar. Inverse Compton scattering processes in the PWN can also generate $\gamma$-ray photons, so
the PWN itself is also detectable. In contrast to the signal from the pulsar itself this emission is
not pulsed.

The magnetic moment of a uniform sphere with surface magnetic field strength $B$ and radius $R$ is
$m = B R^3$. If the magnetic and rotational axes are inclined by $\alpha$, the perpendicular
component of the magnetic moment is $m_\bot = m \sin{\alpha}$. With the period of rotation $P = 2
\pi / \omega$ the radiated power of the dynamo is:

\begin{equation}
  \label{eq:pulsar-radiated-power}
  P_{\mathrm{rad}}
  = \frac{2}{3} \frac{\ddot{m}_\bot^2}{c^3}
  = \frac{2}{3 c^3} (B R^3 \sin{\alpha})^2 \left( \frac{2 \pi}{P} \right)^4 \,,
\end{equation}

where $c$ is the speed of light. The rotational energy of the pulsar is
$E_{\mathrm{rot}} = \frac{1}{2} I \omega^2$ where $I = \frac{2}{5} M R^2$ is the moment of inertia
of a solid uniform sphere, which is approximately universally constant since both mass ($M$) and
radius ($R$) of pulsars do not vary much. The time derivative of the rotational energy is:

\begin{equation}
  \label{eq:pulsar-e-rot}
  \dot{E}_{\mathrm{rot}} = \frac{\mathrm{d}}{\mathrm{d}t}\left( \frac{1}{2} I \omega^2\right)
  = I \omega \dot{\omega} = -4 \pi^2 I \frac{\dot{P}}{P^3} \,,
\end{equation}

with $\omega = 2 \pi / P$ the angular frequency. This results in a huge number, for the Crab nebula
the change of the rotational energy (the power loss) is approximately
$\SI{4e38}{\erg\per\second} = \SI{4e31}{\watt}$, for example. The pulsar's rotation slows down with
time as it loses energy. For a rotation powered pulsar, where all of the rotational energy is lost
by radiation ($P_{\mathrm{rad}} = -\dot{E}_{\mathrm{rot}}$), it is possible to estimate the minimum
magnetic field strength at the pulsar's surface:

\begin{equation}
  \label{eq:pulsar-field-strength}
  B > \sqrt{\frac{3 c^3 I}{8 \pi^2 R^6} P \dot{P}}
  \Rightarrow \left(\frac{B}{\mathrm{\si{\gauss}}}\right) \approx \num{3.2e19} \sqrt{\left( \frac{P
        \dot{P}}{\si{\second}} \right)} \,.
\end{equation}

The inequality is a result of setting $\sin{\alpha} = 1$, since $\alpha$ is generally unknown. For a
typical pulsar this results in field strengths between $\SI{e8}{\gauss}$ and
$\SI{e14}{\gauss}$.

Assuming the magnetic field strength does not change with time,
equation~(\ref{eq:pulsar-field-strength}) can be rearranged to show that the product $P\dot{P}$ is
constant. Thus the characteristic age $\tau$ of the pulsar can be estimated:

\begin{align*}
  &P \mathrm{d}P = P \dot{P} \mathrm{d}t \\
  \Rightarrow \quad &\int_{P_0}^{P}{P' \,\mathrm{d}P'} = \int_0^{\tau}{P \dot{P} \, \mathrm{d}t} = P \dot{P}
                \int_0^{\tau}{\mathrm{d}t} \\
  \Rightarrow \quad &\frac{P^2 - P_0^2}{2} = P \dot{P} \tau \\
  \Rightarrow \quad &\tau \approx \frac{P}{2 \dot{P}} \numberthis \label{eq:pulsar-age} \,,
\end{align*}

which assumes that the original period is much smaller than the current period $P_0 \ll P$. For the
\begin{figure}[t]
  \centering
  \includegraphics[width=0.7\linewidth]{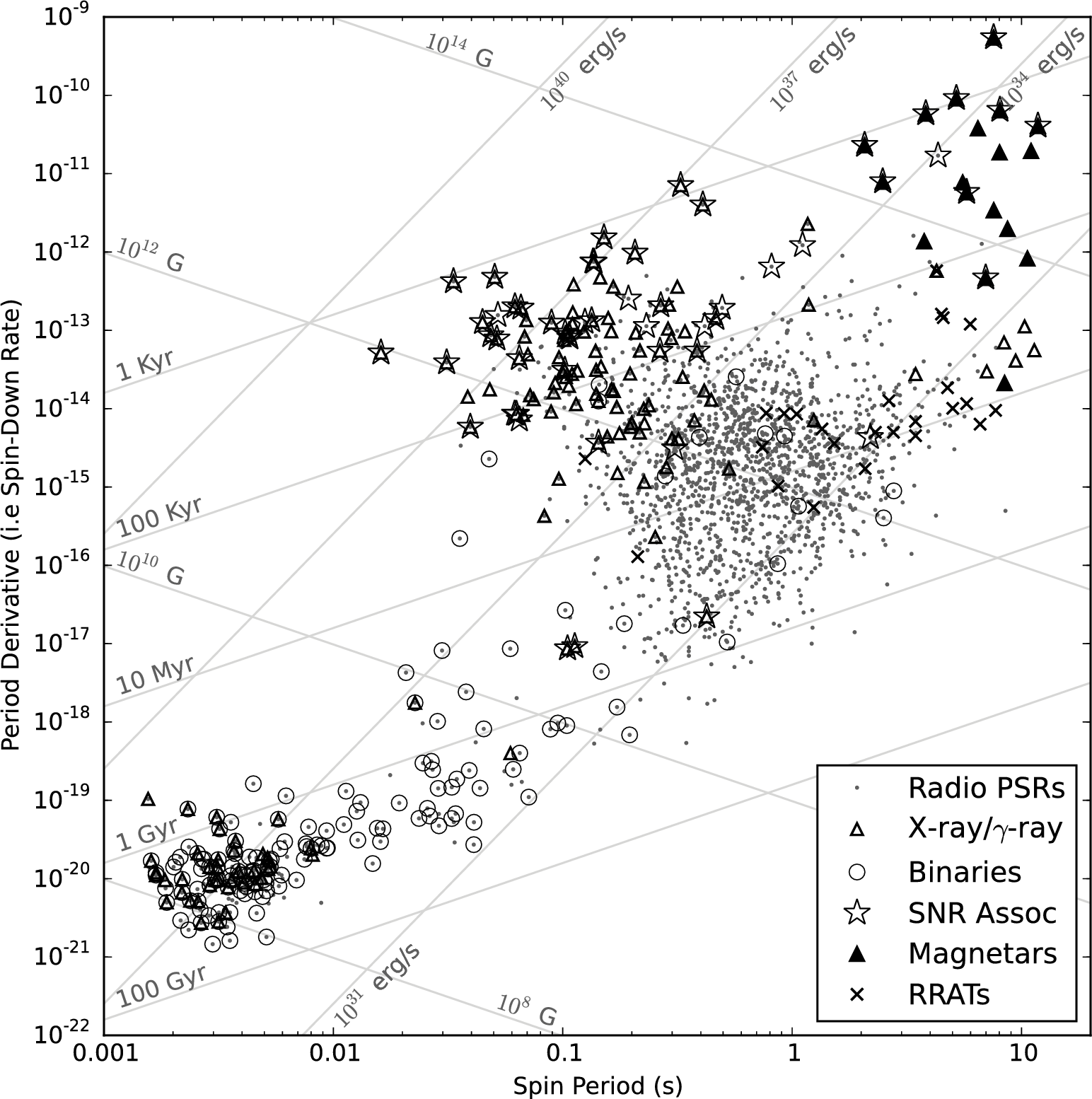}
  \caption{$P\dot{P}$ diagram~\cite{Condon_EssentialRadioAstronomy_2016,ERA_WWW}, showing the
    relation between period $P$ and spindown $\dot{P}$ for a large set of pulsars. The dashed lines
    running from bottom left to top right correspond to various pulsar ages. The dashed-dotted lines
    running from top left to bottom right correspond to various magnetic field strengths. Pulsars
    shown with open triangles have been identified in $\gamma$-rays or X-rays. The encircled dots
    correspond to binary systems.}
  \label{fig:pulsars-p-pdot}
\end{figure}

Crab Nebula ($P = \SI{0.033}{\second}$, $\dot{P} = 10^{-12.4})$~\cite{Crab_Period_1971} this
estimate results in $\tau \approx \SI{1300}{\year}$, which is not too far away from the known age of
\SI{965}{\year}.

Because the masses and radii of pulsars are relatively confined, they are characterized almost
entirely by their period of rotation $P$ and its time derivative $\dot{P}$. Measuring both these
properties enables the estimation of the magnetic field strength $B$ at the surface and the pulsar's
age $\tau$ as shown in equations~(\ref{eq:pulsar-field-strength}) and~(\ref{eq:pulsar-age}).

Figure~\ref{fig:pulsars-p-pdot} shows the distribution of known pulsars in the $P\dot{P}$
diagram. Most regular pulsars have periods between \SI{100}{\milli\second} and \SI{3}{\second} and a
spindown rate of approximately \num{e-17} to \num{e-13}. They form a densely populated blob in the
diagram. Pulsars with period well below \SI{100}{\milli\second} are referred to as milli-second
pulsars. These objects spin very rapidly and are almost always part of a binary system. Young
pulsars such as Crab, Vela and Geminga are found in the top left region. Almost all young pulsars
are located inside Supernova Remnants. Some of these pulsars (Geminga is a prominent example) are
radio quiet: They were identified in X-ray or $\gamma$-rays, but do not pulse in the radio
band~\cite{SAS_2_Geminga_1992,Fermi_Geminga_2010} reason for this effect is not yet understood,
since almost all other pulsars do produce pulsed radio emissions. Magnetars, pulsars with the
strongest magnetic fields, are found in the top right corner. The (empty) bottom right section
corresponds to the ``graveyard'' - the region in which the pulsar is no longer capable of producing
radio emission, since the curvature radiation is not strong enough to generate particle cascades.

The period of rotation of pulsars is generally extremely stable and can be measured with high
precision. Therefore, Pulsar timing can be used to construct astronomical clocks. A network of many
pulsars can also be used to search for signals of gravitational waves, which would be observable due
to their systematic effect on the timing measurements of the ensemble. A dedicated project to study
signals of gravitational waves with pulsars is the NANOGrav project~\cite{NANOGrav_2013}.

Another interesting phenomenon are timing glitches, which are sudden changes in the pulsar's period
or its spindown. In a popular model these glitches are caused by microquakes (release of surface
tension) in the pulsar's outer crust~\cite{Pulsar_Glitches_1975}. Glitches are often found in the
timing of young pulsars, such as Vela and Crab~\cite{ATNF_Glitches_WWW}.

\subsection{Active Galactic Nuclei and Blazars}
\label{sec:sources-agn}

In some galaxies the central Super Massive Black Hole (SMBH) produces enormous amounts of radiation
across the entire electromagnetic spectrum. These central regions of galaxies are called Active
Galactic Nuclei (AGN). In the standard model of AGNs~\cite{AGN_1995} the SMBH is powered by an
accretion disk which surrounds the black hole. During the accretion the matter in the disk is heated
and produces electromagnetic radiation. In addition, relativistic jets are formed in directions
perpendicular to the accretion and rotation of the black hole. In these jets particles are
accelerated to enormous energies.

\begin{figure}[t]
  \centering
  \includegraphics[width=0.75\linewidth]{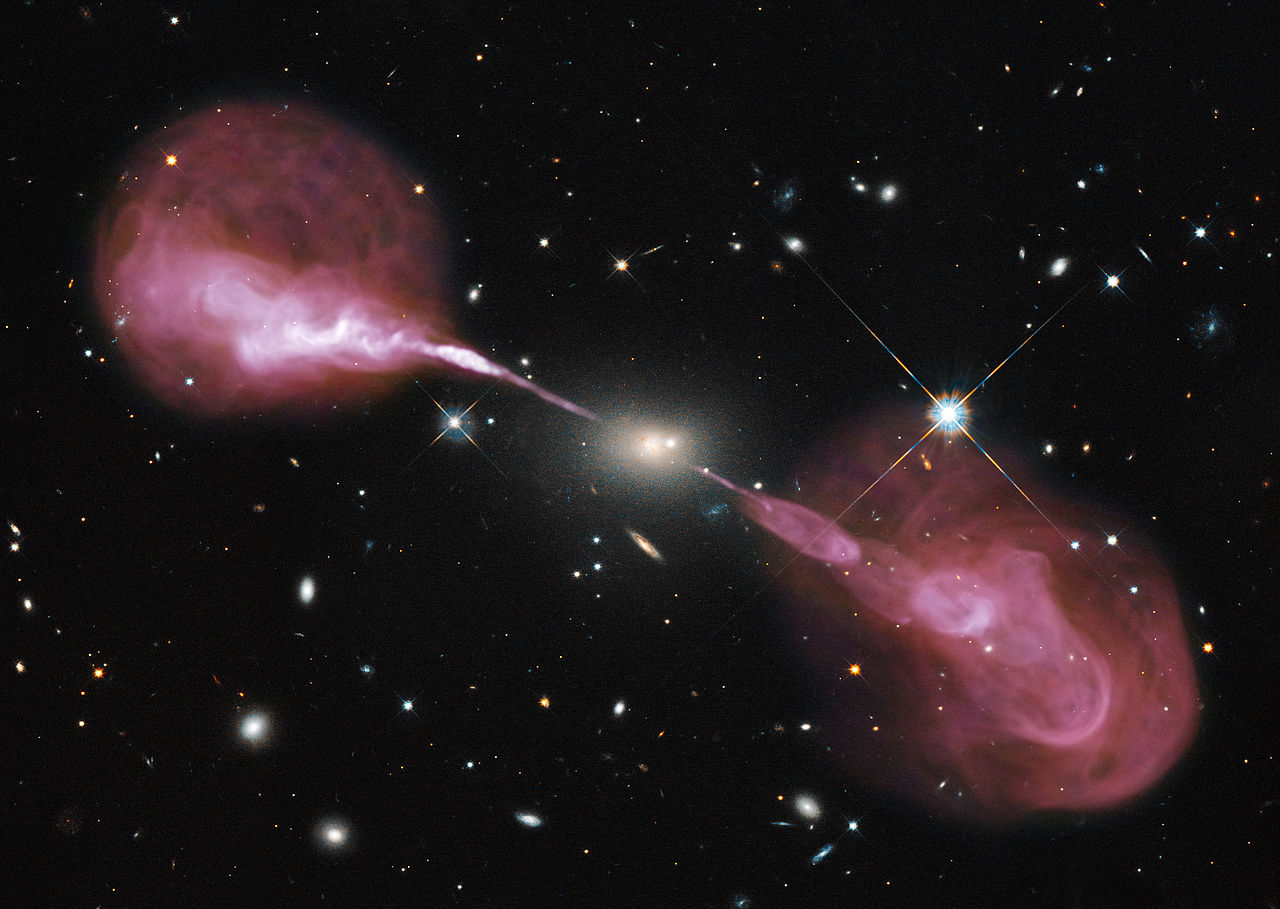}
  \caption{Multi Wavelength view of the radio galaxy \mbox{Hercules
      A}~\cite{Hubble_VLA_HerculesA_2012}. Optical data from the Hubble Space Telescope is combined
    with a radio image (shown in red), recorded with the VLA.}
  \label{fig:agn-image}
\end{figure}

Figure~\ref{fig:agn-image} shows a composite image of the radio galaxy \mbox{Hercules A}. At radio
wavelengths the two jets are clearly distinguishable. At the end of the two jets giant radio lobes
are observed, which are luminous at radio wavelengths. Some AGNs have one-sided jets and the radio
lobes can appear more or less pronounced.

\begin{figure}[t]
  \centering
  \includegraphics[width=0.8\linewidth]{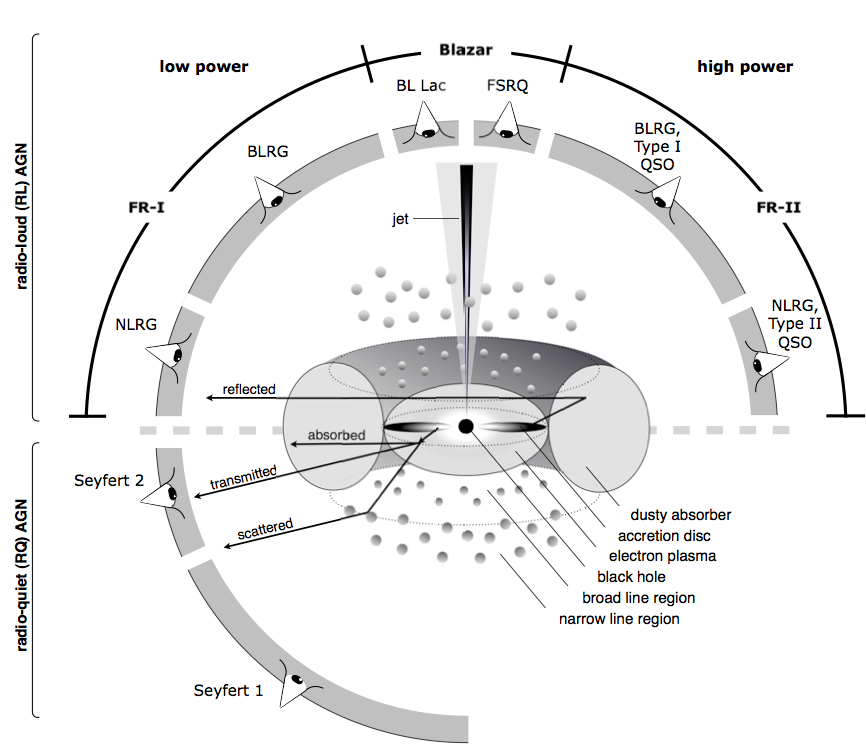}
  \caption{Schematic of different types of AGN~\cite{Beckmann_AGN_2012}.}
  \label{fig:agn-schematic}
\end{figure}

\begin{figure}[t]
  \centering
  \includegraphics[width=0.9\linewidth]{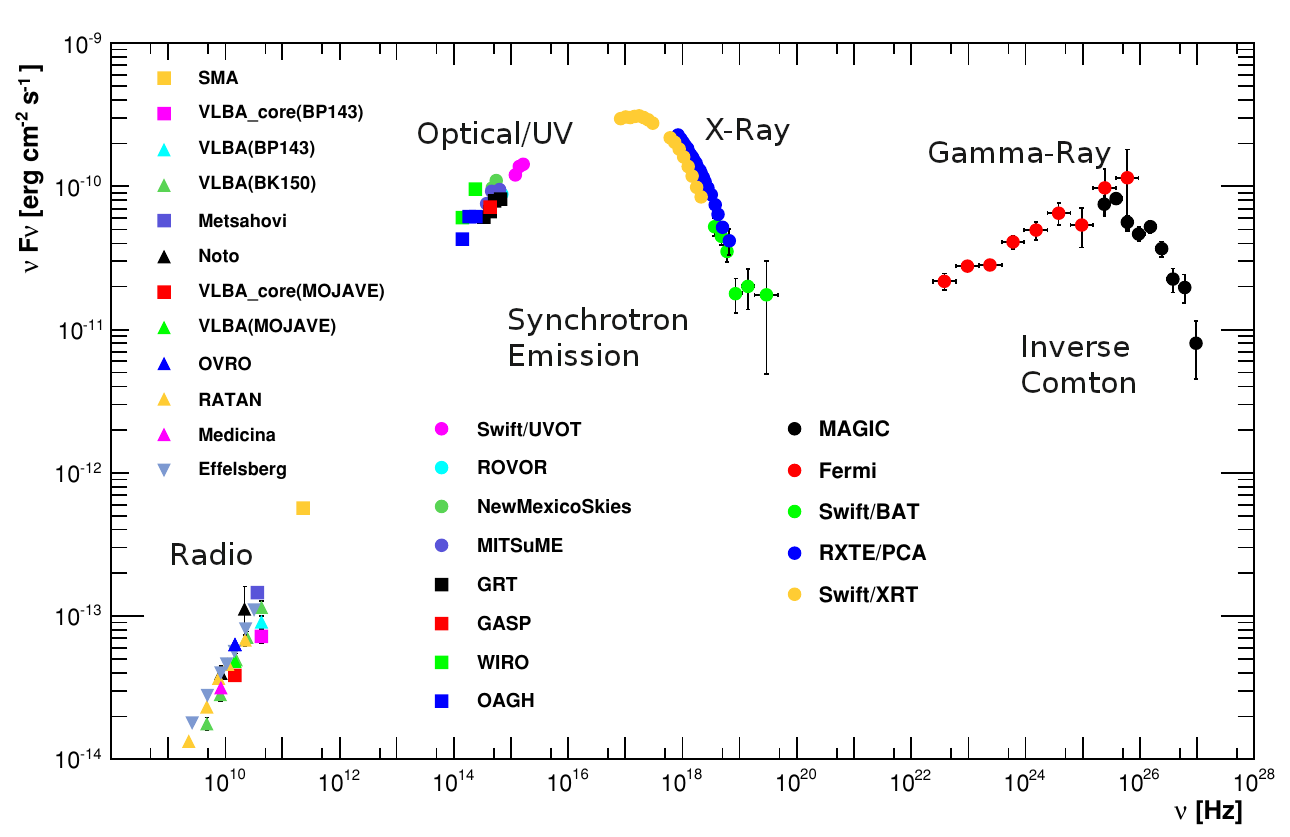}
  \caption{Spectral energy density of the AGN \mbox{Mrk
      421}~\cite{FermiLAT_Mrk421_2011}. Measurements from various telescopes and observatories,
    covering almost 20 orders of magnitude in frequency, are shown as indicated in the legend.}
  \label{fig:agn-sed}
\end{figure}

A schematic overview of various types of AGN is shown in figure~\ref{fig:agn-schematic}. These
classes of objects were historically introduced separately, and only later unified in the AGN
model. In the current understanding the various classes are manifestations of AGN observed under
different angles.

AGNs can be divided into radio-loud and radio-quiet objects, depending on whether or not radio
emissions are observed. The former are radio galaxies and blazars, depending on the observation
angle, and the latter are referred to as Seyfert galaxies.

The optical spectrum of AGNs often contains emission lines. Depending on the width of those lines
one differentiates between Narrow Line Radio Galaxies (NLRG) and Broad Line Radio Galaxies
(BLRG). The same distinction can be used to subdivide Seyfert galaxies into two classes: Seyfert 1
and Seyfert 2. Depending on the orientation radio galaxies can appear very bright (high power), in
which case they outshine the entire host galaxy and appear so bright that they appear to be ``quasi
stellar'' and are referred to as quasars.

In the AGN subclass of blazars, the relativistic jet is oriented directly towards the observer. Due
to relativistic beaming blazars appear extremely bright. Two prominent sub-types of blazars are BL
Lacertae (BL Lac) type objects and Flat Spectrum Radio Quasars (FSRQ). The latter are sometimes also
referred to as Optically Violent Variable (OVV) quasars. The main difference between FSRQ and BL Lac
type objects is that broad emission lines are observed in FSRQ, whereas BL Lac spectra only contain
weak lines, if any.


One common feature of blazars is that they are extremely variable. Variations of the observed
spectrum both on short (minutes to days) and long timescales (weeks to years) have been
observed. This limits the size of the emission region:

\begin{equation}
  \label{eq:agn-variability}
  R < c \Delta t_{\mathrm{min}} \frac{\delta}{1 + z} \,,
\end{equation}

where $R$ is the size of the emitting region, $c$ is the speed of light, $\Delta t_{\mathrm{min}}$
is the variability time scale, $\delta$ is the relativistic Doppler factor and $z$ is the red shift
of the source. For AGNs with variability $\Delta t_{\mathrm{min}} \sim \SI{1}{\day}$ this results in
$R \sim \SI{e-3}{\parsec}$. In fact the limit on $R$ due to the variability time scale, is one of
the strongest arguments for SMBH jets as emission regions.

Figure~\ref{fig:agn-sed} shows the spectral energy density (SED) of Markarian
421~\cite{FermiLAT_Mrk421_2011}, a BL Lac blazar in the constellation Ursa Major. The spectrum shows
a double peak structure, which is typical for AGN spectra~\cite{Konopelko_2003}. In leptonic models
synchrotron emission from electrons and positrons is responsible for the observed intensities from
the radio band all the way to X-ray energies. The position of the synchrotron peak is an important
observable in the characterization of blazars. The second peak, in the $\gamma$-ray energy range, is
assumed to be due to inverse Compton scattering. However, hadronic emission models in which protons
in the jet produce pions, were also proposed to explain the emission.

\section{Modeling the Gamma-Ray Sky}
\label{sec:modeling}

\subsection{Diffuse Gamma Ray Emission}
\label{sec:modeling-diffuse}

The 3D distribution of gas (in its various forms) in the Milky Way is only one of the required
components in the computation of gamma ray maps. Other required ingredients are the cross sections
for the various production channels and the cosmic ray fluxes. These also determine the energy
spectrum of the resulting gamma ray emission.

It is useful to subdivide the galaxy into galactocentric rings, which are commonly referred to as
galactocentric annuli. Then the total observed gamma ray flux from a given location and for a given
production channel (for example $\pi^{0}$ decay) can be calculated as follows:

\begin{displaymath}
  \Phi_{\gamma}\left(E_{\gamma},l,b\right)
  \propto \sum_i{n_{H,i}\left(l,b\right) \int{\frac{\mathrm{d}\sigma}{\mathrm{d}E_{\gamma}}
      \left(T_P,E_\gamma\right)\Phi_{P,i}\left(T_P\right)\mathrm{d}T_P}} \\
  = \sum_i{n_{H,i}\left(l, b\right) q_i\left(E_\gamma\right)} \,,
\end{displaymath}

where $\frac{\mathrm{d}\sigma}{\mathrm{d}E_\gamma}$ is the differential cross section for production
of a $\gamma$-ray with energy $E_\gamma$ for the given channel, $i$ enumerates the galactocentric
rings (galactocentric annuli), $n_{H,i}\left(l,b\right)$ is the column density of the target
material (for example {\Hone} gas) along the intersection of the $\left(l,b\right)$ line of sight
and the ring. In addition $\Phi_{P,i}$ is the projectile flux in the annulus with index $i$ and
$T_P$ is the projectile kinetic energy. Performing the integration over the kinetic energy of the
projectile yields the gamma ray emissivity $q$, which depends on the photon energy and on the
annulus number.

The gamma-ray flux prediction can thus be written as the sum over products of column densities and
emissivities. Since the column densities can be calculated from the gas maps, measuring the diffuse
gamma-ray flux enables an indirect estimation of the average cosmic ray flux of the projectile
species as a function of the galactic radius. Direct measurements of the cosmic ray flux can only be
performed at the location of the solar system (where $R \approx \SI{8.5}{\kilo\parsec}$).

\begin{figure}[t]
  \centering
  \includegraphics[width=0.9\linewidth]{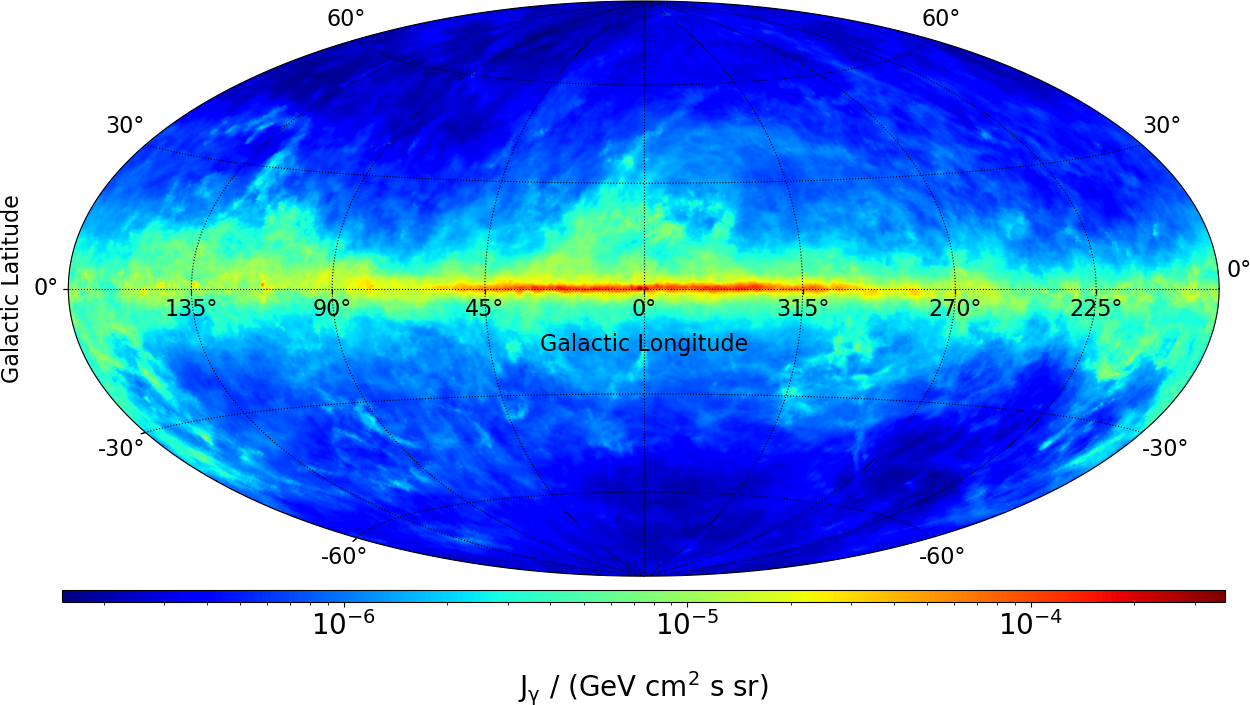}
  \caption{Gamma ray flux from $\pi^0$ decays in galactic coordinates at \SI{1}{\giga\electronvolt}
    as calculated with the GALPROP ``SA0-2D gas'' model from~\cite{Johannesson_APJ_856}. The
    projection is a Hammer-Aitoff projection. Note that galactic longitude increases to the left.}
  \label{fig:modeling-galprop-examples-pi0}
\end{figure}

Assuming that the gas density distribution, the ISRF and the cross sections are known one can
calculate the diffuse emission of photons in Milky Way propagation programs such as
GALPROP~\cite{Strong1998,Galprop_WWW}. In this method the fluxes of the various cosmic ray species
are computed by solving the propagation equations in the Milky Way. The free parameters of the
propagation model are tuned in order to reproduce various measurements of the charged cosmic rays
which were obtained at the location of the solar system.

After propagation the cosmic ray fluxes for the various galactocentric annuli and CR species are
available, which makes it possible to construct predictive gamma-ray maps, based on the measured gas
column densities and interstellar radiation fields. The methods finally yields the diffuse gamma-ray
flux $\Phi_{\gamma,\mathrm{diffuse}}\left(E_{\gamma},l,b\right)$, separately for each of the three
important production channels ($\pi^0$ decay, bremsstrahlung and inverse Compton emission).

\begin{figure}[p]
  \centering
  \includegraphics[width=0.9\linewidth]{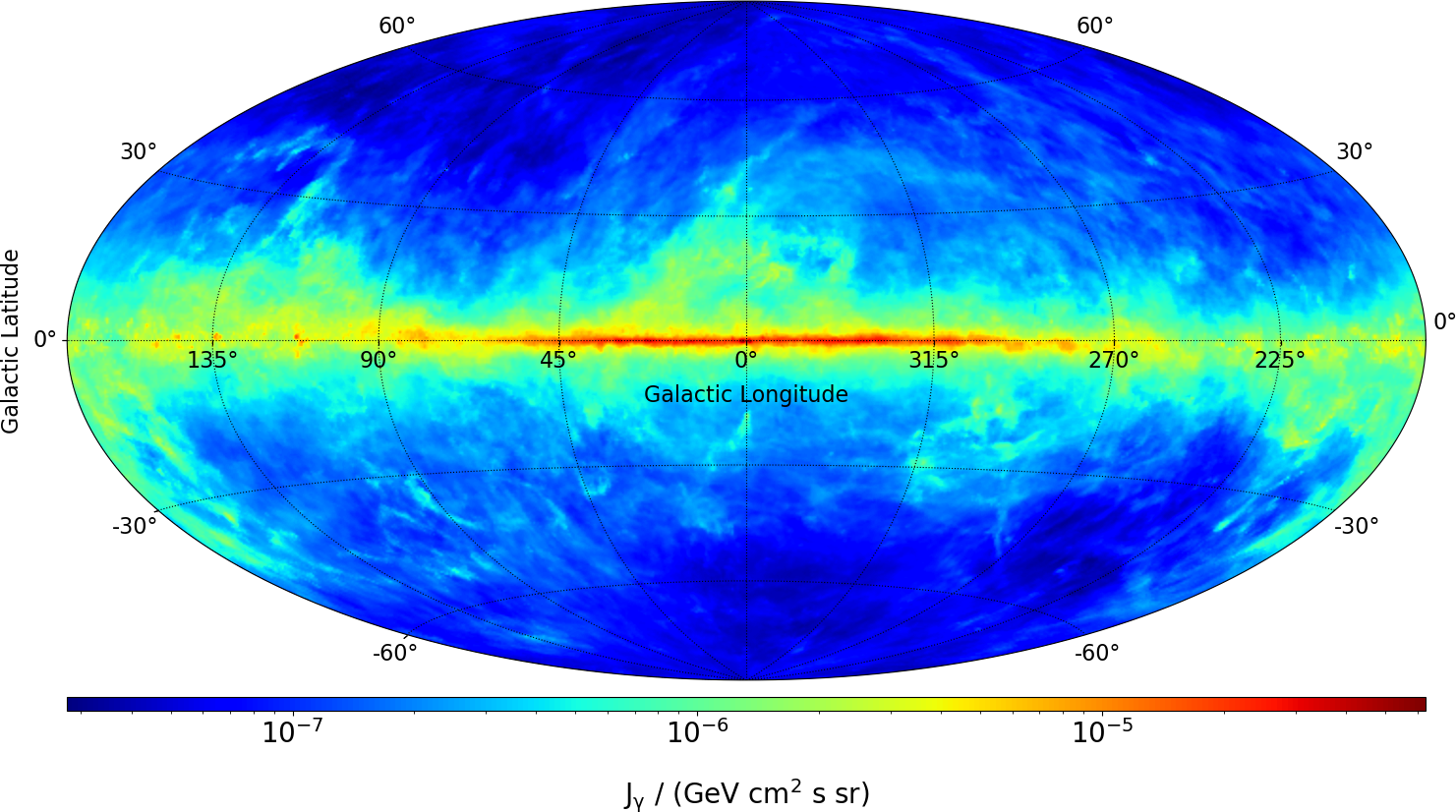}
  \caption{Gamma ray flux from bremsstrahlung in galactic coordinates at \SI{1}{\giga\electronvolt}
    as calculated with the GALPROP ``SA0-2D gas'' model from~\cite{Johannesson_APJ_856}. The
    projection is a Hammer-Aitoff projection. Note that galactic longitude increases to the left.}
  \label{fig:modeling-galprop-examples-brems}

  \vspace*{3\floatsep}

  \centering
  \includegraphics[width=0.9\linewidth]{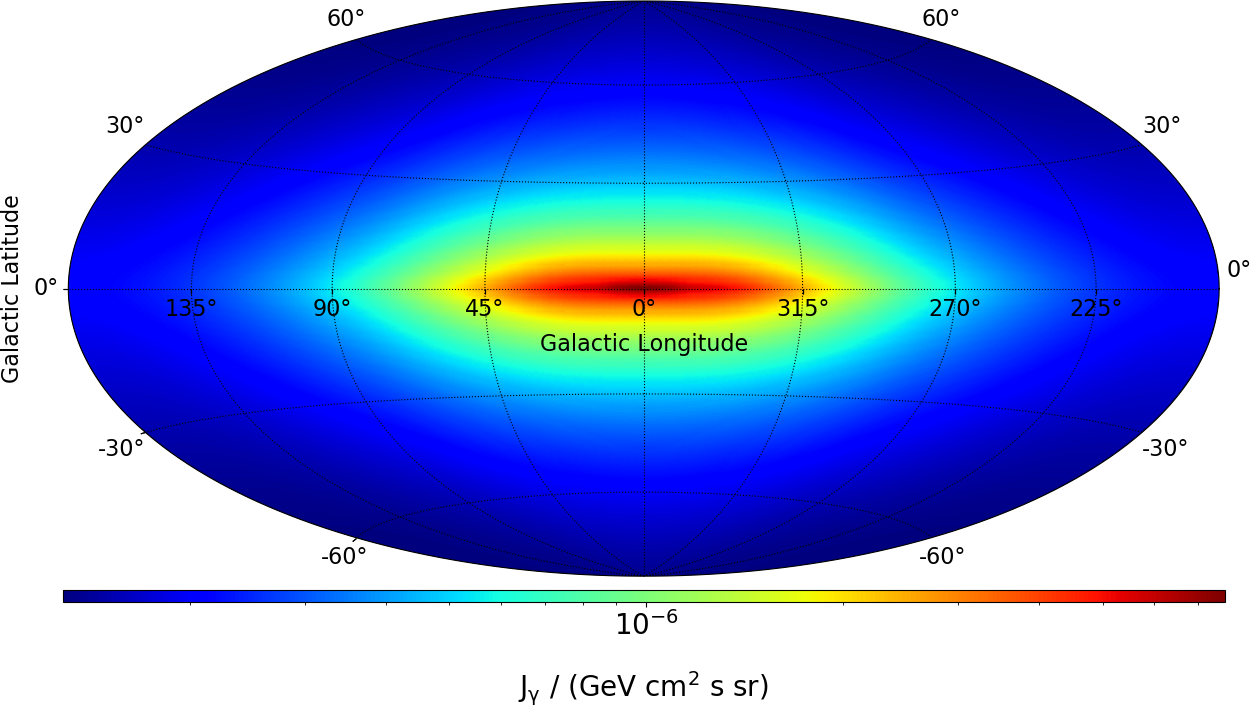}
  \caption{Gamma ray flux from inverse Compton emission in galactic coordinates at
    \SI{1}{\giga\electronvolt} as calculated with the GALPROP ``SA0-2D gas'' model
    from~\cite{Johannesson_APJ_856}. The projection is a Hammer-Aitoff projection. Note that
    galactic longitude increases to the left.}
  \label{fig:modeling-galprop-examples-ics}
\end{figure}

Figures~\ref{fig:modeling-galprop-examples-pi0} to \ref{fig:modeling-galprop-examples-ics} show
examples for gamma ray predictions from GALPROP. The model is a reference case model
from~\cite{Johannesson_APJ_856} (referred to as ``SA0-2D gas''). The predictions shown in the
figures were obtained by running the GALPROP software (version 56) with the models
from~\cite{Johannesson_APJ_856}, which are available from the GALPROP website~\cite{Galprop_WWW}.

Figure~\ref{fig:modeling-galprop-examples-pi0} shows that $\gamma$-rays from pion decays are
strongly correlated with the distribution of the interstellar gas (compare
figure~\ref{fig:lab-hone-emissivity}), which is why the $\gamma$-ray prediction is highly
structured.

The flux of photons from bremsstrahlung emission is shown in
figure~\ref{fig:modeling-galprop-examples-brems}, which also correlates with the gas structure, but
does not depend on the proton density since the $\gamma$-rays are produced in interactions of
electrons and positrons with the gas. Compared to the pion decay component the flux is lower, and a
bit more enhanced for latitudes slightly outside of the galactic plane
($\SI{5}{\degree} < |b| < \SI{10}{\degree}$) and in the third and fourth sector
($\SI{90}{\degree} < l < \SI{270}{\degree}$).

Photons from the inverse Compton process on the other hand do not show such structure as can be seen
from figure~\ref{fig:modeling-galprop-examples-ics}: The gamma rays are correlated with the
structure of the ISRF and with the (local) cosmic ray electron and positron density in the Milky Way
instead.

\begin{figure}[t]
  \begin{minipage}{0.48\linewidth}
    \centering
    \includegraphics[width=1.0\linewidth]{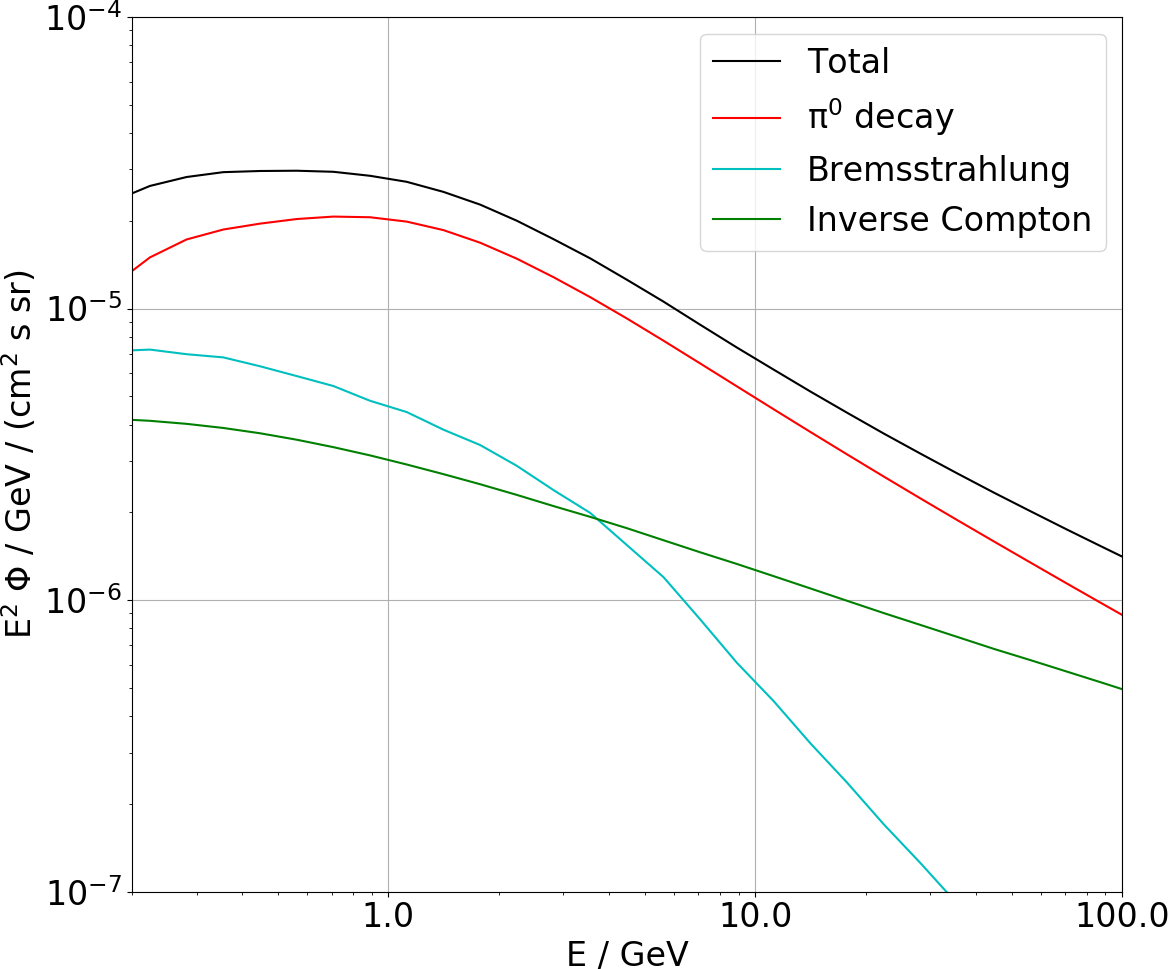}
  \end{minipage}
  \hspace{0.01\linewidth}
  \begin{minipage}{0.48\linewidth}
    \centering
    \includegraphics[width=1.0\linewidth]{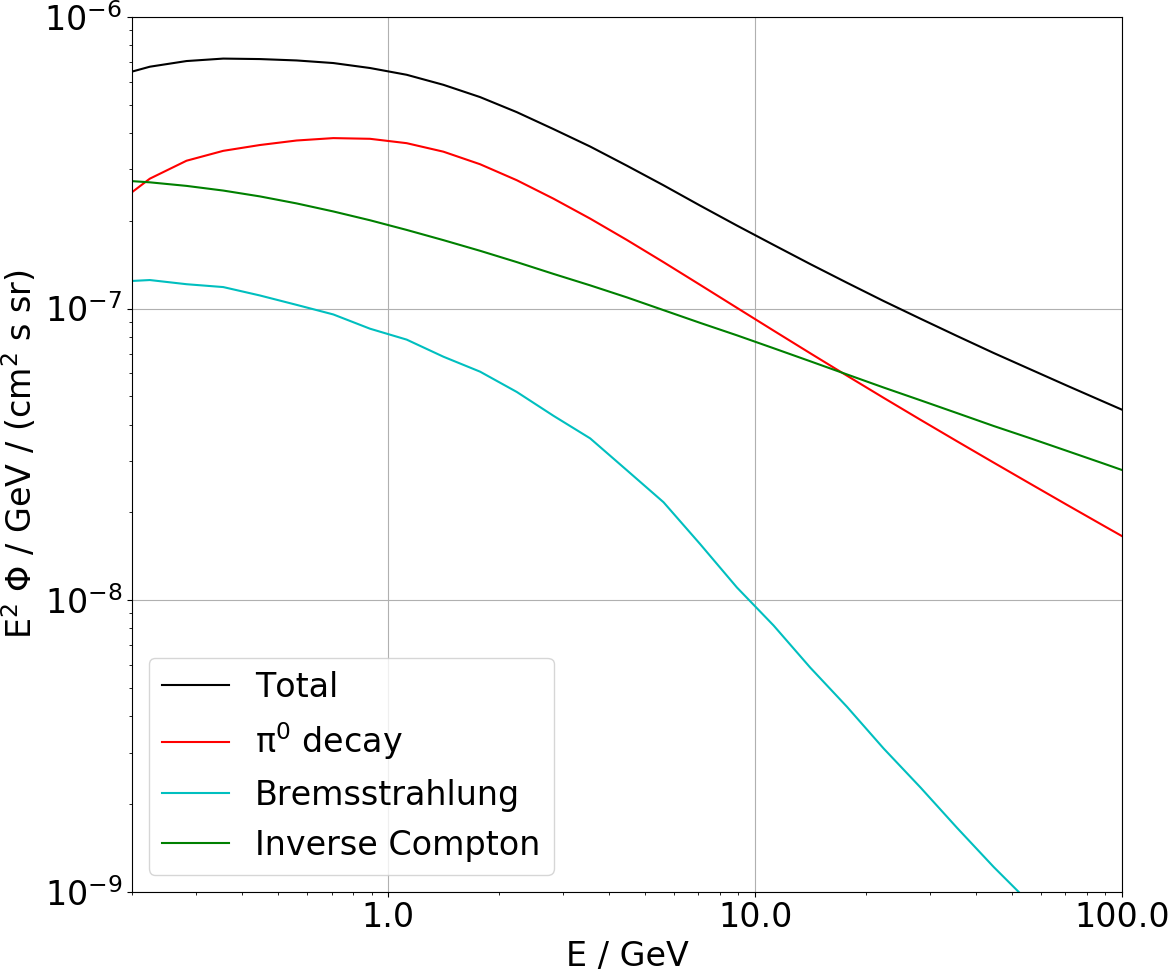}
  \end{minipage}
  \caption{The average gamma ray flux spectrum as predicted by the GALPROP reference model described
    in the text. Left: Spectrum in the inner galaxy ($|l| < \SI{80}{\degree}$ and
    $|b| < \SI{8}{\degree}$). Right: High-latitude spectrum ($b > \SI{60}{\degree}$).}
  \label{fig:modeling-galprop-flux}
\end{figure}

Figure~\ref{fig:modeling-galprop-flux} shows the flux spectrum predicted by GALPROP for the same
model for two different regions of the sky. Since protons are the most abundant cosmic ray species
the gamma ray flux from $\pi^0$ decays dominates the diffuse emission in the inner galaxy, shown on
the left. At low energies photons from bremsstrahlung emission form an important contribution. At
higher energies the bremsstrahlung component falls faster than the pion decay component, which is a
consequence of the softer spectrum of electrons (spectral index $\gamma \approx -3.2$) compared to
protons (spectral index $\gamma \approx -2.8$). The inverse Compton process becomes more and more
important at higher photon energies.

Near the galactic north pole (shown on the right) the inverse Compton process is more important
overall, because the low gas density limits the emission from pion decays and bremsstrahlung. In
both figures the pion decay component exhibits a maximum close to \SI{700}{\mega\electronvolt}. This
characteristic feature is due to the pion bump, which is dictated by the process kinematics as
discussed in section~\ref{sec:gamma-ray-production}.

Diffuse emission models which were obtained in the way described above provide a solid foundation
for analysis of experimental $\gamma$-ray spectra from sources. They contributed immensely to the
identification of regions of excess emission, such as the Fermi bubbles~\cite{Fermi_Bubbles_2010}
for example.

Modeling of the diffuse component from first principles with tools such as GALPROP makes it possible
to study the diffuse emission in a desirable way, since it is directly possible to relate the
building blocks of the galactic model to the gamma ray predictions. It also enables the construction
of a self consistent description of the entire galaxy including predictions for charged cosmic
rays. A comparison with Fermi-LAT data of such an approach was done in
2012~\cite{Fermi_Diffuse_2012}, although the Fermi-LAT data was reprocessed and its understanding
was improved since then.

Although the spatial distribution of gamma ray emission on the sky can be predicted very well by
GALPROP models, the spectral shape of the fluxes (such as the ones in
figure~\ref{fig:modeling-galprop-flux}) often disagrees with the data rather strongly, in particular
for high energies. It also turns out to be very hard to reproduce the entire set of observed cosmic
ray data in a coherent way. Finally, several large scale structures of diffuse emission have been
identified, which are not reproduced by GALPROP models. This includes the Fermi bubbles and the
Loop-I excess~\cite{Fermi_Diffuse_IEM_2016}.

Alternative methods to construct diffuse emission models are therefore needed. One such alternative
method is to leave the gamma-ray emissivities free and determine them by fitting a linear
combination of the various gas column density maps to the gamma-ray data itself. This method will
inherently produce a better fit to the data, but does not necessarily ensure self-consistency with
measurements of charged cosmic ray fluxes.

For the reasons outlined above the primary diffuse model which will be used for the analysis and
comparison with \mbox{AMS-02} data, is based on the Fermi-LAT interstellar emission model (IEM)
which has been constructed for the derivation of the fourth source catalog~\footnote{This version of
  the interstellar emission model is available through the Fermi Science Support Center as
  ``gll\_iem\_v07.fits''}~\cite{Fermi_4FGL_2019, Fermi_Diffuse_IEM_2019_WWW}. Since this model is
derived from the LAT gamma ray data itself it is more difficult to draw physical conclusions from
it. Therefore gamma ray predictions by GALPROP models continue to provide an important tool to study
the diffuse emission and will be provided for several models of the Milky Way.

\begin{figure}[p!]
  \centering
  \includegraphics[width=0.8\linewidth]{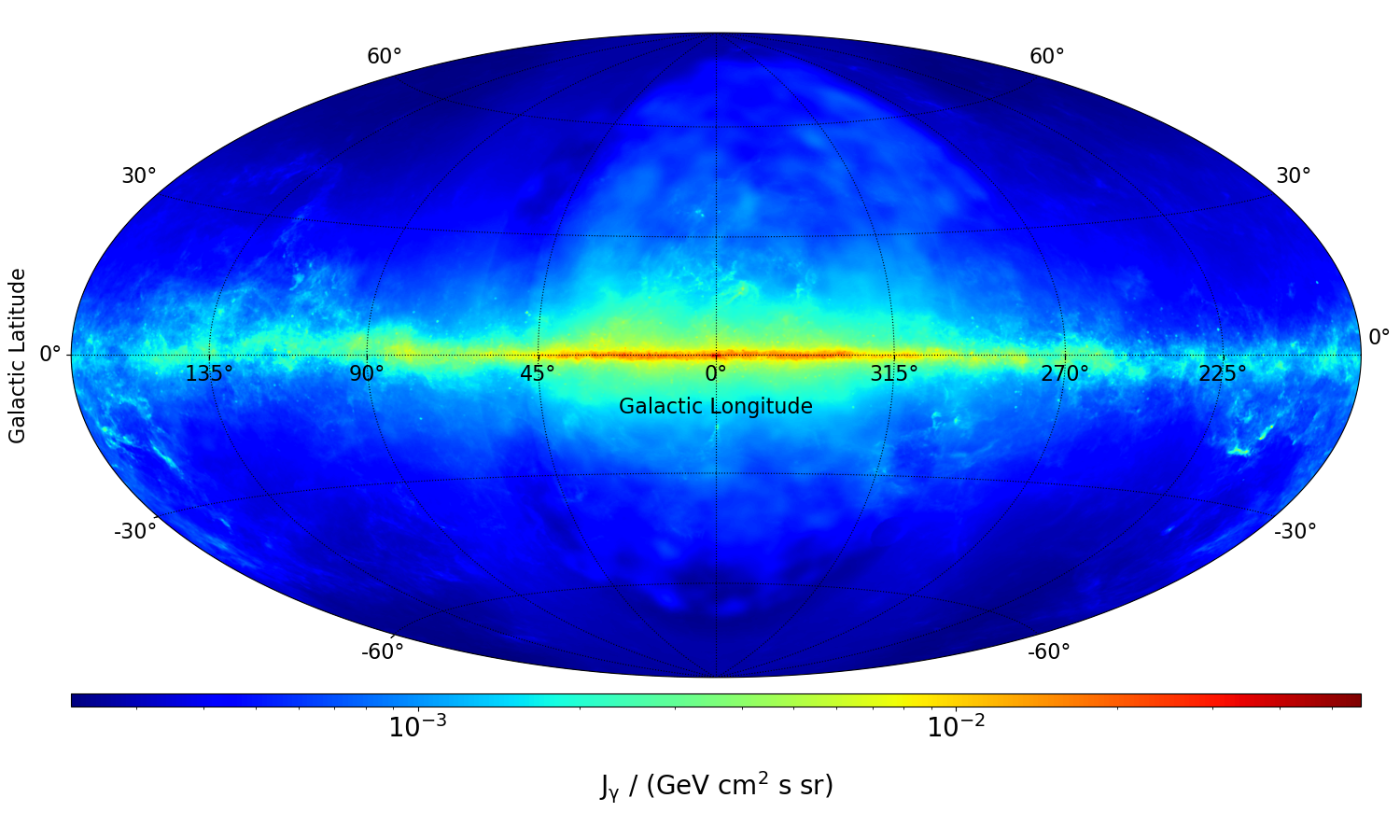}
  \caption{Fermi IEM diffuse $\gamma$-ray flux at \SI{50}{\mega\electronvolt}.}
  \label{fig:fermi-iem-50-mev}

  \vspace*{\floatsep}

  \centering
  \includegraphics[width=0.8\linewidth]{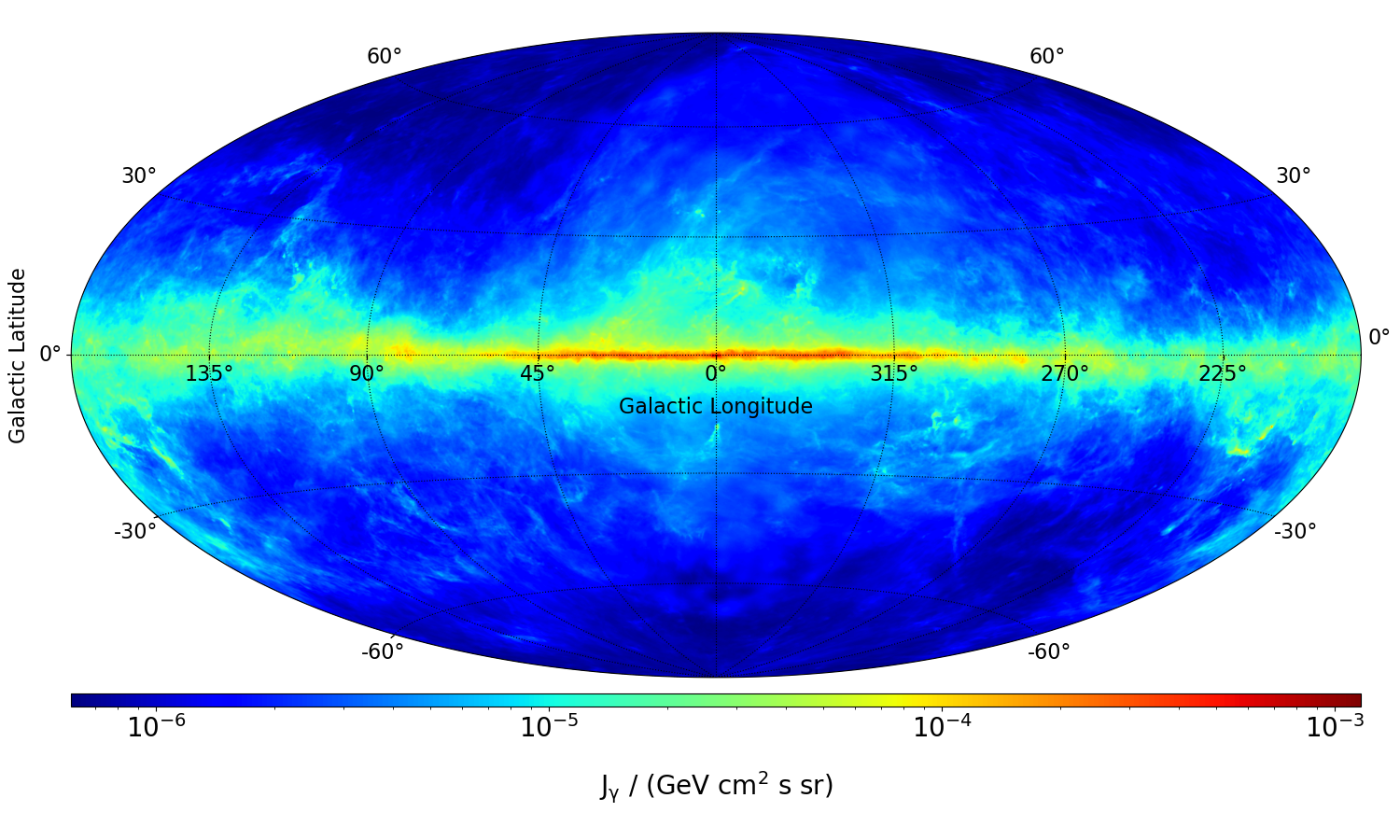}
  \caption{Fermi IEM diffuse $\gamma$-ray flux at \SI{1}{\giga\electronvolt}.}
  \label{fig:fermi-iem-1-gev}

  \vspace*{\floatsep}

  \centering
  \includegraphics[width=0.8\linewidth]{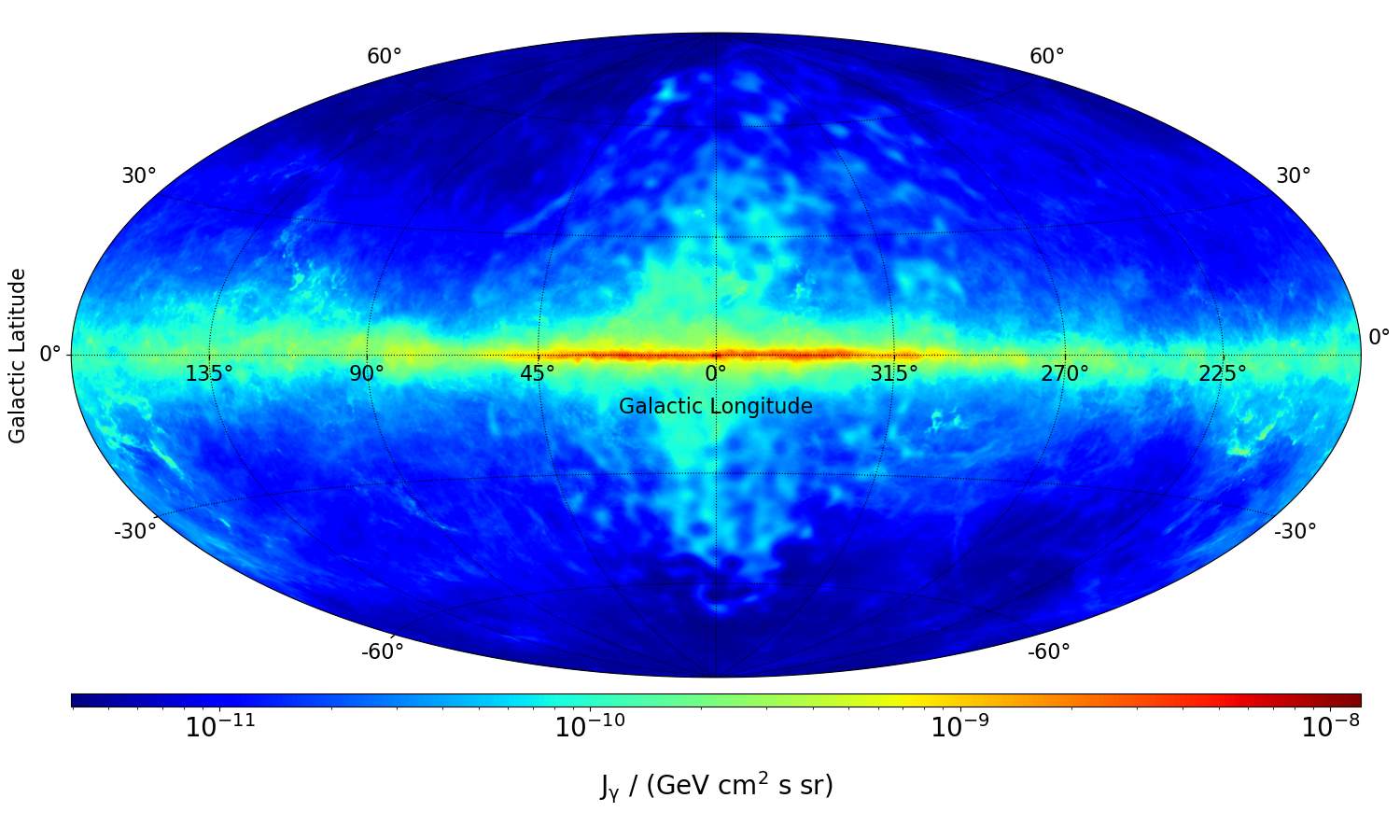}
  \caption{Fermi IEM diffuse $\gamma$-ray flux at \SI{100}{\giga\electronvolt}.}
  \label{fig:fermi-iem-100-gev}
\end{figure}

The Fermi-LAT diffuse emission model is constructed in a similar way as its predecessor, the 4 year
model for the 3FGL~\cite{Fermi_Diffuse_IEM_2016}, which already incorporates extended regions of
emission such as the Fermi Bubbles and the Loop-I excess. These are added ad-hoc without any firm
physical motivation, since the model is primarily designed as a model for gamma ray source detection
and fitting. It is also important to note that the inverse Compton emission is particularly
difficult to model and is calculated with GALPROP in the Fermi diffuse emission model. The IC
emission depends on the cosmic ray electron density in the galaxy, which in turn depends on the
distribution of the cosmic ray sources and on the structure of the (difficult to measure)
ISRF. Recent developments for the modeling of the structure of the IC component and its relation to
various regions of excess in the diffuse model are discussed
in~\cite{Porter_APJ_846,Johannesson_APJ_856}.

The 4FGL version of the Fermi IEM is valid from \SI{50}{\mega\electronvolt} to
\SI{1}{\tera\electronvolt}. This is an improvement over the 3FGL version, which was given up to
approximately \SI{513}{\giga\electronvolt} photon energy. Another major improvement is that the
effect of the \mbox{Fermi-LAT} energy dispersion was included in the fitting procedure to derive the
4FGL Fermi IEM~\cite{Fermi_Diffuse_IEM_2019_WWW}. In the prior version this was not done, which
meant that the energy spectra in the IEM had to be interpreted as functions of \mbox{Fermi-LAT}
measured energy, rather than true photon energy.

Figures~\ref{fig:fermi-iem-50-mev} to~\ref{fig:fermi-iem-100-gev} show the model predictions of the
Fermi IEM for three different energies (\SI{50}{\mega\electronvolt}, \SI{1}{\giga\electronvolt} and
\SI{100}{\giga\electronvolt}, from top to bottom). The strong contribution of the non-structured
inverse Compton emission to the flux at \SI{50}{\mega\electronvolt} is clearly visible in
figure~\ref{fig:fermi-iem-50-mev}. At \SI{100}{\giga\electronvolt} the emission associated with the
Fermi bubbles is particularly visible, which is a result of the hard spectrum ($\Phi \sim E^{-2}$)
of the bubbles.

The extra-galactic isotropic diffuse emission is not observable by \mbox{AMS-02} as it is very
faint~\cite{Fermi_Isotropic_2015}. Therefore it is not included in the constructed model for diffuse
emission.

\subsection{Photons from Gamma Ray Sources}
\label{sec:modeling-sources}

In order to obtain a complete model for the gamma-ray sky one also needs to incorporate gamma-ray
sources into the model. The spectrum and magnitude of the gamma-ray flux depends on the specifics of
each individual source. One way to add them to the model is to simply use a catalog of all the known
gamma-ray sources, which includes their locations and (time-averaged) fluxes. If one assumes the
sources to be point-like, the $\gamma$-ray flux they contribute is:

\begin{displaymath}
  \Phi_{\gamma,\mathrm{sources}}\left(E_{\gamma},l,b\right) =
  \sum_{i}{\Phi_{i}\left(E_{\gamma}\right) \delta\left(l-l_i\right) \delta\left(b-b_i\right)} \,,
\end{displaymath}

where $i$ enumerates the sources, with locations given by $(l_i,b_i)$ and spectra
$\Phi_{i}\left(E_{\gamma}\right)$ given in
$\si{photons\per\centi\meter\squared\per\second\per\giga\electronvolt}$. The Fermi-LAT fourth source
catalog (4FGL)~\cite{Fermi_4FGL_2019} is the most comprehensive list of gamma-ray sources and
includes 5065 objects together with their locations and spectra. Many of these objects were
successfully associated with counterparts in other parts of the electromagnetic spectrum (X-ray,
optical, ...), which allowed a determination of the type of the source. The catalog contains
Pulsars, Supernova-Remnants (SNR), Active Galactic Nuclei (AGN) and other types of objects.

The 4FGL is based on 8 years of Fermi-LAT data (collected from August 2008 to August 2016) and
supersedes the prior third catalog which was based on 4 years of data and included 3034
sources~\cite{Fermi_3FGL_2015}.

More than 3000 of the 5065 listed sources in the 4FGL are blazars (either BL Lac or FSRQ type). The
locations of these extra-galactic sources do not correlate with the galactic plane, which makes
their detection easier, since the background of diffuse emission is much lower. About 230 were
identified as pulsars, with pulsations detected in the $\gamma$-ray band.

To parameterize the flux spectra of the sources three different spectral shapes are used in the
catalog~\cite{Fermi_4FGL_2019}:

\begin{enumerate}
\item \textbf{Power Law}:

  \begin{equation}
    \label{eq:power-law}
    \Phi(E) = K \left(\frac{E}{E_0}\right)^{-\gamma} \,,
  \end{equation}

  where $K$ is the flux normalization, $E$ is the photon energy, $E_0$ is the pivot energy and
  $\gamma$ is the spectral index.

\item \textbf{Log Parabola}:

  \begin{equation}
    \label{eq:log-parabola}
    \Phi(E) = K \left(\frac{E}{E_0}\right)^{-\alpha - \beta \log{E / E_0}} \,,
  \end{equation}

  where $K$ is the flux normalization, $E$ is the photon energy, $E_0$ is the pivot energy and the
  spectral index changes with energy, based on the parameters $\alpha$, and $\beta$.

\item \textbf{Power Law with (Super) Exponential Cutoff}:

  \begin{equation}
    \label{eq:pl-exp-cutoff}
    \Phi(E) = K \left(\frac{E}{E_0}\right)^{-\gamma} e^{a \left(E_0^b - E^b\right)} \,,
  \end{equation}

  where $K$ is the flux normalization, $E$ is the photon energy, $E_0$ is the pivot energy and
  $\gamma$ is the spectral index of the power law component. The spectrum is cut off exponentially,
  regulated by the parameters $a$ and $b$.

\end{enumerate}

Power laws are used for sources whose spectra are not significantly curved, or if statistics only
allows for a crude estimation of the spectral shape. Most pulsars are parameterized by the
exponentially cutoff spectral shape. The catalog lists the spectral parameters used in the
corresponding spectrum type for each individual source. Thus, estimations of all the source spectra
are available in analytical form.

A total of 75 sources in the catalog have been resolved as spatially extended by the LAT. For
simplicity, these sources are also treated as point-like in the \mbox{AMS-02} model. Windows for
flux measurements will be chosen such that even the most extended sources will be fully
contained. In addition, in many cases limited statistics prevents resolving spatial substructure
with the \mbox{AMS-02} data.

\begin{figure}[t]
  \begin{minipage}{0.48\linewidth}
    \centering
    \includegraphics[width=1.0\linewidth]{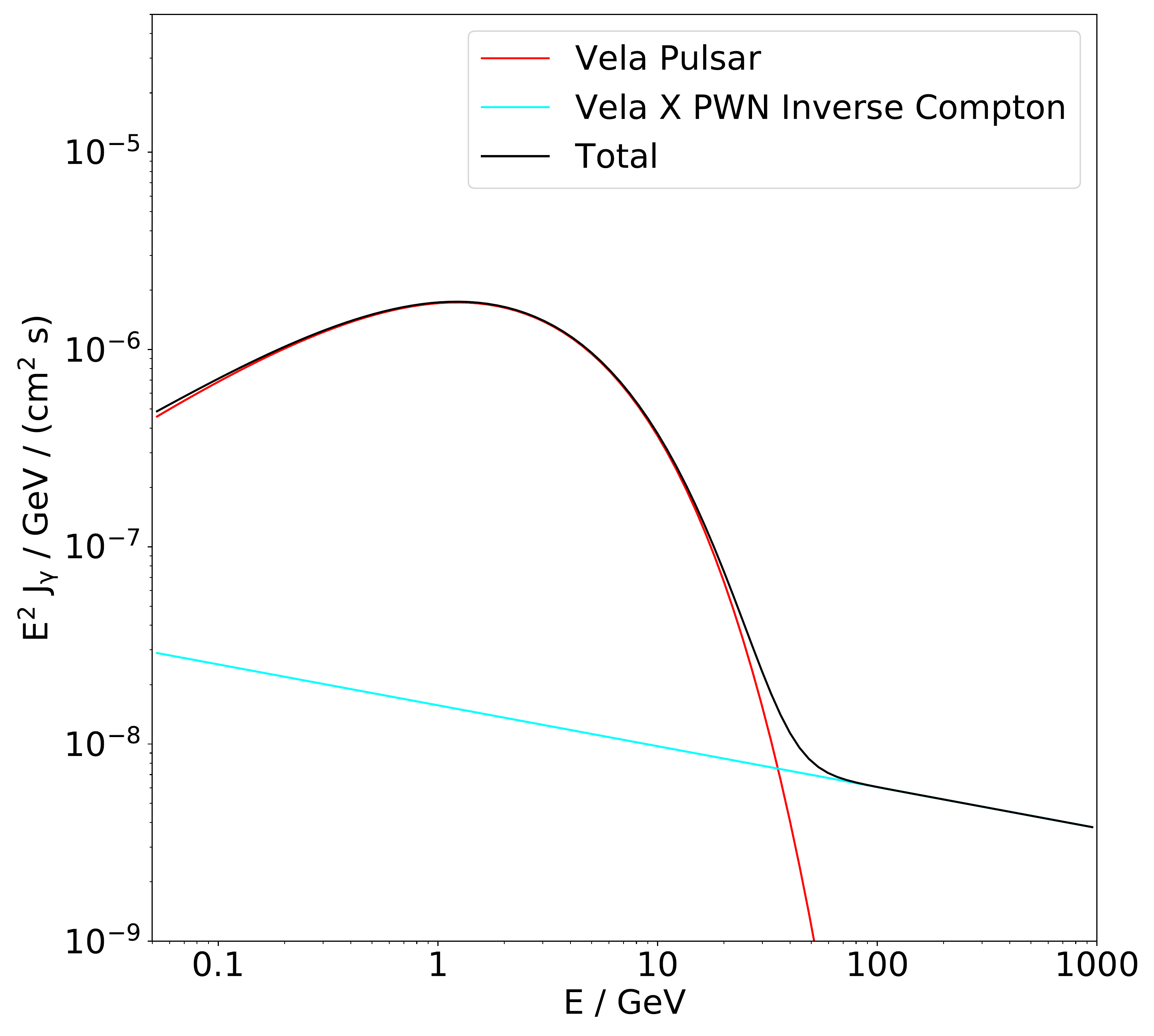}
  \end{minipage}
  \hspace{0.01\linewidth}
  \begin{minipage}{0.48\linewidth}
    \centering
    \includegraphics[width=1.0\linewidth]{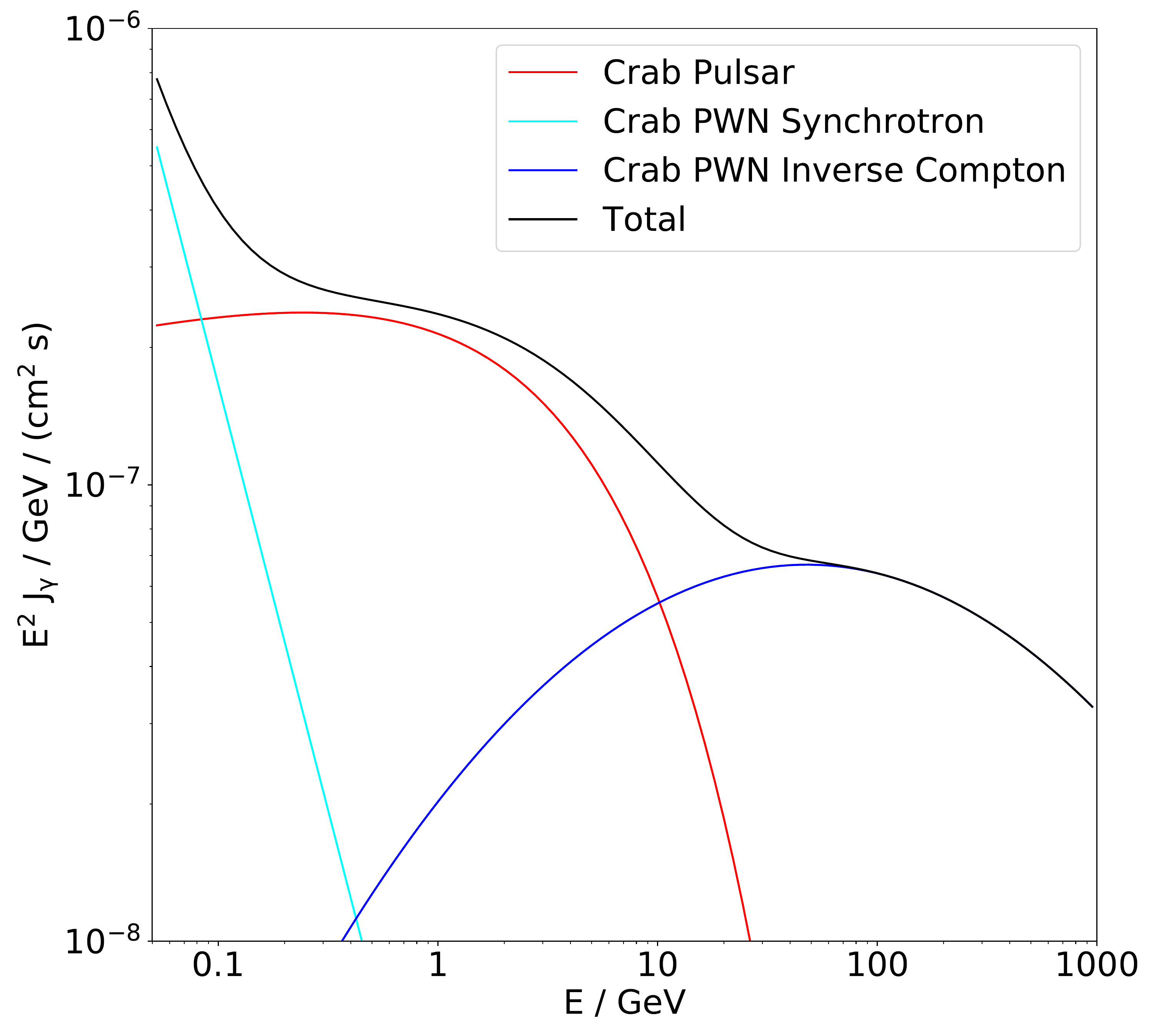}
  \end{minipage}
  \caption{The gamma ray flux of the Vela (left) and Crab (right) pulsars and PWNe, according to the
    Fermi-LAT 4FGL catalog~\cite{Fermi_4FGL_2019}.}
  \label{fig:spectra-4fgl-vela-crab}
\end{figure}

Figure~\ref{fig:spectra-4fgl-vela-crab} shows the spectra of the Vela and Crab pulsars as
examples. Both pulsar spectra are exponentially cutoff at approximately
\SI{10}{\giga\electronvolt}. The Vela X PWN inverse Compton component dominates at high energies,
although the flux is low compared to the pulsar flux (modeled with a pure power law). In the case of
the Crab PWN the IC component is sizable and significantly curved and modeled with a log parabola
spectrum. The Crab PWN also shows a steeply falling power law component at low energies which is
produced by synchrotron radiation of electrons.

\emptypage


%% file: detector.tex

\chapter{Experimental Setup}
\label{sec:experimental-setup}

\begin{figure}[t!]
  \centering
  \includegraphics[width=0.75\textwidth]{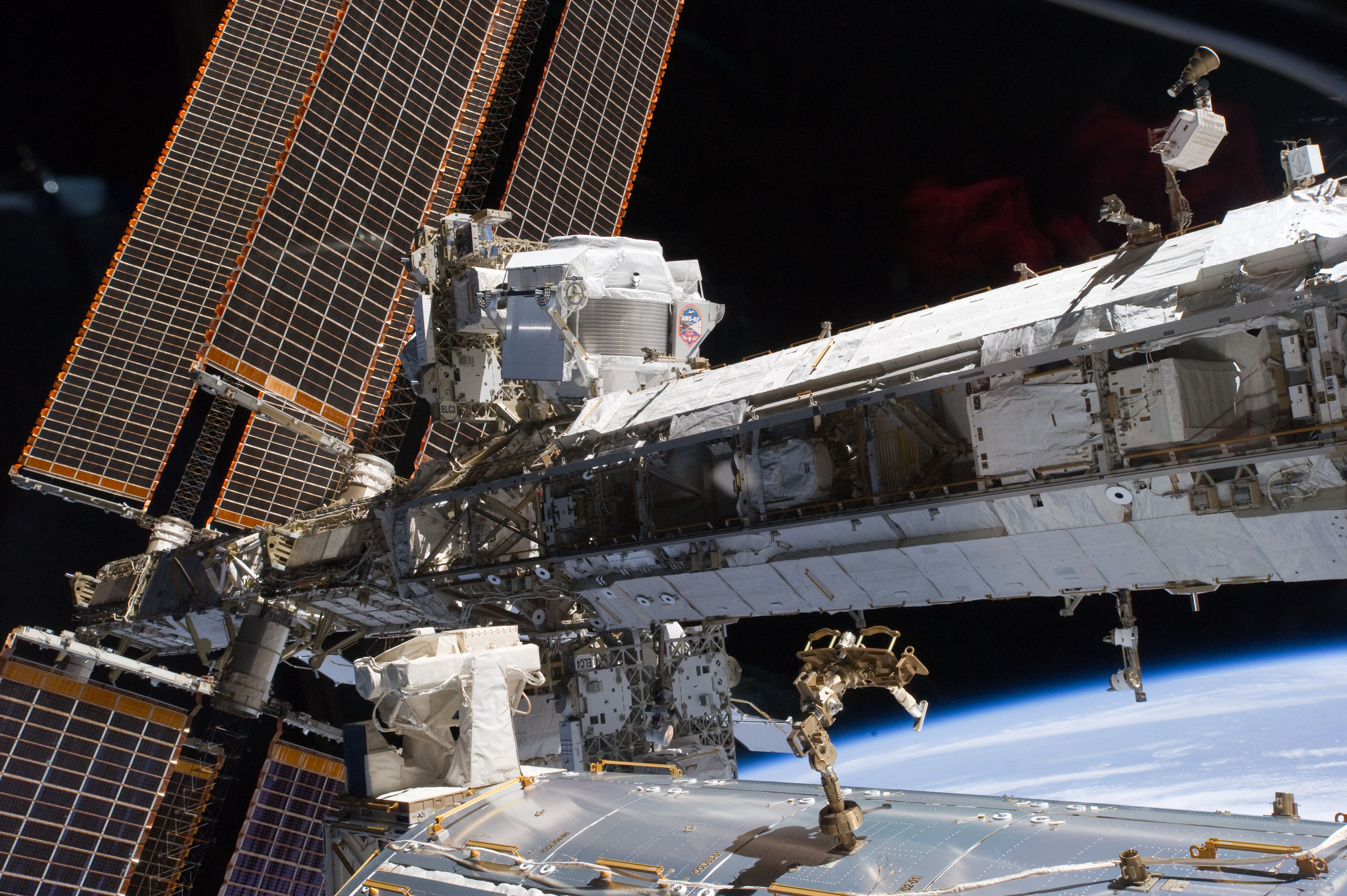}
  \caption{The \mbox{AMS-02} detector installed on the ISS~\cite{NASA_ISS_WWW}.}
  \label{fig:ams02-on-iss-installed}
\end{figure}

The data analyzed in this thesis was collected by the \mbox{AMS-02}
detector~\cite{AMS02_Detector_Kounine} which was installed as an external payload on the
International Space Station (ISS) starboard truss on May 19th 2011 and is operational sincep
(figure~\ref{fig:ams02-on-iss}). This chapter gives an overview of the detector and its various
components, with particular emphasis on the analysis of $\gamma$-rays in the
\SI{100}{\mega\electronvolt} to \SI{1}{\tera\electronvolt} energy range.

Since \mbox{AMS-02} was designed as a general purpose detector it is able to measure all charged
cosmic ray fluxes from protons ($Z=1$) to iron ($Z=26$) and beyond with excellent precision. Due to
its ability to accurately measure particle velocities it can also separate isotopes for Hydrogen,
helium, lithium and beryllium nuclei.

In addition, the detector is capable of measuring $\gamma$-rays in two complementary modes: In case
the photon converts in (or before) the first Time-of-Flight layer it is possible to reconstruct it
fully by analyzing the trajectories of the electron and positron in the tracker. A second method is
to select photons which pass through most of the detector without interacting and initiate an
electromagnetic shower in the calorimeter, which features a standalone trigger for the detection of
these events.

The photon analysis profits from the excellent charged particle detection efficiency, which provides
reliable vetos. Also, the fact that two complementary modes can be used to measure $\gamma$-rays is
a great advantage, since it allows to cross check one result with the other, which can be used to
exclude many sources of systematic uncertainties.

\section{The International Space Station}
\label{sec:iss}

\begin{figure}[t!]
  \centering
  \includegraphics[width=0.9\textwidth]{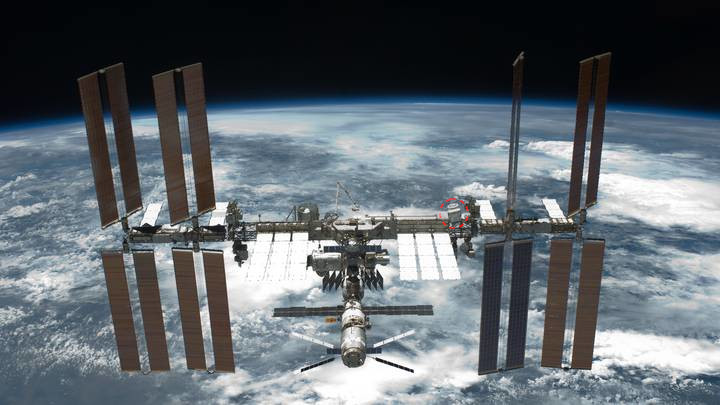}
  \caption{The International Space Station with the \mbox{AMS-02} detector located within the dashed
    red circle on the starboard side S3 truss segment~\cite{NASA_ISS_WWW}.}
  \label{fig:ams02-on-iss}
\end{figure}

The International Space Station (ISS) is a multinational laboratory in near Earth orbit which is
jointly operated by NASA, Roskosmos, ESA and JAXA. It is primarily a science laboratory, housing a
large number of experiments from various countries. Although its construction started in 1998, its
is being continuously extended with new modules even today. In its present form the ISS houses six
astronauts. The crew compartments are subdivided into several modules with cylindrical shape, such
as the US Destiny module, the ESA operated Columbus module and the Japanese JEM Kibo module.

The ISS life support system provides fresh air, clean water and adequate temperature to support
human life in the interior of the station. Debris protection systems protect the astronauts from
debris which could potentially damage the hull of the station, resulting in a loss of cabin
pressure.

Power for the station is provided by eight large solar panels, four on each side of the truss. These
panels are continuously reoriented with rotary joints in order to track the motion of the sun. Each
panel is about \SI{34}{\meter} long and \SI{11}{\meter} wide. The total surface area of the solar
arrays is approximately \SI{2500}{\square\meter}. The solar arrays generate between 84 and
\SI{120}{\kilo\watt} of power on average. Batteries are employed to store the energy when the ISS is
not exposed to direct sunlight. The system allows for sufficient power to operate even large
scientific payloads such as \mbox{AMS-02}.

Communication with the station is possible through UHF and VHF radio links as well as S-band and
KU-band antennas which relay their data through the geostationary ``Tracking and Data Relay
Satellite System'' (TDRSS) to NASA ground stations in White Sands, Goddard and Guam. The available
downlink rate in the KU-band is currently limited to 600 MBit/s as of August 2019. S-band and
KU-band antennas require line of sight connection to the TDRS satellites in order to transmit data,
which is not always available. Therefore, depending on the satellite coverage the signal in the S
and KU bands is interrupted frequently. The data is eventually processed in NASA's Johnson Space
Center (JSC) in Houston, Texas, and in the Huntsville Operations Support Center (HOSC) in the
Marshall Space Flight Center (MSFC) in Huntsville, Alabama. The \mbox{AMS-02} payload specific data
is then forwarded to the \mbox{AMS-02} Payload Operations and Control Center (POCC) in Geneva,
Switzerland.

\mbox{AMS-02} was added as an external payload to the ISS in May 2011 and is installed on the
starboard side of the main truss in the S3 segment, see figure~\ref{fig:ams02-on-iss}.

\subsection{The Orbit}
\label{sec:iss-orbit}

The ISS orbit is a prograde orbit with an inclination angle of \SI{51.6}{\degree}. The inclination
angle is the angle between the orbital plane and the Earth's equatorial plane. It is prograde
because the ISS rotates around the Earth in the same direction as the Earth itself rotates, namely
in the eastern direction. Each revolution around the Earth takes approximately 92 minutes. The
altitude of the ISS is approximately \SI{400}{\kilo\metre} above the Earth's surface. With time the
altitude slowly declines due to the drag of the station in the residual atmosphere, which causes a
loss of velocity and consequently a drop in altitude. From time to time reboosts are employed which
push the station back to higher altitudes. These are performed either with the station's own
thrusters, or with the help of externally docked vehicles.

\begin{figure}[t!]
  \centering
  \includegraphics[width=0.98\linewidth]{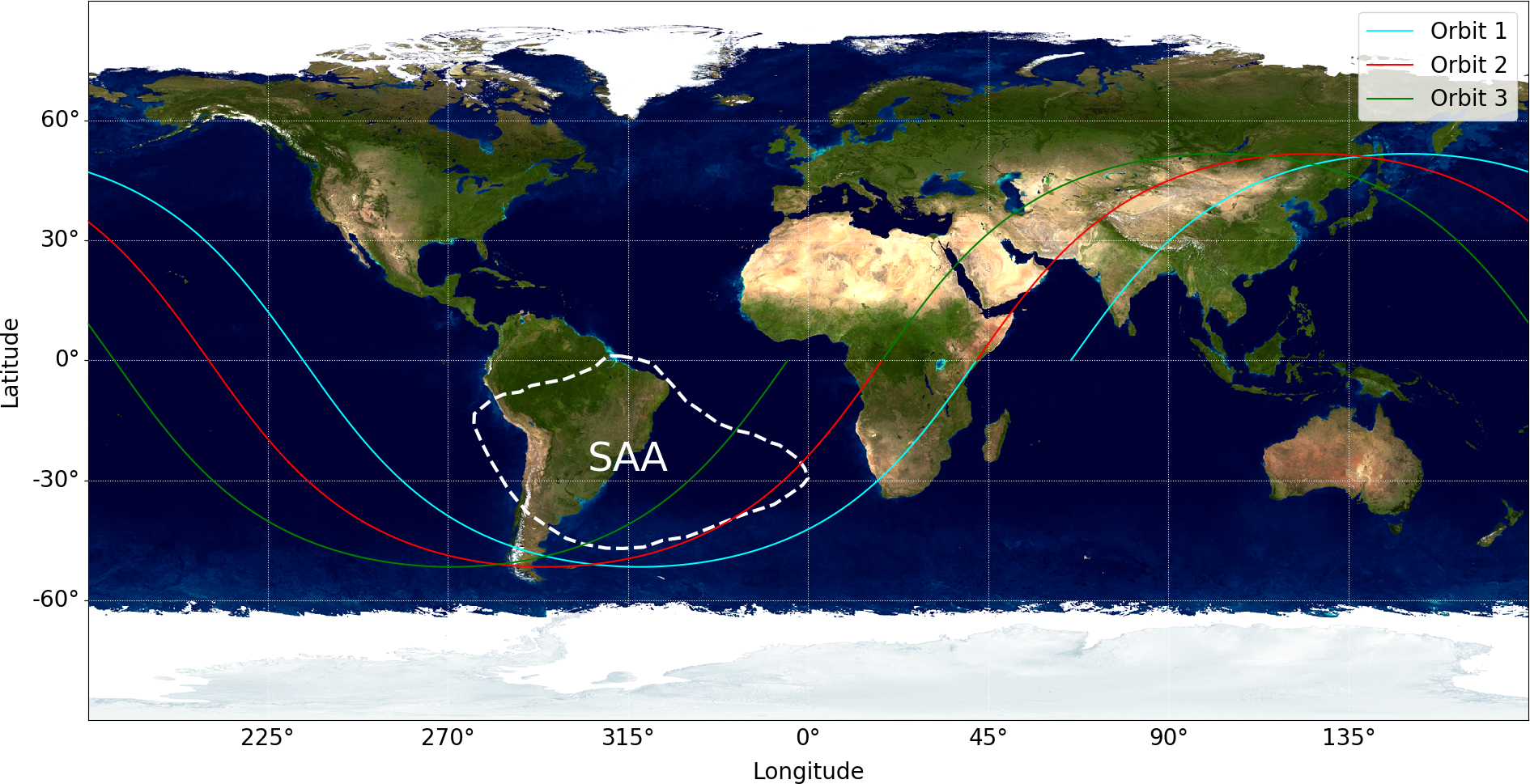}
  \caption{Three example ISS orbits projected onto the surface of the Earth. Background image
    adapted from NASA's ``Blue Marble'' series, available in the Visible Earth
    project~\cite{NASA_BlueMarble_WWW_2002}.}
  \label{fig:ams02-iss-orbit}
\end{figure}

Figure~\ref{fig:ams02-iss-orbit} shows three example orbits of the ISS as it revolves around the
Earth. Each orbit (by convention) begins when the ISS crosses the equator from south to north. Since
the inclination angle is \SI{51.6}{\degree} the maximum latitudes reached are $\SI{51.6}{\degree}$
North and South. Although the ISS moves in a nearly perfect circle around the Earth, the next
south-to-north equator crossing is located approximately \SI{23}{\degree} to the west, because the
Earth itself has rotated by the same amount to the east within the 92 minutes it took to complete
the orbit.

The centrifugal force exerted on the Earth due to its rotation gives rise to its equatorial bulge,
causing the Earth to resemble an oblate spheroid. Because of the slight change in gravitational pull
as the ISS moves, its orbital plane precesses about the Earth's rotational axis with a rate of
approximately \SI{4.5}{\degree} per day. The precession is directed towards the west for the
prograde ISS orbit and completes one full turn after approximately 80 days. It is important to note
that because of this effect the ISS zenith axis (and hence the \mbox{AMS-02} field of view) is not
limited to a fixed path on the sky, even in the default ISS attitude configuration (see
section~\ref{sec:iss-orientation}).

As the space station moves along the orbit the cosmic ray particle rate varies strongly, depending
on the position of the station in the in the Earth's magnetic field. Near the geomagnetic poles the
flux of primary cosmic rays is enhanced at low energies, because the geomagnetic cutoff rigidity is
lower. In addition, secondary cosmic rays spiral along the magnetic field lines and populate the Van
Allen belts. These secondary particles cause an additional increase of the detection rate near the
poles. These conditions are further complicated by external phenomena such as solar flares.

Due to the particular configuration of the Earth's magnetic field there is a region located over
South America and the southern Atlantic Ocean, in which the secondary particle radiation reaches
particularly high intensities at relatively low altitudes. This region is known as the ``South
Atlantic Anomaly'' (SAA) and marked with a gray dashed contour in figure~\ref{fig:ams02-iss-orbit},
although the boundary is not sharp and depends on the ISS altitude. When the ISS passes through the
SAA the rate of secondary particles reaches very high levels, which can damage electronic components
by ionizing radiation. In addition the enormous particle rate can cause particle detector trigger
systems to saturate.

\subsection{Orientation of the Space Station}
\label{sec:iss-orientation}

\begin{figure}[t]
  \begin{minipage}{0.48\linewidth}
    \centering
    \includegraphics[width=1.0\linewidth]{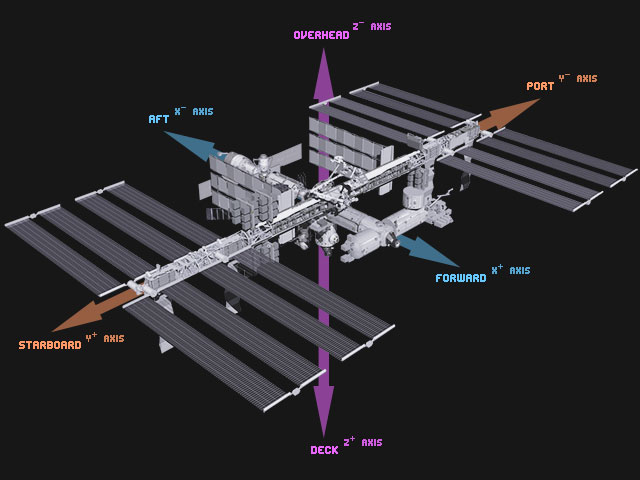}
  \end{minipage}
  \hspace{0.01\linewidth}
  \begin{minipage}{0.48\linewidth}
    \centering
    \includegraphics[width=1.0\linewidth]{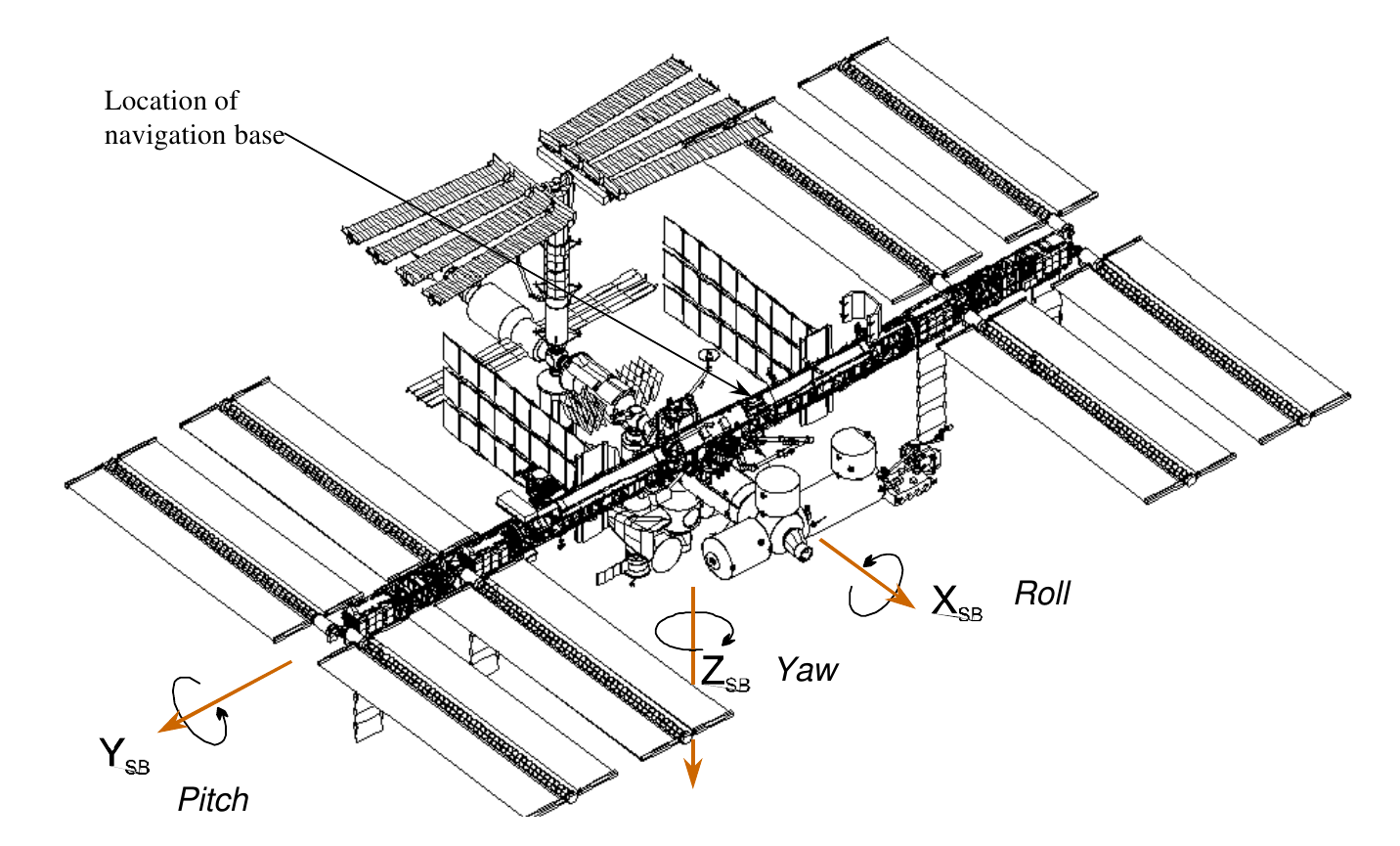}
  \end{minipage}
  \caption{Left: The ISS coordinate system~\cite{NASA_ISS_Coordinates_WWW}. Right: Definition of the
    yaw, pitch and roll Euler angles~\cite{NASA_ISS_SSRCS_WWW}.}
  \label{fig:ams02-iss-coordinate-system}
\end{figure}

In normal circumstances the ISS is oriented such that the camera in figure~\ref{fig:ams02-on-iss} is
looking along the velocity vector. This orientation is the so called ``X axis in velocity vector''
(XVV) attitude. The coordinate system of the space station is depicted on the left hand side of
figure~\ref{fig:ams02-iss-coordinate-system}. The X-axis points towards the European Columbus
module, marked with ``forward'' in the figure. This is the usual direction of flight. The positive
Y-axis points towards the starboard side, on which \mbox{AMS-02} is also installed. The Z-axis
points towards the Nadir, i.e. downwards towards the Earth.

Orientations other than the nominal XVV attitude frequently occur for short periods of time, in
particular in case of spacecraft dockings. When a spacecraft such as a Soyuz rocket approaches, the
station turns backwards, changing its orientation from +XVV to -XVV. This allows for easier docking,
since the two velocity vectors of the spacecrafts are then aligned. Thermal considerations can also
cause changes of the station's orientation.

Rotations of the station are defined in terms of the Euler yaw, pitch and roll angles as depicted on
the right hand side of figure~\ref{fig:ams02-iss-coordinate-system}.

\section{Coordinate Systems}
\label{sec:coordinate-systems}

Gamma ray arrival directions necessarily need to be transformed into one of the established
astronomical frames of reference in order to fully enable a meaningful analysis of the data. It is
therefore useful to summarize the transformations required to convert these arrival directions.

\subsection{Common Reference Systems}
\label{sec:coordinate-systems-list}

When specifying positions on the Earth, a terrestrial reference system is required. Earth centered,
Earth fixed (ECEF) frames have their origin in the Earth's center of mass, while the axes are fixed
to the Earth. It is customary to define zero degrees longitude to be the longitude of the Greenwich
Prime Meridian, and zero degrees latitude to coincide with the Earth's conventional equator. It is
worth noting that the z-axis (and correspondingly the equator) in this system does not exactly
coincide with the Earth's rotational axis, because the latter is subject to a slight ``wobbling''
effect, known as polar motion. This motion is monitored by the International Earth Rotation and
Reference Systems Service (IERS)~\cite{IERS_2004} and published as part of the Earth Orientation
Parameters (EOP) \footnote{The EOP are published in the form of bulletins. The data is also made
  available by the United States Naval Observatory and by the Observatoire de Paris. See:\\
  \url{https://www.iers.org/IERS/EN/Publications/Bulletins/bulletins.html}\\
  \url{https://maia.usno.navy.mil}\\ \url{https://hpiers.obspm.fr/eop-pc/index.php}}.

The International Terrestrial Reference System (ITRS)~\cite{IERS_Conventions_2010} is the current
standard reference frame for precision measurements on the Earth. It is realized in the
International Terrestrial Reference Frame (ITRF)~\cite{ITRF2014_2016}, which is based on the
precisely measured locations and velocities of approximately 400 points on the Earth. Another ECEF
reference system is the World Geodetic System (WGS)~\cite{WGS84_2014} established in 1984, which is
used in the Global Positioning System (GPS).

The IERS also measures and publishes the rate of rotation of the Earth as part of the EOP, which is
an important parameter required to connect the ITRS to celestial coordinates. This is done in terms
of the difference between the UTC and UT1 timescales, since UTC is based on the ticking of atomic
clocks on Earth, whereas UT1 is non-uniform and defined such that one full rotation of the Earth
always corresponds to 86400 seconds. In order to prevent the UTC timescale from drifting away from
UT1 the IERS is also in charge of inserting leap seconds into the UTC time scale.

\begin{figure}[t]
  \begin{minipage}{0.48\linewidth}
    \centering
    \includegraphics[width=1.0\linewidth]{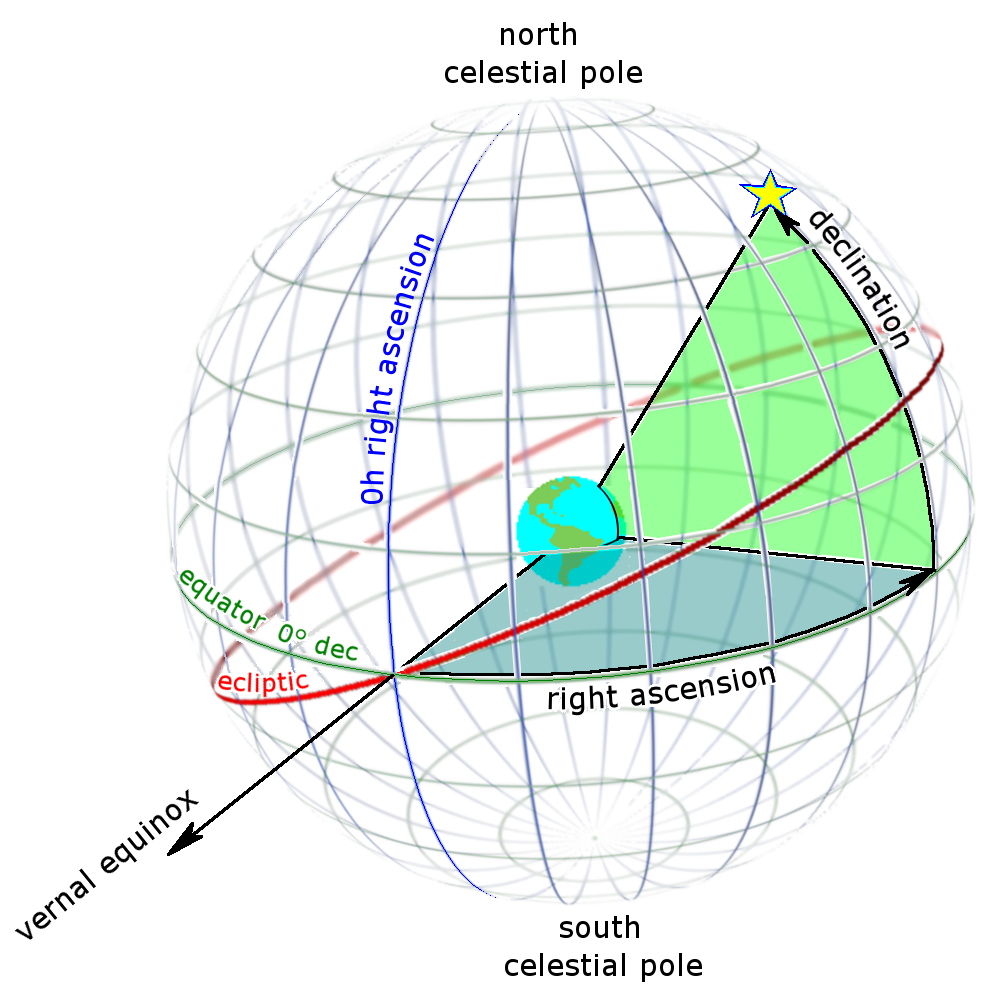}
  \end{minipage}
  \hspace{0.01\linewidth}
  \begin{minipage}{0.48\linewidth}
    \centering
    \includegraphics[width=1.0\linewidth]{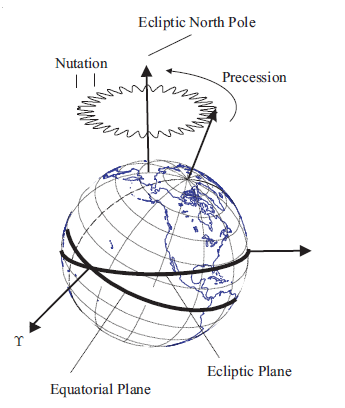}
  \end{minipage}
  \caption{Left: The equatorial coordinate system and the definition of right ascension and
    declination~\cite{RaDec_Schematic_WWW}. Right: Precession and Nutation of the Earth's
    rotational axis around the ecliptic pole axis~\cite{Precessesion_Schematic_WWW}.}
  \label{fig:coordinate-systems-sketch}
\end{figure}

The most important astronomical reference systems are equatorial coordinate systems. The left hand
side of figure~\ref{fig:coordinate-systems-sketch} shows a schematic illustrating the definition of
such a system. The z-axis is aligned with the Earth's rotational axis, the XY-plane is normal to it
and approximately coincides with the Earth's equatorial plane. The X-axis points towards the vernal
equinox, which is the point at which the ecliptic (and hence the Sun) crosses with the equatorial
plane in March. The Y-axis completes the right handed system. Positions in equatorial coordinates
are specified using right ascension ($\alpha$) and declination ($\delta$). Declination is the angle
to the equatorial plane, ranging from \SI{-90}{\degree} to \SI{90}{\degree}, while right ascension
is the azimuth angle between the given point and the vernal equinox in the equatorial plane,
increasing towards the East.

If both the Sun and the Earth were perfect spheres and there were no other bodies in the solar
system the coordinates thus defined would form an adequate inertial system for the measurement of
astronomical phenomena. However, because of the specifics of the gravitational pull of other bodies
in the solar system the Earth's equatorial plane is not fixed with respect to distant
stars. Instead, precession and nutation cause the Earth's rotational axis to constantly
move. Precession causes the Earth's rotational axis to slowly rotate around the ecliptic pole due
the oblate shape of the Earth. This effect corresponds to the large circular motion of the Earth's
pole shown on the right hand side of figure~\ref{fig:coordinate-systems-sketch}. The pole moves on a
cone with an opening angle of approximately \SI{23}{\degree} and completes one full revolution after
approximately 26000 years. This corresponds to a slow movement of the celestial pole on a smooth arc
at a rate of approximately 20 arcseconds per year~\cite{USNO_Circ179_2005}.

\begin{figure}[t]
  \centering
  \includegraphics[width=0.7\linewidth]{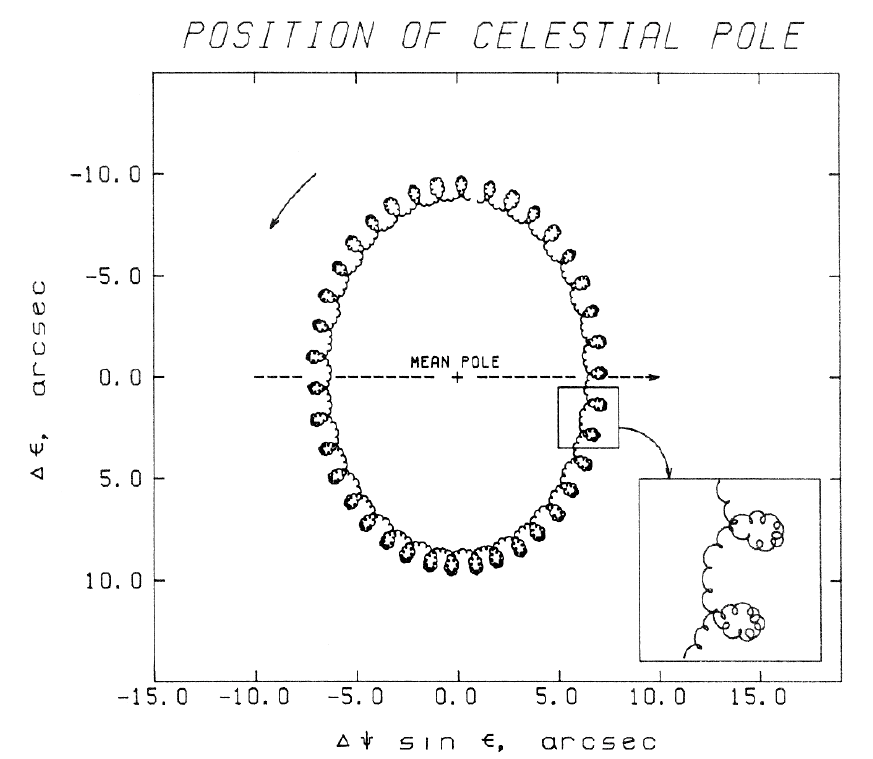}
  \caption{Effect of nutation on the position of the celestial pole with respect to the mean
    pole~\cite{USNO_Circ179_2005} (the position of the pole as predicted by precession theory only),
    over a period of 18 years. The inset shows the details of one year's motion.}
  \label{fig:nutation-sketch}
\end{figure}

In addition, gravitational pull from the Moon and other bodies in the solar system cause the Earth's
polar axis to nutate around the mean path given by precession. This effect is responsible for the
small periodic wobbling of the axis position in figure~\ref{fig:coordinate-systems-sketch}, which
approximately repeats every 18 years. A more detailed illustration of the effect of nutation on the
path of the celestial pole is given in figure~\ref{fig:nutation-sketch}.

Because of these effects and the resulting changes of the orientation of the Earth's equatorial
plane with respect to distant stars it is necessary to specify the instant at which the equatorial
coordinate system is to be defined. Observations at other times can then be converted, by using an
appropriate model for precession and nutation. This point in time is the epoch and is chosen by
convention among astronomers such that they can compare their results. The current standard epoch as
defined by the International Astronomical Union (IAU) is J2000~\cite{IAU_Resolution_1976_1}, which
is the Julian date 2451545.0 TT, the 1st of January 2000 12:00:00 TT (Terrestrial Time),
corresponding to the 1st of January 2000 11:58:55.816 UTC. Before 1984 the standard epoch was B1950
corresponding to the beginning of the Besselian year 1950, which is the 31st of December 1949 22:09
UT.

A coordinate system defined by the actual rotational axis at the epoch $t$ as its z-axis, with the
x-axis pointing towards the intersection of the true equatorial plane at that time with the ecliptic
is known as a ``True of Date'' (TOD) coordinate system for the epoch $t$. Instead, if one does not
consider the effect of nutation, ``Mean of Date'' (MOD) systems are obtained. MOD systems at
different epochs are connected by precession. In a second step a MOD system for the epoch $t$ can be
connected to the TOD system for the same epoch by accounting for nutation.

Although equatorial coordinates are well defined after specifying the epoch, they are a theoretical
concept, since it is impossible to paint the Earth's equatorial plane or the position of the vernal
equinox on the sky. In practice it is necessary to measure and catalogue the positions, proper
motions and parallaxes of stars, in order to form a coordinate frame which realizes the theoretical
concept. Astronomers can then use those stars to orient themselves.

The most important catalogues in the past were the Fourth and Fifth Fundamental Catalogue (FK4 and
FK5)~\cite{FK4_1963,FK5_1988}, which contain the accurate positions of 1535 fundamental stars. The
FK4 positions were specified with respect to the B1950 epoch, whereas the FK5 catalogue is expressed
in terms of the J2000 epoch. The corresponding coordinate systems are also referred to as FK4 and
FK5. Nowadays these catalogues are superseded by the Hipparcos~\cite{Hipparcos_1997} and FK6
catalogues~\cite{FK6_1999}, but the FK4 and FK5 coordinate systems remain relevant.

At the present the most important coordinate system is the ICRS~\cite{ICRS_1995}, which was
theoretically established by a series of specifications published by the IAU between 1997 and
2006~\cite{USNO_Circ179_2005}. Its origin lies in the solar system barycenter and its axes are not
defined by the kinematics of the Earth. Instead they are fixed with respect to distant
stars. However, as a matter of convenience, the actual orientation of the axes was chosen to
coincide almost perfectly with FK5, making the two coordinate systems almost identical for many
practical applications.

The ICRS was first realized by observations of a set of 608 extragalactic radio sources using Very
Long Baseline Interferometry (VLBI)~\cite{Ma_ICRF_1998}, which allowed to improve the precision with
respect to the FK5 by several orders of magnitude. In the optical band the primary realization of
the ICRS is the Hipparcos reference frame~\cite{IAU_Resolution_1997_B2}. Mignard and Froeschle
studied the differences between the ICRS (as realized by Hipparcos) and the FK5 and found the latter
to be non-inertial on the \SI{0.5}{\milli\arcsecondtext\per\year} level~\cite{Mignard_1998}. They
also determined the global rotation required to convert coordinates between the two frames, although
local differences are as large as \SI{150}{\milli\arcsecondtext} and cannot be overcome with a
global rotation.

\begin{figure}[t]
  \centering
  \includegraphics[width=0.7\linewidth]{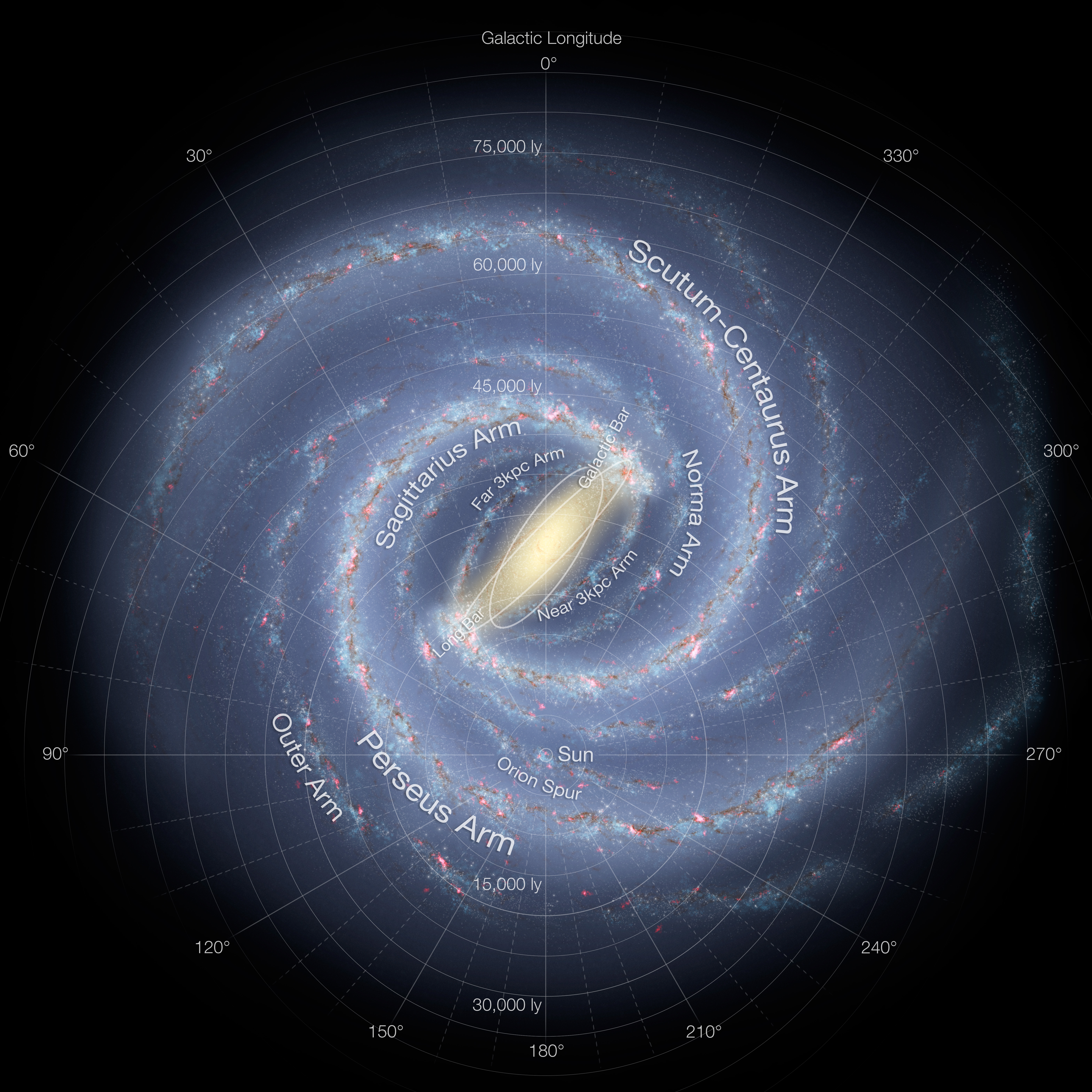}
  \caption{Artist's impression of the Milky Way, together with two dimensional view of galactic
    coordinates in the galactic plane~\cite{GalacticCoordinates_WWW}.}
  \label{fig:galactic-coordinate-systems}
\end{figure}

Another important coordinate system is the galactic coordinate system, first used by William
Herschel in 1785. Figure~\ref{fig:galactic-coordinate-systems} shows an Artist's impression of the
plane of the Milky Way together with orientation of the galactic coordinate system axes. The origin
of galactic coordinates is the sun. The xy-plane coincides with the plane of the Milky Way and the
x-axis points towards the Galactic Center (GC). The z-axis points towards the Northern Galactic Pole
(NGP). Positions in galactic coordinates are measured using galactic longitude ($l$), the azimuth
angle in the galactic plane, and galactic latitude ($b$), the angle with respect to the plane.

An exact definition of galactic coordinates was provided by the IAU in
1958~\cite{IAU_Galactic_1958}, based on the measurements of neutral hydrogen gas, by specifying the
FK4 B1950 positions of the NGP (defining the galactic plane) and the ascending node (defining
longitude zero) to:

\begin{align}
  \alpha_{\mathrm{NGP}} &= \hms{12;49;} = \ang{192;15;} = \ang{192.25}\\
  \delta_{\mathrm{NGP}} &= \ang{27;24;} = \ang{27.4}\\
  l_{\mathrm{NCP}} &= \ang{123;;} \,.
  \label{eq:galactic-coordinates-fk4}
\end{align}

Unfortunately it is not possible to rigorously transform these definitions to the FK5 or ICRS
coordinate system~\cite{Murray_1989,Hipparcos_1997}, which has led some authors to suggest a
redefinition of galactic coordinates based on ICRS coordinates~\cite{Liu_Galactic_2011}. This
suggestion is supported by precise measurements of the radio source Sagitarrius A*~\cite{Reid_2004},
the best physical marker for the center of the Milky Way, which have shown that the position of Sgr
A* deviates by \ang{0.07} from the center of the IAU 1958 galactic coordinates. In addition,
galactic coordinates are rotating, because they were defined based on FK4, which is now shown to be
non-inertial. Reid and Brunthaler also give locations of the NGP and the zero longitude position for
the J2000 epoch, which allow conversion from FK5 to Galactic coordinates.

In this thesis the approach chosen by the Astropy project~\cite{Astropy_2018} is followed, where the
IAU 1958 definitions are translated to FK5 J2000 coordinates (neglecting the FK4 E-terms of
aberration~\cite{Murray_1989}) and the longitude of the ascending node was found by optimizing for
self consistency, such that a circular chain of transformations cancels. In this approach the FK5
coordinates of the northern galactic pole and the galactic longitude of the celestial pole are:

\begin{align}
  \label{eq:galactic-coordinates-fk5-1}
  \alpha_{\mathrm{NGP},\mathrm{J2000}} &= \ang{192.8594812065348}\\
  \label{eq:galactic-coordinates-fk5-2}
  \delta_{\mathrm{NGP},\mathrm{J2000}} &= \ang{27.12825118085622}\\
  \label{eq:galactic-coordinates-fk5-3}
  l_{\mathrm{NCP},\mathrm{J2000}} &= \ang{122.9319185680026} \,.
\end{align}

\subsection{Coordinate Transformations}
\label{sec:coordinate-system-transformations}

\begin{figure}[p]
  \centering
  \includegraphics[width=1.0\linewidth]{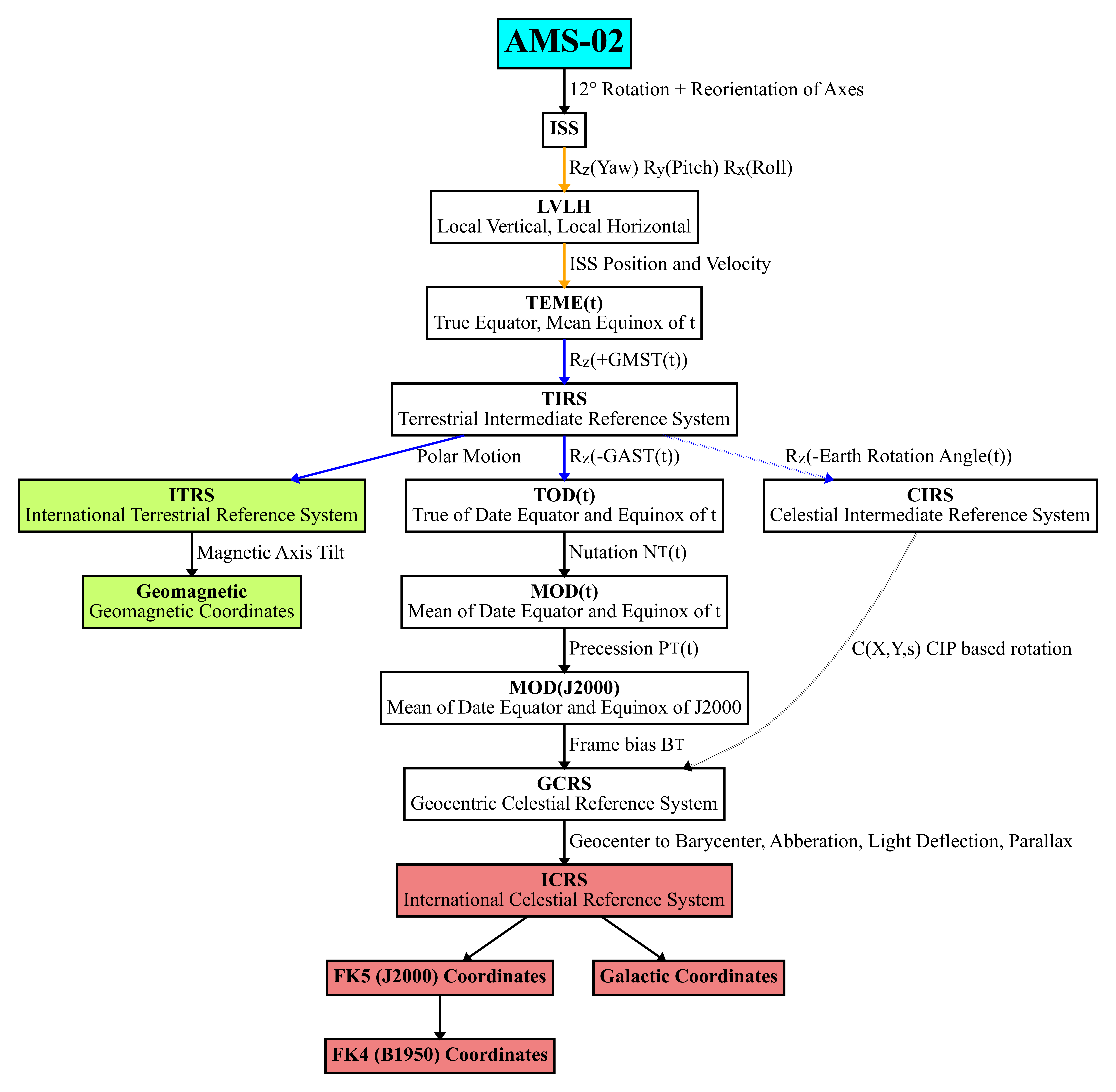}
  \caption{Schematic overview of \mbox{AMS-02} (cyan), terrestrial (green), astronomical (red) and
    intermediate coordinate systems together with associated transformations. Transformations shown
    in orange require ISS position, velocity and attitude information. Operations shown in blue
    require IERS Earth Orientation Parameter information.}
  \label{fig:coordinate-system-transformation}
\end{figure}


In this section the concrete chain of transformations used to convert an arrival direction from
\mbox{AMS-02} internal coordinates to either ICRS (equatorial) or galactic coordinates will be
discussed. Figure~\ref{fig:coordinate-system-transformation} illustrates these transformations
together with all the intermediate coordinate systems. In the following the conventions for rotation
matrices are:

\begin{align*}
  R_{x}(\varphi) &=
  \begin{pmatrix}
    1              & 0              & 0 \\
    0              & \cos{\varphi}  & -\sin{\varphi} \\
    0              & \sin{\varphi}  & \cos{\varphi}
  \end{pmatrix},\\ \\
  R_{y}(\varphi)   &=
  \begin{pmatrix}
    \cos{\varphi}  & 0              & \sin{\varphi} \\
    0              & 1              & 0 \\
    -\sin{\varphi} & 0              & \cos{\varphi}
  \end{pmatrix},\\ \\
  R_{z}(\varphi)   &=
  \begin{pmatrix}
    \cos{\varphi}  & -\sin{\varphi} & 0 \\
    \sin{\varphi}  & \cos{\varphi}  & 0 \\
    0              & 0              & 1
  \end{pmatrix} \,.
\end{align*}

\textbf{1. AMS-02 $\rightarrow$ ISS:} In a first step the direction vector needs to be converted
from the \mbox{AMS-02} internal coordinates to ISS coordinates. This step includes a rotation of
$\delta = \SI{12}{\degree}$ towards the starboard side of the ISS, to account for the fact that the
\mbox{AMS-02} zenith is rotated with respect to the ISS zenith. In a second step the axes are
reoriented such that the new y-axis points along the truss towards the starboard side, the x-axis
points towards the ram side and the z-axis points towards the nadir:

\begin{equation}
  \label{eq:coord-trans-ams-iss}
  \vec{d}_{\mathrm{ISS}} = R_x(\pi) R_z(\frac{\pi}{2}) R_y(\delta)
  \enspace \vec{d}_{\mathrm{AMS}} \,.
\end{equation}

The center of the coordinate system is also shifted from the \mbox{AMS-02} center to the ISS center,
but it is important to note that this shift is omitted for gamma ray direction vectors, which are
assumed to point to a position at infinite distance.

\textbf{2. ISS $\rightarrow$ LVLH:} In the next step the coordinates associated with the ISS body
axes need to be converted to the LVLH (Local Vertical / Local Horizontal) reference frame. For a
given set of yaw ($\alpha$), pitch ($\beta$) and roll ($\gamma$) angles the rotation to go from the
ISS coordinate system to the LVLH reference frame is:

\begin{equation}
  \label{eq:coord-trans-iss-lvlh}
  \vec{d}_{\mathrm{LVLH}} = R_z(\alpha) R_y(\beta) R_x(\gamma)
  \enspace \vec{d}_{\mathrm{ISS}} \,.
\end{equation}

The LVLH reference frame is defined by the current ISS orbital plane. The z-axis points towards the
center of the Earth, the x-axis points in the direction of the projection of the ISS velocity vector
onto the plane normal to the z-axis. The y-axis completes the right handed system. Tabulated values
for the yaw, pitch and roll angles are provided by the Aerospace Logistics Technology Engineering
Company (ALTEC), which is owned by the Italian Space Agency. When the ISS is in the normal XVV
attitude, the typical values of yaw, pitch and roll are only of the order of a few degrees. In that
case the local ISS coordinate system almost coincides with the LVLH reference frame.

\textbf{3. LVLH $\rightarrow$ TEME:} To go from LVLH to any Earth-centered frame of reference
requires knowledge of the station's position and velocity. These are derived from Two-Line Element
sets (TLEs) which are routinely published by Celestrak~\cite{Vallado_2006} or the United States Air
Force Space Track website. The TLE format was defined by the North American Aerospace Defense
Command (NORAD), which monitors the positions and velocities of satellites and debris in Near Earth
Orbit. TLE files specify the Keplerian orbit parameters of satellites in the TEME (True Equator,
Mean Equinox), reference frame. In this coordinate system the z-axis coincides with the true
rotational axis of the Earth at the given time, but the x-axis points towards the mean position of
the vernal equinox. Calculation of the position and velocity vectors based on TLEs are performed by
the Simplified perturbation model SGP4~\cite{STR3_1980}, as implemented in the PREDICT satellite
tracking software~\cite{PREDICT_2011}.

The position ($\vec{r}$) and velocity ($\vec{v}$) of the ISS in the TEME frame are connected to the
axes ($\vec{x}, \vec{y}, \vec{z}$) of the LVLH frame as follows:

\begin{align}
  \label{eq:lvlh-teme-x}
  \vec{x} &= \frac{\vec{r} \times (\vec{v} \times \vec{r})}{|\vec{r} \times (\vec{v} \times \vec{r})|} \\
  \label{eq:lvlh-teme-y}
  \vec{y} &= \frac{\vec{v} \times \vec{r}}{|\vec{v} \times \vec{r}|} \\
  \label{eq:lvlh-teme-z}
  \vec{z} &= \frac{-\vec{r}}{|\vec{r}|} \,.
\end{align}

This defines the matrix which carries vectors in the LVLH frame to the TEME frame of reference:

\begin{displaymath}
  \vec{d}_{\mathrm{TEME}} =
  \begin{pmatrix}
    x_1 & y_1 & z_1 \\
    x_2 & y_2 & z_2 \\
    x_3 & y_3 & z_3
  \end{pmatrix}
  \enspace \vec{d}_{\mathrm{LVLH}} \,,
\end{displaymath}

with the components of the vectors $\vec{x}$, $\vec{y}$ and $\vec{z}$ given by
equations~(\ref{eq:lvlh-teme-x}) to~(\ref{eq:lvlh-teme-z}). This rotation would be again followed by
a change of the center of the coordinate system from the ISS center to the Earth's center of
gravity, but it is unnecessary for astronomical distances.

\textbf{4. TEME $\rightarrow$ TIRS:} In the next step TEME coordinates must be converted into
terrestrial intermediate coordinates. In order to go from celestial to terrestrial coordinates it is
required to account for the rotation of the Earth around its rotational axis. The time variation of
the rotation of the Earth is precisely measured by the IERS. In particular the IERS publishes the
difference between the UTC and UT1 timescales in seconds. The UT1 time is by definition proportional
to the rotational angle of the Earth. Therefore a linear expression can be used to calculate the
Earth Rotation Angle ($\theta$)~\cite{Capitaine_ERA_2000} for any given moment
using~\cite{IAU_Resolution_2000_B18}:

\begin{displaymath}
  \theta(t_{\mathrm{UT1}}) =
  2 \pi (0.7790572732640 + 1.00273781191135448 \cdot (t_{\mathrm{UT1}} - 2451545.0)) \,,
\end{displaymath}

where $t_{\mathrm{UT1}}$ is the time represented as a Julian Date (in the UT1 timescale). The angle
between the Greenwich prime meridian (defining longitude zero of terrestrial coordinates) and the
mean vernal equinox (longitude zero of mean of date equatorial coordinates) at the time $t$ is the
Greenwich Mean Sidereal Time ($\mathrm{GMST}(t)$). It can be calculated from the Earth Rotation
Angle using polynomial expressions~\cite{Capitaine_2003_Precession}:

\begin{align*}
  \mathrm{GMST}(t_{\mathrm{UT1}},t_{\mathrm{TT}}) = \theta(t_{\mathrm{UT1}}) +
  (&0.014506\\
   &+ 4612.156534 \cdot t_{\mathrm{TT}}\\
   &+ 1.3915817 \cdot t_{\mathrm{TT}}^2\\
   &- 0.00000044 \cdot t_{\mathrm{TT}}^3\\
   &- 0.000029956 \cdot t_{\mathrm{TT}}^4\\
   &- 0.0000000368 \cdot t_{\mathrm{TT}}^5) \cdot \pi / 180 / 3600 \,.
\end{align*}

In this expression the time is required in both TT and UT1 timescales. The time argument
$t_{\mathrm{TT}}$ is expressed as the time since the J2000 epoch in Julian centuries. The angle
between the true vernal equinox and the Greenwich prime meridian is the Greenwich Apparent Sidereal
Time ($\mathrm{GAST}(t)$), which differs from the GMST by the ``equation of the equinoxes''
($\mathcal{E}_{\text{\Aries}}$), a term that can be calculated using nutation
theory~\cite{USNO_Circ179_2005}:

\begin{displaymath}
  \mathrm{GAST} =
  \mathrm{GMST} + \mathcal{E}_{\text{\Aries}} = \mathrm{GMST} + \Delta \psi \cos{\epsilon} + C \,,
\end{displaymath}

where $\Delta \psi$ is the nutation in longitude, $\epsilon$ is the mean obliquity of the ecliptic
and $C$ is a small correction, the so called complementary terms.

Applying a rotation about the TEME z-axis by the GMST angle yields coordinates in the Terrestrial
Intermediate Reference System (TIRS):

\begin{displaymath}
  \vec{d}_{\mathrm{TIRS}} = R_z(\mathrm{GMST}(t)) \enspace \vec{d}_{\mathrm{TEME}} \,.
\end{displaymath}

This system is co-rotating with the Earth and largely coincides, apart from a small rotation to
account for the effect of polar motion, with the International Terrestrial Reference System
(ITRS).

\textbf{5. TIRS $\rightarrow$ GCRS:} The transformation to an established celestial coordinate
system must account for the time dependent precession and nutation of the Earth's rotational
axis. The traditional sequence of transformations, which is also used here, proceeds as follows:

\begin{enumerate}
\item Rotate back by the Greenwich Apparent Sidereal Time (-$\mathrm{GAST}(t)$) from TIRS to True of
  Date (TOD) equatorial coordinates at the epoch $t$.
\item Account for nutation by rotating $\mathrm{TOD}(t)$ coordinates to Mean of Date (MOD)
  coordinates at the same epoch, using the matrix $N(t)$.
\item Account for precession by converting MOD coordinates at the epoch $t$ to MOD coordinates at
  the J2000 epoch, using the matrix $P(t)$.
\item Apply a small static frame bias correction (matrix $B$) to go to the Geocentric Celestial
  Reference System (GCRS)~\cite{IAU_Resolution_2000_B13}.
\end{enumerate}

The vector in the terrestrial intermediate system is thus transformed as follows:

\begin{align}
  \vec{d}_{\mathrm{TOD}(t)} &= R_z(-\mathrm{GAST}(t)) \enspace \vec{d}_{\mathrm{TIRS}} \\
  \vec{d}_{\mathrm{MOD}(t)} &= N^{T}(t) \enspace \vec{d}_{\mathrm{TOD}(t)} \\
  \vec{d}_{\mathrm{MOD}(J2000)} &= P^{T}(t) \enspace \vec{d}_{\mathrm{MOD}(t)} \\
  \vec{d}_{\mathrm{GCRS}} &= B^{T} \enspace \vec{d}_{\mathrm{MOD}(J2000)} \,.
\end{align}

In these equations the transpose of the matrices $N$, $P$ and $B$ are used in order to stay
consistent with the literature~\cite{USNO_Circ179_2005} which commonly defines:

\begin{displaymath}
  \vec{d}_\mathrm{TOD(t)} = N(t) P(t) B \enspace \vec{d}_{\mathrm{GCRS}} \,.
\end{displaymath}

Expressions for the matrices $B$, $P(t)$ and $N(t)$ are given in~\cite{USNO_Circ179_2005}. These
matrices as well as the angles $\theta(t)$, $\mathrm{GMST}(t)$ and $\mathrm{GAST}(t)$ are calculated
according to the IAU 2000A~\cite{IAU_Resolution_2000_B16,Mathews_2002} nutation and IAU 2006
precession~\cite{IAU_Resolution_2006_B1,Capitaine_2003_Precession,Hilton_2006} models, which
replaced previous models~\cite{Lieske_1977_Precession,IAU_1980_Nutation} in 2003 and 2006
respectively.

An alternative transformation is based on the Earth Rotation Angle and the Celestial Intermediate
Reference System (CIRS), whose origin points towards the Celestial Intermediate Origin (CIO):

\begin{align}
  \vec{d}_{\mathrm{CIRS}} &= R_z(-\theta(t)) \enspace \vec{d}_{\mathrm{TIRS}} \\
  \vec{d}_{\mathrm{GCRS}} &= C(X,Y,s)^{T} \enspace \vec{d}_{\mathrm{CIRS}} \,.
\end{align}

The Matrix $C(X,Y,s)$ combines the effects of precession, nutation and frame bias and can be
expressed in terms of the position of the celestial pole ($X$, $Y$) in the GCRS reference frame, the
small angle $s$ is the CIO locator~\cite{IERS_Conventions_2010}.

\textbf{6. GCRS $\rightarrow$ ICRS:} The geocentric GCRS coordinates are then transformed into the
International Celestial Reference System (ICRS): The origin is moved to the barycentre of the solar
system and the axis directions are corrected for the annual aberration and annual parallax
effects. A tiny correction for gravitational lensing in the solar system is also applied. In this
step only star independent astrometry parameters are used. This step makes use of astrometry related
functions from the IAU's SOFA software library~\cite{SOFA_2019-07-22}. For distant objects the
difference between GCRS and ICRS coordinates is very small, but it is important for bodies inside
the solar system, due to the shift of the origin to the solar system barycentre.

\textbf{7. ICRS $\rightarrow$ Galactic:} Since most of the gamma rays which reach the \mbox{AMS-02}
experiment on the ISS are produced within the Milky-Way it is also common to show distributions in
galactic coordinates. Given the defining coordinates of the galactic coordinate system according to
the IAU 1958 definition from equations~(\ref{eq:galactic-coordinates-fk5-1})
to~(\ref{eq:galactic-coordinates-fk5-3}) the rotation from ICRS/FK5 to Galactic coordinates is:

\begin{displaymath}
  \vec{d}_{\mathrm{Galactic}} =
  R_{z}(l_{\mathrm{NCP}} - \pi)
  R_{y}(\delta_{\mathrm{NGP}} - \frac{\pi}{2})
  R_{z}(-\alpha_{\mathrm{NGP}})
  \vec{d}_{\mathrm{FK5}} \,.
\end{displaymath}

\section{The AMS-02 Detector}
\label{sec:ams-02-detector}

\begin{figure}[t]
  \centering
  \includegraphics[width=0.5\textwidth]{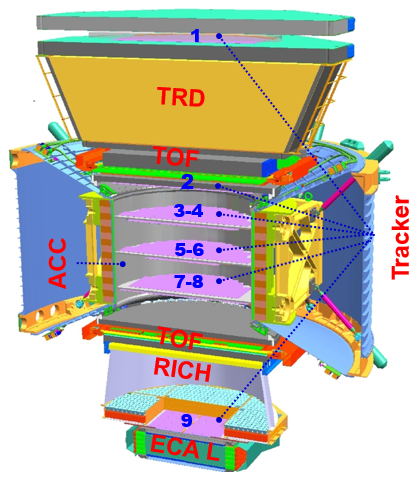}
  \caption{Overview of the \mbox{AMS-02} detector~\cite{AMS_WWW}. From top to bottom the components
    are: The silicon tracker plane 1, the Transition Radiation Detector (TRD), the upper
    Time-of-Flight detector (TOF), the silicon tracker planes 2-8 (the inner tracker), the lower
    Time-of-Flight detector, the Ring Imaging Cherenkov Detector (RICH) radiator and detection
    plane, the silicon tracker plane 9 and the electromagnetic calorimeter (ECAL). The inner tracker
    is surrounded by the permanent magnet as well as the Anti Coincidence Counter (ACC).}
  \label{fig:ams02-detector}
\end{figure}

The \mbox{AMS-02} detector consists of six major subdetectors. Figure~\ref{fig:ams02-detector} shows
an overview of the experiment. The individual subdetectors are explained in detail in the following.

The silicon tracker measures the trajectories of charged particles bent in the magnetic field,
allowing the reconstruction of their momenta and charge signs. The Time-of-Flight system measures
particle velocities, distinguishes up- from down-going particles and provides the main trigger. The
Anti Coincidence Counter surrounds the inner silicon tracker and vetos the trigger in case charged
particles enter the experiment from the sides. The Transition Radiation Detector measures the
trajectories of particles in the upper detector and is able to distinguish light from heavy
particles by detecting the transition radiation X-rays emitted by light particles in the fleece
radiator. The electromagnetic calorimeter reconstructs electromagnetic showers which are induced by
electrons, positrons and photons when they enter the calorimeter volume. By measuring the shower
properties the particle energy as well as its direction are inferred. The calorimeter also provides
the possibility to discriminate between hadronic and leptonic showers, due to their very different
shapes. Furthermore, the calorimeter has a standalone trigger logic for the measurement of photons
using the calorimeter only. Finally the Ring Imaging Cherenkov Detector is able to measure the
velocity of particles by reconstructing the cherenkov cone opening angle.

All detectors are also capable of measuring the particle charge by measuring the energy loss due to
ionization ($\mathrm{d}E/\mathrm{d}x$). The silicon tracker, Time-of-Flight and RICH detectors have
particularly good $\mathrm{d}E/\mathrm{d}x$ resolution, but all other detectors also
contribute. Measuring the energy loss makes it possible to determine the isotope number at low
energies.

Most importantly, the multitude of subdetectors provides redundancy in the measurement of the
particle properties. As an example, both the tracker and the calorimeter measure the particle's
energy, the velocity is measured by the Time-of-Flight and RICH detectors and lepton/hadron
separation is achieved independently by the TRD and ECAL. The particle charge is measured by all
subdetectors. This makes it possible to cross check one subdetector with the others, which is
important for calibration and determination of selection efficiencies from ISS data.

\begin{figure}[t]
  \begin{minipage}{0.48\linewidth}
    \centering
    \includegraphics[width=1.0\linewidth]{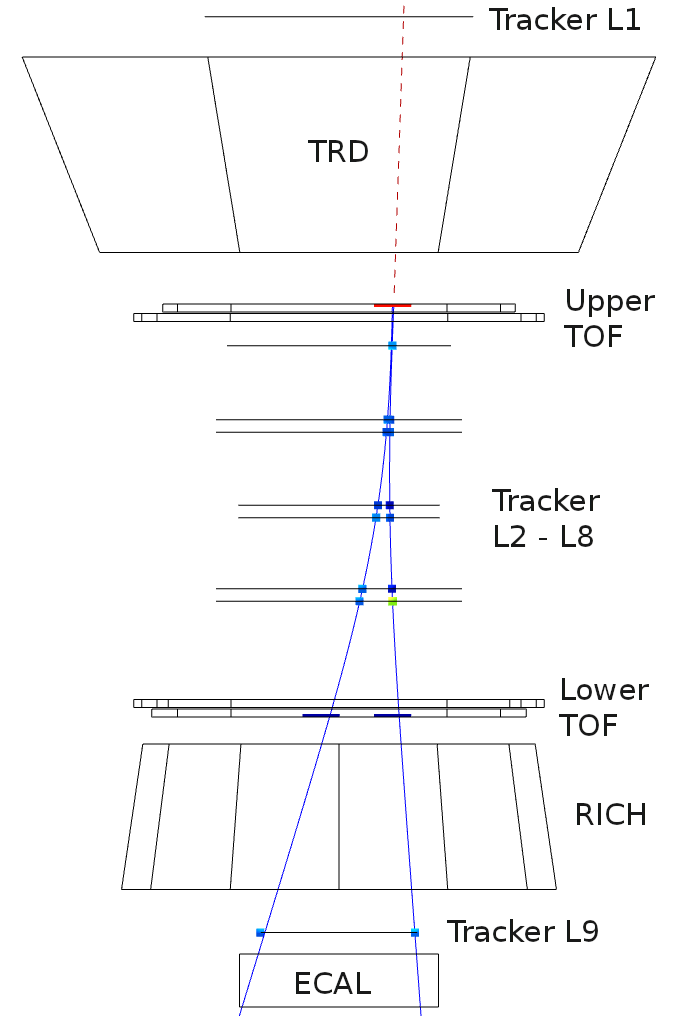}
  \end{minipage}
  \hspace{0.01\linewidth}
  \begin{minipage}{0.48\linewidth}
    \centering
    \includegraphics[width=1.0\linewidth]{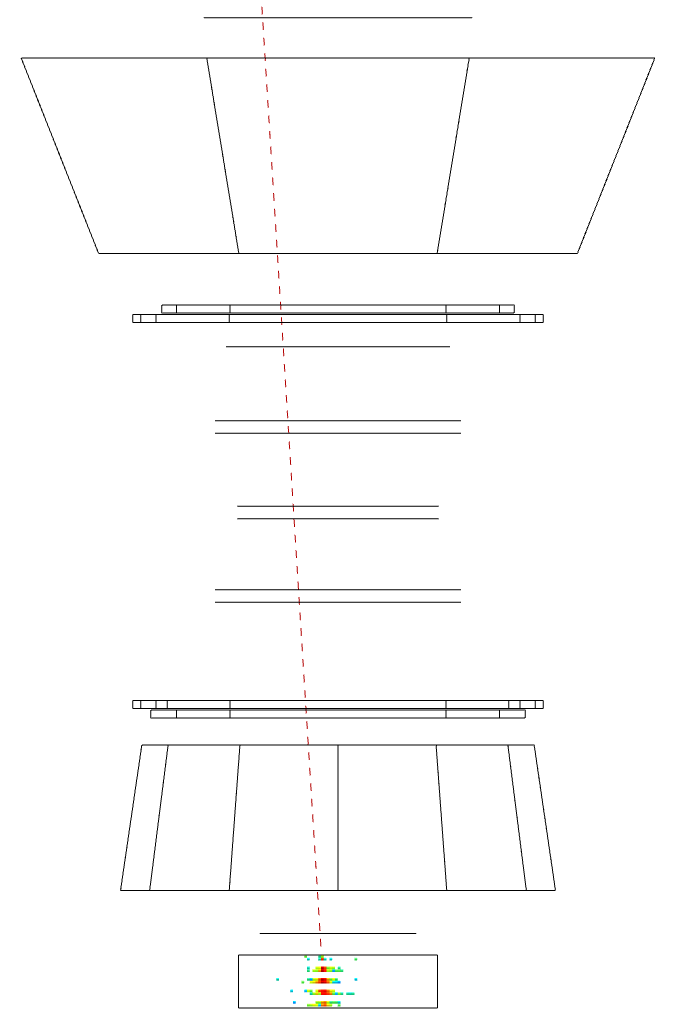}
  \end{minipage}
  \caption{Left: Event display of an event with a \SI{650}{\mega\electronvolt} photon converting in
    the upper TOF. Electron and Positron are reconstructed in the inner tracker. Right: Event
    display of a calorimeter photon event with a \SI{20}{\giga\electronvolt} shower, but no signals
    in the upper detector.}
  \label{fig:ams02-event-dislays}
\end{figure}

Figure~\ref{fig:ams02-event-dislays} illustrates the measurement principle for the two ways in which
$\gamma$-rays are reconstructed in \mbox {AMS-02}. In the conversion mode the photon converts into
an electron / positron pair in the upper Time-of-Flight detector, as show on the left hand side for
an example event. The two charged trajectories are then separated by the magnetic field and are
reconstructed by the inner tracker. The absence of signal in the TRD and the tracker layer 1 is a
key signature of these events. For these events, the trigger is generated by the TOF system and the
energy and direction of the $\gamma$-ray are measured by the tracker.

An alternative approach is the measurement of calorimeter photons. In this class of events the
photon does not convert before the calorimeter. Instead, the conversion only happens when the photon
enters the calorimeter and an electromagnetic shower is produced. The right hand side of
figure~\ref{fig:ams02-event-dislays} shows an example for such an event. In this case the trigger of
the event, as well as the reconstruction of the photon energy and direction are done with the
calorimeter.

Two separate analyses are developed in this thesis. The two approaches are entirely complementary:
Different subdetectors are used for the trigger and measurement, and the event samples are
completely disjunct. They are also suitable for different energy ranges: While the converted photons
can be measured from \SI{100}{\mega\electronvolt} to approximately \SI{10}{\giga\electronvolt} with
the tracker, the energy range for calorimeter photons ranges from \SI{1}{\giga\electronvolt} to
approximately \SI{1}{\tera\electronvolt}. In the overlap region one analysis can be used to
cross-check the other, which allows to exclude systematic uncertainties.

\subsection{Magnet}
\label{sec:detector-magnet}

\begin{figure}[t]
  \begin{minipage}{0.48\linewidth}
    \centering
    \includegraphics[width=0.8\linewidth]{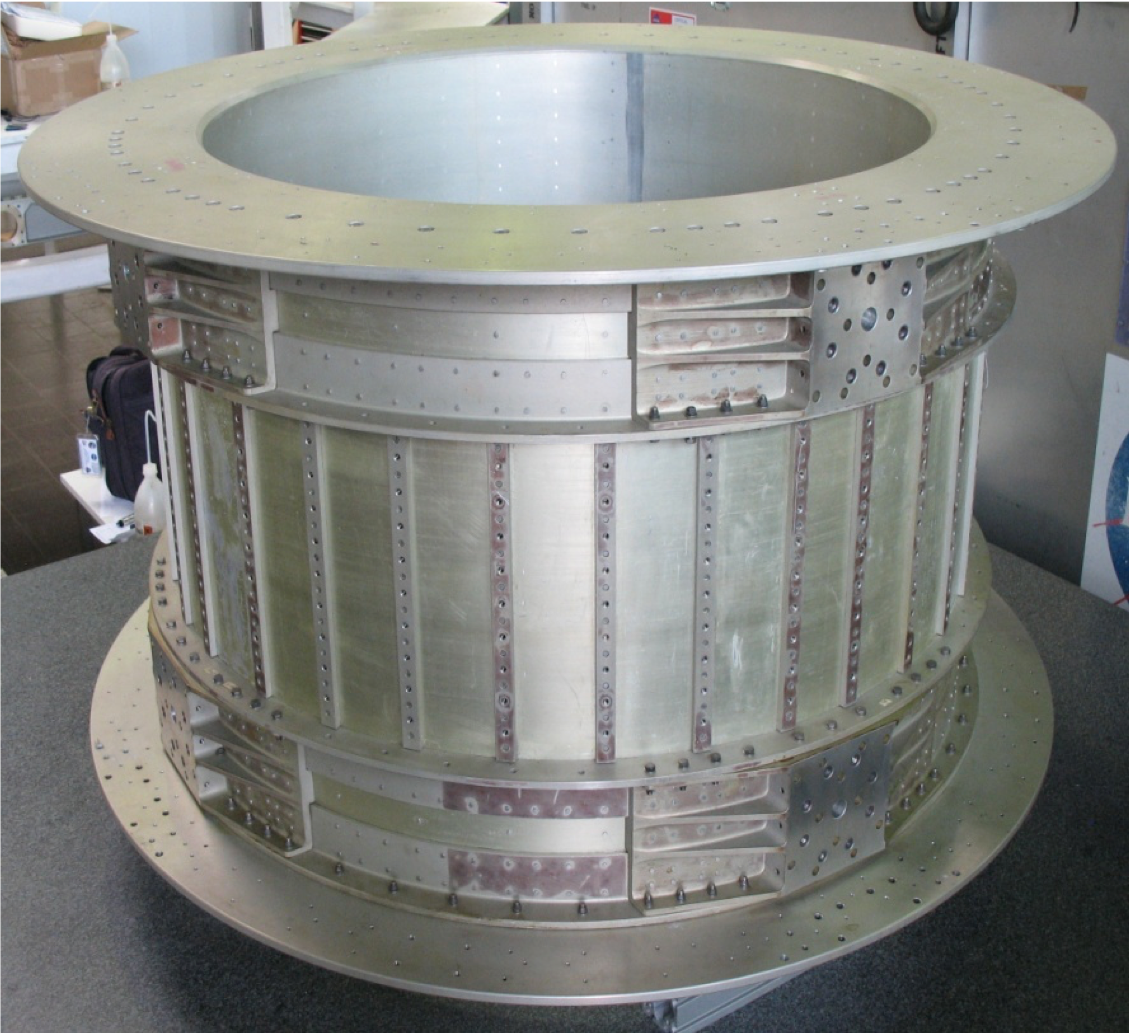}
  \end{minipage}
  \hspace{0.01\linewidth}
  \begin{minipage}{0.48\linewidth}
    \centering
    \includegraphics[width=1.0\linewidth]{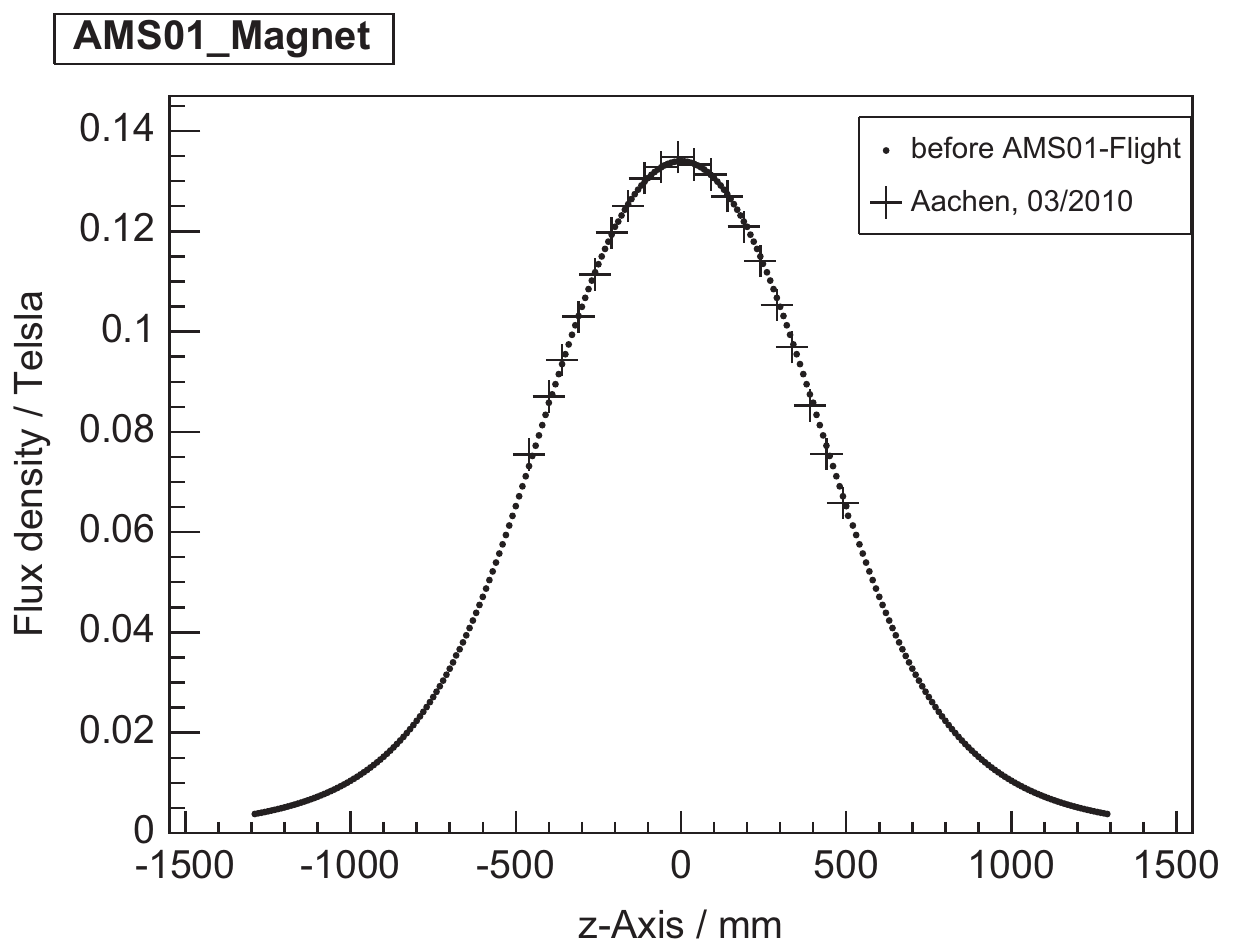}
  \end{minipage}
  \caption{Left: The permanent magnet before integration into AMS~\cite{AMS_WWW}. Right: Magnetic
    field of the permanent magnet in the center as a function of the z-coordinate as measured in
    1997 and in 2010~\cite{AMS02_Detector_Redesign}.}
  \label{fig:ams02-magnet-photo}
\end{figure}

Together with the silicon tracker the magnet is vital for the determination of the rigidity of
charged particles traversing AMS. The requirements on the magnet are:

\begin{itemize}
\item A strong magnetic field in the interior as a stronger magnetic field results in a better
  rigidity resolution.
\item Homogeneity of the magnetic field in the interior, which simplifies the track reconstruction.
\item Stability of the magnetic field as a function of time and temperature since changes in the
  magnetic field impact the rigidity scale systematic uncertainty.
\item Long lifetime to ensure up to 20 years of operations on the ISS.
\item No exterior dipole moment, as that would impose torques due to the interaction with the
  Earth's residual magnetic field on the spacecraft delivering AMS to the space station as well as on
  the station itself.
\end{itemize}

The permanent magnet used in the AMS spectrometer is a cylindrical magnet with a height and diameter
of approximately \SI{1}{\meter} (see figure~\ref{fig:ams02-magnet-photo}). It is made of over 6000
Neodimium-Iron-Boron (Nd-Fe-B) magnetized blocks glued with Epoxy and arranged in such a way that
the resulting total magnetic field is homogeneous along the X-axis of the AMS coordinate system. The
magnetic field strength in the center of the detector is approximately \SI{0.14}{\tesla}.

The AMS permanent magnet was already used successfully in the 1998 precursor flight AMS-01 on board
space shuttle Discovery (STS-91). After the return of the magnet to Earth AMS-01 was disassembled
and the magnet was recovered. Before integration into \mbox{AMS-02} in 2010 the magnetic field
strength was measured again and compared to the 1997 measurement, resulting in no observable
deviations over the course of 12 years, as shown in figure~\ref{fig:ams02-magnet-photo} in the right
panel. The magnetic field is therefore shown to be stable over time and the magnet is suitable for
long term use within \mbox{AMS-02} on the space station.

On orbit changes in temperature have a small, but observable effect on the field. These changes are
therefore corrected for by observing the time evolution of the $^4\mathrm{He}$ mass peak, since any
change in the rigidity scale directly impacts the reconstructed mass. In addition, the rigidity
scale is verified by comparing the rigidity reconstructed for electrons and positrons with the
energy measurement in the electromagnetic calorimeter~\cite{AMS02_Detector_RigidityScale}.

\subsection{Silicon Tracker}
\label{sec:detector-tracker}

\begin{figure}[t!]
  \centering
  \includegraphics[width=0.98\linewidth]{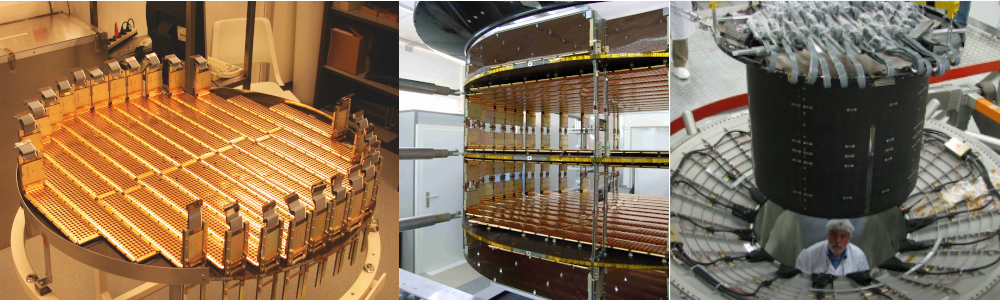}
  \caption{The \mbox{AMS-02} silicon tracker during construction and
    integration~\cite{AMS_Geneve_WWW}. Left: One plane of silicon ladders for the inner tracker
    mounted on the honeycomb support plane. Middle: Side-view of the fully equipped inner
    tracker. Right: Integration of the tracker into \mbox{AMS-02}.}
  \label{fig:ams02-tracker-photo}
\end{figure}

The primary purpose of the tracker is to measure the coordinates of charged particles passing the
detection planes with high accuracy. Together with the magnetic field of the permanent magnet this
provides an accurate estimation of the rigidity and charge sign of the particle. It consists of nine
roughly circular planes equipped with double sided silicon strip detectors. The first plane is
located at the top of the instrument above the TRD, followed by 7 planes in the inner tracker. In
the inner tracker layers (3,4), (5,6) and (7,8) are double layers, located on opposite sides of the
same support plane. The last plane is located directly above the calorimeter and below the RICH
detection plane. In each plane, between 16 and 26 double-sided silicon ``ladders'' are mounted on
top of a aluminum honeycomb support structure with carbon fiber skins, placed next to each other in
order to cover the entire surface of the plane. In total there are 192 ladders in the tracker with
1024 channels each, resulting in almost 200,000 active channels.

\begin{figure}[t]
  \centering
  \includegraphics[width=0.5\linewidth]{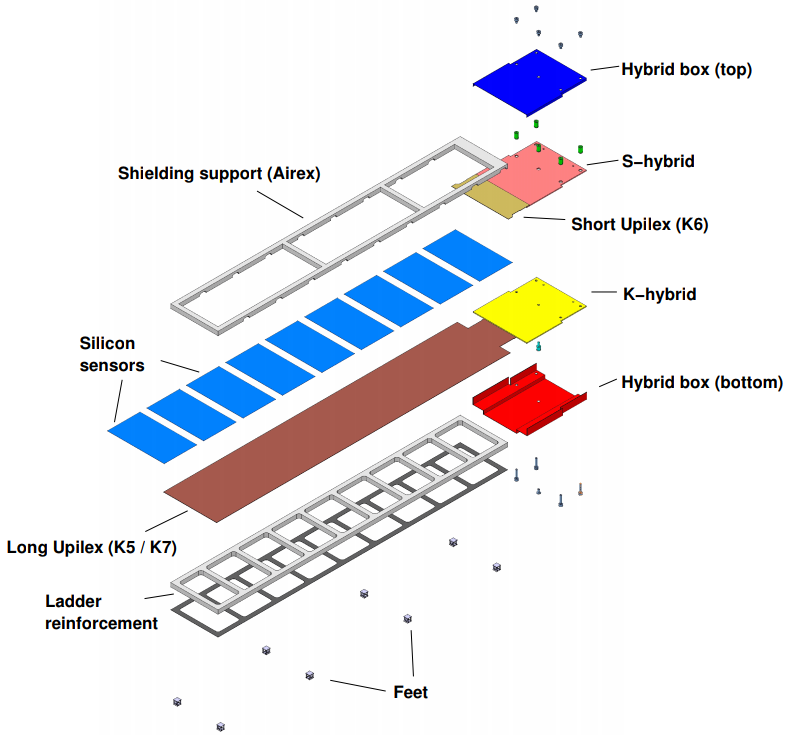}
  \caption{Schematic view of the components of a tracker silicon ladder~\cite{AMS02_PHD_Azzarello}.}
  \label{fig:ams02-tracker-ladder-schematic}
\end{figure}

Silicon ladders are the primary components of the
tracker. Figure~\ref{fig:ams02-tracker-ladder-schematic} shows a schematic of the components of a
ladder. At the heart of the ladder there are between 9 and 15 n-doped silicon sensors, placed next
to each other. The number of sensors depends on the length of the ladder, which varies according to
geometric requirements. The readout preamplifier chips are located at the end of each ladder within
the hybrid box. Upilex cables are used to connect the signals from the bonding wires to the
hybrids. Above and below the silicon sensors support structures reinforce the mechanical stability
of the ladder and aluminum tracker feet are used to establish the connection between the ladder and
the support plane.

The surface area of a silicon sensor is approximately \SI{72.045}{\milli\meter} $\times$
\SI{41.360}{\milli\meter} and each sensor is approximately \SI{300}{\micro\meter} thick. On both
surfaces of each silicon sensor there are doped implant strips ($p^{+}$ on the ``p-side'', $n^{+}$
on the ``n-side''). The implant strip pitch is \SI{27.5}{\micro\meter} for the p-side and
\SI{104}{\micro\meter} for the n-side which results in 2568 p-strips and 384 n-strips. However, in
order to reduce the number of readout channels, the number of strips which are actually read out is
lower, 640 for the p-side and 192 for the n-side per sensor, resulting in a readout pitch of
\SI{110}{\micro\meter} for the p-side and \SI{208}{\micro\meter} for the n-side.

The p-side strips are used to measure the Y-coordinate of the passing particle whereas the n-side
strips measure the X-coordinate. Because of the orientation of the magnetic field, the YZ-plane
corresponds to the bending plane of the particle trajectory, which means that the Y-coordinate is
used to reconstruct the rigidity and charge sign of the track.

In order to improve the spatial resolution capacitive charge sharing is employed, which forwards the
electric signals from strips which are not read out to the connected strips. The capacities reflect
the relative distance between the strips. This technique allows to improve the spatial resolution
from $\SI{110}{\micro\meter} / \sqrt{12} \approx \SI{32}{\micro\meter}$ to \SI{10}{\micro\meter} for
the p-side. Because the readout pitch on the n-side is coarser, and because only two implant strips
are capacitively coupled, the spatial resolution is slightly worse and amounts to approximately
\SI{30}{\micro\meter}. These numbers are valid for protons and electrons, but since the spatial
resolution also depends on the amount of deposited charge it improves for heavier
nuclei~\cite{AMS02_Detector_SpatialRes}.

\begin{figure}[t!]
  \centering
  \includegraphics[width=0.6\linewidth]{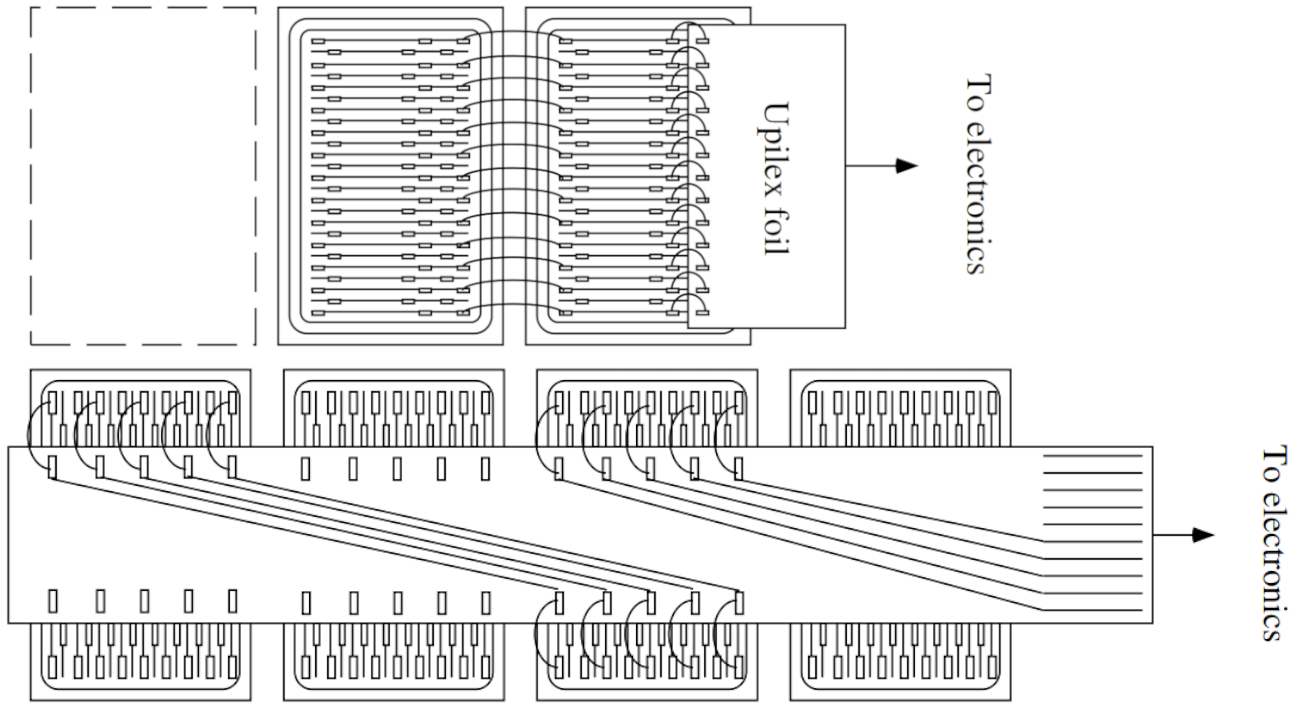}
  \caption{Daisy chaining for the readout channels of the tracker ladders (upper panel: Y-side,
    lower panel: X-side)~\cite{AMS02_PHD_Azzarello}.}
  \label{fig:ams02-tracker-daisy-chaining}
\end{figure}

On the p-side the signals from 640 readout strips of each sensor are daisy-chained and connected to
640 channels in the p-side hybrid (see figure~\ref{fig:ams02-tracker-daisy-chaining}, upper
panel). Effectively, this means that the coordinate along the strips is not restricted to the width
of the sensor which was traversed by the particle, each strip effectively extends along the whole
length of the ladder.

On the n-side sensors 1, 3, 5, 7,... are daisy-chained so that the number of readout-channels is
reduced further (see figure~\ref{fig:ams02-tracker-daisy-chaining}, lower panel). Similarly sensors
2, 4, 6, 8,... are also daisy-chained. This results in a total number of 384 readout channels which
are connected to the n-side hybrid. Because of the daisy-chaining there is an
\SI{8.27}{\centi\meter} ambiguity in the reconstruction of the X-coordinate: It is not possible to
measure the exact (X,Y) point of passing for a particle using only a single ladder. This ambiguity
must be externally resolved, by taking into account the trajectory of the particle in the XZ-plane
as measured by the TRD and/or TOF. The daisy-chaining scheme described above is employed for all
inner tracker ladders (``K5'' ladders), for ladders in layers 1, 2, and 9, a more complicated scheme
with a different pitch is employed (``K7'' ladders) - which is not described in detail here, but
helps to resolve the ambiguities~\cite{AMS02_PHD_Azzarello}.

The alignment of the inner tracker was performed on the sensor level based on beamtest data using
\SI{400}{\giga\electronvolt} protons and is continuously monitored using cosmic ray protons on
orbit. A dedicated laser alignment system also monitors movements of the tracker planes 2-8 and
ensures the time stability. Overall the alignment of the sensors in the inner-tracker is static and
controlled at the submicron level.

The external layers 1 and 9 are moving on the \SI{1}{\milli\meter} scale due to thermal expansion of
the support structure as a result of temperature variations. Cosmic ray protons are used to correct
this effect in a dynamic alignment of the external planes. After this calibration the residual
misalignment is on the level of \SI{3}{\micro\meter}, well below the single point spatial resolution
of \SI{10}{\micro\meter}. Consistency between two different alignment methods is required in order
to minimize the impact of any residual misalignment.

In order to reduce the electronic noise, to minimize thermal expansion and to ensure stable
operating conditions within thermal safety limits there is a dedicated tracker cooling system based
on a two-phase $\mathrm{CO}_2$ loop which controls the temperature of tracker layers 2 to 9. Pumps
are used to circulate the $\mathrm{CO}_2$ which evaporates when absorbing the heat load of the
tracker electronics and condenses on external radiator planes which face outer space. This system
keeps the inner tracker temperature constant at the \SI{1}{\celsius} level.

The reconstruction of the curvature of charged particle tracks in the magnetic field with the
tracker makes it possible to estimate the particle rigidity. The rigidity reconstruction resolution
depends on the single-point spatial resolution of the tracker, on the strength of the magnetic field
along the trajectory of the particle, on the number of hits used in the track fitting, on the
position of the used hits, and on the inclination angle of the trajectory. Most importantly the
resolution depends strongly on the hit pattern, namely which layers participate in the
reconstruction of the track. Hits on the external layers 1 and 9 are particularly helpful in order
to constrain the rigidity of the particle because they extend the lever arm of the
trajectory. Because the spatial resolution of the tracker improves for heavier nuclei it also
depends implicitly on the particle nuclear charge Z.

\begin{figure}[t]
  \begin{minipage}{0.48\linewidth}
    \centering
    \includegraphics[width=1.0\linewidth]{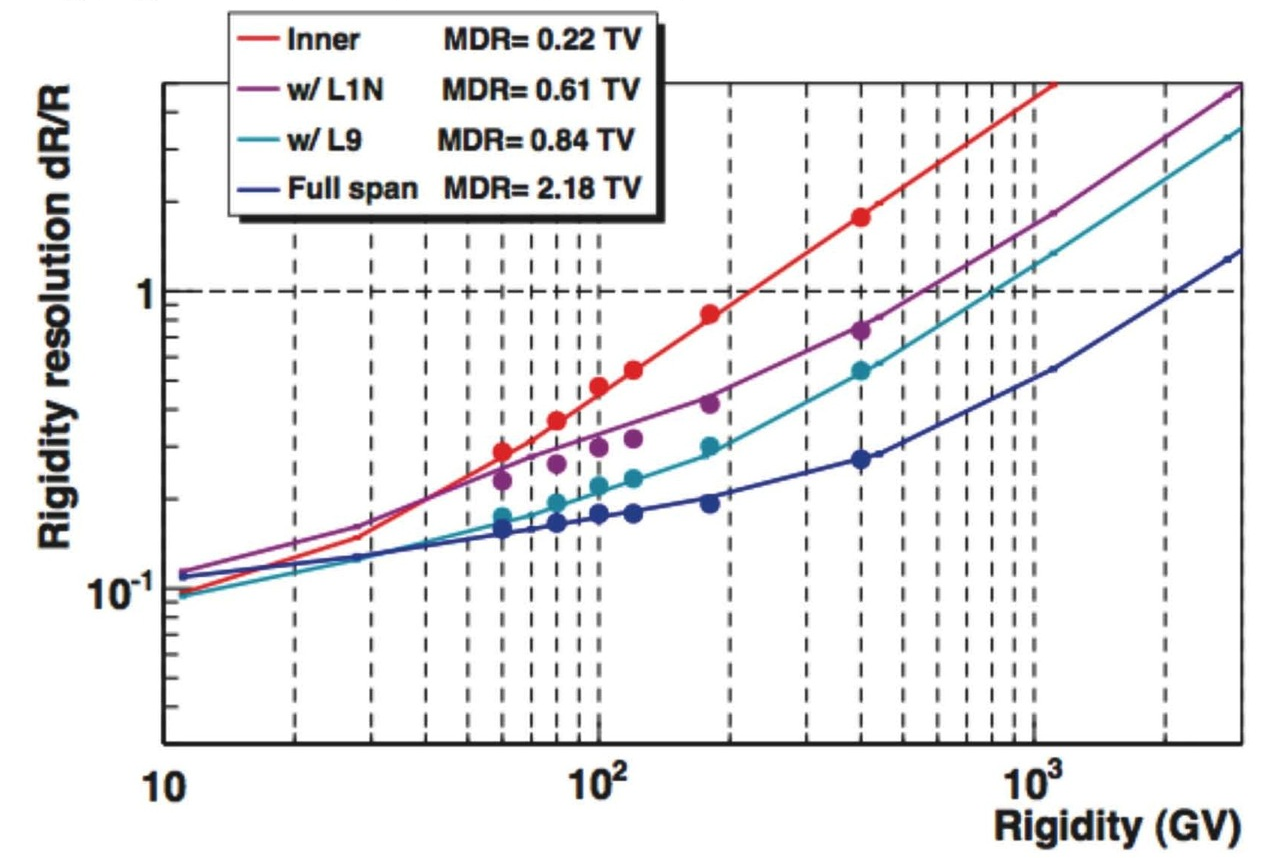}
  \end{minipage}
  \hspace{0.01\linewidth}
  \begin{minipage}{0.48\linewidth}
    \centering
    \includegraphics[width=1.0\linewidth]{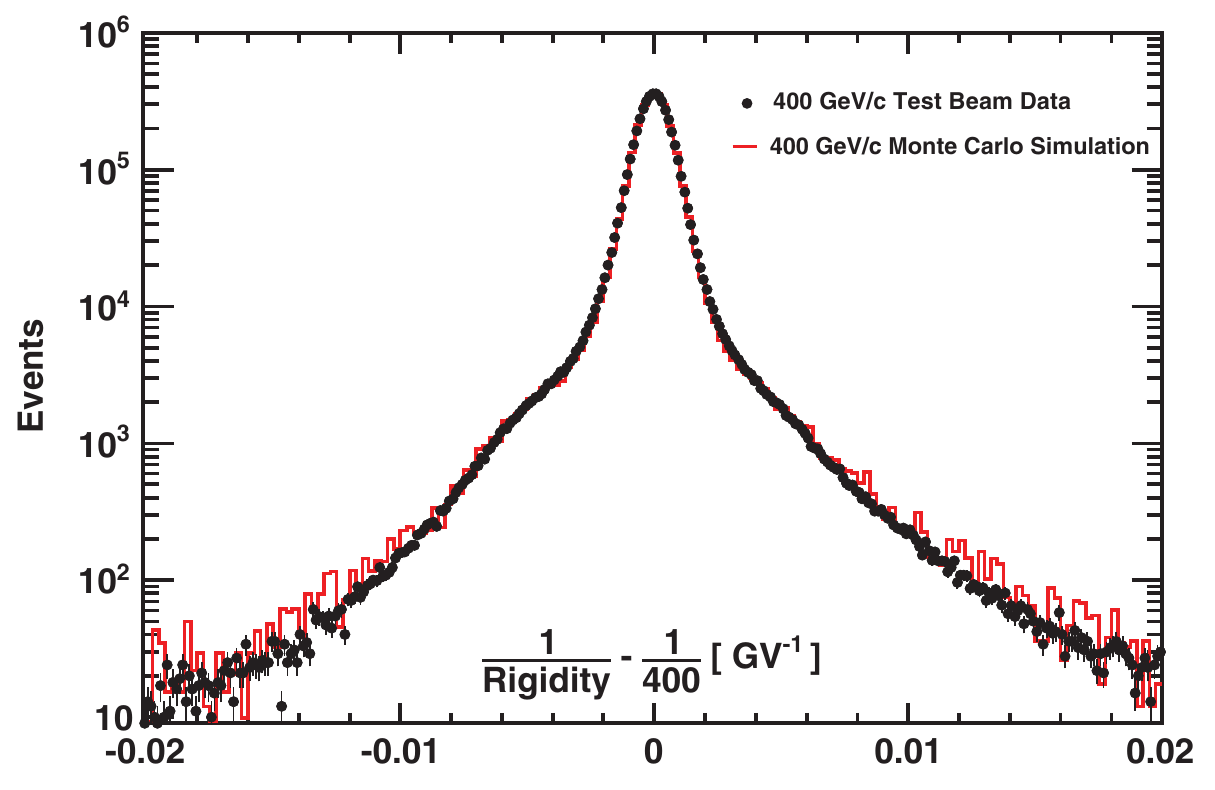}
  \end{minipage}
  \caption{Left: The spectrometer rigidity resolution as measured in the beamtest at CERN's SPS
    using protons and pions with energies 60, 80, 100, 120, 180 and \SI{400}{\giga\electronvolt}
    (points) compared with Monte-Carlo predictions (lines). Right: Detailed comparison of the
    curvature reconstructed by the tracker for the \SI{400}{\giga\electronvolt} beamtest and the
    Monte-Carlo simulation for the same energy~\cite{AMS02_Proton_2015}.}
  \label{fig:ams02-tracker-rigidity-resolution}
\end{figure}

Figure~\ref{fig:ams02-tracker-rigidity-resolution} shows results of the analysis of beamtest proton
and pion data in the 2010 campaign at CERN with predictions from Monte-Carlo simulations. A figure
of merit for the resolution function is the maximum detectable rigidity (MDR), the rigidity at which
the relative rigidity resolution reaches \SI{100}{\percent}. For the inner tracker configuration
(using layers 2 to 8) the MDR is approximately \SI{220}{\giga\volt}, while for the full tracker
using all planes the MDR is \SI{2.18}{\tera\volt}. At low rigidities the proton resolution reaches a
plateau at approximately \SI{10}{\percent}, where it is dominated by the multiple scattering off the
detector material. The material budget of a tracker plane amounts to approximately
\SI{0.4}{\percent} of a radiation length, most of which is due to the silicon itself.

The right panel in figure~\ref{fig:ams02-tracker-rigidity-resolution} shows the excellent agreement
between the Monte-Carlo simulation of the tracker's rigidity reconstruction and the
\SI{400}{\giga\electronvolt} beamtest data, even in the far tails. This excellent matching is
important, since migration matrices obtained from simulations are used to unfold the measured event
counts which is necessary due to bin-to-bin migration.

The inner tracker layers L2-L8 are important for the analysis of gamma conversions, since the
relevant conversions happen in the upper TOF and most of the electron and positron tracks produced
in the conversion do not pass through tracker layer 9. The expected MDR for either one of the two
tracks is therefore around \SI{200}{\giga\volt}, but it is worth noting that the energy resolution
for the photon, which must be reconstructed from both trajectories simultaneously, will be slightly
worse.

The single layer charge resolution for the silicon tracker is approximately \SI{10}{\percent} for
protons and improves for higher charges. For carbon nuclei it is approximately \SI{4}{\percent} and
for Iron ions it is \SI{3}{\permille}. A combination of multiple tracker layers improves the
resolutions further, since each layer provides an independent measurement of the charge.

\subsection{Time-of-Flight System}
\label{sec:detector-tof}

\begin{figure}[t]
  \centering
  \includegraphics[width=0.5\linewidth]{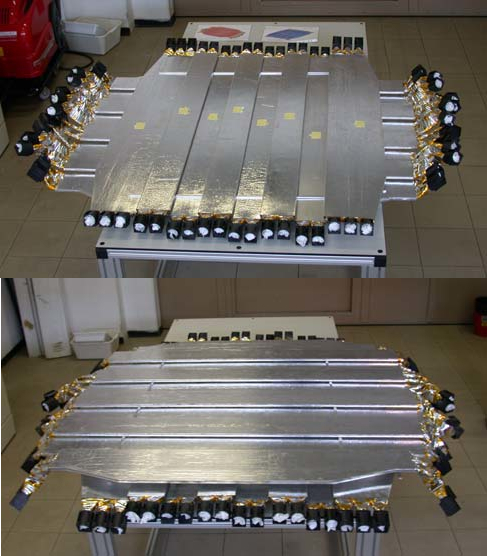}
  \caption{The \mbox{AMS-02} Time-of-Flight detector. The upper photo shows the two upper TOF
    layers, the lower photo shows the lower TOF system~\cite{AMS_WWW}.}
  \label{fig:ams02-tof-photo}
\end{figure}

The Time-of-Flight (TOF) system consists of four layers of polyvinyl-toluene scintillator counters
read out by Photomultiplier-tubes (PMTs) connected to fast electronics. Two layers are located above
the inner tracker and two are located below (see figure~\ref{fig:ams02-detector}). Its main purpose
is to provide the trigger for charged particles when they traverse AMS. The TOF measures the
velocity $\beta = v/c$ of charged particles by comparing the digitized timestamps of the signal in
the lower TOF layers with those collected in the upper layers. The sign of the velocity is important
to discriminate up-going from down-going particles, which is in general not possible with the other
AMS subdetectors. Finally, the TOF has excellent charge resolution, which facilitates the
identification of the nuclear charge Z of the passing particle.

Scintillation light is produced when a charge particle crosses the counters. The photons are
reflected internally and arrive at the end of the bars from where they are transported with the help
of light guides to the PMTs. The attenuation length of the counters is approximately
\SI{3.8}{\meter}, which is much larger than the length of the counters ($\approx
\SI{80}{\centi\meter}$).

Figure~\ref{fig:ams02-tof-photo} shows a photo of the fully assembled Time-of-Flight detector. The
bars in layers 1 and 4 are oriented along the X-axis whereas bars in layers 2 and 3 are oriented
perpendicularly along the Y-axis. Layers 1, 2 and 4 are constructed from 8 scintillator paddles,
while 10 paddles are used for layer 3. The thickness of the scintillator paddles is approximately
\SI{1}{\centi\meter}, the central bars are rectangular with a width of approximately
\SI{12}{\centi\meter}. The light from each bar is detected by four Hamamatsu R5964 PMTs, two on each
side. A particular advantage of these specific PMTs is that they can be used within a strong
magnetic field, such as generated by the AMS magnet. The outer bars in each layer are wider and have
a trapezoidal shape. For these paddles six PMTs (three per side) are used. In each layer the bars
are staggered in height (see figure~\ref{fig:ams02-tof-photo}), with an overlap of
\SI{0.5}{\centi\meter} in order to reduce the impact of inefficiencies at the borders of the bars,
which is important for the trigger efficiency.

\begin{figure}[t]
  \centering
  \includegraphics[width=0.6\linewidth]{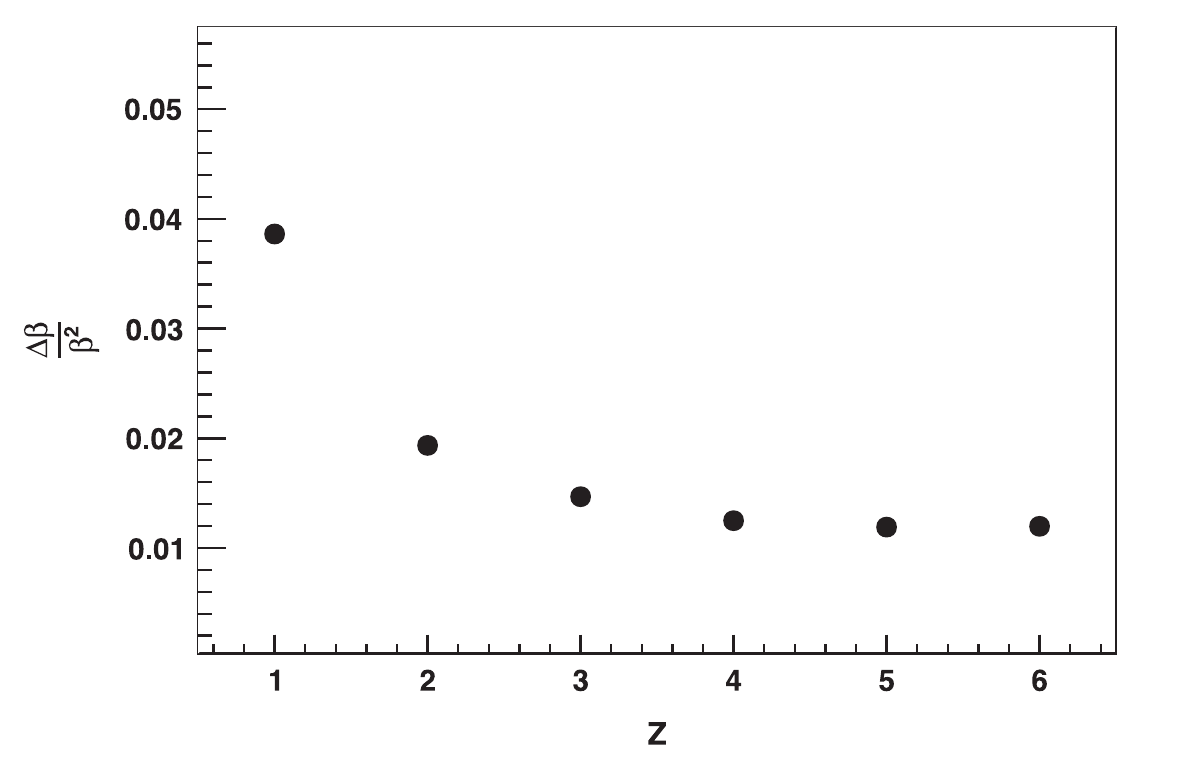}
  \caption{The Time-of-Flight velocity resolution as a function of the nuclear charge Z, measured
    using events with rigidity larger than \SI{20}{\giga\volt} in ISS
    data~\cite{AMS02_Detector_TOF_Calibration}.}
  \label{fig:ams02-tof-velocity-resolution}
\end{figure}

The time resolution for each counter is approximately \SI{160}{\pico\second} for protons, electrons
and positrons, which corresponds to a velocity resolution $\sigma_\beta/\beta$ of approximately
\SI{4}{\percent}~\cite{AMS02_Detector_TOF_Calibration}. This number improves for heavier nuclei,
reaching a lower limit of about \SI{50}{\pico\second} ($\sigma_\beta/\beta = \SI{1.2}{\percent}$)
which is dominated by electronic noise, see figure~\ref{fig:ams02-tof-velocity-resolution}. Since
the distance between the upper and the lower TOF is approximately \SI{1}{\meter}, a relativistic
particle will show a time delay between the signal in the lower and upper TOF of
$\approx \SI{3.3}{\nano\second}$. The probability to mistake an up-going particle for a down-going
one is therefore negligibly small (< $10^{-9}$), which is especially important for the antimatter
searches in AMS.

The charge resolution ($\sigma_Z/Z$) for a single TOF counter is approximately \SI{6}{\percent} for
protons, \SI{2.6}{\percent} for carbon and \SI{1.5}{\percent} for
iron~\cite{AMS02_Detector_TOF_Calibration}.

The material budget of each TOF counter is approximately \SI{2.4}{\percent} of a radiation
length. In addition the \SI{10}{\centi\meter} aluminum honeycomb support structure above the first
TOF layer adds \SI{2.95}{\percent} of a radiation length~\cite{AMS02_Detector_TOF}. This
comparatively large number is the reason why many photons which enter AMS from the top convert
within the upper TOF. In the analysis of converted photons the upper TOF will therefore be used as
the primary converter.

In addition, in case of a photon conversion in the TOF, the deposited energy will be approximately
twice that of a |Z| = 1 particle, because two particles are passing through the same bar. Because
the reconstructed charge Z is proportional to the square root of the deposited energy, this yields a
reconstructed Z of $\sim \sqrt{2}$, which provides a reliable way to identify such events.

\subsection{Anti Coincidence Counter}
\label{sec:detector-acc}

\begin{figure}[t]
  \centering
  \includegraphics[width=0.95\linewidth]{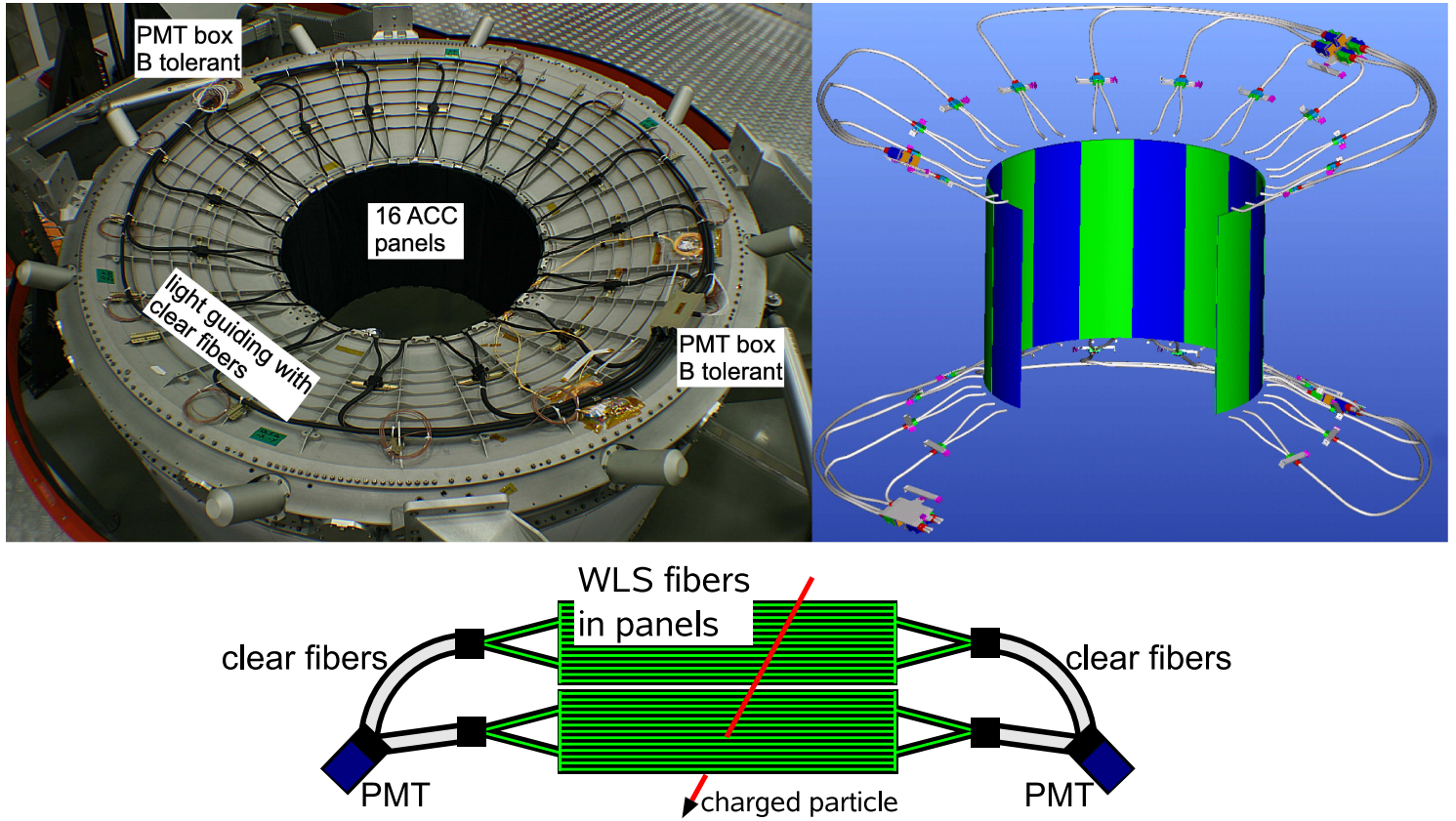}
  \caption{The \mbox{AMS-02} Anti Coincidence Counter (ACC). The upper left panel shows a photo of
    the ACC after integration, the upper right and lower panels show schematics of the detector
    arrangement~\cite{AMS02_Detector_ACC}.}
  \label{fig:ams02-acc-photo-schematic}
\end{figure}

The Anti Coincidence Counter (ACC) consists of 16 scintillation counters surrounding the inner
tracker (see figure~\ref{fig:ams02-acc-photo-schematic}). Its purpose is to veto charged particles
entering AMS from the sides. It also vetoes events in which hadronic interactions cause the primary
particle to break up and produce secondaries which pass through the sides. In case the event is not
vetoed (see section~\ref{sec:detector-trigger}) the amplitude and time information from the ACC can
be used to estimate the amount of backsplash from showers in the calorimeter.

The veto from the ACC is important in order to keep the rate of very low energy secondary particles,
which are not of interest in the analysis of AMS data, at a tolerable level, particularly near the
geomagnetic poles and in the South Atlantic Anomaly. This ensures an acceptable DAQ efficiency for
the measurement of galactic cosmic rays, even if the rate of secondary particles becomes too
high. Because signals from secondary particles entering through the sides could potentially confuse
the reconstruction of particle trajectories in the tracker, the veto from the ACC is particularly
important in the anti-matter searches with AMS.

The 16 scintillation panels are arranged in a cylinder with a diameter of \SI{1.1}{\meter}, and
placed just inside the magnet bore. Each panel is approximately \SI{83}{\centi\meter} long and
\SI{8}{\milli\meter} thick, the material is a Bicron BC-414 polyvinyl-toluene plastic. A tongue and
groove system is used to connect adjacent panels in the cylinder. As can be seen from the lower
panel in figure~\ref{fig:ams02-acc-photo-schematic}, two panels are read out by one photomultiplier
tube on each side (top and bottom). Therefore, the ACC cylinder can be subdivided into 8 independent
sectors, which are redundantly read out on the top and on the bottom of the cylinder. The
ultraviolet scintillation light is picked up by wavelength shifting fibers which are embedded into
the panels and transform the wavelength. At the end of the panels the wavelength shifting fibers
couple to clear fibers which transport the light to the Hamamatsu R5946 PMTs which are located in
PMT boxes, two boxes at the top and two boxes at the bottom of AMS. The clear fiber light guides are
required because the PMTs can not be operated at maximum efficiency within a strong magnetic
field. Each box houses four PMTs.

The signals from the ACC PMTs are forwarded to the AMS trigger electronics where the coincidence
from the two PMTs reading out each sector is used for the veto decision. The average inefficiency
for the ACC system was measured in the beamtest to be smaller than $2.7 \cdot 10^{-5}$ at
\SI{95}{\percent} confidence level~\cite{AMS02_Diplom_Goerres}. The inefficiency is even better for
the bulk of the events which do not pass near the borders of the panels.

In the calorimeter photon analysis the ACC is important because it provides a reliable charged
particle veto in case they enter from the sides of AMS, where the TRD and tracker cannot be used.

\subsection{Transition Radiation Detector}
\label{sec:detector-trd}

The transition radiation detector (TRD) is located at the top of AMS, between the first tracker
layer and the upper TOF. It is a gas detector made of proportional straw tubes filled with a
$\mathrm{Xe/CO_{2}}$ mixture, interleaved with fleece mats. The TRD's main purpose is the
identification of cosmic ray electrons and positrons, by measuring the transition radiation (TR)
photons they emit when passing through the fleece radiator. The TRD also has good tracking
capabilities in both projections, excellent efficiency and a very good signal to noise ratio.

The excellent tracking efficiency of the TRD is used in the photon analysis in order to ensure a
reliable charged particle veto. In the analysis of photons which convert in the upper TOF, the
non-existence of a track in the TRD is essential to suppress the large background of electrons,
protons and helium nuclei.

\begin{figure}[t]
  \centering
  \begin{minipage}{0.48\linewidth}
    \centering
    \includegraphics[width=1.0\linewidth]{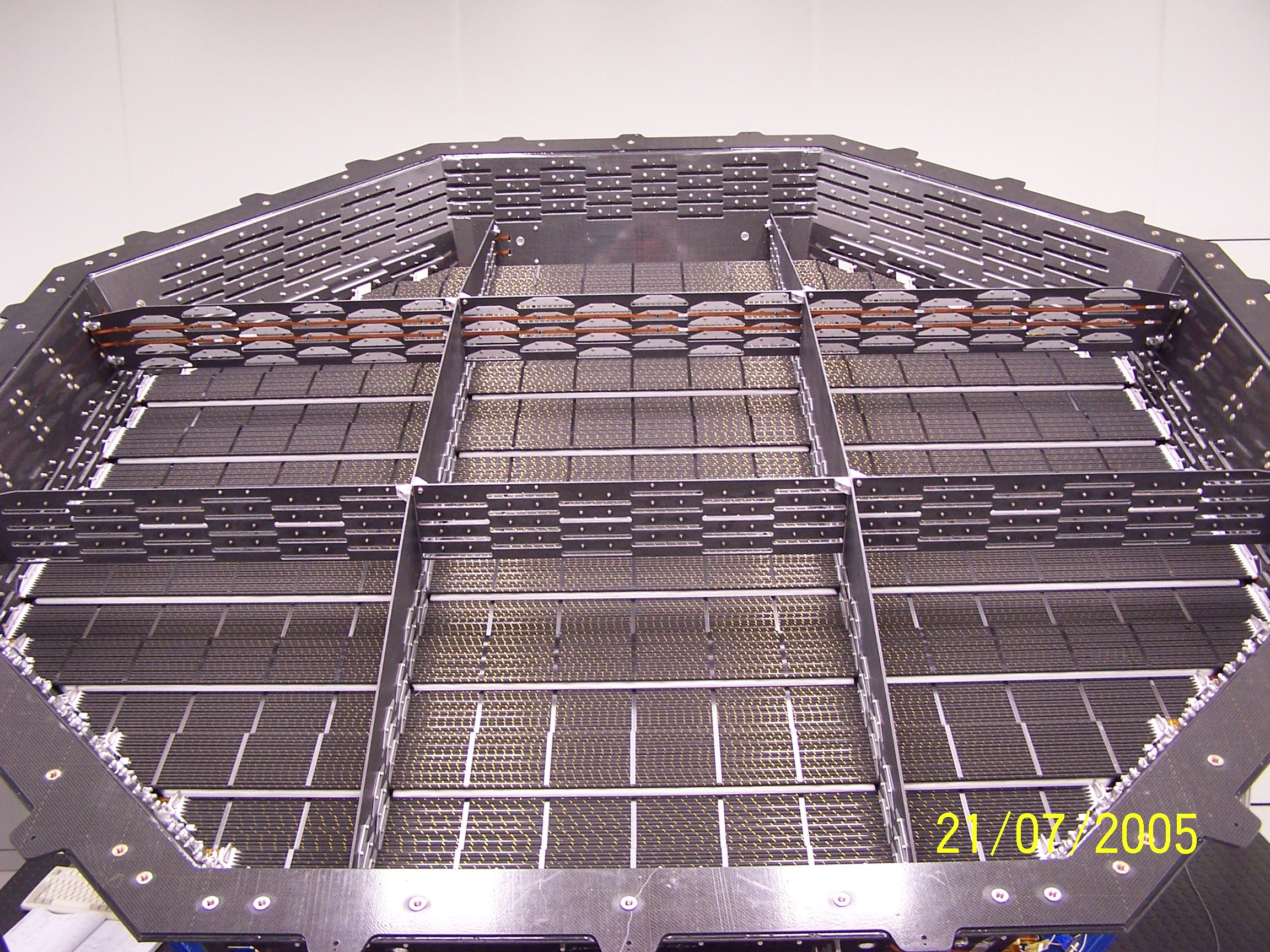}
  \end{minipage}
  \hspace{0.01\linewidth}
  \begin{minipage}{0.48\linewidth}
    \centering
    \includegraphics[width=1.0\linewidth]{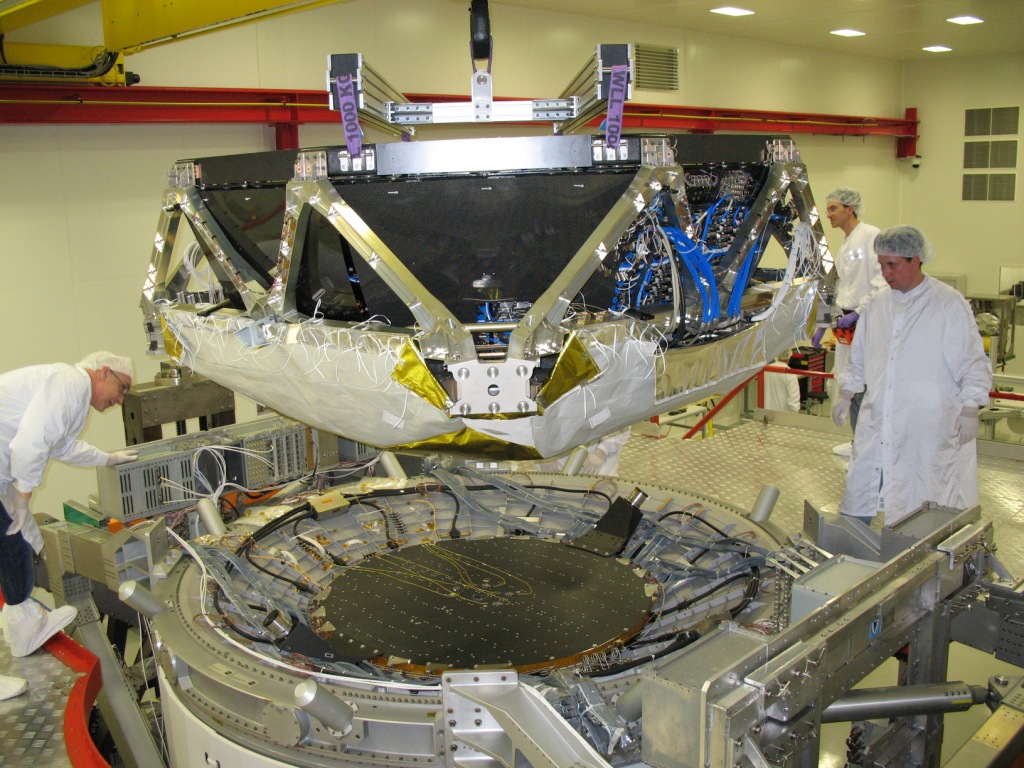}
  \end{minipage}
  \caption{Left: The partially assembled transition radiation
    detector~\cite{AMS_TRD_Schael_WWW}. The fleece radiator above the visible layer of tubes, as
    well as the four uppermost layers have yet to be installed. Right: The installation of the TRD
    on top of AMS~\cite{AMS_TRD_Schael_WWW}.}
  \label{fig:ams02-trd-photo}
\end{figure}

Figure~\ref{fig:ams02-trd-photo} shows photos of the TRD during its construction and
installation. The TRD support structure is a conical octagon (with larger circumference at the top
of AMS). Vertical carbon fiber walls with cutouts (the TRD bulkheads) are installed on the inside of
the octagon for additional support of the modules. The bulkheads can partially be seen in
figure~\ref{fig:ams02-trd-photo} at the top of the detector where the modules are not yet
installed. Lateral and longitudinal stiffeners reinforce the mechanical stability of the TRD
modules.

\begin{figure}[t]
  \centering
  \includegraphics[width=0.5\linewidth]{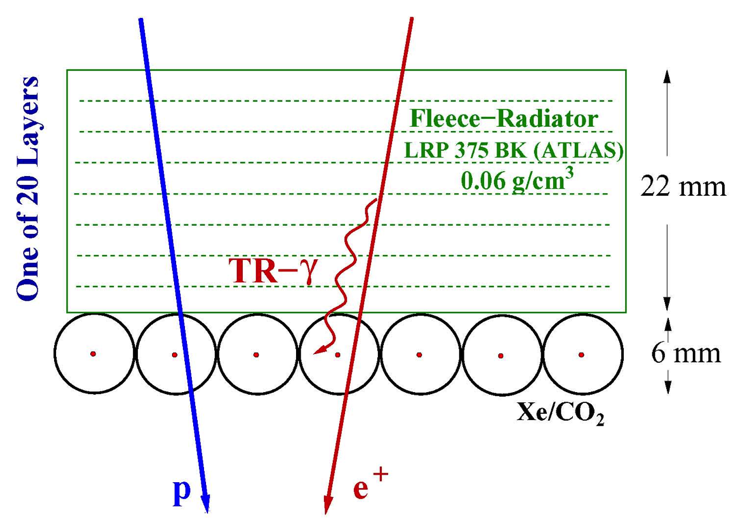}
  \caption{Schematic depiction of a section of a TRD straw tube module, including the fleece
    radiator above~\cite{AMS_TRD_Schael_WWW}.}
  \label{fig:ams02-trd-module-schematic}
\end{figure}

In total the TRD contains 328 straw tube modules, arranged in 20 layers with \SI{22}{\centi\meter}
of fleece radiator on top of each (see figure~\ref{fig:ams02-trd-module-schematic}). Within each
layer the modules are staggered in height in order improve the tracking capabilities. Each module
consists of 16 straw tubes with a diameter of \SI{6}{\milli\meter} filled with $\mathrm{Xe/CO_{2}}$
gas. The relative proportions of the two gas components varies with time, but is typically between
95/5 and 93/7. The length of the modules ranges from \SI{92}{\centi\meter} to \SI{2}{\meter},
depending on where they are placed in the octagon. In the center of each straw a tungsten anode wire
carries high voltage of approximately \SI{1400}{\volt} with respect to the outer multilayer aluminum
kapton foil which makes up the tubes themselves. The correct positioning of the central wire was
verified to an accuracy of \SI{100}{\micro\meter} using a CT scanner in a hospital in Aachen.

A charged particle passing through the straw tubes will ionize the gas along its track. The
liberated electrons will then drift towards the central wire on which the high voltage is
applied. In the vicinity of the wire the electrons will become energetic enough to ionize the gas
themselves, and an avalanche cascade sets in. The gas amplification transforms the charge carried by
the primary ionization into a measurable signal, proportional to the originally deposited
energy. The TRD is operated at a gas gain factor of approximately 3000. The $\mathrm{CO}_{2}$ is
used as a quenching gas in order to stop the avalanche process before the electrical field near the
wire is strong enough to cause corona discharges, which would cause high voltage trips. It also
allows for a faster recovery of the gas, which is important in case particles impinge with a high
rate.

The fleece radiators are made of many irregularly placed polyethylene / polypropylene fibers with
vacuum in between (since the TRD is operated in space) resulting in many boundaries with different
dielectric constants on each side. A particle passing through these boundaries may emit X-ray
transition radiation, depending on its Lorentz boost factor $\gamma = E / m$. The threshold for the
onset of transition radiation emission is located near $\gamma \backsim 200$. For higher values of
$\gamma$ up to approximately 10000 the probability to emit a TR photon is roughly proportional to
$\gamma$, after which it reaches a plateau.

Since protons are approximately 2000 times heavier than electrons and positrons the rate at which TR
photons are produced is strongly suppressed. The X-ray photons emitted by electrons and positrons
can be detected on top of the ionization signal leading to larger energy deposits in the
proportional tubes. This principle is illustrated in
figure~\ref{fig:ams02-trd-module-schematic}. The amplitude information from all 20 layers can then
be incorporated into a likelihood estimator in order to reliably identify the particle type.

The exact strength of the gas amplification depends on the following factors:

\begin{itemize}
\item The density of the gas: A lower density corresponds to a longer mean free path which
  corresponds to higher gas gain.
\item The fraction of $\mathrm{CO}_{2}$: A higher fraction corresponds to stronger damping of the
  gas amplification and thus lower gas gain.
\item The voltage between the anode wire and the tube wall: Larger HV values correspond to higher
  gas gain.
\end{itemize}

These gas amplification parameters vary with time, because of gas losses due to diffusion through
the tube walls. Monthly gas refills are employed to compensate these losses, which also keep the gas
mixture at the desired level. For this purpose a dedicated gas system containing refill supply
vessels with \SI{49}{\kilo\gram} of Xe and \SI{5}{\kilo\gram} of $\mathrm{CO}_{2}$ as well as a
mixing vessel, valves, heaters and pumps is mounted on the side of AMS.

Daily HV adjustments based on the observed gas gain are used to compensate the changes in the
amplification in order to keep the gain at an approximately constant level. In this process
variations between the gain in different parts of the detector are also equalized.

To exploit the full potential of the TRD a time dependent alignment procedure corrects movements and
rotations of the TRD octagon with respect to the tracker due to temperature variations. This is
important in order to correctly estimate the pathlength traversed by particles in each tube. A gain
calibration method based on cosmic ray proton ionization signals fully equalizes all channels and
corrects for time dependent gas amplification variations with an accuracy of approximately
\SI{1}{\percent}.

\begin{figure}[t!]
  \centering
  \includegraphics[width=0.6\linewidth,height=7cm]{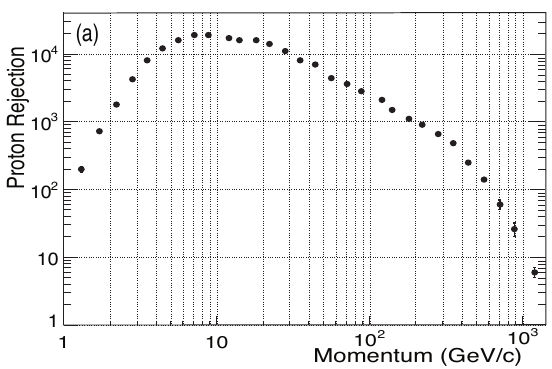}
  \caption{The proton rejection for the TRD alone after calibrations~\cite{AMS02_PosFrac1_2013}.}
  \label{fig:ams02-trd-proton-rejection}
\end{figure}

After alignment and gain calibration the TRD on orbit rejection power (defined as the electron
selection efficiency divided by the proton efficiency) at \SI{90}{\percent} electron efficiency over
the energy range of \SIrange{2}{200}{\giga\electronvolt} is estimated to be better than $10^{3}$,
exceeding the design specifications, as shown in figure~\ref{fig:ams02-trd-proton-rejection}.

\subsection{Electromagnetic Calorimeter}
\label{sec:detector-ecal}

\begin{figure}[t]
  \centering
  \includegraphics[width=0.7\linewidth]{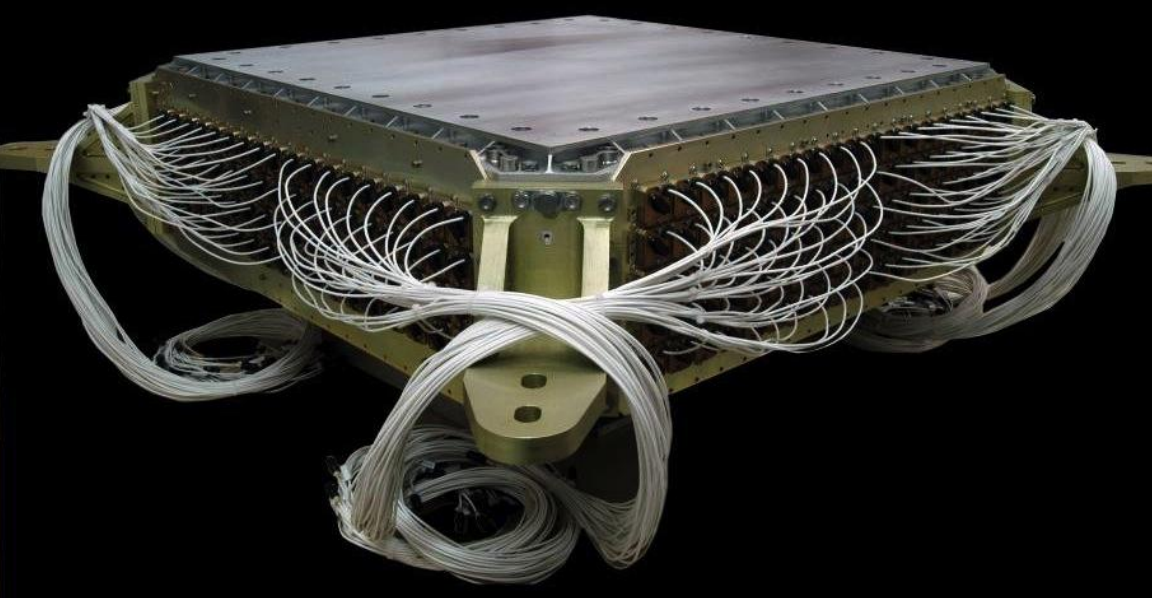}
  \caption{The \mbox{AMS-02} electromagnetic calorimeter~\cite{Schael_Talk_2018_WWW}.}
  \label{fig:ams02-ecal-photo}
\end{figure}

The \mbox{AMS-02} electromagnetic calorimeter (ECAL) is a fine grained lead-scintillating fiber
sampling calorimeter located at the bottom of the experiment, below tracker layer~9 with a total
depth of 17 radiation lengths. It was designed to accurately measure the energy and shower shape of
electrons, positrons and photons up to several \si{\tera\electronvolt}. The analysis of the shower
shape provides strong discrimination power to distinguish electrons and photons from protons. It is
also possible to reconstruct the shower axis which coincides with the direction of the incoming
particle. Figure~\ref{fig:ams02-ecal-photo} shows a picture of the fully assembled calorimeter
including the readout cabling.

\begin{figure}[t!]
  \centering
  \includegraphics[width=0.8\linewidth]{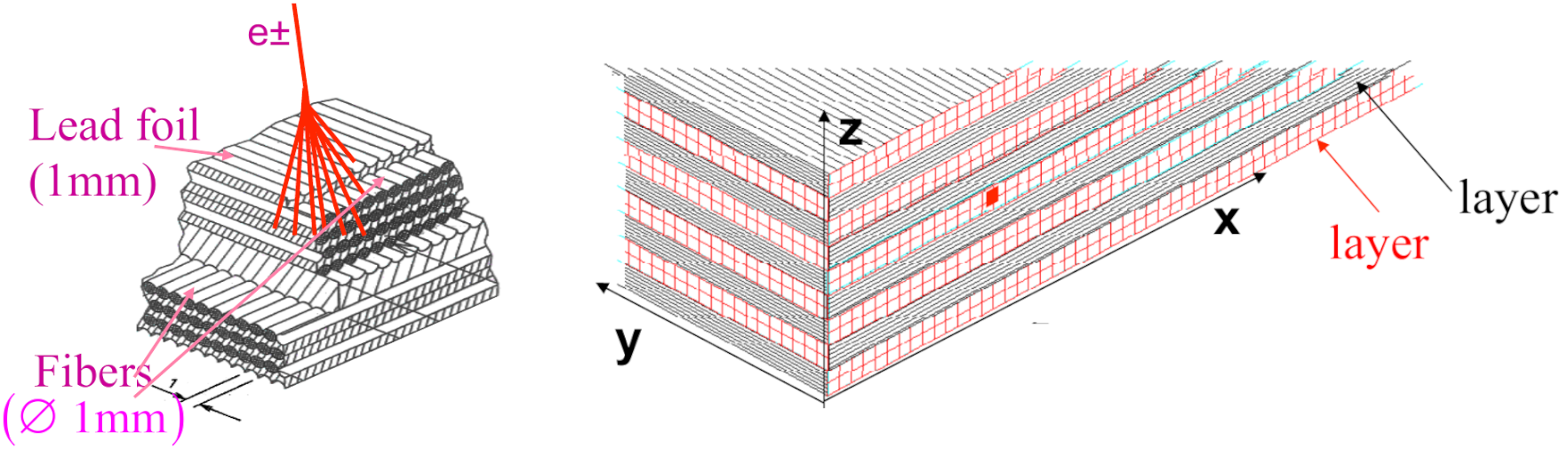}
  \caption{Left: Layout of the lead-fiber structure for two ECAL
    superlayers~\cite{AMS02_Detector_ECAL}. Right: Superlayer and cell arrangement in the
    calorimeter.}
  \label{fig:ams02-ecal-schematic}
\end{figure}

The active area of the calorimeter surface is \SI{64.8}{\centi\meter} $\times$
\SI{64.8}{\centi\meter} and the total thickness is \SI{16.6}{\centi\meter}. The total weight
including mechanical structure and read out cables is \SI{638}{\kilo\gram}. It is composed of 9
superlayers with a thickness of \SI{18.5}{\milli\meter}, each one consisting of 11 grooved,
\SI{1}{\milli\meter} thin lead foils, interleaved with scintillating fibers with a diameter of
\SI{1}{\milli\meter} (Figure~\ref{fig:ams02-ecal-schematic}).

The superlayers are alternatingly rotated along X (5 layers) and Y (4 layers) as shown in
figure~\ref{fig:ams02-ecal-schematic}. The light in the scintillating fibers is read out at the
sides of each superlayer by 36 four anode Hamamatsu R7600-00-M4 PMTs. Each PMT reads out 4 cells,
arranged in a 2 $\times$ 2 matrix. Therefore there are 72 $\times$ 2 cells per superlayer and each
superlayer can be subdivided into two distinct layers along the $z$-direction.

When an electron enters the calorimeter there is a high probability to emit a bremsstrahlung photon
because of the high Z value of the lead foil. This photon will then quickly convert into an
electron/positron pair. Both particles will then again emit bremsstrahlung photons, which leads to
the development of an electromagnetic shower. The total deposited energy in the fibers is then
proportional to the primary electron energy. The shower development has very characteristic shapes,
both in the longitudinal and in the lateral direction.

The above is also true for photons entering the calorimeter, which need to convert into an
electron/positron pair before the shower begins to develop. Therefore photon showers are almost
indistinguishable from electron showers~\footnote{Unless the electron radiates a hard photon in the
  upper detector, in which case the shower shape can be significantly different.}, except for the
fact that their longitudinal development is slightly displaced (on average by half a radiation
length).

Protons on the other hand usually pass through the calorimeter as minimum ionizing particles (MIPs),
leaving a trace only in the fibers through which they pass. Therefore a proton typically deposits
only a few hundred \si{\mega\electronvolt} of energy in the fibers compared to electrons which are
fully absorbed. A proton signal is also typically much narrower compared to an electromagnetic
shower and does not feature the characteristic longitudinal shower shape.

\begin{figure}[t]
  \centering
  \includegraphics[width=0.6\linewidth]{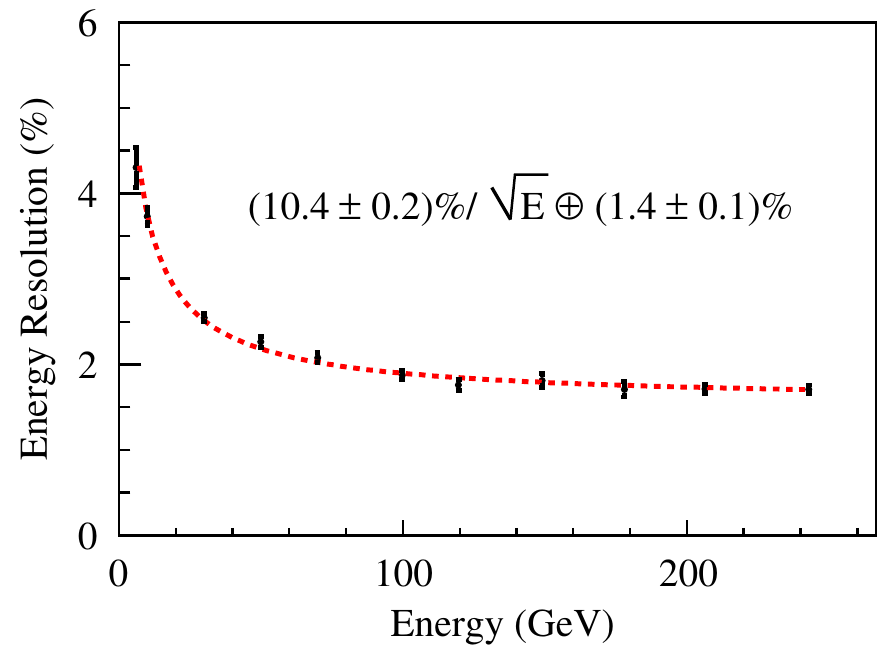}
  \caption{The ECAL energy resolution for electrons measured in the 2010 beamtest at CERN's
    SPS~\cite{AMS02_Detector_ECAL}.}
  \label{fig:ams02-ecal-energy-resolution}
\end{figure}

The calorimeter was fully tested and calibrated together with the rest of AMS in a beamtest in 2010
at CERN. Figure~\ref{fig:ams02-ecal-energy-resolution} shows the resulting energy resolution for
electrons~\cite{AMS02_Detector_ECAL}. It can be described by:

\begin{displaymath}
  \frac{\sigma_E}{E} = \frac{\SI{10.4}{\percent}}{\sqrt{E}} \oplus \SI{1.4}{\percent} \,,
\end{displaymath}

which shows the excellent resolution at high energies. In addition to the event-by-event uncertainty
in the determination of the energy it is important to know the absolute energy scale. Within the
energy range covered by the beamtest (\SIrange{10}{290}{\giga\electronvolt}) the associated
systematic uncertainty on the energy scale is \SI{2}{\percent} and increases to \SI{5}{\percent} at
\SI{0.5}{\giga\electronvolt} and to \SI{4}{\percent} at
\SI{700}{\giga\electronvolt}~\cite{AMS02_ElecPos_2014}. The energy scale is verified using flight
data by the $E / R$ peak position for electrons, positrons and minimum ionizing particles, where the
rigidity is measured by the tracker.

Combining the discriminating power of the ECAL shower shape analysis with the matching of energy and
momentum measured by the tracker a proton rejection above $10^4$ between \SI{3}{\giga\electronvolt}
and \SI{500}{\giga\electronvolt} was achieved, which allows to select very pure electron and
positron samples when combined with the TRD.

\begin{figure}[t!]
  \centering
  \includegraphics[width=0.6\linewidth, height=7cm]{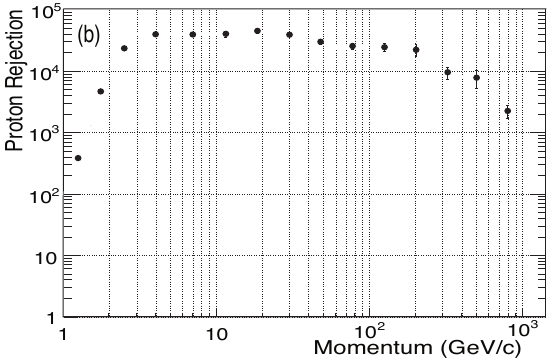}
  \caption{The proton rejection for the combination of ECAL and tracker~\cite{AMS02_PosFrac1_2013}.}
  \label{fig:ams02-ecal-proton-rejection}
\end{figure}

The calorimeter is also equipped with a standalone trigger for the measurement of photons. These
events pass the entirety of AMS and only convert into an electron / positron pair in the
calorimeter. Therefore the trigger from the TOF system is
inadequate. Section~\ref{sec:detector-trigger-calorimeter} discusses the photon specific calorimeter
trigger in detail.

In the analysis of photons the calorimeter is crucial, since it provides all observable quantities,
most notably the energy of the photon and its incoming direction.

\subsection{Ring Imaging Cherenkov Detector}
\label{sec:detector-rich}

\begin{figure}[t!]
  \centering
  \begin{minipage}{0.48\linewidth}
    \centering
    \includegraphics[width=1.0\linewidth]{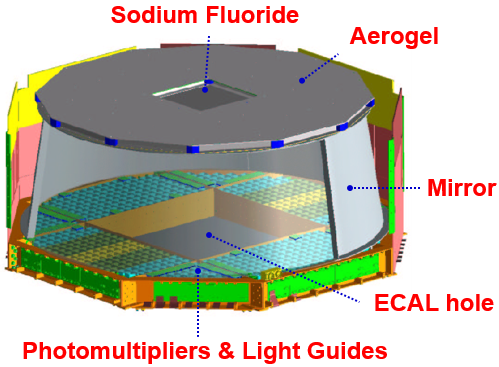}
  \end{minipage}
  \hspace{0.01\linewidth}
  \begin{minipage}{0.48\linewidth}
    \centering
    \includegraphics[width=0.8\linewidth]{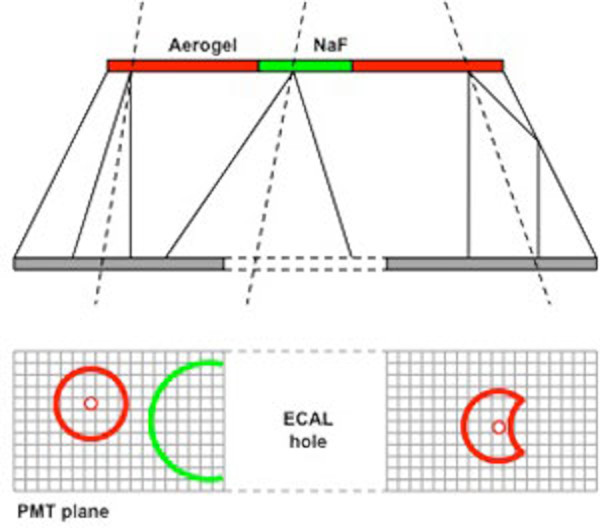}
  \end{minipage}
  \caption{Left: Overview of the components of the RICH detector\cite{Li_2017}. Right: Illustration
    of the various forms of Cherenkov rings, depending on where the particle passes through the
    radiator.}
  \label{fig:ams02-rich-schematic}
\end{figure}

The Ring Imaging Cherenkov Detector is located between the lower TOF and the tracker layer 9. The
left panel of figure~\ref{fig:ams02-rich-schematic} shows a schematic view of the detector. At the
top of the RICH a plane of \SI{0.5}{\centi\meter} thick radiator tiles causes relativistic particles
to emit Cherenkov light when they pass through, if their velocity is above the Cherenkov threshold
$\beta > 1/n$. At a vertical distance of \SI{47}{\centi\meter} below the radiator tiles the
detection plane consists of an array of 680 4x4 multi-anode photomultiplier tubes which detect the
Cherenkov light cone as well as the passage of the primary particle itself. A
\SI{64x64}{\centi\meter\squared} hole in the center of the detection plane assures that particles
can reach tracker layer 9 and the calorimeter undisturbed. The PMTs are arranged in rectangular and
triangular blocks around the central gap. Surrounding the expansion volume, between the radiator and
the detection plane, is a high-reflectivity conical mirror which ensures that the Cherenkov light is
reflected on to the detection plane even for particles which pass near the borders of the radiator.

The radiator plane consists of 4x4 tiles of sodium fluoride (NaF) in the center and 92 surrounding
silica aerogel (Agl) tiles with refractive indices of 1.33 and 1.05 respectively. Due to the higher
refractive index of the sodium fluoride tiles the opening angle of the Cherenkov cone is larger,
which means that rings are detected even if the particle itself passes through the central hole.

\begin{table}[b]
  \centering
  \caption{Cherenkov threshold rigidity for various light isotopes.}
  \vspace*{5mm}
  \label{tab:ams02-rich-thresholds}
  \begin{tabular}[h]{c|cc|cc|cc|ccc}
    & $^1\mathrm{H}$ & $^2\mathrm{H}$
    & $^3\mathrm{He}$ & $^4\mathrm{He}$
    & $^6\mathrm{Li}$ & $^7\mathrm{Li}$
    & $^7\mathrm{Be}$ & $^9\mathrm{Be}$ & $^{10}\mathrm{Be}$ \\
    \hline
    NaF: Rigidity / GV
    & 1.07 & 2.14
    & 1.60 & 2.13
    & 2.13 & 2.48
    & 1.86 & 2.39 & 2.66 \\
    Agl: Rigidity / GV
    & 2.93 & 5.86
    & 4.39 & 5.82
    & 5.83 & 6.80
    & 5.10 & 6.56 & 7.28 \\
  \end{tabular}
\end{table}

Table~\ref{tab:ams02-rich-thresholds} lists the Cherenkov rigidity thresholds for the relevant light
isotopes in cosmic rays. In practice the rigidity needs to be slightly larger than the theoretical
threshold, otherwise the opening angle is too small.  The corresponding thresholds in kinetic energy
per nucleon are \SI{0.48}{\giga\electronvolt\per\nucleon} (NaF) and
\SI{2.13}{\giga\electronvolt\per\nucleon} (Agl) for all isotopes.

The panel on the right hand side of figure~\ref{fig:ams02-rich-schematic} illustrates this
principle. The first particle from the left radiates Cherenkov light in an aerogel tile, which is
reconstructed as a ring. The central particle passes through the sodium fluoride radiator, which
causes emission of light with a larger opening angle (green partial ring). On the right side a
particle passes near the border of the RICH radiator plane with a larger inclination angle, the
Cherenkov photons are reflected on the mirror, and a partially inverted ring is reconstructed in the
detection plane.

The opening angle of the Cherenkov cone is directly accessible after the reconstruction of the ring,
which allows to measure the velocity:

\begin{displaymath}
  \beta = \frac{1}{cos(\theta) n} \,.
\end{displaymath}

\begin{figure}[t]
  \centering
  \begin{minipage}{0.48\linewidth}
    \centering
    \includegraphics[width=1.0\linewidth]{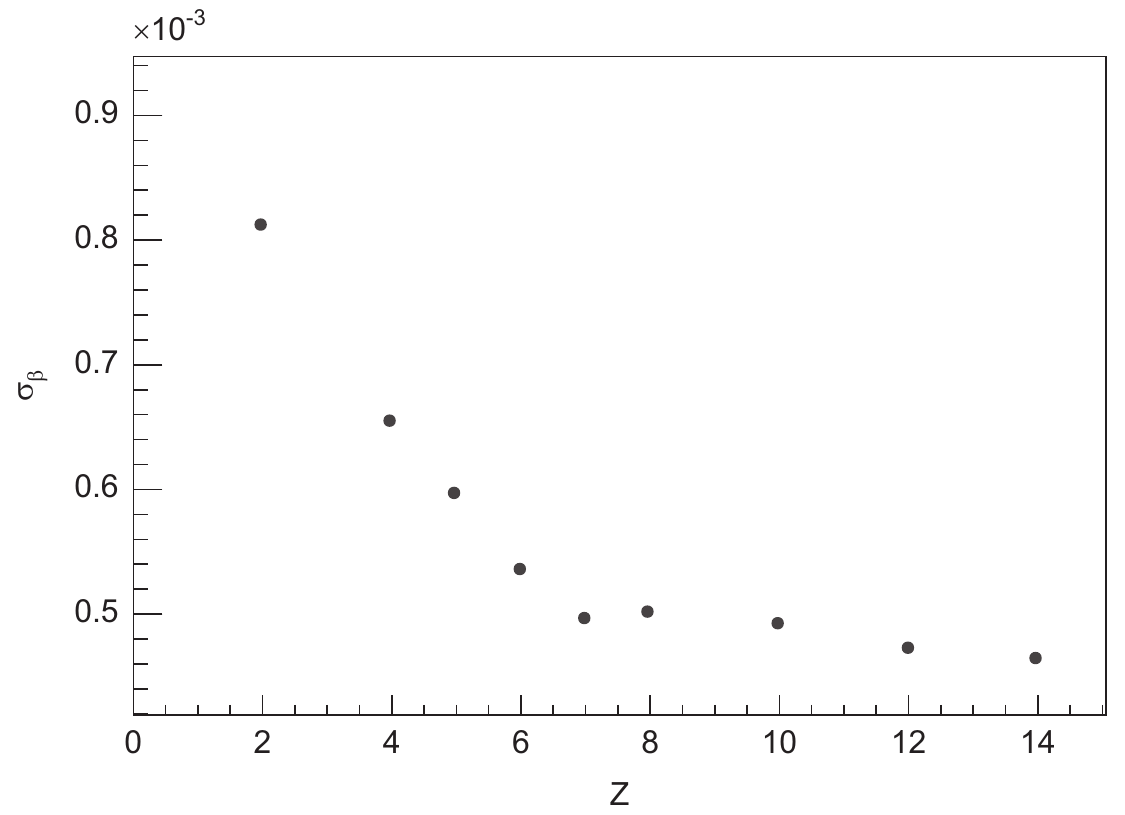}
  \end{minipage}
  \hspace{0.01\linewidth}
  \begin{minipage}{0.48\linewidth}
    \centering
    \vspace{2mm}
    \includegraphics[width=1.0\linewidth]{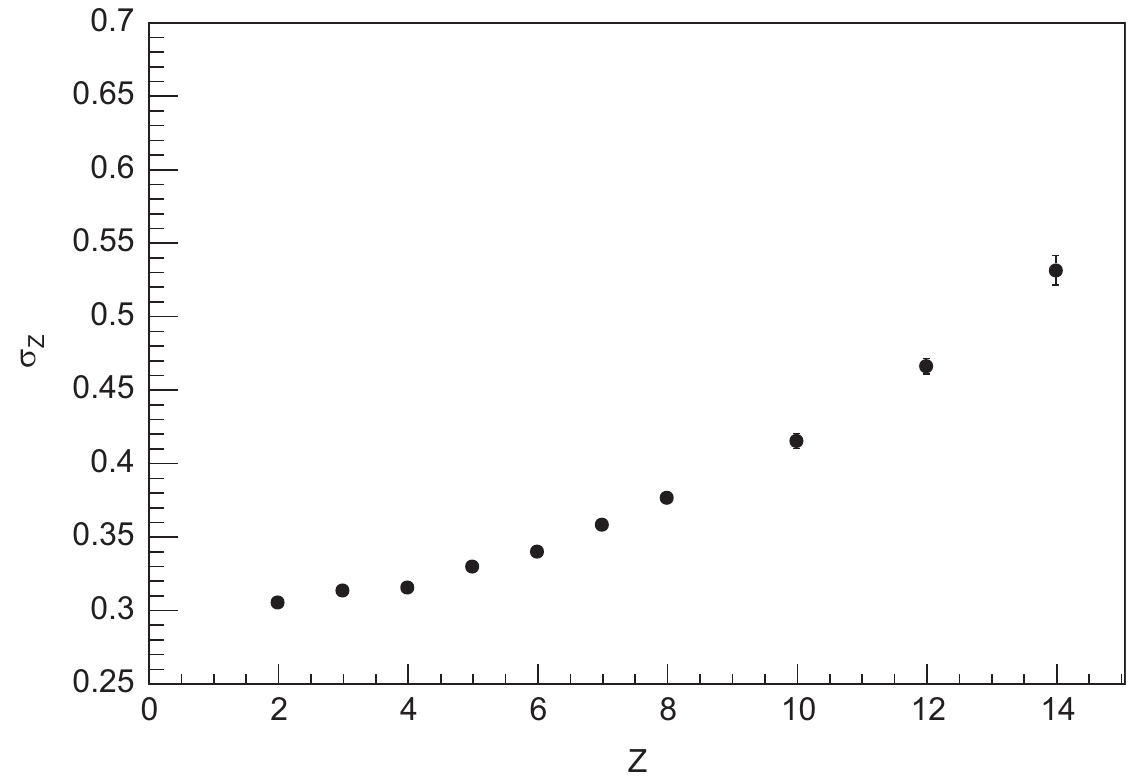}
  \end{minipage}
  \caption{Left: Velocity resolution of the RICH detector after calibrations. Right: Charge
    resolution of the RICH detector after calibrations~\cite{AMS02_Detector_RICH}.}
  \label{fig:ams02-rich-resolution}
\end{figure}

The velocity resolution of the RICH is better than \SI{1}{\permille} for protons and improves for
heavier nuclei, for oxygen and above it is better than \SI{0.5}{\permille}, as shown in
figure~\ref{fig:ams02-rich-resolution}. The RICH detector is, due to its excellent velocity
resolution, essential in the reconstruction of nuclear isotopes. The number of detected
photo-electrons on the ring, as well as the amplitude of the PMT anode signals allow to estimate the
particle charge Z with a relative resolution of \SI{30}{\percent} for protons and \SI{3.7}{\percent}
for silicon, see right panel of figure~\ref{fig:ams02-rich-resolution}.

\section{Trigger and Data Acquisition}
\label{sec:detector-trigger}

The AMS detector has multiple triggers which can cause the recording of a given event, each one is
optimized for the analysis of a specific class of physics events. The trigger board forms a decision
on whether or not to record the event based on three input subdetectors: The TOF, the ACC and the
ECAL.

There are three amplitude thresholds for each TOF counter:

\begin{enumerate}
\item The low threshold (LT) at approximately \SI{20}{\percent} of the MIP signal, used for the time
  measurement.
\item The high threshold (HT) at approximately \SI{50}{\percent} of the MIP signal, used for the
  trigger of |Z| = 1 particles.
\item The super-high threshold (SHT) at approximately four times the MIP signal, used for the
  trigger of |Z| >= 2 particles.
\end{enumerate}

These per-counter signals are then combined within the TOF electronics to form three output signals
per TOF layer:

\begin{enumerate}
\item The ``charged particle'' (CP) signal:\\
  At least one counter with either side exceeding the HT.
\item The ``central track'' (CT) signal:\\
  At least one counter with either side exceeding the HT, but counters 1 and 10 in TOF layer 3 are
  not considered.
\item The ``big-Z'' (BZ) signal:\\
  At least one counter with either side exceeding the SHT.
\end{enumerate}

The four CP, CT and BZ signals are then sent to the AMS trigger board where the logic computes four
different outputs, based on lookup tables, which combine the information from the four layers:

\begin{enumerate}
\item FTCP0: CP signal in at least three out of the four layers.
\item FTCP1: CP signal in all four layers.
\item FTCT0: CT signal in at least three out of the four layers.
\item FTCT1: CT signal in all four layers.
\end{enumerate}

The FTC fast trigger is generated if either FTCP0 or FTCP1 is set, which effectively means that the
FTC fast trigger decision is identical to FTCP0 for the settings deployed in the AMS flight
configuration. The FTZ fast trigger for ions is generated from the coincidence within
\SI{640}{\nano\second} of the BZ signals from all four layers


There also is a fast trigger from the electromagnetic calorimeter (FTE), which is generated if the
last-dynode amplitudes in the calorimeter superlayers 2 to 7 fulfill the following requirements:
Because the readouts of the two projections (XZ for superlayers 2, 4 and 6 and YZ for superlayers 3,
5 and 7) of the calorimeter are physically disconnected, each projection makes an independent
decision first. For each projection the amplitude in each superlayer is compared with a
layer-dependent threshold. The thresholds were optimized for \SI{90}{\percent} efficiency for
\SI{2}{\giga\electronvolt} photons using the Monte-Carlo simulation. If at least two out of the
three superlayers exceed the threshold the XF (or YF) signal is sent to the trigger board. The
trigger board then generates the fast trigger FTE by building the logical OR of XF and YF, i.e. if
either projection FTE is generated if either projection has two out of three layers above
thresholds.

The global fast trigger FT is generated if either FTC, FTZ or FTE is set, which opens a
\SI{240}{\nano\second} gate to latch all the signals from TOF, ACC and ECAL for the final level 1
trigger decision. The input signals to form the trigger decisions are:

\begin{enumerate}
\item FTCP0: Signal in at least three out of the four TOF layers (as above)
\item FTCT0: Signal in at least three out of the four TOF layers, but excluding the edge counters in
  TOF layer 3 (as above)
\item FTCP1: Signal in all four TOF layers (as above)
\item FTCT1: Signal in all four TOF layers, but excluding the edge counters in TOF layer 3 (as
  above)
\item FTZ: Fast trigger for nuclei with \SI{640}{\nano\second} gate (as above)
\item FTE: Fast trigger from the calorimeter (as above)
\item ACC0: Number of ACC hits is zero.
\item ACC1: Number of ACC hits is less than 5/8 \footnote{The setting was less than 5 ACC hits until
    26th of February 2016 when it was changed to less than 8 hits to improve the trigger efficiency for
    very heavy nuclei such as iron.}.
\item BZ: TOF nuclei trigger (BZ)
\item ECALF|: Signal above threshold for at least 2 out of 3 superlayers in either XZ or YZ plane
\item ECALF\& Signal above threshold for at least 2 out of 3 superlayers in both XZ and YZ planes
\item ECALA|: Shower zenith angle less than \SI{20}{\degree} in either XZ or YZ plane
\item ECALA\&: Shower zenith angle less than \SI{20}{\degree} in both XZ and YZ planes
\end{enumerate}

On the basis of these inputs the following physics trigger branches are defined:

\begin{enumerate}
\item Single charge trigger for |Z| = 1 particles as well as light ions.\\
  This trigger is generated if FTCT1 and ACC0 are set.
\item Fast ion trigger for analysis of heavier nuclei.\\
  This trigger is generated if BZ and ACC1 are set.
\item Slow ion trigger for analysis of heavier nuclei, with a longer gate.\\
  This trigger is generated if FTZ is set.
\item Electron trigger: For analysis of electrons and positrons.\\
  This trigger is generated if FTCT1 and ECALF\& are set.
\item Photon trigger: For photons which enter the calorimeter.\\
  This trigger is generated if ECALA\& is set.
\item Unbiased TOF trigger: For TOF trigger efficiency measurement.\\
  This trigger is generated if FTCP0 is set and the prescaling condition is fulfilled.\\
  A prescaling factor of 100 is applied.
\item Unbiased ECAL trigger: For ECAL trigger efficiency measurement.\\
  This trigger is generated if ECALF| is set and the prescaling condition is fulfilled.\\
  A prescaling factor of 1000 is applied.
\end{enumerate}

These are generated from the signals of the four TOF planes and from the calorimeter.

\subsection{Trigger for Converted Photons}
\label{sec:detector-trigger-conversions}

For the majority of the events the trigger for converted photons in \mbox{AMS-02} is the single
charge trigger. The two ion triggers have TOF energy deposition thresholds which are generally not
reached, and the two triggers which involve the calorimeter do not contribute substantially due to
the smaller calorimeter acceptance and due to the energy threshold of the calorimeter trigger.

The single charge trigger fires if there is signal in all four TOF layers, signal in the central 8
layers of TOF layer 3 and no hit in the ACC counters. This configuration has several implications:

\begin{itemize}
\item A photon conversion in the second TOF layer will not trigger the experiment.
\item In case either the electron or the positron is bent out by the magnet and hits the ACC, the
  trigger is vetoed.
\item At least one of the two tracks must pass through one of the 8 central TOF counters in layer 3.
\end{itemize}

In the analysis the focus will therefore be on events which convert in the first TOF layer or just
above in the support material.

Measurements of the trigger efficiency from ISS data are possible due to the availability of the
unbiased TOF trigger. In addition, large parts of the inefficiency due to the geometric effects
(tracks need to pass through the central layers in TOF layer 3, and must not hit the ACC) can be
easily studied using the Monte-Carlo simulation.

\subsection{Trigger for Calorimeter Photons}
\label{sec:detector-trigger-calorimeter}

The trigger for photons which enter the calorimeter is the dedicated photon trigger, all others
require signal in all four TOF layers. The energy thresholds imposed by the calorimeter trigger
therefore limit the energy range which is accessible with this class of events to about
\SI{1}{\giga\electronvolt} and above.

The angular cut on the trigger is important to reduce the trigger rate for events which do not pass
through the upper detector.

It is possible to measure the trigger efficiency using the unbiased ECAL trigger. However the
prescaling factor of 1000 limits the available statistics. An alternative approach is to exploit the
fact that showers initiated by electrons are almost identical to photon showers. This allows to
measure the photon trigger efficiency using electrons, which are triggered by the TOF.


\emptypage


%% file: analysis.tex

\addtocontents{toc}{\protect\vspace{1cm}}
\chapter{Data Analysis}
\label{sec:data-analysis}

In this chapter the datasets on which the analysis is based will be briefly explained. The selection
steps for photons in both detection modes will be presented. Based on these selections the
Instrument Response Functions (IRFs) will be derived: The effective area, the point-spread function
and energy resolution relations will be determined from a simulated set of Monte-Carlo events for
both conversion and calorimetric mode. The exposure maps will be presented, which directly relate
the photon flux to the observed number of events. Based on the exposure a full sky model for
$\gamma$-rays will be constructed. A few important corrections for the simulation will be presented
and systematic uncertainties will be discussed.

\section{Datasets}
\label{sec:analysis-datasets}

\subsection*{ACQt File Format and ACsoft Software}

The \mbox{AMS-02} experiment produces enormous amounts of scientific data which require significant
computing resources to store, process and analyze. The primary data format contains the full
information and allows for flexible re-calibration and re-fitting of the data as needed. At the same
time physicists often only need access to high level information, such as the estimated particle
rigidity from a track fit.

The continuous cycle of making changes, reprocessing the data and evaluation of the result is at the
heart of the day-to-day analysis work. Therefore it is vitally important that the delay between
changes in the analysis chain and the appearance of the reprocessed final result is kept as short as
possible, enabling active experimentation and also encouraging creativity. With this ultimate goal
in mind a custom file format for the \mbox{AMS-02} data was developed in Aachen by Nikolas
Zimmermann, Thorsten Siedenburg, Henning Gast and myself. The ACQt data format and the accompanying
ACsoft software package for analysis is used by the entire AMS Aachen group and was the primary data
source for multiple important publications~\cite{AMS02_ElecPos_2014,AMS02_ElecPosTime_2018}.

The ACQt data format is highly flexible, efficiently compressed and was optimized for parallel
processing in large computing clusters. It was shown to scale linearly to thousands of cores, which
makes it possible to complete a full reprocessing of the entire AMS-02 data within a few hours for a
typical analysis. At the same time the original AMS data can be stored at only \SI{10}{\percent} of
its original size.

The ACsoft software package contains all the tools required for the analysis of ACQt data including
a complete framework for the implementation of cuts, tools for automatic calculation of acceptance
and tag and probe efficiencies, template fitting and unfolding. It also contains tools to fully
automate the entire process of analysis, from batch job submission to the creation of the final
figures, which ensures that the results are reproducible and minimizes the required effort.

\subsection*{ISS Data}

The data which is analyzed in this thesis was collected by \mbox{AMS-02} on the International Space
Station between May 19th 2011 and November 12th 2017, spanning approximately 6.5 years in total. In
this period the detector recorded more than 106 billion events.

The events were reconstructed with AMS software versions ranging from B950 to B961, depending on the
exact period in which they were recorded. The differences between the versions from B950 to B961 are
marginal.

The data was processed with Aachen AMS software package ACsoft in version 7.6.0.0, 7.6.0.1 and
7.6.0.6, producing ACQt files with version 7.6. These ACQt files form the basis for data
analysis. The data is subsequently processed and converted into ROOT~\cite{ROOT_1997} trees, which
contain only a few variables relevant to the analysis of $\gamma$-rays.

\subsection*{Monte-Carlo Data}

The Monte-Carlo simulations used to determine the effective area, point-spread functions and energy
resolution matrices were done using a full detector simulation based on the
Geant4~\cite{Agostinelli2003} package (version 10.3 patch 3). In the simulation the entire
\mbox{AMS-02} detector is modeled in detail, including support structure and sensitive detector
elements. The electromagnetic physics processes used in the simulation correspond to the default
Geant4 electromagnetic processes, with some minor tunings applied by the AMS collaboration, in order
to minimize differences between data and simulation.

The AMS Monte-Carlo software version is B1133 and the same version was used to reconstruct the
simulated data. Photons were generated uniformly on a plane of size
$\SI{3.9}{\meter} \times \SI{3.9}{\meter}$ located \SI{1.95}{\meter} above the center of the
experiment. Their angular distribution is isotropic. Trajectories with large zenith angles would not
pass the detector. They are registered for the correct calculation of the effective area
(section~\ref{sec:analysis-effective-area}), but are immediately discarded in case
$\cos{\theta} < -0.7$ which corresponds to an angle of approximately \SI{45}{\degree}.

After the simulation is performed the signals in the various detector elements are digitized in
order to mimic the detector response. This step also includes the simulation of the \mbox{AMS-02}
trigger. The software then discards all events which did not trigger the experiment.

Three disjunct photon datasets were simulated:

\begin{itemize}
\item Low Energy: \SIrange{0.05}{0.25}{\giga\electronvolt}, approximately $52.6 \cdot 10^{9}$
  generated events.
\item Mid Energy: \SIrange{0.25}{10.0}{\giga\electronvolt}, approximately $35.8 \cdot 10^{9}$
  generated events.
\item High Energy: \SIrange{10.0}{1000.0}{\giga\electronvolt}, approximately $5.5 \cdot 10^{9}$
  generated events.
\end{itemize}

\begin{figure}[t]
  \centering
  \includegraphics[width=0.9\linewidth]{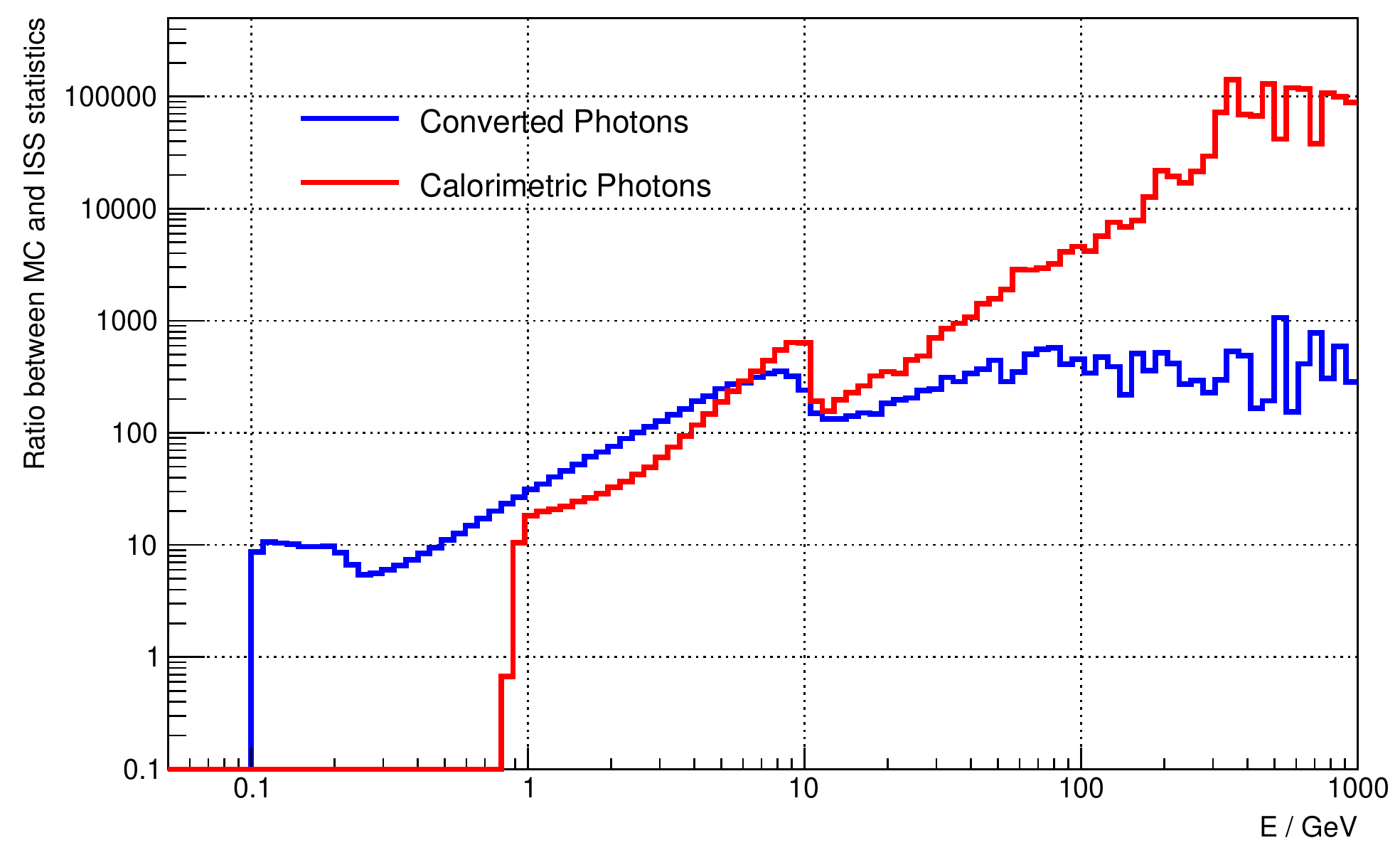}
  \caption{Ratio of the number of events in the Monte-Carlo simulation to the ISS data after
    applying the full set of selection cuts for converted (blue) and calorimetric (red) photons.}
  \label{fig:mc-statistics-ratio}
\end{figure}

In each dataset the energy distribution is flat in $\log(E)$, which corresponds to a power law flux
with spectral index of -1: $\Phi(E) \mathrm{d}E \thicksim E^{-1} \mathrm{d}E$. The available
Monte-Carlo statistics exceeds the number of gamma events in the ISS data by at least one order of
magnitude for almost all photon energies as shown in figure~\ref{fig:mc-statistics-ratio}.

The AMS Monte-Carlo version B1133 incorporates the latest understanding of the detector and its
behavior into the simulation. In particular this version includes effects such as fiber level
saturation in the calorimeter digitization procedure~\cite{AMS02_Detector_Ecal3D}, which is critical
for the correct estimation of the photon energy from calorimeter showers at high energies (E $\gg$
\SI{100}{\giga\electronvolt}). Other parameters relevant to the simulation of electromagnetic
showers were tuned in accordance with both flight and beamtest data.

Extensive tuning of the material budget in order to correctly estimate the multiple scattering of
charged particles has been performed. The material budget was also tuned to ISS nuclei data by
measuring the rate of hadronic interactions in the detector. In the TRD a direct comparison of the
number of interactions between data and Monte-Carlo was carried out and the simulation subsequently
adjusted~\cite{AMS02_Detector_TRD_MC_2017}. The correct description of the material budget is vital
for the photon conversion analysis, since it directly influences the rate with which photons convert
in the relevant parts of the detector.

In addition the tracker resolution as well as elastic and inelastic cross sections were extensively
checked and subsequently optimized to obtain a good agreement with ISS proton, electron and nuclei
data~\cite{AMS02_Proton_2015,AMS02_Helium_2015}.

Similarly to the ISS data the simulated events were processed with the Aachen software package
ACsoft (version 7.7.0.0), resulting in ACQt files with version 7.7 which were used in the subsequent
data analysis steps.

\section{Data Selection}
\label{sec:analysis-selection}

Clean samples of high energy $\gamma$-rays are difficult to obtain because of the large isotropic
background of charged particles. Even in the galactic center, where the flux of $\gamma$-rays is
large, the ratio between photons and galactic cosmic ray protons is smaller than $\num{e-4}$. Close
to the Earth's geomagnetic poles and in the vicinity of the SAA there is an additional background
from secondary protons and electrons, which reduces the signal to background ratio further.

A small contamination of background events in the dataset can be subtracted, because the background
flux is mostly isotropic, which is very different from the highly structured photon signal
flux. Nevertheless, in order to keep the background contamination of the selected event sample below
\SI{10}{\percent} a background rejection of $\num{e6}$ or better is needed. In some cases this
strict requirement implies a small loss in photon signal efficiency.

In this section two complementary selections will be presented: The first one aims to identify
$\gamma$-rays which convert in the upper TOF or directly above it. The second one targets photons
which pass through the detector and shower in the calorimeter, leaving no signal in the upper
detector.

\begin{figure}[t]
  \centering
  \includegraphics[width=0.8\linewidth]{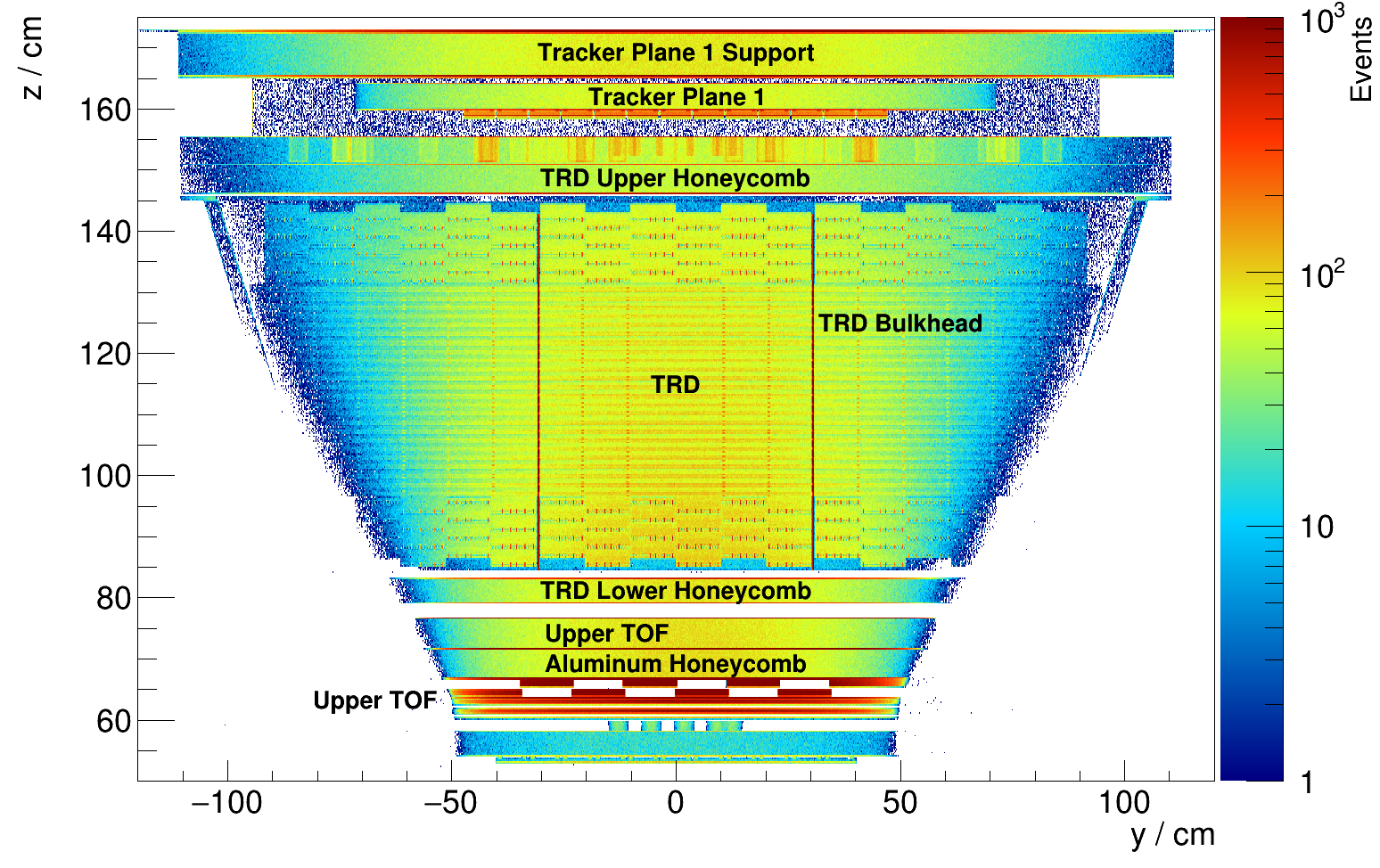}
  \caption{Positions of photon conversion vertices in the YZ plane according to the Monte-Carlo
    simulation.}
  \label{fig:analysis-conversion-points-all-yz}
\end{figure}

Figure~\ref{fig:analysis-conversion-points-all-yz} shows the YZ-distribution of photon conversion
vertices in the upper detector according to the Monte-Carlo simulation. The detector component with
the most photon conversions is the upper TOF, followed by the support structure directly
above. However, there are also quite a few photon conversions in the TRD, in particular close to the
vertical support structures (the TRD bulkheads).

\begin{figure}[t]
  \begin{minipage}{0.48\linewidth}
    \centering
    \includegraphics[width=1.0\linewidth]{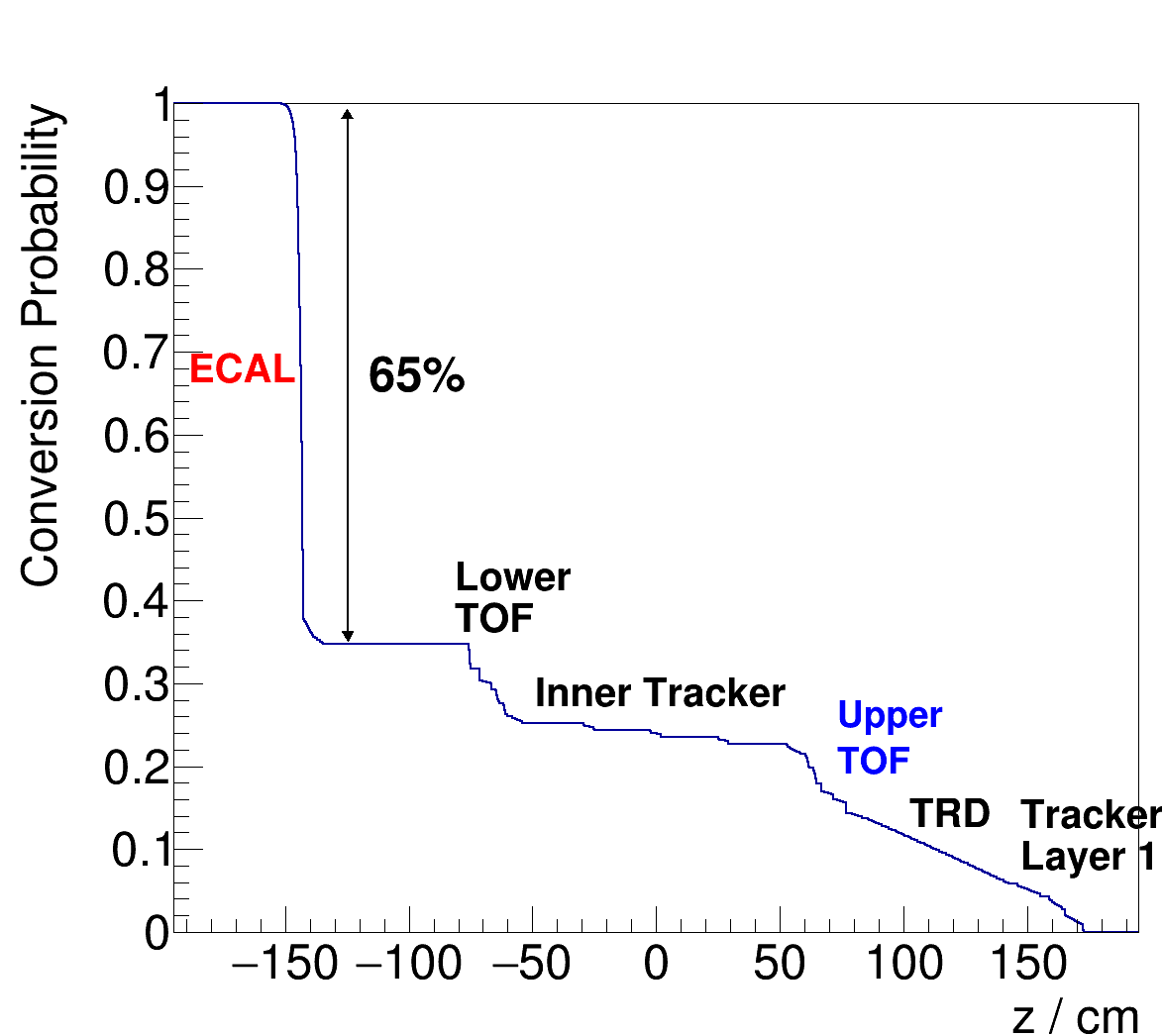}
  \end{minipage}
  \hspace{0.01\linewidth}
  \begin{minipage}{0.48\linewidth}
    \centering
    \includegraphics[width=1.0\linewidth]{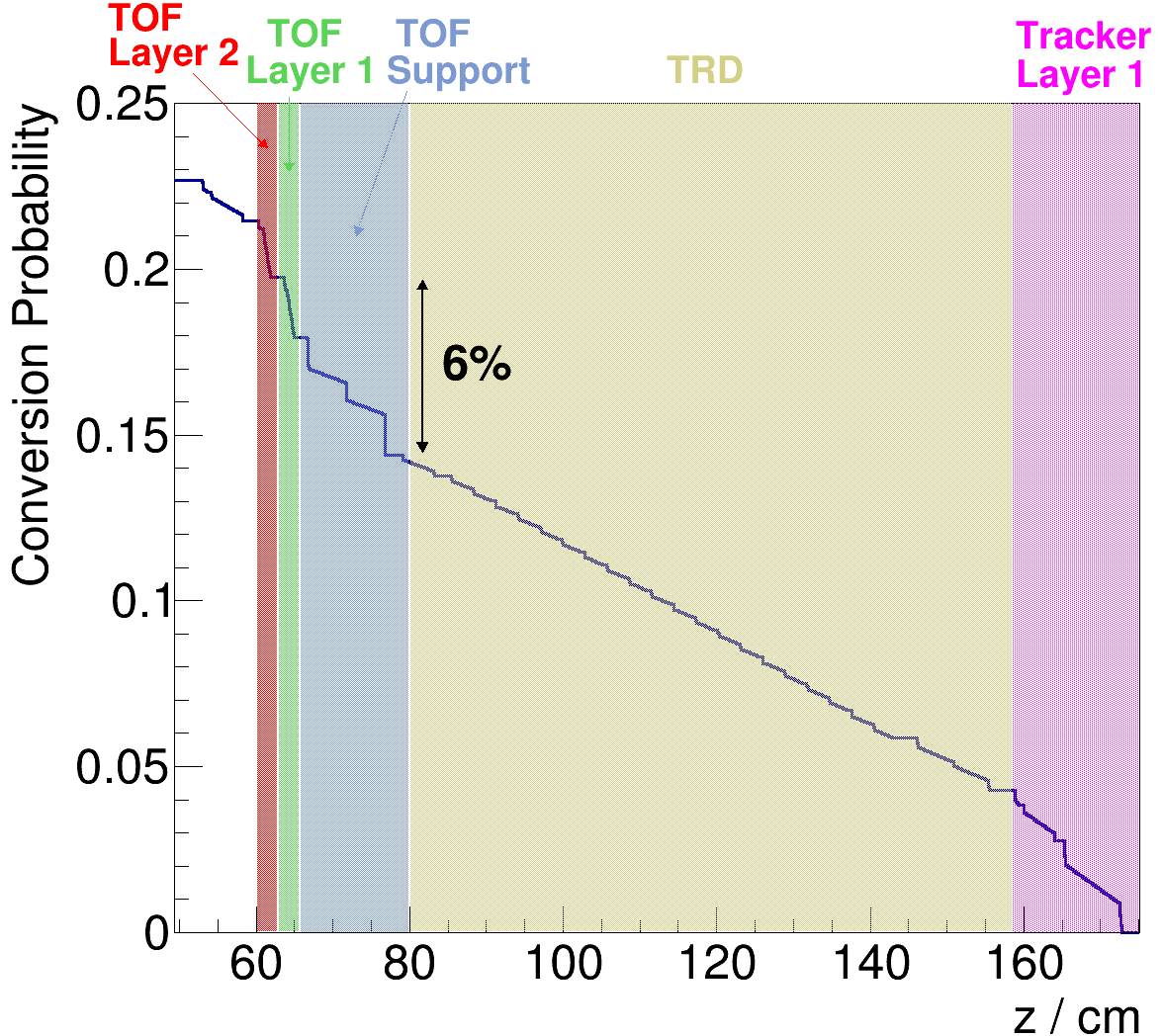}
  \end{minipage}
  \caption{Integrated probability for photon conversions, based on the $\frac{x}{X_0}$ distribution
    in the Monte-Carlo simulation for a perpendicular trajectory passing through
    $x = y = \SI{5}{\centi\meter}$. Left: Entire detector, Right: Zoom into the upper detector
    region.}
  \label{fig:analysis-conversion-probability}
\end{figure}

The probability for a photon conversion to occur directly correlates with the material distribution,
in particular with the integrated radiation length, as described by
equation~(\ref{eq:pair-production-probability}). Figure~\ref{fig:analysis-conversion-probability}
shows the integrated probability for photon conversion to occur at the given z position or above,
for an example perpendicular trajectory passing through $x = y = \SI{5}{\centi\meter}$. The small
displacement from the center was chosen in order to avoid passage through the overlap regions of the
TOF scintillator bars, which would produce a skewed picture. The conversion probability along the
example trajectory is fairly representative of the average.

The left hand side figure illustrates that about \SI{65}{\percent} of the photons pass through the
detector without converting and produce a shower in the calorimeter. This is the event sample
targeted in the calorimeter analysis mode. The right hand side figure illustrates that approximately
\SI{6}{\percent} of the photons convert in the first TOF layer or in the support material above
it. The conversion mode analysis is designed to select these events.

\subsection{Converted Photons}
\label{sec:analysis-selection-vertex}

The goal of the selection in this mode is to select photon events which convert in the first upper
TOF layer (TOF Layer 1) into an electron/positron pair. A conversion in the second upper TOF layer
(TOF Layer 2) would not suffice, because that would violate the 4/4 TOF trigger condition. Because
the absence of hits in the TRD provides the strongest veto against charged particles the conversion
must not happen in the TRD active volume. The conversion target material is therefore the upper TOF
Layer 1 scintillator and the support material between the TRD and the upper TOF.

Because of the 4/4 TOF trigger condition it is required that at least one of the two charged
particles traverses the entire inner detector and passes through both lower TOF layers. In addition,
the other track must not hit the ACC as that would generate an ACC veto for the trigger as described
in section~\ref{sec:detector-trigger}.

The target event signature can thus be summarized as follows:

\begin{itemize}
\item Two tracks with opposite rigidity sign in the inner tracker, possibly extending to tracker
  layer 9.
\item Signal in all four TOF layers.
\item No signal in the TRD or first tracker layer.
\item No signal in the ACC.
\end{itemize}

\begin{figure}[t]
  \centering
  \includegraphics[width=0.9\linewidth]{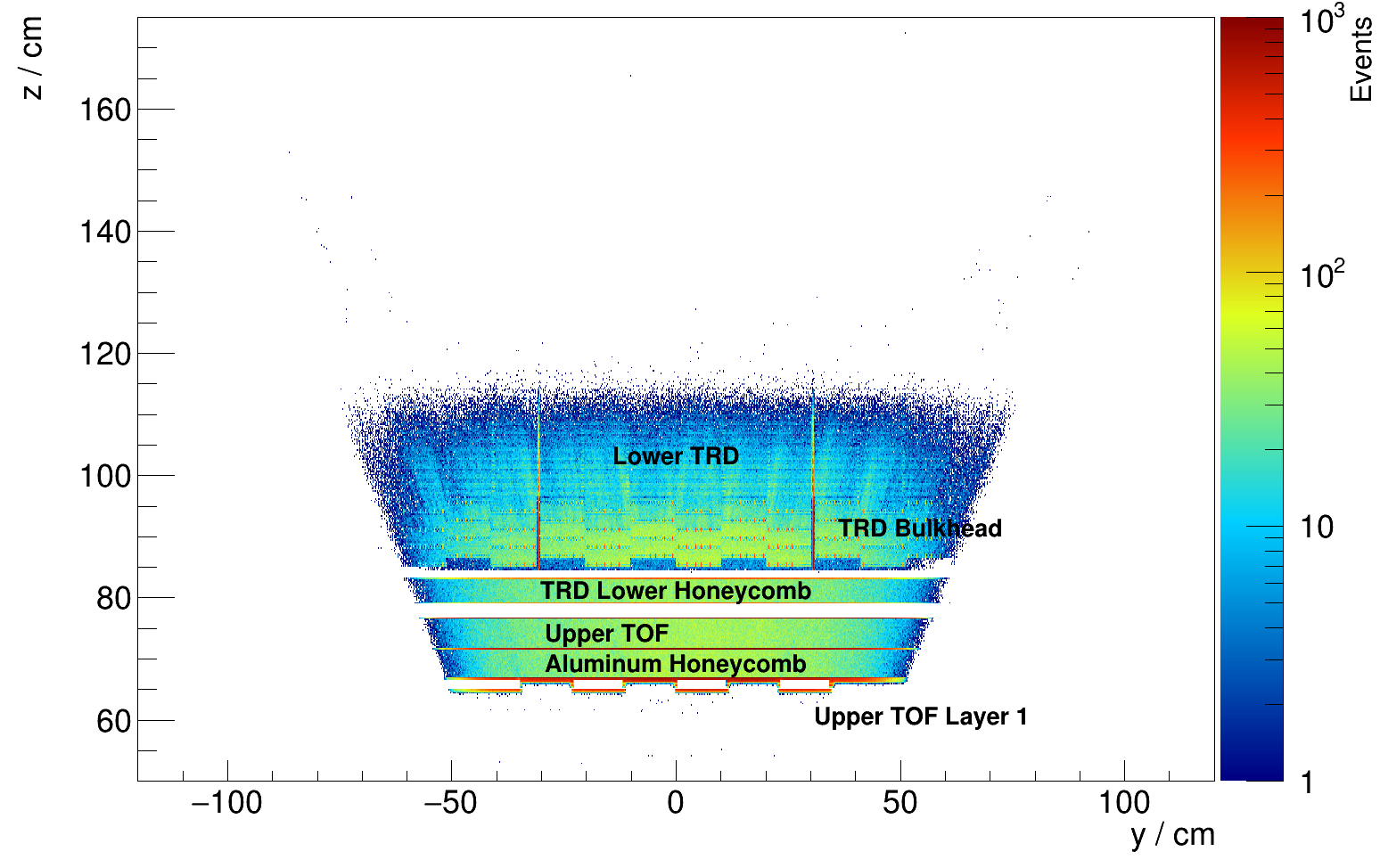}
  \caption{Positions of photon conversion vertices after applying the full selection for converted
    photons.}
  \label{fig:analysis-conversion-points-vertex-selected}
\end{figure}

Figure \ref{fig:analysis-conversion-points-vertex-selected} shows the distribution of photon
conversion positions after applying the selection which is detailed in the following. Photons which
convert in the central part of the TRD or above are removed, because the electron and positron
create tracks in the TRD. Photons which convert below the very first TOF layer are also removed,
because they do not fulfill the 4/4 TOF trigger requirement. Thus it is possible to identify the
first TOF layer, the upper TOF support structure, and the lowest part of the TRD including the
vertical TRD bulkheads as prominent converter materials.

\subsubsection*{Tracker}
\label{sec:analysis-selection-vertex-tracker}

The trajectories measured by the silicon tracker are the cornerstones of the converted photon
analysis. It is therefore necessary to identify tracks in the inner tracker with high
efficiency. The standard track reconstruction in AMS, while adequate for single charged particle
events traversing the entire detector, including the TRD, is not sufficient for this task. Because
of its known deficiencies an alternate track finding algorithm for \mbox{AMS-02} was developed in
2016 by Z. Qu~\cite{AMS02_Detector_TrReconQ}. The new track finding is more efficient, especially
for tracks left by nuclei in the silicon tracker, but also for photon conversions. In some rarer
cases however, tracks are only identified in the old algorithm. The idea employed here is therefore
to combine the results of both algorithms in a merging scheme in order to maximize the track finding
efficiency.

The merging scheme begins by adding all tracks from the old track reconstruction algorithm to the
set of selected tracks. Afterwards tracks found by the new reconstruction algorithm are added, but
only if they share at most one Y-cluster with any of the already selected tracks. This scheme avoids
track duplication, while keeping the efficiency high.

After merging the results of the two track finding algorithms the analysis requires at least two
identified tracker tracks, for which the Choutko track fit algorithm, using the electron mass
hypothesis for the treatment of multiple scattering, is required to converge. Among all possible
pairs of tracks there must be at least one pair with opposite rigidity signs according to the track
fit. In case there is more than one such pair the selection chooses the pair which minimizes:

\begin{displaymath}
  \chi^{2} = \left(\frac{\Delta Y}{\SI{0.25}{\centi\meter}}\right)^{2} +
  \left(\frac{\Delta X}{\SI{0.2}{\centi\meter}}\right)^{2} + \left(\frac{\Delta
      \alpha}{\SI{0.02}{\radian}}\right)^{2} \,,
\end{displaymath}

where $\Delta Y$ and $\Delta X$ are the differences in the estimated position at
$z = \SI{63.65}{\centi\meter}$ which corresponds to the position of the upper TOF. $\Delta \alpha$
is the angle between the two tracks at the same z position. The numbers in the denominators are
meant to roughly normalize the contributions of the various terms, they were determined from
inspection of the relevant distributions in the Monte Carlo simulation. The exact values are
unimportant, since this criterion is only required in case three tracks or more are found, which
happens for less than one permille of the events.

In case such a pair is found the track with the negative rigidity sign is referred to as the
``electron'' and the track with the positive rigidity sign is referred to as the ``positron'' in the
following, otherwise the event is discarded.

The two tracks must not be separated by more than $\SI{5}{\centi\meter}$ in the $Y$-coordinate at
$z = \SI{90.35}{\centi\meter}$ which corresponds to the position of the lower end of the TRD. In
addition, neither track is allowed to point into the ACC panels when extrapolated through the
magnetic field.

In order to suppress photons produced by hadronic interactions of nuclei with the material at the
very top of AMS the selection requires that there is no tracker hit in \mbox{layer 1} with a
measured charge larger than 1.5.

The kinematical properties of the photon itself are reconstructed from the two tracks as follows:

\begin{align*}
  \vec{x}_{\gamma} &= 0.5 \cdot \left(\vec{x}_{e^{-}} + \vec{x}_{e^{+}}\right) \\
  \vec{d}_{\gamma} &= \left|R_{e^{-}}\right| \vec{d}_{e^{-}} + \left|R_{e^{+}}\right| \vec{d}_{e^{+}}
                     \numberthis \label{eq:vertex-direction} \\
  E_{\gamma} &= E_{e^{-}} + E_{e^{+}} = \sqrt{R_{e^{-}}^{2}
               + m_{e^{-}}^{2}} + \sqrt{R_{e^{+}}^{2} + m_{e^{+}}^{2}} \approx \left|R_{e^{-}}\right| +
               \left|R_{e^{+}}\right|
               \numberthis \label{eq:vertex-energy}
\end{align*}

where $\vec{x}_{e^{-/+}}$ is the position of the electron / positron track at the reference z
coordinate at the upper TOF, $\vec{d}_{e^{-/+}}$ is the respective track direction and $R_{e^{-/+}}$
its measured rigidity. The vectors $\vec{x}_{\gamma}$ and $\vec{d}_{\gamma}$ define a straight line
on which the conversion point lies and which provides the reconstructed photon direction. It is
difficult to pinpoint the exact location of the conversion along this line because the two tracks
are essentially parallel close to the vertex, but this information is not required to reconstruct
the photon direction and energy.

\subsubsection*{Time-of-Flight System}
\label{sec:analysis-selection-vertex-tof}

The TOF is important for three reasons in the analysis. First and foremost it provides the trigger
for the rest of the detector. In addition, its excellent $\mathrm{d}E/\mathrm{d}x$ measurement
enables discrimination of photon conversions from ordinary cosmic rays such as protons, electrons
and helium nuclei. Finally its measurement of the time of flight allows to discriminate against slow
particles such as low energy protons.

At least one TOF cluster in each of the four TOF layers is required in the analysis. In the first
two TOF layers the number of clusters must not be larger than 2.

The $\mathrm{d}E/\mathrm{d}x$ measured in the TOF is proportional to the squared elementary charge
of the particle. As a result an electron/positron pair deposits energy in the TOF which corresponds
to an effective particle charge of $Q_{\mathrm{eff}} \approx \sqrt{2}$, which is very different from
the charge deposited by protons or electrons ($Q_{p/e-} \approx 1$) and Helium nuclei
($Q_{He} \approx 2$).

\begin{figure}[t]
  \begin{minipage}{0.48\linewidth}
    \centering
    \includegraphics[width=1.0\linewidth]{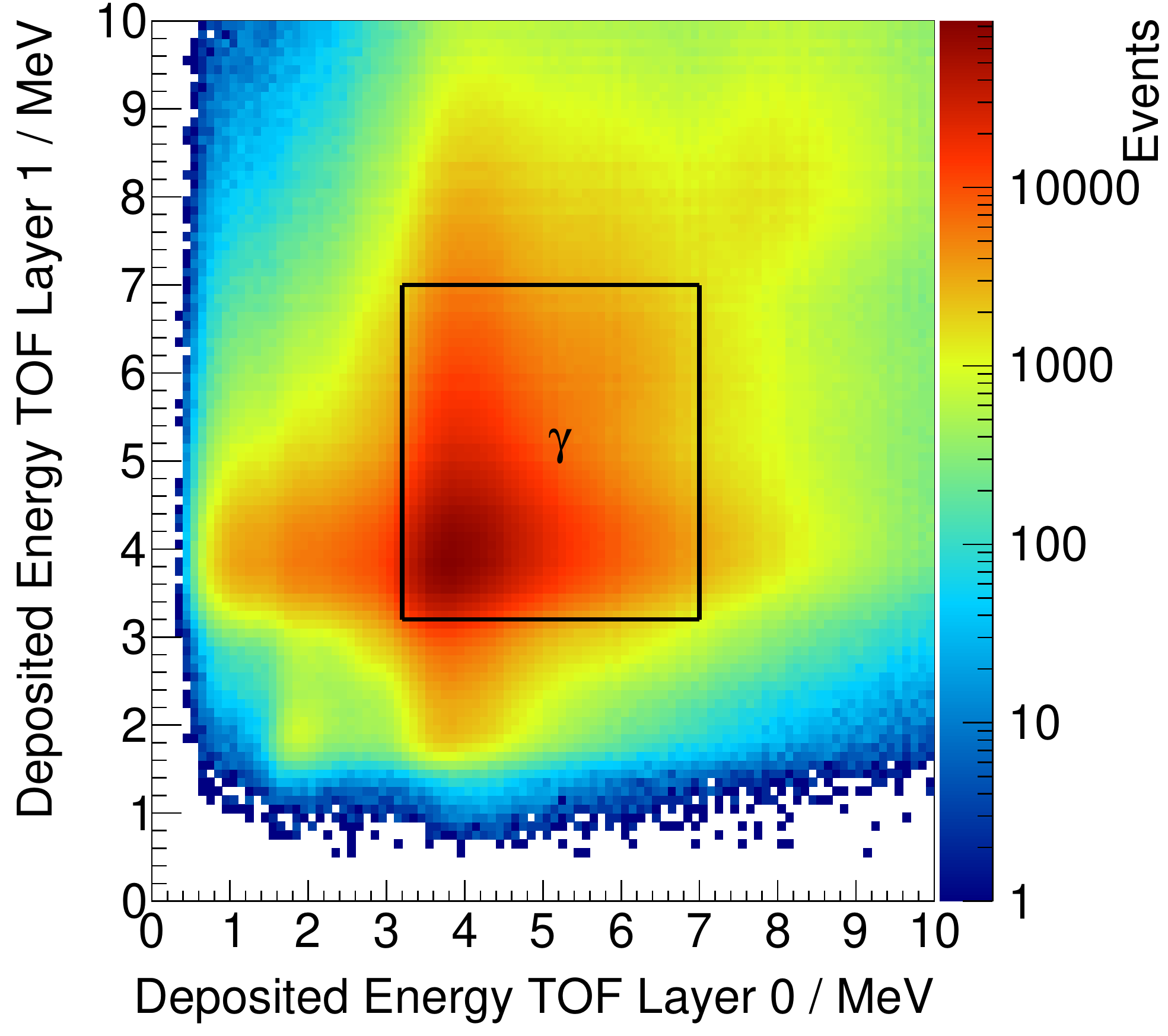}
  \end{minipage}
  \hspace{0.01\linewidth}
  \begin{minipage}{0.48\linewidth}
    \centering
    \includegraphics[width=1.0\linewidth]{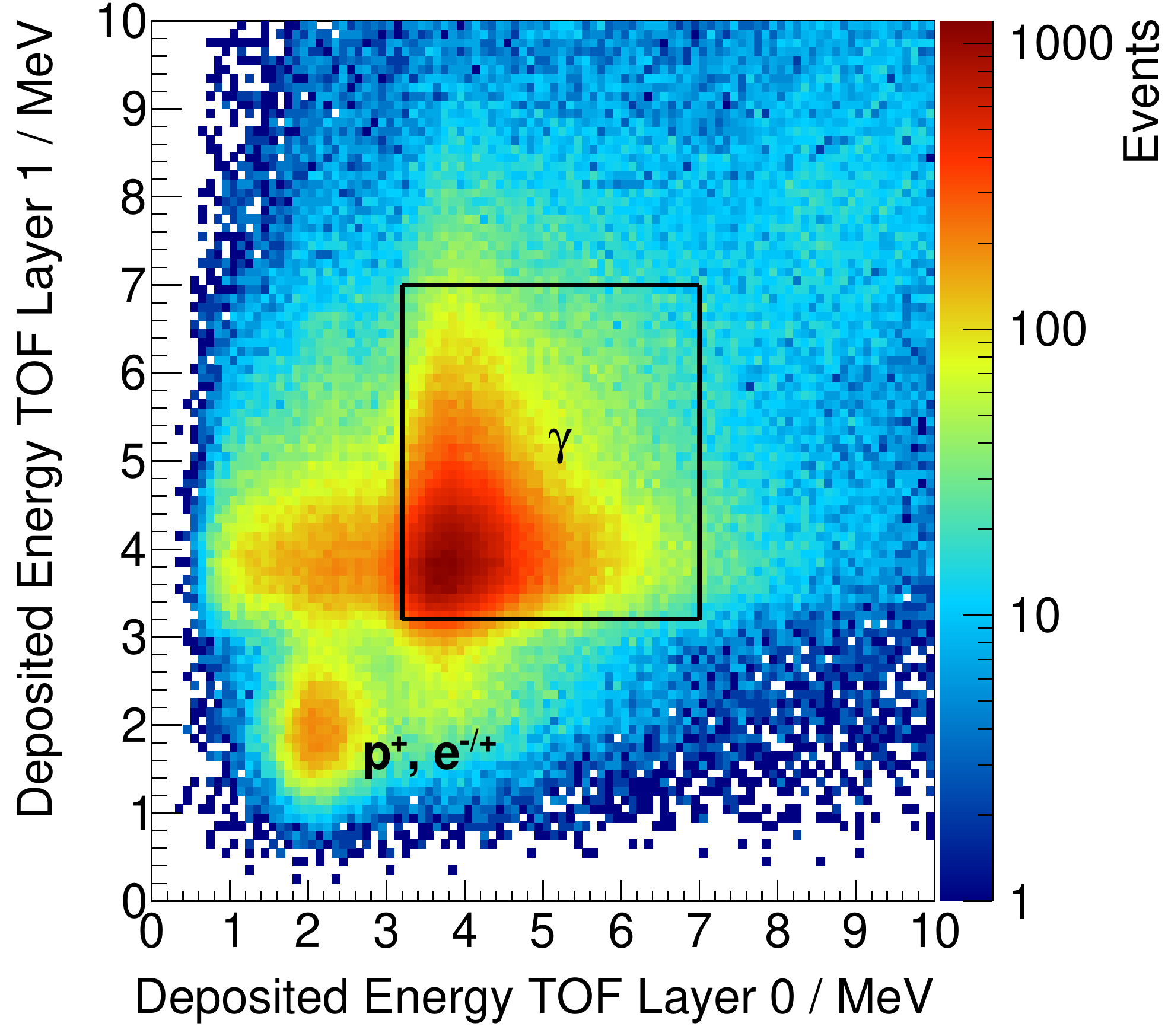}
  \end{minipage}
  \caption{The deposited energy in the two upper TOF layers for Monte-Carlo (left) and ISS (right)
    data, together with the region selected by the cuts shown in black. The distributions were
    obtained by applying all selection cuts, except for those concerning the deposited energy in the
    TOF.}
  \label{fig:control-plots-vertex-upper-tof}
\end{figure}

Therefore the total energy deposited in the first TOF layer divided by the number of TOF clusters is
required to be between \SI{3.2}{\mega\electronvolt} and \SI{7.0}{\mega\electronvolt} which
corresponds to the energy deposited by two singly-charged relativistic particles. The same cut is
placed on the energy deposition in the second TOF layer. These numerical cut values were found by
inspection of the one-dimensional distribution of deposited energy in each TOF layer separately.

Figure~\ref{fig:control-plots-vertex-upper-tof} shows the distribution of the deposited energy in
the first two TOF layers for events selected by the full set of cuts, except for those concerning
the deposited energy in the TOF. The majority of the signal events are situated within the marked
region and thus selected as is apparent by the Monte-Carlo distribution on the left. The ISS data
distribution shows an additional peak, close to \SI{2}{\mega\electronvolt} deposited energy, which
is due to residual proton and electron contamination. These background events are removed by the
cut. The small peak in the Monte-Carlo distribution at the same position is due to photon events
which convert above the upper TOF, and produce an electron or positron which either misses the TOF
or is absorbed before entering it.

For the third TOF layer the same cut is also used, but only if there is exactly one TOF cluster in
that layer. It is not used for the fourth TOF layer because due to the magnetic field the particle
trajectories are bent and the two tracks do not pass through the same TOF bar unless the photon
energy is high enough. At low energies one of the two tracks often does not pass through the last
TOF layer at all, in which case the energy deposit in that layer is identical to that of proton or
electron events.

In the next step the two tracker tracks are matched to clusters in the TOF. A cluster is defined to
``match'' the tracker track if the track passes geometrically through the associated TOF
scintillator bar. Both tracks must have matching TOF clusters in the two upper TOF layers. In the
two lower TOF layers only one of the two tracks is required to match with the clusters, because due
to the magnetic field one of the tracks might be bend out and miss the TOF plane.

For the two upper TOF layers the clusters matched to the two tracks must be in the same bar, or
adjacent to each other. Two bars with signal are not uncommon because of the overlap of the TOF
scintillator bars. In TOF Layer 3 at least one of the two matched bars must be one of the central
scintillator bars (bar number = 2..9). This is because a hit in one of the edge paddles in the third
TOF layer is insufficient for the 4/4 TOF trigger signal.

The velocity $\beta$ as measured by the TOF clusters matched to the track must be greater than 0.9
for both tracks. The individual tracker tracks are used for the pathlength estimation when
calculating the time of flight. This cut removes slow particles such as protons or alpha particles
at low rigidities and makes sure that the event is downward going.

\subsubsection*{Transition Radiation Detector}
\label{sec:analysis-selection-vertex-trd}

The reconstructed photon line of passage must be fully within the active volume of the TRD. In
addition, at the top of the TRD, the X-coordinate of the photon line must be within $\left|X\right|
< \SI{80}{\centi\meter}$. These cuts ensure that the reconstructed particle passed through the TRD
sensitive volume, which is required to guarantee the reliability of the TRD veto.

In order to suppress charged particles the number of TRD hits is required to be less than 10. For a
charged trajectory approximately 18-20 TRD hits are expected. In addition it is required that no TRD
track segments are found in either of the two projections.

It should in principle suffice to reject only those events in which the TRD track approximately
matches with the direction of the reconstructed photon. While such an approach would not result in
an increase of effective area, it would still be worthwhile to pursue, since it would reduce the
magnitude of the TRD pileup correction discussed in
section~\ref{sec:analysis-trd-pileup-weight}. Unfortunately it was found that the background of
protons and electrons is currently too high to manage in such types of selections.

Also, it is in principle possible to include photon events which convert in the TRD in the
analysis. These events exhibit the unique signature of a partial TRD track which begins in the
middle of the TRD. Also, the amplitude of the TRD tube signals is special, since the electron and
positron, both of which are able to produce transition radiation, pass through the same tubes.

Although special selections designed to identify these events were developed, it was found that the
gain in effective area and statistics of approximately \SI{30}{\percent} was accompanied by a
substantial increase in the number of background events. Ultimately, it was decided not to include
these sets of events.

However, given more time, it should be feasible to improve the analysis with respect to both of the
above points.

\subsubsection*{Combined Signal Efficiency}
\label{sec:signal-efficiency-vertex}

\begin{figure}[t]
  \centering
  \includegraphics[width=0.65\linewidth]{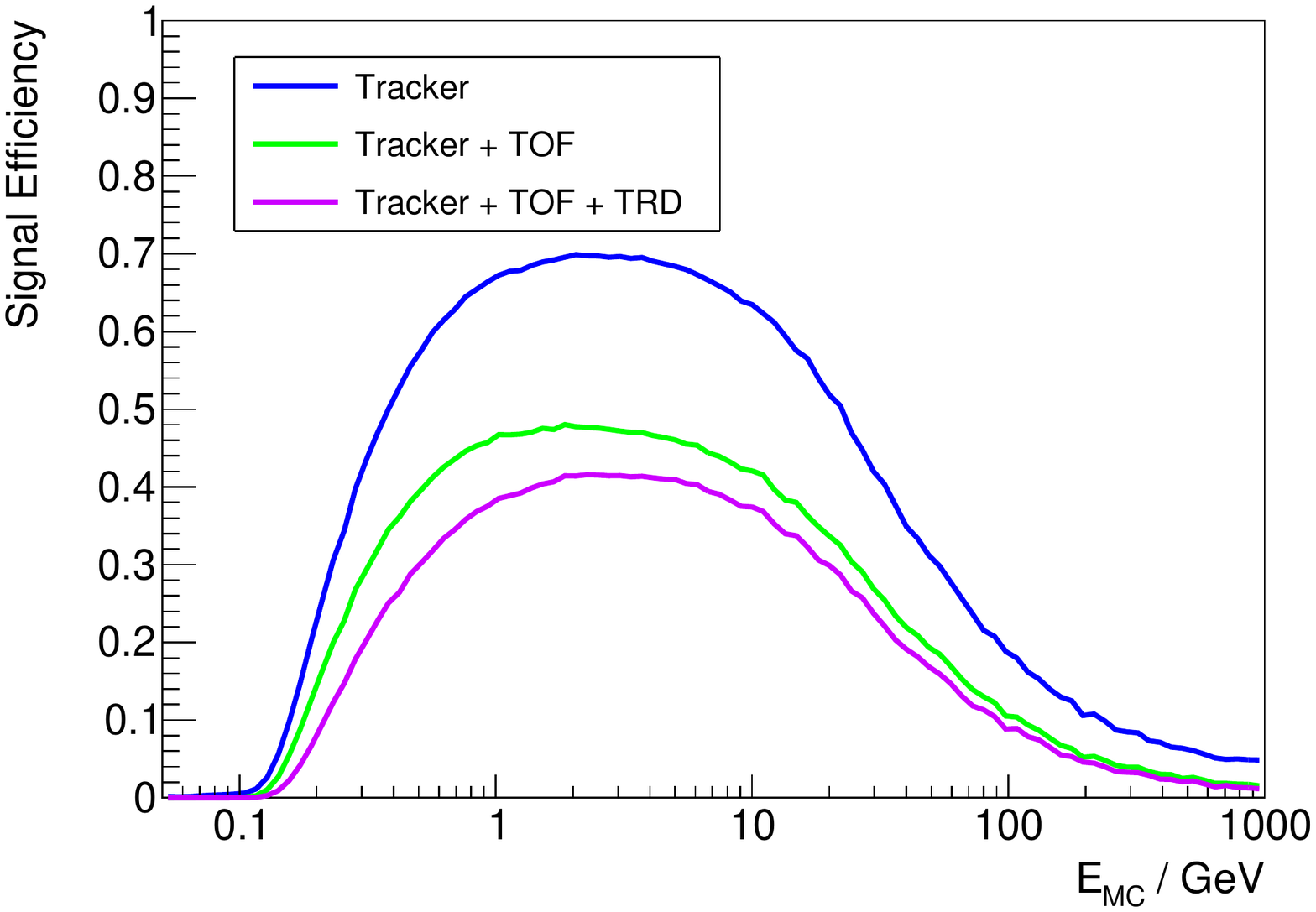}
  \caption{The combined conversion mode photon selection efficiency for perpendicular incidence in
    the Monte-Carlo simulation, based on a sample of events in which the photon converts in the
    first upper TOF layer or just above it, according to the Monte-Carlo truth. Blue: Tracker
    selection efficiency. Green: Efficiency after applying both tracker and TOF cuts. Magenta: Final
    selection efficiency, which also includes the TRD cuts.}
  \label{fig:signal-efficiency-vertex}
\end{figure}

Figure~\ref{fig:signal-efficiency-vertex} shows an estimate of the signal efficiency for photons
which impinge from the zenith, for the selection outlined above. This efficiency was obtained using
only photons which convert in the target region according to the Monte-Carlo truth. This
precondition is required: It is not meaningful to study the efficiency of the selection above for
photons which convert in the lower TOF for example. However, for the calculation of photon fluxes
such a precondition must not be applied, instead the effective area should be used, which is
determined and discussed in section~\ref{sec:analysis-effective-area}.

The efficiency of the selection criteria relating to the tracker alone are shown in blue. The
efficiency reaches a maximum of approximately \SI{70}{\percent} around \SI{3}{\giga\electronvolt}
and slowly drops towards both higher and lower energies. The blue curve indicates that the shape of
the final combined signal efficiency is governed by the tracker selection. At lower energies the
momentum of either the electron or positron may be too small, so that the particle can easily be
absorbed in the detector material or swept away by the magnetic field. This means that the tracker
will not reconstruct two tracks with opposite charge sign. At higher energies the two tracks begin
to overlap, which interferes with the ability to reconstruct two independent trajectories in the
tracker.

Cuts on the TOF signal amplitudes (see figure~\ref{fig:control-plots-vertex-upper-tof}) cause the
signal efficiency to drop to a combined value of about \SI{45}{\percent} in the maximum, as
indicated by the green curve. The selection requirements for the TOF amplitudes is indeed strict,
which is a result of the need for a very high background rejection. The curve shown in magenta in
the figure shows the final combined selection efficiency is around \SI{40}{\percent} in the maximum
at around \SI{3}{\giga\electronvolt}.

\subsection{Calorimeter Photons}
\label{sec:analysis-selection-calorimeter}

Photons within the ECAL acceptance which do not convert in the material above the calorimeter will
pass most of the detector unnoticed and finally produce a shower in the ECAL. The unique signature
is thus the existence of an isolated calorimeter shower of electromagnetic shape, without any tracks
in the rest of the detector.

The veto for charged particles is provided by the absence of tracks in the TRD and tracker. While
the upper TOF is typically also empty for these events, some activity in the lower TOF and tracker
layer 9 is expected from backsplash from the calorimeter shower, in particular for high energy
photons.

The target event signature can thus be summarized as follows:

\begin{itemize}
\item Electromagnetic shower in the calorimeter with shower axis pointing to the top of the
  instrument.
\item No activity in the TRD, Tracker, upper TOF and ACC.
\end{itemize}

\subsubsection*{Trigger}
\label{sec:analysis-selection-ecal-trigger}

Because these photons do not interact with the Time-of-Flight system they will not fire the regular
charged particle trigger. Therefore a special trigger based on ECAL information only is required to
record these events as described in section~\ref{sec:detector-trigger-calorimeter}.

For the event selection the special calorimeter trigger is required to fire, furthermore the
Time-of-Flight trigger must be absent.

\subsubsection*{Electromagnetic Calorimeter}
\label{sec:analysis-selection-ecal-ecal}

In the calorimeter the existence of exactly one reconstructed particle shower as identified and
reconstructed with the new 3D shower reconstruction method developed by the MIT group in
AMS~\cite{AMS02_Detector_Ecal3D} is required. The longitudinal shower shape must be compatible with
that of a downward going electron or photon according to the longitudinal shower fit in order to
discriminate against upgoing events which stop in the calorimeter. The total deposited energy in the
calorimeter must be at least \SI{1}{\giga\electronvolt} before corrections, because of the
calorimeter trigger threshold.

\begin{figure}[t]
  \begin{minipage}{0.48\linewidth}
    \centering
    \includegraphics[width=1.0\linewidth]{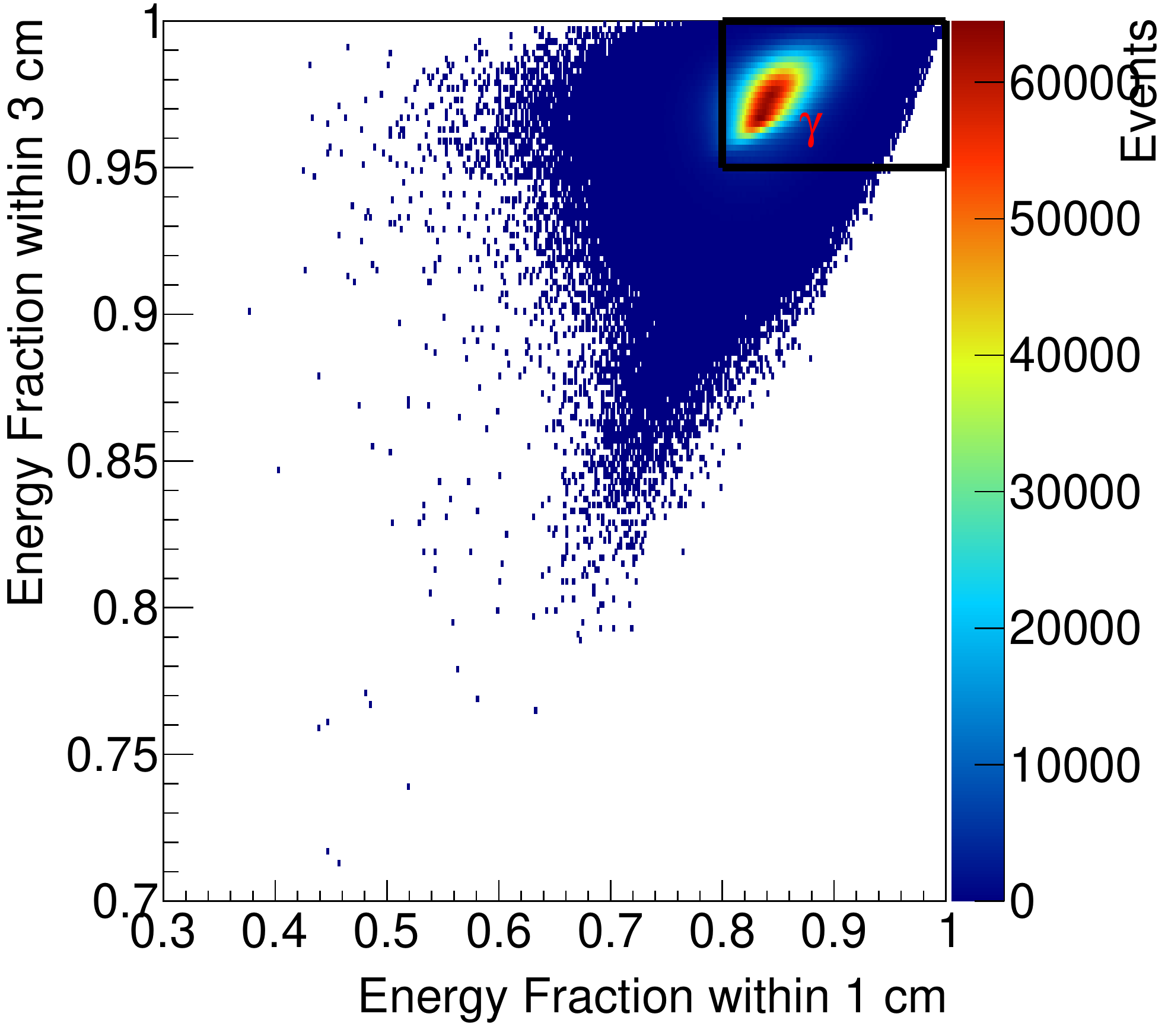}
  \end{minipage}
  \hspace{0.01\linewidth}
  \begin{minipage}{0.48\linewidth}
    \centering
    \includegraphics[width=1.0\linewidth]{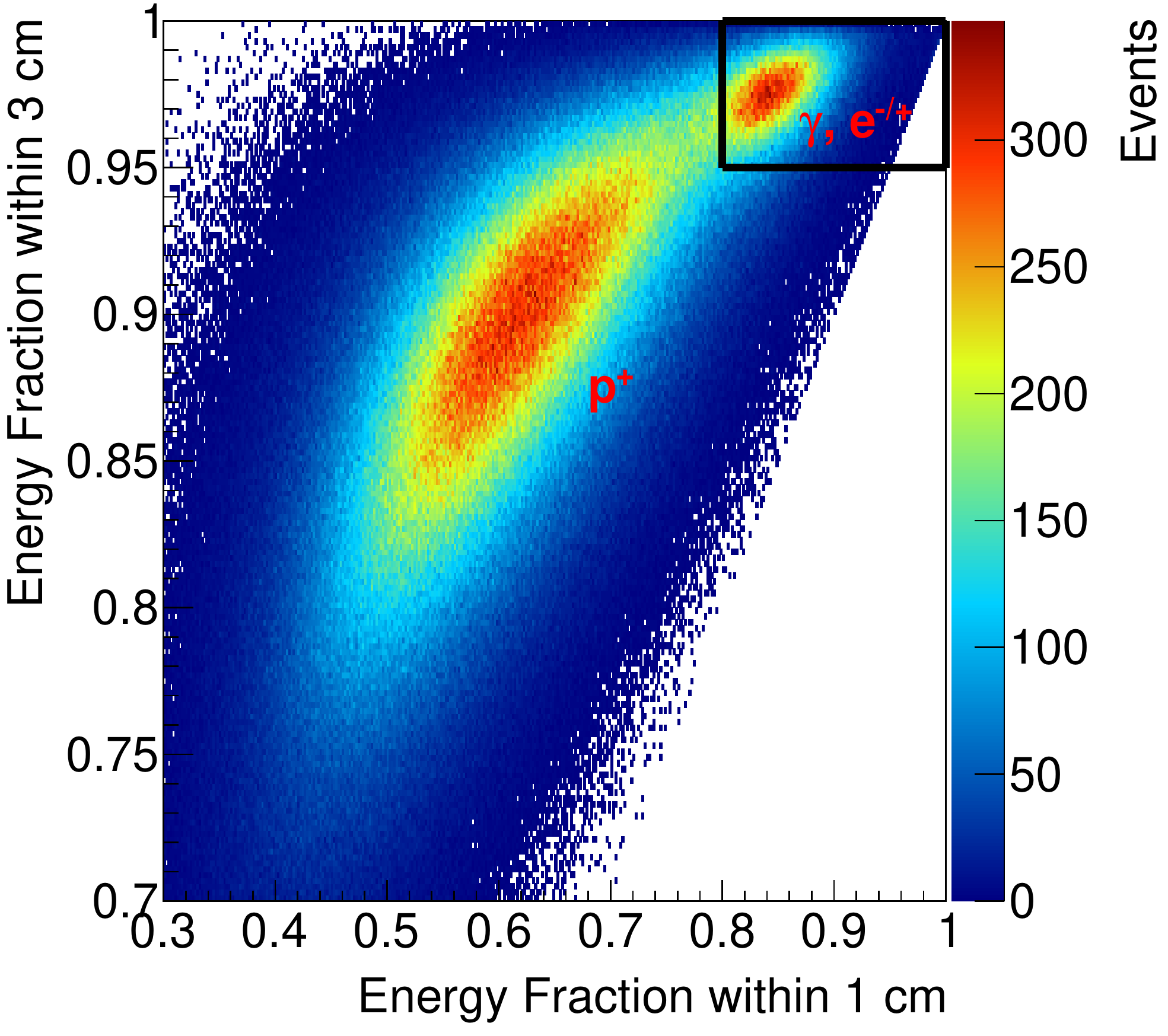}
  \end{minipage}
  \caption{Left: Distribution of two important ECAL shower shape parameters in the photon
    Monte-Carlo. The black box indicates the region from which events are accepted by the
    cuts. Right: The same distribution for ISS data, obtained after applying all other cuts.}
  \label{fig:control-plots-ecal-energy-ratios}
\end{figure}

The shower shape is required to be electromagnetic with the help of the following restrictions: The
energy contained within cylinders with radii of \SI{1}{\centi\meter} and \SI{3}{\centi\meter} around
the shower core must be greater than \SI{80}{\percent} and \SI{95}{\percent} respectively. Figure
\ref{fig:control-plots-ecal-energy-ratios} shows the distribution of these two shower shape
parameters for both photon Monte-Carlo and ISS data. In the ISS data photons can clearly be
separated from background protons. The agreement between data and simulation for the photon
component in the distributions is very good.

The $\chi^2$ of the longitude shower profile must be less than 20. The value of the reweighted layer
likelihood estimator~\cite{AMS02_Detector_Ecal3D} of the 3D shower reconstruction must be less than
3.2, which removes proton events.

The shower axis is reconstructed by several different methods and it is required that these
reconstructions match within \SI{8}{\degree}. The primary shower reconstruction method is the 3D
shower fit which provides the photon direction and axis.

A zenith angle cut of $\cos{\theta} > 0.9$ on the shower axis ensures that all other subdetectors
can be used for a reliable charged particle veto. For the same reason the reconstructed shower axis
is explicitly required to pass through the entire active TRD volume.

The energy of the photon is estimated by the energy reconstruction method of the 3D shower fit
routine as described in~\cite{AMS02_Detector_Ecal3D}. Although this estimator was designed to
measure the energy of electrons and positrons it is used without modification. The small differences
between photons and electrons are corrected for in the unfolding procedure described in
section~\ref{sec:corrections-unfolding}.

\subsubsection*{Tracker and Transition Radiation Detector}
\label{sec:analysis-selection-ecal-trd}

Events with tracker tracks found by either the standard reconstruction or the new development
reconstruction~\cite{AMS02_Detector_TrReconQ} in AMS are removed. In addition there must not be any
TRD segments in either projection and the total number of TRD hits must be less than 10.

The discussion in section \ref{sec:analysis-psf} will show that it is hard to reconstruct the photon
direction from the calorimeter shower axis, in particular at low energies. For this reason other
approaches in which events with TRD segments which do not match with the shower axis are allowed
were not pursued further.

\subsubsection*{Time-of-Flight and Anti-Coincidence-Counter}
\label{sec:analysis-selection-ecal-tof-acc}

\begin{figure}[t]
  \begin{minipage}{0.48\linewidth}
    \centering
    \includegraphics[width=1.0\linewidth]{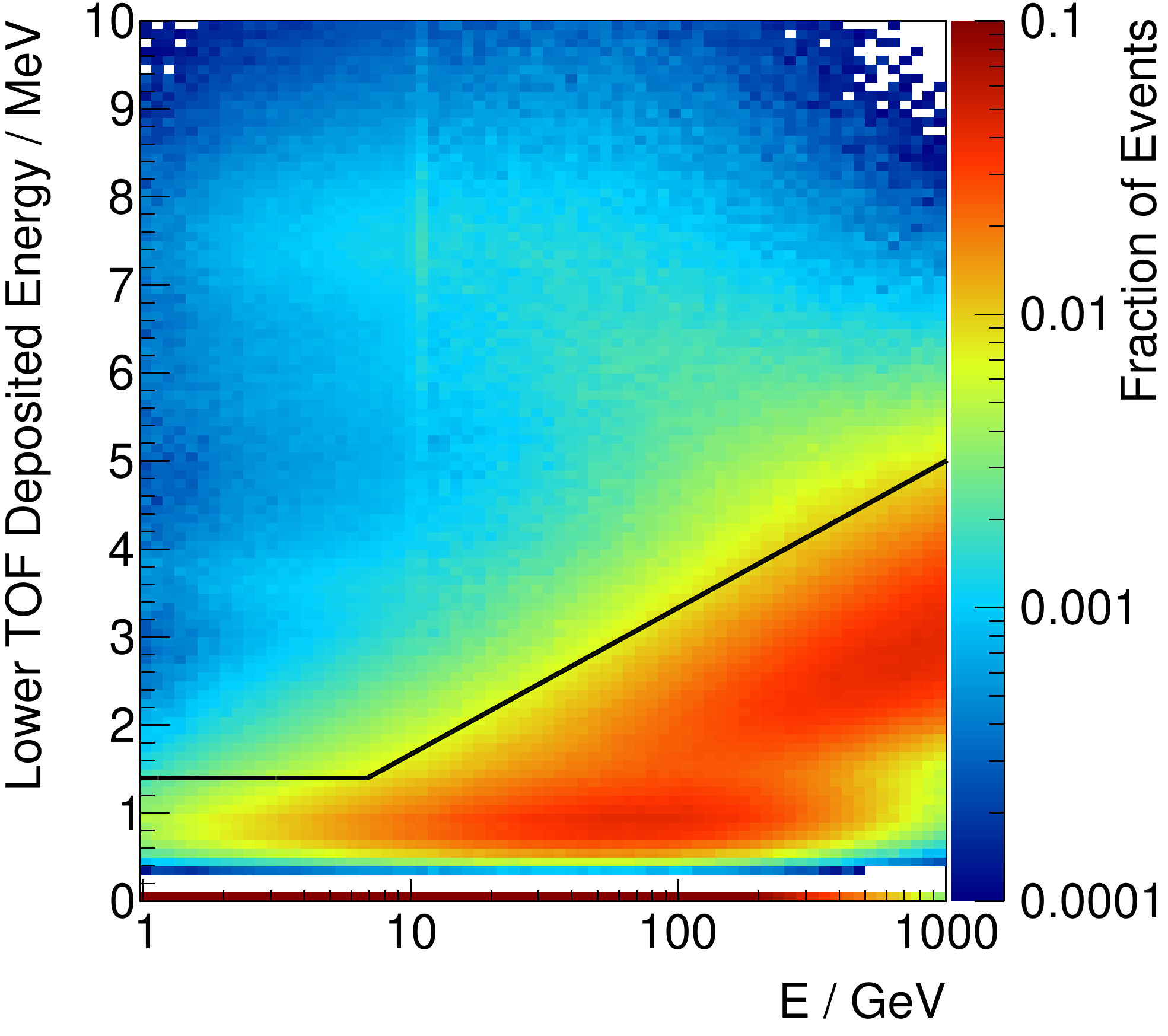}
  \end{minipage}
  \hspace{0.01\linewidth}
  \begin{minipage}{0.48\linewidth}
    \centering
    \includegraphics[width=1.0\linewidth]{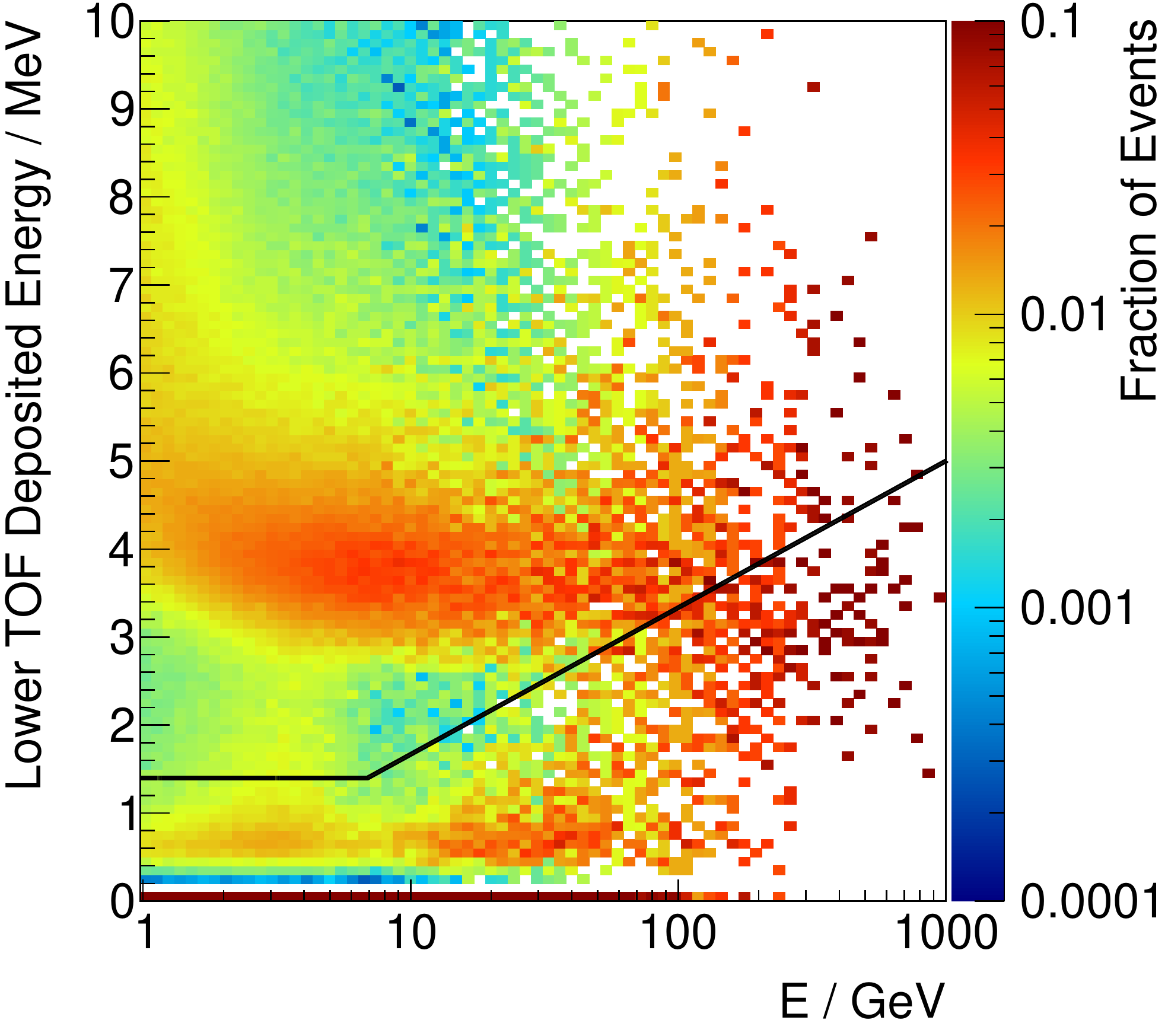}
  \end{minipage}
  \caption{Left: Average deposited energy in the lower TOF as a function of the calorimeter energy
    for the photon Monte-Carlo simulation. Events from below the black line are accepted by the
    cuts. Right: The same distribution for ISS data, after applying all other cuts.}
  \label{fig:control-plots-ecal-lower-tof-edep}
\end{figure}

For a photon which traverses the entirety of the detector and converts only in the calorimeter one
naively would not expect any signal in the TOF or ACC detectors. However, at higher energies, there
can be non negligible energy deposits in the TOF and ACC due to the high number of backsplash
particles from the calorimeter shower. Therefore there the analysis requires either no clusters in
the lower TOF, or if there are any, it is required that the average deposited energy must not exceed
a threshold which rises logarithmically with energy:

\begin{displaymath}
  \bar{E}_{dep} < \SI{5}{\mega\electronvolt} \cdot \log10(E / \si{\giga\electronvolt})
\end{displaymath}

Figure \ref{fig:control-plots-ecal-lower-tof-edep} shows distributions for the average deposited
energy in the lower TOF as a function of energy. In the ISS data the charged particle background is
clearly visible at low energies as an additional component above the black line. The cut function
was optimized for \SI{95}{\percent} signal efficiency according to the Monte-Carlo simulation. Note
that in most photon events there actually is no signal in the lower TOF, such that the average
deposited lower TOF energy is zero, i.e. at the lower edge of the distribution.

\begin{figure}[t]
  \begin{minipage}{0.48\linewidth}
    \centering
    \includegraphics[width=1.0\linewidth]{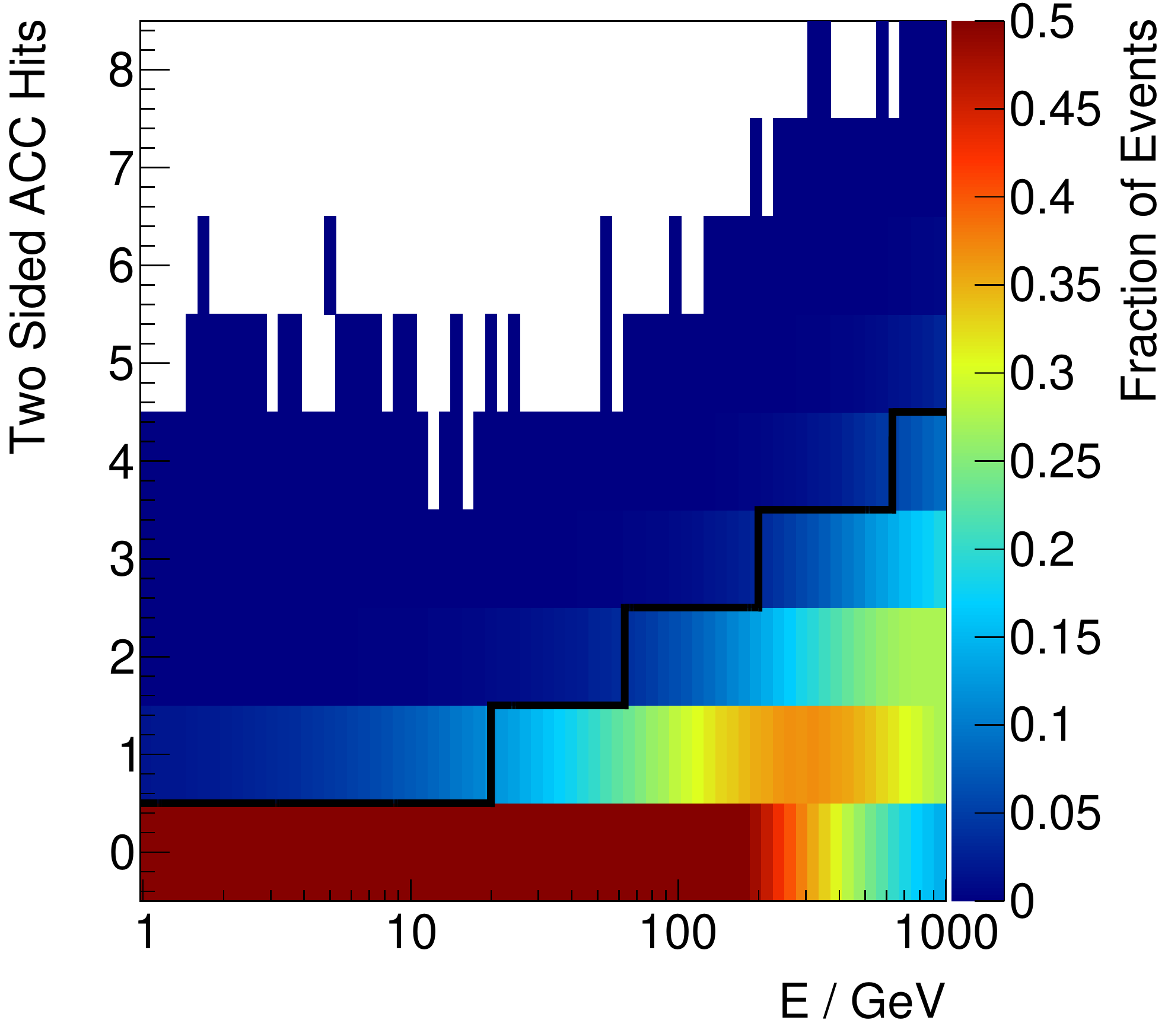}
  \end{minipage}
  \hspace{0.01\linewidth}
  \begin{minipage}{0.48\linewidth}
    \centering
    \includegraphics[width=1.0\linewidth]{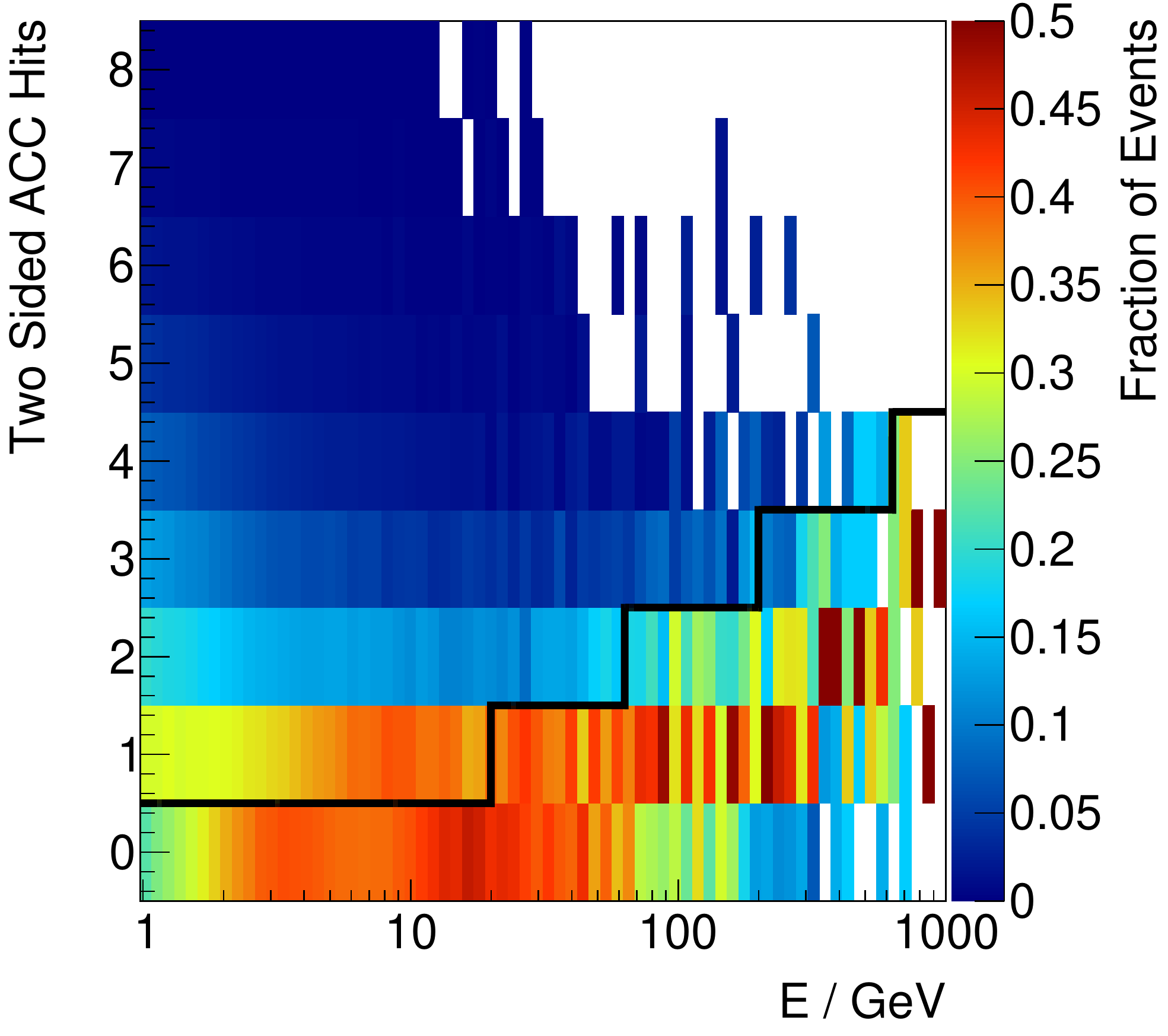}
  \end{minipage}
  \caption{Left: The number of two sided ACC hits as a function of energy for the photon Monte-Carlo
    simulation. The black line indicates the cut threshold, events below are accepted. Right: The
    same distribution for the ISS data.}
  \label{fig:control-plots-ecal-acc-hits}
\end{figure}

For the same reason it is required that the number of ACC hits with coincident signal in both
photomultipliers is limited:

\begin{equation}
  \label{eq:ecal-acc-cut}
  N_{\mathrm{ACC}} < 2 \cdot \left(\log10(E / \si{\giga\electronvolt}) - 0.8\right)
\end{equation}

These formulas were found by inspecting the relevant distribution for events with backsplash in the
Monte-Carlo simulation, shown in figure \ref{fig:control-plots-ecal-acc-hits}. The charged particle
background in the ISS distribution is visible above the black line, typically producing a single ACC
cluster, in particular at low energies. This component is not present in the photon Monte-Carlo and
removed by the selection cuts.

As the photon energy increases the number of observed two sided ACC clusters increases as well, due
to backsplash particles produced in interactions in the calorimeter. Therefore, the selection cut
must allow for an increasing number of ACC hits as indicated in the figure.

\subsubsection*{Combined Signal Efficiency}
\label{sec:signal-efficiency-ecal}

\begin{figure}[t]
  \centering
  \includegraphics[width=0.65\linewidth]{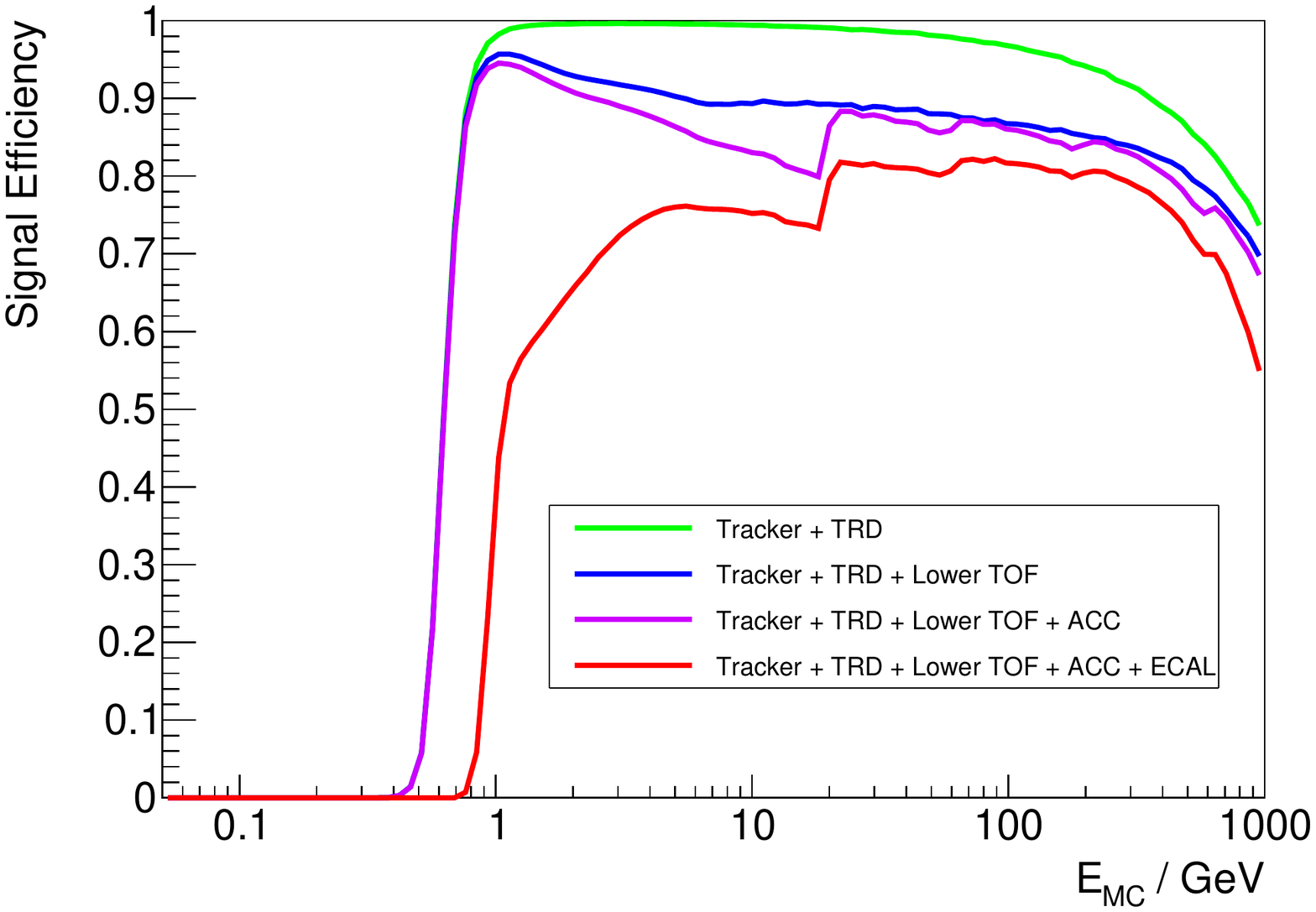}
  \caption{The combined calorimeter mode photon selection efficiency for perpendicular incidence in
    the Monte-Carlo simulation, based on a sample of events in which the photon passes through AMS
    without converting and produces a calorimeter shower according to the Monte-Carlo truth. Green
    line: Efficiency of selection cuts relating to Tracker and TRD vetos. Blue line: Combined
    efficiency of Tracker, TRD and Lower TOF cuts. Magenta line: Combined efficiency of Tracker,
    TRD, Lower TOF and ACC cuts. Red: Final selection efficiency.}
  \label{fig:signal-efficiency-ecal}
\end{figure}

Similarly to figure~\ref{fig:signal-efficiency-vertex}, the calorimeter mode selection efficiency is
detailed in figure~\ref{fig:signal-efficiency-ecal} for photons from the zenith, based on a subset
of Monte-Carlo events which fall into the target region according to the Monte-Carlo truth.

The combined efficiency of the selection (shown in red) is above \SI{70}{\percent} between
\SI{3}{\giga\electronvolt} and approximately \SI{500}{\giga\electronvolt}. The step at around
\SI{20}{\giga\electronvolt} is due to the ACC requirement, as can be seen from the magneta curve.
The selection criterion begins to allow events with one ACC hit at this energy, see
equation~(\ref{eq:ecal-acc-cut}).

The drop in the overall efficiency at \SI{1}{\giga\electronvolt} is due to the calorimeter trigger
efficiency, which will be derived in section~\ref{sec:analysis-trigger-efficiency}. At the highest
energies the selection is not well optimized to account for backsplash from calorimeter showers in
the upper detector. The green line indicates that the drop in efficiency at the highest energy is
mostly due to backsplash particles which are energetic enough to form tracks in the tracker and
TRD. They also produce additional energy depositions in the lower TOF.

Overall these results are encouraging, since the signal efficiency is quite high over a large energy
range in the calorimetric photon selection. However, for a full flux analysis it is more meaningful
to study the effective area, which is derived in section~\ref{sec:analysis-effective-area}.

\section{Instrument Response Functions}
\label{sec:analysis-irf}

The Instrument Response Functions (IRFs) describe the response of the \mbox{AMS-02} detector to
photons. The measurement of photons with the \mbox{AMS-02} detector is subject to inaccuracies due
to detector resolution effects. After a triggered event is identified as a photon candidate there
are two quantities in which one is primarily interested: The direction from which the photon arrived
and its energy. Both of these quantities can deviate from their true values due to detector
resolution.

Uncertainties in the reconstruction of the direction give rise to the Point Spread Function (PSF),
which describes the observed spread of an ideal point source in the sky. The PSFs for the two
\mbox{AMS-02} photon detection modes are obtained and described in section~\ref{sec:analysis-psf}.

Misreconstruction of the photon energy and the associated energy resolution leads to bin-to-bin
migration of events when determining the photon spectrum. The effect is described in
section~\ref{sec:analysis-energy-resolution}, together with results for the \mbox{AMS-02}
resolution.

In addition, the efficiency to detect a photon needs to be estimated in order to measure photon
fluxes. The exact required quantity for the calculation of non-isotropic fluxes is the product of
all selection and detector efficiencies with the apparent geometric size of the apparatus. Therefore
it is customary to directly determine the product, which is referred to as the ``effective area'',
typically from Monte-Carlo simulations of the experiment. The effective area is determined and
discussed in section~\ref{sec:analysis-effective-area}.

Finally the trigger efficiency is determined from the simulation in
section~\ref{sec:analysis-trigger-efficiency}.

\subsection{Point Spread Function}
\label{sec:analysis-psf}

The angular resolution of the two photon detection modes in \mbox{AMS-02} can be studied in the
Monte-Carlo simulation. The sample of events is obtained by requiring the full set of selection cuts
described in sections~\ref{sec:analysis-selection-vertex}
and~\ref{sec:analysis-selection-calorimeter} for the conversion and calorimeter mode respectively.

The angles between the reconstructed photon direction and the true direction in the two detector
planes are defined as follows:

\begin{displaymath}
  \alpha_{x/y} = \arctan{\left(\frac{d_{x/y}}{d_{z}}\right)} - \arctan{\left(\frac{v_{x/y}}{v_{z}}\right)},
\end{displaymath}

where $\vec{d}$ is the reconstructed photon direction and $\vec{v}$ is the true incoming direction
according to the Monte-Carlo truth.

\begin{figure}[t]
  \begin{minipage}{0.48\linewidth}
    \centering
    \includegraphics[width=1.0\linewidth]{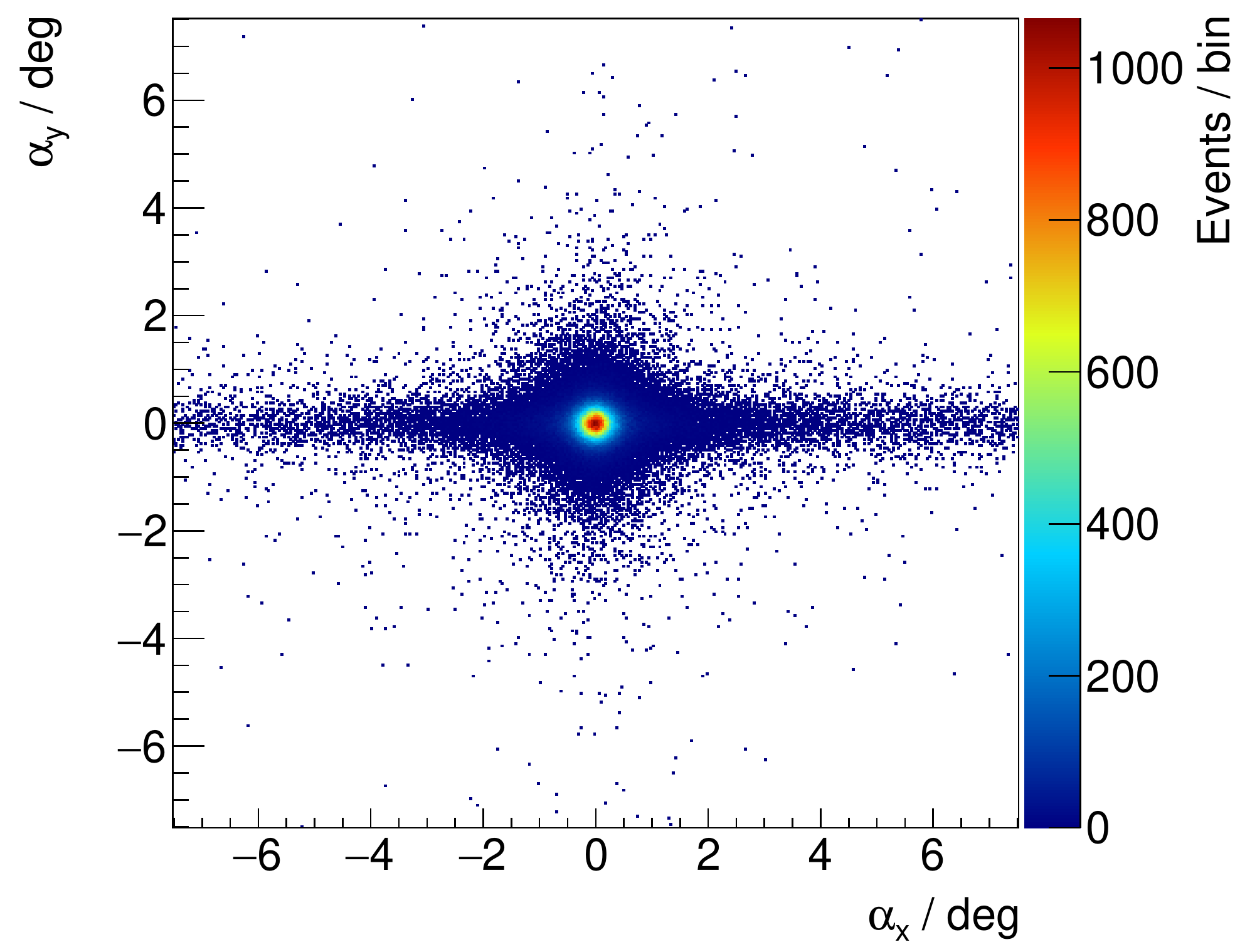}
  \end{minipage}
  \hspace{0.01\linewidth}
  \begin{minipage}{0.48\linewidth}
    \centering
    \includegraphics[width=1.0\linewidth]{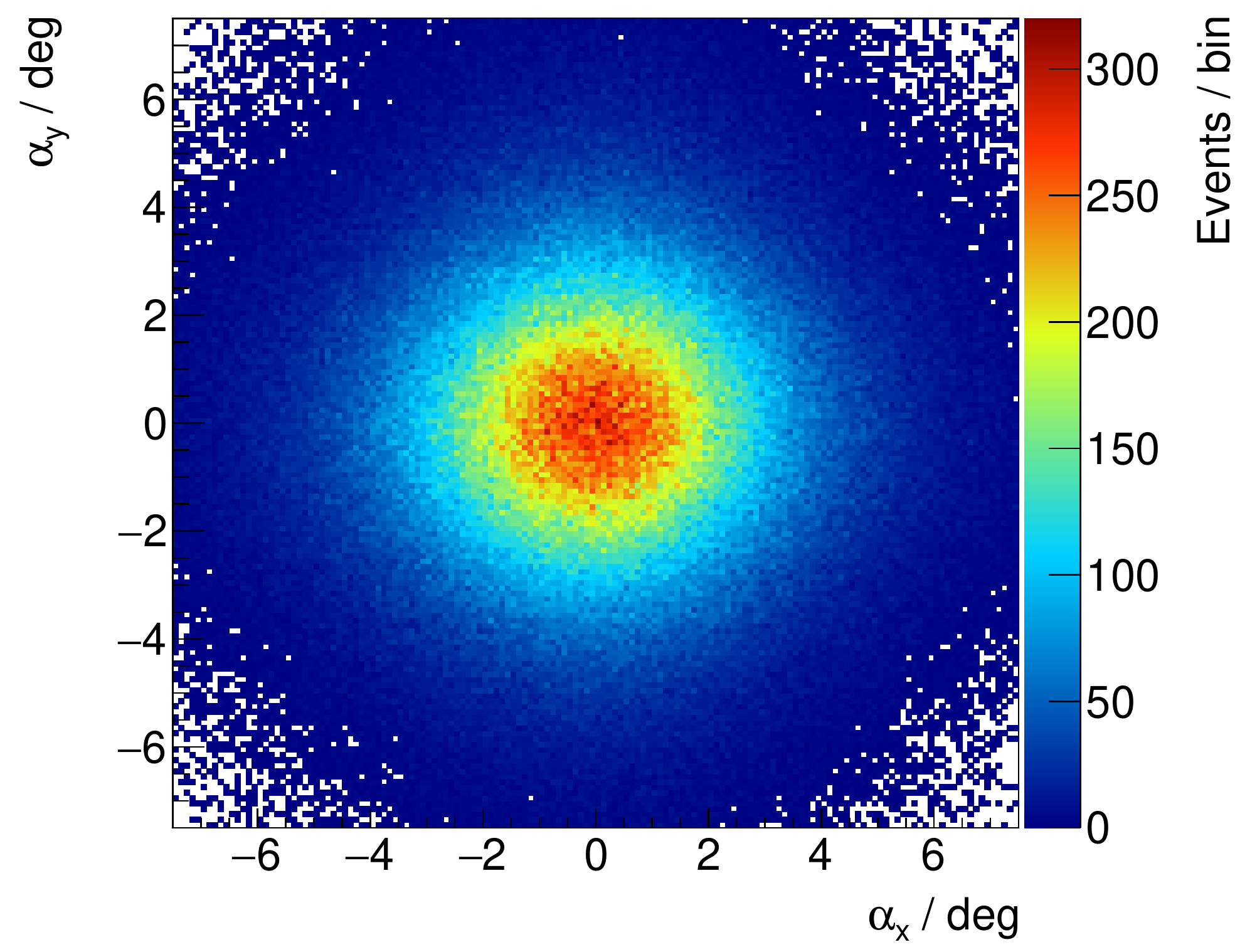}
  \end{minipage}
  \caption{The point spread function for the vertex (left) and calorimeter (right) analysis for
    \SI{2}{\giga\electronvolt} photons.}
  \label{fig:psf-2d}
\end{figure}

The left hand side of figure~\ref{fig:psf-2d} shows the two dimensional distribution of $\alpha_y$
versus $\alpha_x$ for \SI{2}{\giga\electronvolt} photons which convert in the upper TOF. The
electron and positron tracks are reconstructed with the tracker and combined to estimate the photon
direction according to equation~\ref{eq:vertex-direction}. There is no correlation between the
reconstructed direction in the two planes and the distribution is approximately Gaussian with some
minor tails, especially in the X direction. These tails stem from rare misreconstructions of the
ambiguity in the X coordinate measurement for some events (see
section~\ref{sec:detector-tracker}). The two dimensional Gaussian distribution is approximately
symmetric, i.e.~the width is almost the same in both directions. The distribution is centered around
the origin which indicates that there is no bias in the direction reconstruction in either
projection.

The right panel in figure~\ref{fig:psf-2d} shows the PSF for \SI{2}{\giga\electronvolt} photons
which enter the calorimeter and are reconstructed by shower shape analysis. The shower axis is
determined from a fit of a 3D shower profile to the calorimeter data. Because of inherent
statistical fluctuations in the longitudinal and lateral development of the shower as well as
coarser granularity of the calorimeter readout cells the reconstruction of the photon direction is
considerably more difficult than in the vertex analysis. The resulting angular resolution is
therefore worse compared to the vertex angular resolution.

As in the vertex analysis the shape of the PSF is approximately Gaussian. It is symmetric, without
bias and without correlation between the angles reconstructed in the two planes as can be seen from
figure~\ref{fig:psf-2d}.

\begin{figure}[t]
  \begin{minipage}{0.48\linewidth}
    \centering
    \includegraphics[width=1.0\linewidth]{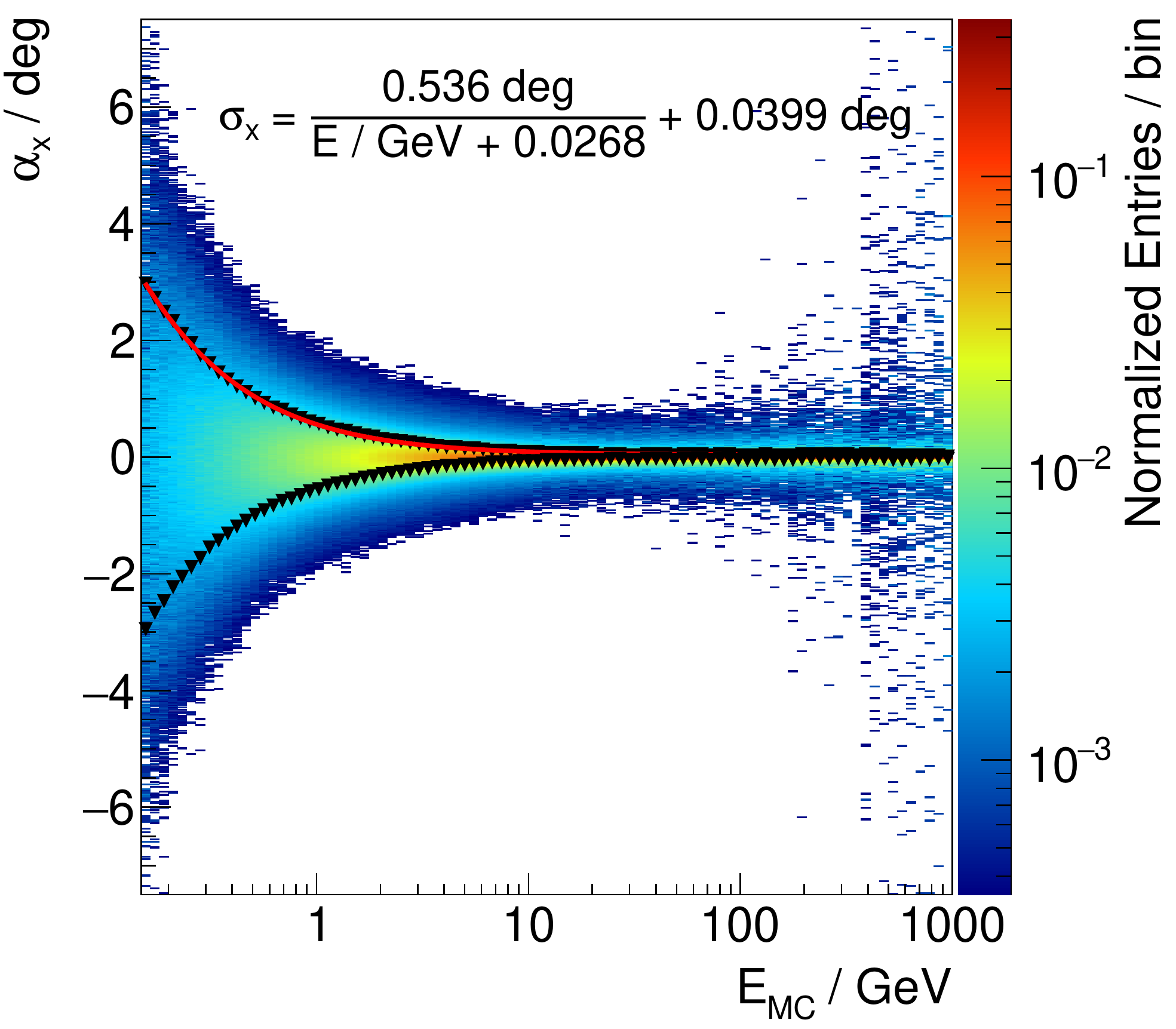}
  \end{minipage}
  \hspace{0.01\linewidth}
  \begin{minipage}{0.48\linewidth}
    \centering
    \includegraphics[width=1.0\linewidth]{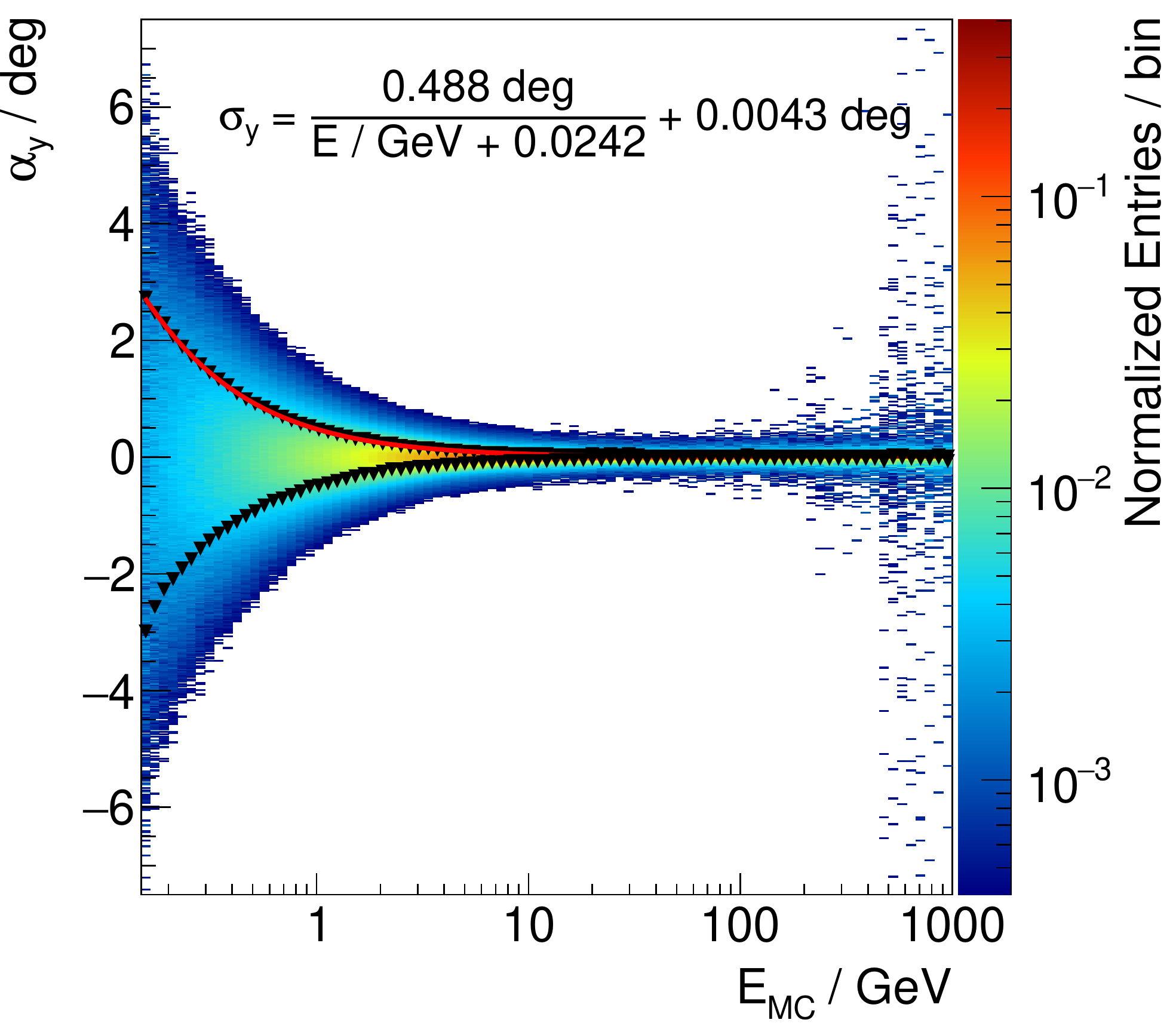}
  \end{minipage}
  \begin{minipage}{0.48\linewidth}
    \centering
    \includegraphics[width=1.0\linewidth]{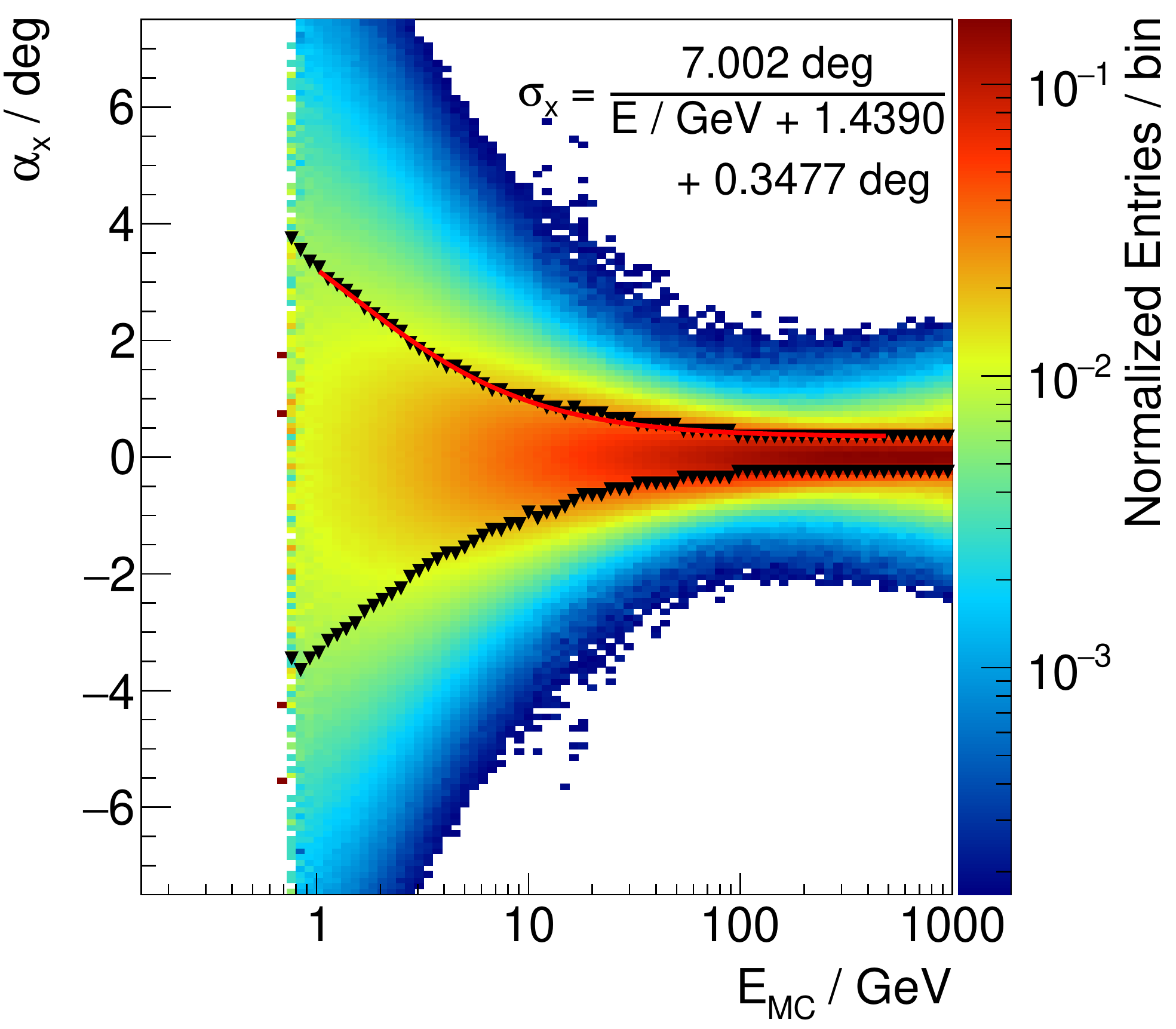}
  \end{minipage}
  \hspace{0.01\linewidth}
  \begin{minipage}{0.48\linewidth}
    \centering
    \includegraphics[width=1.0\linewidth]{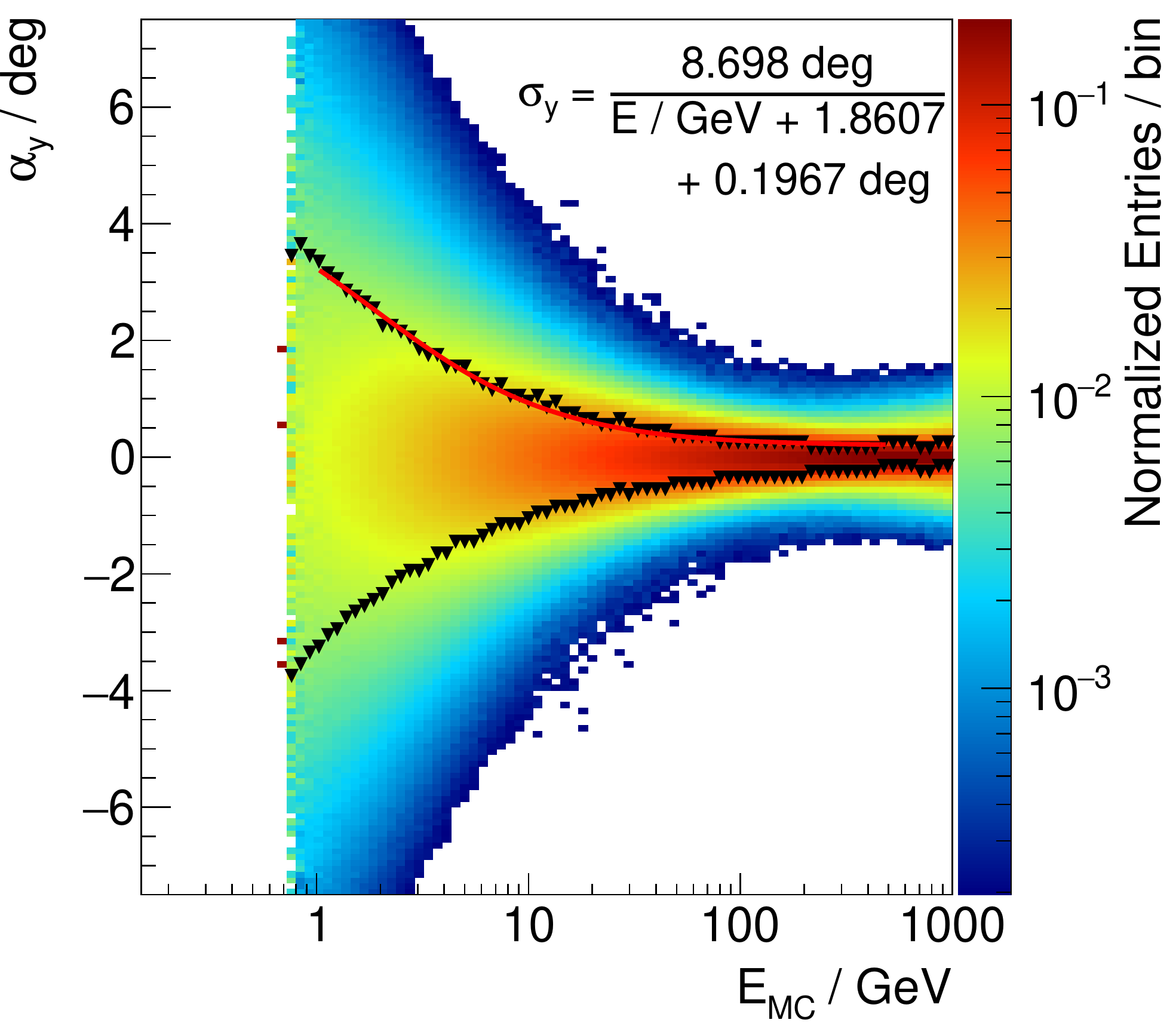}
  \end{minipage}
  \caption{The evolution of the PSF with energy. Upper row: Conversion mode, XZ (top left) and YZ
    (top right) projections. Lower row: Calorimetric mode, XZ (bottom left) and YZ (bottom right)
    projections. The black markers correspond to the Gaussian $\sigma$ in each vertical slice. The
    red lines are parametrizations of the evolution of $\sigma_{x/y}$ with energy according to the
    formulas given in the figures.}
  \label{fig:psf-energy}
\end{figure}

Because of the absence of correlations it is meaningful to look at the one dimensional projection
onto each axis separately. The top row of figure~\ref{fig:psf-energy} shows the evolution of the
$\alpha_x$ (left) and $\alpha_y$ (right) angles for converted photons with energy. The resolution is
approximately \SI{3}{\degree} at \SI{150}{\mega\electronvolt} and improves with energy. It is
approximately \SI{0.5}{\degree} at \SI{1}{\giga\electronvolt} and better than \SI{0.1}{\degree}
above \SI{10}{\giga\electronvolt} for both projections.

At low energies the resolution is dominated by multiple scattering of the electron and positron on
the material of the upper TOF and the first tracker planes. As the energy increases the magnitude of
the scattering decreases and the resolution improves. At the highest energies the resolution
approaches a constant which is related to the single-point spatial resolution of the
tracker. Because of the good spatial resolution of the AMS tracker, multiple scattering is the
dominant effect over almost the entire energy range. This is also the main reason why the resolution
in the XZ and YZ planes is very similar even though the tracker spatial resolution in the YZ
direction is better.

For the calorimeter analysis the evolution of the PSF with energy is shown in the bottom row of
figure~\ref{fig:psf-energy}. The resolution is about \SI{3}{\degree} at \SI{1}{\giga\electronvolt}
and improves with energy to \SI{1}{\degree} at \SI{10}{\giga\electronvolt}. At
\SI{100}{\giga\electronvolt} and above it is better than \SI{0.3}{\degree}. The resolution in the XZ
plane is slightly worse than the YZ resolution at high energies because only four superlayers are
available for the reconstruction in this view instead of five.

The PSF is important in the development of the model for the gamma-ray sky (see
chapter~\ref{sec:modeling}). Because the resolutions in XZ and YZ are very similar in both analysis
modes the average of the two resolution functions is used in the following. This also greatly
simplifies the modeling, because the AMS coordinate axes do not correspond to fixed directions in
the sky.

\begin{figure}[t]
  \centering
  \includegraphics[width=0.9\textwidth]{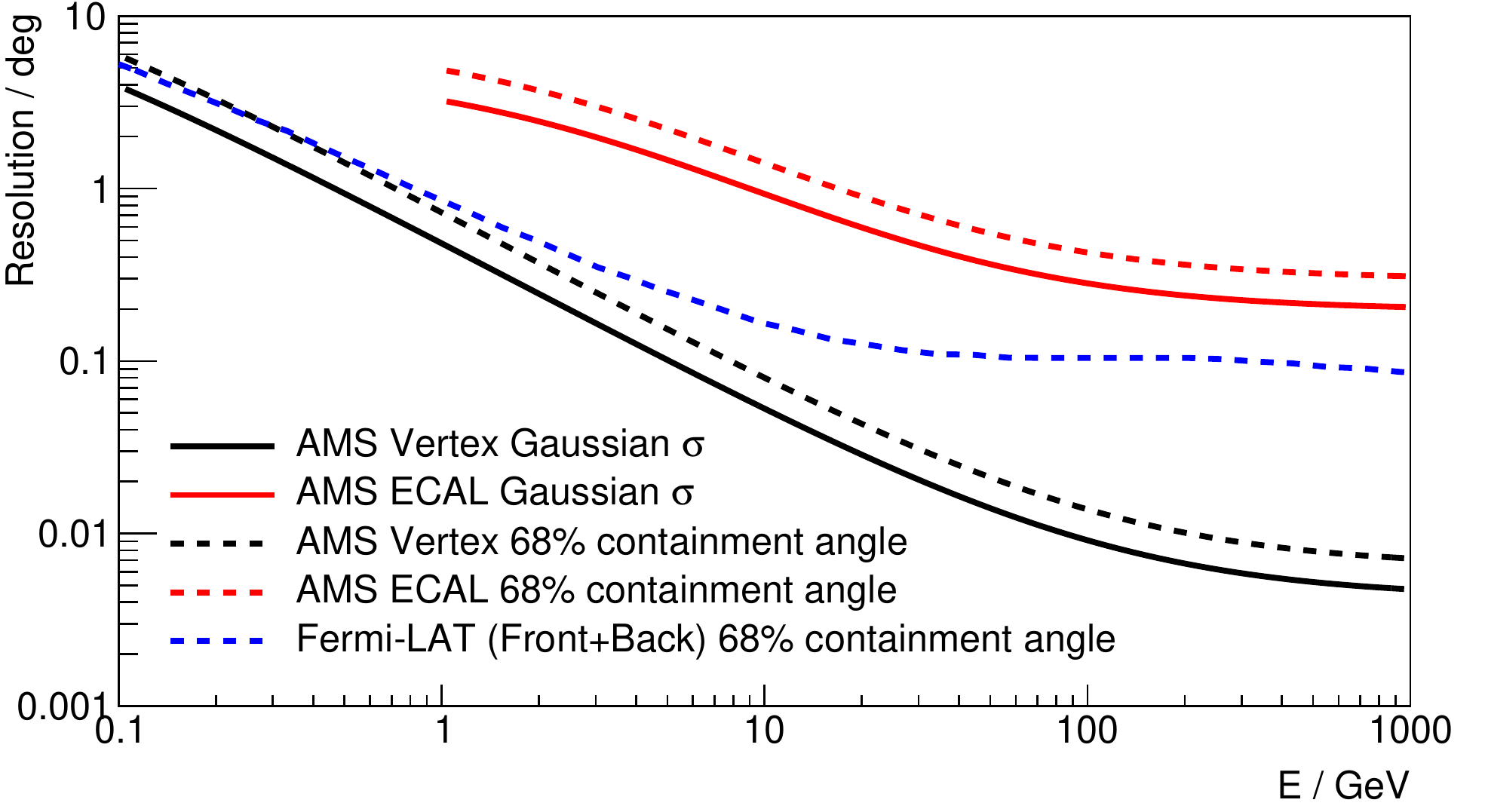}
  \caption{Comparison of the pointing resolution for the AMS Vertex and Calorimeter mode. Also
    included is the Fermi-LAT \SI{68}{\percent} containment angle for the ``SOURCE'' event class
    from the pass 8 instrument response functions~\cite{Fermi_Performance_WWW_2016}.}
  \label{fig:psf-comparison}
\end{figure}

Figure~\ref{fig:psf-comparison} shows a comparison of the angular resolution for the vertex and
calorimeter reconstruction methods. Also shown are the \SI{68}{\percent} containment angles
($\alpha_{68}$) for the two modes. For a given photon energy the reconstructed direction will differ
from the true direction by at most $\alpha_{68}$ for \SI{68}{\percent} of the events. It is defined
as:

\begin{displaymath}
  \int_{0}^{2\pi}{\int_{0}^{\alpha_{68}}{ \mathrm{PSF}(\theta) \sin{\theta} \, \mathrm{d}\theta \,
      \mathrm{d}\varphi} } = 0.68 \,,
\end{displaymath}

where $\theta$ is the angle with respect to the true photon direction. It is related to the $\sigma$
parameter of the two dimensional Gaussian distribution by:

\begin{displaymath}
  \alpha_{68} = \sqrt{-2 \ln{(1 - \alpha)}} \sigma \stackrel{\alpha = 0.68}{\approx} 1.51 \sigma \,.
\end{displaymath}

Also shown in the figure is the Fermi-LAT PSF \SI{68}{\percent} containment angle for the SOURCE
class in the pass 8 reconstruction~{\cite{Fermi_Performance_WWW_2016}} when the front and back
converting event classes are combined.

The vertex angular resolution is excellent over the entire energy range. It is better than the
Fermi-LAT resolution except at the lowest energies where it is approximately the same. At high
energies it is significantly better, due to the better spatial resolution of the AMS tracker. This
enables resolving of fine structure in the galactic diffuse gamma ray emission and to study the
morphology of extended gamma ray sources, if statistics allows.

The calorimeter angular resolution is significantly worse, which is an unavoidable consequence of
the difficulty to reconstruct the axis from an electromagnetic shower. Above approximately
\SI{5}{\giga\electronvolt} the resolution suffices to study point sources and their fluxes in
detail. When measuring the diffuse photon flux the angular resolution becomes irrelevant as long as
the size of region of the sky is large enough.



\subsection{Energy Resolution and Migration}
\label{sec:analysis-energy-resolution}

The energy of the photon has to be reconstructed from the traces it leaves in the detector. For the
vertex conversions this means that the energy is reconstructed from the curvature of the electron
and positron track, for calorimeter photons the energy is estimated from the electromagnetic
shower. Both methods are subject to statistical fluctuations and detector resolution effects,
leading to a potential mismeasurement of the photon energy.

To study the energy resolution one can compare the reconstructed energy $E_{\mathrm{rec}}$ with the
true photon energy in the Monte-Carlo simulation $E_{\mathrm{MC}}$. The sample of simulated data for
the study is the one which is obtained after applying the full set of selection cuts for the two
respective analyses.

For the vertex analysis the energy is calculated from the rigidities of the two tracks according to
formula~\ref{eq:vertex-energy}. The two tracks are typically measured with the inner tracker layers
2 to 8. The rigidity resolution of the tracker was discussed in section~\ref{sec:detector-tracker}
and specifically shown in figure~\ref{fig:ams02-tracker-rigidity-resolution} for protons. Because
the rigidities of the two tracks need to be combined, one can expect the resolution for the photon
energy to be worse than the single-track resolution by at least a factor of $\sqrt{2}$.

In addition, electrons and positrons often emit bremsstrahlung photons. These photons can escape the
detector without being detected and can carry away significant parts of the electron or positron
energy. This results in an additional difficulty to measure the photon energy. Finally, at high
energies above \SI{30}{\giga\electronvolt}, the tracks of the electron and positron begin to
overlap, which results in hit association difficulties when reconstructing the two tracks and has
negative implications for the track fitting procedure.

\begin{figure}[t]
  \begin{minipage}{0.48\linewidth}
    \centering
    \includegraphics[width=1.0\linewidth]{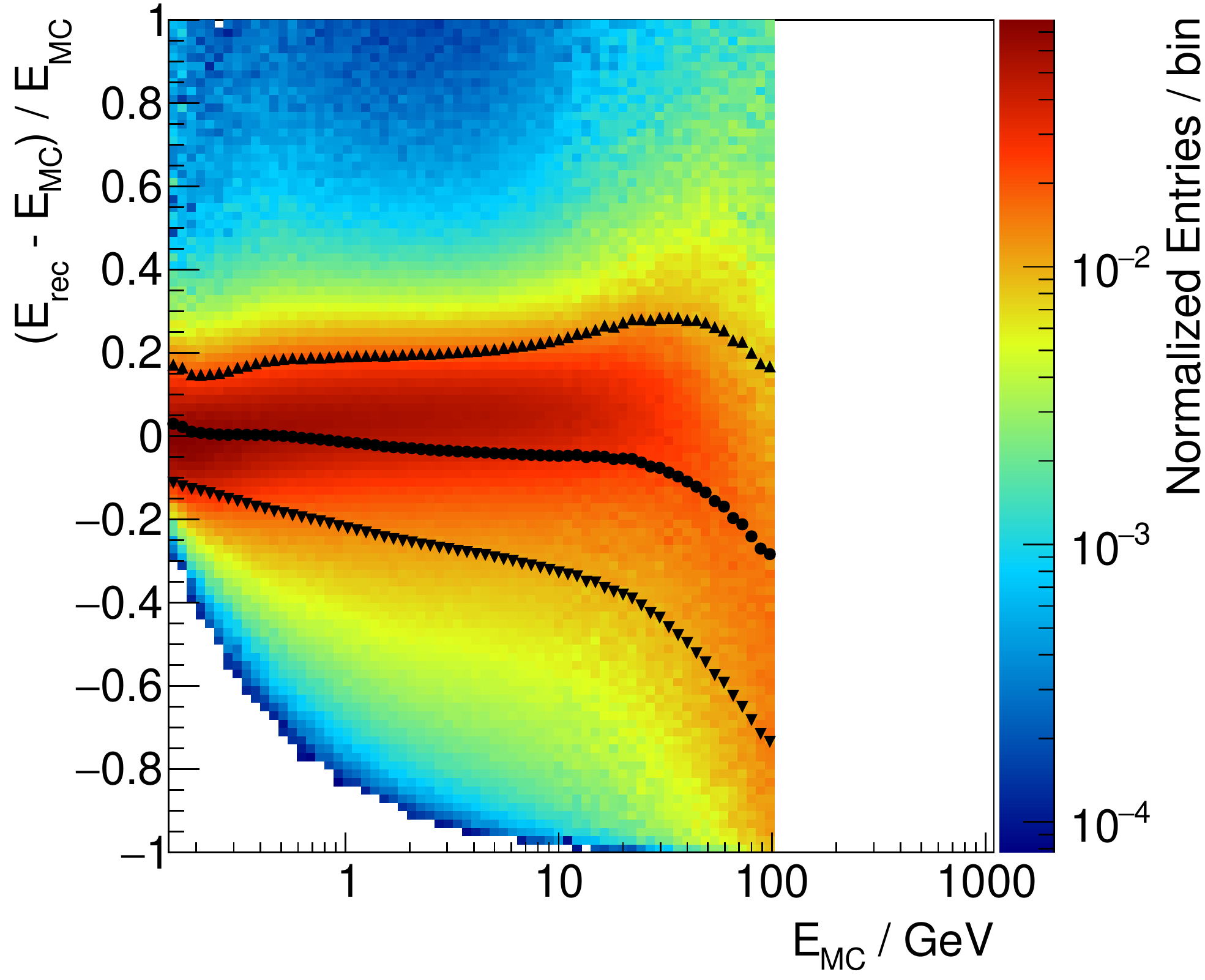}
  \end{minipage}
  \hspace{0.01\linewidth}
  \begin{minipage}{0.48\linewidth}
    \centering
    \includegraphics[width=1.0\linewidth]{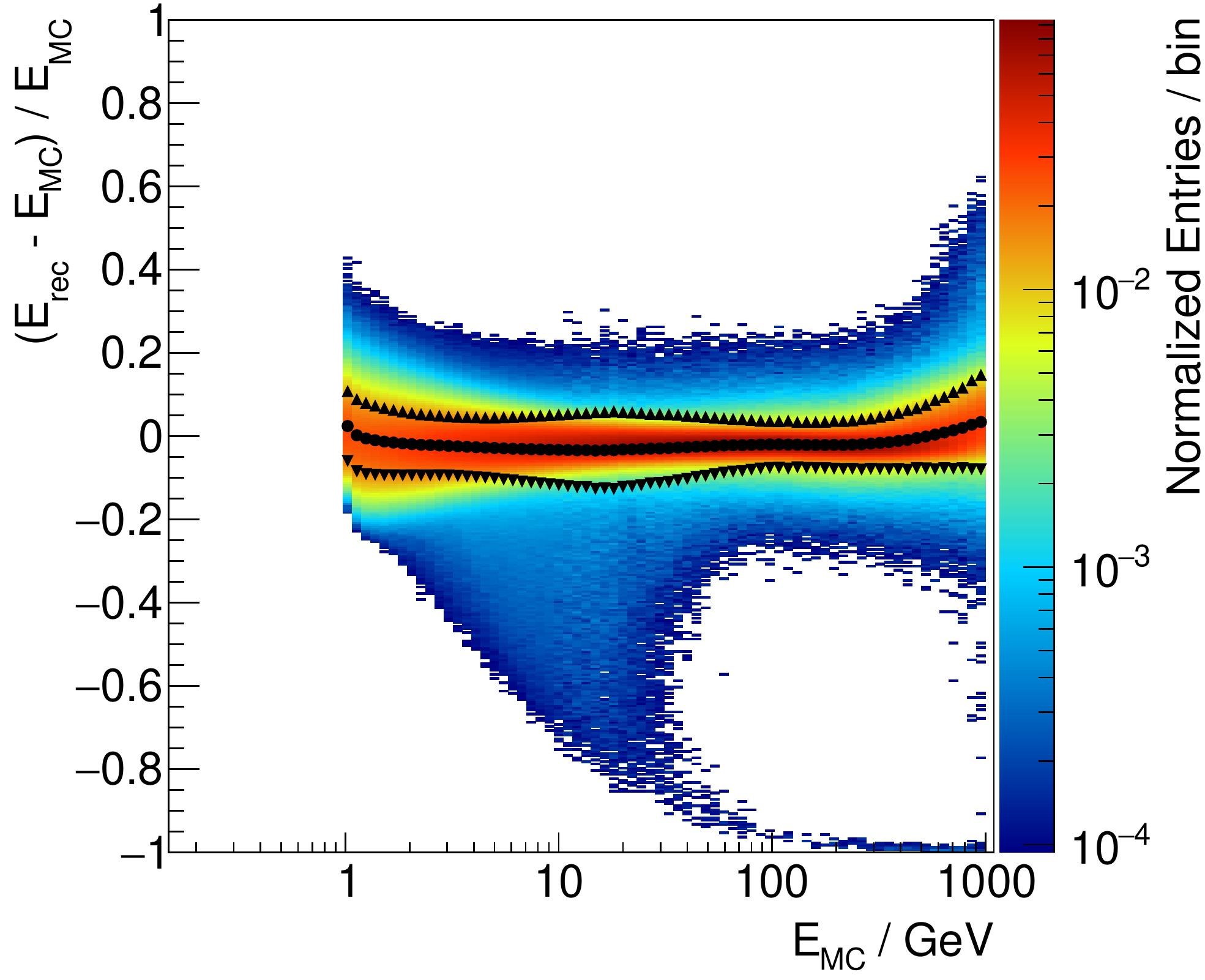}
  \end{minipage}
  \caption{The relative energy resolution $(E_{\mathrm{rec}} - E_{\mathrm{MC}})/E_{\mathrm{MC}}$ as
    a function of the true photon momentum for conversions (left) and calorimeter photons
    (right). The circular markers correspond to the mean value of the distributions in each vertical
    slice. The triangular markers corresponds to the mean plus RMS and mean minus RMS positions and
    are indicative of the energy resolution.}
  \label{fig:energy-resolution}
\end{figure}

Figure~\ref{fig:energy-resolution} depicts the relative energy resolution functions for the vertex
and calorimeter methods. The vertex resolution on the left side is approximately \SI{15}{\percent}
at \SI{200}{\mega\electronvolt} and increases with energy to \SI{28}{\percent} at
\SI{10}{\giga\electronvolt} and approximately \SI{45}{\percent} at \SI{100}{\giga\electronvolt}. The
mean of the distribution has a small bias of a few percent between \SI{1}{\giga\electronvolt} and
\SI{30}{\giga\electronvolt}. Above that point the bias increases and the photon energy is estimated
too low systematically, mainly because of overlapping tracks. Emission of bremsstrahlung causes an
asymmetric shape of the resolution functions, which means that the mean of each vertical slice does
not necessarily coincide with the maximum position. Overall the energy resolution is sufficient to
measure fluxes of photons, although resolving fine structures in the spectra can be challenging.

For the calorimeter analysis on the other hand the energy resolution is very good as can be seen
from the right hand side figure. The difficulties mentioned above do not apply to the energy
measurement with the calorimeter. Photons enter the calorimeter without converting before, so the
entire energy is deposited in the electromagnetic shower. It is better than \SI{10}{\percent} for
all energies between \SI{1}{\giga\electronvolt} and \SI{1}{\tera\electronvolt}. In addition the bias
is relatively small, only a few percent at most. The energy estimator used for the calorimeter
analysis is based on a 3D shower shape fit with leakage corrections. Rear leakage corrections become
important if the shower energy approaches \SI{1}{\tera\electronvolt}, which is the reason why the
energy resolution begins to worsen above \SI{200}{\giga\electronvolt}. The temporary increase of the
RMS at approximately \SI{10}{\giga\electronvolt} is due to the population of events with low
reconstructed energies in the tail of the distribution. This tail is a result of edge effects in the
calorimeter geometry, it disappears when only considering showers which are well within the central
part of the calorimeter.

\begin{figure}[t]
  \begin{minipage}{0.48\linewidth}
    \centering
    \includegraphics[width=1.0\linewidth]{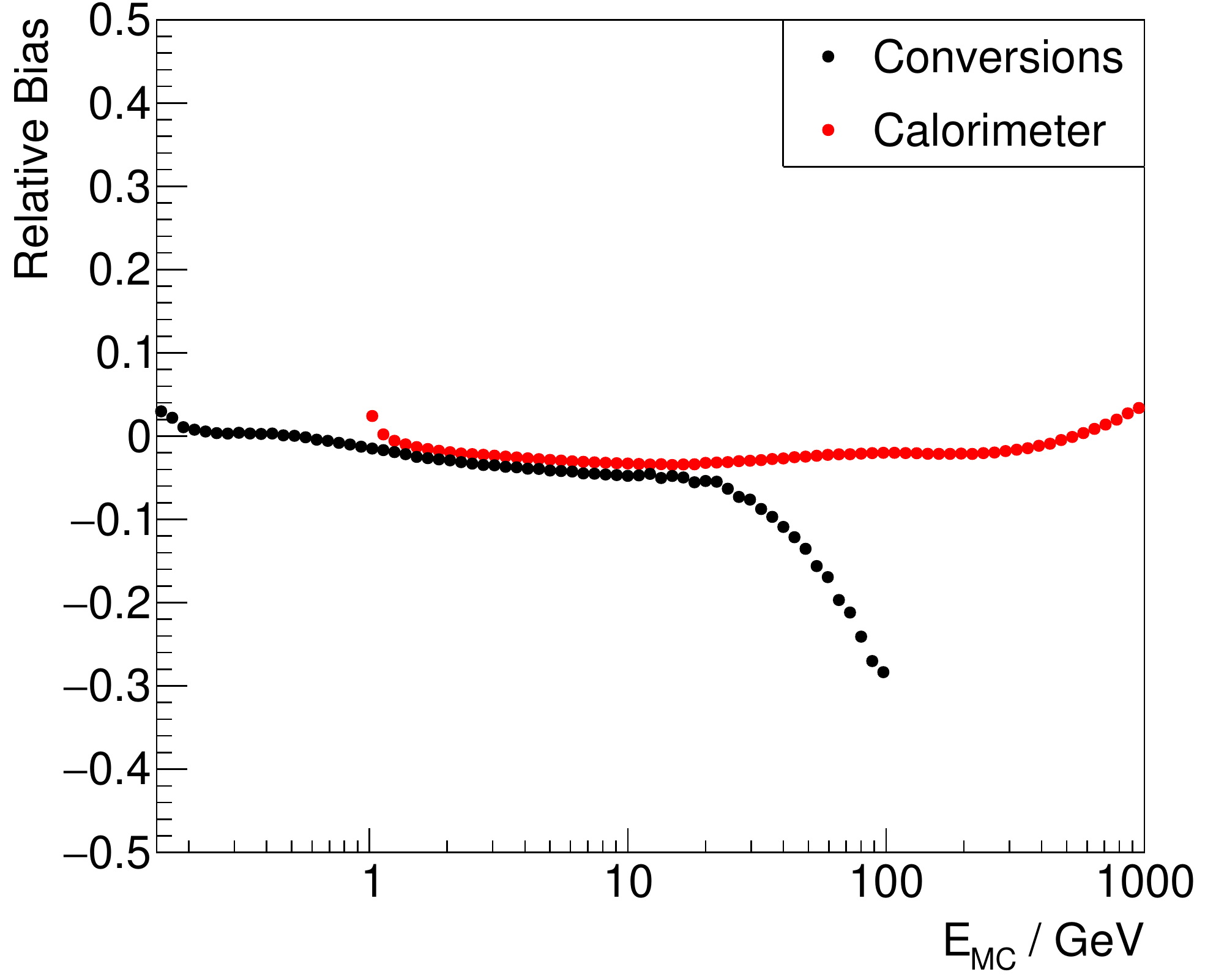}
  \end{minipage}
  \hspace{0.01\linewidth}
  \begin{minipage}{0.48\linewidth}
    \centering
    \includegraphics[width=1.0\linewidth]{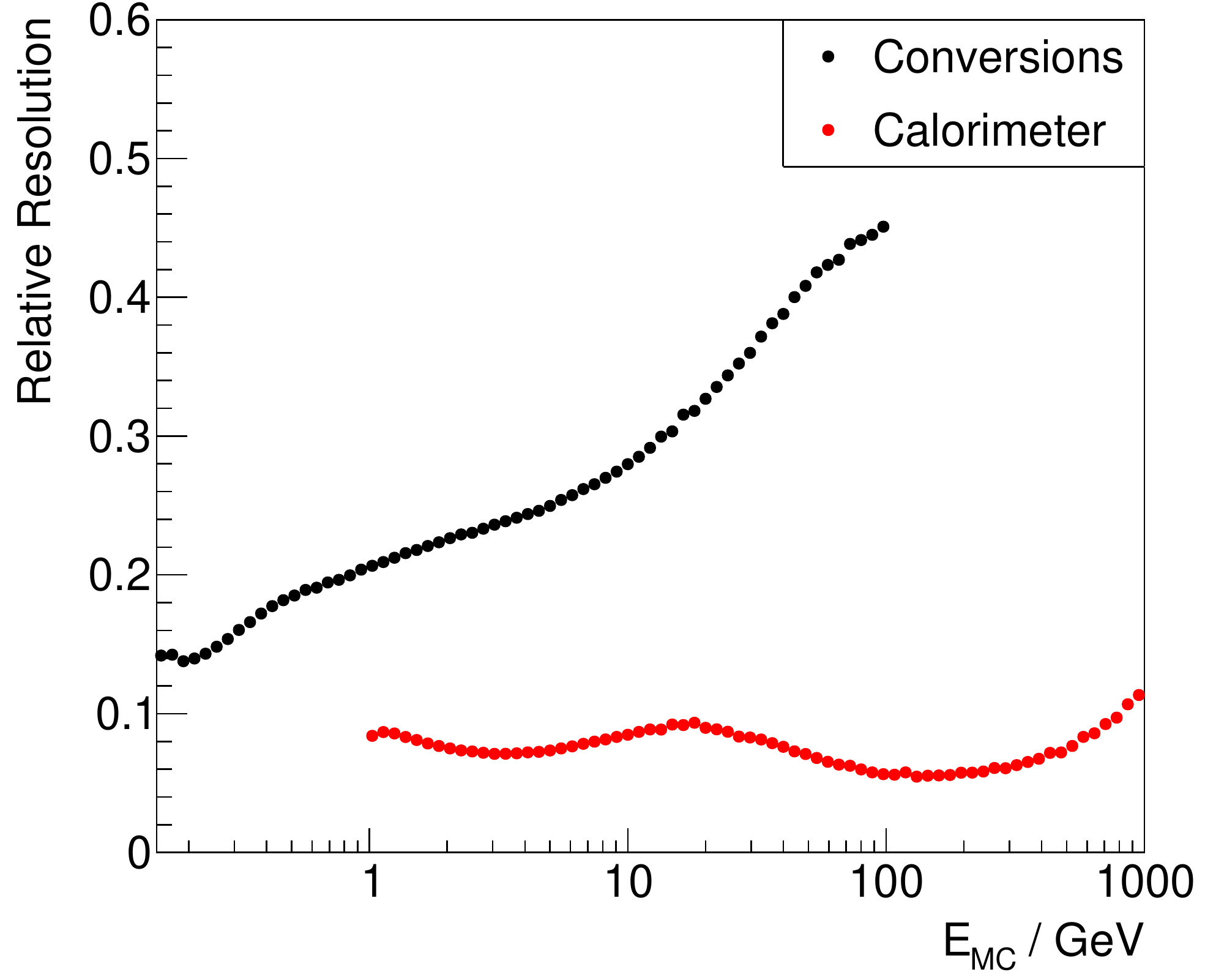}
  \end{minipage}
  \caption{The mean (left) and RMS (right) of the energy resolution distributions for the two
    analyses as a function of the true photon momentum.}
  \label{fig:energy-resolution-mean-and-rms}
\end{figure}

Figure~\ref{fig:energy-resolution-mean-and-rms} summarizes the bias and resolution of the two
reconstruction methods, which illustrates again that photons reconstructed with the calorimeter have
a significantly better energy resolution with less bias.

A mismeasurement of the photon energy results in bin-to-bin migration when measuring fluxes. In
order to correct for this effect one needs to know the probability with which a photon of energy
$E_{\mathrm{true}}$ is reconstructed in a bin corresponding to the energy $E_{\mathrm{rec}}$. For
the case of $n$ energy bins the total number of photons observed in bin $i$ is then:

\begin{displaymath}
  N_{i} = \sum_{j=1}^{n}{p(i|j) N^{\mathrm{true}}_{j}} \,,
\end{displaymath}

where $p(i|j)$ is the probability for an event to migrate from bin $j$ to bin $i$ and
$N^{\mathrm{true}}_{j}$ is the true number of photons in bin $j$. The $n \times n$ matrix defined by
the coefficients $p(i|j)$ is referred to as the migration matrix. This formula can be understood as
``forward folding the true event counts with the migration matrix''. The probabilities $p(i|j)$
depend on the energy resolution and on the binning used in the analysis and can be determined from
the Monte-Carlo simulation - provided the simulation accurately models the detector resolution.

\begin{figure}[t]
  \begin{minipage}{0.48\linewidth}
    \centering
    \includegraphics[width=1.0\linewidth]{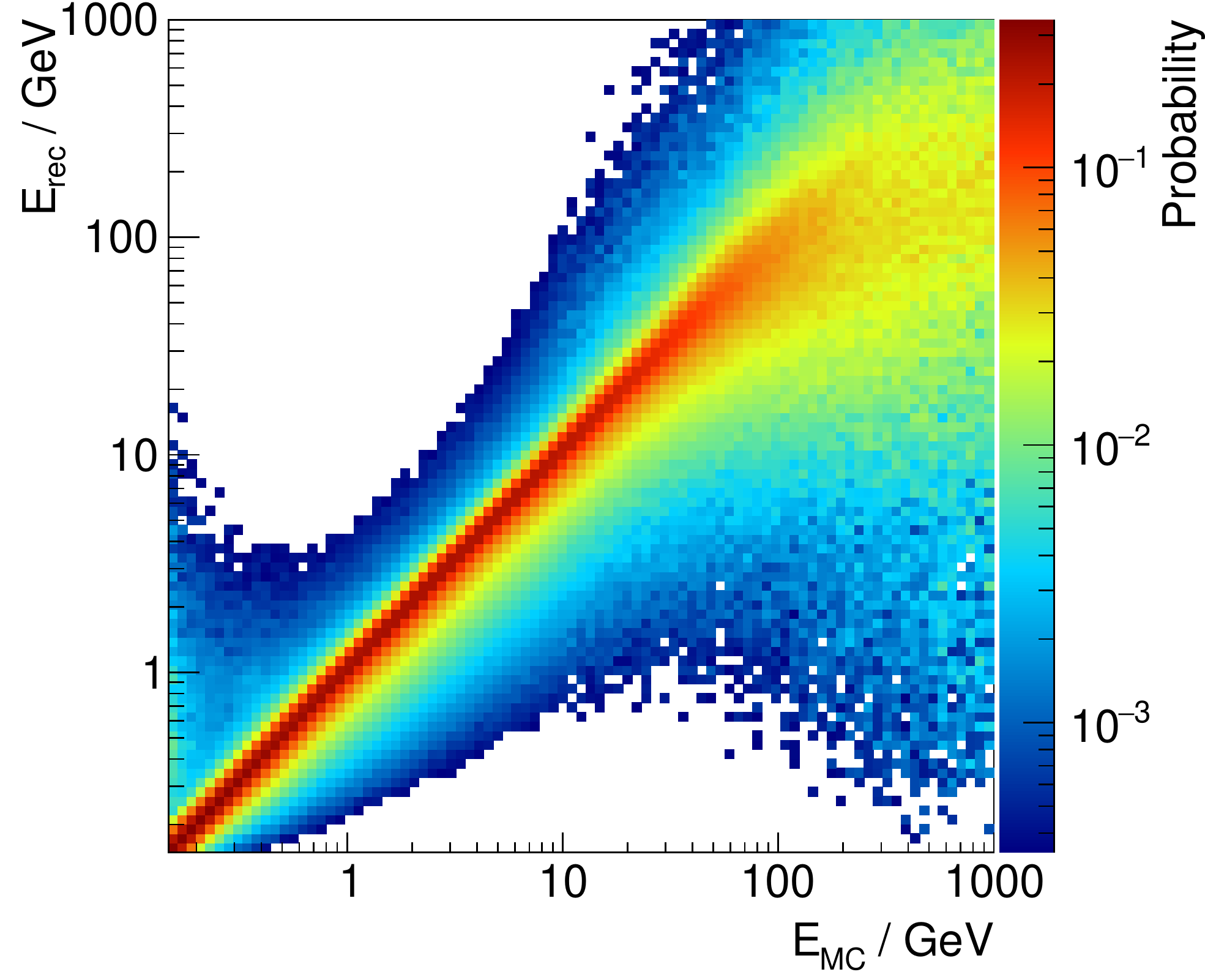}
  \end{minipage}
  \hspace{0.01\linewidth}
  \begin{minipage}{0.48\linewidth}
    \centering
    \includegraphics[width=1.0\linewidth]{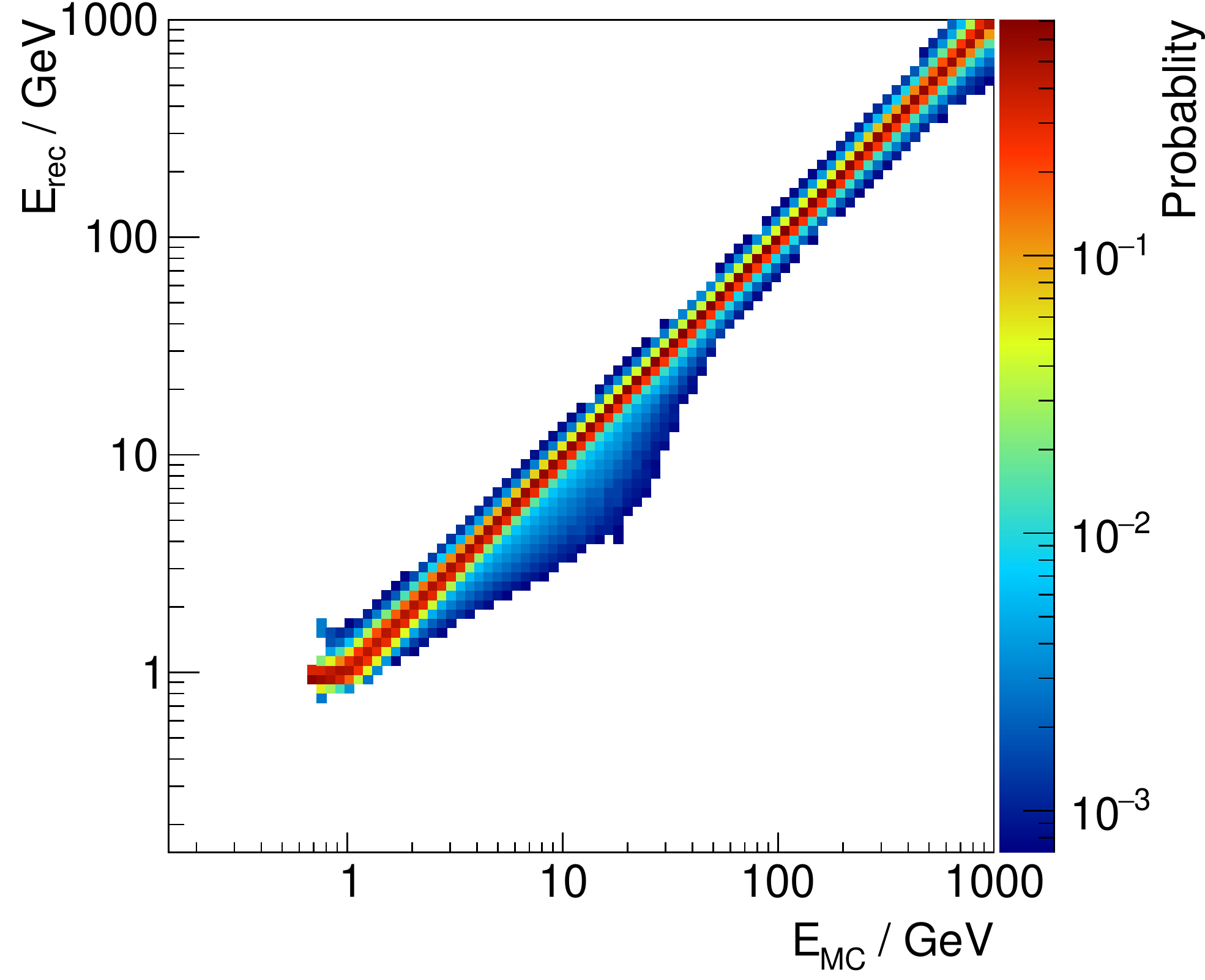}
  \end{minipage}
  \caption{The migration matrices for the analysis binning in the vertex (left) and calorimeter
    (right) selections.}
  \label{fig:energy-resolution-migration}
\end{figure}

Figure~\ref{fig:energy-resolution-migration} shows the energy migration matrices for the two
analyses as determined from the Monte-Carlo simulation. The vertex migration matrix shows that
events can sometimes migrate far away from the bin in which they would have been reconstructed if
the resolution were perfect. Above \SI{100}{\giga\electronvolt} the resolution becomes insufficient
for a flux measurement.  In contrast, the nearly diagonal matrix for the calorimeter analysis shows
that, for the same binning, migration is a minor effect when measuring the energy with the
calorimeter.

The effect of the migration needs to be corrected for using an unfolding procedure, which is
discussed in section~\ref{sec:corrections-unfolding}.

\subsection{Effective Area}
\label{sec:analysis-effective-area}

The effective area ($A_{\mathrm{eff}}$) is the most vital ingredient for the calculation of photon
fluxes. It depends on the photon energy ($E$) as well the photon arrival direction in detector
coordinates ($\cos{\theta}$, $\varphi$) and is measured in units of area. It represents the
projected size of the experiment for a photon with incident direction with respect to the
\mbox{AMS-02} zenith given by $\cos\theta$ and $\varphi$, multiplied by the selection efficiency for
photons from that direction. It can be thought of as the virtual size of the experiment, if the
experiment and the selection were \SI{100}{\percent} efficient in the collection of photons.

The effective area factorizes into a geometric part ($A_{\mathrm{geom}}$), sometimes referred to as
the directional response function~\cite{Sullivan1971}, and the detection and selection efficiency
$\epsilon_{\mathrm{selection}}(E)$:

\begin{equation}
  \label{eq:effective-area-response-function}
  A_{\mathrm{eff}}\left(E,\cos{\theta},\varphi\right) =
  A_{\mathrm{geom}}\left(\cos{\theta}, \varphi\right) \cdot \epsilon_{\mathrm{selection}}(E) \,.
\end{equation}

For a single plane detector of any shape, collecting photons with an efficiency $\epsilon(E)$, the
effective area is:

\begin{equation}
  \label{eq:effective-area-single-plane}
  A_{\mathrm{eff}}(E_,\cos{\theta},\varphi) = \epsilon(E) \cos{\theta} A \,,
\end{equation}

where $A$ is the surface area of the plane and $\cos{\theta}$ is the angle between the source and
the plane's normal. Note that it is independent of $\varphi$, regardless of the shape of the plane,
which is due to the fact that this hypothetical detector is infinitely thin. For a more realistic
experimental configuration the following factors contribute to the effective area for the
measurement of photons:

\begin{itemize}
\item The geometric configuration of the detector.
\item The probability for a photon to convert in (a certain part of) the detector.
\item The efficiency with which the electron/positron pair is triggered.
\item The efficiency with which the electron/positron pair is selected by the selection cuts.
\end{itemize}

The geometric part of the effective area can be calculated analytically~\cite{Sullivan1971}, but for
complex setups it can be difficult to perform the integration. Instead, a numeric method using
Monte-Carlo data is usually employed, which allows an estimation of the contributions from all
detector and selection efficiencies at the same time. In a binned approach (where the indices $i$,
$j$ and $k$ enumerate the bins in energy and the two angles) the effective area can be computed from
the simulation of the experiment using:

\begin{equation}
  \label{eq:effective-area-phi}
  A_{\mathrm{eff}}(E_i,\cos{\theta}_j,\varphi_k) =
  \frac{1}{\Delta\varphi_k\Delta\cos{\theta}_j}
  \frac{N_{\mathrm{passed}}(E_i,\cos{\theta}_j,\varphi_k)}{N_{\mathrm{generated}}(E_i)}
  \mathcal{A}_{\mathrm{gen}} \,,
\end{equation}

where $\Delta\cos{\theta}_j$ and $\Delta\varphi_k$ are the bin widths in $\cos{\theta}$ and
$\varphi$ respectively, $N_{\mathrm{passed}}$ is the number of events passing all selection cuts in
the given bin, $N_{\mathrm{generated}}$ is the number of simulated events in the energy bin and
$\mathcal{A}_{\mathrm{gen}}$ is the geometric acceptance of the surface from which particle
trajectories are generated.

For the \mbox{AMS-02} Monte-Carlo simulation the plane from which all particle trajectories
originate is a square with a side-length of \SI{3.9}{\meter}, located \SI{1.95}{\meter} above the
center of the experiment. The geometric acceptance of the generation surface is therefore:

\begin{equation}
  \label{eq:acceptance-generator-plane}
  \mathcal{A}_{\mathrm{gen}} = A_{\mathrm{gen}} \pi = \left(\SI{3.9}{\meter}\right)^{2} \pi \,,
\end{equation}

where $A_{\mathrm{gen}} = \SI{3.9}{\meter} \times \SI{3.9}{\meter}$ is the area of the plane from
which particles are generated. In case multiple planes of the cube surrounding the detector are used
to generate particles in the Monte-Carlo simulation the formula becomes:

\begin{equation}
  \label{eq:acceptance-generator-plane-nsides}
  \mathcal{A}_{\mathrm{gen}} = N_{\mathrm{sides}} \cdot
  A_{\mathrm{gen}} \pi = N_{\mathrm{sides}} \cdot \left(\SI{3.9}{\meter}\right)^{2} \pi \,,
\end{equation}

with $N_{\mathrm{sides}} = \numrange[range-phrase=..]{1}{6}$. Integration of the effective area over
solid angle yields the effective acceptance ($\mathcal{A}_{\mathrm{eff}}$):

\begin{align}
  \label{eq:effective-acceptance}
  \mathcal{A}_{\mathrm{eff}}(E_i)
  &= \int_{\Omega}{A_{\mathrm{eff}}\left(E_i, \cos{\theta}, \varphi\right) \mathrm{d}\Omega} =
    \int_{0}^{2\pi}{\int_{-1}^{1}{A_{\mathrm{eff}}\left(E_i, \cos{\theta},
    \varphi\right)\mathrm{d}\cos{\theta}}\mathrm{d}\varphi}\\
  &= \sum_{j}{ \sum_{k}{A_{\mathrm{eff}}\left(E_i,
    \cos{\theta}_j, \varphi_k\right) \Delta\varphi_k\Delta\cos{\theta}_j} }\\
  &= \frac{N_{\mathrm{passed}}(E_i)}{N_{\mathrm{generated}}(E_i)} \mathcal{A}_{\mathrm{gen}} \,.
\end{align}

The effective acceptance is commonly used when measuring isotropic fluxes ($\Phi$), such as those of
charged galactic cosmic rays:

\begin{equation}
  \label{eq:isotropic-flux}
  \Phi(E_{i}) = \frac{N(E_{i})}{\mathcal{A}_{\mathrm{eff}}(E_{i})
    \epsilon_{\mathrm{trigger}}(E_{i}) \Delta T \Delta E_{i}} \,,
\end{equation}

where $N(E_i)$ is the number of observed events in bin $i$, $\epsilon_{\mathrm{trigger}}$ is the
trigger efficiency, $\Delta T$ is the observation time and $\Delta E_i$ is the energy bin width.

For the non-isotropic gamma ray flux exposure maps are used, which are constructed from the
effective area and the path of the detector's zenith axis on the celestial sphere in
section~\ref{sec:analysis-exposure-maps}.

Another quantity which is sometimes quoted is the field-of-view (FOV):

\begin{equation}
  \label{eq:effective-area-fov}
  \mathrm{FOV}(E_i) = \frac{\mathcal{A}_{\mathrm{eff}}(E_i)}{A_{\mathrm{eff}}(E_i, 0, 0)} \,,
\end{equation}

which corresponds to the ratio of the effective acceptance to the on-axis effective area.

For geometries which are either very thin or feature cylindrical symmetry the effective area does
not depend on $\varphi$. To a good approximation this is also true for the \mbox{AMS-02} geometry,
even though some of the detector planes are rectangular and not circular. In this case it makes
sense to express the effective area as a function of $E$ and $\cos{\theta}$, only, by averaging over
$\varphi$:

\begin{equation}
  \label{eq:effective-area-averaged}
  A_{\mathrm{eff}}(E_i,\cos{\theta}_j) =
  \frac{1}{2 \pi \Delta\cos{\theta}_j}
  \frac{N_{\mathrm{passed}}(E_i,\cos{\theta}_j)}{N_{\mathrm{generated}}(E_i)}
  \mathcal{A}_{\mathrm{gen}} \,.
\end{equation}

This approximation is particularly adequate for the \mbox{AMS-02} calorimeter selection, since the
ECAL can be considered a thin detector. In the case of the vertex analysis, since the inner tracker
layers are roughly circular it is also a good approximation, but not perfect due to the rectangular
shape of the TOF planes. This approach allows to bin the data sample in only two dimensions (energy
and $\cos{\theta}$), which significantly reduces the fluctuations due to finite Monte-Carlo
statistics. However, a correction to this approximation will be derived in
section~\ref{sec:analysis-effective-area-phi-correction}.

\begin{figure}[t]
  \begin{minipage}{0.48\linewidth}
    \centering
    \includegraphics[width=1.0\linewidth]{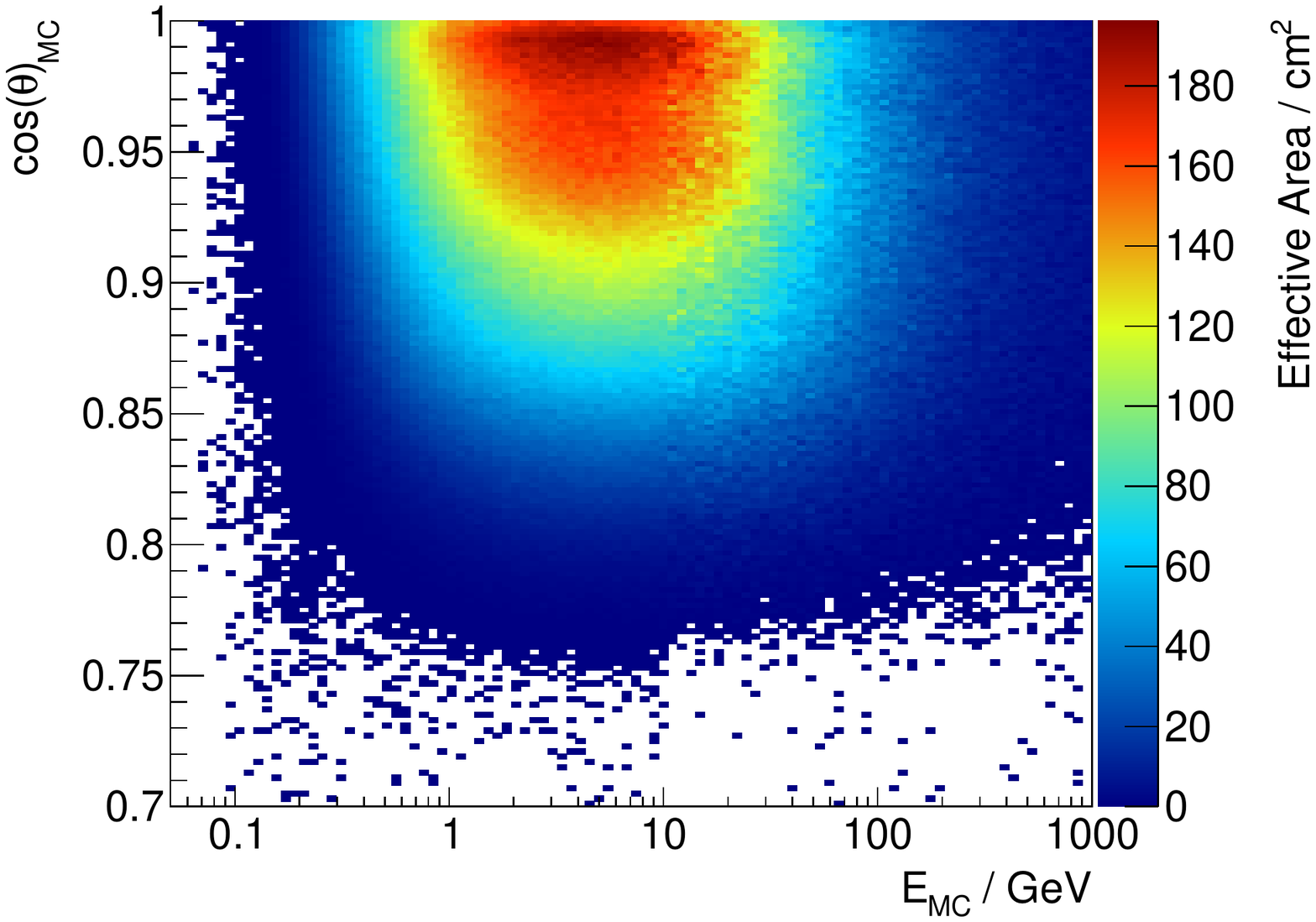}
  \end{minipage}
  \hspace{0.01\linewidth}
  \begin{minipage}{0.48\linewidth}
    \centering
    \includegraphics[width=1.0\linewidth]{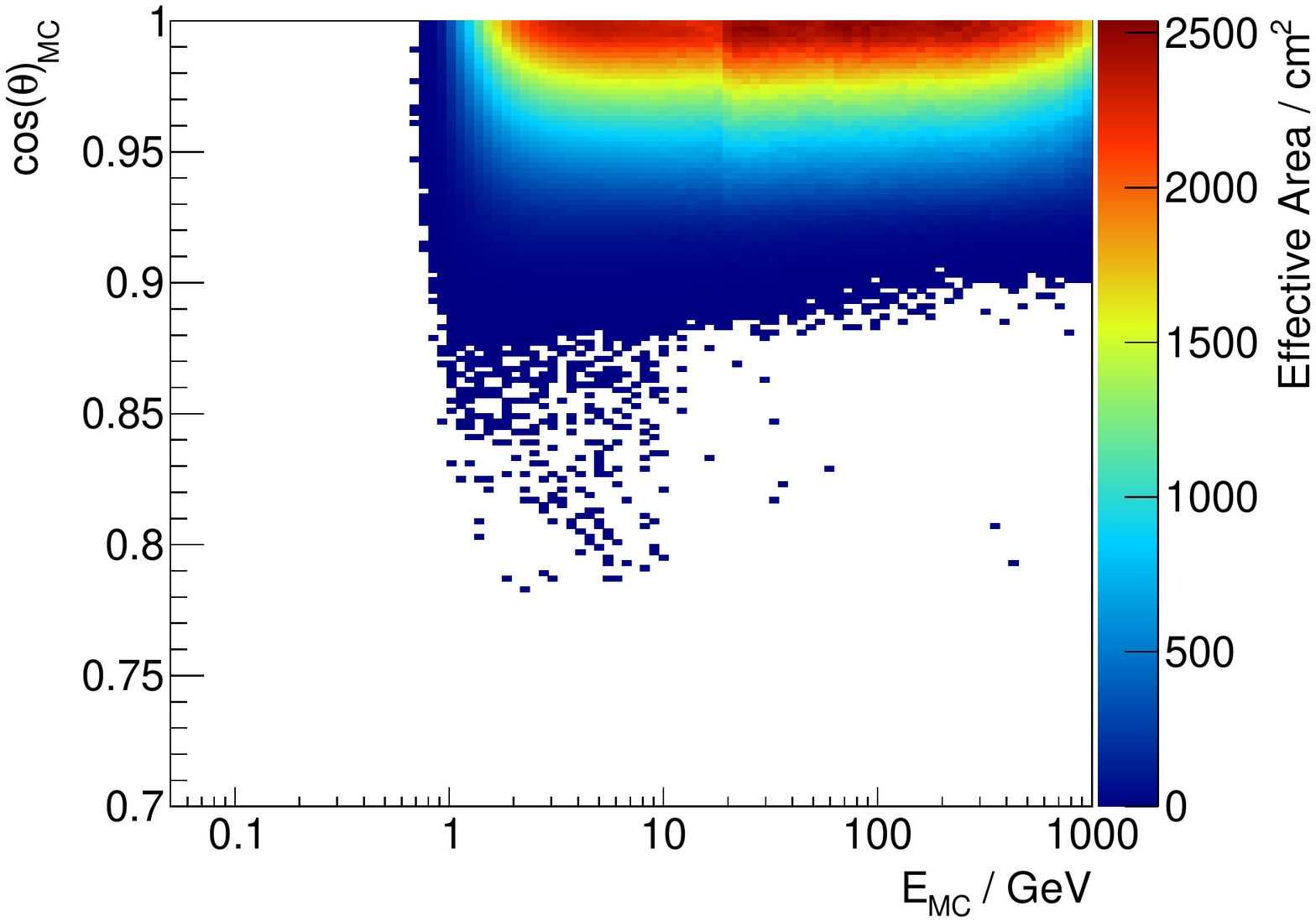}
  \end{minipage}
  \caption{The effective area as a function of energy and $\cos{(\theta)}$ for the vertex (left) and
    calorimeter (right) analysis.}
  \label{fig:effective-area-2d}
\end{figure}

Figure~\ref{fig:effective-area-2d} shows the resulting effective area for the two selections as a
function of the photon energy and zenith angle. The vertex effective area, shown on the left, shows
that events with zenith angles $\cos{\theta}$ greater than 0.76 are selected. This corresponds to an
angular acceptance cone size of approximately \SI{40}{\degree}. It rises from approximately
\SI{100}{\mega\electronvolt}, reaches its maximum around \SI{2}{\giga\electronvolt} and slowly
declines from \SI{10}{\giga\electronvolt} onwards. The maximum of the effective area is
approximately $\SI{180}{cm^{2}}$ for photon sources close to the zenith of AMS.

At low energies the effective area is low because there are many events in which either the electron
or positron is swept away by the magnetic field and deflected to the ACC which vetoes the trigger
decision. It is also possible for the electron or positron to stop in the detector material, in case
the pair production is asymmetric, producing a particle with extremely low energy. Above
\SI{10}{\giga\electronvolt} the two tracks begin to overlap also in the bending plane, in which case
the track reconstruction fails to identify two distinct tracks.

For the calorimeter, shown on the right, the effective area is non zero for energies above
approximately \SI{1}{\giga\electronvolt}, and for angles above $\cos{\theta} = 0.9$, which
corresponds to an angular acceptance cone size of approximately \SI{25}{\degree}. Below
\SI{1}{\giga\electronvolt} the calorimeter trigger is too inefficient for photons to be registered
by the calorimeter. At the highest energies, above \SI{500}{\giga\electronvolt}, calorimeter
backsplash and leakage become important, which cause a decline in the effective area.

\begin{figure}[t]
  \begin{minipage}{0.48\linewidth}
    \centering
    \includegraphics[width=1.0\linewidth]{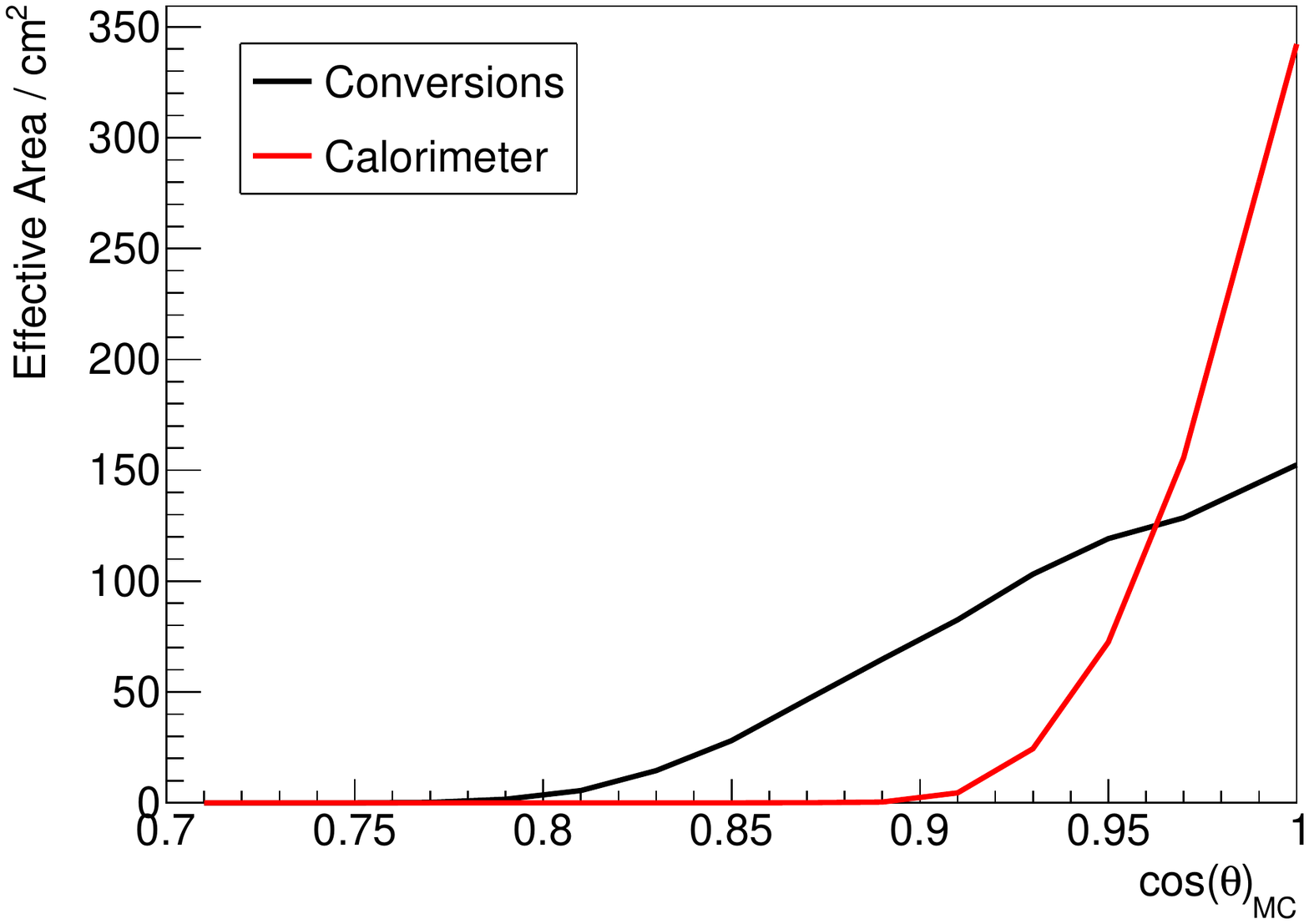}
  \end{minipage}
  \hspace{0.01\linewidth}
  \begin{minipage}{0.48\linewidth}
    \centering
    \includegraphics[width=1.0\linewidth]{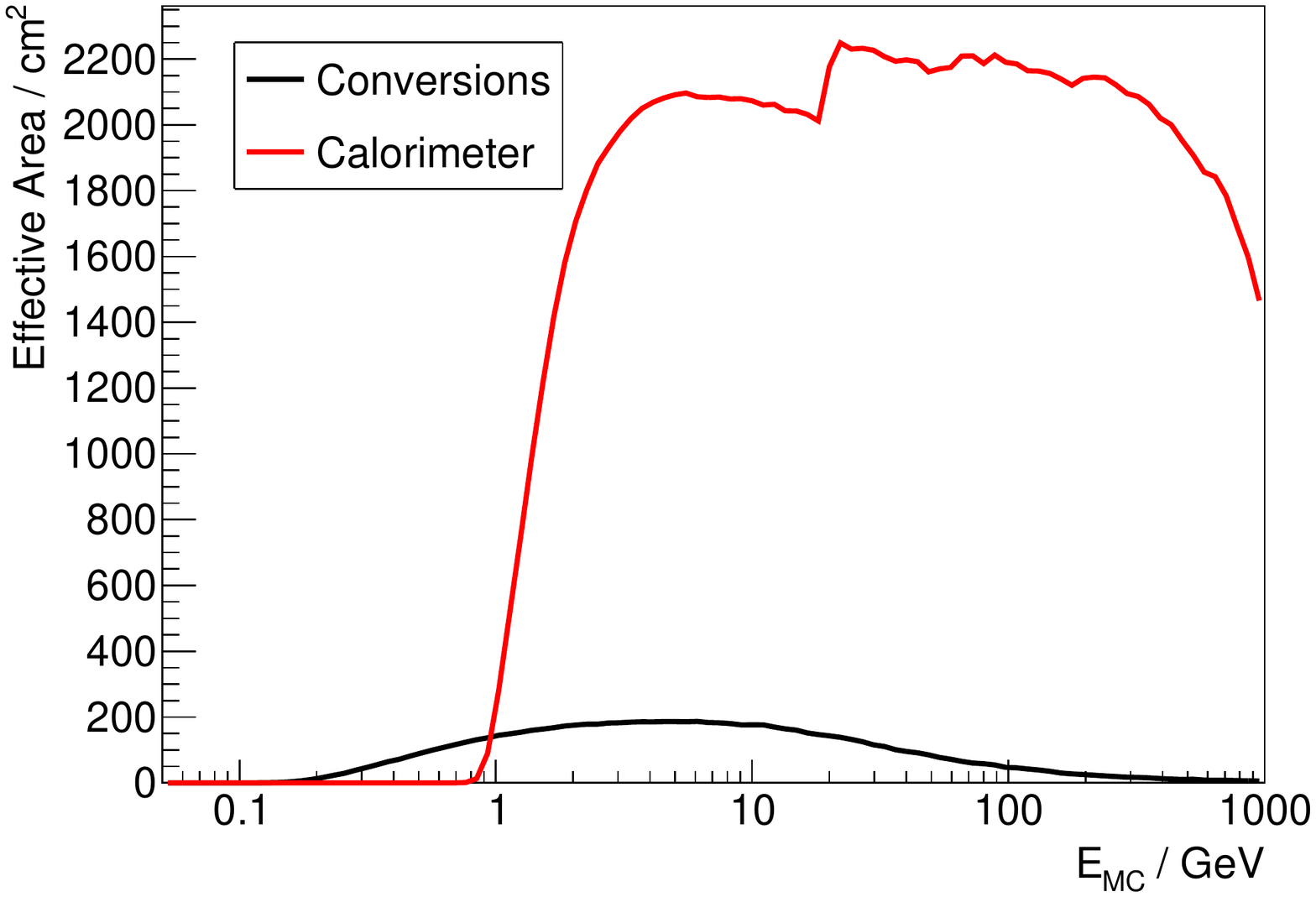}
  \end{minipage}
  \caption{The effective area as a function of $\cos({\theta})$ for \SI{1}{\giga\electronvolt}
    photons on the left, and for perpendicular incidence as a function of the photon energy on the
    right.}
  \label{fig:effective-area-fixed-energy-fixed-angle}
\end{figure}

The left hand side of figure~\ref{fig:effective-area-fixed-energy-fixed-angle} shows the effective
area for \SI{1}{\giga\electronvolt} photons for the two selections as a function of the zenith angle
$\cos{\theta}$. These curves correspond to slices of the two-dimensional effective area
distributions in figure~\ref{fig:effective-area-2d} for $\cos{\theta} \thicksim 1$. Comparing the
two analysis modes, an advantage of the conversion mode is the ability to collect photons impinging
with larger zenith angles.

The zenith effective area is shown on the right hand side of
figure~\ref{fig:effective-area-fixed-energy-fixed-angle} as a function of the photon energy. It
shows that photons between \SI{100}{\mega\electronvolt} and \SI{1}{\giga\electronvolt} can only be
studied with the vertex analysis. The calorimeter analysis features a better effective area above
\SI{1}{\giga\electronvolt} for photons impinging perpendicularly. The step in calorimeter effective
area at approximately \SI{20}{\giga\electronvolt} is a result of the cut on the number of two-sided
ACC hits, see equation~(\ref{eq:ecal-acc-cut}), when the allowed number of ACC hits changes from
zero to one.

The zenith effective area can be compared to the geometric expectation, in order to understand the
order of magnitude of the selection efficiencies.

The vertex analysis selection is designed to select photons which convert in the first upper TOF
layer, or in the support material directly above. The probability for a photon to convert in the
desired region can be estimated from the Monte-Carlo simulation, taking into account the material
distribution ($X / X_0$) in the conversion region. It is also important to consider the amount of
material above the conversion region, because photons converting too early are also rejected in the
analysis. The result is that about \SI{6}{\percent} of all photons convert in the target
region. This number is largely independent of energy, because the pair production cross section does
not vary with energy above \SI{100}{\mega\electronvolt}.

In order to estimate the ``surface area'' of the detector, as seen from the zenith, one needs only
to consider the surface area of the smallest plane in the detector. Since the electron and positron
track need to pass through the inner tracker in order to be measured, these are the inner tracker
layers 3 to 8 whose active surface area is approximately \SI{6700}{\centi\meter\squared}. Thus, the
theoretical upper limit of the effective area for perpendicular incidence is approximately
\SI{400}{\centi\meter\squared}. The maximum of the observed effective area, including the selection
efficiencies, is \SI{180}{\centi\meter\squared}, which means that the combined selection efficiency
of all selection cuts is approximately \SI{45}{\percent} for \SIrange{2}{10}{\giga\electronvolt}
photons. This rough estimate is in agreement with the signal efficiency determined in
figure~\ref{fig:signal-efficiency-vertex}.

For the calorimeter case one needs to consider the surface area of the ECAL, which is roughly
$\SI[product-units=brackets-power,output-product=\times]{60 x 60}{\centi\meter} =
\SI{3600}{\centi\meter\squared}$. According to the Monte-Carlo simulation about \num{2 / 3} of the
photons enter the calorimeter without converting before. So the theoretical upper bound for the
effective area in the calorimeter analysis is approximately
$\num{2 / 3} \cdot \SI{3600}{\centi\meter\squared} = \SI{2400}{\centi\meter\squared}$. The full
effective area of the selection is very close to this geometrical estimation for energies above
\SI{3}{\giga\electronvolt}, which shows that the selection efficiency is high (compare also
figure~\ref{fig:signal-efficiency-ecal}).

Above \SI{300}{\giga\electronvolt} the selection efficiency drops because backsplash particles from
the shower can mimic charged particle signals in the TOF, ACC and tracker.

\begin{figure}[t]
  \centering
  \includegraphics[width=0.6\linewidth]{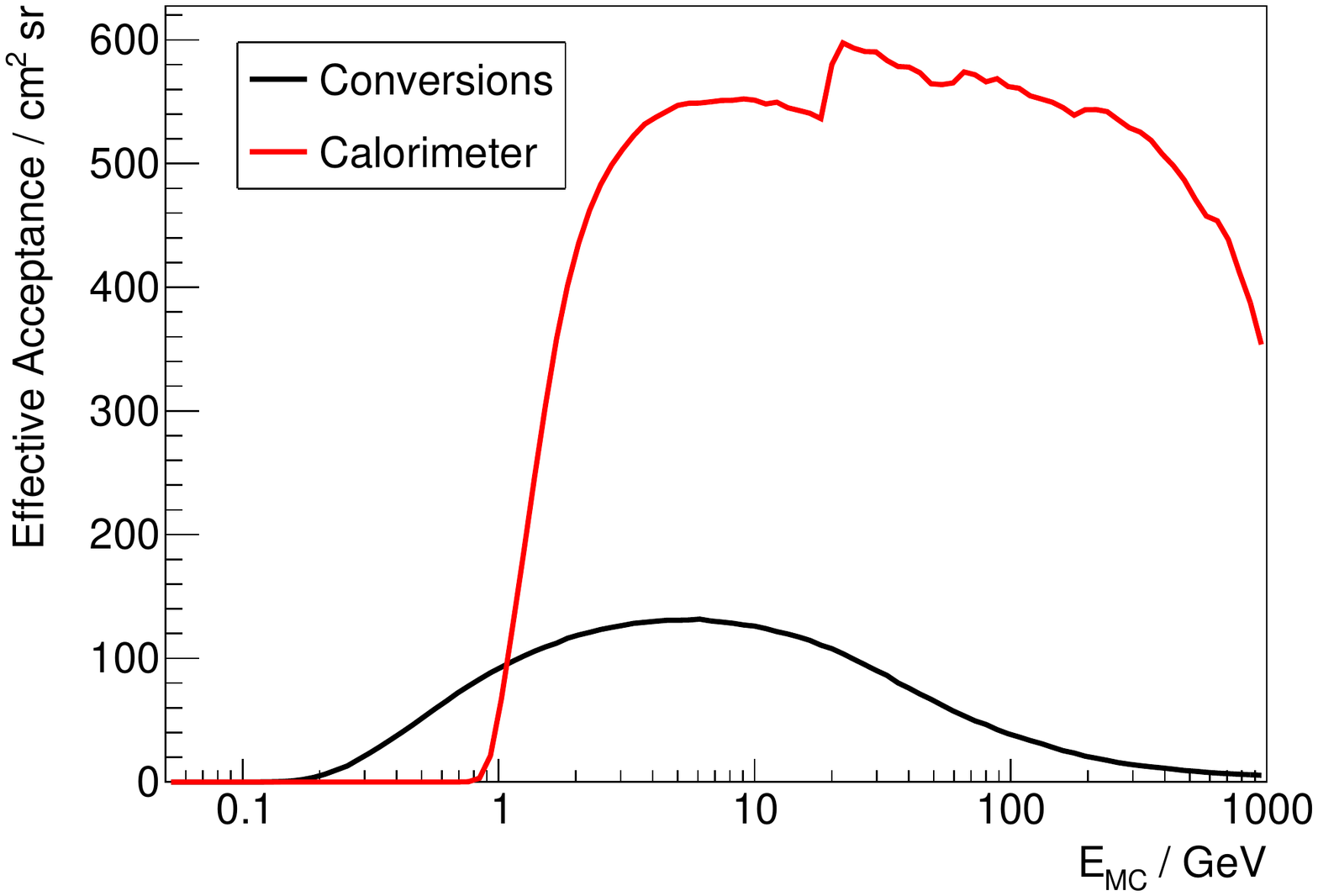}
  \caption{The effective acceptance as a function of energy for the vertex (black) and calorimeter
    (red) analysis.}
  \label{fig:effective-area-acceptance}
\end{figure}

The effective acceptance is the integral of the effective area performed over solid angle as
expressed by equation~(\ref{eq:effective-acceptance}). It is shown in
figure~\ref{fig:effective-area-acceptance}. It depends only on the photon energy and has units of
$\si{\centi\meter\squared\steradian}$. Compared to the zenith effective area the large difference
between vertex and calorimeter analyses is reduced, due to the larger angular acceptance cone size
of the vertex analysis. The maximum effective acceptance is roughly
\SI{140}{\centi\meter\squared\steradian} for the vertex analysis and
\SI{660}{\centi\meter\squared\steradian} for the calorimeter analysis.

\subsubsection{Phi Correction}
\label{sec:analysis-effective-area-phi-correction}

In order to account for the $\varphi$-dependence of the effective area the full effective area is
factorized as follows:

\begin{equation}
  \label{eq:effective-area-phi-correction}
  A_{\mathrm{eff}}\left(E_i,\cos{\theta}_j,\varphi_k\right)
  = A_{\mathrm{eff}}\left(E_i,\cos{\theta}_j\right) \cdot C\left(\cos{\theta}_j, \varphi_k\right) \,,
\end{equation}

where $A_{\mathrm{eff}}\left(E_i,\cos{\theta}_j\right)$ was calculated above and
$C\left(\cos{\theta}_j, \varphi_k\right)$ is a geometric correction function of order 1. The
underlying assumption is that the correction does not depend on energy, but purely on the geometry
of the detector.

The correction function is calculated from the Monte-Carlo simulation as follows:

\begin{equation}
  \label{eq:phi-correction-function}
  C\left(\cos{\theta}_j, \varphi_k\right) = \frac{2 \pi}{\Delta\varphi_k} \cdot
  \frac{N_{\mathrm{passed}}\left(\cos{\theta}_j,\varphi_k\right)}{N_{\mathrm{passed}}\left(\cos{\theta}_j\right)} \,.
\end{equation}

\begin{figure}[t]
  \begin{minipage}{0.48\linewidth}
    \centering
    \includegraphics[width=1.0\linewidth]{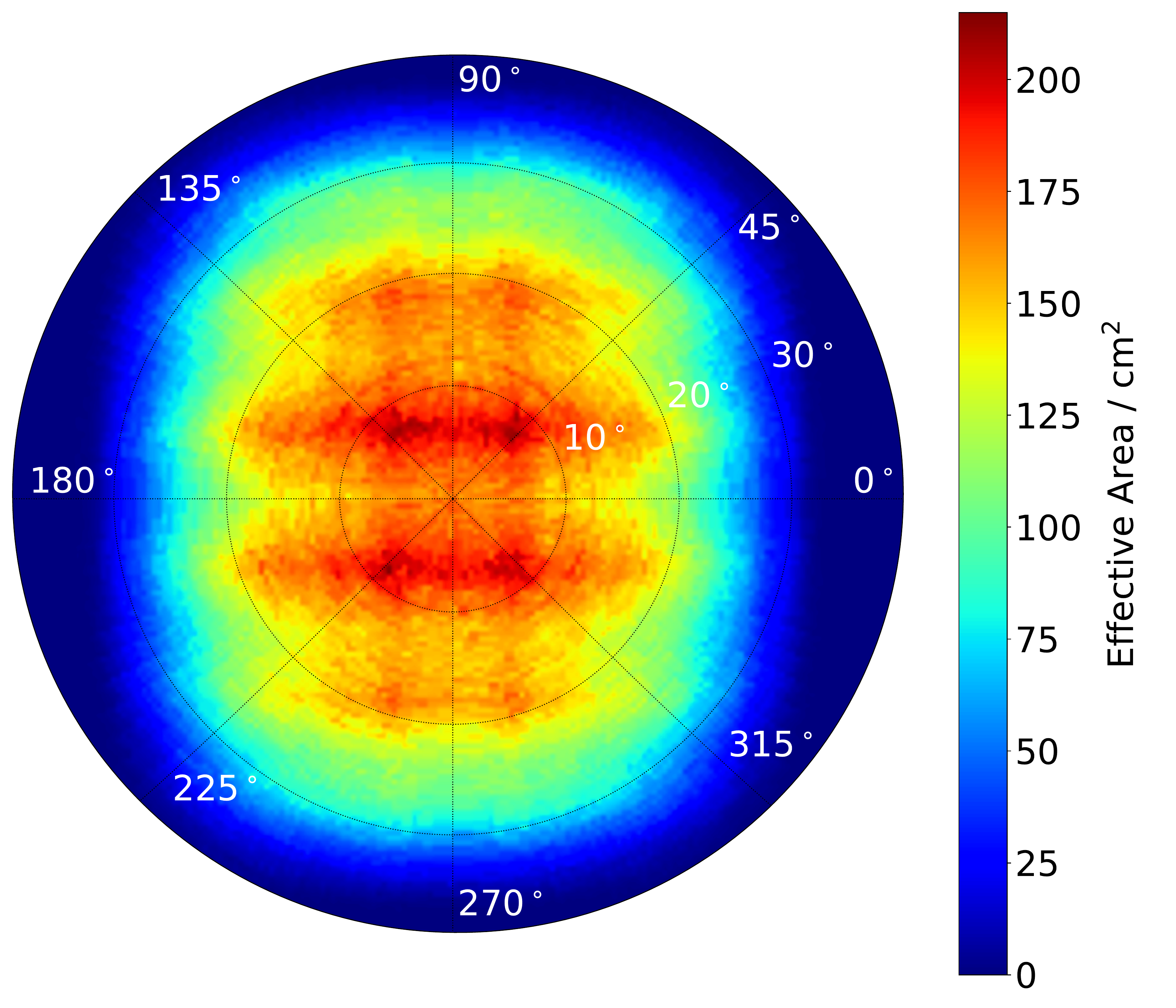}
  \end{minipage}
  \hspace{0.01\linewidth}
  \begin{minipage}{0.48\linewidth}
    \centering
    \includegraphics[width=1.0\linewidth]{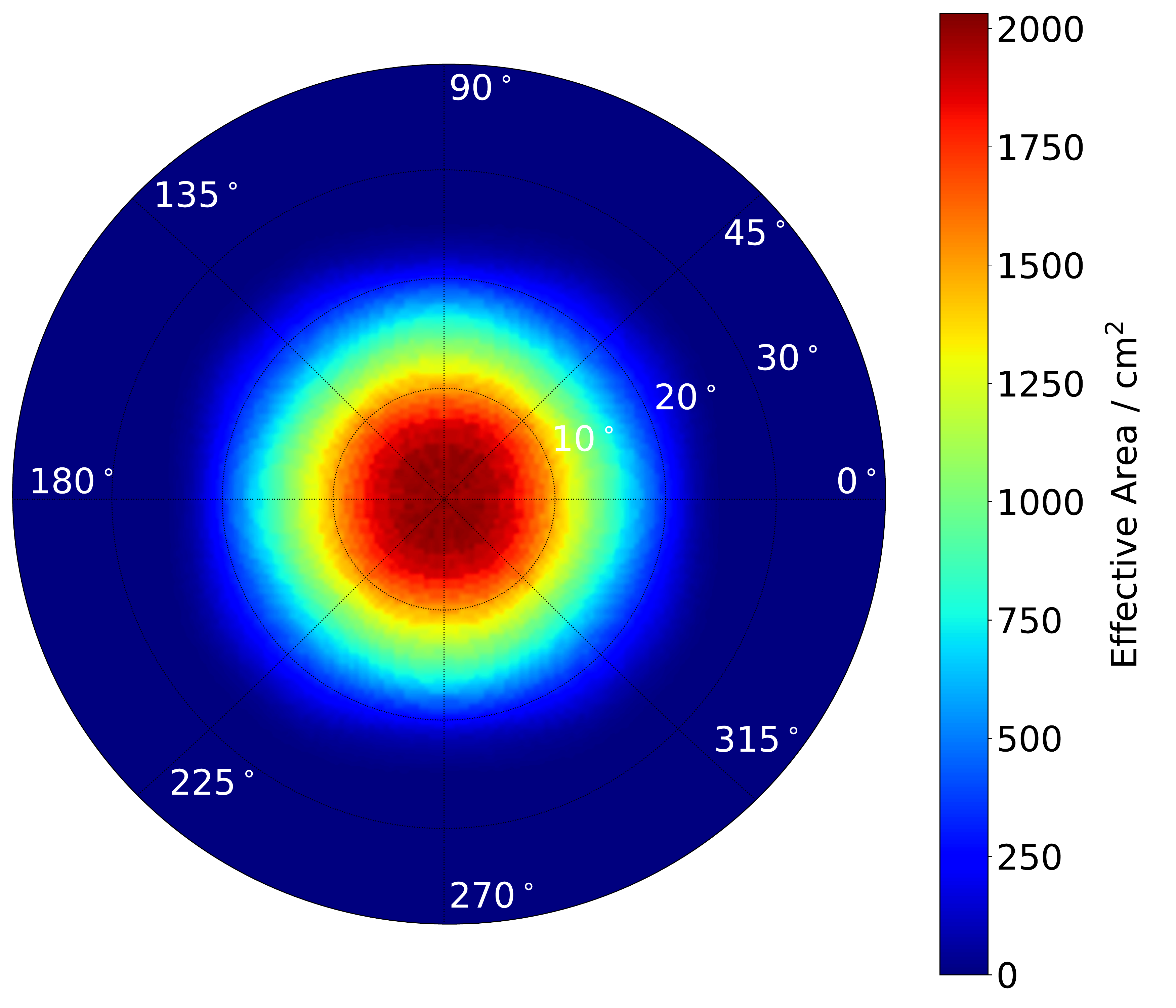}
  \end{minipage}
  \caption{The effective area for the vertex (left) and calorimeter (right) analyses for
    \SI{2}{\giga\electronvolt} photons as a function of the source position with respect to the AMS
    zenith, including the $\varphi$ correction.}
  \label{fig:effective-area-phi-2d}
\end{figure}

Figure~\ref{fig:effective-area-phi-2d} shows the resulting effective area as a function of the two
angles for \SI{2}{\giga\electronvolt} photons, including the $\varphi$ correction. The calorimeter
effective area on the right hand side depends only very weakly on $\varphi$, it is almost perfectly
symmetric under rotations around the center. The vertex effective area on the left shows vertical
and horizontal structures. These structures arise, because of photon conversions in the bulkhead
support material in the lower part of the TRD. These are dense enough for photons to convert in
them. For polar angles around \SI{30}{\degree} the rectangular structure of the TOF planes becomes
visible.

\subsection{Trigger Efficiency}
\label{sec:analysis-trigger-efficiency}

The efficiency for the AMS trigger to induce the recording of a photon with a given energy is not
necessarily part of the effective area, which was estimated in
section~\ref{sec:analysis-effective-area}. Whether or not the effective area includes the trigger
efficiency depends on two factors:

\begin{itemize}
\item Is there a request for a specific (physics) trigger in the list of selection cuts?
\item Does the Monte-Carlo simulation store events without positive trigger decision?
\end{itemize}

The list of selection cuts does not include a request for the presence of a physics trigger decision
in neither the vertex nor the calorimeter analysis. However, in the case of the photon Monte-Carlo
used, the simulation only stores events which have a positive simulated trigger decision. Therefore
events without a positive trigger decision can never enter the numerator of
equation~(\ref{eq:effective-area-phi}), and any effective area determined from the simulation in
principle already includes the trigger efficiency.

One major difference between the Monte-Carlo trigger simulation and the actual ISS trigger
configuration is that the unbiased trigger paths are not prescaled. Therefore, any event featuring a
positive unbiased trigger decision is stored in the simulation. Since the assumed unbiased TOF
trigger efficiency is \SI{100}{\percent}, any event in which either the electron or positron passes
at least three TOF layers is stored, which is always the case for the events passing the vertex
selection. This means that for the vertex analysis, the trigger efficiency is not part of the
effective area and needs to be calculated separately.

For the calorimeter analysis the situation is more complex, because the unbiased ECAL trigger is not
fully efficient in neither data nor simulation. In addition, in the simulation the unbiased ECAL
trigger overlaps completely with the physics calorimeter trigger. As a result, the simulation only
records events with positive physics trigger decision, so the trigger efficiency is included in the
effective area, but will be corrected for differences between data and simulation in
section~\ref{sec:corrections-ecal-trigger}.

For the vertex analysis the trigger efficiency is calculated by comparing the number of events
triggered with any physics trigger branch to the ``all'' sample composed of those with physics
trigger ($N_P$) and those without ($N_{\bar{P}}$):

\begin{displaymath}
  \epsilon_{\mathrm{trigger}}(E_i) = \frac{N_{P}(E_i)}{N_{P}(E_i) + N_{\bar{P}}(E_i)} \,.
\end{displaymath}

The number of events without physics trigger ($N_{\bar{P}}$) is not directly measurable in data. But
counting the number of events which are only triggered by the two unbiased triggers it is possible
to approximate it. Looking at the sample of events without physics triggers, defining the following
names for the events:

\begin{itemize}
\item T: The unbiased TOF trigger fired for the event
\item E: The unbiased ECAL trigger fired for the event
\item S: The event is actually recorded
\end{itemize}

then the number of observable, recorded events, depending on T and E are related to the total number
of events without physics triggers $N_{\bar{P}}$ as follows:

\begin{align*}
  N_{ST\bar{E}} &= p(T) \cdot (1 - p(E|T)) \cdot p(S|T,\bar{E}) \cdot N_{\bar{P}} \\
  N_{S\bar{T}E} &= (1 - p(T)) \cdot p(E|\bar{T}) \cdot p(S|\bar{T},E) \cdot N_{\bar{P}} \\
  N_{STE} &= p(T) \cdot p(E|T) \cdot p(S|T,E) \cdot N_{\bar{P}} \\
  N_{S\bar{T}\bar{E}} &= (1 - p(T)) \cdot (1 - p(E|\bar{T})) \cdot p(S|\bar{T},\bar{E}) \cdot
                        N_{\bar{P}} \,,
\end{align*}

where (conditional) probabilities are denoted by $p$. Since the unbiased TOF and ECAL triggers share
no logic and are based on different subdetector signals it is reasonable to assume that the
statistical events E and T are independent:

\begin{align*}
  N_{ST\bar{E}} &= p(T) \cdot (1 - p(E)) \cdot p(S|T,\bar{E}) \cdot N_{\bar{P}} \\
  N_{S\bar{T}E} &= (1 - p(T)) \cdot p(E) \cdot p(S|\bar{T},E) \cdot N_{\bar{P}} \\
  N_{STE} &= p(T) \cdot p(E) \cdot p(S|T,E) \cdot N_{\bar{P}} \\
  N_{S\bar{T}\bar{E}} &= (1 - p(T)) \cdot (1 - p(E)) \cdot p(S|\bar{T},\bar{E}) \cdot N_{\bar{P}} \,.
\end{align*}

Events triggered by the unbiased TOF trigger, but not the ECAL trigger will be recorded if the
prescaling condition for the unbiased TOF trigger is fulfilled. Vice versa events triggered by the
unbiased ECAL trigger, but not the TOF trigger will be recorded if the prescaling condition for the
unbiased ECAL trigger is fulfilled. If both unbiased trigger branches fire the event will be
recorded if either one of the prescaling conditions is fulfilled. If $f_{T}$ is the prescaling
factor for the unbiased TOF triggers and $f_{E}$ is the prescaling factor for the unbiased ECAL
triggers:

\begin{align*}
  p(S|T,\bar{E}) &= 1 / f_{T} \\
  p(S|\bar{T},E) &= 1 / f_{e} \\
  p(S|T,E) &= 1 - (1 - 1 / f_{T}) \cdot (1 - 1 / f_{E}) \\
  p(S|\bar{T},\bar{E}) &= 0 \,.
\end{align*}

The actual number of triggered events can then be estimated, based on these probabilities:

\begin{align*}
  N_{T\bar{E}} &= f_{T} \cdot N_{ST\bar{E}} \\
  N_{\bar{T}E} &= f_{E} \cdot N_{S\bar{T}E} \\
  N_{TE} &= 1 / (1 - (1 - 1 / f_{T}) \cdot (1 - 1 / f_{E})) \cdot N_{STE} \,.
\end{align*}

The number of events without any unbiased trigger $N_{\bar{T}\bar{E}}$ cannot be estimated in this
way since they are never recorded. In general this leaves the following system of equations:

\begin{align*}
  N_{T\bar{E}} &= p(T) \cdot (1 - p(E)) \cdot N_{\bar{P}} \\
  N_{\bar{T}E} &= (1 - p(T)) \cdot p(E) \cdot N_{\bar{P}} \\
  N_{TE} &= p(T) \cdot p(E) \cdot N_{\bar{P}} \\
  N_{\bar{T}\bar{E}} &= (1 - p(T)) \cdot (1 - p(E)) \cdot N_{\bar{P}} \,,
\end{align*}

from which it is possible to solve for the four unknowns $N_{\bar{P}}$, $p(T)$, $p(E)$,
$N_{\bar{T}\bar{E}}$.

The situation simplifies if one of two unbiased trigger efficiencies is near unity. For AMS this is
the case for the unbiased TOF trigger, but not for the unbiased ECAL trigger. In the vertex analysis
$p(E)$ also includes a geometrical factor: There are many events in which neither the electron nor
the positron enters the calorimeter.

\begin{align*}
  N_{T\bar{E}} &\approx (1 - p(E)) \cdot N_{\bar{P}} \\
  N_{\bar{T}E} &\approx 0 \\
  N_{TE} &\approx p(E) \cdot N_{\bar{P}} \\
  N_{\bar{T}\bar{E}} &\approx 0
\end{align*}

and thus:

\begin{align*}
  N_{\bar{P}}
  &= N_{T\bar{E}} + N_{\bar{T}E} + N_{TE} + N_{\bar{T}\bar{E}} \\
  &\approx N_{T\bar{E}} + N_{TE} \\
  &= f_{T} \cdot N_{ST\bar{E}} + \frac{1}{(1 - (1 - 1 / f_{T}) \cdot (1 - 1 / f_{E}))} \cdot N_{STE} \,.
\end{align*}

In the AMS ISS data the prescaling factors are $f_{T} = 100$ and $f_{E} = 1000$, while in the
Monte-Carlo simulation no prescaling is applied, so $f_{T} = 1$ and $f_{E} = 1$.

Once the number of events without physics triggers is estimated the efficiency of any physics
trigger branch to fire is:

\begin{displaymath}
  \epsilon_{\mathrm{trigger}}(E_i) = \frac{N_{P}(E_i)}{N_{P}(E_i) + f_{T} \cdot
    N_{ST\bar{E}}(E_i) + \frac{1}{(1 - (1 - 1 / f_{T}) \cdot (1 - 1 / f_{E}))} \cdot N_{STE}(E_i)} \,.
\end{displaymath}

Using this equation it is also possible to measure the efficiency of individual physics triggers, by
replacing the numerator with the number of events triggered by the individual trigger.

\begin{figure}[t]
  \centering
  \includegraphics[width=0.8\linewidth]{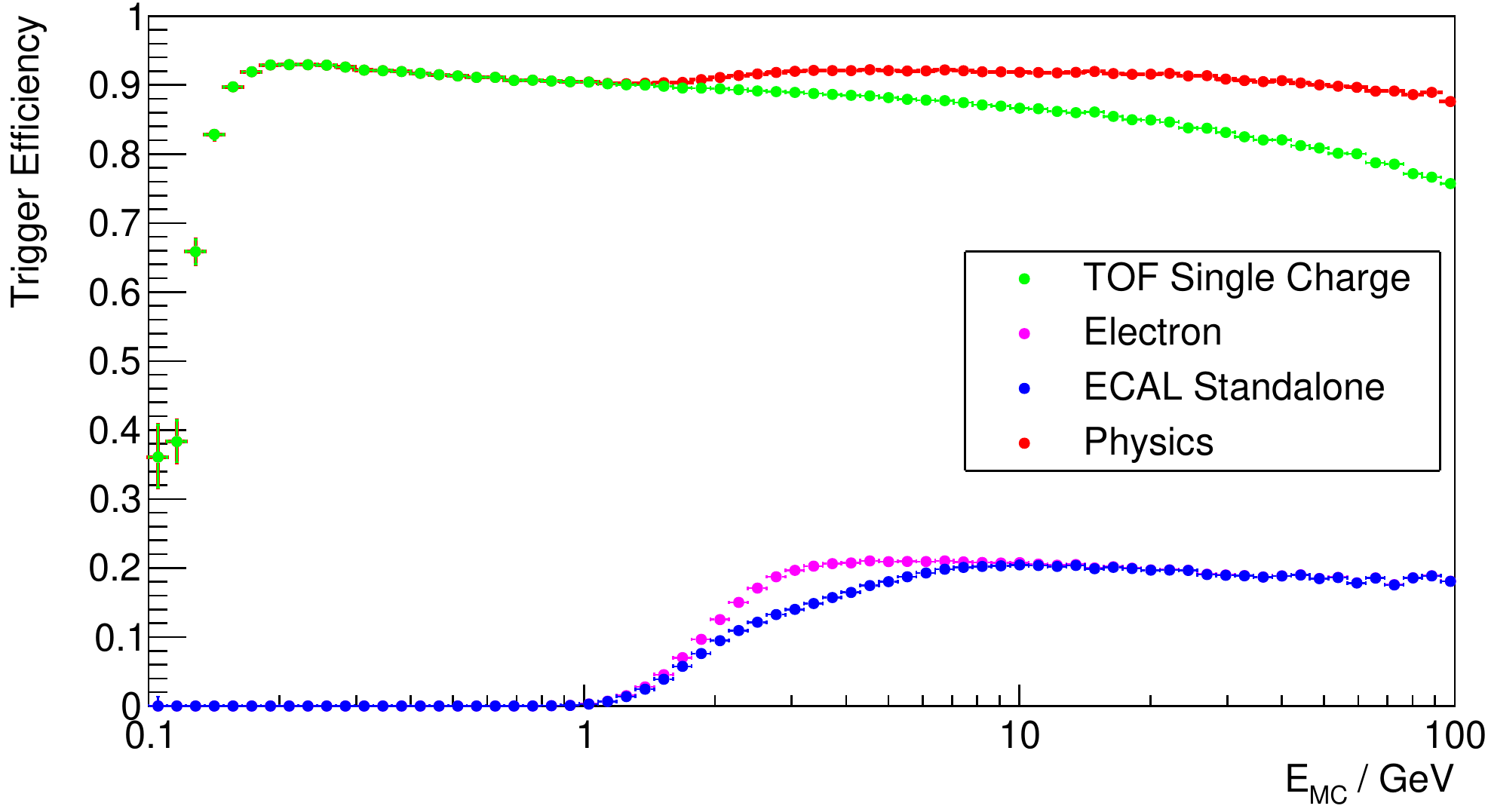}
  \caption{The trigger efficiency for the vertex selection from the Monte-Carlo simulation.}
  \label{fig:vertex-mc-trigger-efficiency}
\end{figure}

Figure~\ref{fig:vertex-mc-trigger-efficiency} shows the resulting trigger efficiency for the vertex
analysis according to the Monte-Carlo simulation. The red points correspond to the efficiency of any
physics trigger to fire. The green curve corresponds to the efficiency of the 4/4 TOF single charge
trigger, which includes the ACC veto. The magenta and blue points correspond to the two physics
trigger branches which involve the calorimeter in the trigger decision: The electron and ECAL
standalone triggers. These individual branches are not exclusive, it is possible to obtain a
positive trigger decision for more than one physics trigger branch.

The overall trigger efficiency is approximately \SI{90}{\percent} for photon energies above
\SI{150}{\mega\electronvolt}. Below that the efficiency drops because one of the two tracks is often
bent out by the magnetic field and hits the ACC. The TOF single charge trigger slowly drops in
efficiency above \SI{200}{\mega\electronvolt}, due to the slow increase of potential calorimeter
backsplash to the ACC counters. From \SI{1}{\giga\electronvolt} onwards some of the efficiency loss
is compensated by the trigger branches which involve the calorimeter. Since only about
\SI{20}{\percent} of the photons produce an electron or positron which enters into the calorimeter
volume, the calorimeter cannot recover all of the inefficiency, though.

The overall efficiency, shown in red in the figure, is applied as an additional energy dependent
correction to the effective area derived in section~\ref{sec:analysis-effective-area}.

Deriving the ECAL standalone trigger efficiency, needed for the ECAL photon analysis, is more
involved. It requires generation of a special Monte-Carlo simulation in which all generated events
are stored in the result file (in regular Monte-Carlo simulations only events with positive trigger
decision are stored). This is because the unbiased ECAL trigger is not correctly modeled in the
simulation. In the second step only those events in which the generated photon passes into the
calorimeter according to the Monte-Carlo truth are kept. Events in which the photon converts before
the calorimeter are also discarded. In addition it is required that the generated photon passes
through all four TOF layers and through the TRD. This selection does not make use of any
reconstructed information from the detector, but makes sure that the sample of photons is adequate
for the study.

The trigger efficiency can then be estimated by counting the number of events which have a positive
ECAL standalone trigger decision (E):

\begin{displaymath}
  \epsilon_{\mathrm{trigger}}(E_i) = \frac{N_{E}(E_i)}{N_{E}(E_i) + N_{\bar{E}}(E_i)} \,.
\end{displaymath}

The number of events without ECAL standalone trigger $N_{\bar{E}}$ are directly accessible and do
not need to be estimated using unbiased trigger branches, because of the special nature of the
Monte-Carlo simulation.

\begin{figure}[t]
  \centering
  \includegraphics[width=0.8\linewidth]{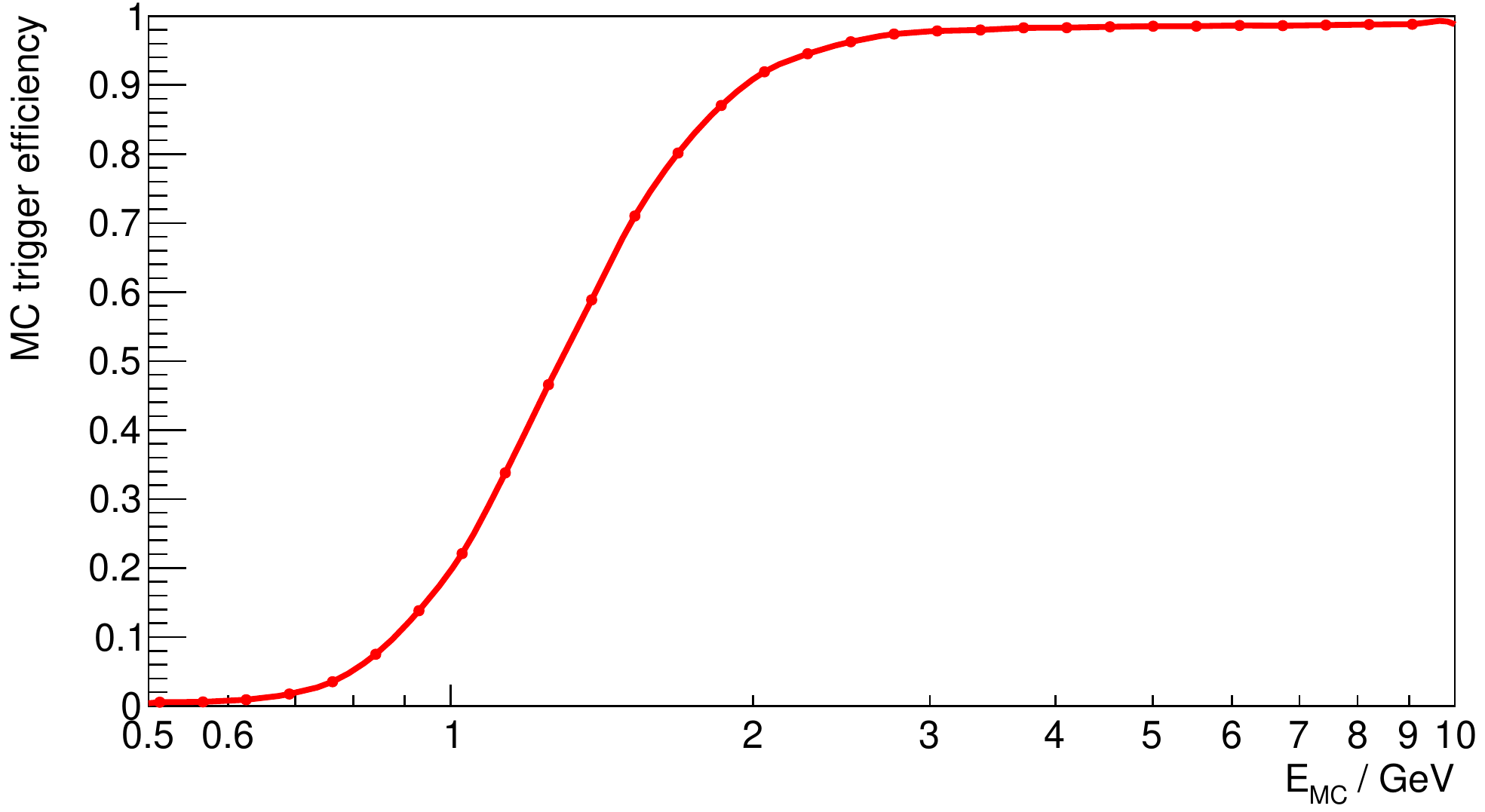}
  \caption{The trigger efficiency for the ECAL selection from the Monte-Carlo simulation.}
  \label{fig:ecal-mc-trigger-efficiency}
\end{figure}

The resulting efficiency is shown in figure~\ref{fig:ecal-mc-trigger-efficiency}. It is close to
zero for photons below \SI{500}{\mega\electronvolt} and quickly rises, reaching a plateau at about
\SI{99}{\percent} around \SI{3}{\giga\electronvolt}. At \SI{1}{\giga\electronvolt} the efficiency is
approximately \SI{20}{\percent}. The inefficiency below \SI{1}{\giga\electronvolt} is the reason why
photons with lower energies cannot be measured with sufficient statistics with the calorimeter.

Unlike in the vertex analysis this efficiency is not used as an additional correction to the
effective area for the calorimeter analysis, as it is already included in it. The reason is the
inefficiency of the unbiased calorimeter trigger in the simulation, which means that all photons
selected with the calorimeter selection on a regular Monte-Carlo already, by definition, have a
positive trigger decision. However, the ECAL standalone photon trigger efficiency is needed in order
to correct the effective area for differences between data and simulation, see
section~\ref{sec:corrections-ecal-trigger}.

\section{TRD Pileup Weight}
\label{sec:analysis-trd-pileup-weight}

There is one instrumental effect which must be considered, but is not modeled properly in the
simulation. When the flux of charged particles is high there is a significant probability for
signals from prior particle crossings to be visible in the TRD at the time of the trigger of the
actual event. This is because the TRD electronics use a rather long pulse integration time of up to
\SI{100}{\micro\second}. Thus, one can expect signals from prior particle crossings in case the rate
of particles crossing the TRD exceeds \SI{10}{\kilo\hertz}. Such prior particle crossings are
referred to as pileup in the following. Because of the TRD pulse shape the tracks of pileup events
in the TRD often feature peculiarly low amplitudes in the associated tubes.

The pileup effect is important because both the vertex selection and the calorimeter selection use
the global absence of charged particle tracks and track segments in the TRD in order to establish a
reliable veto. So the presence of a track in the TRD from a prior particle crossing can spoil the
selection of genuine photon events in both selections. Since the Monte-Carlo simulation does not
treat pileup from secondary events, the corresponding efficiency correction needs to be determined
from ISS data. In addition, the pileup probability is completely unrelated to the properties of the
actual photon passing AMS. Instead it depends on the absolute rate of charged particles, which
varies according to the geomagnetic position of the ISS due to the geomagnetic cutoff.

Because the effect is instrumental and only depends on the flux of charged particles it is not
necessary to study it using photons. Instead protons or electrons which are much more abundant than
photons can be used. However, for protons and electrons passing the detector from top to bottom the
presence of a TRD track from the primary particle itself is a problem. Such tracks do not exist for
photons which convert only in the upper TOF or in the calorimeter. Although it is possible to count
only the number of excess hits and tracks the method is inaccurate because interactions of the
primary in the TRD volume could create additional segments and tracks.

Instead a better way is to use electrons which enter the calorimeter from below. These events can be
triggered with the calorimeter standalone trigger. If the selection ensures that the primary
electron is fully absorbed in the calorimeter, any additional hits and track segments in the TRD
must be due to pileup from prior particle crossings. The full selection for upgoing calorimeter
electrons for the pileup study is presented in appendix~\ref{sec:appendix-trd-pileup-selection}.

\begin{figure}[t]
  \centering
  \includegraphics[width=0.98\linewidth]{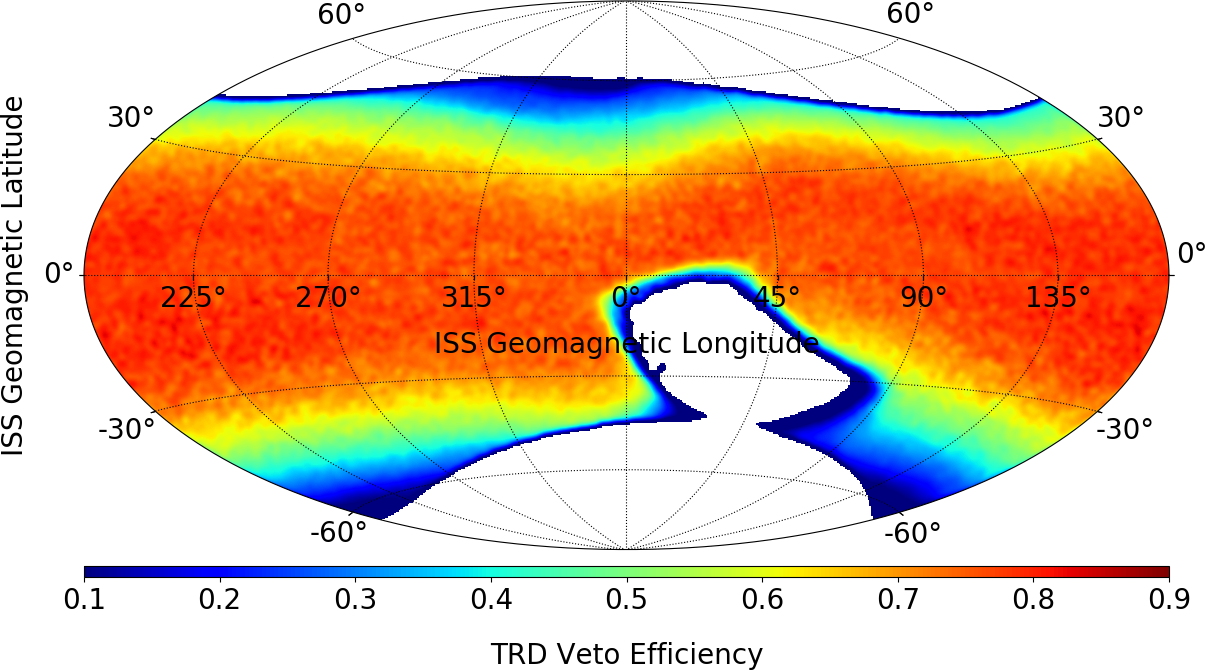}
  \caption{The TRD veto cut efficiency for electron events entering the calorimeter from below as a
    function of the ISS geomagnetic position. The hole in the center of the image corresponds to the
    location of the SAA.}
  \label{fig:trd-pileup-correction}
\end{figure}

In this specific event sample, in which the primary particle enters from below, stops in the
calorimeter, and thus causes a trigger and subsequent readout of the entire experiment, one does not
expect any signal in the TRD, so the TRD veto cuts should always pass. However, because of the
aforementioned pileup of events this naive assumption is incorrect, and the TRD veto efficiency
(corresponding to the TRD pileup weight $w_{\mathrm{pileup}}$) is:

\begin{displaymath}
  w_{\mathrm{pileup}}(t) =
  \epsilon_{\mathrm{veto}}(\Phi_{\mathrm{ISS}}(t),\lambda_{\mathrm{ISS}}(t)) =
  \frac{N_{\mathrm{pass}}\left(\Phi_{\mathrm{ISS}},\lambda_{\mathrm{ISS}}\right)}
  {N_{\mathrm{total}}\left(\Phi_{\mathrm{ISS}},\lambda_{\mathrm{ISS}}\right)} \,,
\end{displaymath}

where $(\Phi_{\mathrm{ISS}}, \lambda_{\mathrm{ISS}})$ are the geomagnetic longitude and latitude of
the ISS position, $N_{\mathrm{pass}}$ is the number of events for which the TRD veto cuts pass and
$N_{\mathrm{total}}$ is the total number of events for which the ISS was located in the bin centered
on $(\Phi_{\mathrm{ISS}}, \lambda_{\mathrm{ISS}})$.

The resulting efficiency is shown in figure~\ref{fig:trd-pileup-correction}. Within a window of
$|\lambda_{\mathrm{ISS}}| < \SI{30}{\degree}$ around the geomagnetic equator the veto efficiency is
better than \SI{80}{\percent}, except for regions close to the border of the SAA. There is no data
from inside the SAA, this region is excluded from the study and from the photon analysis. However,
close the magnetic poles the veto efficiency drops to \SI{10}{\percent} due to increased pileup of
secondary particles.

The TRD pileup weight is an important correction for the exposure maps derived in the next section.

In the future it should be possible to improve the analysis, such that the problem of the TRD pileup
is partially alleviated. If events which feature TRD track segments which do not appear to be
related to the actual reconstructed photon were allowed in the selection, the overall size of the
pileup effect and the corresponding correction would be reduced. As a result, more photons would be
selected (which would contain spurious TRD tracks) and the correction to the effective area would be
lower.

\section{Exposure Maps}
\label{sec:analysis-exposure-maps}

Exposure maps are required to convert the observed counts from a photon source in a given location
in the sky into a photon flux. They combine the effective area results derived in
section~\ref{sec:analysis-effective-area} with the effective measuring time for any location in the
sky. The exposure $\mathcal{E}(E_{i},l,b)$ is a key quantity as it combines all efficiencies,
geometry factors and also includes the observation time for each point in the sky. By convention it
does not include the trigger efficiency.

For a point in the sky with coordinates $(l,b)$ the exposure $\mathcal{E}(E_i,l,b)$ can be
calculated by integration over time:

\begin{equation}
  \label{eq:exposure}
  \mathcal{E}(E_{i},l,b) =
  \int_{t_{\mathrm{start}}}^{t_{\mathrm{end}}}{A_{\mathrm{eff}}(E_i,\cos{\theta}_{l,b}(t),\varphi_{l,b}(t)) \cdot
    \epsilon_{\mathrm{DAQ}}(t) \cdot w_{\mathrm{pileup}}(t) \mathrm{d}t} \,.
\end{equation}

Here $\cos{\theta}_{l,b}(t)$ and $\varphi_{l,b}(t)$ are the angles of the given point in the sky
with coordinates $(l,b)$ with respect to the \mbox{AMS-02} zenith at the time $t$ and
$t_{\mathrm{start}}$ and $t_{\mathrm{end}}$ are the start and end times of the observation period.

The efficiency $\epsilon_{\mathrm{DAQ}}(t)$ is the data acquisition efficiency at the given point in
time. The \mbox{AMS-02} detector electronics require a small but noticeable time to digitize
signals, to store and read out events from the electronics buffers and to assemble the final event
from the various pieces of information from the individual subdetectors. During this time the
detector is ``busy'' and can not record further events. The ratio of the non-busy time to the total
time in a second is called the DAQ livetime, and reflected in $\epsilon_{\mathrm{DAQ}}(t)$.

The quantity
$w_{\mathrm{pileup}}(t) =
\epsilon_{\mathrm{veto}}(\Phi_{\mathrm{ISS}}(t),\lambda_{\mathrm{ISS}}(t))$ is the TRD veto
efficiency (as shown in figure~\ref{fig:trd-pileup-correction}). Even though the location given by
$(l,b)$ is fixed, the angles $\cos{\theta}_{l,b}$ and $\varphi_{l,b}$ are functions of time because
the \mbox{AMS-02} zenith moves over the sky according to the ISS orbit position and orientation.

Time intervals which correspond to periods with hardware problems, reduced data quality or other bad
detector operating conditions are removed from the analysis and from the integration based on a
purely time based selection. This selection also removes periods in which the ISS was inside the SAA
and in which AMS was not taking data. Overall about \SI{20}{\percent} of the total time is removed
in this step, most of which is removed because of TRD refill operations, ISS SAA passings and two
months of operations in 2014 in which only part of the tracker was powered.

\begin{figure}[t]
  \begin{minipage}{0.48\linewidth}
    \centering
    \includegraphics[width=1.0\linewidth]{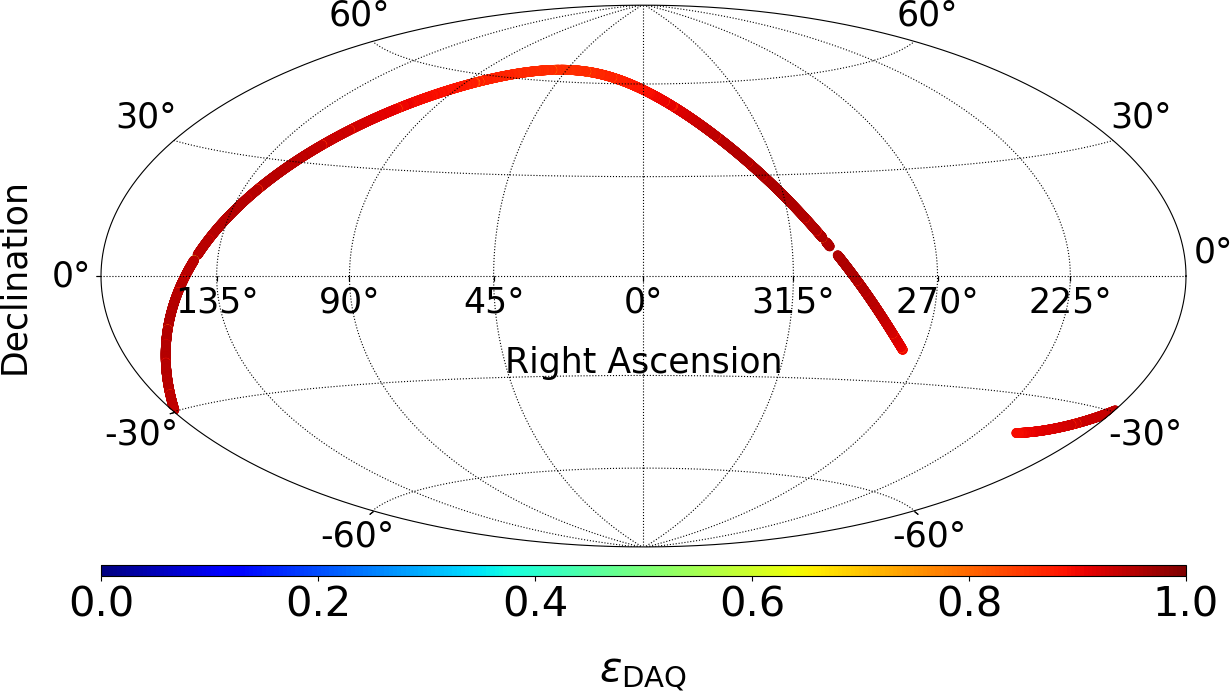}
  \end{minipage}
  \hspace{0.01\linewidth}
  \begin{minipage}{0.48\linewidth}
    \centering
    \includegraphics[width=1.0\linewidth]{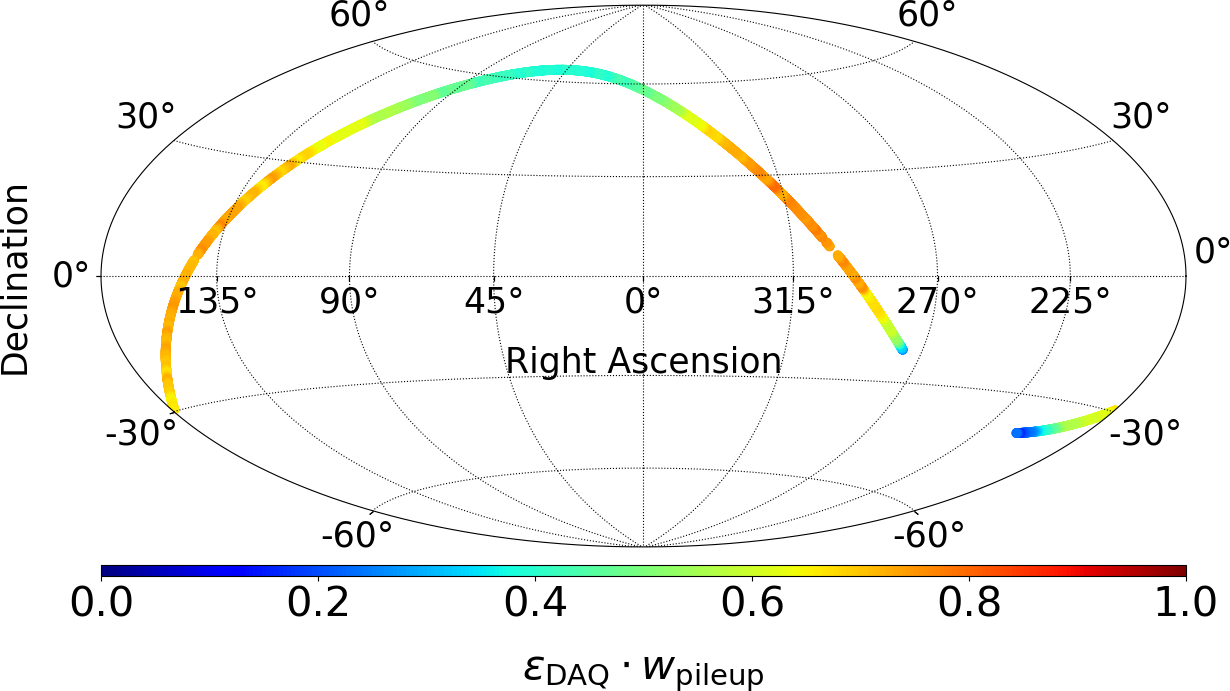}
  \end{minipage}
  \caption{Trajectory of the AMS zenith over the equatorial sky for one example orbit (starting at
    July 16th 2014 20:49:02 UTC). Left: The color code depicts the DAQ efficiency for each
    second. Right: The color code depicts the product of the DAQ efficiency with the TRD pileup
    weight for each second.}
  \label{fig:exposure-map-zenith}
\end{figure}

Figure~\ref{fig:exposure-map-zenith} shows the trajectory of the AMS zenith projected on to the sky
for one example orbit. On the left the color corresponds to the data acquisition efficiency
$\epsilon_{\mathrm{DAQ}}$ which is generally better than \SI{90}{\percent} except at the poles where
it drops to \SI{80}{\percent}. The gaps in the trajectory are due to the removal of time intervals,
such as the passage of the ISS through the SAA. In the right hand side figure the TRD pileup weight
factor is included. The effective measuring time is between \SI{40}{\percent} and \SI{80}{\percent}
of a second for most of the orbit, except for a few seconds in which the ISS was close to the
geomagnetic south pole and the SAA border.

\begin{figure}[p]
  \centering
  \includegraphics[width=0.98\linewidth]{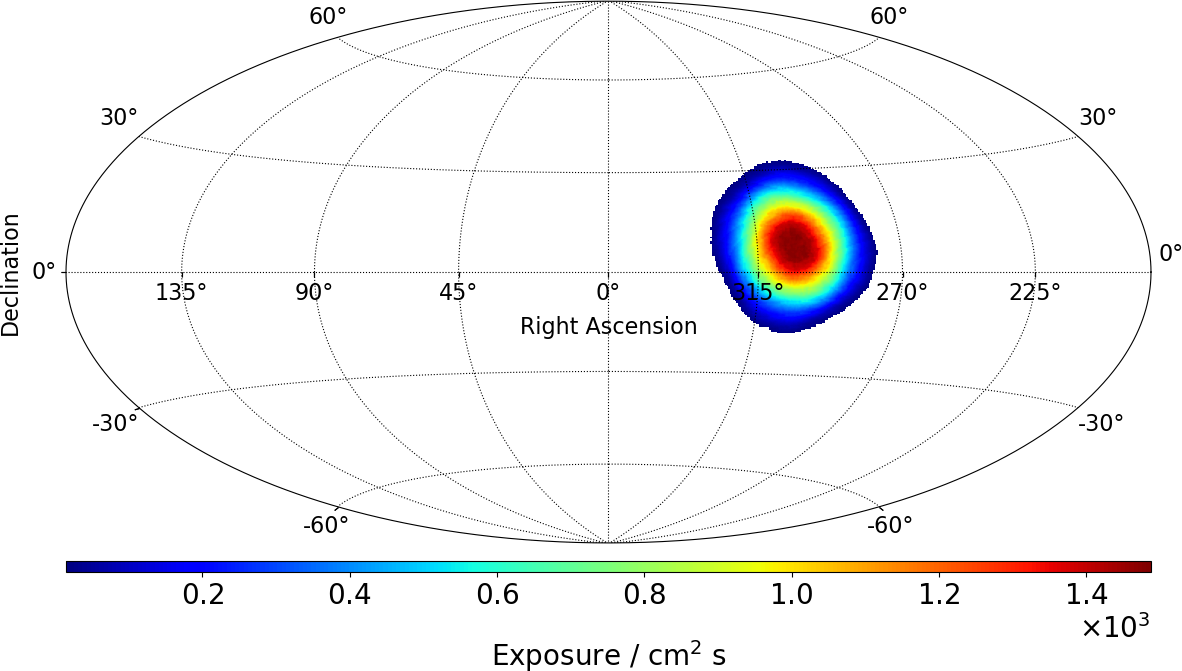}
  \caption{Projection of the calorimeter mode effective area at \SI{2}{\giga\electronvolt} onto the
    sky, centered on the AMS zenith position at July 16th 2014 20:49:02 UTC and rotated according to
    the AMS orientation, weighted by the effective observation time in that second.}
  \label{fig:exposure-map-one-second}

  \vspace*{\floatsep}

  \centering
  \includegraphics[width=0.98\linewidth]{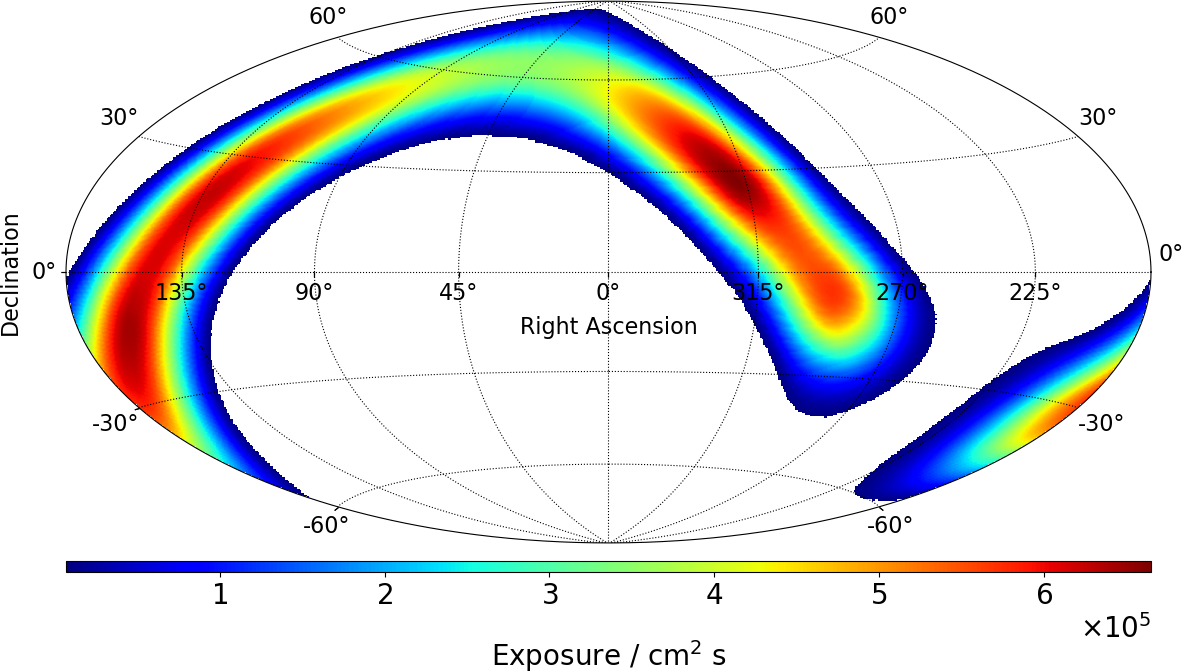}
  \caption{Calorimeter mode exposure map at \SI{2}{\giga\electronvolt} for the example orbit
    starting at July 16th 2014 20:49:02 UTC.}
  \label{fig:exposure-map-one-orbit}
\end{figure}

For any given second the exposure map is constructed by projecting the effective area onto the sky,
weighted by the DAQ efficiency and TRD pileup, according to equation~(\ref{eq:exposure}).
Figure~\ref{fig:exposure-map-one-second} illustrates this principle for calorimeter photons at
\SI{2}{\giga\electronvolt}. In this example $t_{\mathrm{start}}$ is July 16th 2014 20:49:02 UTC and
$t_{\mathrm{end}}$ is set to $t_{\mathrm{end}} = t_{\mathrm{start}} + \SI{1}{\second}$ . The
effective area is centered around the AMS zenith position and rotated according to the current
rotation of \mbox{AMS-02} around its zenith in equatorial coordinates.

Figure~\ref{fig:exposure-map-one-orbit} shows the \SI{2}{\giga\electronvolt} exposure map for the
calorimeter analysis for the full example orbit. This corresponds to setting $t_{\mathrm{start}}$ to
July 16th 2014 20:49:02 UTC, $t_{\mathrm{end}} = t_{\mathrm{start}} + \SI{92}{\minute}$ and
performing the integral in equation~(\ref{eq:exposure}). The exposure near the north pole is reduced
because the DAQ efficiency drops slightly and the TRD pileup weight drops significantly. Near the
SAA there is a gap, the size of which is determined by the angular acceptance cone size of the
calorimeter selection.

\begin{figure}[p]
  \centering
  \includegraphics[width=0.98\linewidth]{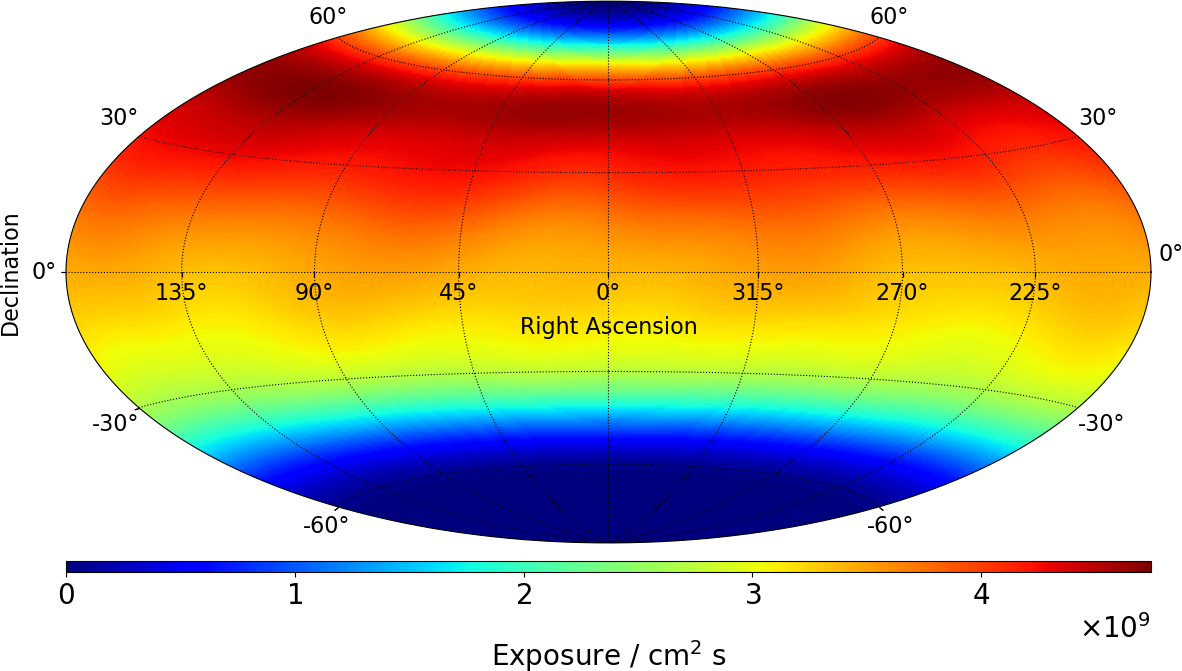}
  \caption{Complete exposure map for \SI{2}{\giga\electronvolt} photons for the calorimeter analysis
    in ICRS coordinates for the time period from May 19th 2011 to November 12th 2017.}
  \label{fig:exposure-map-ecal-equatorial}

  \vspace*{\floatsep}

  \centering
  \includegraphics[width=0.98\linewidth]{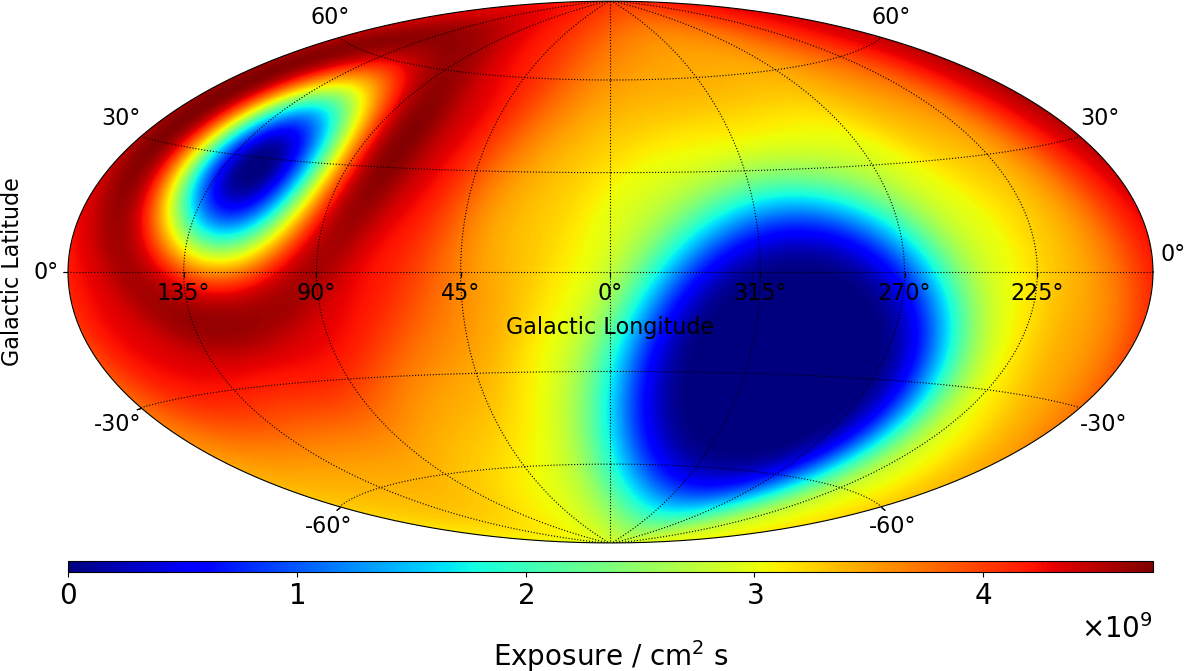}
  \caption{Complete exposure map for \SI{2}{\giga\electronvolt} photons for the calorimeter analysis
    in galactic coordinates for the time period from May 19th 2011 to November 12th 2017.}
  \label{fig:exposure-map-ecal-galactic}
\end{figure}

\begin{figure}[p]
  \centering
  \includegraphics[width=0.98\linewidth]{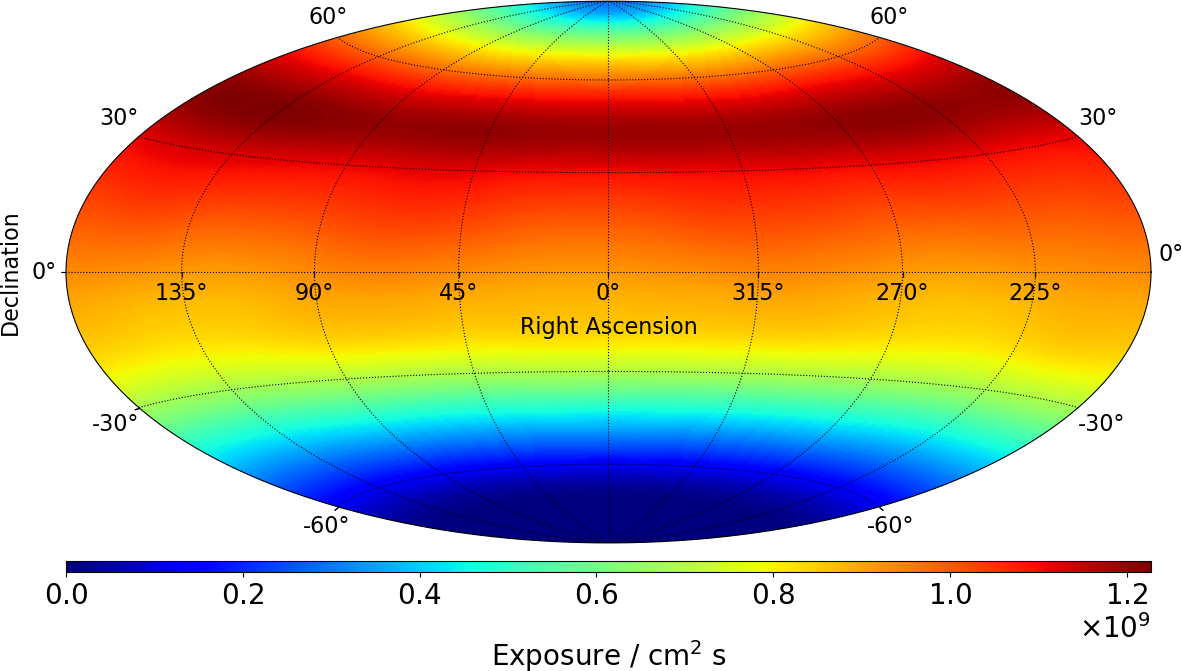}
  \caption{Complete exposure map for \SI{2}{\giga\electronvolt} photons for the conversion analysis
    in ICRS coordinates for the time period from May 19th 2011 to November 12th 2017.}
  \label{fig:exposure-map-vertex-equatorial}

  \vspace*{\floatsep}

  \centering
  \includegraphics[width=0.98\linewidth]{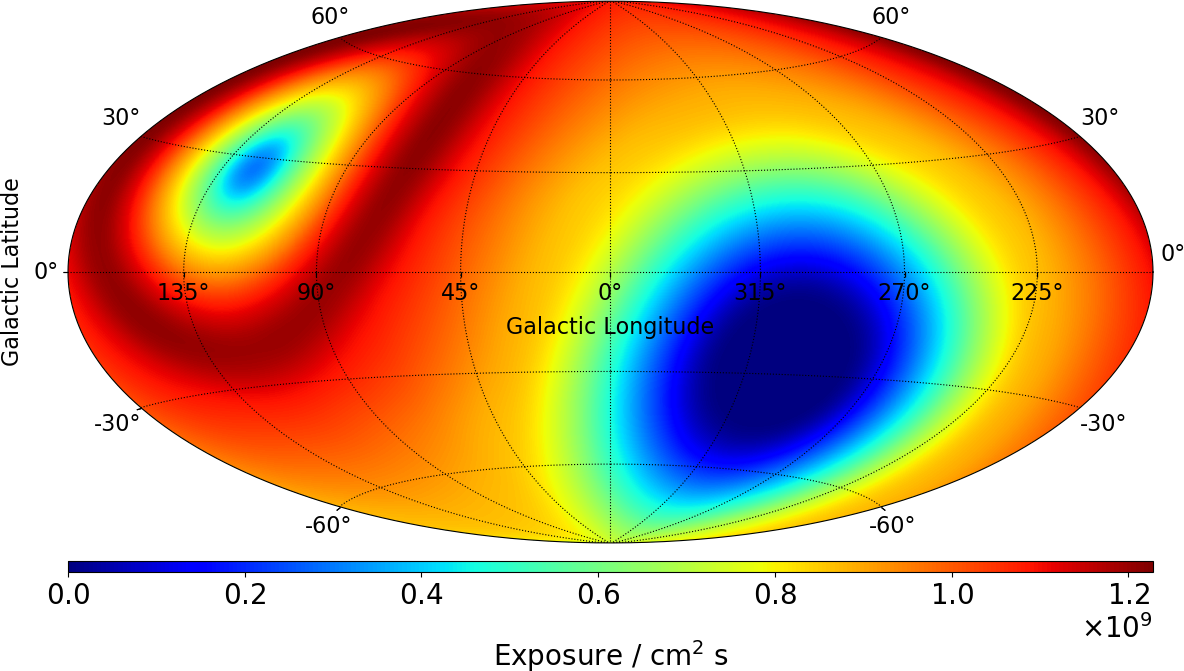}
  \caption{Complete exposure map for \SI{2}{\giga\electronvolt} photons for the conversion analysis
    in galactic coordinates for the time period from May 19th 2011 to November 12th 2017.}
  \label{fig:exposure-map-vertex-galactic}
\end{figure}

The full exposure maps are obtained by integrating equation~(\ref{eq:exposure}) from May 19th 2011
to November 12th 2017, which is the full time period
analyzed. Figures~\ref{fig:exposure-map-ecal-equatorial} and~\ref{fig:exposure-map-ecal-galactic}
show the result for \SI{2}{\giga\electronvolt} photons for the calorimeter analysis in ICRS
equatorial and galactic coordinates respectively.

The exposure for the northern sky is larger compared to the exposure for the southern sky. This is a
result of three factors:

\begin{itemize}
\item The \mbox{AMS-02} zenith is tilted \SI{12}{\degree} towards the port side of the ISS, which
  corresponds to a northern tilt of up to \SI{12}{\degree}.
\item The exposure on the southern side is reduced because seconds in which the ISS passed through
  the SAA are excluded.
\item The TRD pileup weight is generally lower when the ISS is in the southern hemisphere.
\end{itemize}

In addition the sky north of $\delta = \SI{80}{\degree}$ cannot be observed with the calorimeter
analysis. This is a result of the ISS orbital plane inclination of \SI{51.6}{\degree}, the
\mbox{AMS-02} zenith tilt and the smaller acceptance cone size of the calorimeter
analysis. Likewise, the sky south of $\delta = \SI{-45}{\degree}$ cannot be observed in the
calorimeter analysis.

Figures~\ref{fig:exposure-map-vertex-equatorial} and~\ref{fig:exposure-map-vertex-galactic} show the
final exposure maps for the analysis of converted photons in ICRS equatorial and galactic
coordinates. Because the acceptance cone size in the conversion analysis is larger compared to the
calorimeter analysis, the north pole region is observable, although with a reduced exposure of
approximately \SI{5e8}{\centi\meter\squared\second}. Similarly the gap around the south pole is
smaller.

Because of the larger effective area, the calorimeter analysis features better exposure for most
regions of the sky for \SI{2}{\giga\electronvolt} photons compared to the converted photon analysis.

Exposure maps for other energy bins are computed analogously. The spatial shapes are similar to
those in figures~\ref{fig:exposure-map-ecal-equatorial} to~\ref{fig:exposure-map-vertex-galactic},
but the normalization differs according to the variation of the effective area with energy.

\section{Construction of Model Predictions}
\label{sec:analysis-expectation-maps}

Once the flux of diffuse photons and source photons is given it is relatively straightforward to
calculate expectation maps for \mbox{AMS-02} using the instrument response functions (IRFs). For
each event selection the exposure map (ref. section~\ref{sec:analysis-exposure-maps}) converts the
diffuse flux into counts for each pixel in the sky:

\begin{equation}
  \label{eq:modeling-diffuse-expected-counts-diffuse}
  N'_{\mathrm{diffuse}} (E_{i},l_j,b_k) =
  \Phi_{\mathrm{diffuse}}(E_{i},l_{j},b_{k})
  \cdot \mathcal{E} (E_{i},l_j,b_k)
  \cdot \epsilon_{\mathrm{trigger}}(E_{i})
  \cdot \Delta\Omega_{jk}
  \cdot \Delta E_{i} \,.
\end{equation}

Here $E_i$, $l_j$ and $b_k$ are the energy, longitude and latitude bin index, $\mathcal{E}$ is the
exposure at the given location and energy, $\epsilon_{\mathrm{trigger}}$ is the trigger efficiency,
$\Delta\Omega_{jk}$ is the solid angle subtended by the bin $(l_j, b_k)$, and $\Delta E_i$ is the
energy bin width.

In addition the count maps need to be convoluted with the PSF
(cf. section~\ref{sec:analysis-psf}). For the diffuse emission this is achieved by computing the
Fourier transform of the unsmeared count map $N'_{\mathrm{diffuse}}$ and of the PSF function,
multiplying the Fourier transforms and transforming the result back:

\begin{equation}
  \label{eq:modeling-diffuse-convolve}
  N_{\mathrm{diffuse}} = \tilde{\mathcal{F}} \bigl( \mathcal{F}
  \left(N'_{\mathrm{diffuse}}\right) \cdot \mathcal{F} \left(PSF\right) \bigr) \,,
\end{equation}

where $\mathcal{F}$ is the Fourier transform and $\tilde{\mathcal{F}}$ is the inverse Fourier
transform. For the sources a slightly different, procedure is used instead. In the first step the
expected events for a given source $s$ are calculated in each energy bin:

\begin{equation}
  \label{eq:modeling-expected-counts-sources}
  N'_{s} (E_{i}) = \Phi_{s}(E_{i}) \cdot \mathcal{E} (E_{i},l_s,b_s) \cdot
  \epsilon_{\mathrm{trigger}}(E_{i}) \cdot \Delta E_{i} \,.
\end{equation}

Because the sources are assumed to be point like there is no solid angle factor here. Since the
probability density of the PSF is normalized to unity it is possible to model the source $s$ by
simply scaling the PSF with the expected event yield in each energy bin and then placing the result
on the sky at the (unbinned) location of the source $(l, b)$ for each energy bin. This procedure is
repeated for all the sources in the 4FGL catalog~\cite{Fermi_4FGL_2019} and the results are summed.

\begin{figure}[p!]
  \centering
  \includegraphics[width=0.8\linewidth]{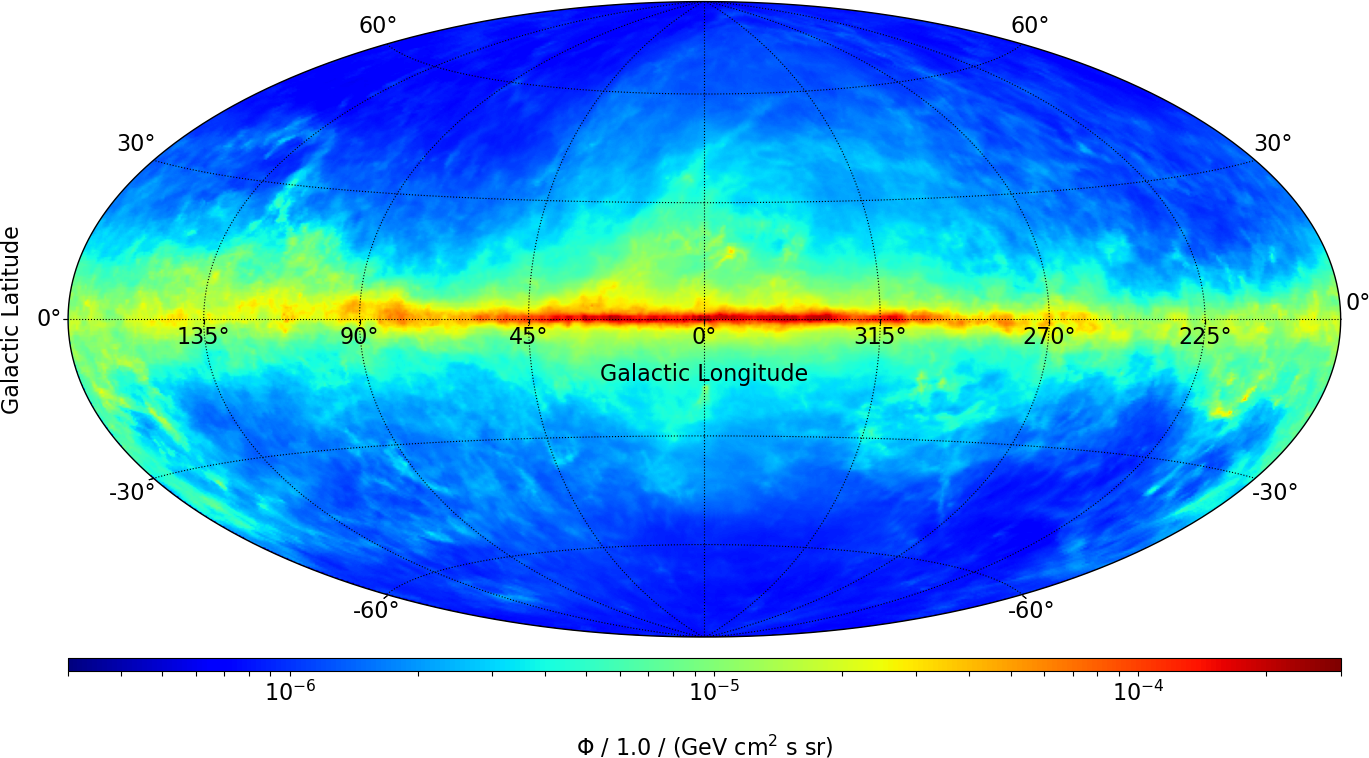}
  \caption{Diffuse photon flux at \SI{1}{\giga\electronvolt} according to the Fermi-LAT diffuse
    emission model.}
  \label{fig:diffuse-model-flux}

  \vspace*{\floatsep}

  \centering
  \includegraphics[width=0.8\linewidth]{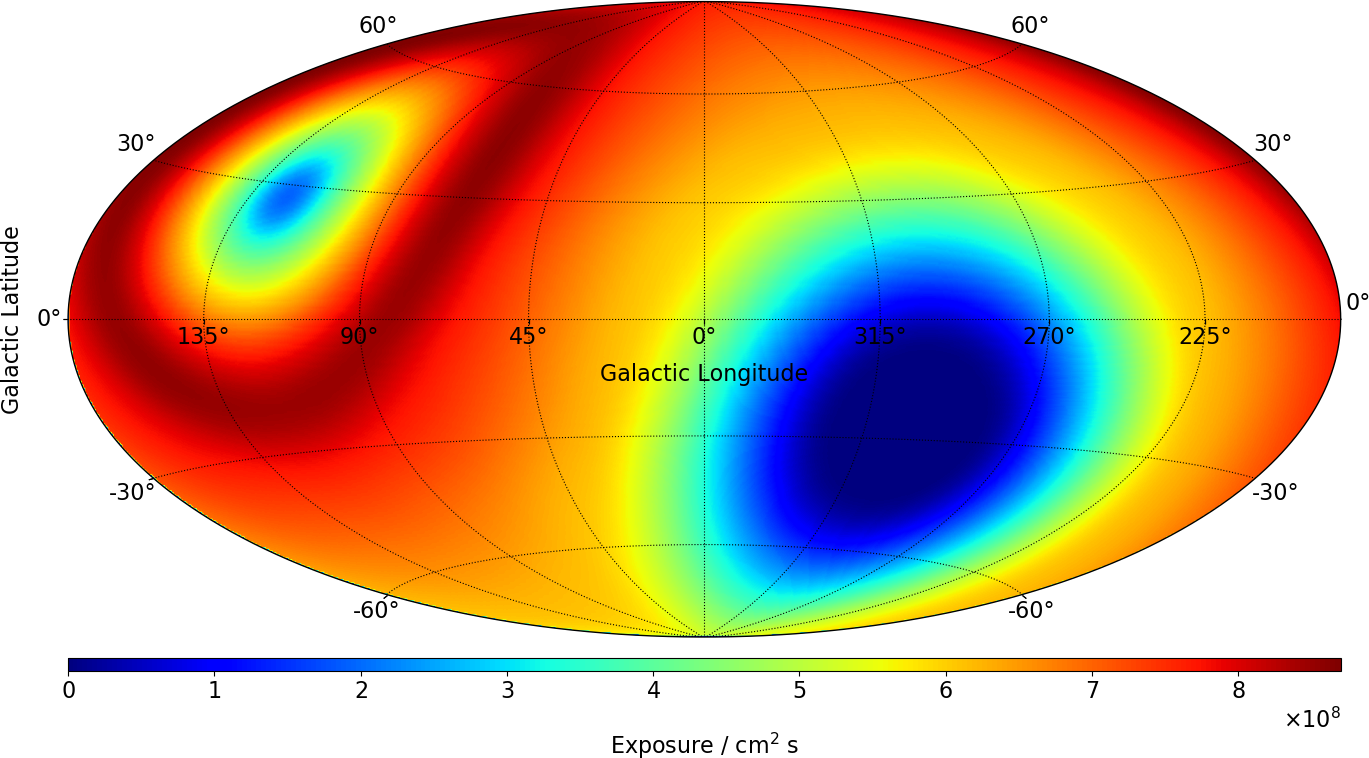}
  \caption{Vertex analysis exposure map at \SI{1}{\giga\electronvolt} as constructed in
    section~\ref{sec:analysis-exposure-maps}.}
  \label{fig:diffuse-model-exposure-map}

  \vspace*{\floatsep}

  \centering
  \includegraphics[width=0.8\linewidth]{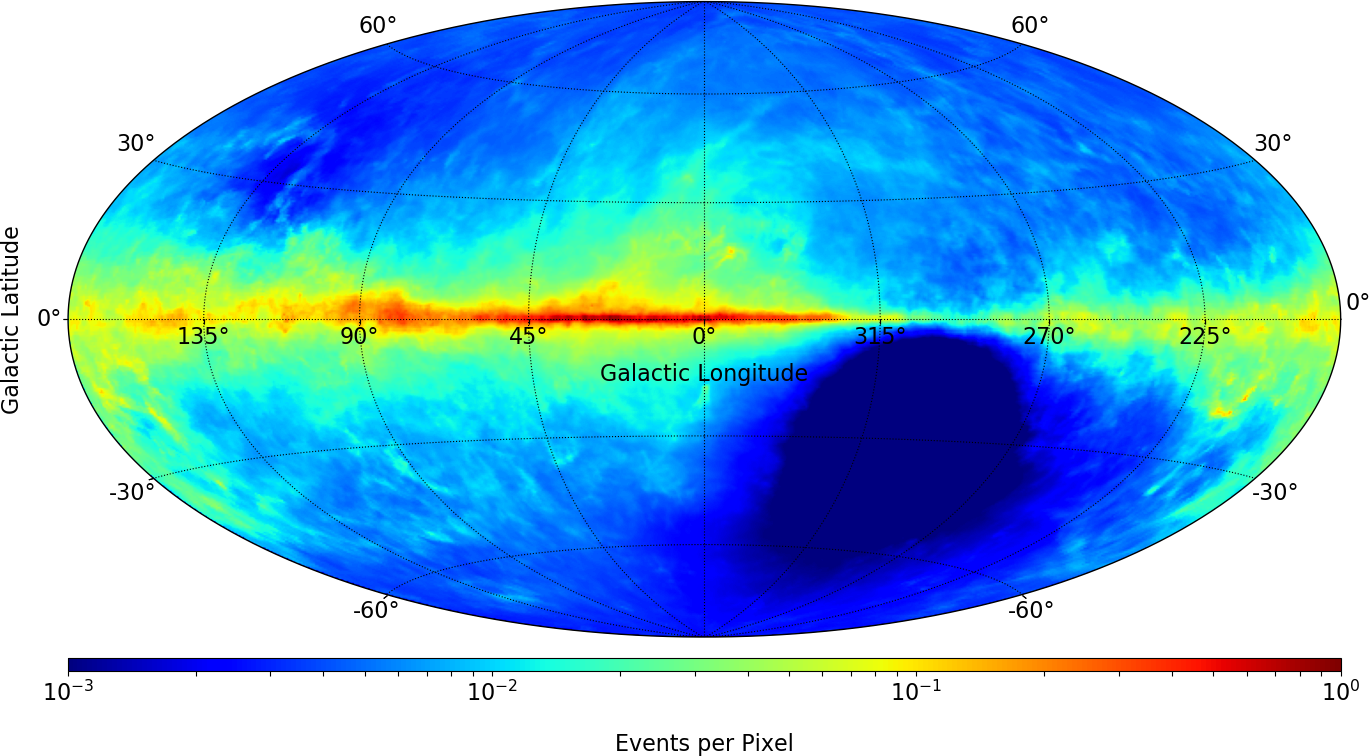}
  \caption{Expected event counts between \SI{0.97}{\giga\electronvolt} and
    \SI{1.07}{\giga\electronvolt} from diffuse emission for the vertex analysis, assuming ideal
    angular resolution.}
  \label{fig:diffuse-model-counts}
\end{figure}

\begin{figure}[p!]
  \centering
  \includegraphics[width=0.8\linewidth]{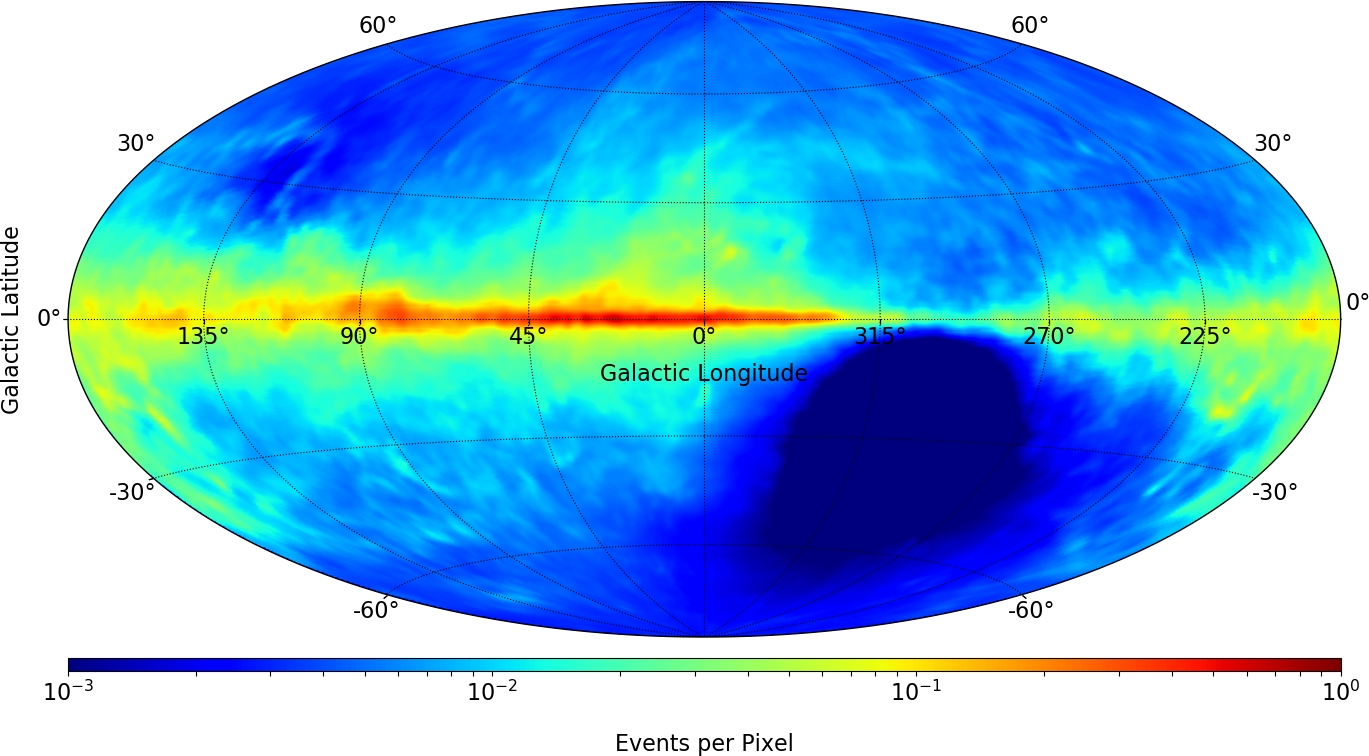}
  \caption{Expected photon counts between \SI{0.97}{\giga\electronvolt} and \SI{1.07}{\giga\electronvolt}
    from diffuse emission for the vertex analysis, after convolution with the PSF.}
  \label{fig:total-model-diffuse-smeared}

  \vspace*{\floatsep}

  \centering
  \includegraphics[width=0.8\linewidth]{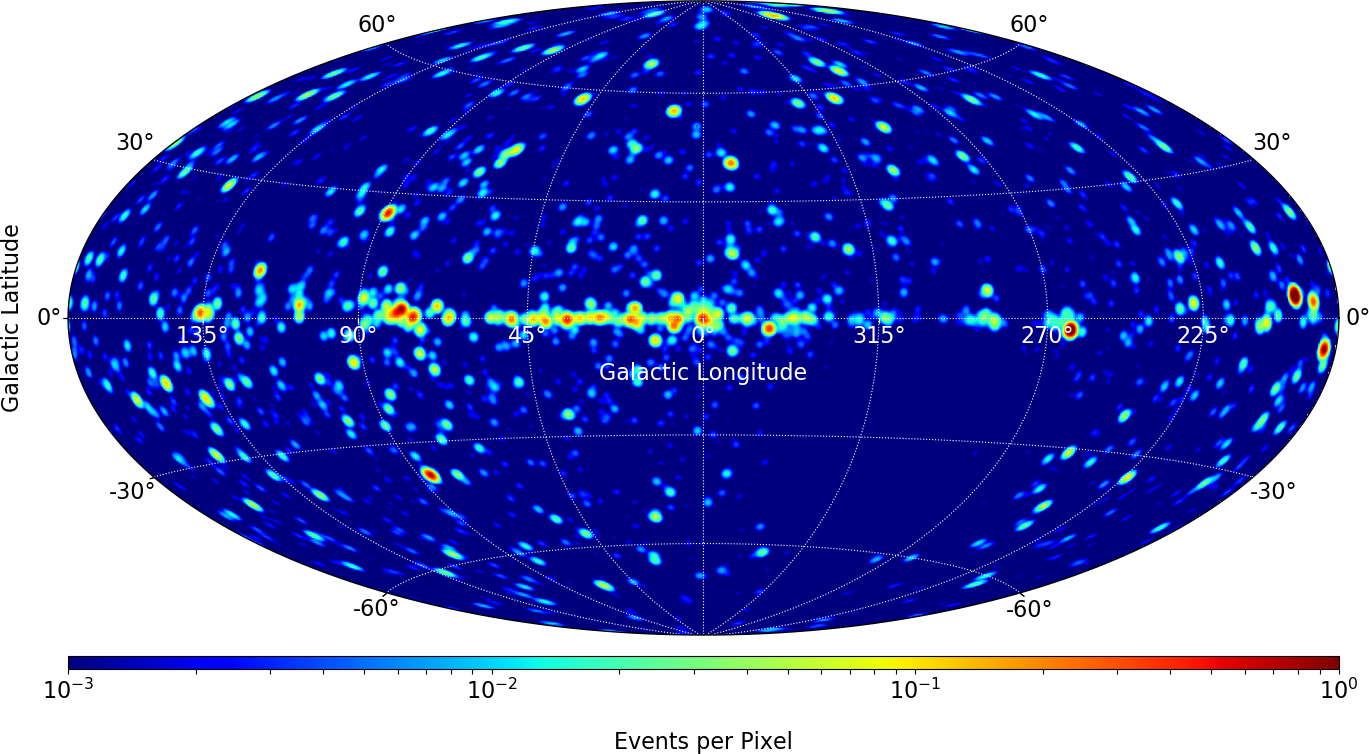}
  \caption{Expected photon counts from sources between \SI{0.97}{\giga\electronvolt} and
    \SI{1.07}{\giga\electronvolt} using the 4FGL and the vertex analysis exposure map, after
    convolution with the PSF.}
  \label{fig:total-model-sources-smeared}

  \vspace*{\floatsep}

  \centering
  \includegraphics[width=0.8\linewidth]{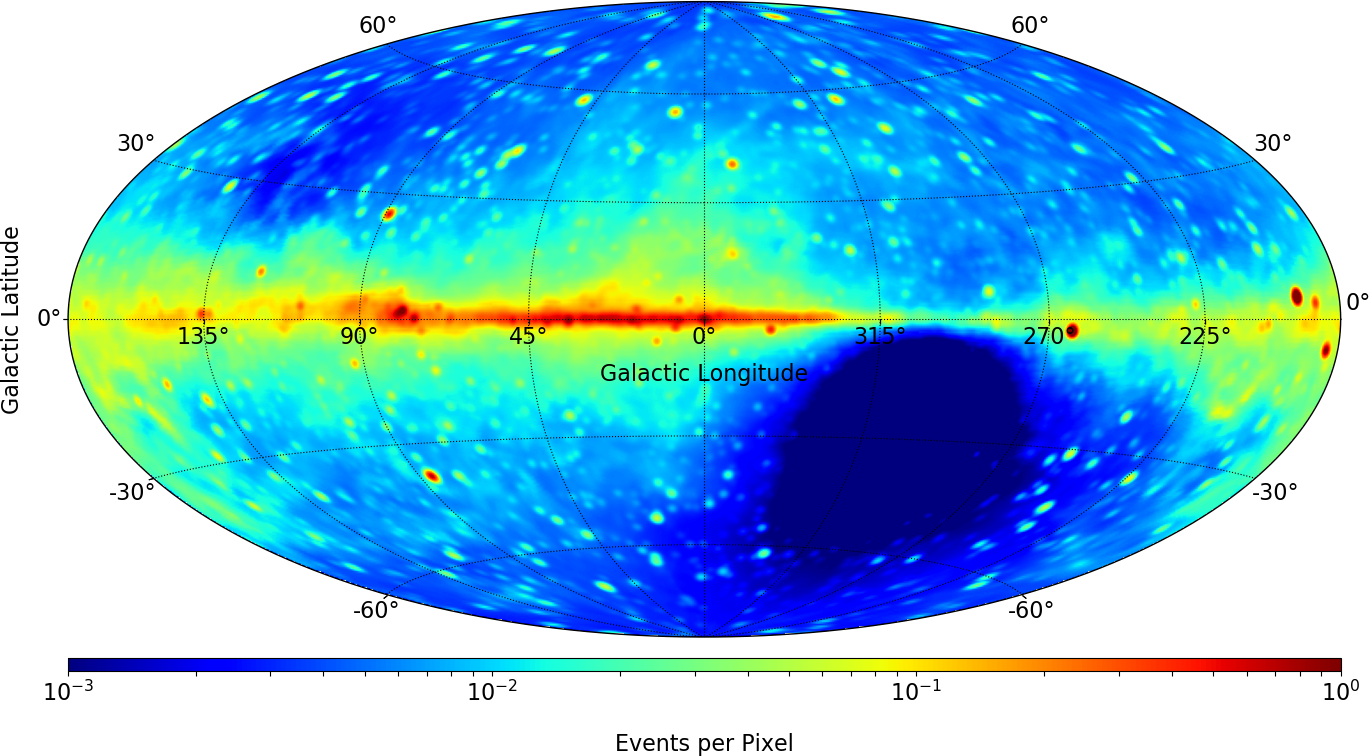}
  \caption{Expected photon counts according to the final model between \SI{0.97}{\giga\electronvolt}
    and \SI{1.07}{\giga\electronvolt} for the vertex analysis.}
  \label{fig:total-model-both}
\end{figure}

Figures~\ref{fig:diffuse-model-flux} to~\ref{fig:total-model-both} illustrate this
procedure. Figure~\ref{fig:diffuse-model-flux} shows the diffuse flux at \SI{1}{\giga\electronvolt}
photon energy according to the Fermi-LAT diffuse emission
model~\cite{Fermi_Diffuse_IEM_2019_WWW}. Figure~\ref{fig:diffuse-model-exposure-map} shows the
exposure map at \SI{1}{\giga\electronvolt} photon energy for the vertex analysis as constructed in
section~\ref{sec:analysis-exposure-maps}.

Figure~\ref{fig:diffuse-model-counts} is the product of the former two scaled by the energy bin
width and solid angle, which yields the expected event counts from diffuse emission following
equation~(\ref{eq:modeling-diffuse-expected-counts-diffuse}). Figure~\ref{fig:total-model-diffuse-smeared}
illustrates the effect of the smearing with the PSF, as constructed in
section~\ref{sec:analysis-psf}. Fine structures in the diffuse emission are blurred according to the
detector angular resolution. Combining the measured fluxes and positions of the sources in the
fourth Fermi-LAT source catalog with the PSF and the exposure map allows to estimate the number of
expected events for each source as shown in figure~\ref{fig:total-model-sources-smeared}. Finally,
the maps for diffuse emission and source contributions can be summed yielding the final model of the
gamma-ray sky in figure~\ref{fig:total-model-both}.



The model thus constructed has no free parameter and is therefore very predictive. However it is
also possible to include scaling factors for the flux of a particular source, for example. These can
then be fitted in a maximum likelihood fit to the \mbox{AMS-02} data in order to measure the source
flux independently from other contributions in the same region of the sky. Similarly it is also
possible to vary the position and spatial extension of a given source. If the diffuse emission is
considered a background it is also customary to include scaling factors for it in fits, which
improves the background description in small regions of interest and allows for a better fit of
source spectra.

Because this model is based on the Fermi-LAT diffuse emission
model~\cite{Fermi_Diffuse_IEM_2019_WWW} and the Fermi-LAT fourth source catalog
(4FGL)~\cite{Fermi_4FGL_2019}, both of which are derived from LAT data, it reproduces the Fermi-LAT
gamma ray sky rather well. Comparing the \mbox{AMS-02} data with the model therefore makes it
possible to indirectly quantify the agreement between \mbox{AMS-02} and Fermi-LAT data, to the
extent of the validity of the model.

It is also important to note that the 4FGL catalog includes a number of sources which are variable
in time. This is in particular true for extragalactic gamma ray sources such as quasars and
AGN. These objects are known to produce strong gamma-ray flares which can substantially impact their
time-averaged spectra when the fluxes are integrated over a long period of time. In addition these
sources can remain in an active (or quiet) state for many months or even years. Examples of such
variable sources are 3C 454.3, 3C 279 and CTA 102.

The 4FGL catalog was created from an analysis of approximately 8 years of Fermi-LAT data (recorded
between August 4th 2008 and August 2nd 2016)~\cite{Fermi_4FGL_2019}. This period overlaps only
partially with the interval in which the \mbox{AMS-02} data was recorded (May 19th 2011 until
November 12th 2017). Hence, discrepancies between the model and the \mbox{AMS-02} data are to be
expected for these types of sources. For many of the bright sources (in particular pulsars such as
Vela, Geminga, Crab, PSR J1836+5925 and others, whose fluxes are remarkably stable in time), the
predictions of the model are transferable to the \mbox{AMS-02} time period without restrictions.
\enlargethispage{1cm}

\section{Background Estimation}
\label{sec:corrections-background}

\begin{figure}[p!]
  \centering
  \includegraphics[width=1.0\linewidth]{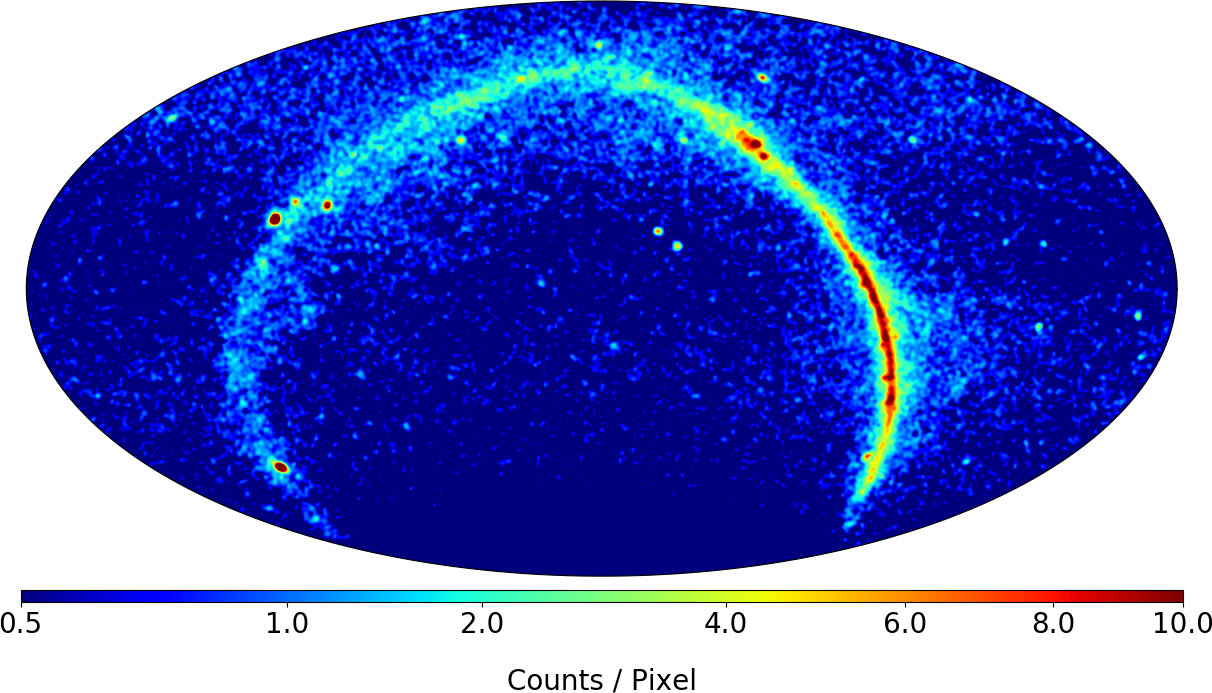}
  \caption{Measured photon counts for the vertex analysis integrated from
    \SI{500}{\mega\electronvolt} to \SI{10}{\giga\electronvolt} in ICRS coordinates, shown with
    square root color scale.}
  \label{fig:counts-vertex-data}

  \vspace*{5\floatsep}

  \centering
  \includegraphics[width=1.0\linewidth]{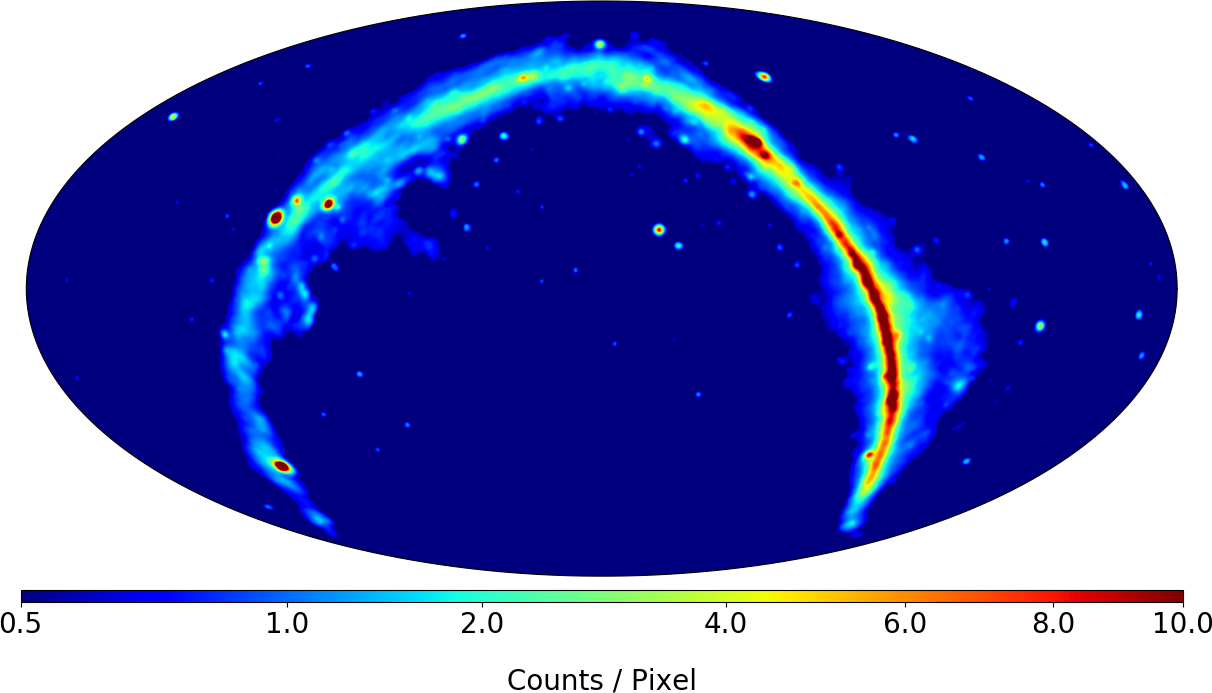}
  \caption{Model photon counts for the vertex analysis integrated from \SI{500}{\mega\electronvolt}
    to \SI{10}{\giga\electronvolt} in ICRS coordinates, shown with a square root color scale.}
  \label{fig:counts-vertex-model}
\end{figure}

Applying the selection described in section~\ref{sec:analysis-selection} to the AMS-02 data yields
231495 photon candidate events in the vertex analysis as well as 323245 photon candidates in the
calorimeter analysis. The two numbers are not directly comparable, because of the different energy
thresholds of the analyses. For each event the position in celestial coordinates can be calculated
based on the reconstructed direction from either tracker tracks or calorimeter shower, the ISS
position and rotation as well as the event time.

Figure~\ref{fig:counts-vertex-data} shows a binned skymap in ICRS equatorial coordinates constructed
from all events between \SI{500}{\mega\electronvolt} and \SI{10}{\giga\electronvolt} in the vertex
analysis. Figure~\ref{fig:counts-vertex-model} below shows the model for the expected event counts
in the same energy range. The projections are full-sky Hammer-Aitoff projections with 720 bins in
longitude and 360 bins in latitude, resulting in a grid of 203588 active bins with a size of
approximately $\SI{0.5}{\degree} \times \SI{0.5}{\degree}$ in the center of the image. Each bin
covers the same solid angle of approximately \SI{6.17e-5}{\steradian}.

Similar figures are shown for the calorimeter analysis in
appendix~\ref{sec:appendix-ecal-background-estimation}.

Overall there is a clear similarity between the measured data and the model prediction. However, in
both the vertex and calorimeter analyses there is an additional component of measured events which
is not present in the model predictions. This component forms ring like structures around lines of
constant declination and is particularly visible for the high and low declination regions.



Because the excess correlates with declination, which corresponds to latitude on the Earth, the
hypothesis is that these events are due to charged particle background. Two major sources of such
background are: Particles which are misidentified as gamma rays and those which produce genuine
gamma rays in the detector material near the top of the experiment.

In the calorimeter analysis protons and electrons (which are more abundant than gamma rays by
several orders of magnitude, in particular near the geomagnetic poles) can enter the calorimeter
from outside the regular acceptance cone, without passing through the upper TOF, TRD and inner
tracker or ACC. At low energies, where the resolution of the shower axis reconstruction is poor, it
is possible that such events are badly reconstructed (in particular in the case of proton
background), such that the reconstructed axis points towards the upper detector. In that case the
absence of signal in the TRD, tracker and upper TOF system is misleading and the event might be
selected based on the set of cuts outlined in section~\ref{sec:analysis-selection-calorimeter}. This
effect was verified with the help of proton and electron Monte-Carlo and cannot be neglected.

\begin{figure}[t]
  \begin{minipage}{0.48\linewidth}
    \centering
    \includegraphics[width=1.0\linewidth]{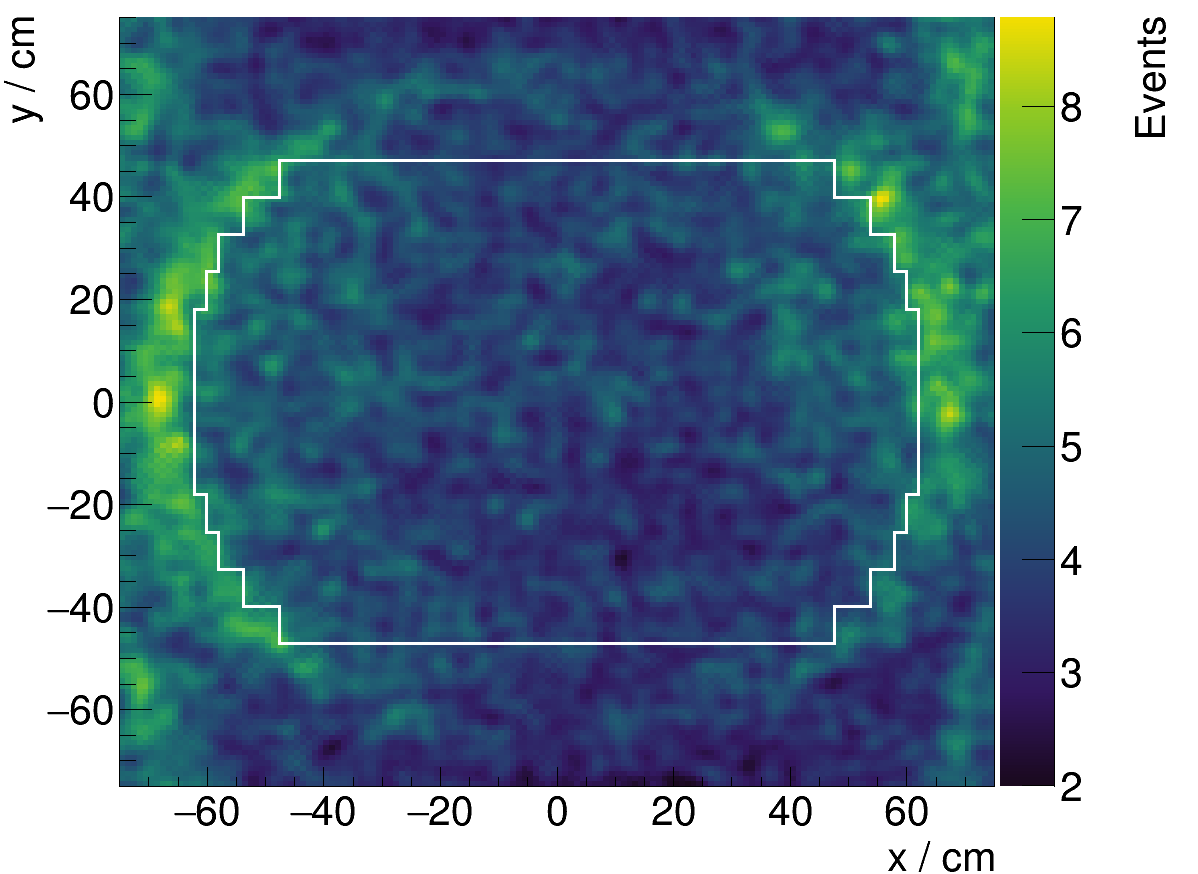}
  \end{minipage}
  \hspace{0.01\linewidth}
  \begin{minipage}{0.48\linewidth}
    \centering
    \includegraphics[width=1.0\linewidth]{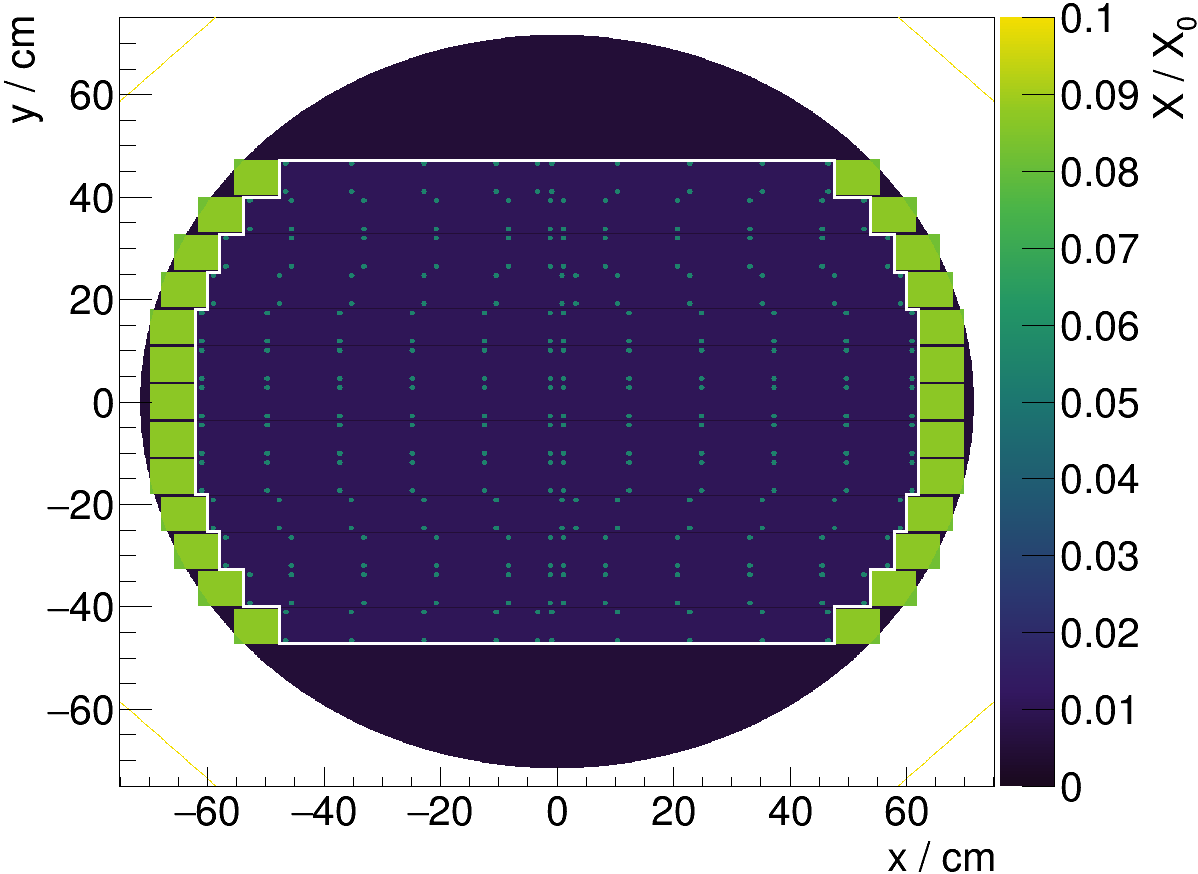}
  \end{minipage}
  \caption{Left: Extrapolation of reconstructed photon directions to the tracker layer 1 plane for
    events outside of the galactic plane ($|b| > \SI{10}{\degree}$). Right: Integrated radiation
    length ($X / X_0$) in the tracker layer 1 material budget according to the AMS-02 model in the
    simulation. The white line marks the circumference of the tracker layer 1 active area in both
    figures.}
  \label{fig:layer1-background-and-material}
\end{figure}

It is also possible for charged particles to produce gamma rays in the material at the top of the
detector, particularly through production and subsequent decay of $\pi^0$ mesons. Such interactions
become visible when the tracks of reconstructed photon candidates are extrapolated towards the top
of the experiment. Such extrapolations are shown in figure~\ref{fig:layer1-background-and-material}
on the left hand side, for events which do not fall into the galactic plane. For genuine gamma rays
a uniform illumination of the top of the instrument is expected (when integrating over long periods
of time). However, the actual observed distribution features structures which correlate with the
material of the detector, such as the tracker layer 1 electronics, as shown on the right hand side
of figure~\ref{fig:layer1-background-and-material}. This effect becomes more apparent if the event
sample is restricted to those regions of the sky in which a significant amount of background is
observed, such as regions at high and low declination.

In both cases it is hard to remove such events. The first case is simply a result of the resolution
of the calorimeter shower axis reconstruction, which cannot be substantially improved at low
energies. In the second case genuine gamma rays are produced, with the actual charged primary
escaping detection. Although it would be possible to remove the second class by vetoing all events
which pass through regions with dense material at the top of the instrument, this would result in a
significant reduction of the effective area and subsequent loss of statistics.

Naively one would assume that the background flux of charged particles is isotropic, in which case
the exposure map would serve as a good spatial template for the distribution of the observed
background events on the sky. Such an approach is inadequate for two reasons: Firstly, the charged
particle flux is not isotropic at low energies, because of trapped secondary particles in the
Earth's geomagnetic field. The additional flux of secondary particles near the Earth's geomagnetic
poles and in the vicinity of the SAA correlates with an substantial increase in background
events. Secondly, the exposure map calculated in section~\ref{sec:analysis-exposure-maps} is not
suitable for this task, since it was calculated with the effective area for photon signal
events. However, both types of background events preferentially arrive at larger zenith-angles,
which means that the $\cos{\theta}$ dependence of the background effective area is very different.

It is also difficult to predict the exact normalization of the expected background from the
Monte-Carlo simulation, because it would require generating vast amounts of protons and electrons in
order to produce a significant sample of interacting events which survive the photon selection. Even
if the probabilities for such events to occur were determined, such a method would require a good
description of the proton and electron particle flux at low energies, in particular below the
geomagnetic cutoff near the poles. However, the primary and secondary particle fluxes are subject to
significant variations with time, due to changes in the solar activity.

A better option is to quantify the residual amount of background from charged particles with a data
driven method directly from the photon data, which is the approach followed here. The principal
observation is that the charged particle background spatially depends only on the declination angle,
but not on right ascension, which is not physical for the $\gamma$-ray signal. Also, the expected
flux of signal photons at high galactic latitudes is negligible, except in the vicinity of a few
strong sources. Therefore it is possible to use large parts of the high galactic latitude region to
directly estimate the background from the data.

The procedure starts by defining a mask which excludes regions with significant amounts of signal
photons from the background determination. All of the regions of interest which will be analyzed in
chapter~\ref{sec:results} are masked completely in this step. The following regions of the sky are
excluded in the background determination procedure:

\begin{itemize}
\item \makebox[5cm][l]{\textbf{The galactic plane}:} |b| < \SI{15}{\degree}
\item \makebox[5cm][l]{\textbf{3C 454.3} and \textbf{CTA-102}:} \SI{73}{\degree} < l <
  \SI{93}{\degree}, \SI{-42}{\degree} < b < \SI{-34}{\degree}
\item \makebox[5cm][l]{\textbf{J1836.2+5925}:} \SI{85}{\degree} < l < \SI{93}{\degree},
  \SI{21}{\degree} < b < \SI{29}{\degree}
\item \makebox[5cm][l]{\textbf{Other bins}:} Bins in which more than 0.5 photon events are\\
  \makebox[5cm][l]{ } expected when integrating from \SI{50}{\mega\electronvolt} to infinity.
\end{itemize}

\begin{figure}[t!]
  \centering
  \includegraphics[width=0.98\linewidth]{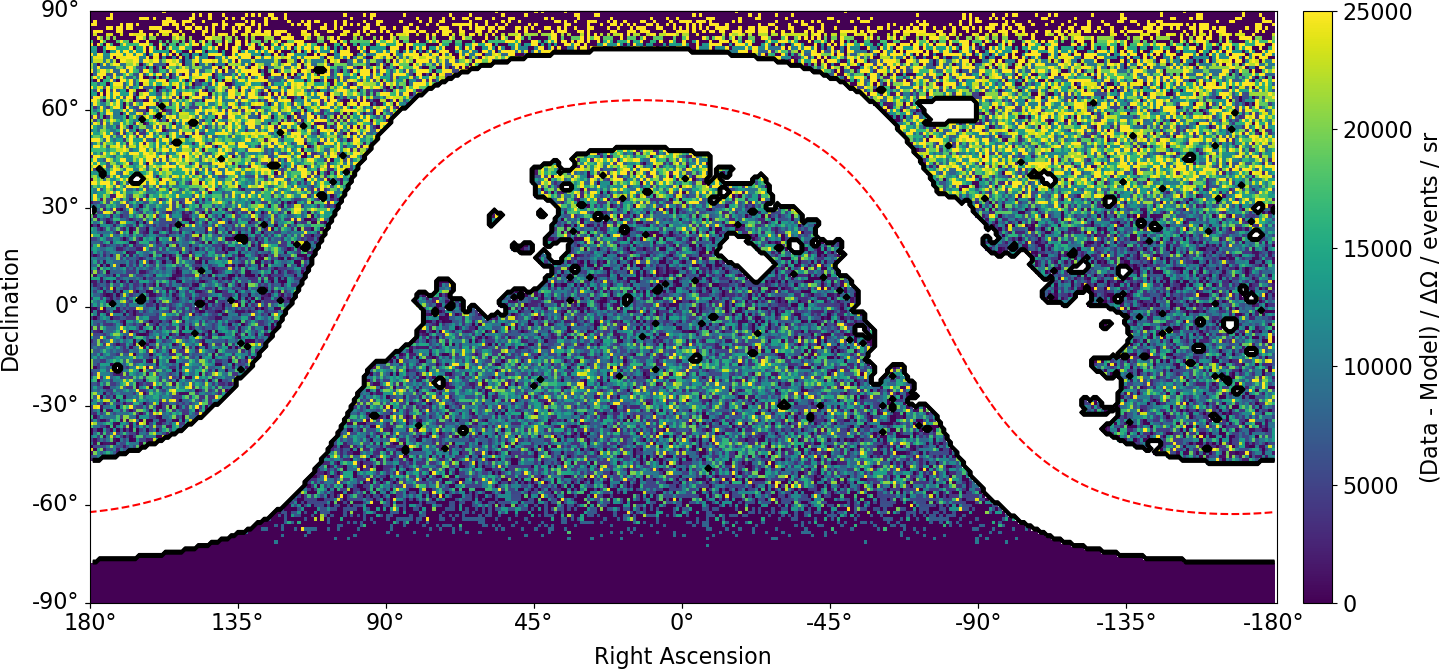}
  \caption{Subtraction of observed and predicted event counts in ICRS equatorial plate carree
    projection, divided by bin solid angle for the vertex analysis, counts integrated from
    \SI{100}{\mega\electronvolt} to \SI{100}{\giga\electronvolt}. The white areas correspond to
    masked pixels, the dashed red line corresponds to the location of the galactic plane.}
  \label{fig:background-fit-vertex-masked-excess}
\end{figure}

In order to fix the spatial shape of the background the observed photons and the photon model are
summed over energy. In the vertex analysis the summation runs over all energy bins from
\SI{100}{\mega\electronvolt} to \SI{100}{\giga\electronvolt}. Pixels which are masked are removed
from both the observation and the prediction. The model prediction for the number of $\gamma$-ray
photons in the remaining bins is very small, but for mathematical correctness the difference of the
observed and predicted photon counts is constructed and divided by the solid angle of each skymap
bin. For the vertex analysis the resulting distribution is shown in
figure~\ref{fig:background-fit-vertex-masked-excess} in ICRS equatorial coordinates. Horizontal
bands corresponding to regions with large background are identifiable. There are no other visible
structures in the excess map.

\begin{figure}[t!]
  \centering
  \includegraphics[width=0.8\linewidth]{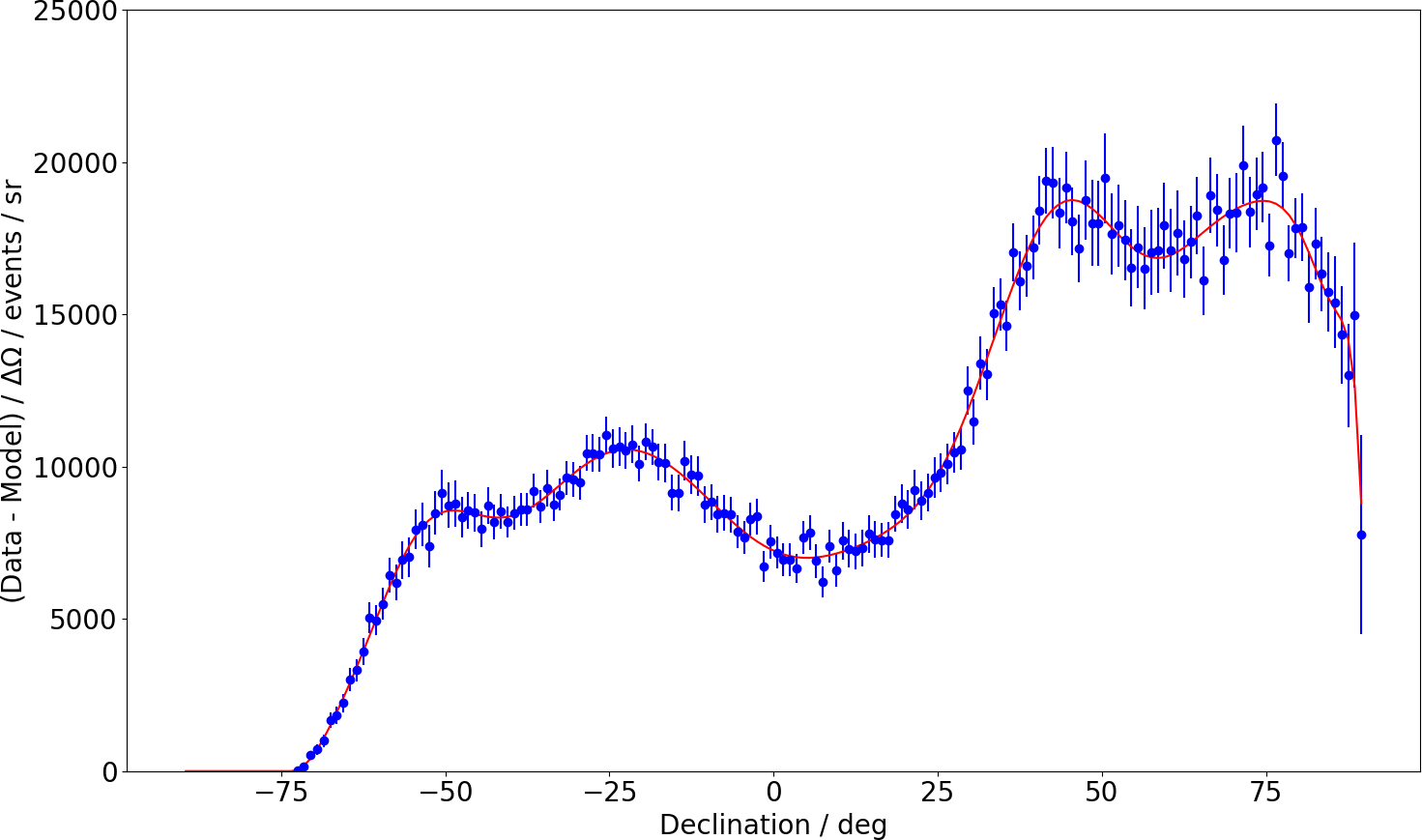}
  \caption{Average excess flux as a function of declination for the vertex analysis. The red line is
    a polynomial of order 20, which is used as an analytical description of the shape.}
  \label{fig:background-fit-vertex-polynomial-fit}
\end{figure}

In order to parameterize the background shape a one dimensional distribution of the average excess
as a function of declination is built. For each declination angle bin, the average excess flux is
constructed, averaging over right ascension for all non-masked pixels. The resulting distribution is
shown for the vertex analysis in blue in figure~\ref{fig:background-fit-vertex-polynomial-fit}. A
polynomial of order 20 is used as an empirical analytical model and fit to the data. The result is
shown in red. The shape of the resulting polynomial roughly represents the declination dependence of
the exposure (see figure~\ref{fig:exposure-map-vertex-equatorial}). The double peak structure is a
result of the preference for large zenith angles in the background.

\begin{figure}[t!]
  \begin{minipage}{0.48\linewidth}
    \centering
    \includegraphics[width=1.0\linewidth]{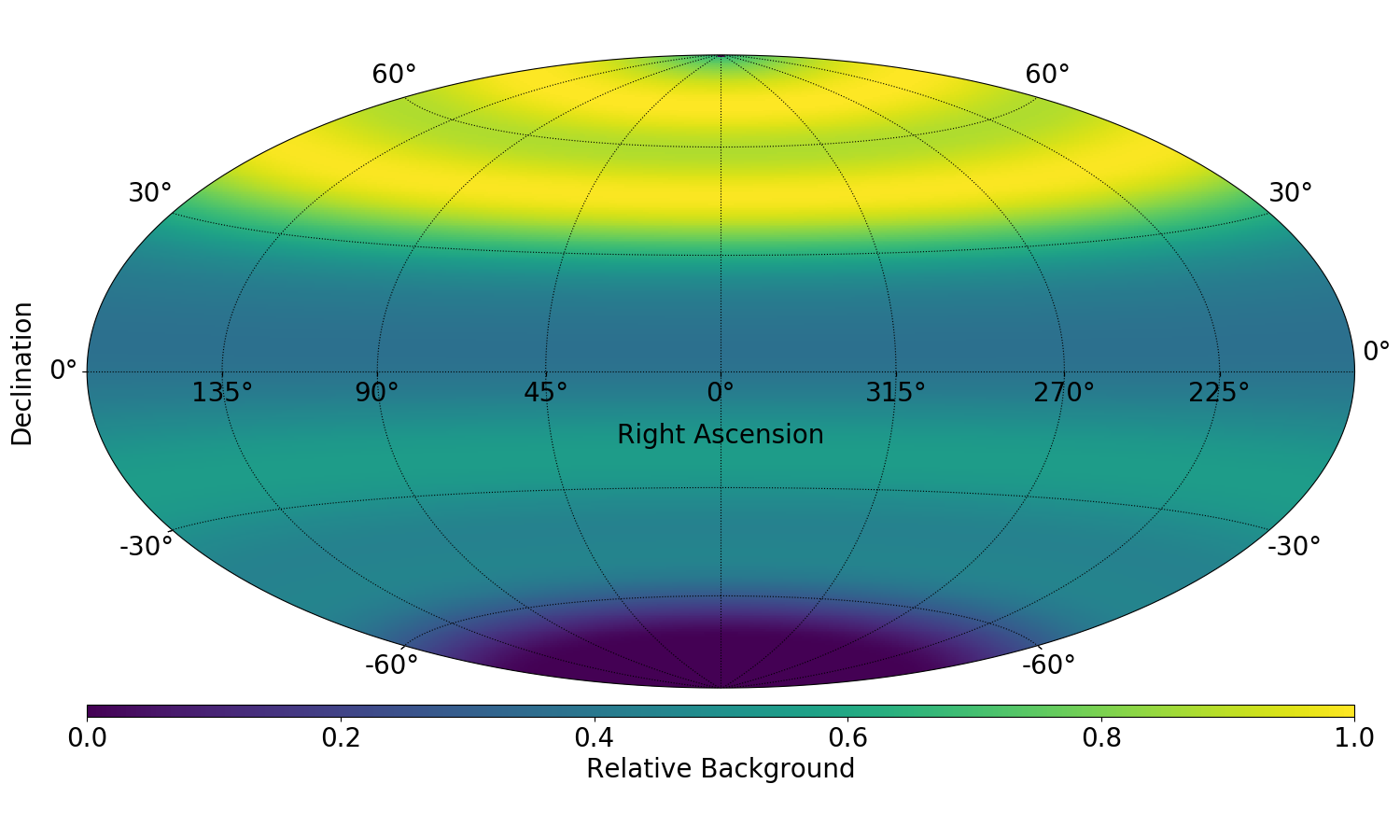}
  \end{minipage}
  \hspace{0.01\linewidth}
  \begin{minipage}{0.48\linewidth}
    \centering
    \includegraphics[width=1.0\linewidth]{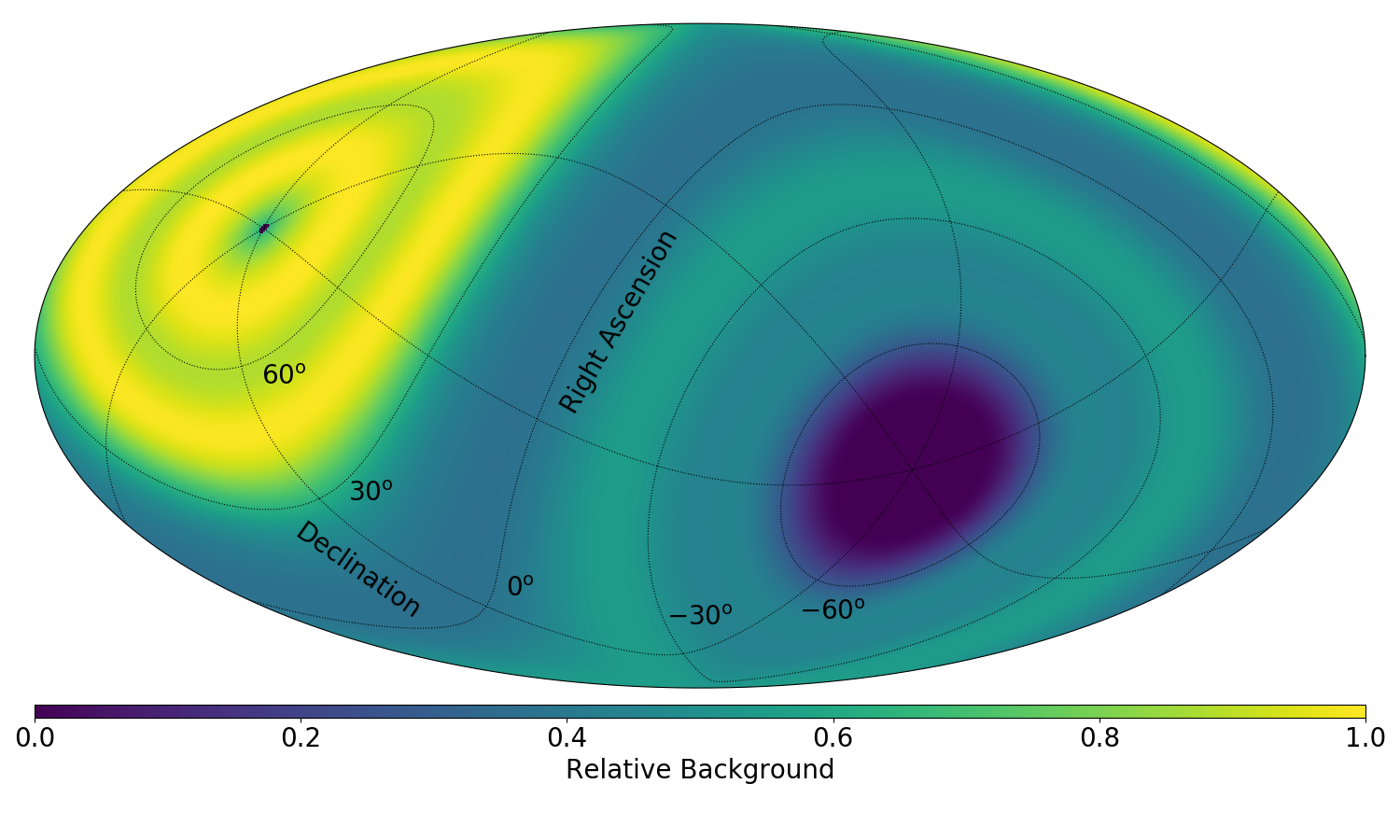}
  \end{minipage}
  \caption{Vertex analysis background template map in ICRS equatorial (left) and galactic (right)
    coordinates.}
  \label{fig:background-fit-vertex-template}
\end{figure}

The analytical function is then used to produce a full sky template map for the background, which
also extends into the masked regions. The result for the vertex analysis is shown in
figure~\ref{fig:background-fit-vertex-template}, for both equatorial and galactic coordinates.

For the vertex analysis no significant spatial variation of the background with energy is observed,
so the spatial template is assumed to be valid at all energies. For the calorimeter there are minor
differences when constructing the template at low energies and at high energies. Therefore two
independent templates, one constructed from data between \SI{1}{\giga\electronvolt} to
\SI{2}{\giga\electronvolt} and one constructed from data between \SI{2}{\giga\electronvolt} and
\SI{1}{\tera\electronvolt} are assembled using the procedure outlined above.

The final energy dependent background maps are linear combinations of the templates, in which the
coefficients depend on the energy bin:

\begin{align}
  b(E_i,\alpha_j,\delta_k) = \sum_{j=1}^{n_{\mathrm{templates}}}{x^j_i \cdot b^j(\delta_k)} \,.
\end{align}

In order to determine the scaling factors $x$ a likelihood fit is employed in each energy bin. For
the case of only one background template the number of observed events in each (non-masked) pixel in
the sky is assumed to follow a Poisson distribution with expectation value given by

\begin{displaymath}
  n^{\mathrm{exp}}(E_i,\alpha_j,\delta_k) =
  n^{\mathrm{model}}(E_i,\alpha_j,\delta_k) + x_i \cdot b(\delta_k) \,,
\end{displaymath}

where $x_i$ is a free scaling factor and $b$ is the background template from the polynomial fit
which depends only on the declination angle. Summing this expression over right ascension yields:

\begin{align}
  \mu_{ik} := N^{\mathrm{exp}}(E_i,\delta_k)
  &= N^{\mathrm{model}}(E_i,\delta_k) + x_i \cdot B(\delta_k) \\
  &= N^{\mathrm{model}}_{ik} + x_i \cdot B_k \,.
\end{align}

\begin{figure}[t!]
  \centering
  \includegraphics[width=0.98\linewidth]{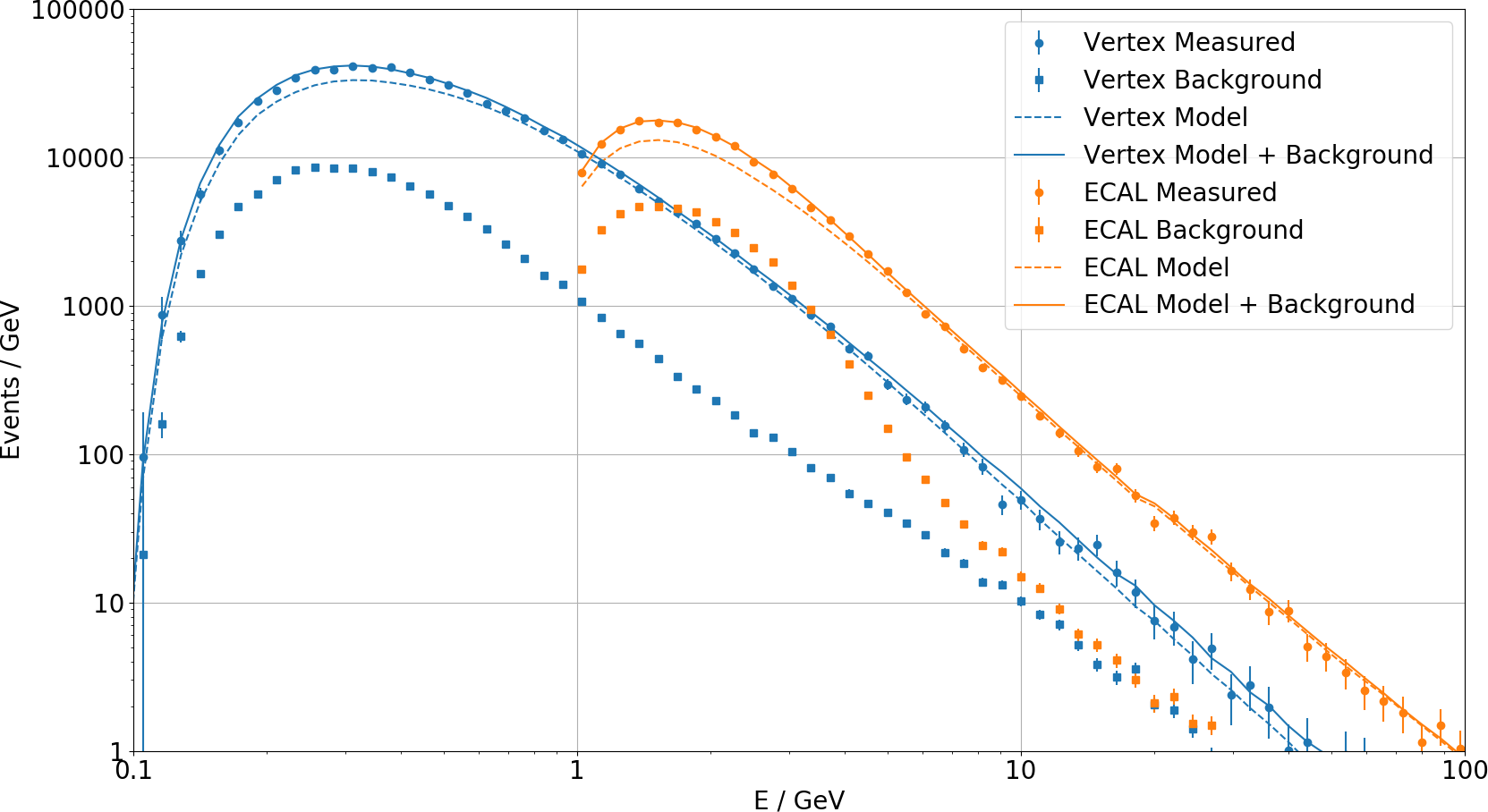}
  \caption{Comparison of the obtained background yield in the inner galaxy
    ($\SI{-20}{\degree} < l < \SI{80}{\degree}, |b| < \SI{8}{\degree}$) with the measured and
    predicted photon counts for the vertex (blue) and calorimeter (orange) analyses.}
  \label{fig:background-norm}
\end{figure}

The sum runs over all pixels which are not masked, so the number of summed right ascension pixels is
not the same for each declination bin. Since all the individual observations in each pixel are
independent $N$ will also follow a Poisson distribution. The likelihood function for a single energy
bin (dropping the energy bin indices) is thus:

\begin{align}
  \mathcal{L}
  &= \prod_{k=1}^{n_{\mathrm{bins}}}{\frac{e^{-\mu_k} \mu_k^{N_k}}{N_k!}} \\
  \Rightarrow -\log{\mathcal{L}}
  &= \sum_{k=1}^{n_{\mathrm{bins}}}{\mu_k - N_k \cdot \log{\mu_k} + \log{N_k!}} \\
  &= \sum_{k=1}^{n_{\mathrm{bins}}}{N^{\mathrm{model}}_k +
    x \cdot B_k - N_k \cdot \log{(N^{\mathrm{model}}_k + x \cdot B_k)} + \log{N_k!}} \,.
\end{align}

Dropping terms which do not depend on $x$ this can be simplified to:

\begin{displaymath}
  -\log{\mathcal{L}} = \sum_{k=1}^{n_{\mathrm{bins}}}{x \cdot B_k - N_k \cdot
    \log{(N^{\mathrm{model}}_k + x \cdot B_k)}} \,.
\end{displaymath}

Finding the minimum of $-\log{\mathcal{L}}$ yields the maximum likelihood estimator for $x$ which is
used as a scaling factor for the background template.

\begin{figure}[t!]
  \centering
  \includegraphics[width=0.98\linewidth]{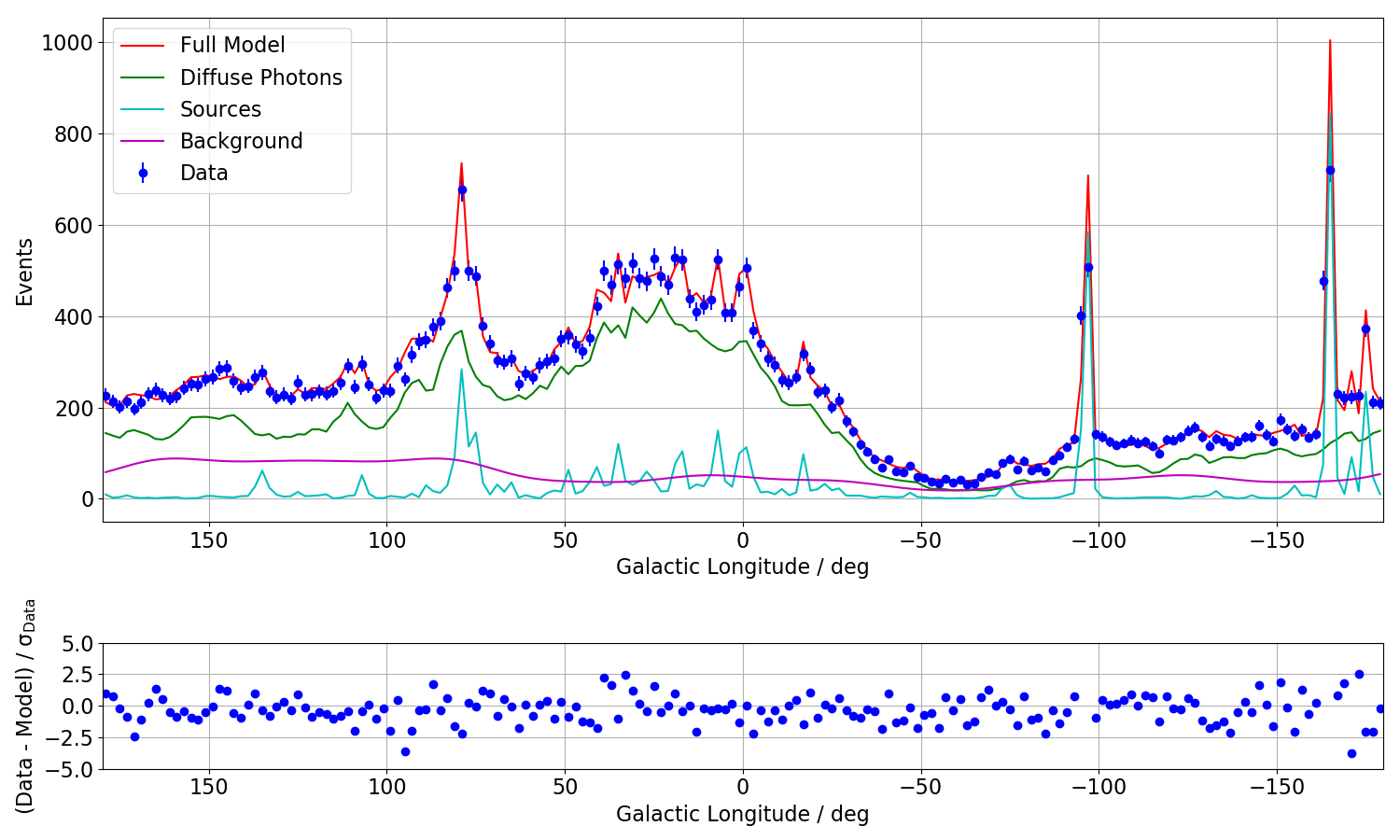}
  \caption{One dimensional comparison of measured photon counts between \SI{500}{\mega\electronvolt}
    and \SI{100}{\giga\electronvolt} in the galactic plane ($|b| < \SI{8}{\degree}$) for the vertex
    analysis with the model prediction including the background component as a function of galactic
    longitude. The lower panel shows the pull distribution, uncertainties are statistical only.}
  \label{fig:compare-counts-vertex-galactic-plane}
\end{figure}

In case of the calorimeter analysis, where two templates are used, the likelihood is

\begin{displaymath}
  -\log{\mathcal{L}} = \sum_{k=1}^{n_{\mathrm{bins}}}{\left(
      \sum_{j=1}^{2}{\left(x^{j} \cdot B^j_k\right)} - N_k \cdot \log{\left(N^{\mathrm{model}}_k +
          \sum_{j=1}^{2}{\left(x^{j} \cdot B^j_k\right)}\right)} \right)} \,,
\end{displaymath}

which requires finding two scaling factors per energy bin and their covariance. Both coefficients
are allowed to vary freely in each energy bin, allowing for a gradual transition from the low energy
template to the high energy template around \SI{2}{\giga\electronvolt}.

Figure~\ref{fig:background-norm} shows the evolution of the obtained background yield after fitting
as a function of energy. Photons are summed over the inner galaxy
($\SI{-20}{\degree} < l < \SI{80}{\degree}, |b| < \SI{8}{\degree}$) and compared with both measured
and predicted photons for both analyses. For the vertex analysis the background is at the
\SI{10}{\percent} level in this window. For the calorimeter analysis the background is at the
\SI{30}{\percent} level at energies below \SI{3}{\giga\electronvolt} but quickly drops below
\SI{10}{\percent} for energies above \SI{10}{\giga\electronvolt}. In both cases the sum of model
prediction and background yield fits the measured data very well over all energies. The uncertainty
of the background fit result is small compared to the statistical uncertainty of the data itself and
will be neglected in subsequent analyses.

Figure~\ref{fig:compare-counts-vertex-galactic-plane} shows a one dimensional comparison obtained by
projecting events from the vertex analysis in the galactic plane ($|b| < \SI{8}{\degree}$) onto the
galactic longitude axis and includes the result for the background component in the model. The
agreement between the data and the model is very good. The structure of both the diffuse emission as
well as the contribution of gamma ray sources are very well described.

\begin{figure}[p!]
  \centering
  \includegraphics[width=1.0\linewidth]{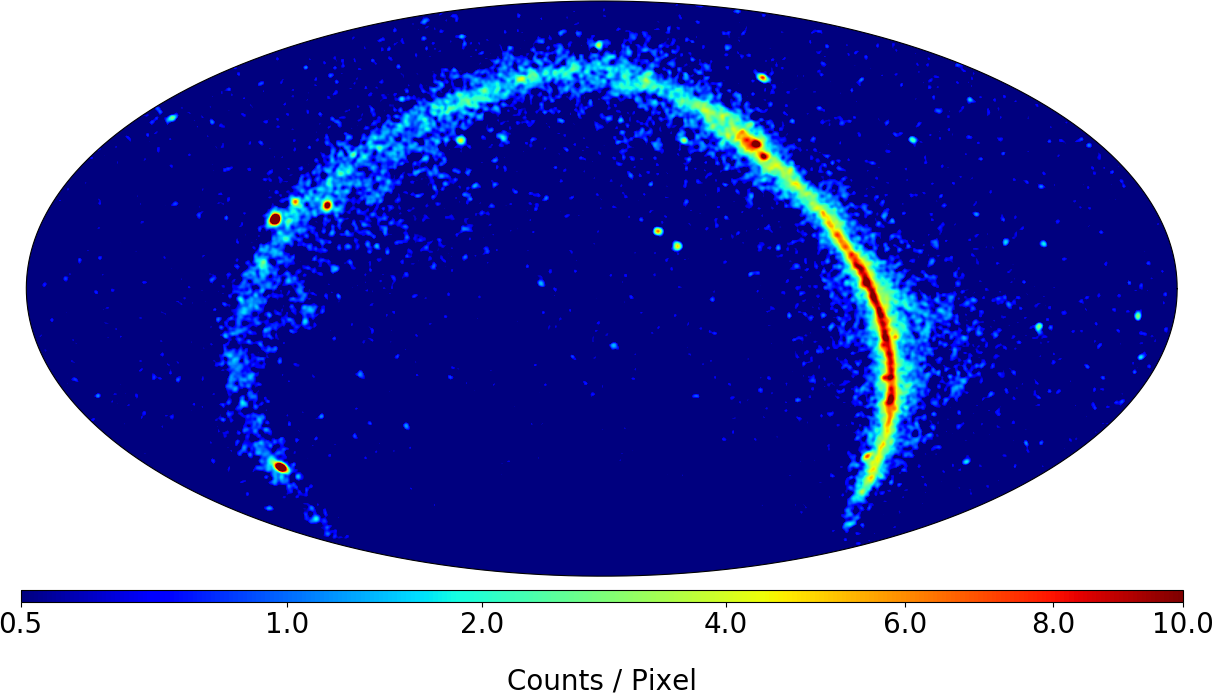}
  \caption{Background subtracted measured photon counts for the vertex analysis between
    \SI{500}{\mega\electronvolt} and \SI{10}{\giga\electronvolt} in ICRS equatorial coordinates,
    shown with a square root color scale.}
  \label{fig:counts-vertex-data-background-removed}

  \vspace*{5\floatsep}

  \centering
  \includegraphics[width=1.0\linewidth]{figures/analysis/Vertex_Counts_Equatorial_Model_0_5_10_jet_sqrt.png}
  \caption{Model photon counts for the vertex analysis between \SI{500}{\mega\electronvolt} and
    \SI{10}{\giga\electronvolt} in ICRS equatorial coordinates, shown with a square root color
    scale.}
  \label{fig:counts-vertex-model-repeated}
\end{figure}

\begin{figure}[p!]
  \centering
  \includegraphics[width=1.0\linewidth]{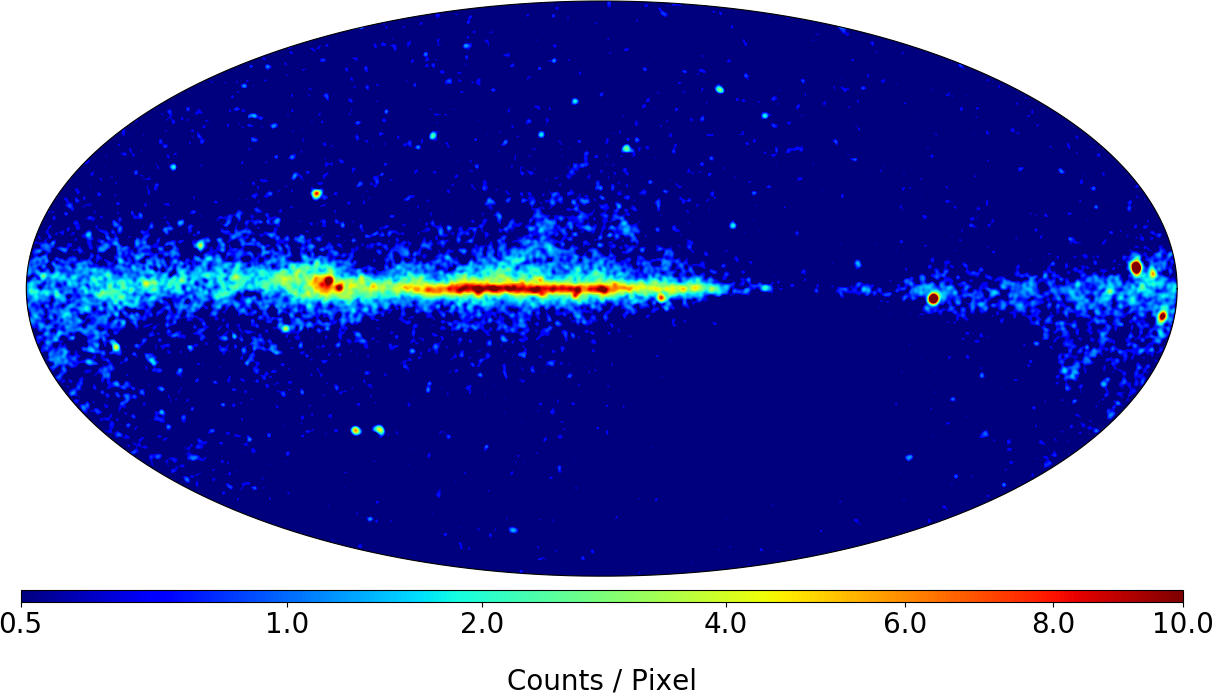}
  \caption{Background subtracted measured photon counts for the vertex analysis between
    \SI{500}{\mega\electronvolt} and \SI{10}{\giga\electronvolt} in galactic coordinates,
    shown with a square root color scale.}
  \label{fig:counts-vertex-data-galactic-background-removed}

  \vspace*{5\floatsep}

  \centering
  \includegraphics[width=1.0\linewidth]{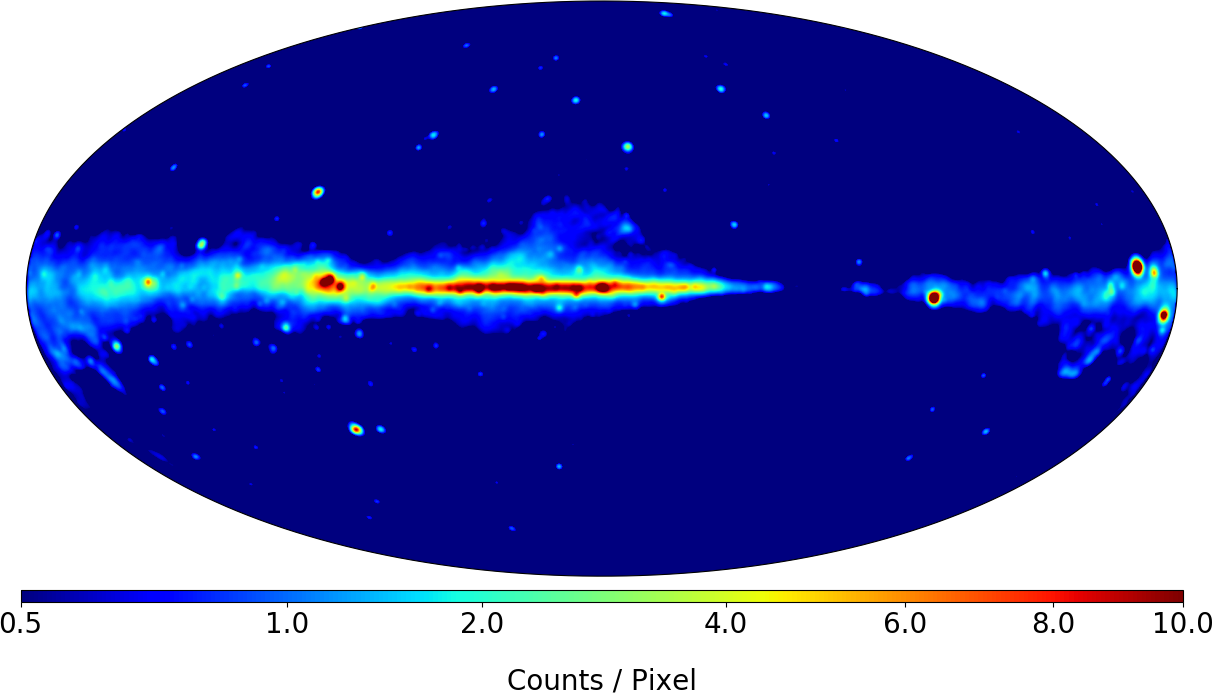}
  \caption{Model photon counts for the vertex analysis between \SI{500}{\mega\electronvolt} and
    \SI{10}{\giga\electronvolt} in gactic coordinates, shown with a square root color
    scale.}
  \label{fig:counts-vertex-model-galactic-repeated}
\end{figure}

Figures~\ref{fig:counts-vertex-data-background-removed}
to~\ref{fig:counts-vertex-model-galactic-repeated} show the background corrected measured photon
counts together with the prediction from the model in ICRS equatorial and galactic coordinates. In
figures~\ref{fig:counts-vertex-data-background-removed}
and~\ref{fig:counts-vertex-data-galactic-background-removed} the predicted background yield was
subtracted from the data. The ring like structures seen in figure~\ref{fig:counts-vertex-data} are
no longer visible, while many sources emerge as predicted by the model.

\section{Calorimeter Trigger Efficiency Correction}
\label{sec:corrections-ecal-trigger}

The AMS-02 trigger system features a dedicated trigger branch for the measurement of photons with
the calorimeter. This trigger branch makes use of special hardware and software and is described in
detail in section~\ref{sec:detector-trigger-calorimeter}. The Monte-Carlo trigger efficiency for
this trigger branch was shown and discussed in section~\ref{sec:analysis-trigger-efficiency}.

However, in the calorimeter trigger case, the ISS trigger efficiency is significantly different from
the Monte-Carlo efficiency. This means that a dedicated correction to the effective area is required
to account for this difference. Such a correction is not required for the trigger in the converted
photon analysis where it is provided by the TOF system. The main reason is that the TOF trigger is
extensively studied, since it must be estimated in all charged cosmic ray flux
measurements. Subsequently the Monte-Carlo was improved until the efficiency matched closely with
the in-flight TOF trigger efficiency.

In contrast, the calorimeter trigger is only needed for the measurement of photons. Parts of it are
also used in the measurement of electrons and positrons, but the details are different: For example
the angular calorimeter shower axis restriction implemented in the photon trigger is not used for
the electron trigger. In addition the electron trigger always requires the coincidence with a 4/4
TOF trigger decision. The simulation of the photon trigger has not been optimized in the software
and differences with respect to the actual efficiency can be expected.

Unfortunately it is not straightforward to estimate the photon calorimeter trigger efficiency using
the photon dataset obtained in this analysis. This is because the photon statistics available at low
energies (roughly \SIrange{1}{5}{\giga\electronvolt}) is inadequate. One particular problem is that
the events which fail the trigger and are recorded with the unbiased calorimeter trigger, are
prescaled with a factor of 1000. In addition, there is a sizable component of primary and secondary
cosmic rays in the dataset at these low energies, which makes it difficult to select a pure sample
of photons. While it is possible to reduce the fraction of background events by using only events
from regions where a large signal to background ratio is expected (such as the galactic plane at
intermediate declination values), this would reduce the available statistics further. Overall only a
qualitative estimation of the trigger efficiency is possible when using photon events.

Instead of using photons one can exploit the fact that electron showers look almost the same as
photon showers in the calorimeter. Since the calorimeter trigger only uses shower information its
efficiency should be almost the same for both photons and electrons. This statement can be checked
with the Monte-Carlo simulation where both photons and electrons are available with sufficient
statistics. This means that the in flight efficiency of the photon calorimeter trigger branch can be
estimated using ISS electrons which are significantly more abundant than photons. The same procedure
can then be carried out for Monte-Carlo electrons in order to estimate the Monte-Carlo to ISS
correction factor.

Another benefit of using electrons to measure the ECAL standalone trigger efficiency is that one
does not need to rely on the unbiased ECAL trigger to estimate the ECAL photon trigger
efficiency. Because most electrons fulfill the 4/4 TOF trigger condition, and because the TOF
trigger and ECAL trigger are statistically independent, one can instead use the TOF trigger as a tag
and calculate the conditional probability $p(E|T) = p(E)$ where E is the ECAL standalone trigger and
T the TOF 4/4 trigger, which is not prescaled. This greatly improves the available statistics.

In order to select electrons in both data and simulation the presence of a single calorimeter shower
is requested. In addition the event must have a single well reconstructed track with negative
rigidity in the tracker. The reduced $\chi^2/ndf$ for the track fit is required to be smaller than
10 in both projections. The measured tracker charge must be between 0.5 and 1.5 charge units. The
tracker track must pass through all four TOF layers and must not point to the ACC. The measured time
of flight velocity must be compatible with a downgoing particle, and must be larger than 0.9 in
order to select relativistic particles. The presence of a single TRD track with more than 14 active
layers and TRD electron/proton likelihood ratio smaller than 0.5 is required. Finally the ratio of
deposited energy in the calorimeter to the absolute rigidity value must be greater than 0.5 and
there is a loose cut on the calorimeter shower shape boosted decision tree of BDT > 0.5 in order to
select electromagnetic showers.

The conditional trigger efficiency of the photon trigger $P$ given that the TOF trigger $T$ fired
can then be calculated as:

\begin{displaymath}
  \epsilon_{\mathrm{trigger}}(E_i) = \frac{N_{\mathrm{PT}}(E_i)}{N_{\mathrm{T}}(E_i)} \,,
\end{displaymath}

where $E_i$ is the energy bin, $N_{\mathrm{PT}}$ is the number of events with both photon and TOF
trigger and $N_{\mathrm{T}}$ is the total number of events with a 4/4 TOF trigger.

\begin{figure}[t]
  \begin{minipage}{0.48\linewidth}
    \centering
    \includegraphics[width=1.0\linewidth]{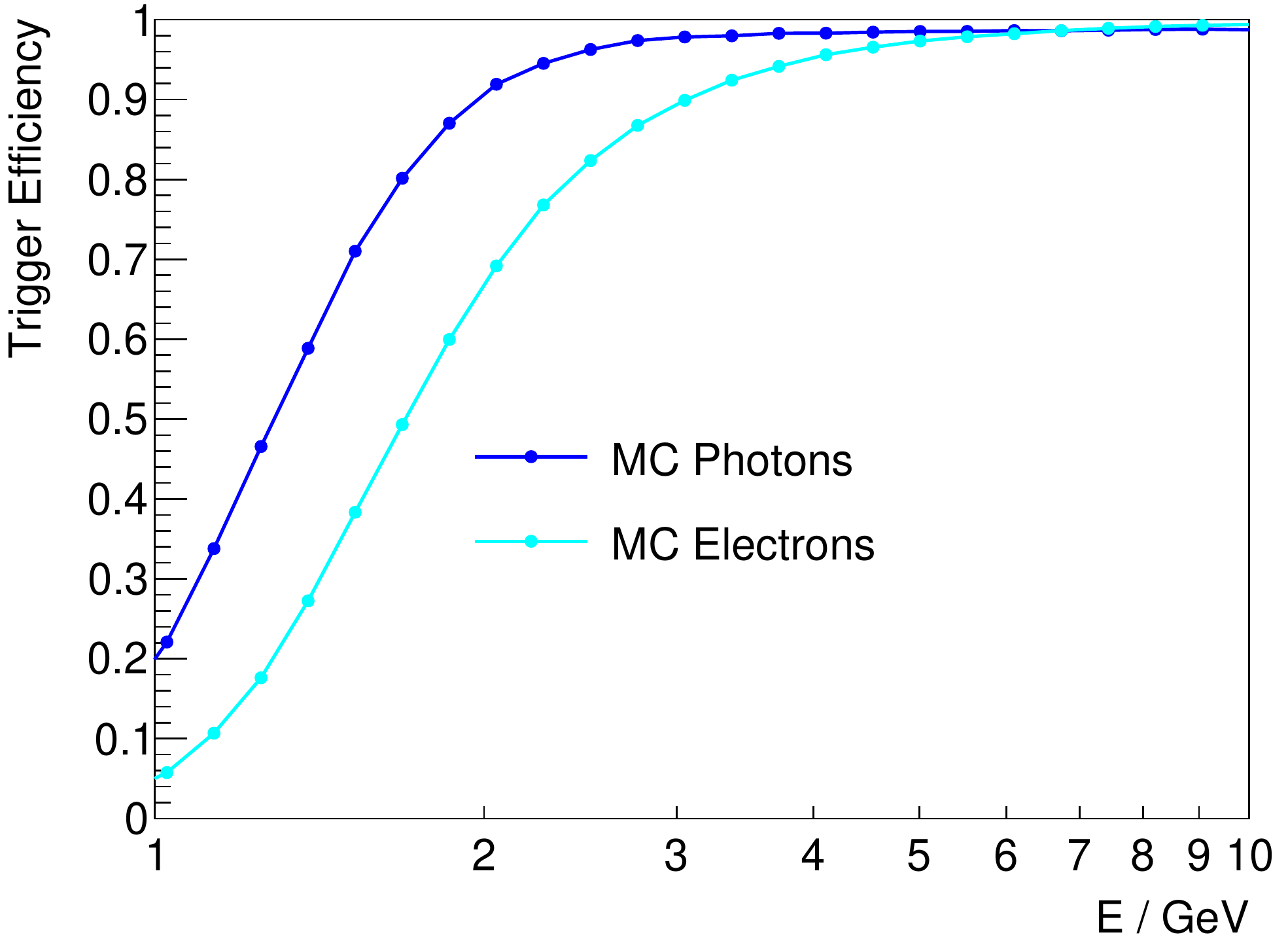}
  \end{minipage}
  \hspace{0.01\linewidth}
  \begin{minipage}{0.48\linewidth}
    \centering
    \includegraphics[width=1.0\linewidth]{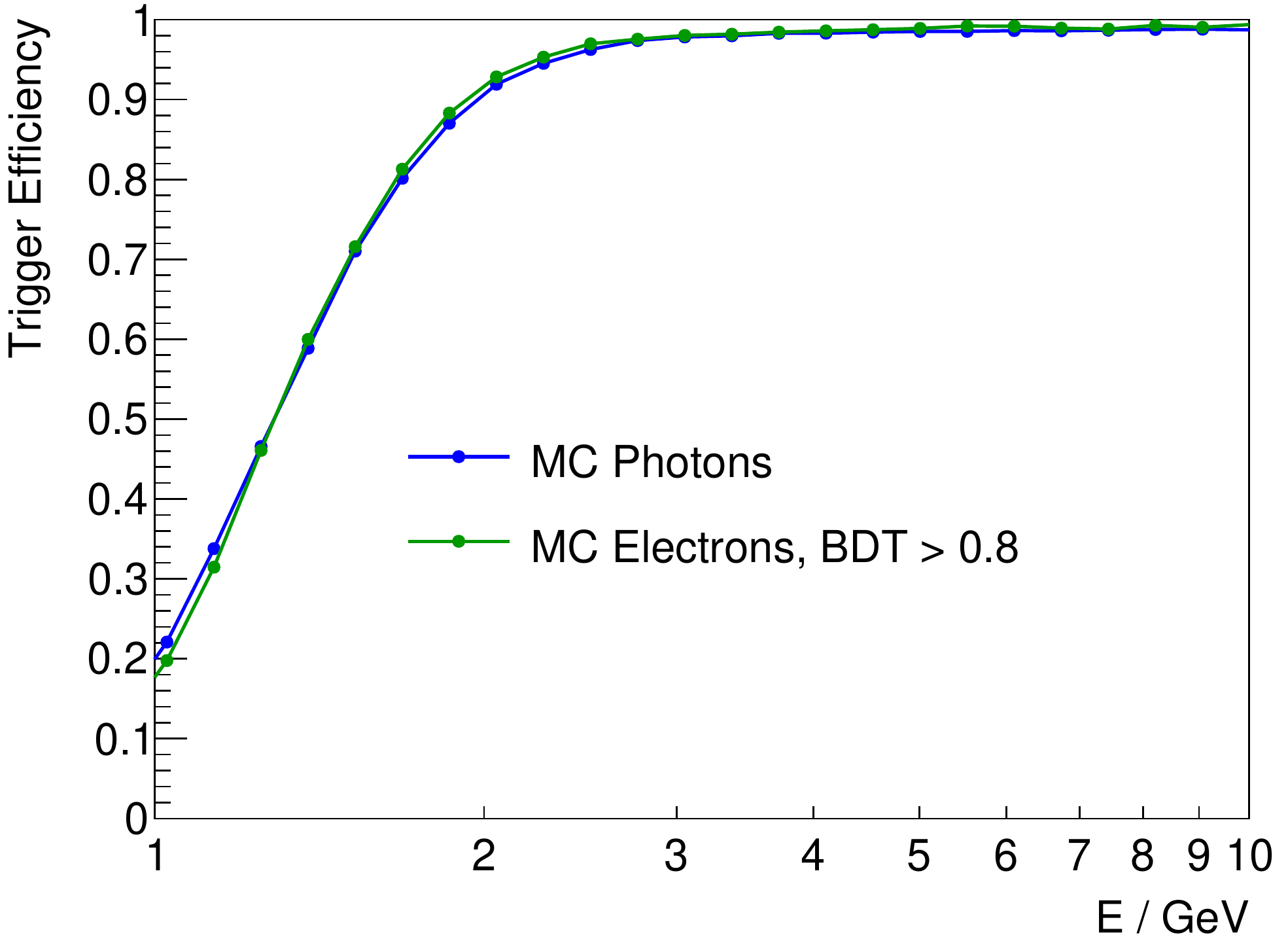}
  \end{minipage}
  \caption{Left: Comparison of the trigger efficiency derived from the simulation for photons (cyan)
    and electrons (blue). Right: Same comparison after applying a cut on the bremsstrahlung BDT.}
  \label{fig:ecal-trigger-correction-photons-electrons}
\end{figure}

The resulting trigger efficiency for Monte-Carlo electrons is shown together with the Monte-Carlo
photon trigger efficiency derived in section~\ref{sec:analysis-trigger-efficiency} on the left hand
side of figure~\ref{fig:ecal-trigger-correction-photons-electrons}. For both species the energy
quantity in the figure is the Monte-Carlo true energy at the top of the instrument. Contrary to the
initial assumption the two efficiencies do not match. The reason is that electrons can emit
energetic bremsstrahlung Photons in the material above the ECAL. If that happens the signal in the
calorimeter will look different: Instead of only a single shower, there will be two (partially)
overlapping showers with lower energy depositions in each one. In that case the two showers will not
penetrate as deeply into the calorimeter and the layer dependent energy thresholds in the lower
calorimeter layers are too high to accept such events, resulting in a lower trigger efficiency for
electrons compared to photons of the same energy.

Since the bremsstrahlung process is a discrete process there are two classes of electron events:
Those which radiate a hard photon and those which do not. In order to identify electrons which do
not undergo bremsstrahlung, or radiate only a small fraction of their energy, a dedicated boosted
decision tree (BDT) classifier was developed within the framework of the TMVA~\cite{TMVA_2007}
toolkit. Figures relating to the input variables and classifier output distribution of the BDT are
available in appendix~\ref{sec:appendix-bremsstrahlung-bdt}.

The BDT classifier uses shower shape variables, exploiting the fact that (at low energies) the
primary electrons and the bremsstrahlung photon are spatially displaced because of the magnetic
field above the calorimeter. In particular, this means that the single reconstructed shower is wider
in case two displaced particles entered the calorimeter. In addition the energy contained in the
shower center of gravity cell, and in various corridors around its axis will be lower, due to the
larger spread of the energy released in the ECAL. In contrast, an electron which enters the
calorimeter without emission of a photon will produce a rather well defined shower, with
characteristic shower shape properties.

The classifier also evaluates the longitudinal position of the shower maximum, which is expected to
be deeper in the calorimeter for electrons which did not emit a hard photon. Finally, the classifier
uses the ratio of measured energy to the absolute value of the rigidity: For electrons which radiate
a photon this ratio can be much larger than 1, depending on the position of the photon emission in
the upper detector.

Finally the TRD track is extrapolated through the magnetic field to the calorimeter surface and the
impact point and impact angle are compared to the tracker track extrapolation. In case a photon was
emitted there can be a sizable displacement (with a well defined sign) in the bending plane. The TRD
track extrapolation is also compared to the shower center of gravity position and to the shower axis
direction in the bending plane.

The signal sample for the BDT training consists of all electrons which retained more than
\SI{90}{\percent} of their energy just before entering into the calorimeter, according to the
Monte-Carlo truth. Conversely, the background sample consists of those electrons which radiated away
more than half of their energy before reaching the calorimeter surface.

After applying a cut on the BDT output variable of BDT > 0.8 the electron trigger efficiency can be
redetermined. This is shown on the right hand side of
figure~\ref{fig:ecal-trigger-correction-photons-electrons}. Overall the efficiency (green) now
matches well with the photon trigger efficiency (blue). The bremsstrahlung BDT classifier is also
used for ISS data, in order to remove electrons which radiated a hard photon.

The trigger efficiency correction must be a function of the true energy at the top of the
instrument, in order to apply it as a correction to the effective area. Unfortunately, the true
energy is not known for measured electrons in the ISS dataset. It is however possible to unfold the
event counts of electrons before determining the trigger efficiency. This requires an estimation of
the migration matrix for electrons, which can be done with the help of the Monte-Carlo. Because the
energy resolution is significantly different for events which have a positive calorimeter trigger
decision compared to those which do not, separate migration matrices need to be evaluated for the
``passed'' and for the ``all'' sample in the efficiency fraction. The unfolding method described in
section~\ref{sec:corrections-unfolding} can then be applied to correct both those samples.

\begin{figure}[t]
  \begin{minipage}{0.48\linewidth}
    \centering
    \includegraphics[width=1.0\linewidth]{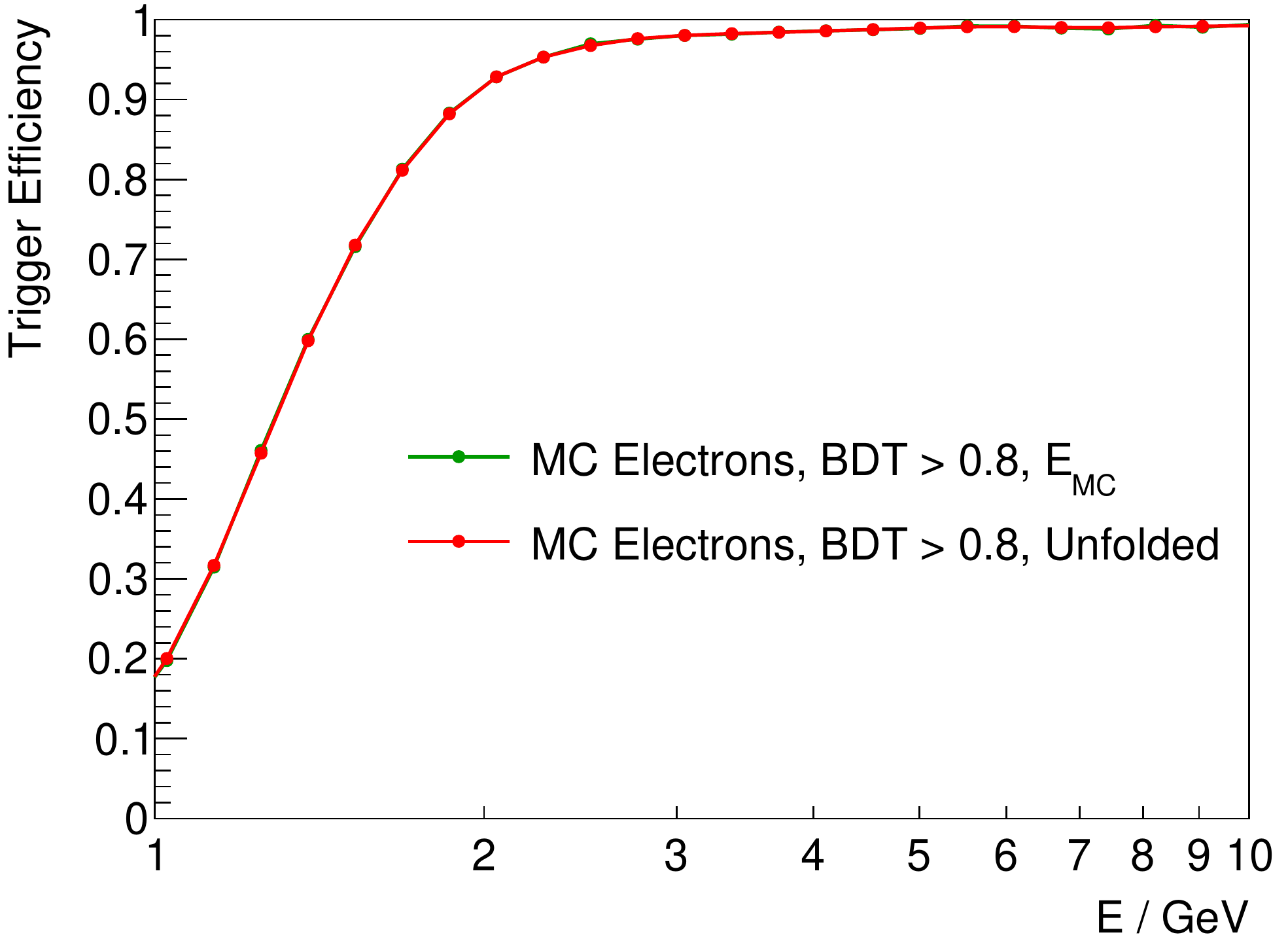}
  \end{minipage}
  \hspace{0.01\linewidth}
  \begin{minipage}{0.48\linewidth}
    \centering
    \includegraphics[width=1.0\linewidth]{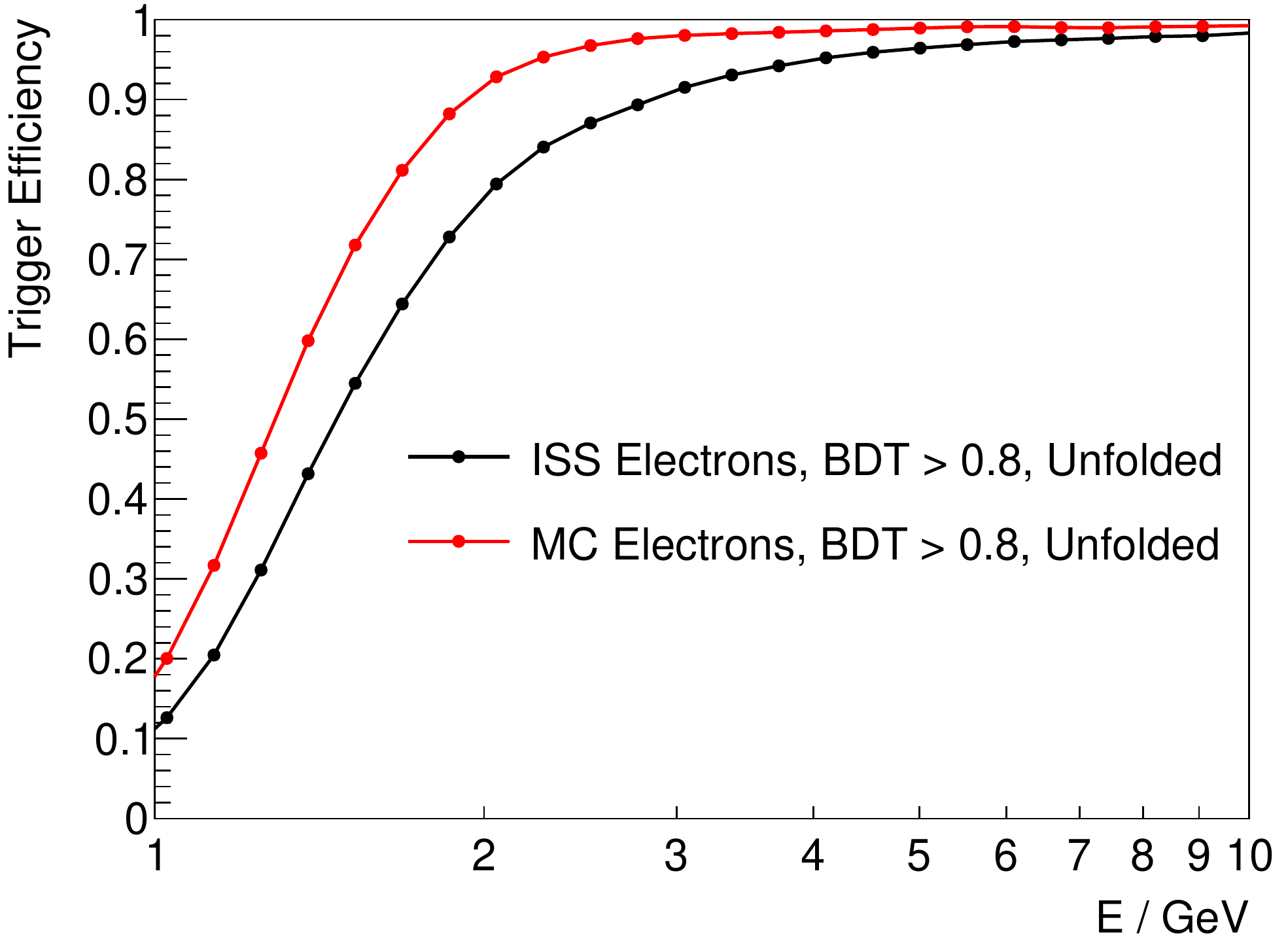}
  \end{minipage}
  \caption{Left: Comparison of the trigger efficiency derived from the simulation for electrons as a
    function of the true energy (green), and using the reconstructed energy and unfolding for
    migration (red). Right: Trigger efficiency as determined from ISS electrons (black) and MC
    electrons (red) after unfolding.}
  \label{fig:ecal-trigger-correction-electrons}
\end{figure}

The unfolding procedure was carried out for both the electrons in data and for those in the
Monte-Carlo. The left hand side of figure~\ref{fig:ecal-trigger-correction-electrons} shows the
trigger efficiency when determined as a function of the true electron energy and when determined as
a function of the reconstructed electron energy and unfolding the numerator and denominator event
counts separately, with the respective migration matrices. The two efficiencies match very well,
which validates the unfolding procedure.

The same technique is also used for electrons in the ISS dataset. The right hand side of
figure~\ref{fig:ecal-trigger-correction-electrons} shows the resulting trigger efficiency for
electrons which did not emit bremsstrahlung photons for both data and simulation after
unfolding. The data efficiency is significantly lower than the simulation, which illustrates the
need for a correction to the photon trigger efficiency.

\begin{figure}[t!]
  \centering
  \includegraphics[width=0.8\linewidth]{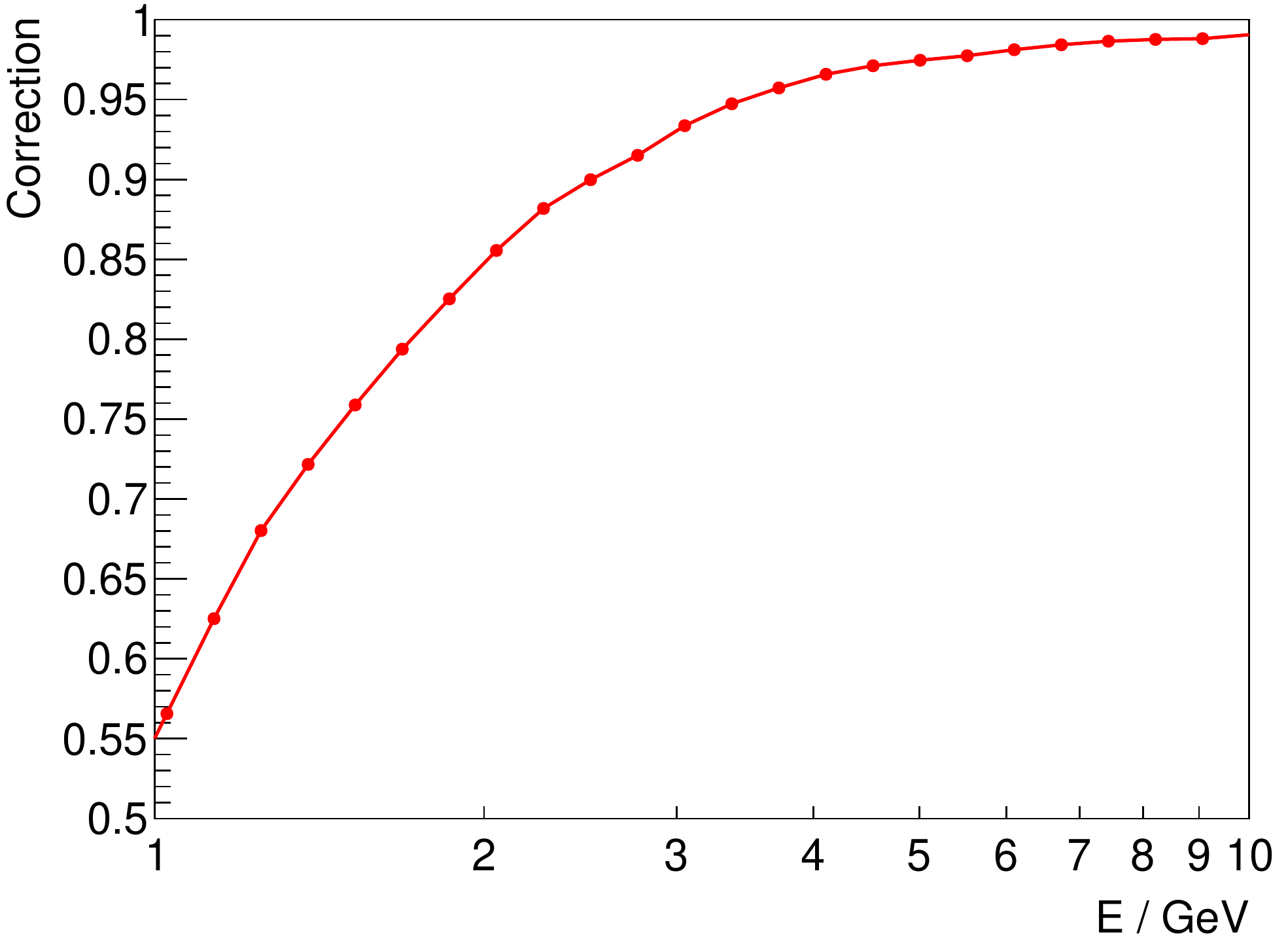}
  \caption{Correction to the calorimeter trigger efficiency defined as the ratio of ISS over MC
    efficiency for electrons.}
  \label{fig:ecal-trigger-correction}
\end{figure}

The final correction which is applied to the effective area is defined as the ratio of data over
simulation efficiency and shown in figure~\ref{fig:ecal-trigger-correction}. The correction is
sizable and is subsequently applied to the effective area of the calorimeter photon analysis.

\newpage
\section{Unfolding}
\label{sec:corrections-unfolding}

The imperfect energy resolution of the experiment results in energy bin-to-bin migration of events,
as discussed in section~\ref{sec:analysis-energy-resolution}. This migration can have a sizable,
systematic effect on the reconstructed flux if it is not corrected for. Mathematically the energy
density of the reconstructed event counts can be expressed as

\begin{equation}
  \label{eq:corrections-unfolding-forward-fold}
  \frac{\mathrm{d}N(E')}{\mathrm{d}E'} =
  \int_{-\infty}^{\infty}{p(E'|E) \frac{\mathrm{d}N(E)}{\mathrm{d}E} \mathrm{d}E} \,,
\end{equation}

where $E'$ is the reconstructed energy, $E$ is the true energy, $\mathrm{d}N/\mathrm{d}E$ is the
differential number of events and $p(E'|E)$ is the probability for an event with energy $E$ to be
reconstructed with energy $E'$. This is equivalent to forward folding of the true event counts with
the detector resolution to obtain the reconstructed event counts in each energy interval. The
reverse operation is the unfolding of the observed event counts with the resolution function, which
results in an estimation of the true distribution of events.

Although unfolding can be challenging and numerically unstable it is required in order to provide
results which are independent of the experimental setup. The alternative is to publish the results
as a function of the measured energy, together with the detector resolution function. Models then
need to be forward folded with the resolution function in order to compare them with the
experimental data. Such a procedure does not allow direct comparison between different experimental
results however. Therefore, unfolding should be applied to the data if possible. The procedure
employed here is based on an iterative Bayesian approach~\cite{Unfolding_Bayes_1995} and is
discussed in this section.

The binned equivalent of equation~(\ref{eq:corrections-unfolding-forward-fold}) is:

\begin{equation}
  \label{eq:corrections-unfolding-forward-fold-matrix}
  N'_i = \sum_{l=1}^{N_{\mathrm{bins}}}{P_{il} N_l} \,, \qquad i \in [1, N'_{\mathrm{bins}}] \,,
\end{equation}

where $N'_i$ is the number of observed events in bin $i$, $N_l$ is the true number of events in bin
$l$ and $P_{il}$ is the transition probability to migrate from bin $l$ to bin $i$. Note that in
general the binning of the true event counts and of the observed counts must not be the same and $P$
is not necessarily a square matrix, although that is often the case. The unfolding task is to find a
matrix $D$ which satisfies

\begin{equation}
  \label{eq:corrections-unfolding-unfolding-matrix}
  N_i \approx \sum_{l=1}^{N'_{\mathrm{bins}}}{D_{il} N'_l} \,, \qquad i \in [1, N_{\mathrm{bins}}] \,.
\end{equation}

The matrix $D$ implements the unfolding. The matrix $D$ is in general not the inverse of the matrix
$P$, since $P$ can be singular or even not a square matrix so that the inverse is not required to
exist. In addition, the columns of $P$ correspond to probabilities. By the same logic the columns of
$D$ should be probabilities: For an event observed in a given energy bin $j$ there exists a
(non-negative) probability that it originated from bin $k$ for every $k$ and the sum of all these
probabilities should be identical to unity. These properties are not fulfilled by the mathematical
inverse of $P$.

Instead the matrix $D$ is estimated in an iterative Bayesian procedure, as described in
ref. ~\cite{Unfolding_Bayes_1995}:

\begin{align}
  \label{eq:corrections-unfolding-unfolding-matrix-calculation}
  D^{k}_{ij} &= \frac{P(E'_j|E_i) P^{k-1}(E_i)}{\left[\sum_{l=1}^{N'_{\mathrm{bins}}}{P(E'_l|E_i)}\right]
               \left[\sum_{l=1}^{N_{\mathrm{bins}}}{P(E'_j|E_l) P^{k-1}(E_l)}\right]} \,,\\
  N^{k}_i &= \sum_{l=1}^{N'_{\mathrm{bins}}}{D^k_{il} N'_l} \,,\\
  P^{k}(E_i) &= \frac{N^k_i}{\sum_{l=1}^{N_{\mathrm{bins}}}{N^k_l}} =
               \frac{1}{N_{\mathrm{true}}} \sum_{l=1}^{N'_{\mathrm{bins}}}{D^k_{il} N'_{l}} \,,
\end{align}

where $k \in [1, N_{\mathrm{iter}}]$ and $P(E'_j|E_i) = P_{ji}$ are the elements of the migration
matrix and $P^0(E_i)$ is the initial probability for an event to be in energy bin $i$ according to
an initial guess of the true distribution, which must be specified from the outside. In the absence
of any a priori knowledge a flat distribution can be used. The quantities $D$ and $N_i$ in
equation~(\ref{eq:corrections-unfolding-unfolding-matrix}) correspond to the final iterations:
$D \vcentcolon = D^{N_{\mathrm{iter}}}, N_i \vcentcolon = N^{N_{\mathrm{iter}}}_i$.

The covariance matrix $U(N)$ of the reconstructed true distribution can be estimated by:

\begin{equation}
  \label{eq:corrections-unfolding-covariance-matrix}
  U(N)_{ij} =
  \sum_{k=1}^{N'_{\mathrm{bins}}}{\sum_{l=1}^{N'_{\mathrm{bins}}}{\tilde{D}_{ik} U(N')_{kl} \tilde{D}^T_{lj}}} \,,
  \qquad
  i,j \in [1, N_{\mathrm{bins}}] \,.
\end{equation}

The matrix $\tilde{D}$ is the matrix which implements the error propagation
($\tilde{D}_{ij} = \mathrm{d}N_i/\mathrm{d}N'_j$) and $U(N')$ is the covariance matrix of the
observed event counts. The matrix $\tilde{D}$ is not the same as $D$ because the observed event
counts were used to derive $D$ in all but the first iteration, which has to be taken into account
when differentiating equation~(\ref{eq:corrections-unfolding-unfolding-matrix}) with respect to
$N'_j$. In addition it is optionally possible to consider the contribution to $U(N)_{ij}$ due to
uncertainties related to the migration matrix itself, such as the limited statistics with which the
migration matrix was estimated from the Monte-Carlo simulation. It is important to note that the
matrix $U(N)$ is usually not diagonal even if $U(N')$ is diagonal. This is because the unfolding
mixes events from several of the observed energy bins, which correlates their uncertainties.

In this analysis the number of iterations is equal to 3. The initial guess of the distribution
$N^0(E)$ is based on the combined model of diffuse emission and source photons in the inner galaxy
($\SI{-20}{\degree} < l < \SI{80}{\degree}$, $\SI{-8}{\degree} < b < \SI{8}{\degree}$) as
constructed in chapter~\ref{sec:modeling}. When unfolding the event counts in a given part of the
sky the events are first summed over all the pixels in the region of interest and then
unfolded. This procedure yields the matrices $D$ and $\tilde{D}$, which implement the unfolding and
the corresponding error propagation. Using these matrices it is possible to estimate the unfolded
distribution in the individual pixels (indexed by $p$) of the region of interest:

\begin{align}
  \label{eq:corrections-unfolding-counts-per-pixel}
  N_{ip} &\approx \sum_{l=1}^{N'_{\mathrm{bins}}}{D_{il} N'_{lp}} \,,\\
  \label{eq:corrections-unfolding-counts-per-pixel-cov}
  U(N)_{ijp} &\approx \sum_{k=1}^{N'_{\mathrm{bins}}}{\sum_{l=1}^{N'_{bins}}{\tilde{D}_{ik} U(N')_{klp}
               \tilde{D}^T_{lj}}} \,.
\end{align}

The matrix $U(N')_{ijp}$ is assumed to be diagonal, the counts follow a Poisson distribution:
$U(N')_{ijp} = \delta_{ij} \sqrt{N'_{ip}}$. These equations constitute an approximation. In general
it would be necessary to unfold each pixel independently, so that the matrices $D$ and $\tilde{D}$
would vary from pixel to pixel. Given the limited statistics in the individual pixels such a
procedure is not practical, though. Instead it will be shown that the approximation is valid with
the help of a suitable toy Monte-Carlo study in the following.

Computing the photon flux and its covariance for each pixel in the region of interest is then
straightforward:

\begin{align}
  \label{eq:corrections-unfolding-flux}
  \Phi_{ip} &= \frac{N_{ip}}{C_{ip}} \,, \\
  \label{eq:corrections-unfolding-flux-cov}
  U(\Phi)_{ijp} &= \frac{U(N)_{ijp}}{C_{ip} C_{jp}} \,,
\end{align}

with the pixel-dependent count to flux conversion factor $C_{ip} = \mathcal{E}_{ip} \epsilon_{i}
\Delta E_{i} \Delta \Omega_p$. Here $\mathcal{E}_{ip}$ is the exposure in the given energy bin and
pixel, $\epsilon_i$ is the trigger efficiency, $\Delta E_i$ is the energy bin width and $\Delta
\Omega_p$ is the solid angle subtended by the pixel $p$. The average photon flux in the region of
interest is:

\begin{align}
  \label{eq:corrections-unfolding-average-flux}
  \Phi_{i} &= \sum_{p=1}^{N_{\mathrm{pixel}}}{\Phi_{ip} \frac{\Delta\Omega_p}{\Delta\Omega_{\mathcal{W}}}} \,, \\
  \label{eq:corrections-unfolding-average-flux-cov}
  U(\Phi)_{ij} &= \sum_{p=1}^{N_{\mathrm{pixel}}}{U(\Phi)_{ijp}
                 \left(\frac{\Delta\Omega_p}{\Delta\Omega_{\mathcal{W}}}\right)^{2}} \,,
\end{align}

where $\Delta \Omega_{\mathcal{W}}$ is the total solid angle subtended by the window on the
sphere. If required these quantities can be rebinned in energy into a coarser binning:

\begin{align}
  \label{eq:corrections-unfolding-average-flux-rebinned}
  \tilde{\Phi}_{i} &= \sum_{l=i_{1}}^{i_{N}}{\Phi_{l} \frac{\Delta E_{l}}{\Delta \tilde{E}_i}} \,, \\
  \label{eq:corrections-unfolding-average-flux-rebinned-cov}
  U(\tilde{\Phi})_{ij} &= \sum_{l=i_{1}}^{i_{N}}{\sum_{m=j_{1}}^{j_{M}}{U(\Phi)_{lm}
                         \frac{\Delta E_{l}\Delta E_{m}}{\Delta \tilde{E}_i \Delta \tilde{E}_j}}} \,.
\end{align}

Here the merged bin $i$ (or $j$) is the union of the original bins $i_{1}, ..., i_{N}$, $\Delta
\tilde{E}_i$ is the energy bin width of the merged bin and $\tilde{\Phi}_i$ is the average photon
flux in the merged bin.

As discussed in section~\ref{sec:analysis-energy-resolution} the energy resolution in the vertex
analysis is significantly worse than that of the calorimeter for the ECAL photon analysis. While
unfolding needs to be applied in both cases, the correction is much more important in the vertex
case. The following figures and studies will therefore focus on the vertex analysis, but they were
also carried out for the calorimeter photons. The complementary set of figures is available in
appendix~\ref{sec:appendix-unfolding-ecal}.

The primary ingredient for the unfolding is the migration matrix. It is directly related to the
energy resolution shown in figure~\ref{fig:energy-resolution}. This matrix must be estimated from
the Monte-Carlo simulation since the true energy is needed for each event.

\begin{figure}[t]
  \centering
  \includegraphics[width=0.8\linewidth]{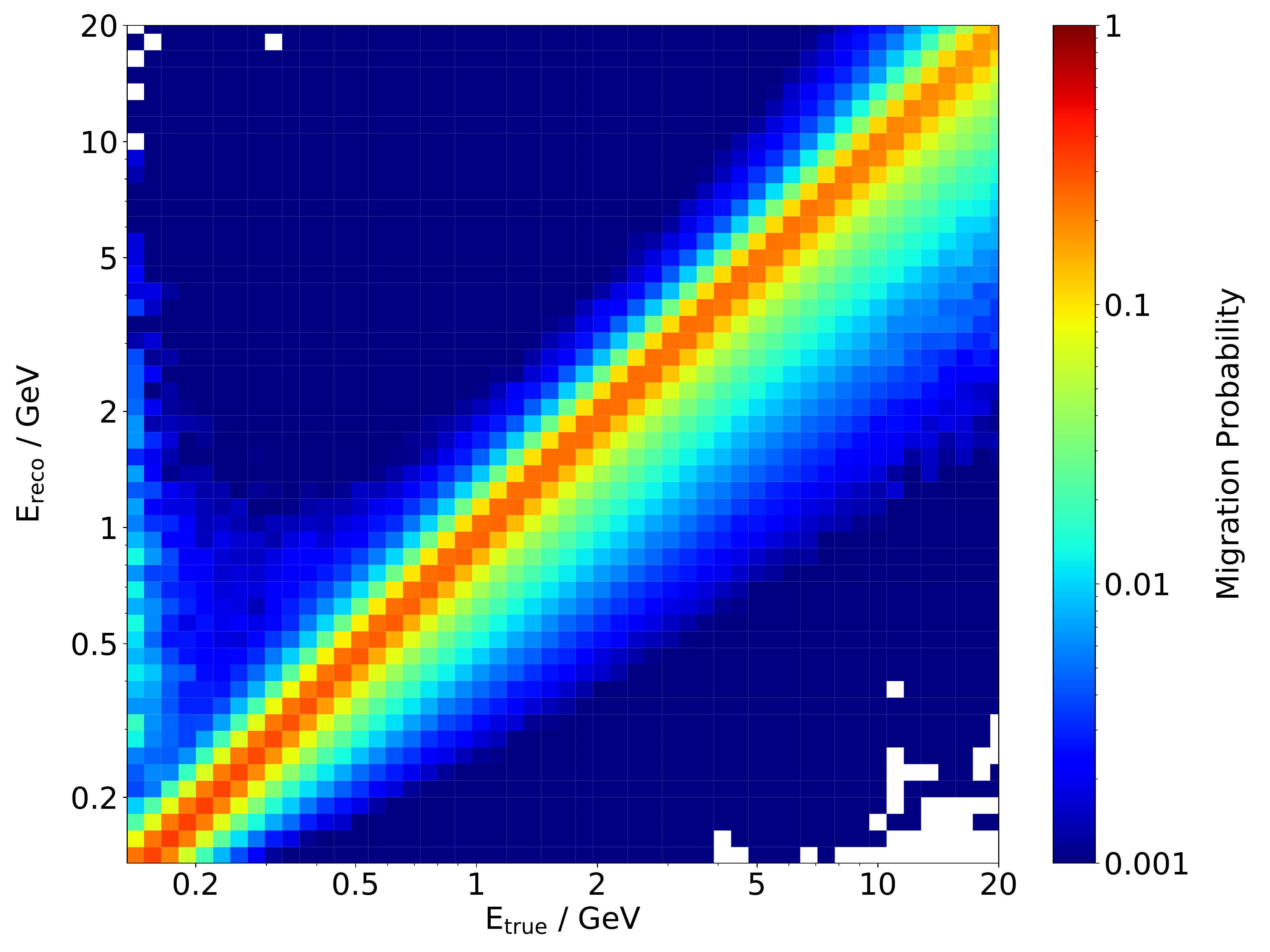}
  \caption{Migration matrix for the vertex analysis, used in the unfolding procedure.}
  \label{fig:unfolding-correction-migration-matrix}
\end{figure}

The migration matrix for the vertex analysis is shown in
figure~\ref{fig:unfolding-correction-migration-matrix}. The energy binning is equidistant in
$\log{E}$ and spans 100 bins from \SI{50}{\mega\electronvolt} to \SI{1}{\tera\electronvolt} on both
axes, but only the relevant portion of the matrix is shown in the figure. In this binning most of
the events are reconstructed on the diagonal: These are events which are reconstructed in the
correct energy bin. However, due to the imperfect resolution there are also substantial
contributions to the neighboring bins, in particular for $E_{\mathrm{reco}} < E_{\mathrm{true}}$.

In order to check the validity of the unfolding procedure a toy Monte-Carlo study was performed in
the following way:

\begin{enumerate}
\item The model of the expected counts based on the galactic diffuse emission and the source catalog
  is used as the ``true photon distribution''.
\item For each toy experiment random event counts are sampled for each energy bin and pixel, based
  on independent Poisson distributions with their means given by the model counts in that energy bin
  and pixel.
\item For each ``event'' the energy is smeared according to the migration matrix by randomly
  assigning a new energy bin, with the probabilities given by the migration matrix. There is no
  pixel to pixel migration, since smearing according to the PSF is already part of the model
  construction. As a result of this step the simulated ``measured'' distribution is obtained in each
  pixel ($N'_{ip}$).
\item The counts in the pixels in the region of interest are summed to obtain the summed spectrum as
  a function of energy.
\item The summed spectrum is unfolded according to
  equations~(\ref{eq:corrections-unfolding-unfolding-matrix})
  and~(\ref{eq:corrections-unfolding-unfolding-matrix-calculation}), which yields the unfolding
  matrix $D$ and the error propagation matrix $\tilde{D}$.
\item The unfolding procedure is applied to each individual pixel as described in
  equations~(\ref{eq:corrections-unfolding-counts-per-pixel})
  and~(\ref{eq:corrections-unfolding-counts-per-pixel-cov}).
\item The average (rebinned) photon flux and its covariance matrix are computed according to
  equations~(\ref{eq:corrections-unfolding-flux})
  to~(\ref{eq:corrections-unfolding-average-flux-rebinned-cov}).
\item The reconstructed event counts and photon flux are compared with their true distributions.
\end{enumerate}

In total 10000 toy experiments were performed. For simplicity the following figures will focus on
the unfolded event counts rather than on the unfolded flux and compare them to the true
distributions. The same set of figures was studied by looking at the photon flux as the final
quantity. In all of the cases shown there is no visible difference in the figures and all of the
statements translate to the flux without restrictions.

\begin{figure}[t]
  \centering
  \includegraphics[width=0.8\linewidth]{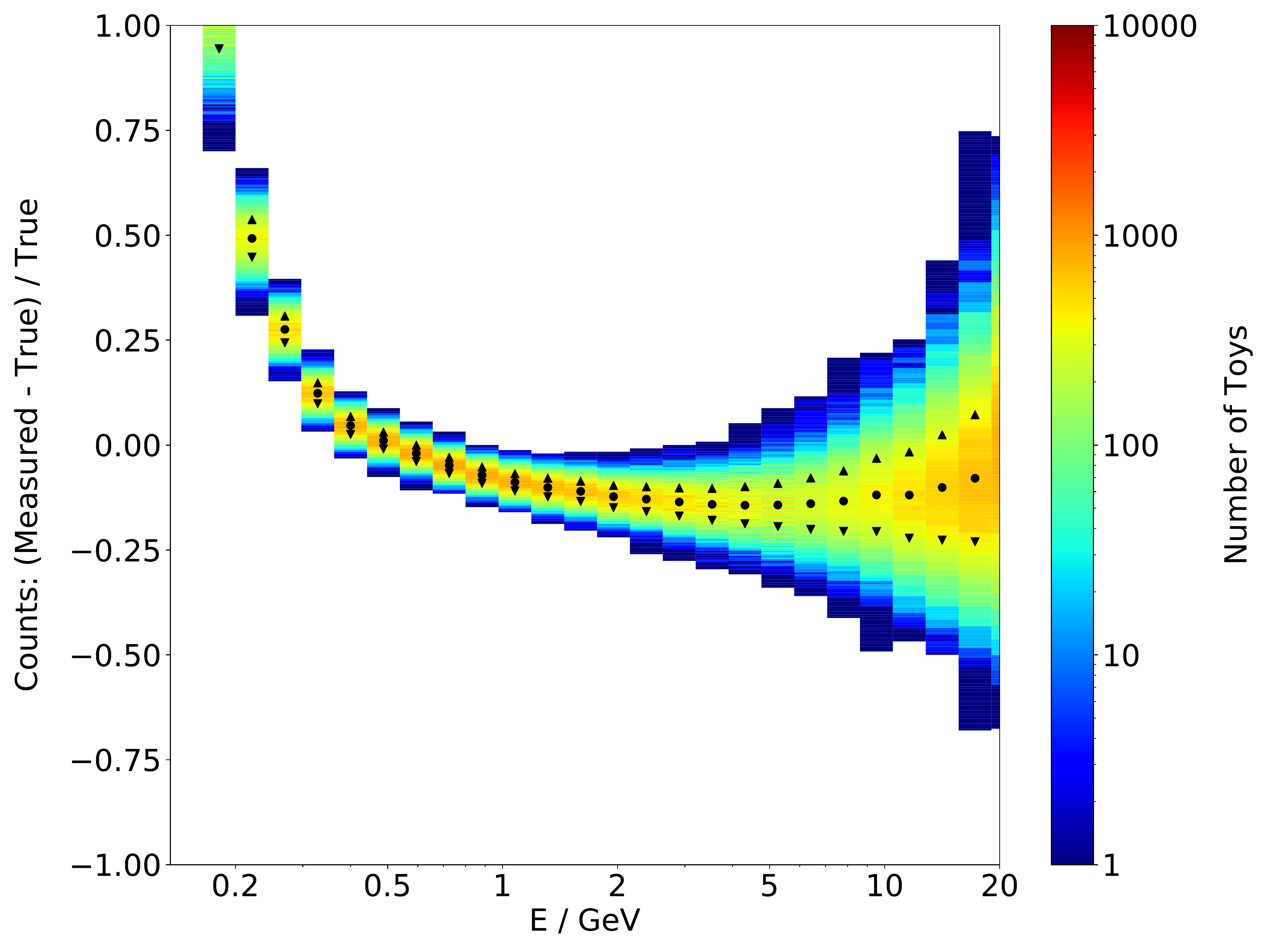}
  \caption{Distribution of the relative difference between the measured event counts and true
    average counts for 10000 toy experiments in the vertex analysis. The black circles correspond to
    the mean in each vertical slice, the triangles corresponds to the mean $\pm$ RMS position.}
  \label{fig:unfolding-correction-difference-measured-true}
\end{figure}

Figure~\ref{fig:unfolding-correction-difference-measured-true} shows the distribution of the
relative difference between the measured (toy) counts and the true average counts in each rebinned
energy bin for the vertex analysis. The distribution illustrates the magnitude of the effect of the
migration. Across all energies the measured distribution systematically differs from the true
distribution. At low energies the measured distribution is too high by almost a factor of 2 and for
energies above approximately \SI{500}{\mega\electronvolt} it is too low, by up to
\SI{15}{\percent}. The width of the distribution in each vertical slice corresponds to the spread of
the different toy experiments, which corresponds to the size of the fluctuations due to the given
statistics. The observed bias is larger than the statistical uncertainty in almost all of the
bins. An adequate unfolding correction is therefore required.

\begin{figure}[t]
  \centering
  \includegraphics[width=0.8\linewidth]{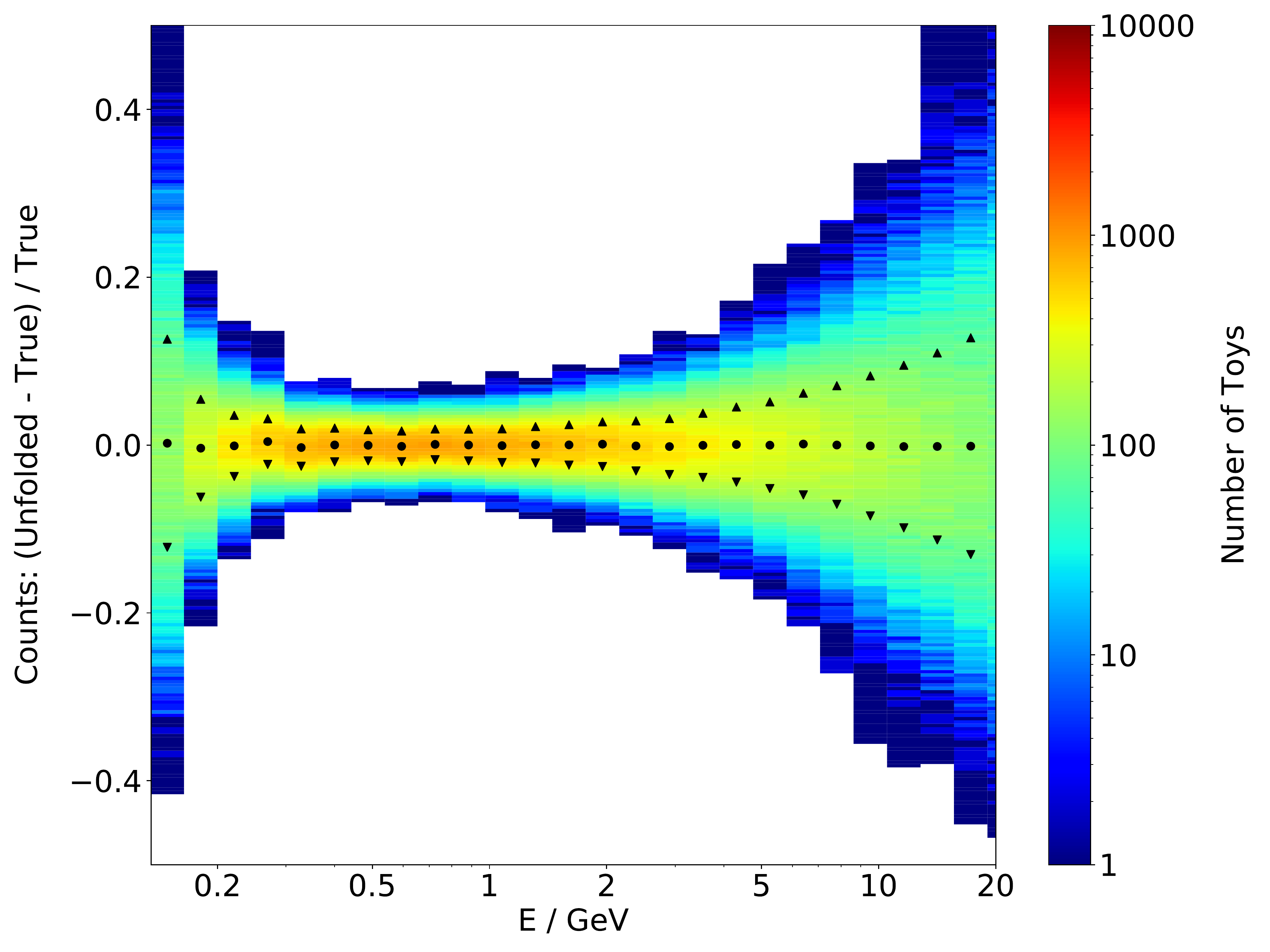}
  \caption{Distribution of the relative difference between the unfolded event counts and true
    average counts for 10000 toy experiments in the vertex analysis. The black markers correspond to
    the mean and RMS positions.}
  \label{fig:unfolding-correction-difference-unfolded-true}
\end{figure}

Figure~\ref{fig:unfolding-correction-difference-unfolded-true} shows the distribution of the
relative difference between the unfolded counts and true average counts for the same set of toy
experiments. After the unfolding the reconstructed distribution no longer systematically differs
from the true distribution. The mean value is compatible with zero at the permille level, except for
the lowest energy bin which is excluded in the final analysis.

This result shows that the unfolding procedure is able to correct for the systematic effects
introduced by the migration. It also shows that there is no inherent bias in the procedure outlined
above. In particular, because the unfolded flux is also bias free on average, the approximation to
use the same matrix $D$ for all the pixels is justified (as long as one is not actually interested
in the unfolded flux of individual pixels).

\begin{figure}[t]
  \centering
  \includegraphics[width=0.8\linewidth]{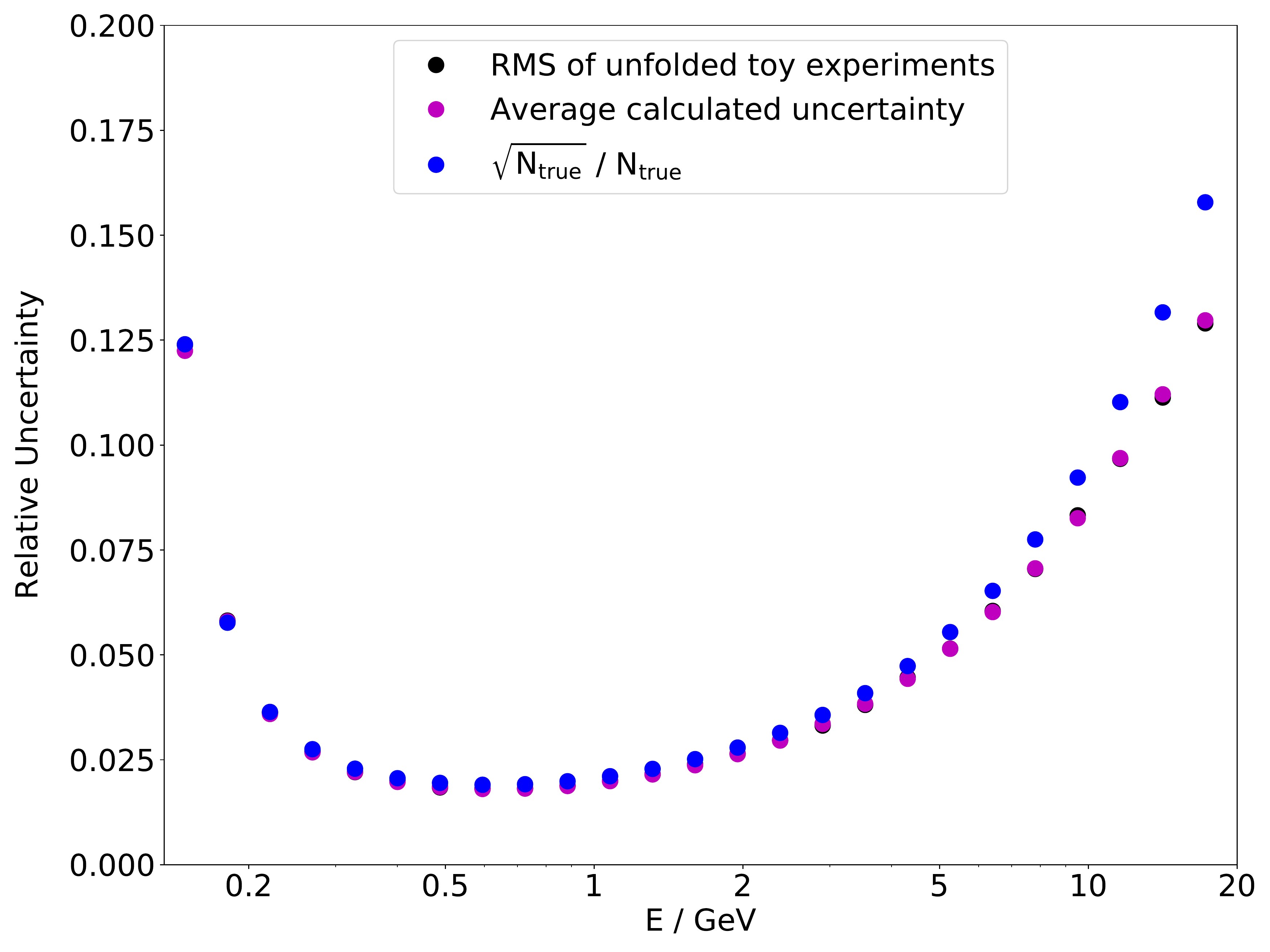}
  \caption{Relative uncertainty of the unfolded event counts compared with the inherent statistical
    uncertainty of the true distribution.}
  \label{fig:unfolding-correction-uncertainty}
\end{figure}

In order to understand the uncertainties of the unfolded result it is useful to compare the
variation of the unfolded toy results with the expectation for the statistical fluctuations
according to the true distribution. This is done in
figure~\ref{fig:unfolding-correction-uncertainty}. The blue markers correspond to the expected
relative statistical uncertainty of the true average counts without any migration. This uncertainty
is based on the assumption that the counts in individual bins fluctuate according to a Poisson
distribution. The black markers were derived from the width of the distribution of the unfolded toy
experiment results in each rebinned energy bin. Finally, the magenta markers correspond to the
uncertainties obtained from the diagonal elements of the covariance matrix as calculated by error
propagation. The covariance matrix used is the average of the individual matrices obtained in each
toy experiment. The variations of the obtained matrices from toy to toy are small.

The uncertainties obtained by error propagation agree with the observed variation of the toy
results. Both of the uncertainties are at high energies smaller than the expected statistical
uncertainty of the true average counts, which is due to the fact that the unfolding mixes events
from neighboring bins in such a way that the reconstructed counts in each bin are actually
calculated from a larger sample size, which implies an effect that is similar to averaging. However,
this does not mean that the actual uncertainties decrease due to the unfolding. Instead bins have a
non-zero correlation with their neighbors after the unfolding, which has to be taken into account
when the unfolded result is used in further analyses such as model fits or when the distribution is
rebinned.

\begin{figure}[t]
  \centering
  \includegraphics[width=0.9\linewidth]{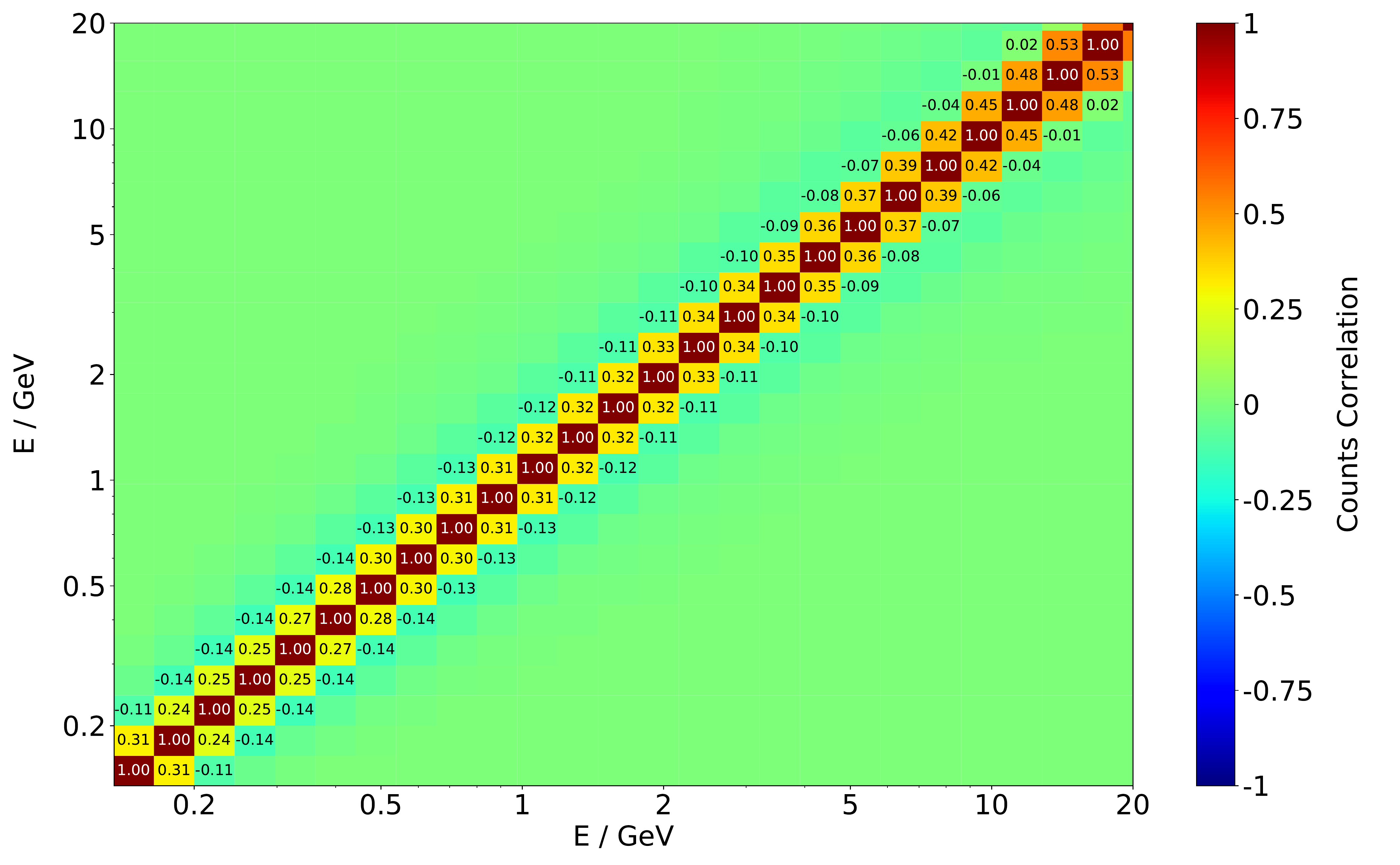}
  \caption{Correlation matrix of the unfolded counts.}
  \label{fig:unfolding-correction-correlation-matrix}
\end{figure}

In order to study the correlation coefficients between neighboring bins it is useful to compute the
average covariance matrix of the unfolded flux (see
equation~(\ref{eq:corrections-unfolding-average-flux-rebinned-cov}). The correlation matrix is
defined based on the covariance matrix as

\begin{displaymath}
  \rho_{ij} = \frac{\sigma^2_{ij}}{\sigma_{i}\sigma_{j}} \in [-1, 1] \,.
\end{displaymath}

The correlation matrix for the unfolded counts is shown in
figure~\ref{fig:unfolding-correction-correlation-matrix}. Between \SI{200}{\mega\electronvolt} and
\SI{10}{\giga\electronvolt} the direct neighbors of each bin show a positive correlation of
\SI{25}{\percent} to \SI{40}{\percent} whereas the neighbors one bin further away show a slight
anti-correlation of up to -\SI{14}{\percent} at low energies.

The correctness of the covariance matrix off-diagonal elements was crosschecked by calculating the
bin-to-bin correlations from the sample of reconstructed toy fluxes:

\begin{displaymath}
  \sigma^2_{ij,\mathrm{sample}} = \frac{1}{N_{\mathrm{toys}} - 1} \sum_{t=1}^{N_{\mathrm{toys}}}{\left(N_i^t
      - \bar{N_i}\right) \left(N_j^t - \bar{N_j}\right)} \,.
\end{displaymath}

Here $N_i^t$ are the unfolded event counts in energy bin $i$ for the toy experiment $t$ and
$\bar{N_i}$ are the average unfolded event counts in energy bin $i$ across all toys. The sample
covariance matrix matches the covariance matrix calculated by
equation~(\ref{eq:corrections-unfolding-covariance-matrix}) within a few percent.

\section{Systematic Uncertainties}
\label{sec:analysis-systematic-uncertainties}

The estimation of systematic uncertainties for the two analyses is not a straightforward task,
because the same event is only ever measured by one sub-detector at a time.

In AMS it is customary to use the tag and probe method to construct a sample of signal events
without using the subdetector under study, for example, a sample of electrons can be selected by the
calorimeter and TRD on which efficiency of the selection requirements regarding the tracker can be
measured in data. The same procedure can be applied to the simulation. The difference between the
two results can be used to correct the simulation and to estimate the systematic uncertainty on the
selection efficiency.

Unfortunately this method is not available here. For example, it is impossible to select a sample of
photons which do not convert in the upper detector for the estimation of calorimeter selection
efficiencies without using the calorimeter itself. For the conversion mode analysis similar
arguments apply, since the only part of AMS which is involved in the measurement is the tracker.

However, the measurement of photons is similar in many ways to other analyses in AMS where the
relevant instrumental effects have been studied in detail.

For the conversion mode analysis the following systematic uncertainties are relevant:

\begin{itemize}
\item \textbf{Differences in selection efficiency between data and simulation}\\
  The differences in the tracker selection cut efficiencies for electron and positron selection were
  studied in detail in the analysis of electrons and
  positrons~\cite{AMS02_ElecPos_2014,ZimmermannPHD_2019}. The agreement between data and Monte-Carlo
  is at the \SI{2}{\percent} to \SI{3}{\percent} level~\cite{ZimmermannPHD_2019}, for the full
  electron selection.

  However, most of the selection cuts which show the largest difference between data and simulation
  are irrelevant in the analysis of converted photons (for example, the matching between tracker and
  ECAL shower, the matching between energy and rigidity, the number of active TRD layers and the
  existence of a TRD track). Therefore the systematic uncertainty related to the agreement between
  data and simulation is estimated to be \SI{1}{\percent} in this analysis and affects the effective
  area.

\item \textbf{Trigger efficiency (TOF)}\\
  The uncertainty on the TOF trigger decision with 4/4 ACC veto was found to be negligible (less
  than \SI{4}{\permille}) in the proton analysis~\cite{AMS02_Proton_2015} in the relevant energy
  range. The converted photon analysis uses the same trigger, so equal arguments apply.

\item \textbf{Unfolding (tracker)}\\
  The uncertainty on the unfolding arises from two separate effects. The first one is the stability
  of the unfolding method. This effect was shown to be minor in the study in
  section~\ref{sec:corrections-unfolding}. The observed bias is at the \SI{5}{\permille} level for
  all relevant energies.

  The second one is the knowledge of the migration matrix which directly corresponds to the
  description of the tracker rigidity resolution in the Monte-Carlo simulation. The tracker
  resolution was studied in detail in the AMS analysis of protons~\cite{AMS02_Proton_2015}, where
  the agreement between data and simulation was shown to be excellent even for
  \SI{400}{\giga\electronvolt} particles, so no further contribution to the uncertainty is
  considered.

\item \textbf{Absolute tracker rigidity scale}\\
  The uncertainty on the absolute rigidity scale is determined by two effects. The first one is
  residual misalignment of the tracker, in particular the external layers. The absolute rigidity
  scale uncertainty due to this effect was estimated to be approximately
  \SI{1/30}{\per\tera\volt}~\cite{AMS02_Detector_RigidityScale}, which shows that this effect is
  only important at the highest energies. Also, the conversion mode photon analysis does not make
  use of external tracker layers.

  The second part of the uncertainty arises from the knowledge of the magnetic field. This
  uncertainty is \SI{0.25}{\percent} (absolute) and \SI{0.1}{\percent} (temperature
  correction)~\cite{AMS02_Proton_2015} and is the major part of the rigidity scale uncertainty.

  It is important to point out that this uncertainty does not translate directly to an uncertainty
  in the measured flux, instead the spectral shape of the flux must be considered when calculating
  the effect of the rigidity scale uncertainty.
\end{itemize}

Overall the total systematic uncertainty for the conversion mode analysis is \SI{1.2}{\percent},
which is the quadratic sum of the contributions listed above and is dominated by the uncertainty of
the effective area due to differences in data and simulation. The rigidity scale uncertainty of
\SI{0.27}{\percent} is considered separately, since its effect depends on the spectral index of the
flux.

\begin{figure}[t]
  \centering
  \includegraphics[width=0.75\linewidth]{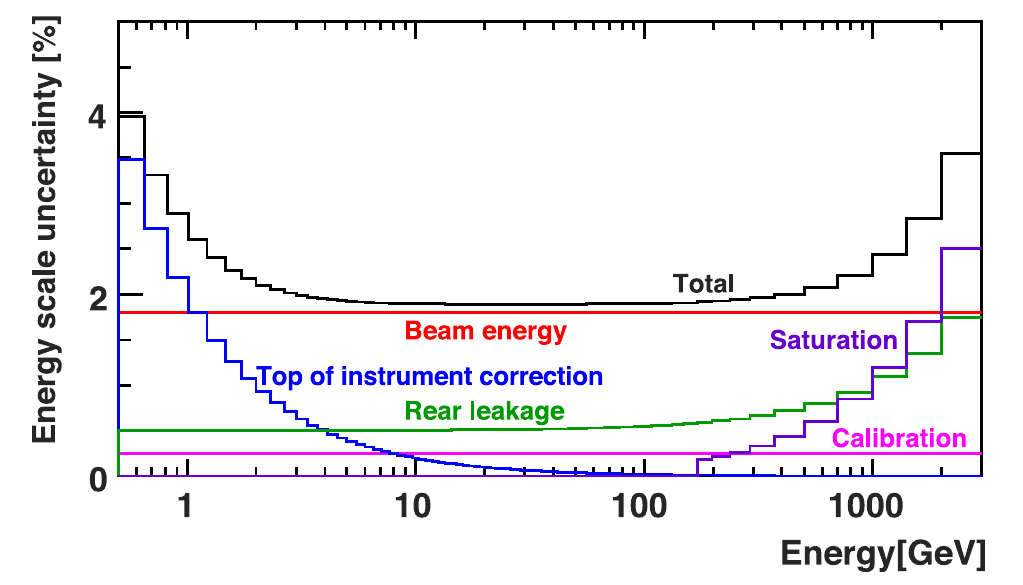}
  \caption{Uncertainty on the ECAL absolute energy scale for the measurement of electrons, together
    with the various components of the uncertainty~\cite{AMS02_Detector_Ecal3D}.}
  \label{fig:ecal-absolute-energy-scale}
\end{figure}

For the calorimeter mode analysis the following systematic uncertainties are relevant:

\begin{itemize}
\item \textbf{Differences in selection efficiency between data and simulation}\\
  The calorimeter electromagnetic shower shape description in the simulation was found to be
  good. This has been verified in comparisons with both beamtest and ISS
  data~\cite{AMS02_Detector_ECAL,AMS02_Detector_Ecal3D,ZimmermannPHD_2019}.

  In the analysis of electrons and positrons, no large discrepancies between the selection
  efficiencies relating to the ECAL selection cuts were identified as part of the acceptance
  uncertainty~\cite{ZimmermannPHD_2019}. The main related systematic uncertainty is given by the
  uncertainty on the ECAL electron likelihood estimator, which was found to be on the permille
  level, except at energies above \SI{200}{\giga\electronvolt} where the cut on the ECAL estimator
  tightens in the lepton analysis~\cite{ZimmermannPHD_2019}.

  In the case of the photon analysis there is no such tightening of the cut value and the estimation
  of the associated systematic uncertainty is \SI{1}{\percent}.

\item \textbf{Trigger efficiency (ECAL)}\\
  The ECAL photon trigger efficiency was corrected for differences between data and simulation in
  section~\ref{sec:corrections-ecal-trigger}. The estimated uncertainty on the derived correction
  factor is \SI{2}{\percent}, which was determined by studying the stability of the method.

\item \textbf{Unfolding (ECAL)}\\
  This effect was studied in detail in the lepton analysis~\cite{ZimmermannPHD_2019}. The associated
  systematic uncertainty was shown to be important at energies below \SI{1}{\giga\electronvolt}
  which are not accessible in the calorimeter mode photon analysis, due to the inefficiency of the
  ECAL trigger. Above \SI{1}{\giga\electronvolt} the systematic uncertainty due to the calorimeter
  energy unfolding is at the permille level and only reaches \SI{1}{\percent} at
  \SI{1}{\tera\electronvolt} energy.

\item \textbf{Absolute ECAL energy scale}\\
  The knowledge of the absolute ECAL energy scale is one of the most important effects. It was
  studied using electrons in a separate publication~\cite{AMS02_Detector_Ecal3D} and is shown in
  figure~\ref{fig:ecal-absolute-energy-scale}.

  For most of the energy range the uncertainty is limited by the knowledge of the beamtest energy
  which is approximately \SI{1.8}{\percent}. Together with leakage and calibration the combined
  uncertainty is about \SI{2}{\percent} from \SI{10}{\giga\electronvolt} to \SI{300}{\giga\electronvolt}.

  At the highest energies rear leakage and the correction of a saturation effect become important
  uncertainties. At \SI{1}{\tera\electronvolt} particle energy the absolute energy scale uncertainty
  is \SI{2.5}{\percent}.

  The uncertainty relating to the top of the instrument correction is needed only for electrons and
  positrons and compensates for bremsstrahlung losses along the particle's trajectory. This
  component of the uncertainty should be disregarded in the context of the photon analysis.

  As with the absolute rigidity scale uncertainty the spectral shape of the flux must be considered
  in order to translate the uncertainty of the absolute ECAL energy scale into an uncertainty of the
  measured flux.
\end{itemize}

Overall the uncertainty in the calorimeter analysis is \SI{2.2}{\percent}, which is dominated by the
trigger efficiency uncertainty and the error on the effective area. The absolute energy scale
uncertainty is also important, but will be considered separately.

Because the two analyses are performed using different subdetectors and should be considered
complementary, a comparison of the results will allow to reduce the uncertainties listed above and
will be discussed in section~\ref{sec:results-diffuse}. However, there are also a few effects which
affect both analyses:

\begin{itemize}
\item \textbf{TRD pileup weight}\\
  The biggest correction to the exposure map is the TRD pileup correction which was discussed in
  section~\ref{sec:analysis-trd-pileup-weight}. This correction is relevant in particular for low
  and high declination angles.

  The associated uncertainty was estimated to be \SI{3}{\percent}, by variation of the selection
  criteria for the upgoing electron event sample, used in the TRD pileup study (see
  section~\ref{sec:analysis-trd-pileup-weight}).

  Because the very same correction is used in both analyses they are equally affected by the
  uncertainty. The pileup correction affects only the normalization of the reconstructed photon
  fluxes as it is independent of photon energy.

\item \textbf{Description of material in the simulation}\\
  The correct material description in the simulation is important for a reliable estimation of the
  effective area, since it directly influences the number and locations of photon
  conversions. Although this uncertainty is relevant for both analysis modes, it typically has
  opposite effects on two respective effective areas. For example: An increase in the material in
  the upper TOF would lead to more photon conversions in the conversion mode analysis and at the
  same time to a reduction of the effective area for the calorimeter analysis.

  The material description in AMS was verified in the analysis of Helium~\cite{AMS02_Helium_2015}
  and other nuclei~\cite{AMS02_BoverC_2016,AMS02_HeCO_2017,AMS02_LiBeB_2018}. The material in the
  upper detector and in particular in the TRD was also verified in a dedicated
  study~\cite{AMS02_Detector_TRD_MC_2017}. As a result the uncertainty related to the material
  budget is negligible.

\item \textbf{Background subtraction}\\
  Because the procedure to fix the charged particle background is entirely data driven it is free of
  uncertainties relating to the simulation. There are two relevant components of the background
  subtraction uncertainty: The spatial shape of the background templates and the normalization
  uncertainty.

  The shape of the background templates was completely fixed by the data itself and no assumptions
  were made. It depends only on the declination angle, so it is simple in structure. As a result the
  uncertainty on the background template shape is negligible.

  The normalization error for the background is determined by the likelihood fit given in
  section~\ref{sec:corrections-background} and was found to be at the permille level for each
  individual energy bin. Overall the uncertainty due to the background subtraction is very small.

\end{itemize}
\emptypage


%% file: fermi.tex

\chapter{Fermi-LAT Analysis}
\label{sec:fermi-lat-analysis}

Comparing the AMS-02 results with physical models provides valuable insights into the physics of
gamma rays, but it is equally interesting to directly compare these results with those obtained from
other gamma ray experiments. In particular, the most sensitive high energy gamma ray detector in
space is the Large Area Telescope (LAT)~\cite{Fermi_LAT_Instrument_2009} on the Fermi satellite.

Comparing the AMS-02 results with those obtained with the LAT instrument on the Fermi satellite
requires performing the analysis of the \mbox{Fermi-LAT} data, since the LAT gamma ray flux was not
directly published in a suitable form. Fortunately both the \mbox{Fermi-LAT} data and the analysis
software (``fermitools'') are publicly available through the Fermi Science Support Center
(FSSC). Although a private analysis cannot be considered an official \mbox{Fermi-LAT} result, it is
worthwhile to pursue, since it provides the opportunity to judge the compatibility of the AMS-02
data with the LAT data without using models for the galactic diffuse emission. In addition, as
mentioned in section~\ref{sec:modeling-sources}, the published \mbox{Fermi-LAT} catalogs of sources
which include their fluxes were derived for specific time intervals, which only partially overlap
with the period of AMS-02 data taking. This is of particular importance for the flux of highly
variable blazars, such as \mbox{3C-454-3} or \mbox{CTA-102}, which can change dramatically over the
course of days or weeks.

The Fermi-LAT detector is described in detail elsewhere~\cite{Fermi_LAT_Instrument_2009}.

\section{Data Selection}
\label{sec:fermi-lat-data-selection}

The version of the \mbox{Fermi-LAT} data is Pass 8 Release 3 (P8R3)~\cite{Fermi_Pass8_2013,
  Fermi_P8R3_2018}, which improves upon the prior release by significantly reducing the charged
particle background, in particular near the ecliptic. Weekly photon event lists are provided through
the FSSC.

In the first step the \mbox{Fermi-LAT} photon events of the ``SOURCE'' event class and the
``FRONT+BACK'' conversion type are selected. Only those \mbox{Fermi-LAT} data files which fall into
the time range of the AMS measurement from the 19th of May 2011 to the 12th of November 2017 are
processed. This is important because otherwise it is impossible to compare the fluxes from variable
sources between the two experiments. The minimum and maximum photon energy in the selection is
\SI{50}{\mega\electronvolt} and \SI{1}{\tera\electronvolt}, respectively.

In the second step ``Good Time Intervals'' (GTI) are assigned, which correspond to the time periods
in which the \mbox{Fermi-LAT} detector was operating under normal conditions. This selection
includes a cut on the angle between the detector's $z$-axis and the spacecraft zenith direction of
$\delta < \SI{52}{\degree}$. This cut is important in order to remove photons coming from the
Earth's limb, which are created in interactions of charged particles with the Earth's
atmosphere. The event lists are correspondingly filtered to only contain events from GTIs.

\begin{figure}[t!]
  \centering
  \includegraphics[width=0.98\linewidth]{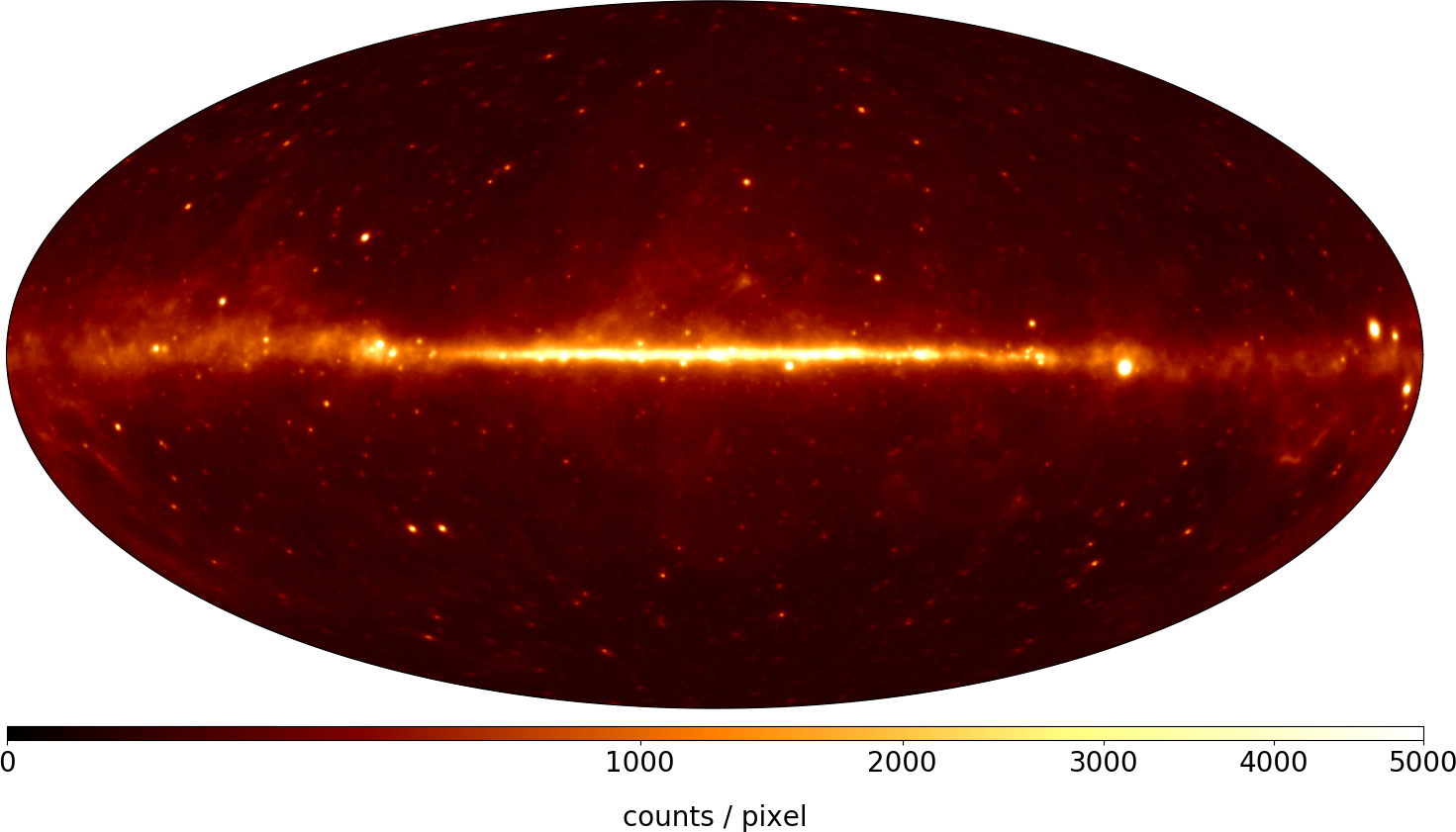}
  \caption{Measured photon counts for the \mbox{Fermi-LAT} experiment between
    \SI{500}{\mega\electronvolt} and \SI{100}{\giga\electronvolt} in galactic coordinates, shown
    with a square root color scale.}
  \label{fig:fermi-lat-counts-galactic}
\end{figure}

In order to create count maps the events are then binned into three-dimensional data cubes (energy,
celestial latitude and longitude). The energy binning used is logarithmic, spanning 100 bins from
\SI{50}{\mega\electronvolt} to \SI{1}{\tera\electronvolt}. The spatial pixel size is
$\SI{0.5}{\degree} \times \SI{0.5}{\degree}$, which defines 720 bins in longitude and 360 bins in
latitude for an all-sky image in the plate carree
projection. Figure~\ref{fig:fermi-lat-counts-galactic} shows an all-sky view in the Hammer-Aitoff
projection of the integrated photon counts from \SI{500}{\mega\electronvolt} to
\SI{100}{\giga\electronvolt}.

\section{Instrument Response Functions}
\label{sec:fermi-lat-irfs}

\begin{figure}[t]
  \begin{minipage}{0.48\linewidth}
    \centering
    \includegraphics[width=1.0\linewidth]{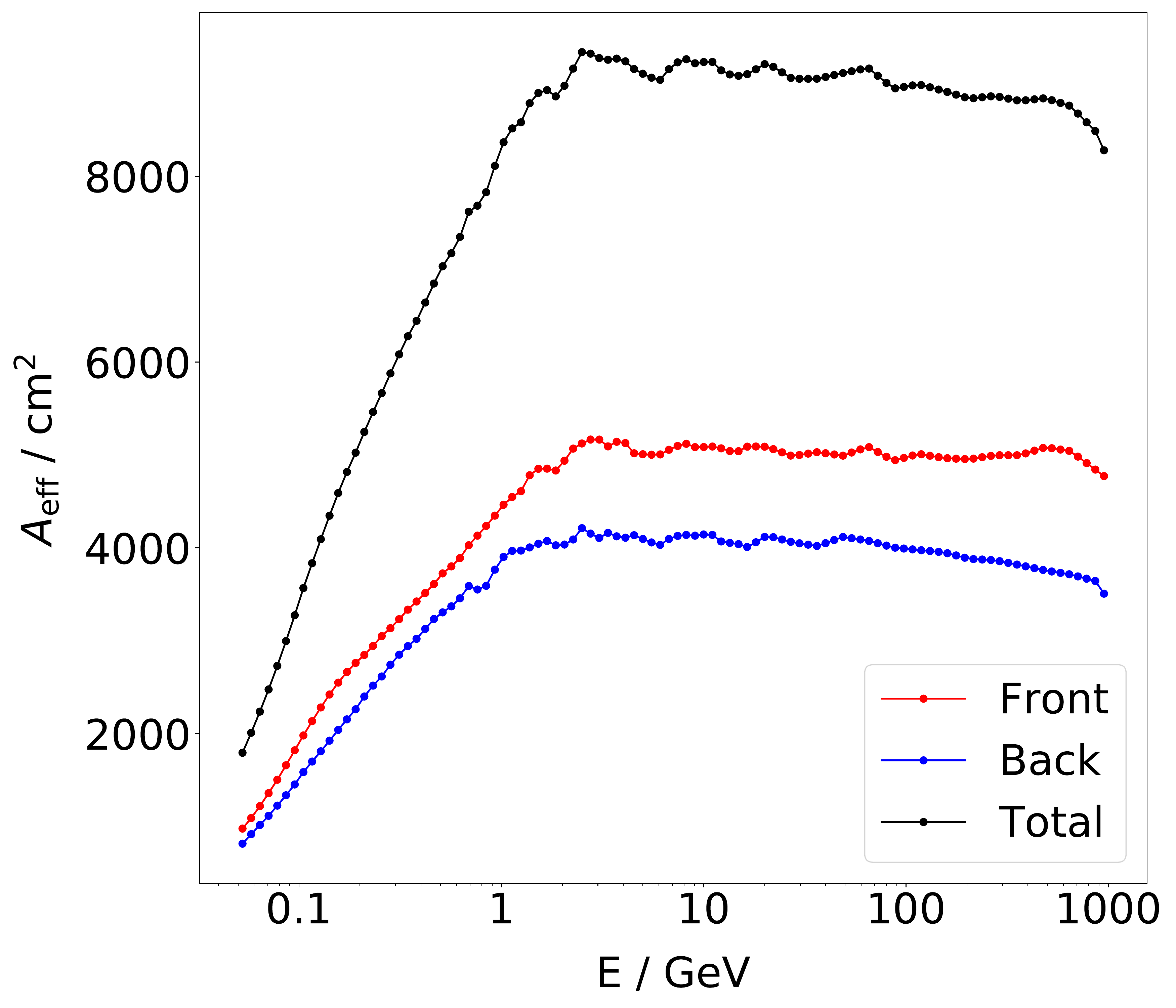}
  \end{minipage}
  \hspace{0.01\linewidth}
  \begin{minipage}{0.48\linewidth}
    \centering
    \includegraphics[width=1.0\linewidth]{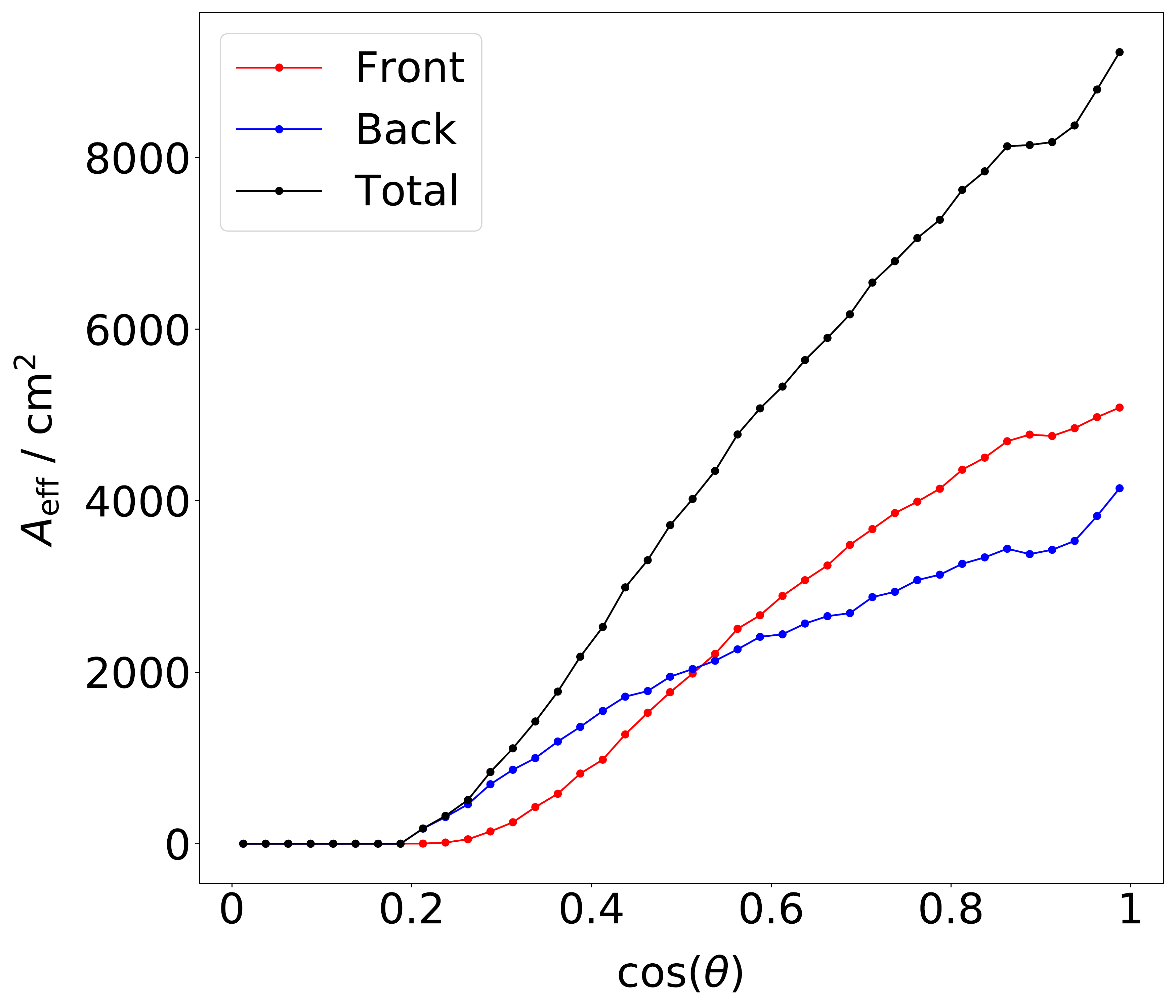}
  \end{minipage}
  \caption{The \mbox{Fermi-LAT} P8R3\_SOURCE\_V2 effective area for perpendicular incidence as a
    function of the photon energy on the left and as a function of $\cos({\theta})$ for
    \SI{10}{\giga\electronvolt} photons on the right. These are slices of the two dimensional
    effective area distribution.}
  \label{fig:fermi-lat-irf-effective-area}
\end{figure}

Since the \mbox{Fermi-LAT} selection cuts are fixed by defining the event class and conversion type,
it is possible to use the official \mbox{Fermi-LAT} IRFs without any further modifications. These
IRFs were pre-derived by the \mbox{Fermi-LAT} team from Monte-Carlo simulations and include
corrections for differences between data and simulation. Given the selection specified above the two
IRF types used are ``P8R3\_SOURCE\_V2::(FRONT|BACK)''. These IRFs include the effective area, as
well as the energy and angular resolution probability density functions. The effective area for the
selection is the sum of the FRONT and BACK effective areas, since both selections are orthogonal.

The effective area is shown in figure~\ref{fig:fermi-lat-irf-effective-area}. For perpendicular
incidence, shown on the left, the effective area rises from \SI{3500}{\centi\meter\squared} at
\SI{100}{\mega\electronvolt} and reaches its maximum of approximately
\SI{9000}{\centi\meter\squared} around \SI{2}{\giga\electronvolt}. It stays approximately constant
up to \SI{500}{\giga\electronvolt} from where it begins to drop, due to the finite size of the LAT
calorimeter. As shown on the right side of figure~\ref{fig:fermi-lat-irf-effective-area}, the
effective area is approximately proportional to $\cos\theta$. The maximum allowed polar angle in the
selection is approximately $\cos\theta = 0.2$ which corresponds to
$\theta \approx \SI{78}{\degree}$. The corresponding acceptance is approximately
\SI{25000}{\centi\meter\squared\steradian}.

\begin{figure}[t]
  \begin{minipage}{0.48\linewidth}
    \centering
    \includegraphics[width=1.0\linewidth]{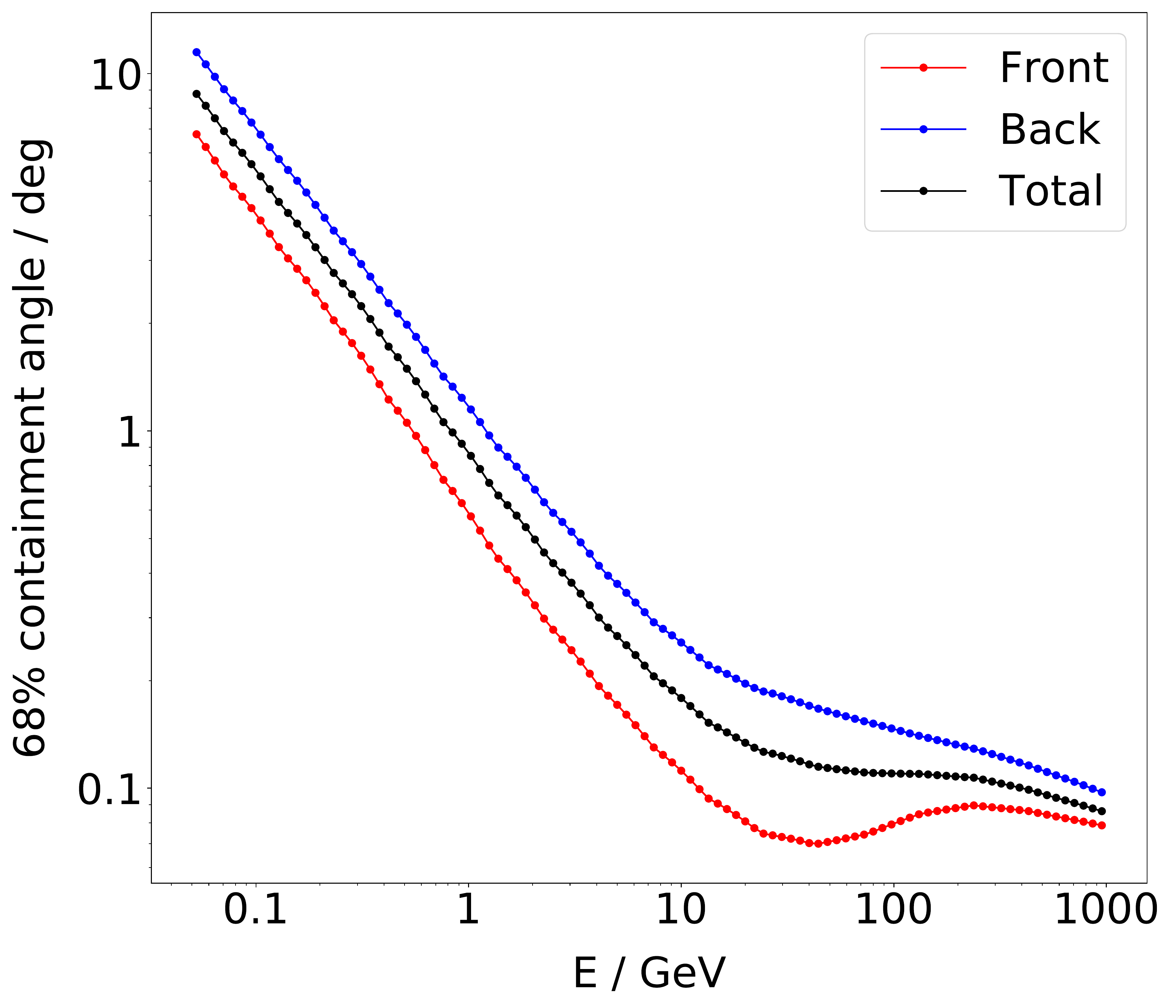}
  \end{minipage}
  \hspace{0.01\linewidth}
  \begin{minipage}{0.48\linewidth}
    \centering
    \includegraphics[width=1.0\linewidth]{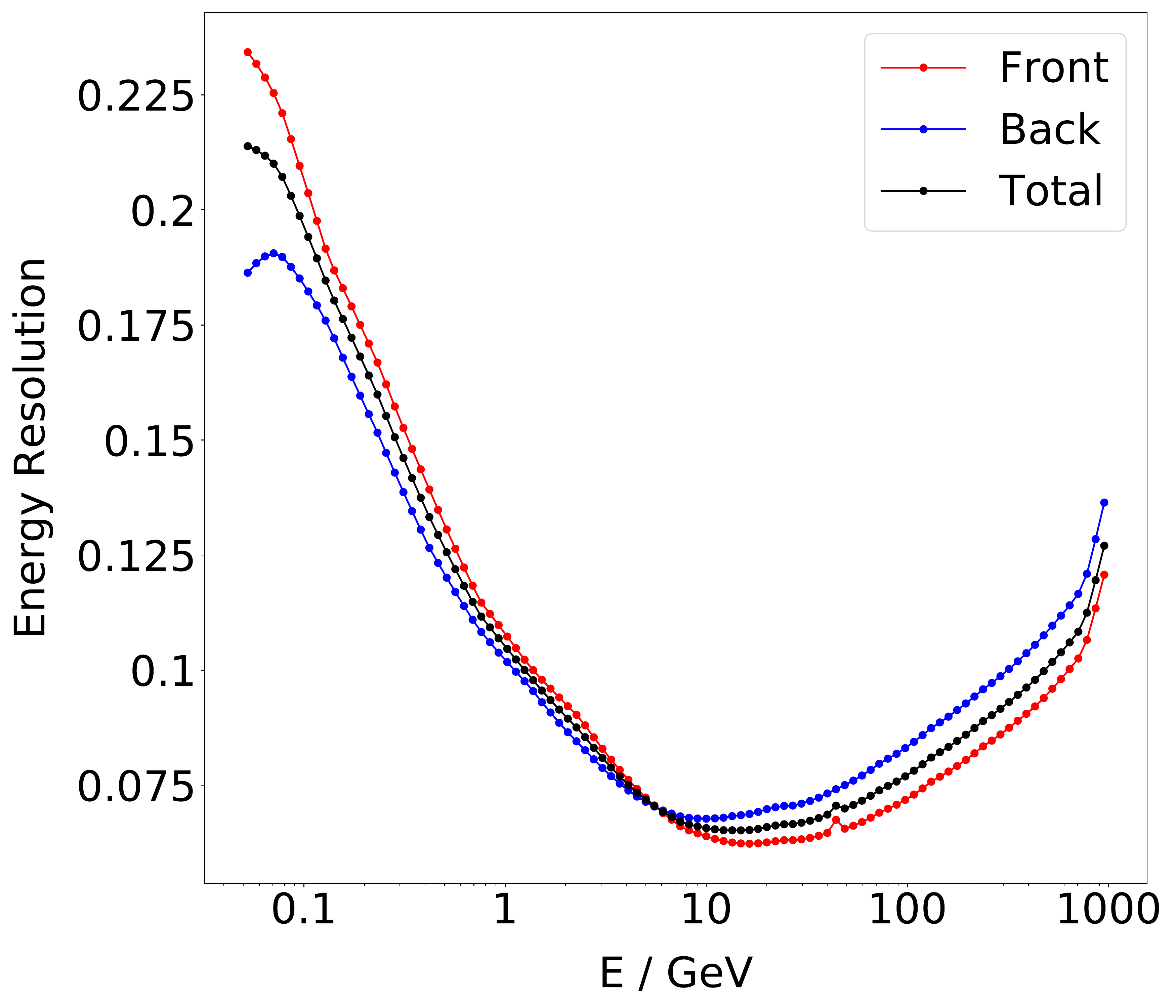}
  \end{minipage}
  \caption{The \mbox{Fermi-LAT} P8R3\_SOURCE\_V2 point spread function \SI{68}{\percent} containment
    angle on the left and the relative energy resolution on the right.}
  \label{fig:fermi-lat-irf-psf-energy-resolution}
\end{figure}

The energy and angular resolution functions are given as a function of energy $E$ and inclination
angle $\cos\theta$. They are provided separately for each of the two conversion types. In order to
calculate the averaged resolution function for the combined selection these functions are weighted
with the corresponding effective area and then integrated over $\cos\theta$:

\begin{displaymath}
  f^{F+B}(E) = \frac{1}{\mathcal{A}^{F+B}(E)}
  \int_{-1}^{1}{(A_{eff}^{F}(E,\cos\theta) f^{F}(E,\cos\theta) + A_{eff}^{B}(E,\cos\theta)
    f^{B}(E,\cos\theta)) \mathrm{d}\cos\theta} \,,
\end{displaymath}

where $\mathcal{A}^{F+B}(E) = \mathcal{A}^{F}(E) + \mathcal{A}^{B}(E)$ is the total acceptance of
the combined FRONT and BACK selection. This factor normalizes the PDF to unity for each energy value
$E$.

The resolution functions are shown in figure~\ref{fig:fermi-lat-irf-psf-energy-resolution}. The left
hand side shows the \SI{68}{\percent} containment angle, which corresponds to the angular resolution
of the LAT for the given photon selections. For energies below approximately
\SI{20}{\giga\electronvolt} the resolution is dominated by multiple scattering of the electron and
positron in the tracking volume and correspondingly improves with energy as $1 / E$. At
\SI{1}{\giga\electronvolt} the containment angle is approximately \SI{0.8}{\degree}. For
\SI{100}{\giga\electronvolt} and above the resolution is limited by the spatial resolution of the
LAT tracker, the containment angle reaches a plateau of approximately \SI{0.1}{\degree}.

The energy resolution of the calorimeter is shown on the right hand side. It is approximately
\SI{20}{\percent} at \SI{100}{\mega\electronvolt} and improves as $1 / E$ with energy. The best
resolution of approximately \SI{7}{\percent} is reached for \SIrange{10}{20}{\giga\electronvolt}
photons. For higher energies the resolution deteriorates, due to the finite thickness of the LAT
calorimeter.

\section{Exposure Maps}
\label{sec:fermi-lat-exposure}

\begin{figure}[t!]
  \centering
  \includegraphics[width=0.98\linewidth]{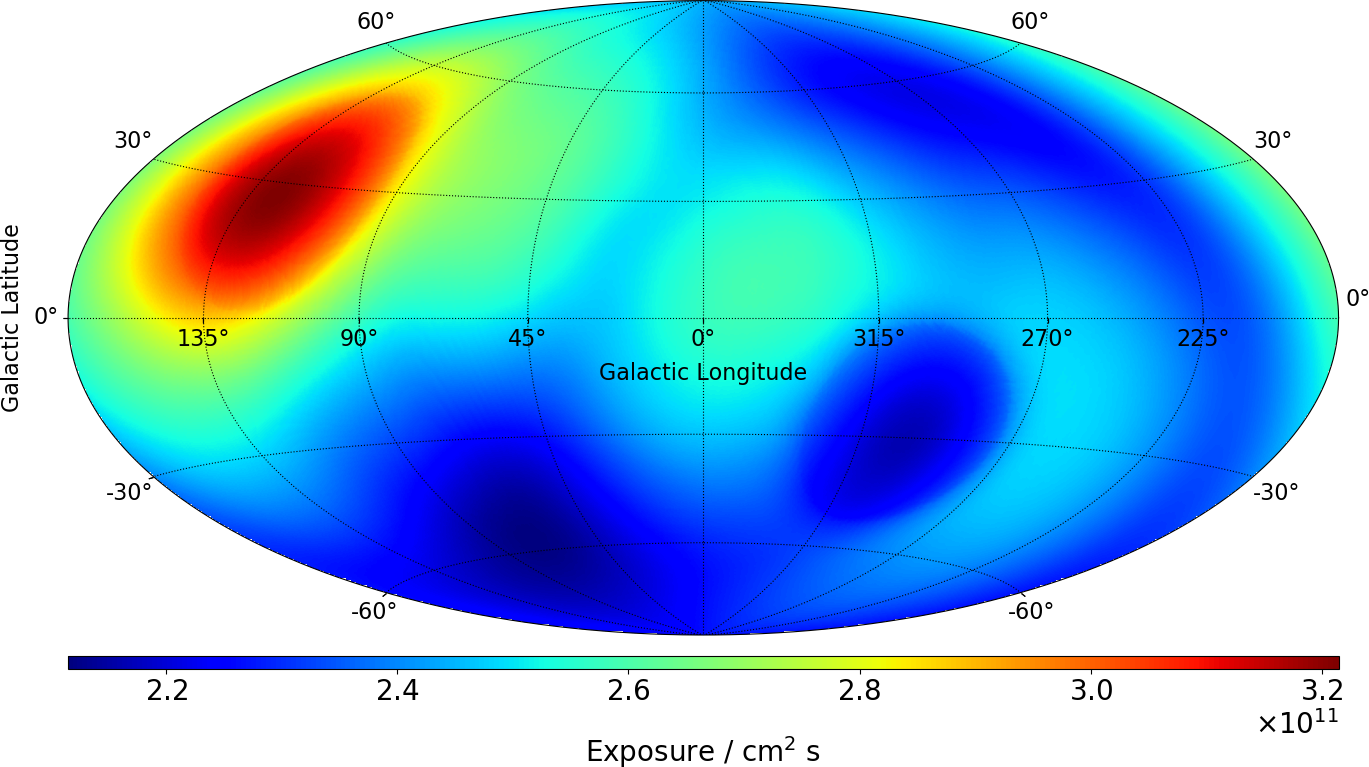}
  \caption{Exposure map at \SI{10}{\giga\electronvolt} for the \mbox{Fermi-LAT} experiment P8R3
    SOURCE selection in galactic coordinates. The integration over time was performed from 19th of
    May 2011 to 12th of November 2017.}
  \label{fig:fermi-lat-exposure-map-galactic}
\end{figure}

Based on the effective area and the GTIs the exposure maps can be computed in the same binning as
the event count maps. As shown in figure~\ref{fig:fermi-lat-exposure-map-galactic} the
\mbox{Fermi-LAT} exposure for \SI{10}{\giga\electronvolt} photons varies only weakly over the sky
due to the large angular acceptance of the experiment. For the same reason there are no ``blind
spots'' on the sky. At \SI{10}{\giga\electronvolt} the maximum exposure value of
\SI{3.2e11}{\centi\meter\squared\second} is reached close to the northern celestial pole. Close to
the southern celestial pole the minimum exposure value is approximately
\SI{2.2e11}{\centi\meter\squared\second}. The southern exposure is generally smaller, due to the
existence of the SAA, in which the experiment is unable to record useful data. Because the LAT
effective area is almost constant between \SI{2}{\giga\electronvolt} and
\SI{500}{\giga\electronvolt} exposure maps for other energy bins look very similar.

\section{Systematic Uncertainties}
\label{sec:fermi-lat-systematics}

\begin{figure}[t]
  \begin{minipage}{0.48\linewidth}
    \centering
    \includegraphics[width=1.0\linewidth]{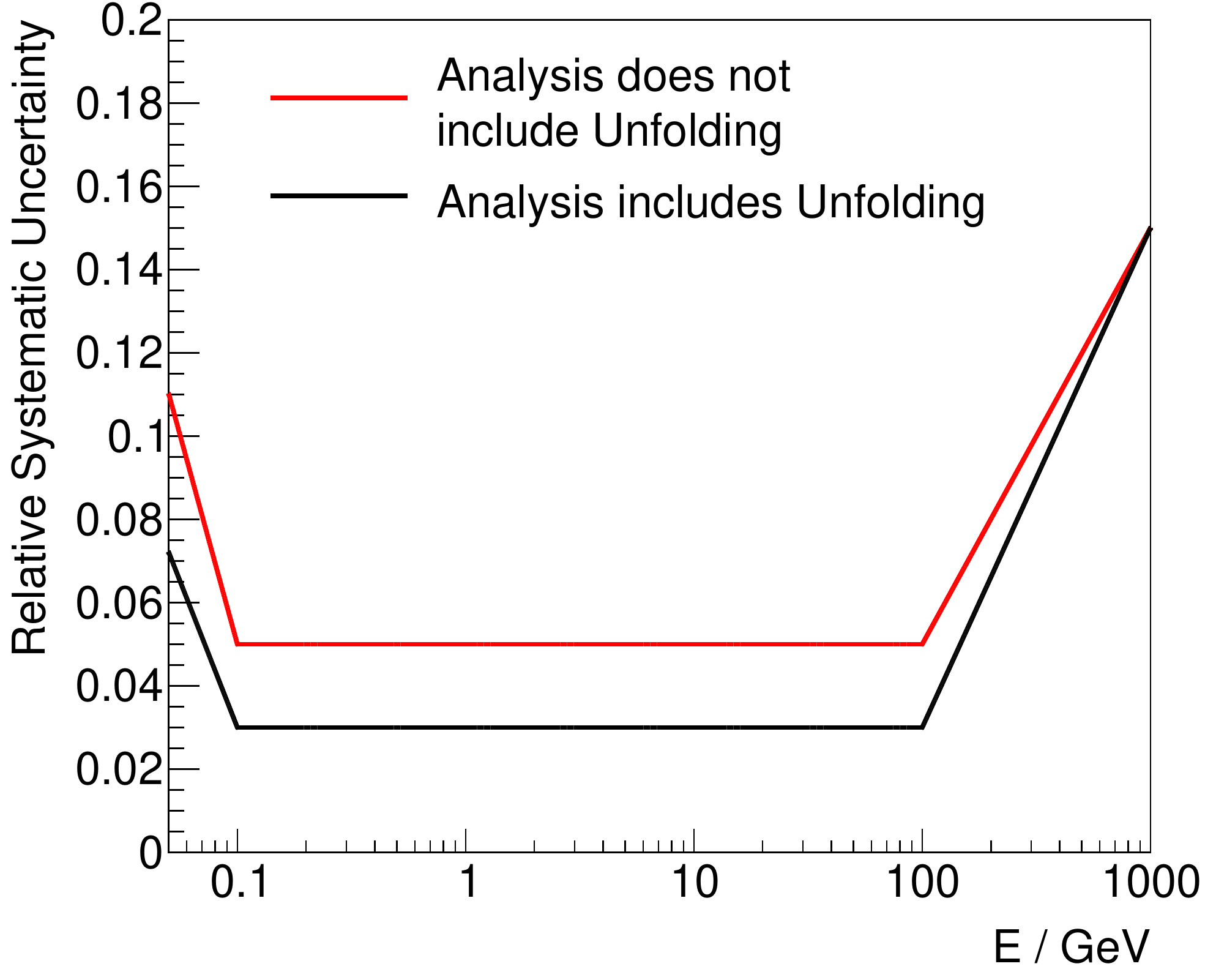}
  \end{minipage}
  \hspace{0.01\linewidth}
  \begin{minipage}{0.48\linewidth}
    \centering
    \includegraphics[width=1.0\linewidth]{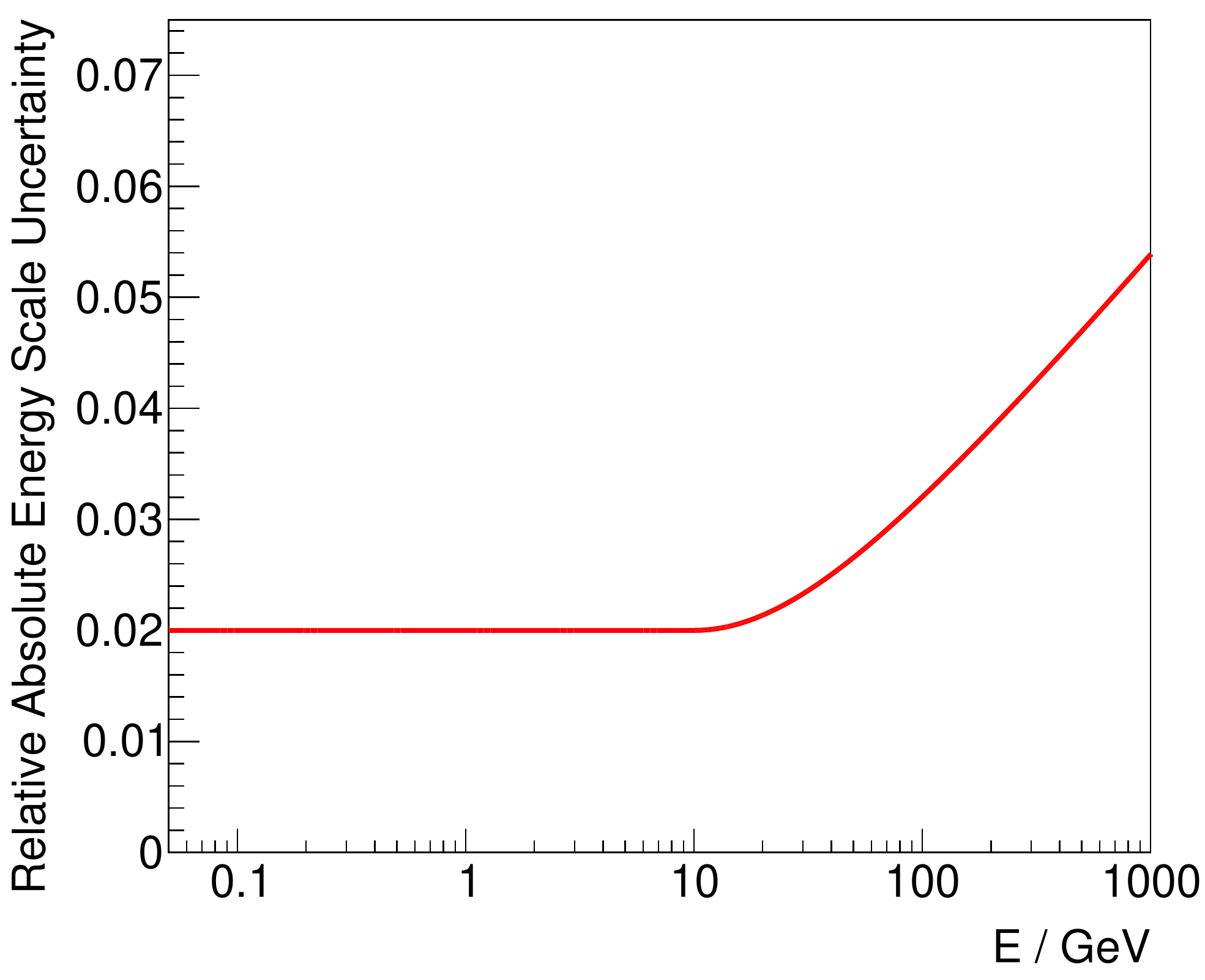}
  \end{minipage}
  \caption{Left: The \mbox{Fermi-LAT} systematic uncertainty on the effective area. Right: The
    absolute energy scale uncertainty.}
  \label{fig:fermi-lat-systematics}
\end{figure}

The main systematic uncertainty relevant to the determination of gamma ray fluxes with the LAT is
the uncertainty of the effective area. The estimate of this uncertainty is shown on the left hand
side of figure~\ref{fig:fermi-lat-systematics}. It was derived by comparing the measured fluxes and
cut efficiencies to the Monte-Carlo predictions for some gamma ray sources using various subsets of
the data~\cite{Fermi_IRFs_2012}. Details are given in the FSSC
\footnote{\url{https://fermi.gsfc.nasa.gov/ssc/data/analysis/LAT\_caveats.html}\\
  \url{https://fermi.gsfc.nasa.gov/ssc/data/analysis/scitools/Aeff\_Systematics.html}}. The black
curve, corresponding to the smaller uncertainty, is to be used in case the effect of bin-to-bin
migration is corrected for using appropriate techniques. In that case the systematic uncertainty on
the effective area is given as \SI{3}{\percent} from \SI{100}{\mega\electronvolt} to
\SI{100}{\giga\electronvolt} and rises logarithmically with energy above and below.

However, the effect of energy resolution and bin-to-bin migration is often neglected in analyses of
\mbox{Fermi-LAT} data~\cite{Fermi_3FGL_2015,Fermi_Diffuse_IEM_2016}, and a larger systematic
uncertainty, shown in red in the same figure is to be used instead. In that case the suggested
systematic uncertainty is \SI{5}{\percent} between \SI{100}{\mega\electronvolt} and
\SI{100}{\giga\electronvolt}, and also increases with $log(E)$ above and below. In the 4FGL
publication, which lists the correction of the energy migration effect as one of the improvements
over the prior catalog, the authors note that considering energy dispersion ``tends to increase the
flux (by \SI{4}{\percent} on average)''~\cite{Fermi_4FGL_2019}, although the specifics depend on the
spectral shape of the source.

The absolute energy scale of the LAT instrument is another important issue, which can systematically
influence the measured photon flux. Although a calibration unit of the LAT calorimeter and tracker
were calibrated and tested in a beam test, the full flight model was
not~\cite{Fermi_EnergyScale_2012}. To verify the absolute energy scale of the instrument a study was
performed which compared the measured position of the geomagnetic cutoff for cosmic ray electrons
with model calculations based on the spacecraft altitude and position in the Earth's magnetic
field~\cite{Fermi_EnergyScale_2012}. Because of the orbital inclination and altitude of the LAT
spacecraft the geomagnetic cutoff for cosmic ray electrons varies from approximately
\SI{6}{\giga\electronvolt} to approximately \SI{13}{\giga\electronvolt}, which is the range in which
such comparisons are possible. The study concludes that the measured cutoff energies exceed the
predicted ones by \SI{2.6}{\percent} on average, but because the systematic uncertainty on this
result is \SI{2.5}{\percent}, no correction to the measured LAT energies is applied. No statement is
made about the validity of the absolute energy scale at lower or higher energies.

In the analysis of cosmic ray electrons plus positrons~\cite{Fermi_AllElec_2017} the geomagnetic
cutoff study is repeated with seven years of data. In the repeated analysis the measured cutoffs
exceed the predicted ones by \SI{3.3}{\percent} on average, with an estimated systematic uncertainty
of \SI{2.0}{\percent}. As a result the energy scale for the electron plus positron analysis is
decreased by \SI{3.3}{\percent}. However, a similar energy scale shift is not commonly used in the
analysis of gamma ray data.

The absolute energy scale uncertainty as proposed in~\cite{Fermi_AllElec_2017} is shown in red on
the right hand side of figure~\ref{fig:fermi-lat-systematics}. At energies below
\SI{10}{\giga\electronvolt} a constant value of \SI{2}{\percent} is proposed, which was estimated on
the basis of various cross checks in the context of the geomagnetic cutoff study. This value also
includes the uncertainty due to the modeling of the geomagnetic field and the corresponding cutoff
rigidity prediction. Above \SI{10}{\giga\electronvolt} leakage becomes an increasingly important
effect, which could also systematically influence the absolute energy scale. At
\SI{1}{\tera\electronvolt} the overall energy scale uncertainty is approximately \SI{5.4}{\percent}.

\section{Corrections and Unfolding}
\label{sec:fermi-lat-corrections}

The LAT photon data contains an isotropic flux component which is predominantly due to misidentified
charged cosmic rays, but also contains extragalactic diffuse photons. The isotropic flux component
was estimated by the LAT team and is available through the FSSC. It depends on the event class
selection and conversion type, as well as the galactic diffuse emission model used to derive it. In
order to account for this charged particle background the isotropic flux is converted into expected
event counts using the exposure map derived above. The result is subsequently subtracted from the
measured data. This corresponds to the AMS-02 background subtraction procedure described in
section~\ref{sec:corrections-background}, however in the LAT case the correction is significantly
smaller.

\begin{figure}[t]
  \begin{minipage}{0.48\linewidth}
    \centering
    \includegraphics[width=1.0\linewidth]{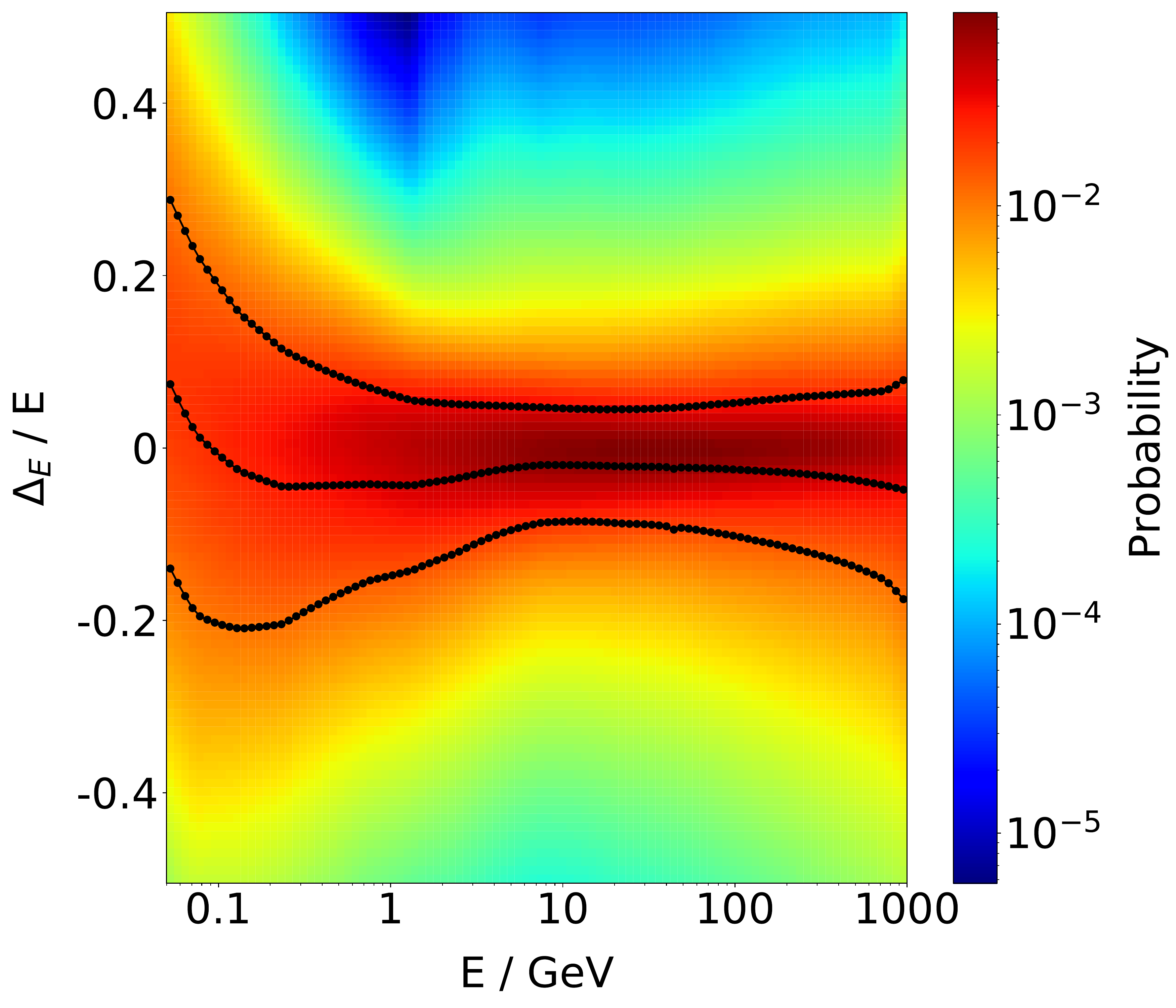}
  \end{minipage}
  \hspace{0.01\linewidth}
  \begin{minipage}{0.48\linewidth}
    \centering
    \includegraphics[width=1.0\linewidth]{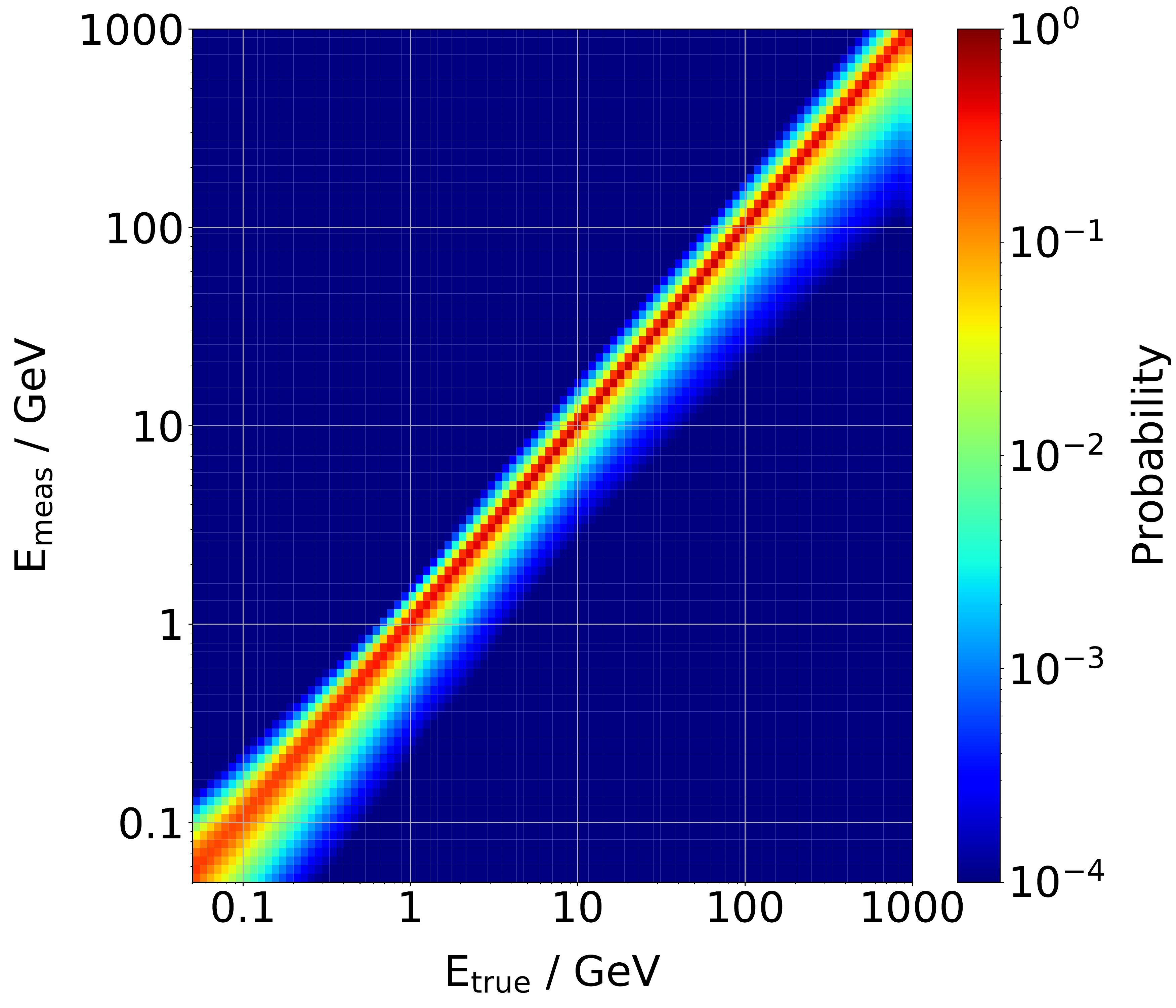}
  \end{minipage}
  \caption{The \mbox{Fermi-LAT} P8R3\_SOURCE\_V2 relative energy resolution on the left, and the
    corresponding migration matrix on the right.}
  \label{fig:fermi-lat-irf-energy-resolution-migration}
\end{figure}

The effect of energy migration is treated by unfolding the background corrected measured event
counts. The procedure is the same as the one described in section~\ref{sec:corrections-unfolding}
for AMS-02. The migration matrix element $(i,j)$ for the unfolding is derived from the energy
resolution PDF $f(E^{\mathrm{rec}}|E^{\mathrm{true}}) \mathrm{d}E^{\mathrm{rec}}$ by integration
over the bins in reconstructed energy:

\begin{displaymath}
  A_{ij} = p\left(E^{\mathrm{rec}}_{j},E^{\mathrm{true}}_{i}\right) =
  \int_{E^{\mathrm{rec}}_{j,\mathrm{low}}}^{E^{\mathrm{rec}}_{j,\mathrm{high}}} { f \left(
      E^{\mathrm{rec}} | \sqrt{E^{\mathrm{true}}_{i,\mathrm{low}} \cdot
        E^{\mathrm{true}}_{i,\mathrm{high}}} \right) \mathrm{d} E^{\mathrm{rec}} } \,,
\end{displaymath}

where the subscripts $\mathrm{low / high}$ refer to the energy bin lower and upper boundaries and
the square root expression corresponds to the logarithmic bin center. This expression assumes that
the PDF $f(E^{\mathrm{rec}}|E^{\mathrm{true}})\mathrm{d}E^{\mathrm{rec}}$ varies only weakly with
true energy in each bin in true energy, so that the logarithmic bin center value can be used in the
integral.

The resulting relative energy resolution and migration matrix are shown in
figure~\ref{fig:fermi-lat-irf-energy-resolution-migration}. The best resolution is reached around
\SI{10}{\percent}. It is noteworthy that although the mean value of the relative energy resolution
curve is typically negative, the most probable value in each slice is close to zero.

As discussed in section~\ref{sec:fermi-lat-systematics}, the latest study of the absolute energy
scale of the LAT concludes that the energy scale in the pass 8 reconstruction is biased by
\SI{3.3}{\percent}~\cite{Fermi_AllElec_2017}. This shift was found within the context of the
analysis of electrons and positrons, but it is reasonable to assume that it applies equally to the
energy reconstruction of photon showers in the LAT calorimeter and to attempt to correct for this
bias. It is important to point out that, unlike in this analysis, the energy scale bias is not
corrected for in published Fermi-LAT photon analyses. The effect of the bias correction will be
studied in section~\ref{sec:results-inner-galaxy}.

One way to achieve the bias correction is to simply multiply all photon event measured energies by
$1.0 / 1.033 \approx 0.968$ in the \mbox{Fermi-LAT} event selection procedure. Another way is to
construct a special migration matrix, in which the true energy axis is scaled with respect to the
previous Monte-Carlo true energy, corresponding to a shift to the left of the band in the migration
matrix in figure~\ref{fig:fermi-lat-irf-energy-resolution-migration}. The latter is the approach
chosen here, but both methods were verified to yield the same result.

\emptypage


%% file: results.tex

\chapter{Results}
\label{sec:results}

\begin{figure}[t!]
  \centering
  \includegraphics[width=0.95\linewidth]{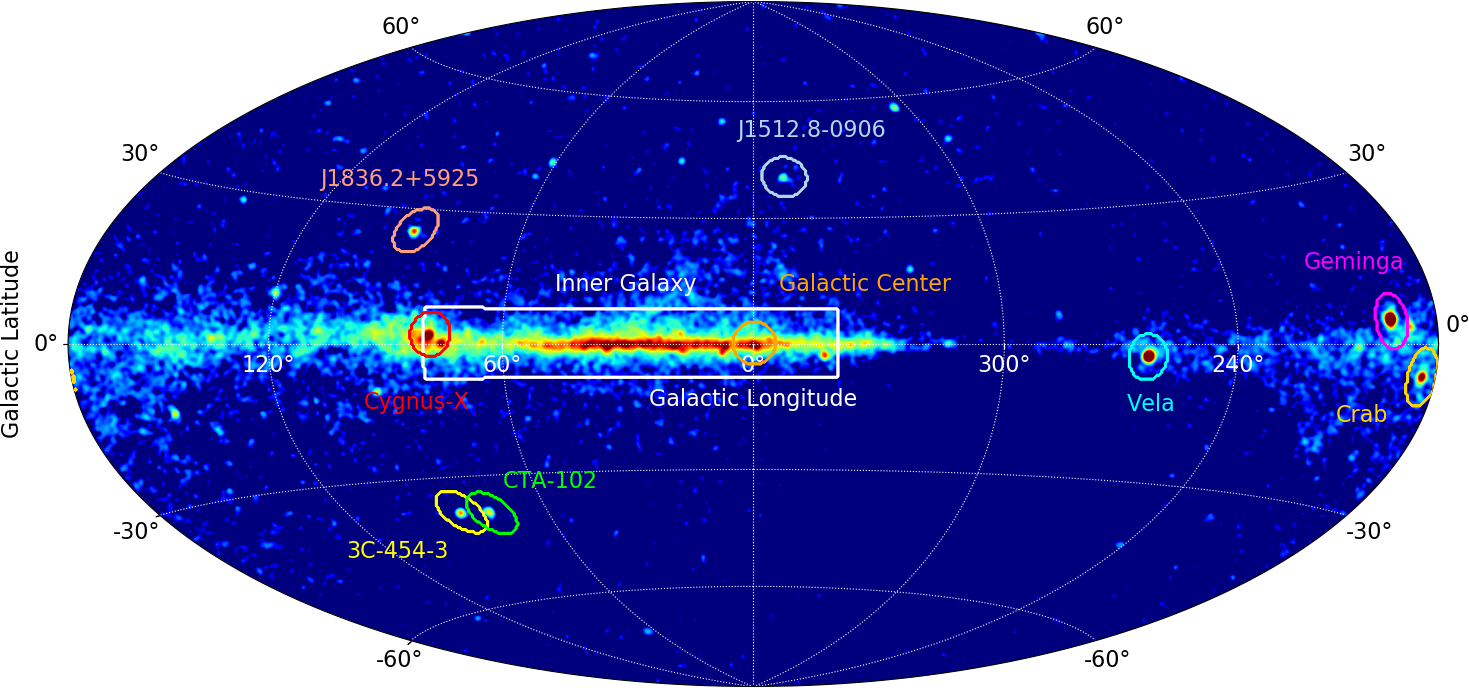}
  \caption{Integrated photon counts in the conversion analysis between \SI{500}{\mega\electronvolt}
    and \SI{100}{\giga\electronvolt} together with the locations of various windows discussed in the
    text.}
  \label{fig:results-counts-window-labels}
\end{figure}

In this chapter the results for photon fluxes for various parts of the sky are presented. A general
description of the formulae for the calculation of non-isotropic fluxes is given in
section~\ref{sec:calculation-flux}.

Figure~\ref{fig:results-counts-window-labels} shows a skymap of the integrated photon counts between
\SI{500}{\mega\electronvolt} and \SI{100}{\giga\electronvolt} in the photon conversion analysis. On
top of the figure various circles, rectangles and labels mark the windows which were analyzed in
this thesis. In this chapter the focus is on a few important regions, results for other regions can
be found in appendix~\ref{sec:appendix-flux-other}.

Results for regions of interest which are dominated by diffuse emission are shown in
section~\ref{sec:results-diffuse}. The white window in the center of the figure is the inner galaxy
region, defined by $\SI{-20}{\degree} < l < \SI{80}{\degree}$ and
$\left|b\right| < \SI{8}{\degree}$, where $l$ and $b$ are the galactic longitude and latitude,
respectively. In this region hundreds of prominent gamma ray sources contribute to the total gamma
ray flux, but overall the diffuse emission dominates. The \mbox{Cygnus} region is particularly
active in $\gamma$-rays: It contains the \mbox{Cygnus-X} star-forming region as well as prominent
Supernova Remnants (the Cygnus Loop and the $\gamma$-Cygni SNR). It also includes two strong pulsars
(PSR J2021+4026 and PSR J2021+3651) and extended regions of emission, such as the Cygnus
cocoon. Results for both regions are presented.

Sources which produce high energy $\gamma$-rays are studied in section~\ref{sec:results-sources}. As
an example for a $\gamma$-ray pulsar results for Geminga are shown and discussed. This includes
pulsar timing with \mbox{AMS-02} data, from which the canonical age and surface magnetic field
strength of the pulsar are estimated. Results for the pulsars Vela and Crab are also shown. Due to
their variability blazars are particularly interesting objects. The photon flux from the Flat
Spectrum Radio Quasar \mbox{CTA-102} is presented. Strong flaring activity on time scales of hours
was detected for this source in the \mbox{AMS-02} data.

\section{Calculation of Fluxes}
\label{sec:calculation-flux}

For a point-like gamma-ray source located at $(l,b)$ in the sky, the exposure map converts the
measured number of events produced by the source into its flux:

\begin{equation}
  \label{eq:flux-point-source}
  \Phi_{\mathrm{source}}(E_{i}) =
  \frac{N_{\mathrm{source}}(E_{i})}
  {\mathcal{E}(E_{i},l,b) \epsilon_{\mathrm{trigger}}(E_{i}) \Delta E_{i}} \,,
\end{equation}

where $N_{\mathrm{source}}(E_{i})$ is the number of collected events from the source,
$\mathcal{E}(E_{i},l,b)$ is the exposure, $\epsilon_{\mathrm{trigger}}$ is the trigger efficiency
and $\Delta E_{i}$ is the energy bin width. Whether or not the trigger efficiency is included in the
exposure is a matter of convention. The flux from point sources has units
$1 / (\si{\giga\electronvolt\square\centi\meter\second})$, or equivalent.

For the measurement of diffuse gamma ray fluxes a different formula is required. It is useful to bin
the sky into a grid, for example, by simply binning in terms of longitude and latitude in a
rectangular plate carree grid, using $N_l$ equidistant bins from \SIrange{-180}{180}{\degree} in
longitude and $N_b$ bins from \SIrange{-90}{90}{\degree} in latitude. In this scheme the solid angle
subtended by each bin varies and is given for the bin $j,k$ by

\begin{equation}
  \label{eq:solid-angle}
  \Delta \Omega_{jk} = \cos{b_k} \Delta l \Delta b \,,
\end{equation}

where $\Delta l = 2\pi / N_l$, $\Delta b = \pi / N_b$ and $b_{k}$ is the galactic latitude of the
bin center. Another binning scheme which is widely used is the HEALPix~\cite{HEALPix2005} binning in
which each bin subtends the same solid angle. The average diffuse photon flux in the bin $j,k$ is
then:

\begin{equation}
  \label{eq:flux-diffuse}
  \Phi_{\mathrm{diffuse}}(E_{i},l_{j},b_{k}) =
  \frac{N(E_{i},l_j,b_k)}
  {\mathcal{E}(E_{i},l_j,b_k)
    \epsilon_{\mathrm{trigger}}(E_{i})
    \Delta\Omega_{jk}
    \Delta E_{i}} \,,
\end{equation}

where $N(E_{i},l_j,b_k)$ is the number of events counted in the bin $j,k$. The diffuse flux has
units $1 / (\si{\giga\electronvolt\square\centi\meter\second\steradian})$, or equivalent.

Computing the average flux for a given region of interest in the sky is done by integration:

\begin{align}
  \label{eq:flux-window-integral}
  \Phi_{\gamma,{\mathcal{W}}}(E_{i})
  &= \frac{1}{\Delta \Omega_{\mathcal{W}}}
    \int_{\mathcal{W}}{\Phi_{\gamma}(E_{i},l,b) \, \mathrm{d}\Omega}\\
  \label{eq:flux-window-sum}
  &= \frac{1}{\Delta \Omega_{\mathcal{W}}}
    \sum_{j,k \in {\mathcal{W}}}{\Phi_{\gamma}(E_{i},l_j,b_k) \Delta \Omega_{jk}} \,,
\end{align}

where $\mathcal{W}$ is the window definition and $\Delta \Omega_{\mathcal{W}}$ is the total solid
angle subtended by that window. Fluxes presented in this chapter will be calculated according to
equation~(\ref{eq:flux-window-sum}).

\section{Flux in Regions Dominated by Diffuse Emission}
\label{sec:results-diffuse}

\subsection*{The Inner Galaxy}
\label{sec:results-inner-galaxy}

\begin{figure}[t!]
  \centering
  \includegraphics[width=0.95\linewidth]{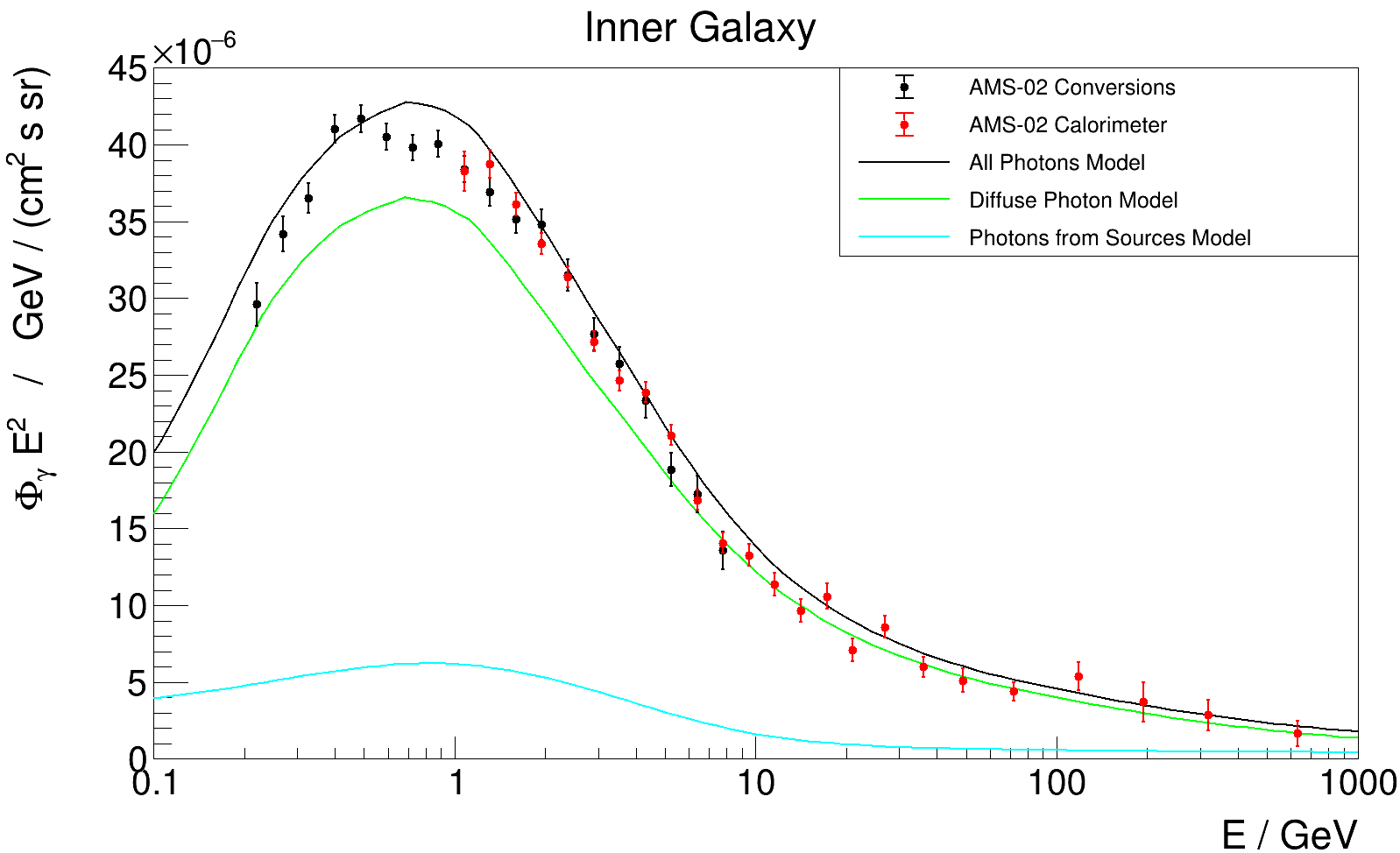}
  \caption{Average photon flux in the inner galaxy multiplied by $E^2$ as a function of the photon
    energy. Black points: AMS-02 photon flux from the conversion analysis. Red points: AMS-02 photon
    flux from the calorimeter analysis. Uncertainties are statistical only. Cyan line: Summed model
    prediction for all sources in the window. Green line: Model prediction for the photon flux from
    diffuse emission. Black line: Sum of the model predictions for sources and diffuse emission.}
  \label{fig:flux-inner-galaxy}
\end{figure}

Figure~\ref{fig:flux-inner-galaxy} shows the measured average photon flux in the inner galaxy for
the two analyses modes. The figure also includes the model prediction for the diffuse model (green),
source model (cyan) and their sum (black line). The data points were obtained from the unfolded and
background corrected distributions and calculated according to
equation~(\ref{eq:flux-window-sum}). Within each bin the position of the data point on the abscissa
has been calculated according to the procedure suggested by Lafferty and Wyatt~\cite{Lafferty_1995},
assuming a spectral index of $\gamma = 2.5$.

The results for the two \mbox{AMS-02} analyses modes (conversions and calorimeter) were obtained
individually, using the respective selections, exposure maps and unfolding corrections. Although the
window for the inner galaxy region was chosen to be large enough, such that the effect of the PSF is
marginal, a very minor correction is applied to the calorimeter analysis. This correction accounts
for event migration in and out of the inner galaxy window, due to the imperfect angular resolution
and was obtained by comparing the reconstructed model flux after convolution with the calorimeter
angular uncertainty to the unsmeared model flux.

The model generally predicts a slightly higher photon flux compared to the data, in particular in
the region from \SI{500}{\mega\electronvolt} to \SI{2}{\giga\electronvolt} and around
\SI{10}{\giga\electronvolt}. The disagreement is not unexpected, because the diffuse model was
primarily optimized for spatial rather than spectral compatibility with the \mbox{Fermi-LAT} data.

\begin{figure}[t!]
  \centering
  \includegraphics[width=0.65\linewidth]{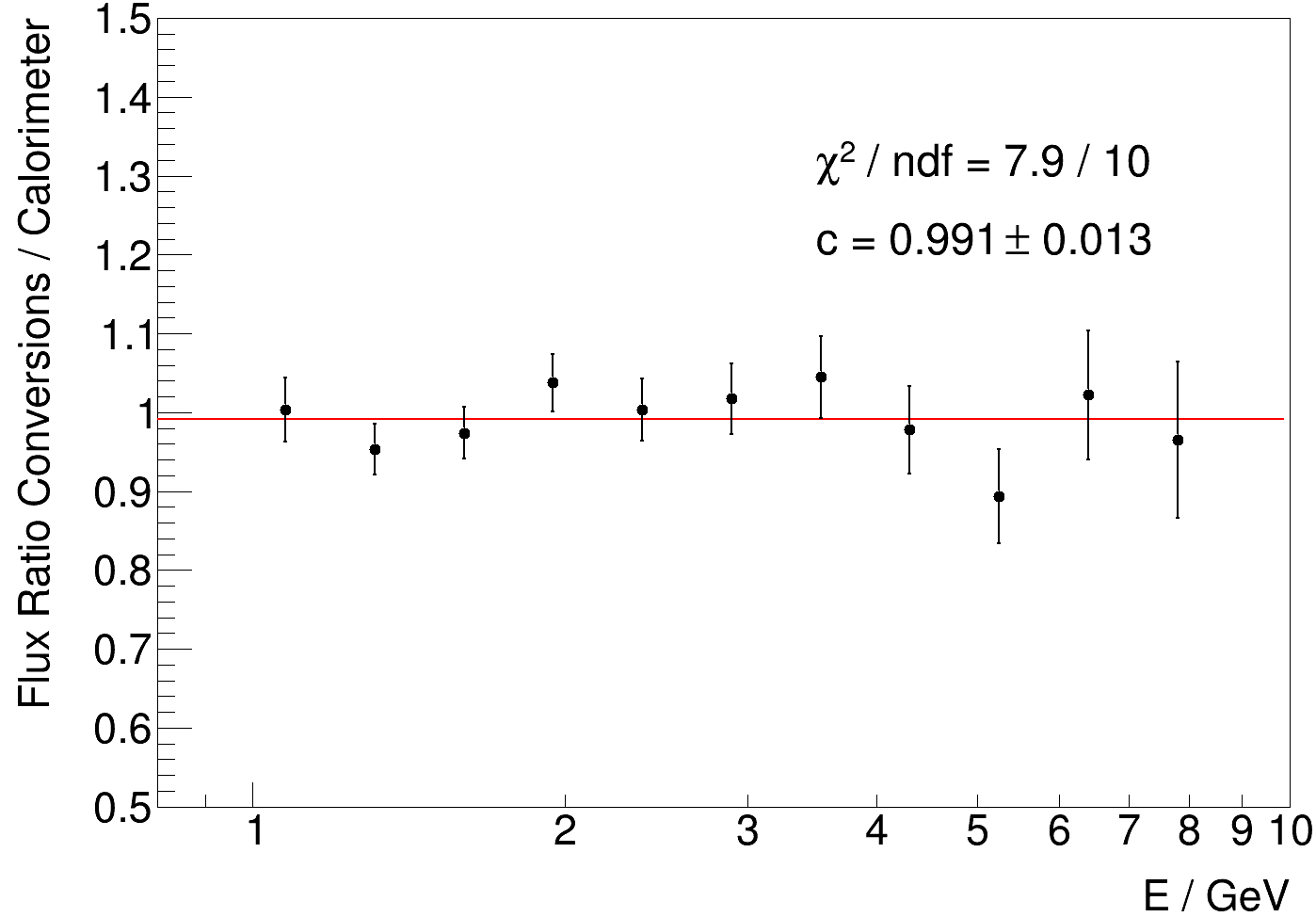}
  \caption{Ratio of fluxes measured with the \mbox{AMS-02} vertex and calorimeter analyses with
    statistical uncertainties. The red line is a constant ($c$) which is fit to the ratio and given
    in the figure together with the $\chi^2/ \mathrm{ndf}$ of the fit.}
  \label{fig:ratio-vertex-ecal}
\end{figure}

In the overlap range between \SI{1}{\giga\electronvolt} and \SI{10}{\giga\electronvolt} the two
complementary \mbox{AMS-02} analysis modes are in excellent agreement with each
other. Figure~\ref{fig:ratio-vertex-ecal} shows the ratio of the fluxes reconstructed with the two
\mbox{AMS-02} analyses modes in the overlap region. The error bars on the ratio were constructed
from the two individual analyses by error propagation and represent statistical uncertainties only.

The average of the ratio is compatible with unity within one standard deviation as indicated by the
constant line fit, which is shown in red in the figure. The $\chi^2 / \mathrm{ndf}$ of the fit is
very good, the fluctuations of the data around the red line are purely statistical.

Because the two analyses were performed using completely different parts of the \mbox{AMS-02}
detector, the agreement between the two results indicates that neither analysis is influenced by
sizable systematic uncertainties. Based on the result of the constant line fit, the associated
systematic uncertainty is \SI{1}{\percent}. This is to be contrasted with the individual
uncertainties of \SI{1.2}{\percent} (conversions) and \SI{2.2}{\percent} (calorimeter),
respectively. However, as discussed in section~\ref{sec:analysis-systematic-uncertainties}, both
analyses are affected equally by the uncertainty on the TRD pileup weight, which introduces an
additional normalization uncertainty of \SI{3}{\percent}.

\begin{figure}[t]
  \centering
  \includegraphics[width=0.95\linewidth]{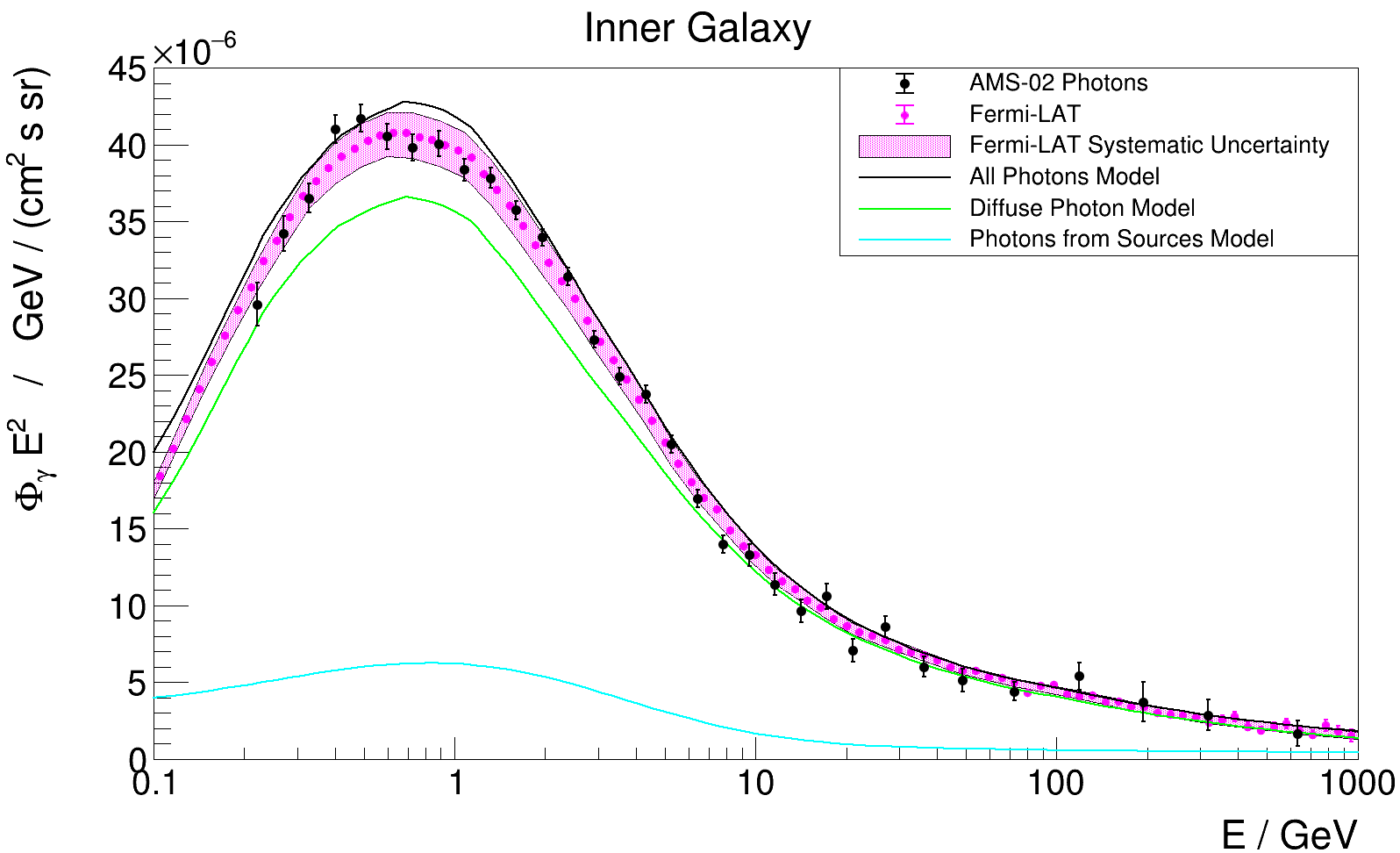}
  \caption{Average photon flux in the inner galaxy multiplied by $E^2$ as a function of the photon
    energy. Black points: Weighted average of two \mbox{AMS-02} analysis modes, uncertainties are
    statistical only. Magenta points: \mbox{Fermi-LAT} measured flux. Magenta band: \mbox{Fermi-LAT}
    systematic uncertainty. The other components are the same as in
    figure~\ref{fig:flux-inner-galaxy}.}
  \label{fig:flux-inner-galaxy-linear-average}
\end{figure}

Given the agreement of the results of the two \mbox{AMS-02} analysis modes within their statistical
uncertainties, it is useful to construct the weighted average, which represents the combined final
\mbox{AMS-02} result, and is shown in figure~\ref{fig:flux-inner-galaxy-linear-average} in black
markers. The figure also includes the flux measured by the \mbox{Fermi-LAT} experiment, which was
obtained using the count map, exposure map, background model and migration matrix presented in
chapter~\ref{sec:fermi-lat-analysis}.

For the \mbox{Fermi-LAT} result (magenta) the absolute energy scale was decreased by
\SI{3.3}{\percent} as suggested in~\cite{Fermi_AllElec_2017}. Also shown is the systematic
uncertainty band for the \mbox{Fermi-LAT} result, which was obtained by first calculating the
uncertainty on the flux due to the absolute energy scale uncertainty:

\begin{equation}
  \label{eq:flux-energy-scale-error}
  \frac{\sigma_\Phi}{\Phi} = \left|\gamma - 1\right| \frac{\sigma_E}{E} \,,
\end{equation}

where $\gamma$ is the local spectral index and $\sigma_E / E$ is shown in
figure~\ref{fig:fermi-lat-systematics}. The local spectral index is estimated using local power law
fits. The final systematic uncertainty band is obtained by adding the systematic uncertainty on the
effective area (shown in figure~\ref{fig:fermi-lat-systematics} in black) in quadrature to the flux
energy scale uncertainty.

The \mbox{AMS-02} results are in excellent agreement with the \mbox{Fermi-LAT} results within their
respective uncertainties. Although AMS can not beat the dedicated \mbox{Fermi-LAT} experiment in
terms of pure photon statistics, valuable information is contained in the new result. The systematic
uncertainty of the \mbox{Fermi-LAT}, indicated by the magenta band, represents an uncertainty which
cannot be reduced unless data from another experiment is considered. Since there are very few
experiments capable of measuring $\gamma$-rays in the Fermi energy range the AMS data is more than
just an independent verification and, when combined with the Fermi data, allows to reduce the
overall uncertainty.

\begin{figure}[t]
  \centering
  \includegraphics[width=0.95\linewidth]{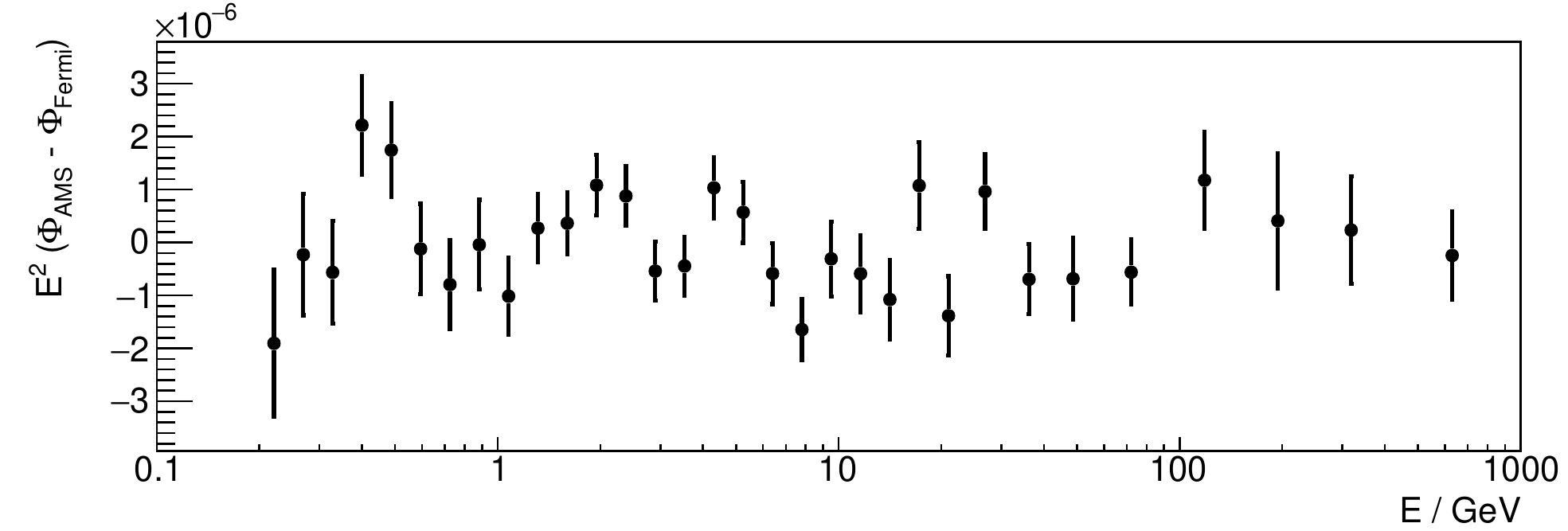}
  \caption{The difference between the measured fluxes by \mbox{AMS-02} and \mbox{Fermi-LAT} in the
    inner galaxy, multiplied by $E^{2}$. The uncertainty on each point is the quadratic sum of the
    respective statistical uncertainties of the individual fluxes.}
  \label{fig:residuals-ams-fermi}
\end{figure}

Figure~\ref{fig:residuals-ams-fermi} shows the residuals of the AMS measurements compared to the
\mbox{Fermi-LAT} result, which also highlights the excellent agreement. The comparison is made using
statistical uncertainties only, since the most important systematic uncertainties affect only the
normalization.

\begin{figure}[t]
  \centering
  \includegraphics[width=0.75\linewidth]{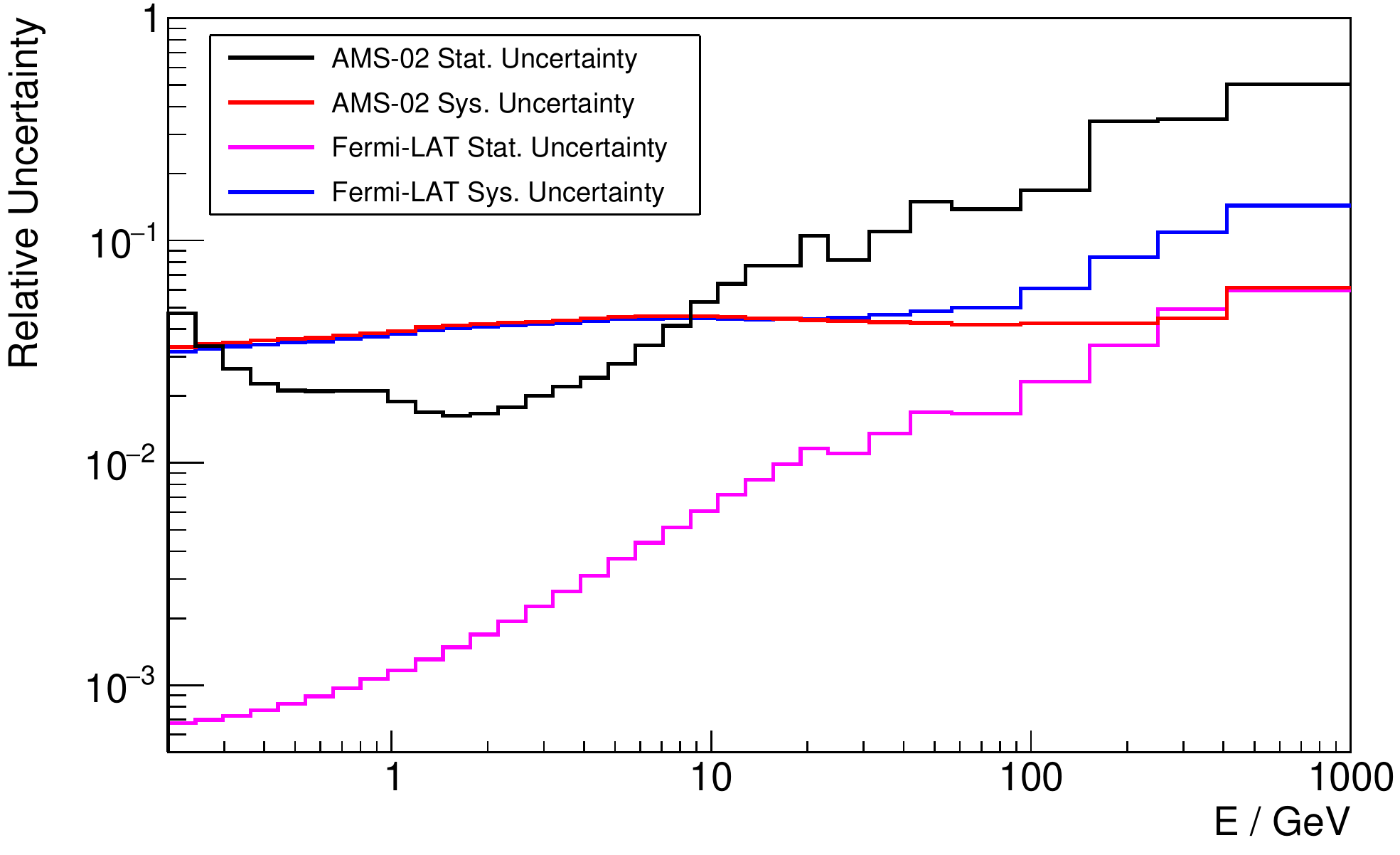}
  \caption{Comparison of the uncertainties of the two \mbox{AMS-02} and \mbox{Fermi-LAT}
    results. Black: Statistical uncertainty of the AMS photon flux in the inner galaxy. Red:
    Corresponding systematic uncertainty, including absolute energy scale uncertainty. Magenta:
    Statistical uncertainty of the \mbox{Fermi-LAT} photon flux (in the AMS binning). Blue: Total
    \mbox{Fermi-LAT} systematic uncertainty, including energy scale.}
  \label{fig:errors-ams-fermi}
\end{figure}

Figure~\ref{fig:errors-ams-fermi} shows a comparison of the statistical and systematical
uncertainties associated with the two results. The \mbox{Fermi-LAT} measurement is dominated by
systematic uncertainties over the entire energy range. In contrast, the \mbox{AMS-02} result is
limited by statistics above \SI{10}{\giga\electronvolt}.

The total systematic uncertainty as well as the absolute energy scale uncertainty are similar at
intermediate energies. At the highest energies the \mbox{AMS-02} energy scale uncertainty is
smaller, due to the thicker calorimeter which suffers less from rear leakage. In addition the
calibration of the calorimeter is better understood, because the flight model was tested extensively
in a beamtest at CERN. This leads to a smaller total systematic uncertainty above
\SI{100}{\giga\electronvolt}. The AMS systematic uncertainty is dominated by the normalization
uncertainty due to the TRD pileup weight of \SI{3}{\percent}.

The systematic uncertainties of the two results are almost the same below approximately
\SI{100}{\giga\electronvolt}. For this reason the magenta band in
figure~\ref{fig:flux-inner-galaxy-linear-average} is also indicative of the AMS systematic
uncertainty. Above \SI{100}{\giga\electronvolt} the AMS result is dominated by statistical
uncertainties.

\begin{figure}[p]
  \centering
  \includegraphics[width=0.95\linewidth]{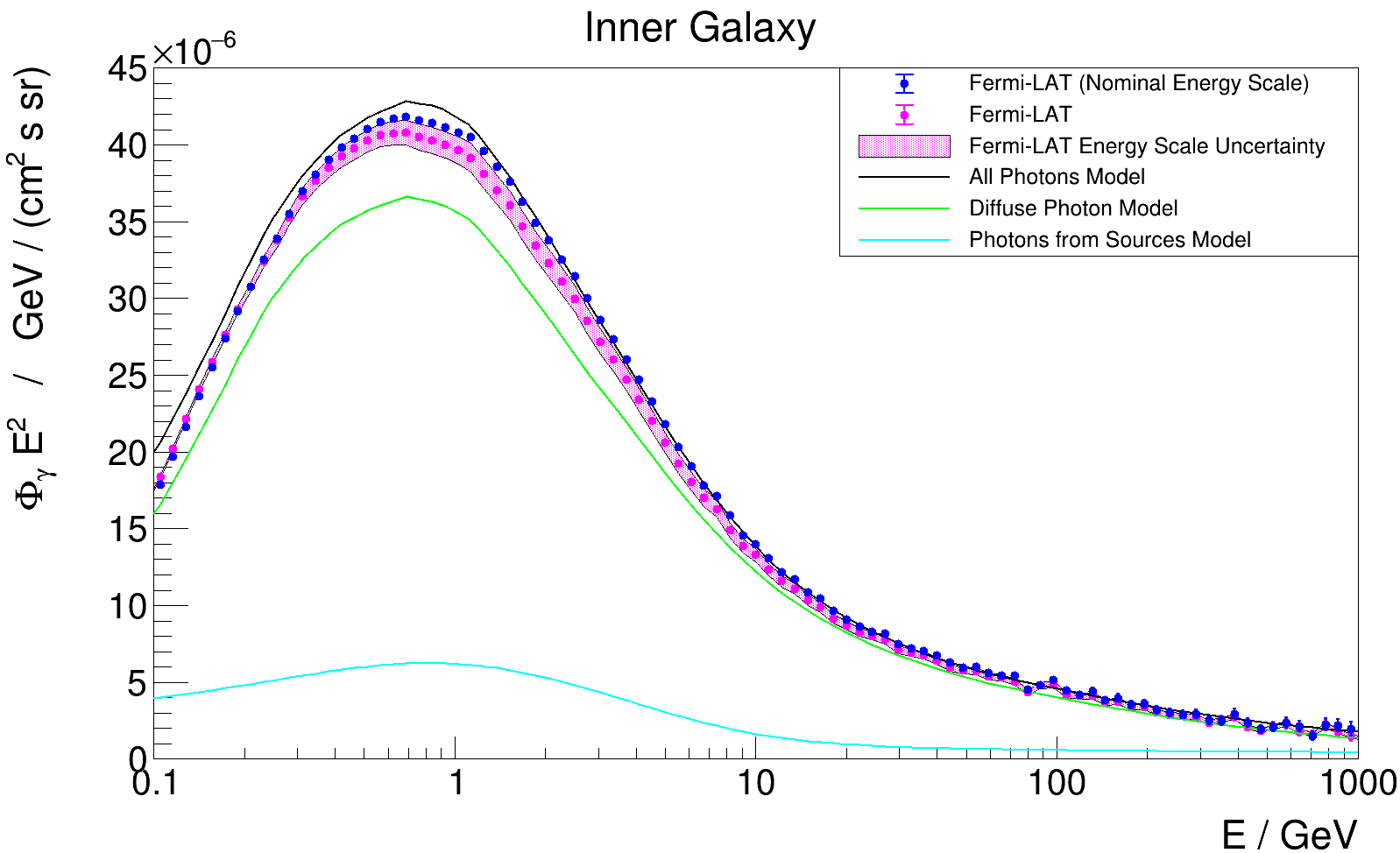}
  \caption{The \mbox{Fermi-LAT} reconstructed flux in the inner galaxy
    ($\SI{-20}{\degree} < l < \SI{80}{\degree}$, $|b| < \SI{8}{\degree}$), using the nominal (blue)
    and corrected (magenta) energy scales. The magenta band is the energy scale systematic
    uncertainty.}
  \label{fig:fermi-lat-flux-unfolding-fluxes}

  \vspace*{3\floatsep}

  \centering
  \includegraphics[width=0.95\linewidth]{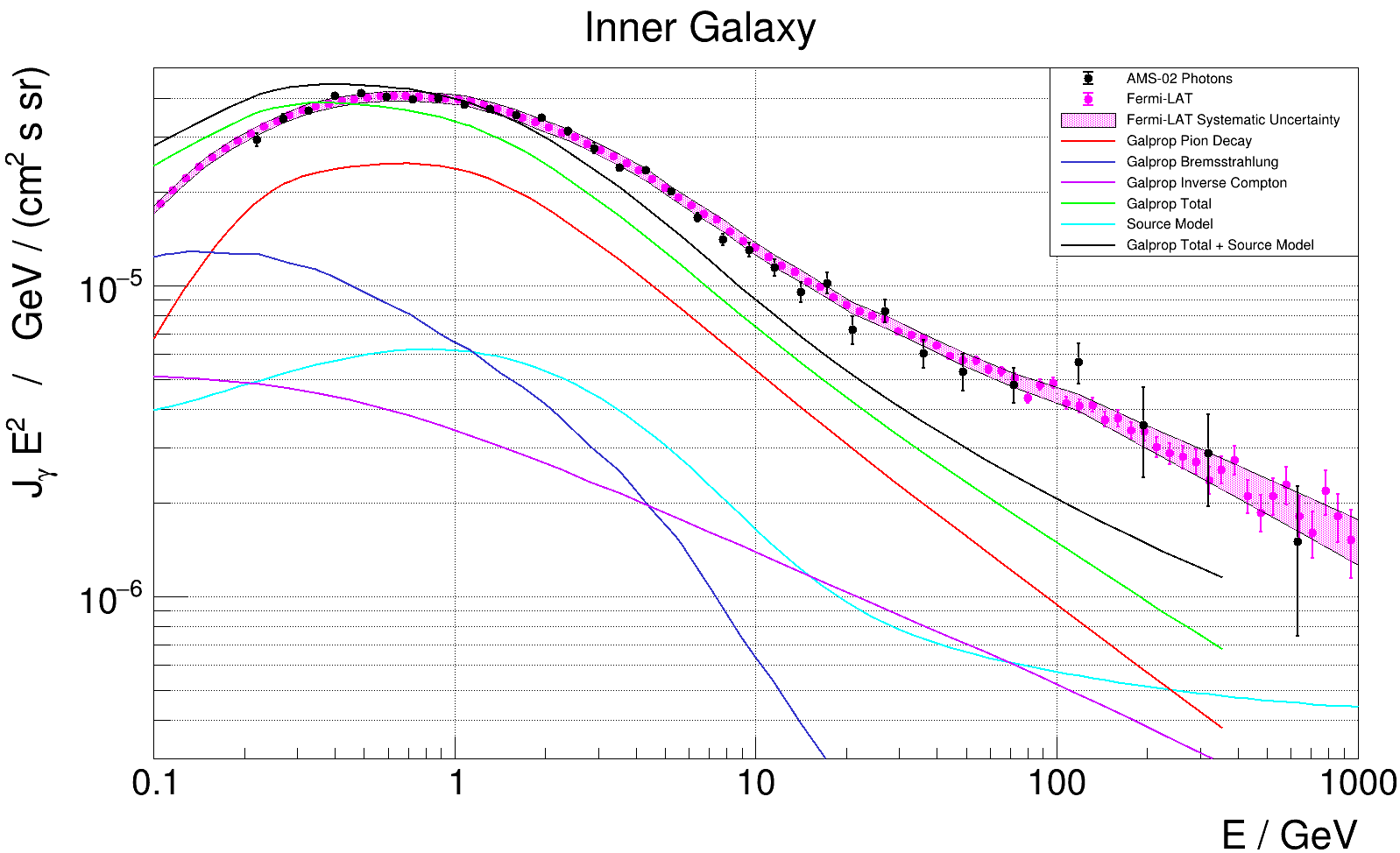}
  \caption{Comparison of the measured $\gamma$-ray flux in the inner galaxy with a recent GALPROP
    model~\cite{Johannesson_APJ_856}. AMS uncertainties are statistical only. GALPROP components:
    Fluxes for $\pi^0$-decays (red), bremsstrahlung (blue) and inverse Compton emission
    (magenta). Green line: Sum of GALPROP predictions for diffuse emission. Cyan line: Summed model
    prediction for $\gamma$-ray sources. Black line: Sum of green and cyan lines, corresponding to
    the total model prediction.}
  \label{fig:flux-galprop-comparison}
\end{figure}

The \mbox{Fermi-LAT} flux in figure~\ref{fig:flux-inner-galaxy-linear-average} included a correction
for the \SI{3.3}{\percent} energy scale shift that was found in the Fermi electron analysis, as
discussed in
section~\ref{sec:fermi-lat-corrections}. Figure~\ref{fig:fermi-lat-flux-unfolding-fluxes} shows the
effect of this energy scale bias correction on the measured flux in the inner galaxy. With the
nominal \mbox{Fermi-LAT} energy scale without correction, the blue flux is obtained. This result is
systematically above the magenta result, which is the same as the one shown in
figures~\ref{fig:flux-inner-galaxy} and~\ref{fig:flux-inner-galaxy-linear-average}. For energies
above \SI{1}{\giga\electronvolt} the difference is approximately \SI{5}{\percent}, on average, and
barely covered by the systematic energy scale uncertainty shown in magenta in the figure.

The \mbox{AMS-02} results are in excellent agreement with the energy scale corrected Fermi results,
but not with the nominal ones. In fact, fitting a constant to the ratio of the AMS flux to the
nominal Fermi-LAT inner galaxy flux yields $c = 0.966 \pm 0.005$, i.e. a deviation from unity by 6.8
standard deviations, whereas the ratio of the flux to the corrected Fermi result is compatible with
unity with $c = 1.002 \pm 0.005$. This, indicates that the energy scale shift is indeed correctly
determined in the \mbox{Fermi-LAT} electron flux publication~\cite{Fermi_AllElec_2017} and needs to
be considered in photon flux measurements. This new result provides valuable information, and if
confirmed, has direct consequences for other \mbox{Fermi-LAT} photon measurements.

This study also shows that the agreement between the Fermi measured flux and the Fermi diffuse model
improves if the nominal energy scale is used, which is not surprising since that is how the diffuse
model was obtained.

Figure~\ref{fig:flux-galprop-comparison} shows a comparison of the measured \mbox{AMS-02} flux with
a GALPROP model calculation~\cite{Johannesson_APJ_856}, in which the new 3D gas model is combined
with CR source distribution, which puts \SI{50}{\percent} of the sources in the spiral arms and
distributes the rest evenly across the galaxy (``SA50-3D gas'' in the publication).

The model was tuned to reproduce the measured \mbox{AMS-02} cosmic ray fluxes of protons, helium and
electrons. This includes a spectral break of both protons and helium nuclei at \SI{200}{\giga\volt}
rigidity in the injection spectrum, which was introduced in order to reproduce the break observed in
the \mbox{AMS-02} data. This break is important for the prediction of $\pi^0$ decays at intermediate
and high energies. The model was also tuned to reproduce the boron over carbon ratio observed by
AMS.

This model is a specific example, but other GALPROP models in
recent~\cite{Johannesson_APJ_856,Porter_APJ_846} as well as older~\cite{Fermi_Diffuse_2012}
publications have qualitatively similar spectra, although the spatial predictions show interesting
variations, which are discussed in the publications. This was verified by calculating the model
predictions with GALPROP versions 54 and 56. The model parameter files and the corresponding input
data files (containing the gas maps, radiation fields and nuclear cross sections) were obtained from
the GALPROP website~\cite{Galprop_WWW}, the supplemental material of the
publications~\cite{Fermi_Diffuse_2012} and from the authors themselves, which allowed to reproduce
the published results.

The GALPROP model includes predictions for the $\pi^0$-decay (red), bremsstrahlung (blue) and IC
(magenta) components of diffuse emission. The green line shows the sum of all three
components. These lines end at approximately \SI{350}{\giga\electronvolt}, since that is the upper
energy limit in the GALPROP model. As before, the $\gamma$-ray flux from sources in the inner galaxy
window is calculated from the 4FGL and shown in cyan. Adding this component to the GALPROP model
yields the total prediction, shown in black.

Although the spectral shape of the data and the model is similar, the model does not fit the
\mbox{AMS-02} data. In particular, the diffuse model peak position appears shifted and its spectral
index above \SI{3}{\giga\electronvolt} indicates a softer spectrum compared to the data. The
predicted model flux can be well approximated by a single power law above
\SI{3}{\giga\electronvolt}, even though the spectral break in the proton spectrum is included in the
model. The measured flux on the other hand hardens at approximately
\SI{30}{\giga\electronvolt}. These spectral discrepancies are the main reason why the model for
galactic diffuse emission derived by \mbox{Fermi-LAT} was used as the primary diffuse model in this
thesis.

\subsection*{The Cygnus-X Star Forming Region}
\label{sec:results-cygnus-x}

\begin{figure}[t]
  \centering
  \includegraphics[width=0.9\linewidth]{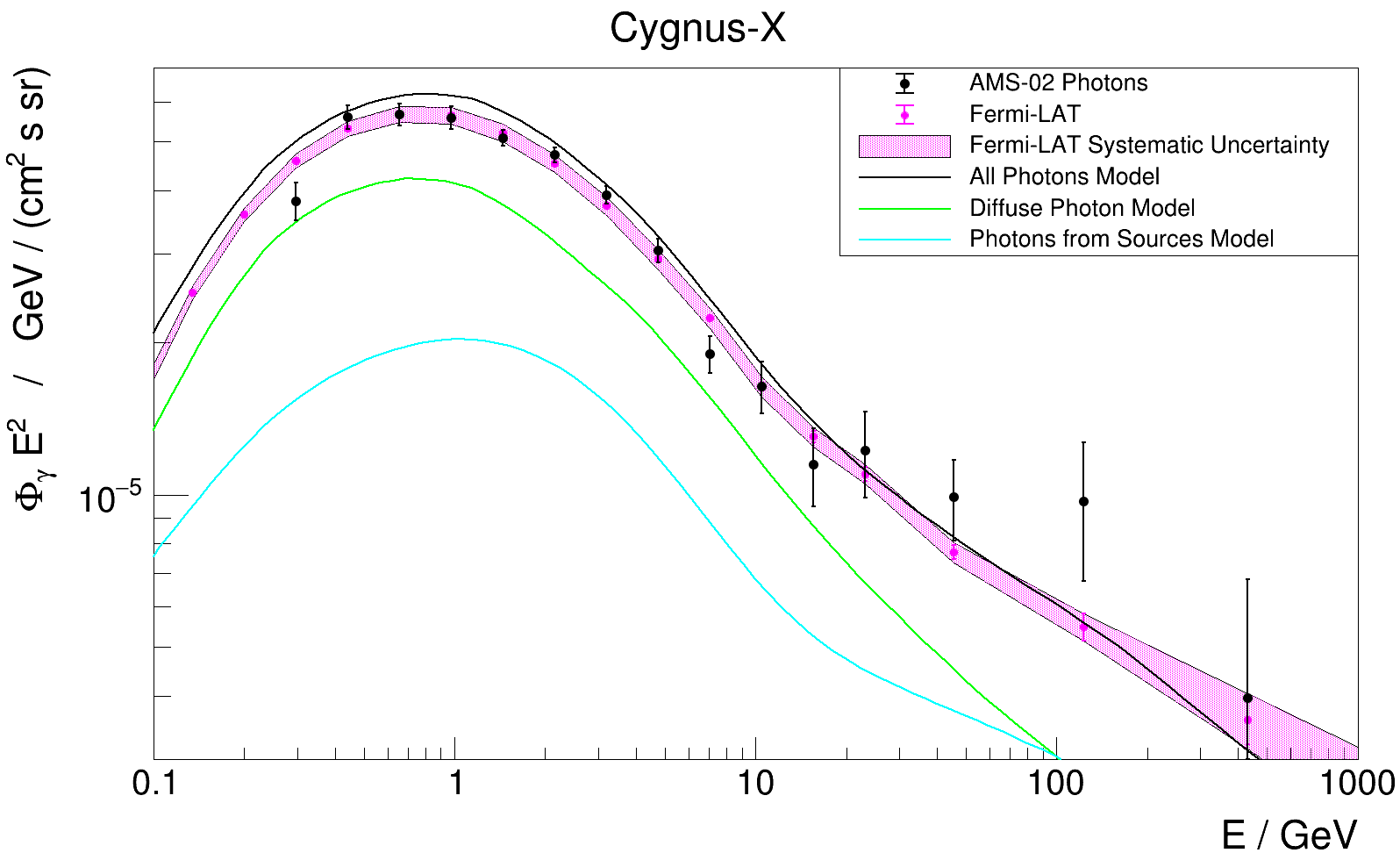}
  \caption{Average photon flux from the \mbox{Cygnus-X} region, multiplied by $E^2$ as a function of
    the photon energy. See figures~\ref{fig:flux-inner-galaxy}
    and~\ref{fig:flux-inner-galaxy-linear-average} for an explanation of the components.}
  \label{fig:flux-cygnus-x-log-average}
\end{figure}

Figure~\ref{fig:flux-cygnus-x-log-average} shows the photon flux in a \SI{5}{\degree} window
enclosing the \mbox{Cygnus-X} star forming region, centered at
$(l, b) = (\SI{78.2}{\degree}, \SI{2.0}{\degree})$. This region is located in the Cygnus
constellation, close to the star $\gamma$-Cygni. It is a strong source of diffuse $\gamma$-ray
emission, due to the abundance of freshly accelerated cosmic rays as well as the special structure
of gas and magnetic fields in this complex region. In addition, multiple $\gamma$-ray producing
sources have been identified in this window, including two pulsars and the $\gamma$-Cygni SNR.

The agreement between the \mbox{AMS-02} data and the \mbox{Fermi-LAT} result is very good. The model
slightly over-predicts the $\gamma$-ray flux in this region.

\section{Spectra from $\gamma$-ray Sources}
\label{sec:results-sources}

\subsection*{The Radio-Quiet Pulsar Geminga}
\label{sec:results-geminga}

Among the strongest sources in the $\gamma$-ray sky are three pulsars: The Vela pulsar (\mbox{PSR
  J0835-4510}), the Geminga pulsar (\mbox{PSR J0633+1746}) and the pulsar at the center of the Crab
Nebula (\mbox{PSR J0534+2200}). As an example, the flux from Geminga will be studied in this
section. Geminga is special, since it is one of only a few radio-quiet pulsars, i.e. up to now no
radio emissions from Geminga have been identified. At the same time Geminga is fairly close to the
solar system: It's distance is estimated to be approximately
\SI{160}{\parsec}~\cite{Geminga_Distance_1996}.

Because of its proximity Geminga was suggested as a possible source of high energy positrons, which
could possibly explain the rise in the positron fraction above
\SI{10}{\giga\electronvolt}~\cite{Yuksel2009}. However, measurements by the HAWC telescope disfavor
this explanation~\cite{HAWC_Pulsars_2017}.

\begin{figure}[t]
  \centering
  \includegraphics[width=0.95\linewidth]{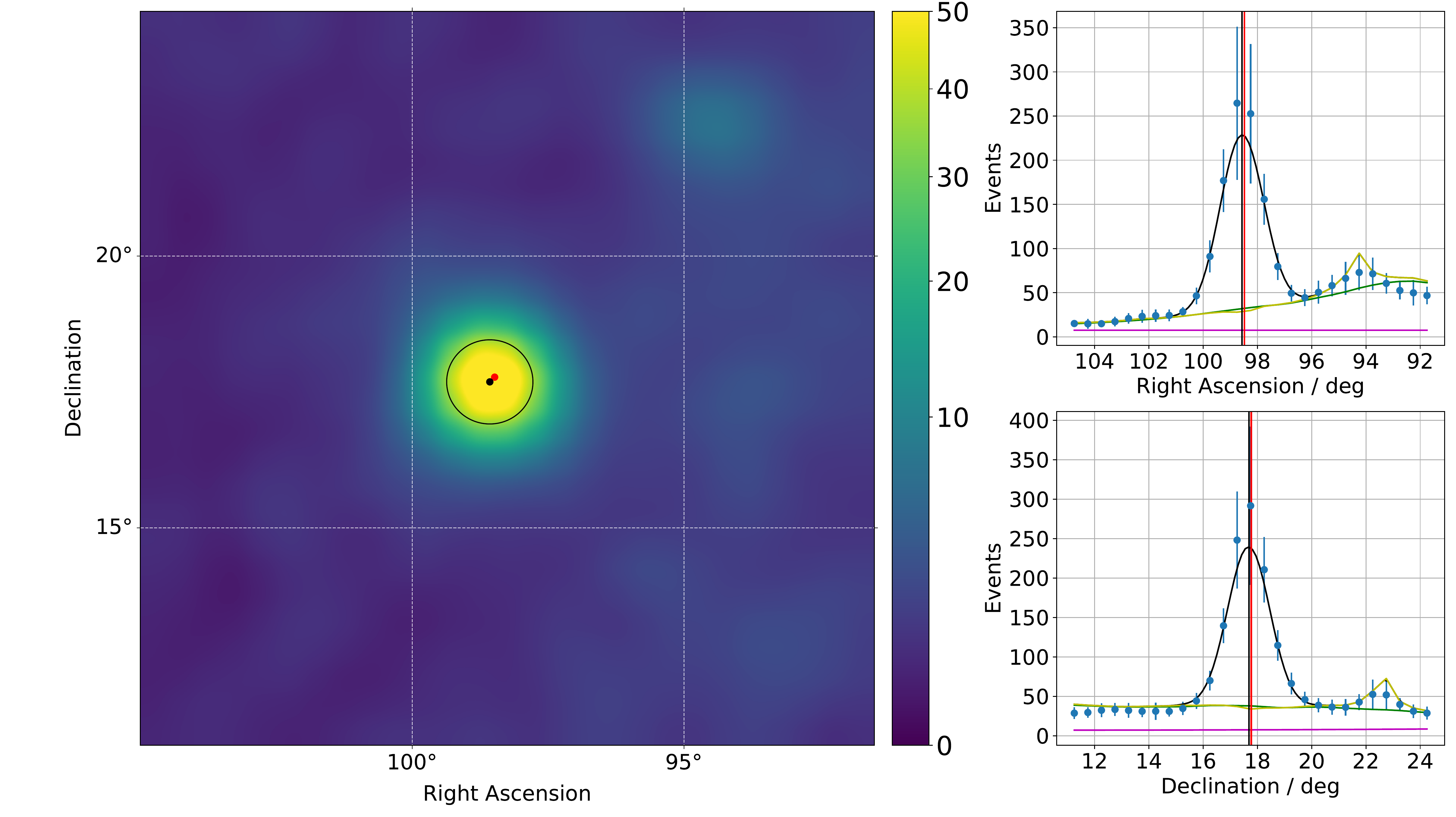}
  \caption{Geminga pulsar in \mbox{AMS-02} conversion data. Left: Skymap in equatorial coordinates
    showing integrated event counts between \SI{0.5}{\giga\electronvolt} and
    \SI{100}{\giga\electronvolt}. Top right: Projection on the right ascension axis. Bottom right:
    Projection on the declination axis. See text for description of components in the figures.}
  \label{fig:flux-geminga-source-position}
\end{figure}

Figure~\ref{fig:flux-geminga-source-position} shows a \mbox{AMS-02} conversion mode count map for a
$\SI{14}{\degree} \times \SI{14}{\degree}$ section of the sky centered on the Geminga pulsar in
equatorial coordinates in the large plot on the left side. The smaller, right hand side, figures
show projections onto the two axes. The 4FGL catalog position of the pulsar is indicated by the red
dot in the center of the left figure and by red vertical lines in the right hand side projections.

The two dimensional distribution, shown in color, is fit with a background plus signal model, in
which the background is held fixed and modeled by the predetermined background component (see
section \ref{sec:corrections-background}), the diffuse emission model and the sum of the fluxes of
all sources in the window except Geminga. The signal model is a two-dimensional Gaussian with width
determined by the energy dependent PSF of the conversion mode analysis. The mean values (the
position) of the Gaussian and it's normalization are left free in the fit.

The background (violet), diffuse emission (green), other sources (yellow) and total model (black)
after the fit are shown in the one dimensional projections. Overall the model agrees very well with
the measured data. The small bump in the yellow curve on the right hand side of the top right figure
is the Supernova Remnant \mbox{IC 443}.

Resulting from the fit is the source position
($\alpha = 6^{\mathrm{h}} 34^{\mathrm{m}} 08^{\mathrm{s}} \pm 14^{\mathrm{s}}, \delta = 17^{\circ}
40^{\prime} \pm 4^{\prime}$), which agrees within uncertainties with the catalog position
($\alpha = 6^{\mathrm{h}} 33^{\mathrm{m}} 56^{\mathrm{s}}, \delta = 17^{\circ}
46^{\prime}$)~\cite{Fermi_4FGL_2019}. This demonstrates that there is no significant bias in the
pointing of the AMS detector or the reconstruction of the photon directions. The reconstructed
position is indicated by the black dot in the two dimensional plot and by the vertical black lines
in the projections.

\begin{figure}[t]
  \centering
  \includegraphics[width=0.95\linewidth]{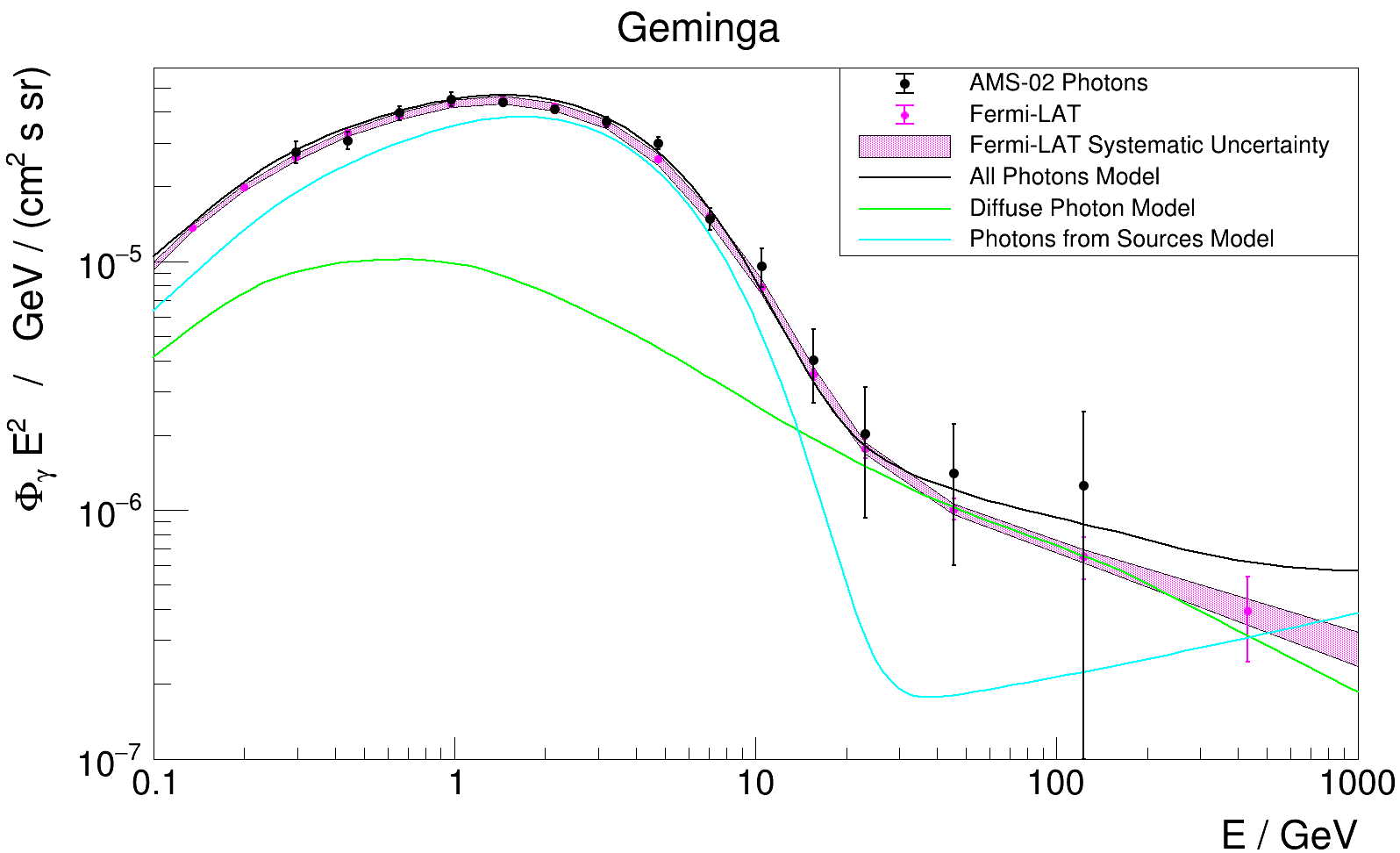}
  \caption{Average photon flux from the Geminga region, multiplied by $E^2$ as a function of the
    photon energy. See figures~\ref{fig:flux-inner-galaxy}
    and~\ref{fig:flux-inner-galaxy-linear-average} for an explanation of the components.}
  \label{fig:flux-geminga-log-average}
\end{figure}

The flux of the pulsar is obtained in a \SI{5}{\degree} window, centered around the 4FGL position of
Geminga, $(l, b) = (\SI{195.0}{\degree}, \SI{3.9}{\degree})$. The results from both \mbox{AMS-02}
and \mbox{Fermi-LAT} are shown in figure~\ref{fig:flux-geminga-log-average}. The spectrum of Geminga
is exponentially cut-off around \SI{5}{\giga\electronvolt}. Above \SI{10}{\giga\electronvolt}
diffuse emission dominates the window flux.

The flux measured by \mbox{AMS-02} agrees well with the \mbox{Fermi-LAT} flux measurement. In
addition, it also agrees well with the model over the entire energy range.

\begin{figure}[t]
  \begin{minipage}{0.48\linewidth}
    \centering
    \includegraphics[width=1.0\linewidth]{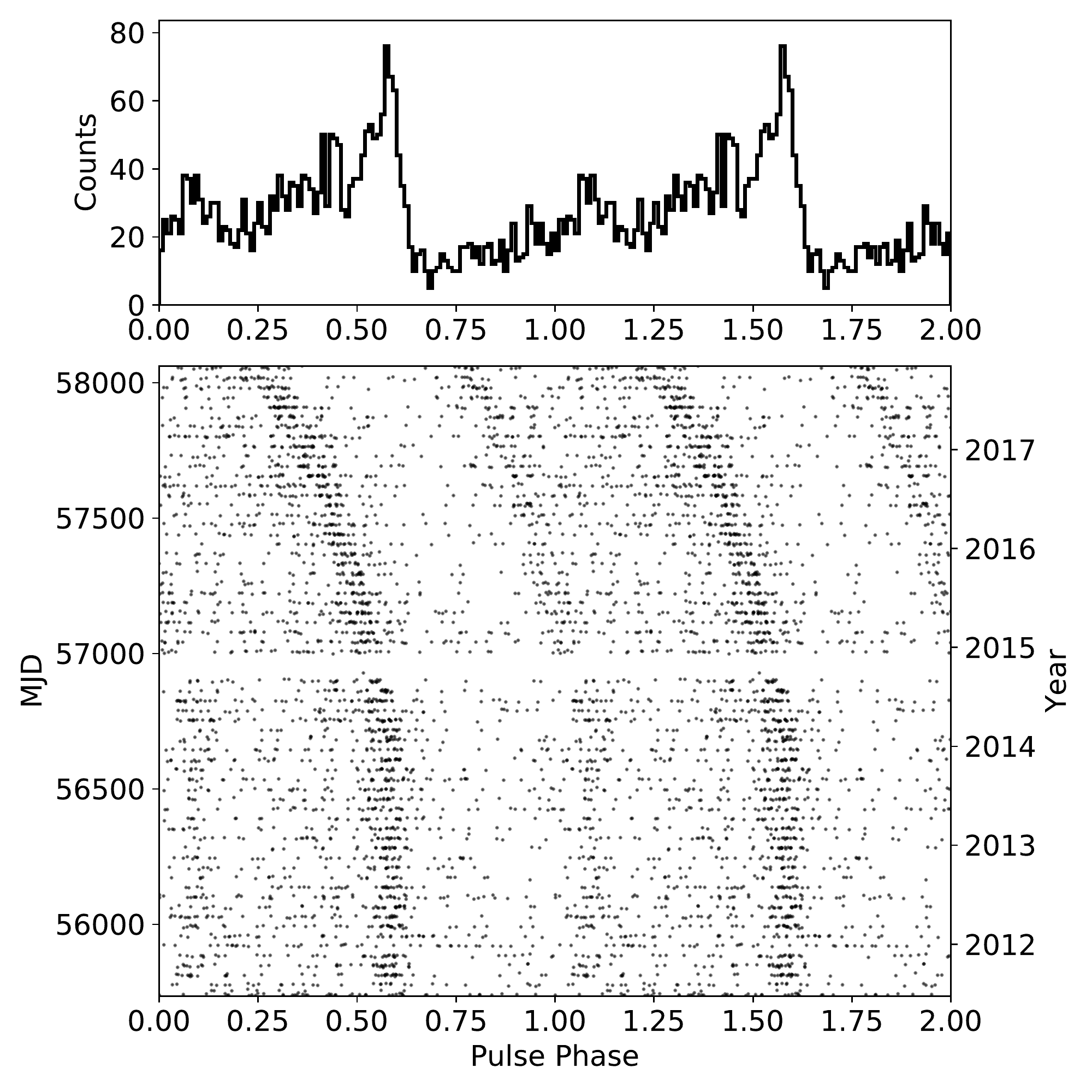}
  \end{minipage}
  \hspace{0.01\linewidth}
  \begin{minipage}{0.48\linewidth}
    \centering
    \includegraphics[width=1.0\linewidth]{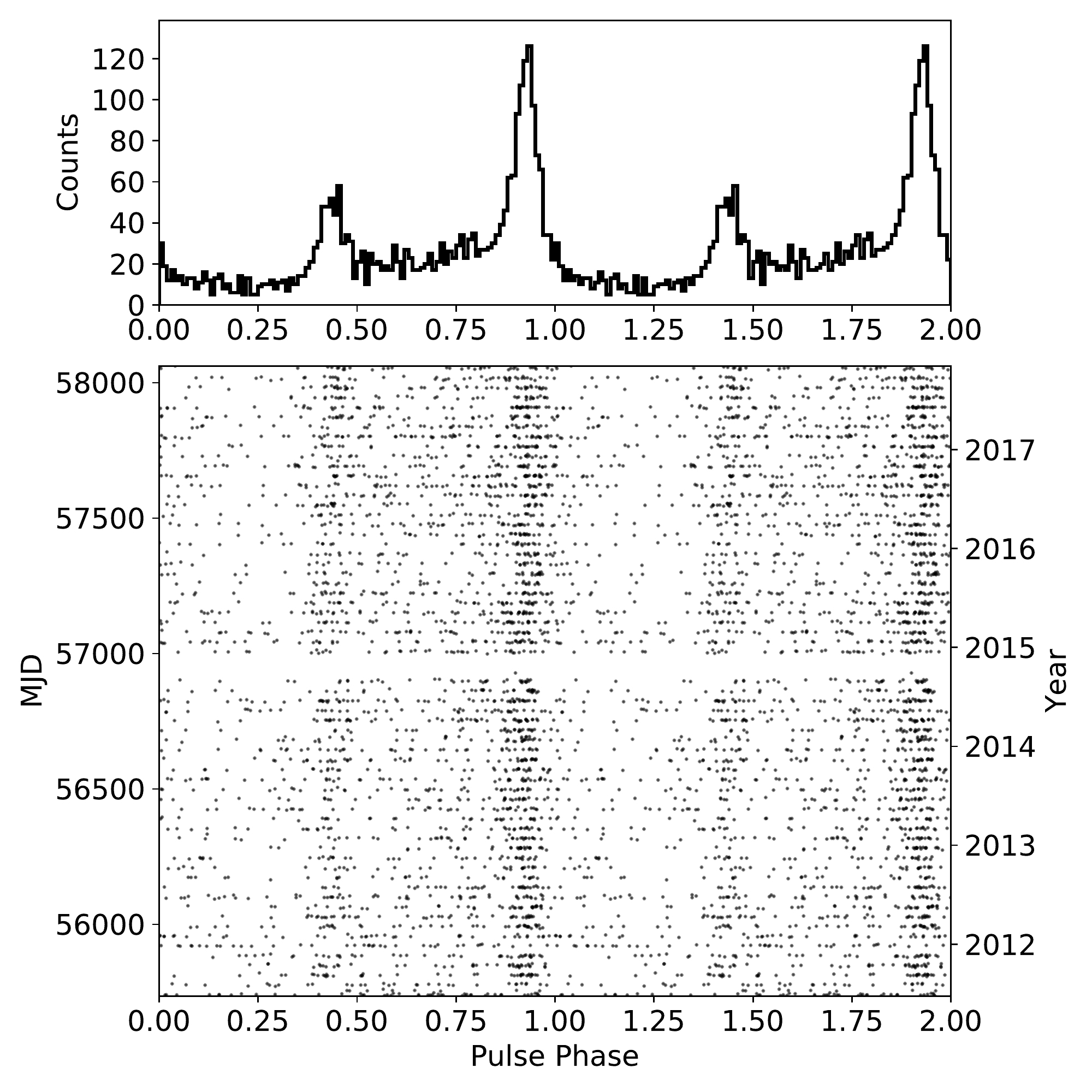}
  \end{minipage}
  \caption{Top: Phase folded light curve of the Geminga pulsar, before fitting (left) and after
    fitting the pulsars period $P$ and spindown $\dot{P}$ (right). Bottom: Evolution of the phase
    folded light curve with time, before fitting (left) and after (right).}
  \label{fig:geminga-phaseogram}
\end{figure}

The Geminga pulsar can also be identified by its pulsed emission of photons. The \mbox{AMS-02} event
timestamps are assigned by the main DAQ system of the experiment. The clock of the DAQ system is
corrected for drift and synchronized with the GPS system of AMS on a daily basis. As a result the
recorded events can be used for precision timing.

In order to correct for the light travel time as well as relativistic effects the photon arrival
times must be converted from the TAI timescale to the TDB timescale~\footnote{TAI (Temps Atomique
  International) is the International Atomic Time standard and TDB (Temps Dynamique Barycentrique)
  is Barycentric Dynamical Time, a time scale which includes relativistic corrections needed to
  convert times to equivalent instants in the Solar System barycenter.}, which can be done with
appropriate software~\cite{Astropy_2018}. To account for the motion of the Earth and the ISS with
respect to Geminga, the JPL DE430 planetary ephemerides~\cite{JPL_Ephemerides_WWW} are used to
calculate the exact position of the solar system barycenter and to convert the photon arrival times
to the barycentric arrival times. This correction includes the R{\o}mer delay for the orbital motion
of the Earth and the Einstein and Shapiro delay to account for relativistic effects.

Pulsar ephemerides are required to account for the pulsar's proper motion, it's dispersion measure
and to provide good estimates of its position and rotational parameters. Such ephemerides are
available from the radio telescope community, for example from the Australia Telescope National
Facility (ATNF)\cite{ATNF_Catalogue_2005,ATNF_Psrcat_WWW}. Since Geminga is radio-quiet, a timing
model from the \mbox{Fermi-LAT} FSSC\cite{Fermi_PulsarTiming_WWW,Ray_2011} is used here. In the used
timing model the white noise of the pulsar's timing is removed using a tabulated function.

\begin{figure}[p]
  \centering
  \includegraphics[width=0.95\linewidth]{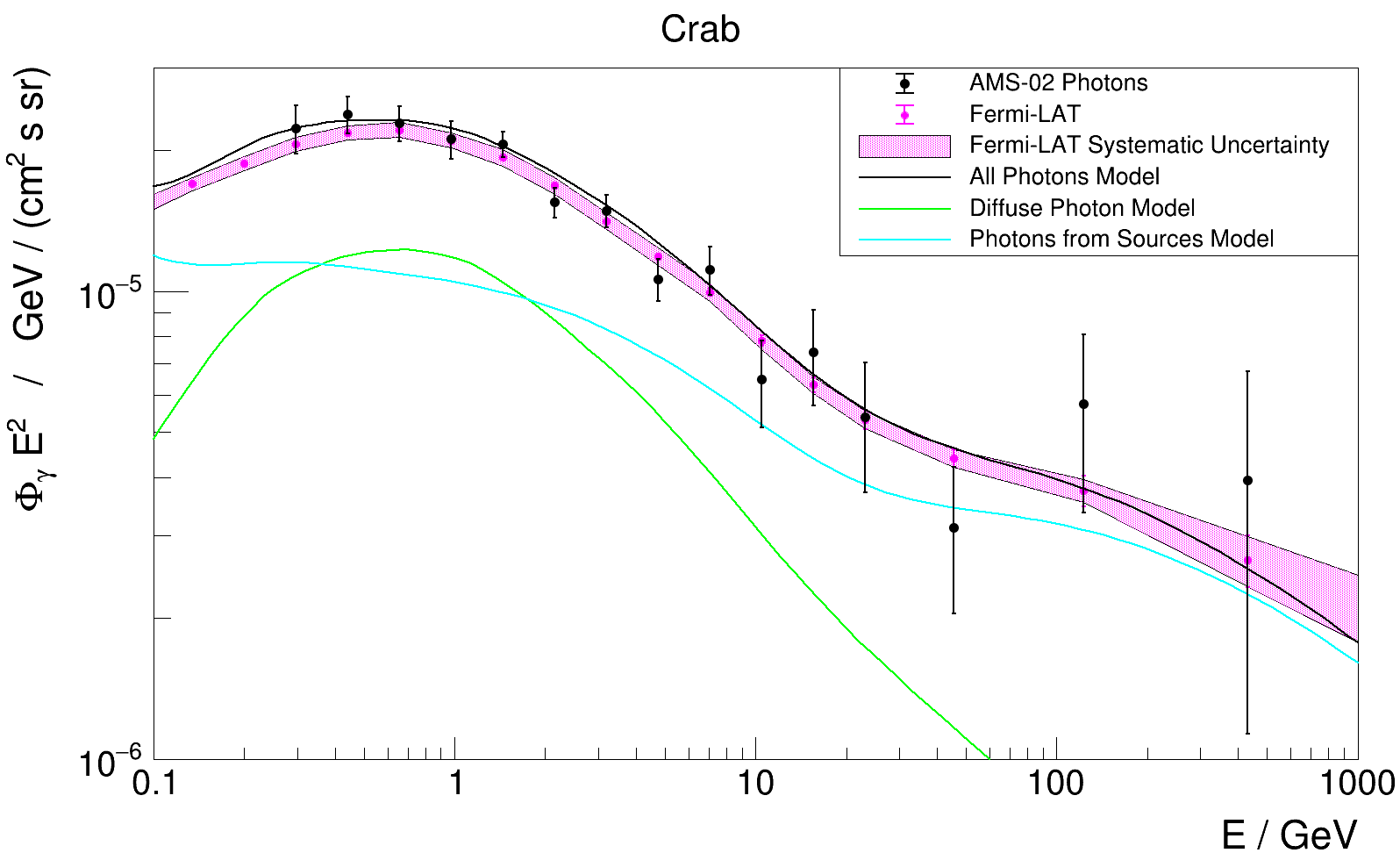}
  \caption{Average photon flux from the Crab region, multiplied by $E^2$ as a function of the photon
    energy. See figures~\ref{fig:flux-inner-galaxy} and~\ref{fig:flux-inner-galaxy-linear-average}
    for an explanation of the components.}
  \label{fig:flux-crab-log-average}

  \vspace*{3\floatsep}

  \centering
  \includegraphics[width=0.95\linewidth]{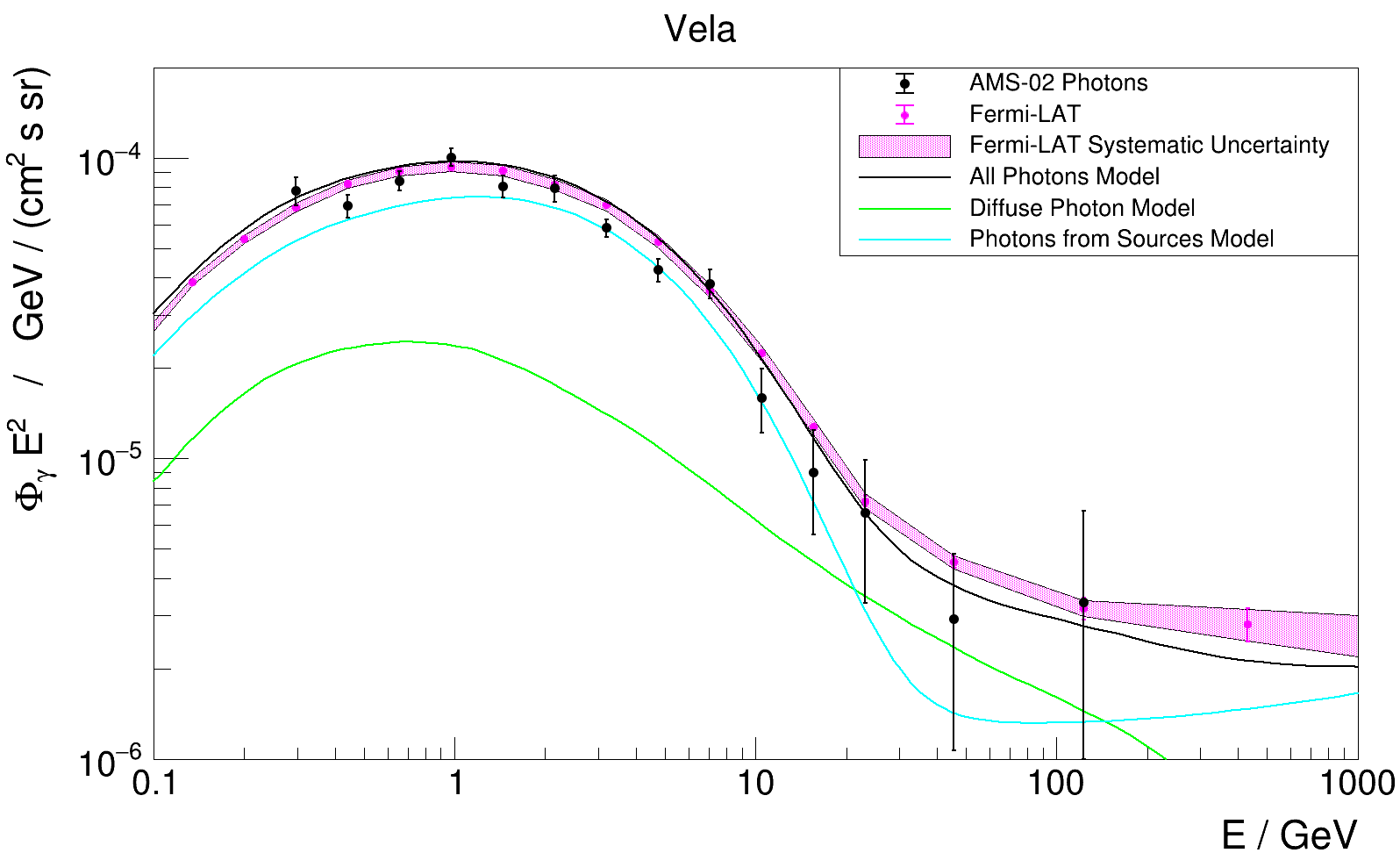}
  \caption{Average photon flux from the Vela region, multiplied by $E^2$ as a function of the photon
    energy. See figures~\ref{fig:flux-inner-galaxy} and~\ref{fig:flux-inner-galaxy-linear-average}
    for an explanation of the components.}
  \label{fig:flux-vela-log-average}
\end{figure}

Figure~\ref{fig:geminga-phaseogram} shows the phase folded light curve of the Geminga pulsar, using
all events (from both \mbox{AMS-02} analysis modes) between \SI{200}{\mega\electronvolt} and
\SI{10}{\giga\electronvolt}, in a \SI{3}{\degree} window around the pulsar. Note that the same data
is shown twice, which avoids the visual problem of wrapping events at the borders near phase 0 and
1. The left hand side shows the pre-fit result, using the pulsar ephemerides without
modifications. For event times before 2015 the phaseogram is approximately constant, but the pulse
begins to drift away after that point, which is a consequence of the fact that the model is being
extrapolated.

In order to fix this problem the timing solution can be fitted, by using the
TEMPO2~\cite{TEMPO2_2006} and PINT~\cite{PINT_2019} software packages. The result is shown on the
right hand side of figure~\ref{fig:geminga-phaseogram}. This reveals a second peak in the pulse
profile, approximately offset by half a pulsar rotation from the strongest peak. The curve in the
phaseogram has disappeared and the pulse profile is clearly visible from the phase folded light
curve. The fit solutions for the pulsar's rotational period $P$ and spindown $\dot{P}$ at the epoch
(MJD 54800) are:

\begin{align*}
  P &= \SI[separate-uncertainty=true, multi-part-units=brackets]{0.2371035208991(59)}{\second} \\
  \dot{P} &= \SI[separate-uncertainty=true,
            multi-part-units=brackets]{1.0973095(69)e-14}{\second\per\second} \,,
\end{align*}

which places thne Geminga pulsar in the main population of pulsars in the $P\dot{P}$ diagram (see
figure~\ref{fig:pulsars-p-pdot}). The canonical age and surface magnetic field strength of the
pulsar can be estimated using equations~(\ref{eq:pulsar-age}) and~(\ref{eq:pulsar-field-strength}):

\begin{align*}
  \tau &\approx \SI{340000}{\year}\\
  B &> \SI{1.63e12}{\gauss} \,,
\end{align*}

which constitutes a young to middle aged pulsar with a strong magnetic field.

\subsection*{The Crab and Vela Pulsars}
\label{sec:results-crab-vela}

In analogy to the flux from the Geminga region, figures~\ref{fig:flux-crab-log-average}
and~\ref{fig:flux-vela-log-average} show the fluxes obtained in windows around the Crab and Vela
pulsars. In the case of the Crab pulsar the agreement between \mbox{AMS-02} and \mbox{Fermi-LAT} is
excellent. At the highest energies the measured \mbox{AMS-02} flux is incompatible with pure diffuse
photon production, which is a result of the high energy inverse Comton component of the Crab PWN.

The measured flux in the Vela region is slightly lower than the \mbox{Fermi-LAT} result on
average. Although it is not excluded that the Fermi measurement is slightly to high, the discrepancy
could also be a result of the southern location of the source and its proximity to the southern
exposure hole in both AMS analysis modes. The overall agreement of the flux shape is still good
however, and the exponential cutoff is well reproduced.

\subsection*{The Flaring Blazar CTA-102}
\label{sec:results-cta-102}

The Flat Spectrum Radio Quasar \mbox{CTA-102} was discovered in the 1960s in the Caltech radio
survey~\cite{Harris_1960}. It is located at redshift $z \approx 1.032$~\cite{Monroe_2016} and was
once proposed to be home to an extraterrestrial civilization, because of its variability in the
radio band (this was before the discovery of pulsars, which were also associated with
extraterrestrial life at first).

\begin{figure}[p]
  \centering
  \includegraphics[width=0.95\linewidth]{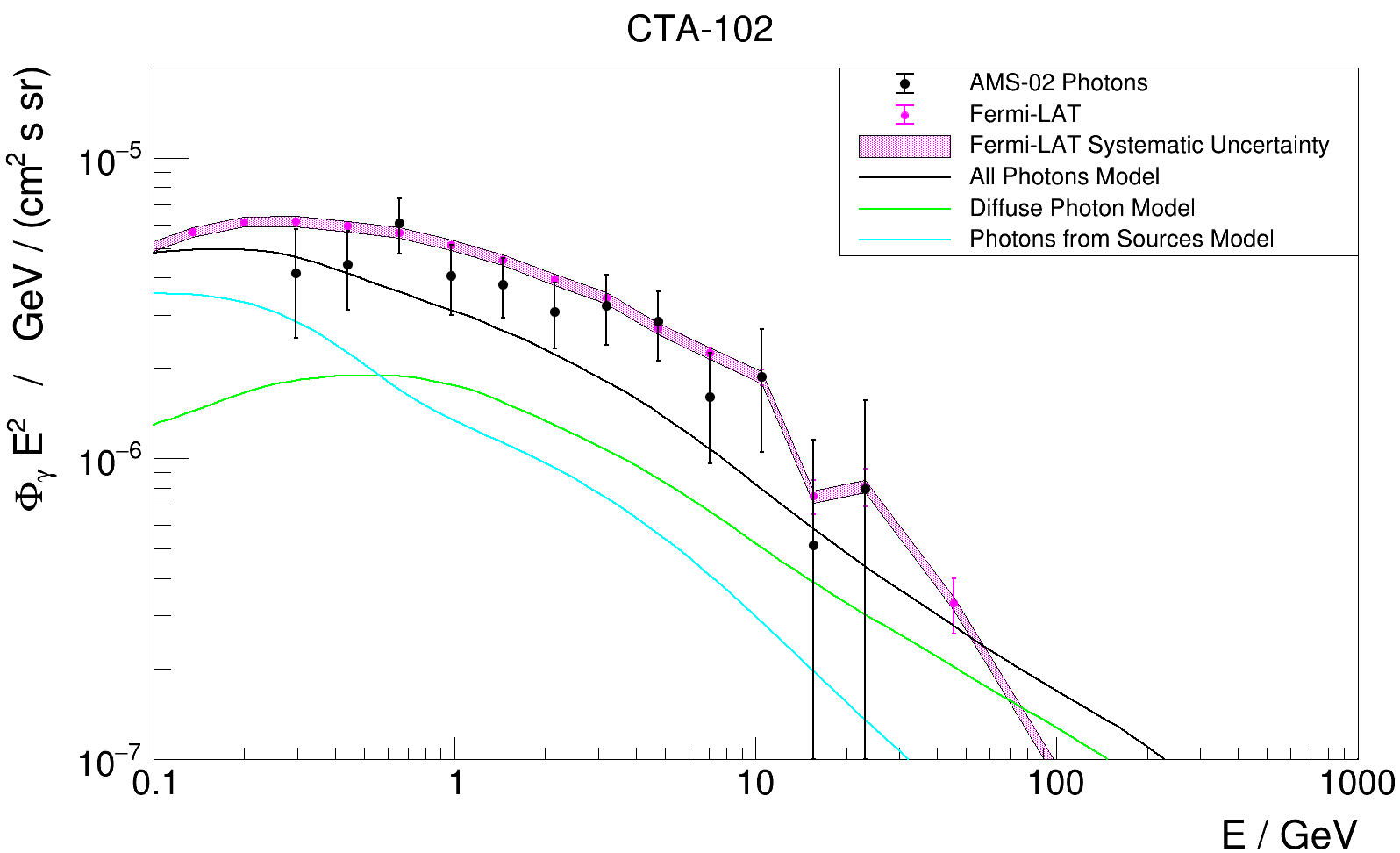}
  \caption{Average photon flux from the \mbox{CTA-102} region, multiplied by $E^2$ as a function of
    the photon energy. See figures~\ref{fig:flux-inner-galaxy}
    and~\ref{fig:flux-inner-galaxy-linear-average} for an explanation of the components.}
  \label{fig:flux-cta-102-log-average}

  \vspace*{3\floatsep}

  \centering
  \includegraphics[width=0.95\linewidth]{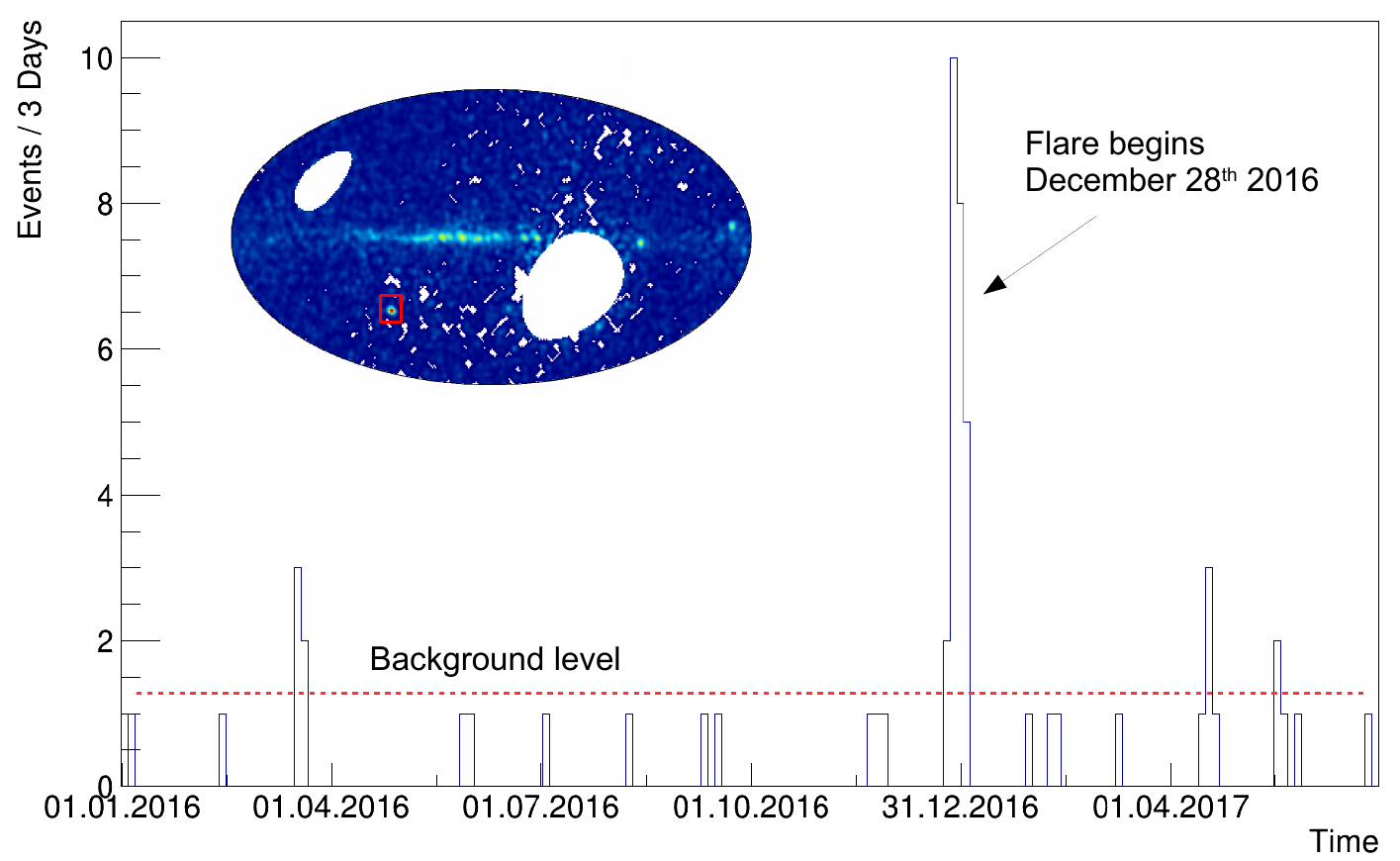}
  \caption{Number of observed events in a \SI{3}{\degree} window centered on \mbox{CTA-102} per 3
    day time bin. The flaring activity begins abruptly on 28th of December 2016 and lasts for a few
    days. The inset shows the measured $\gamma$-ray flux in December 2016 and January 2017 during
    which period CTA-102 was the brightest object in the sky.}
  \label{fig:flux-cta-102-time-flare}
\end{figure}

Figure~\ref{fig:flux-cta-102-log-average} shows the measured flux for both \mbox{AMS-02} and
\mbox{Fermi-LAT}. The AMS flux is systematically slightly below the measured Fermi flux. However,
both measured fluxes are significantly higher than the 4FGL catalog prediction. This is because the
source is highly variable. The catalog was constructed based on data collected between August 2008
and August 2016~\cite{Fermi_4FGL_2019}. The \mbox{AMS-02} flux, as well as the \mbox{Fermi-LAT} flux
presented here, were derived from photon data collected between May 2011 and November 2017.

The blazar \mbox{CTA-102} entered into a very active flaring state in the second half of
2016~\cite{Fermi_CTA_102_2018}, with multiple outburst in that period. The strongest flares were
observed first in the middle of December 2016 and in particular in a period which started on
December 28th 2016 and lasted for a few days. The flaring activity which began on December 28th was
also registered by the \mbox{AMS-02} detector.

Figure~\ref{fig:flux-cta-102-time-flare} shows the distribution of observed event counts in
\SI{3}{\degree} window around the \mbox{CTA-102} source as a function of time. Each bin contains 3
days of collected data. The 12 days beginning on December 28th 2016 contain 25 events, which is the
result of an enormous increase of flux. During December 2016 and January 2017 this source was
measured to be the strongest $\gamma$-ray sources in the sky.

The source entered another period of strong flaring activity in April 2017. In this period, the flux
was shown to be variable on timescales as short as 5 minutes, which leads to an estimation of the
emission region which is smaller than the light travel time across the black hole at the center of
the blazar ($\approx 70$ light minutes)~\cite{Shukla_2018}. As a consequence, a compact emission
region inside the jet itself was proposed.

\emptypage


%% file: summary.tex

\chapter{Summary}
\label{sec:summary}

The \mbox{AMS-02} measurement of $\gamma$-rays presented in this thesis is the first independent
test of the \mbox{Fermi-LAT} data in the energy range from \SI{200}{\mega\electronvolt} to
\SI{1}{\tera\electronvolt}.

The results show that the \mbox{AMS-02} detector contributes valuable information to the measurement
of high energy $\gamma$-rays in this energy range.

Two complementary analysis methods were developed and successfully applied to the \mbox{AMS-02}
data. The two measurements are almost entirely independent, because different parts of the
experiment are used in their derivation.

The conversion mode analysis is suitable for photon energies between \SI{200}{\mega\electronvolt}
and \SI{10}{\giga\electronvolt}. It features a very good angular resolution of \SI{0.5}{\degree} at
\SI{1}{\giga\electronvolt} and improves with energy. The on-axis effective area is approximately
\SI{180}{\square\centi\meter} for \SI{2}{\giga\electronvolt} photon energy.

In the calorimeter analysis the pointing resolution is worse than in the conversion mode at low
energies, but is still better than \SI{1}{\degree} above \SI{5}{\giga\electronvolt}. The energy
resolution of the calorimeter is excellent and allows photon reconstruction up to
\si{\tera\electronvolt} energies. The peak on-axis effective area is \SI{2200}{\square\centi\meter}.

A predictive model of the \mbox{AMS-02} sky, including diffuse emission and $\gamma$-ray sources was
developed and compared with the data, showing excellent agreement.

A dedicated \mbox{Fermi-LAT} analysis of high energy photons was carried out, which included a
correction for the absolute energy scale shift of \SI{3.3}{\percent} found by the LAT team.

The results for the $\gamma$-ray flux in the inner galaxy
($\SI{-20}{\degree} < l < \SI{80}{\degree}, |b| < \SI{8}{\degree}$), which is dominated by diffuse
emission, show that two \mbox{AMS-02} analysis modes are in excellent agreement, which excludes
sizable systematic uncertainties.

The results are also in good agreement with the measured \mbox{Fermi-LAT} flux, if the absolute
energy scale correction is applied to the LAT data. In case it is not applied the disagreement is at
the level of 6.8 standard deviations, which shows that the energy scale correction must be
considered in the LAT photon analysis. This new result highlights the importance of the
\mbox{AMS-02} measurement.

The measured emission from the \mbox{Cygnus-X} is also in good agreement with the LAT data. The high
energy fluxes of the pulsars Geminga, Crab and Vela were presented. In all cases the \mbox{AMS-02}
results are in excellent agreement with the \mbox{Fermi-LAT} result. A pulsar timing analysis for
Geminga was presented, which allowed to determine the pulsar's period of rotation and its spindown
with excellent precision.

Finally \mbox{AMS-02} observed the outburst of a flaring blazar, \mbox{CTA-102}, whose measured flux
is also in agreement with the \mbox{Fermi-LAT} result.

\emptypage


%% file: bibliography.tex

\thispagestyle{empty}
\clearpage
\addcontentsline{toc}{chapter}{Bibliography}
\bibliographystyle{thesis}
\bibliography{./papers/db}



%% file: appendix.tex

\thispagestyle{empty}
\clearpage

\begin{appendices}

\addtocontents{toc}{\protect\setcounter{tocdepth}{-5}}


\chapter{TRD Pileup Study Electron Selection}
\label{sec:appendix-trd-pileup-selection}

In order to select electrons which enter the calorimeter from below and are fully absorbed there for
the TRD pileup study in section~\ref{sec:analysis-trd-pileup-weight} the following cuts are used:

\begin{itemize}
\item Positive physics trigger from the calorimeter.
\item Exactly one reconstructed ECAL shower.
\item Up-going longitudinal shower shape from a longitudinal shower fit.
\item Angle between shower axis determination methods using the shower center of gravity in each
  layer and using ratios of cell amplitudes smaller than \SI{6}{\degree}.
\item No significant leakage of the shower to any side
\item Ratio of energy release within \SI{1}{\centi\meter} around the shower core > 0.8.
\item Ratio of energy release within \SI{3}{\centi\meter} around the shower core > 0.95.
\item At most 1 lower TOF cluster.
\item Deposited energy in each lower TOF layer smaller than \SI{10}{\mega\electronvolt}.
\end{itemize}

\KOMAoptions{open=any}
\chapter{ECAL Background Estimation}
\label{sec:appendix-ecal-background-estimation}

\begin{figure}[h]
  \centering
  \includegraphics[width=0.8\linewidth]{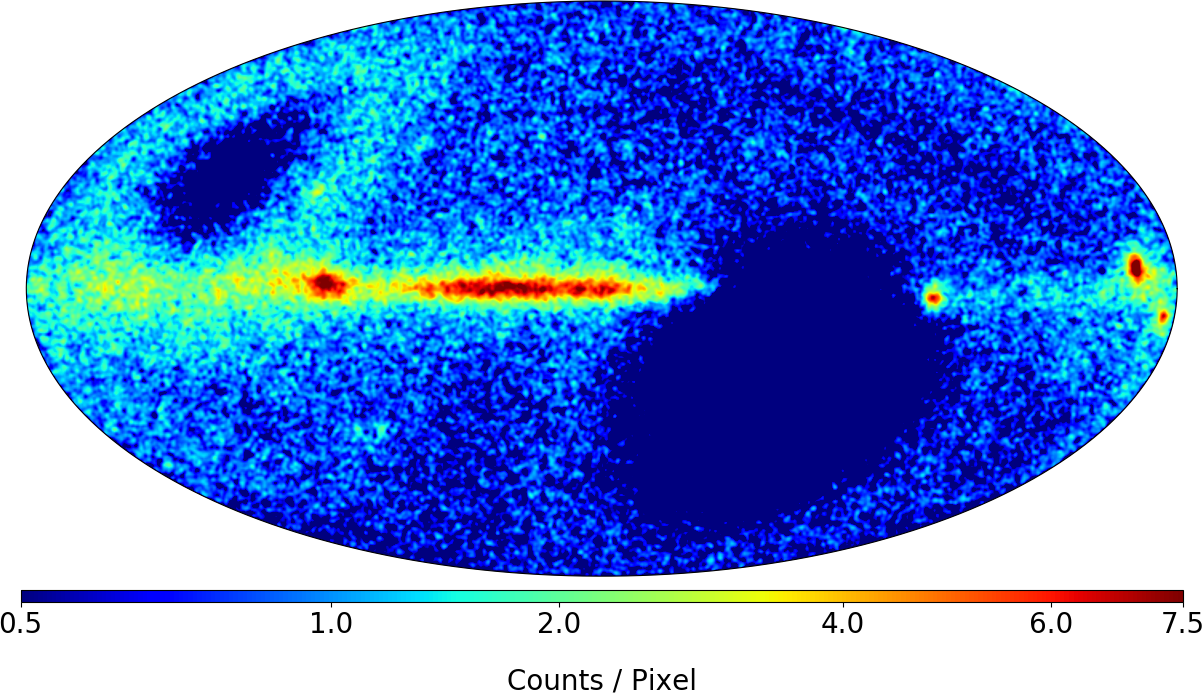}
  \caption{Measured photon counts for the calorimeter analysis between
    \SI{2}{\giga\electronvolt} and \SI{1}{\tera\electronvolt} in galactic
    coordinates, shown with a square root color scale.}
  \label{fig:counts-ecal-data}
\end{figure}

\begin{figure}[h]
  \centering
  \includegraphics[width=0.8\linewidth]{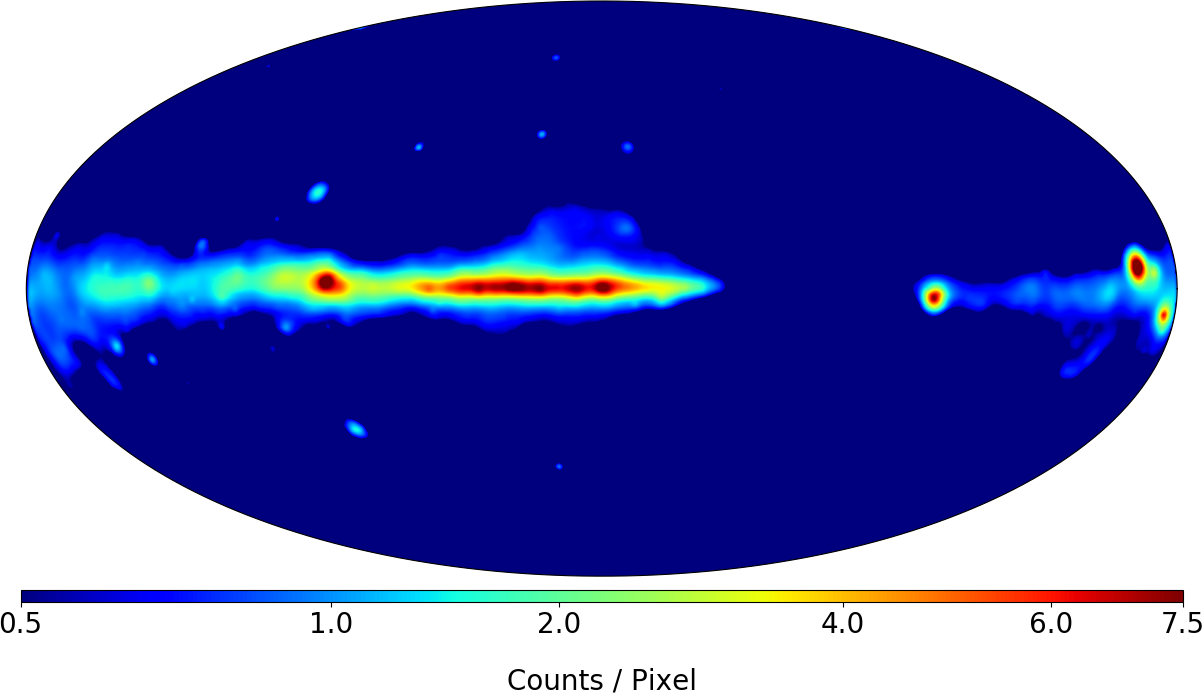}
  \caption{Model photon counts for the calorimeter analysis between
    \SI{2}{\giga\electronvolt} and \SI{1}{\tera\electronvolt} in galactic
    coordinates, shown with a square root color scale.}
  \label{fig:counts-ecal-model}
\end{figure}

\begin{figure}[t!]
  \centering
  \includegraphics[width=0.8\linewidth]{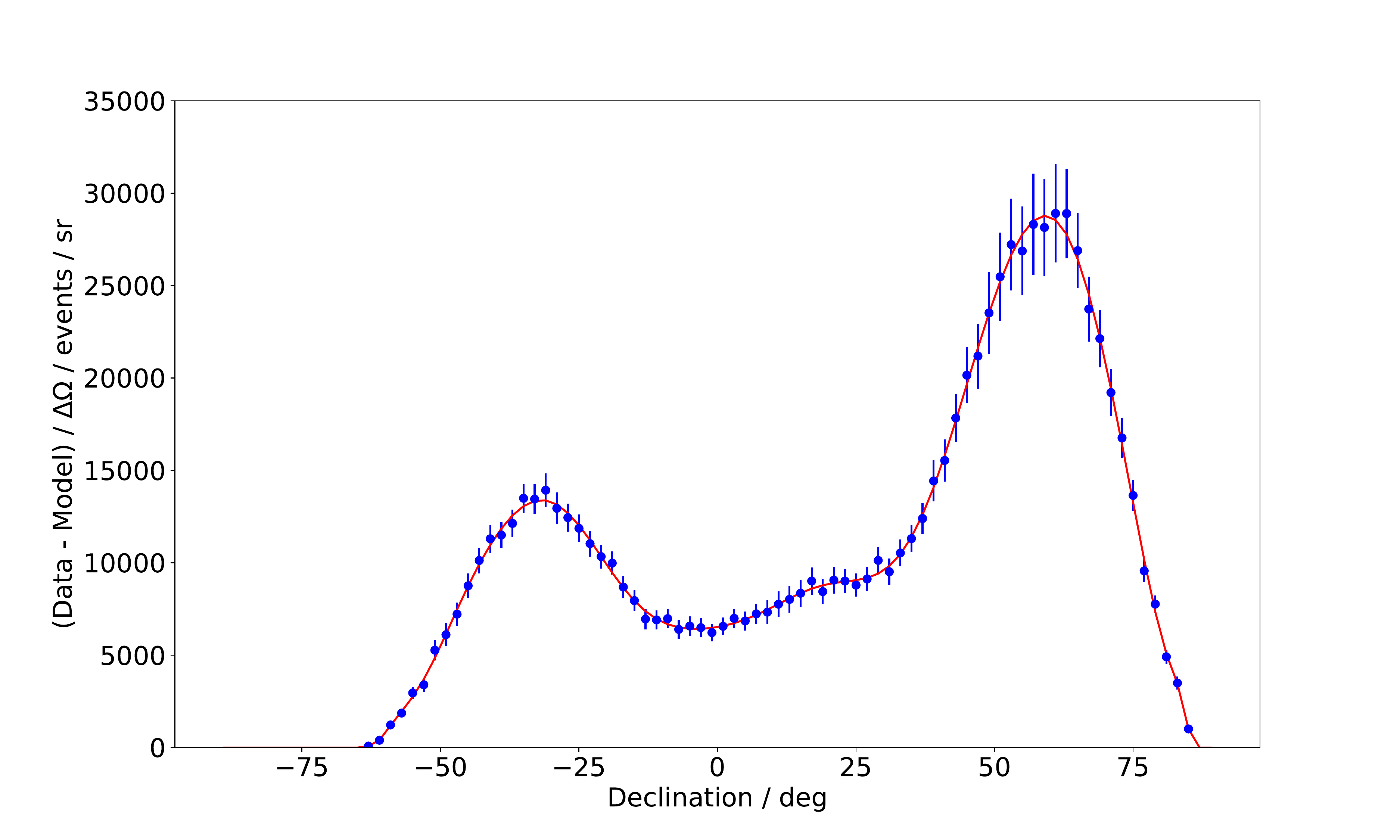}
  \caption{Average excess flux as a function of declination for the ECAL analysis, determined from
    data below \SI{2}{\giga\electronvolt}. The red line is a polynomial of order 20, which is used
    as an analytical description of the shape.}
  \label{fig:background-fit-ecal-le-polynomial-fit}
\end{figure}

\begin{figure}[t!]
  \centering
  \includegraphics[width=0.8\linewidth]{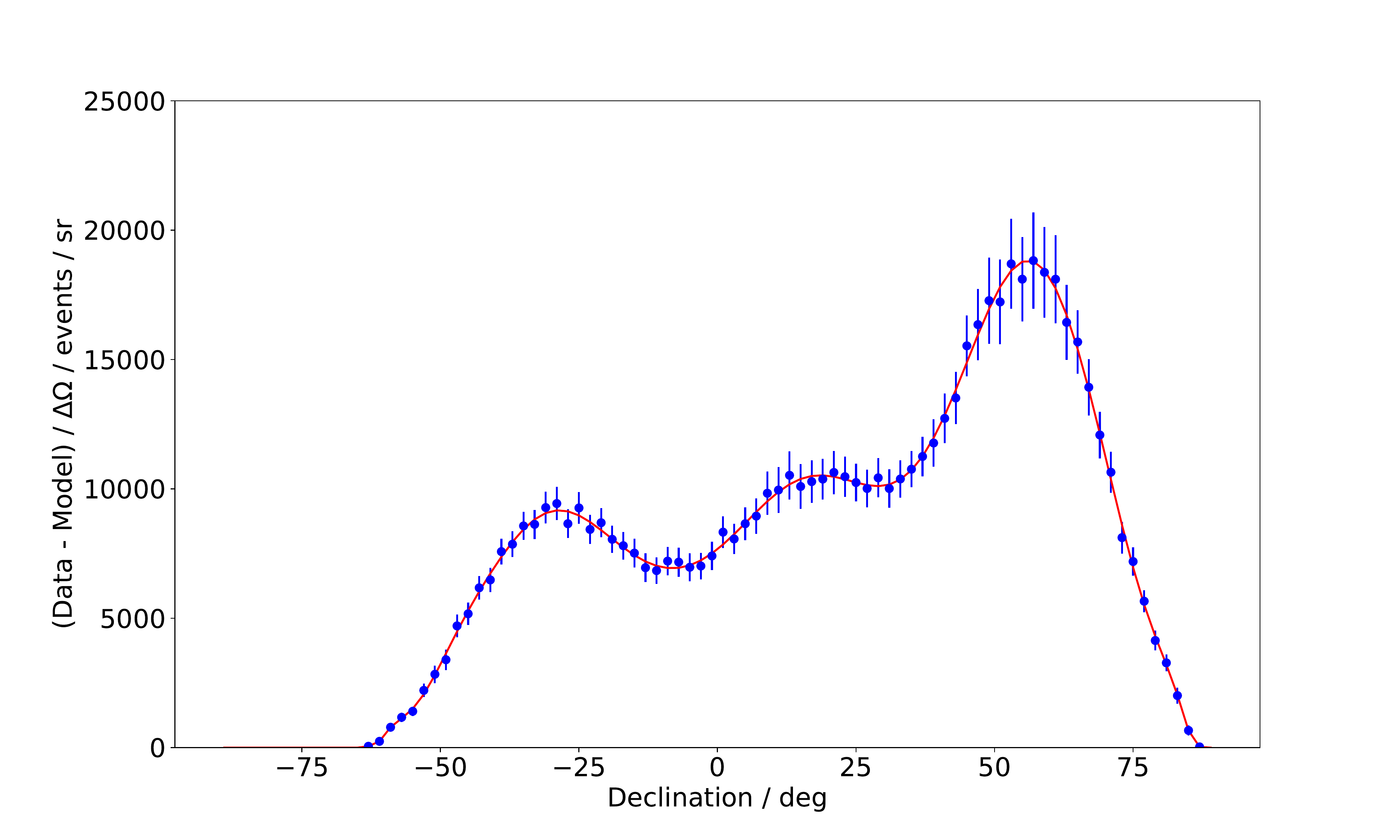}
  \caption{Average excess flux as a function of declination for the ECAL analysis, determined from
    data above \SI{2}{\giga\electronvolt}. The red line is a polynomial of order 20, which is used
    as an analytical description of the shape.}
  \label{fig:background-fit-ecal-he-polynomial-fit}
\end{figure}

\begin{figure}[t!]
  \begin{minipage}{0.48\linewidth}
    \centering
    \includegraphics[width=1.0\linewidth]{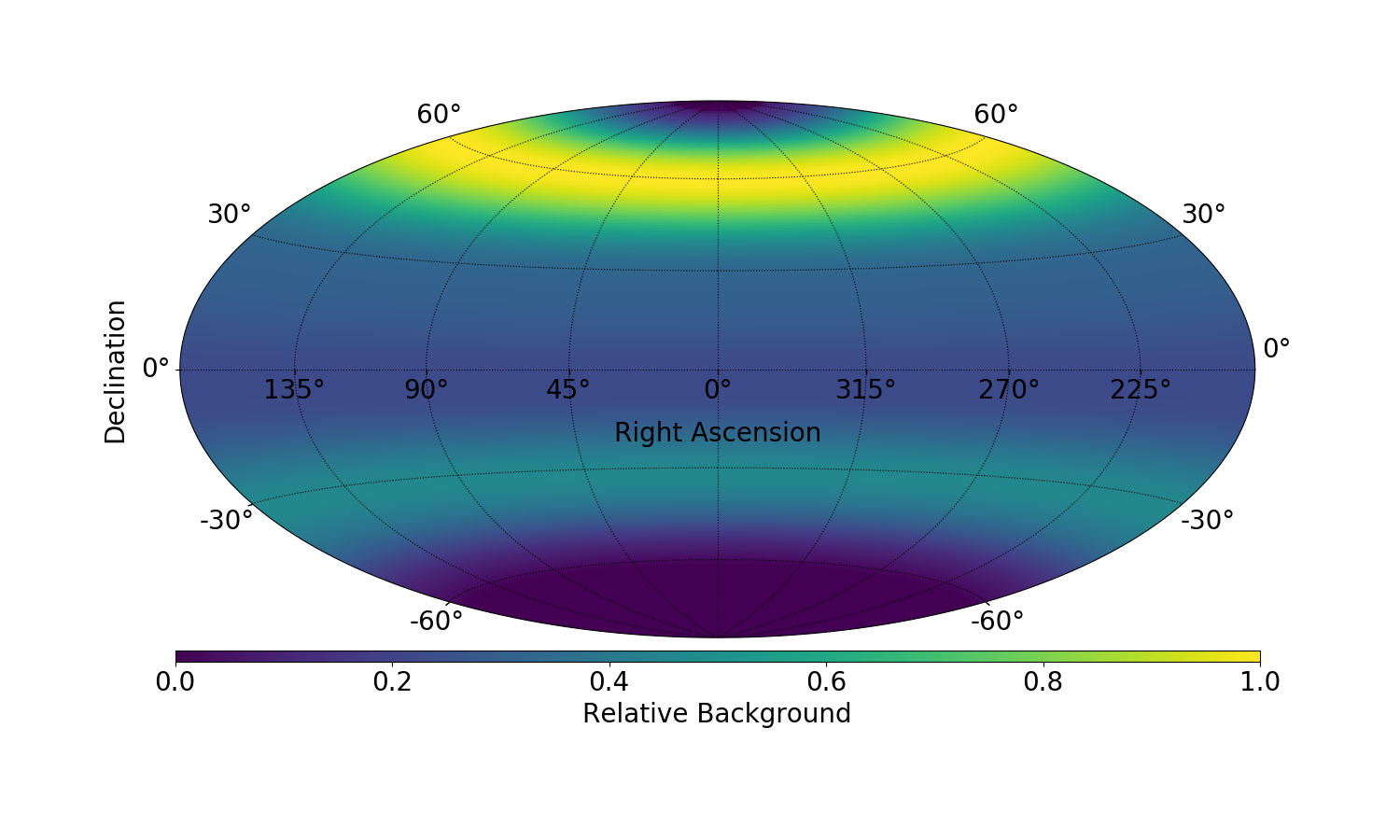}
  \end{minipage}
  \hspace{0.01\linewidth}
  \begin{minipage}{0.48\linewidth}
    \centering
    \includegraphics[width=1.0\linewidth]{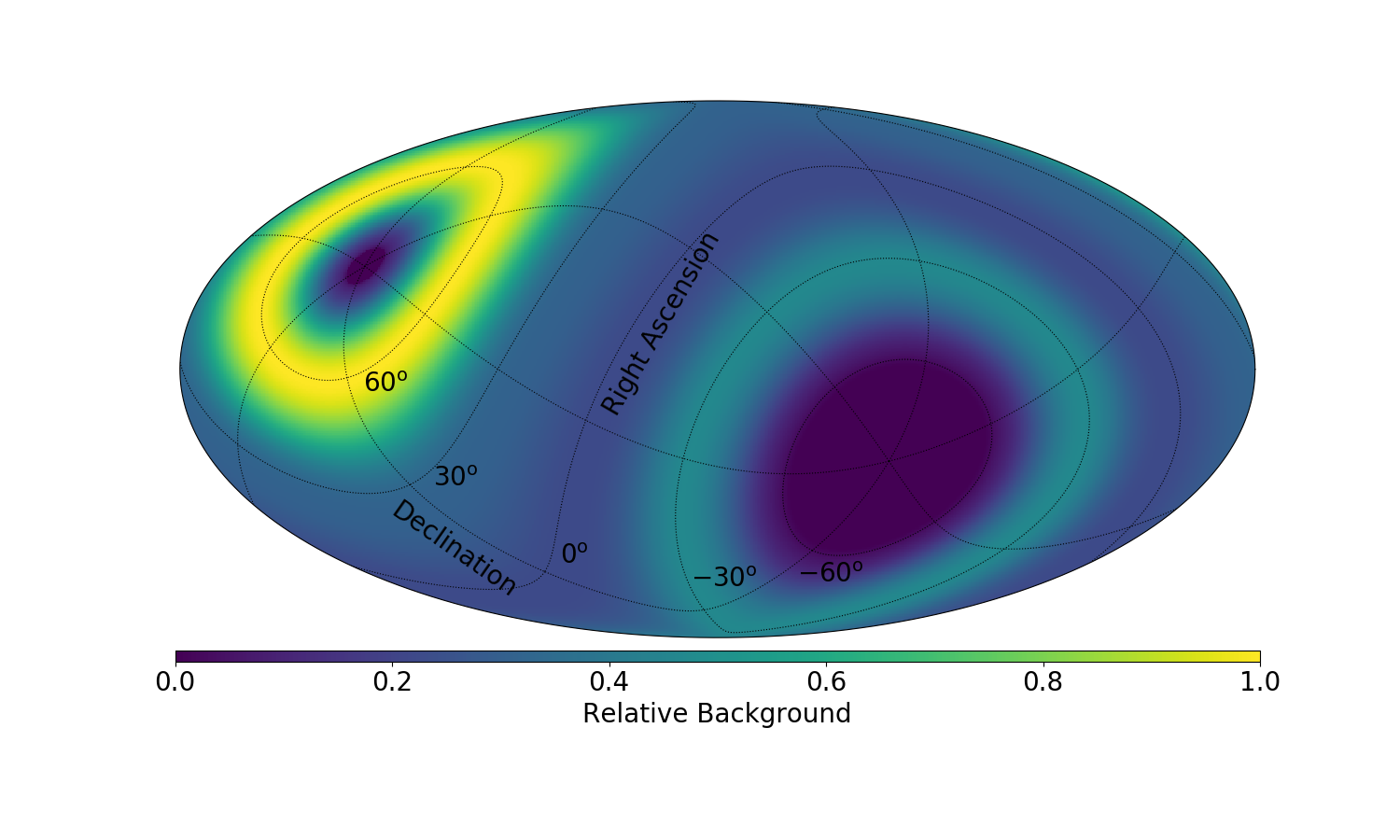}
  \end{minipage}
  \caption{ECAL analysis low energy background template map in ICRS equatorial (left) and galactic (right)
    coordinates.}
  \label{fig:background-fit-ecal-le-template}
\end{figure}

\begin{figure}[t!]
  \begin{minipage}{0.48\linewidth}
    \centering
    \includegraphics[width=1.0\linewidth]{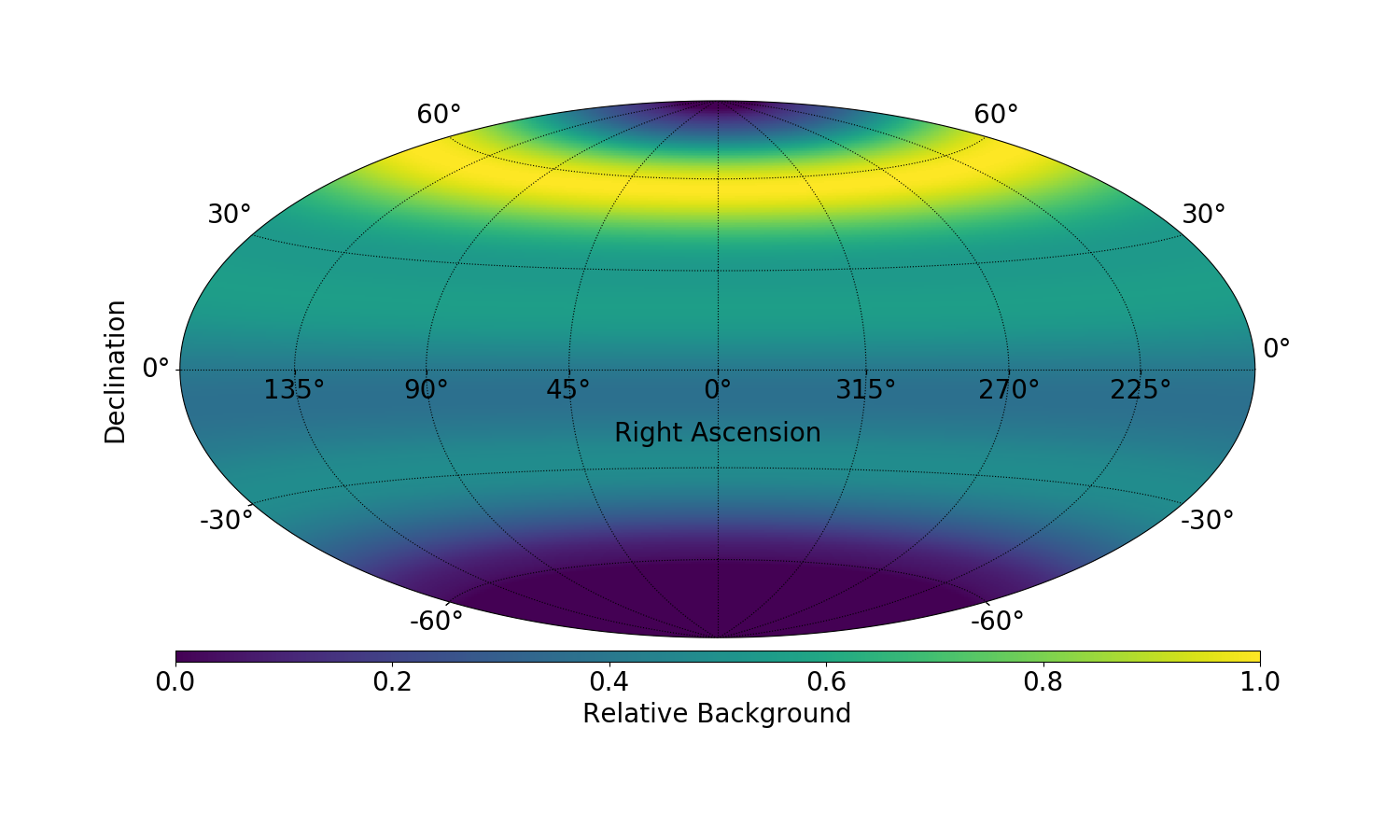}
  \end{minipage}
  \hspace{0.01\linewidth}
  \begin{minipage}{0.48\linewidth}
    \centering
    \includegraphics[width=1.0\linewidth]{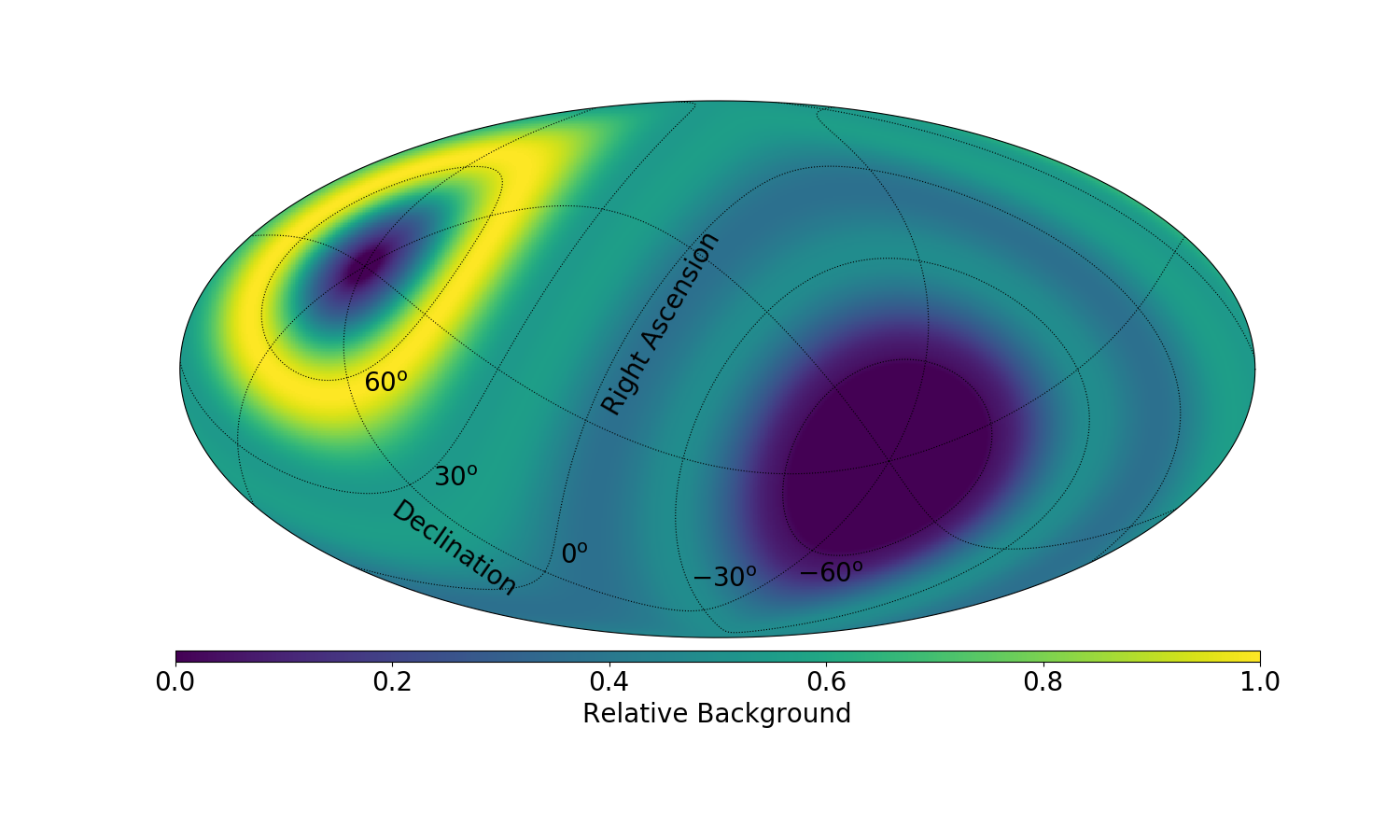}
  \end{minipage}
  \caption{ECAL analysis high energy background template map in ICRS equatorial (left) and galactic (right)
    coordinates.}
  \label{fig:background-fit-ecal-he-template}
\end{figure}


\begin{figure}[h]
  \centering
  \includegraphics[width=0.98\linewidth]{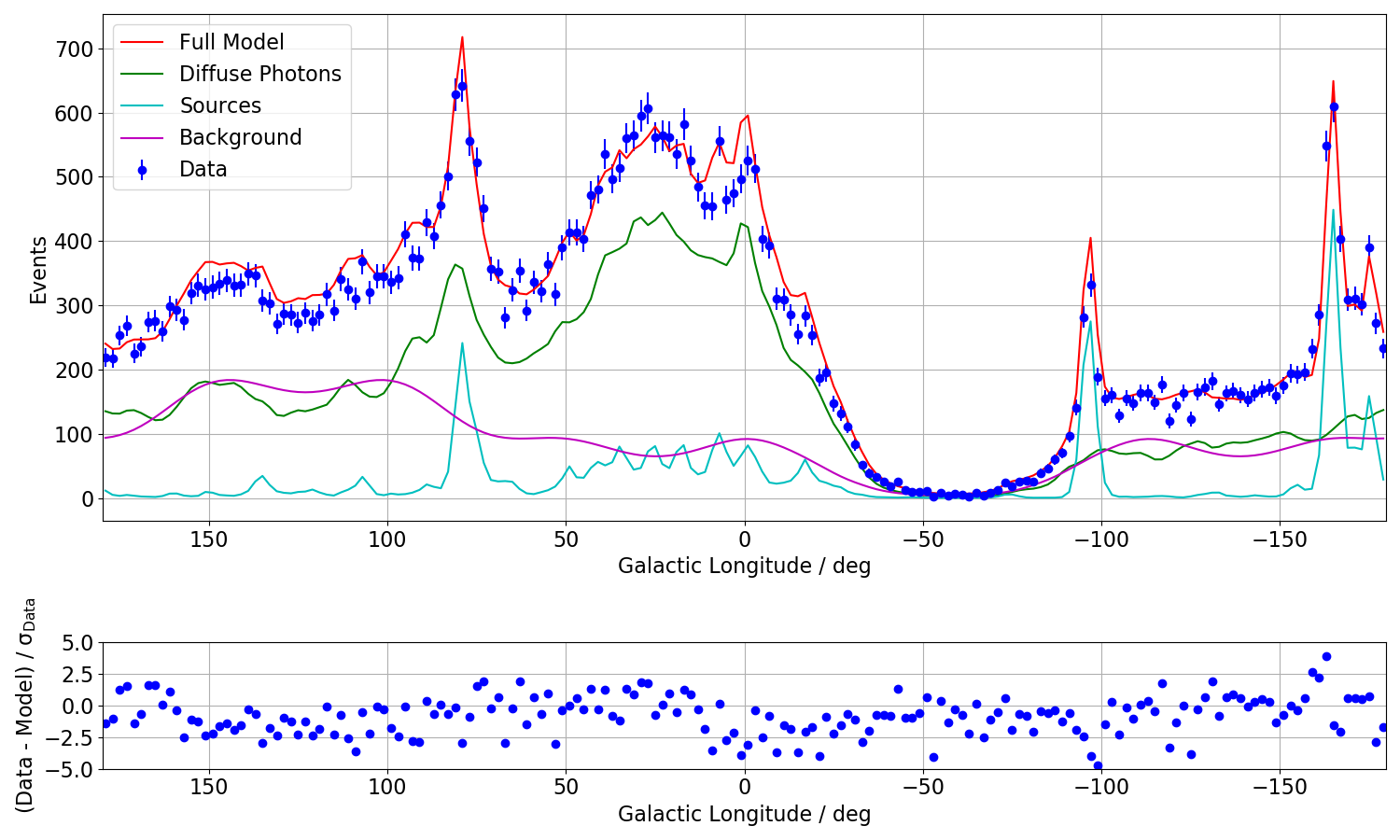}
  \caption{Comparison of measured photon counts between \SI{2}{\giga\electronvolt} and
    \SI{1}{\tera\electronvolt} in the galactic plane ($|b| < \SI{8}{\degree}$) as a function of
    galactic longitude, for the calorimeter analysis together with the full model prediction
    including background.}
  \label{fig:compare-counts-ecal-galactic-plane}
\end{figure}

\begin{figure}[h]
  \centering
  \includegraphics[width=0.98\linewidth]{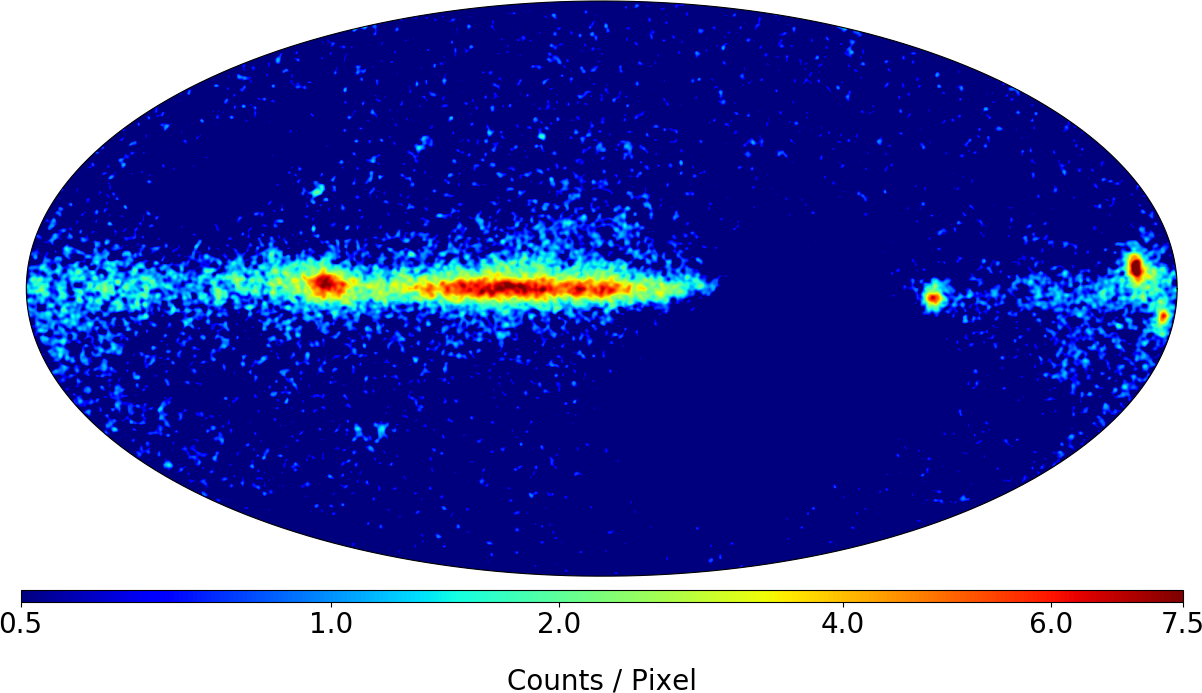}
  \caption{Background subtracted measured photon counts for the calorimeter
    analysis between \SI{2}{\giga\electronvolt} and \SI{1}{\tera\electronvolt}
    in galactic coordinates, shown with a square root color scale.}
  \label{fig:counts-ecal-data-background-removed}
\end{figure}


\KOMAoptions{open=any}
\chapter{MVA for Electron Bremsstrahlung Identification}
\label{sec:appendix-bremsstrahlung-bdt}

\begin{figure}[h]
  \centering
  \includegraphics[width=0.95\linewidth]{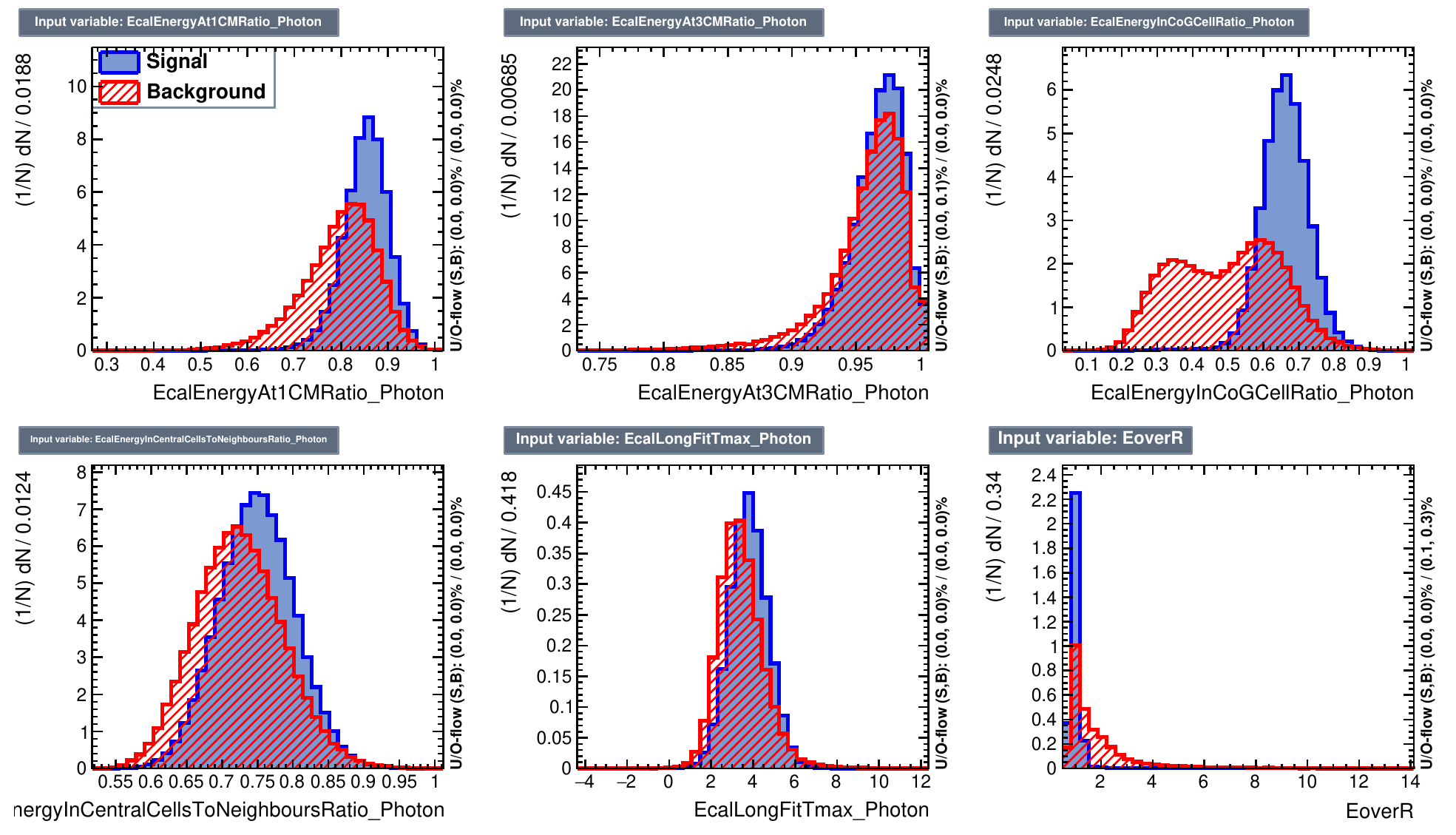}
  \includegraphics[width=0.95\linewidth]{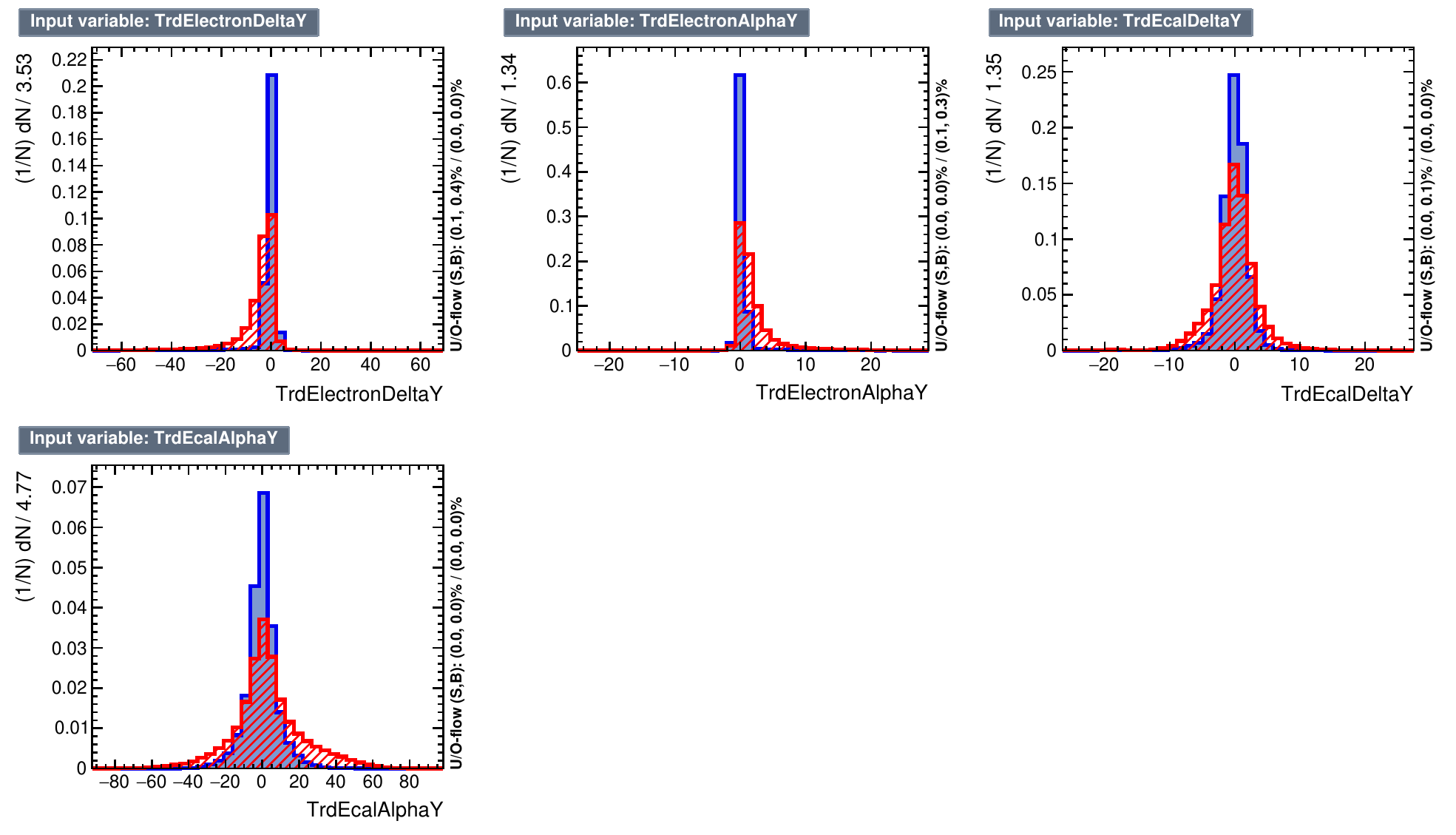}
  \caption{Input variables for the BDT classifier to identify bremsstrahlung in electron events.}
  \label{fig:brems-bdt-input}
\end{figure}

\begin{figure}[h]
  \centering
  \includegraphics[width=0.8\linewidth]{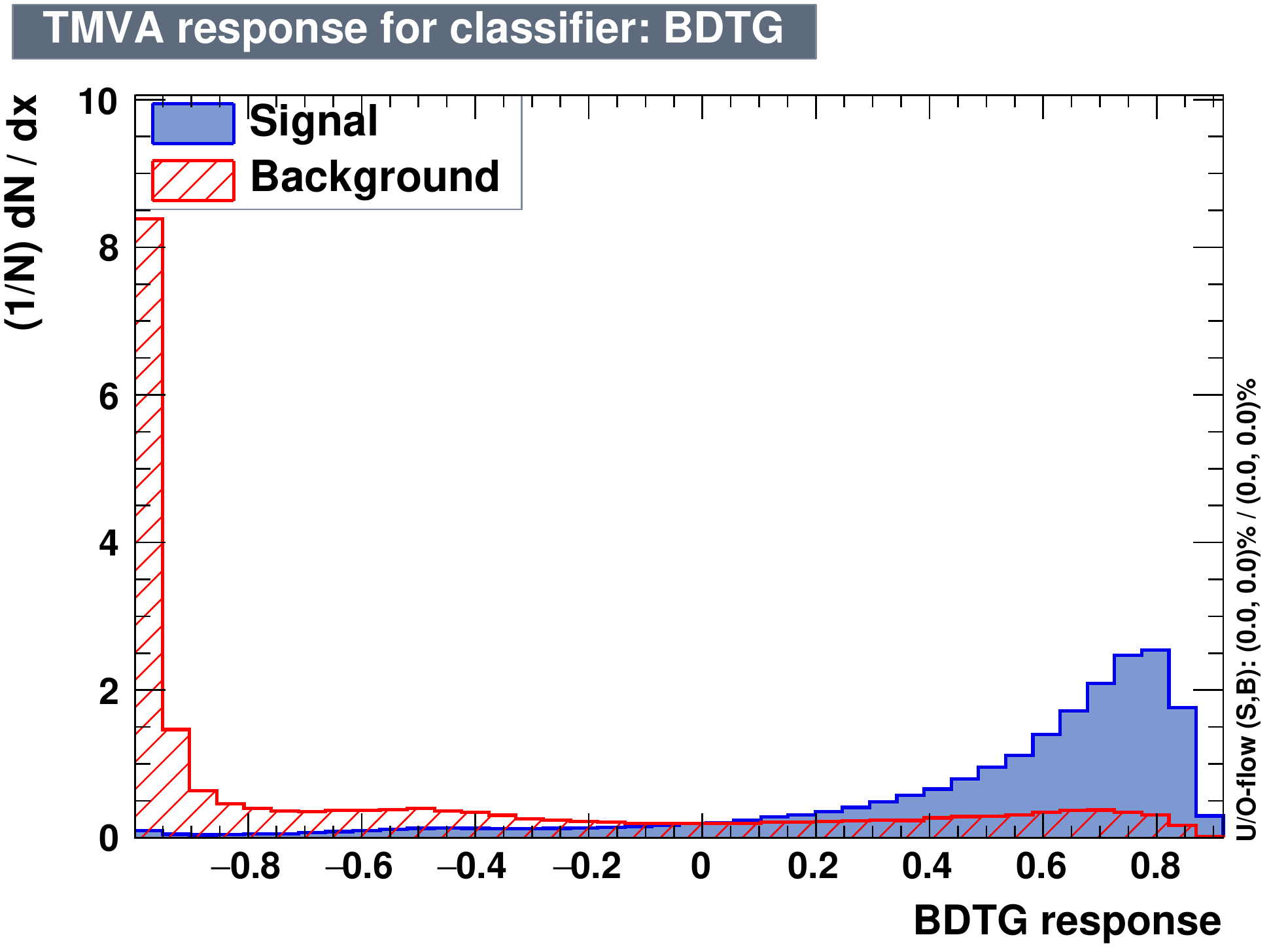}
  \caption{Distribution of the trained BDT classifier for signal and background events from the test
    set in the BDT training.}
  \label{fig:brems-bdt-output}
\end{figure}

\KOMAoptions{open=any}
\chapter{Unfolding Study for ECAL Analysis}
\label{sec:appendix-unfolding-ecal}

\begin{figure}[h]
  \centering
  \includegraphics[width=0.8\linewidth]{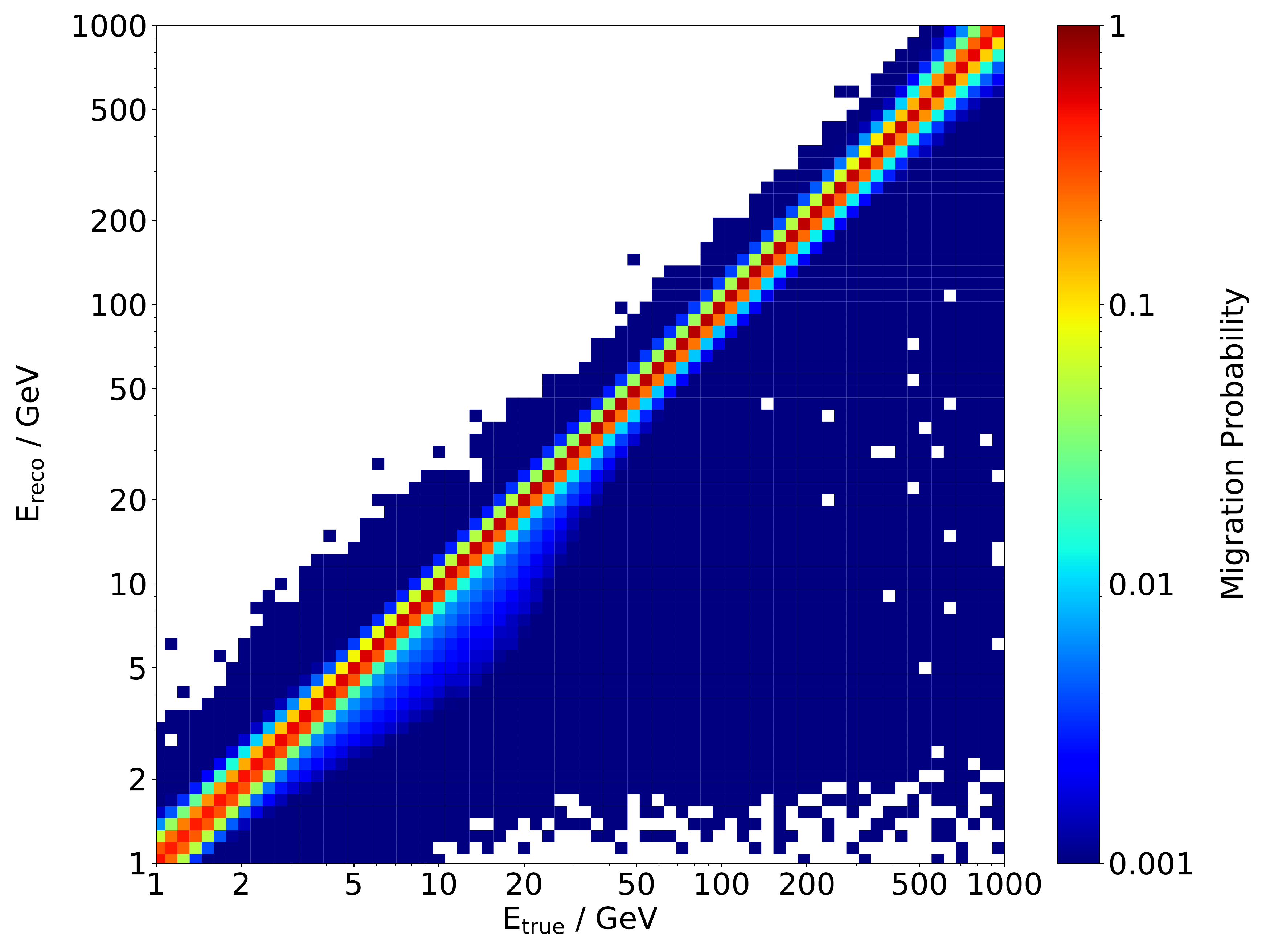}
  \caption{Migration matrix for the calorimeter analysis, used in the unfolding procedure.}
  \label{fig:unfolding-correction-migration-matrix-ecal}
\end{figure}

\begin{figure}[h]
  \centering
  \includegraphics[width=0.8\linewidth]{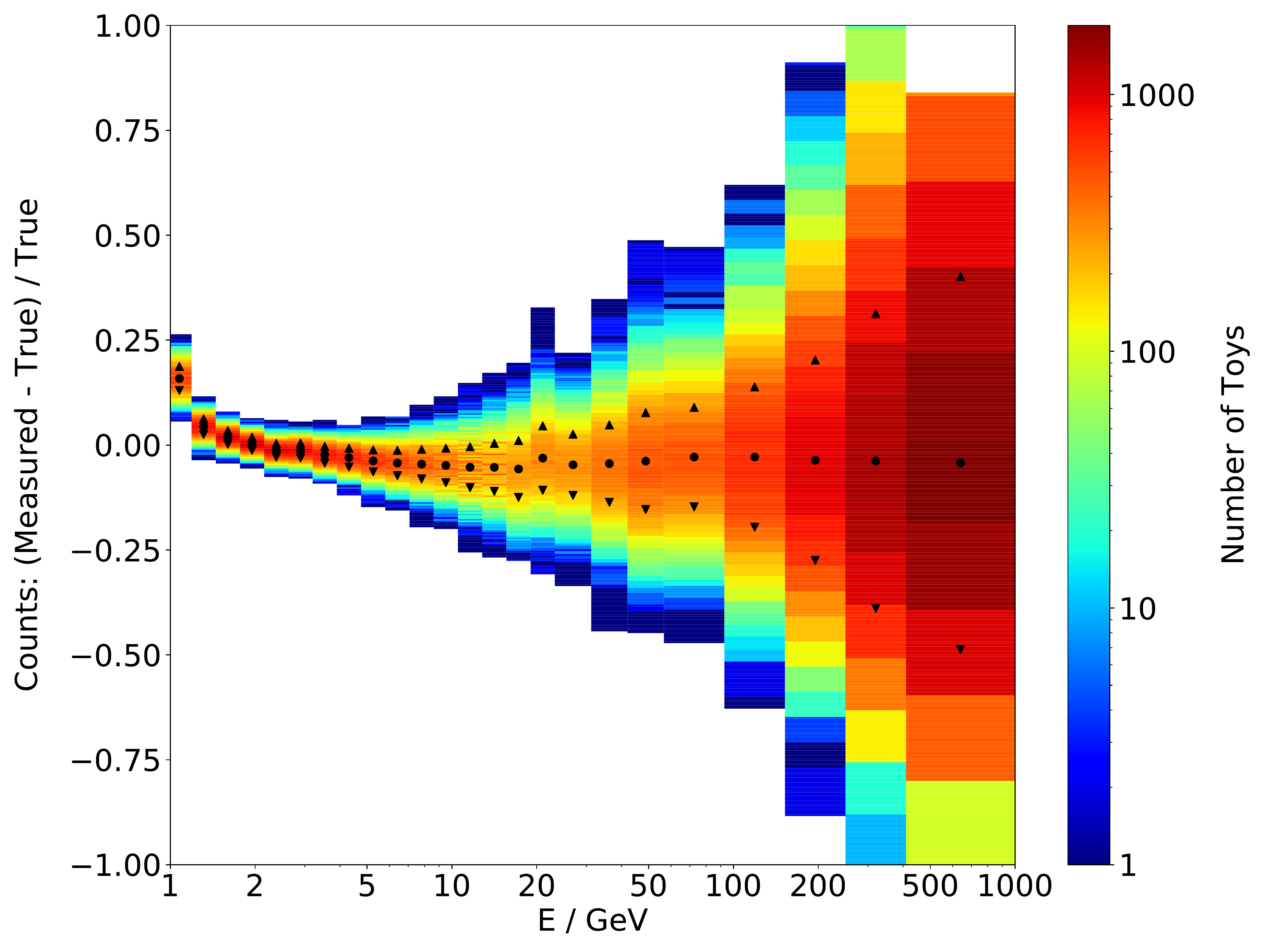}
  \caption{Distribution of the relative difference between the measured event counts and true
    average counts for 10000 toy experiments in the calorimeter analysis. The black circles
    correspond to the mean in each vertical slice, the triangles corresponds to the mean $\pm$ RMS
    position.}
  \label{fig:unfolding-correction-difference-measured-true-ecal}
\end{figure}

\begin{figure}[h]
  \centering
  \includegraphics[width=0.8\linewidth]{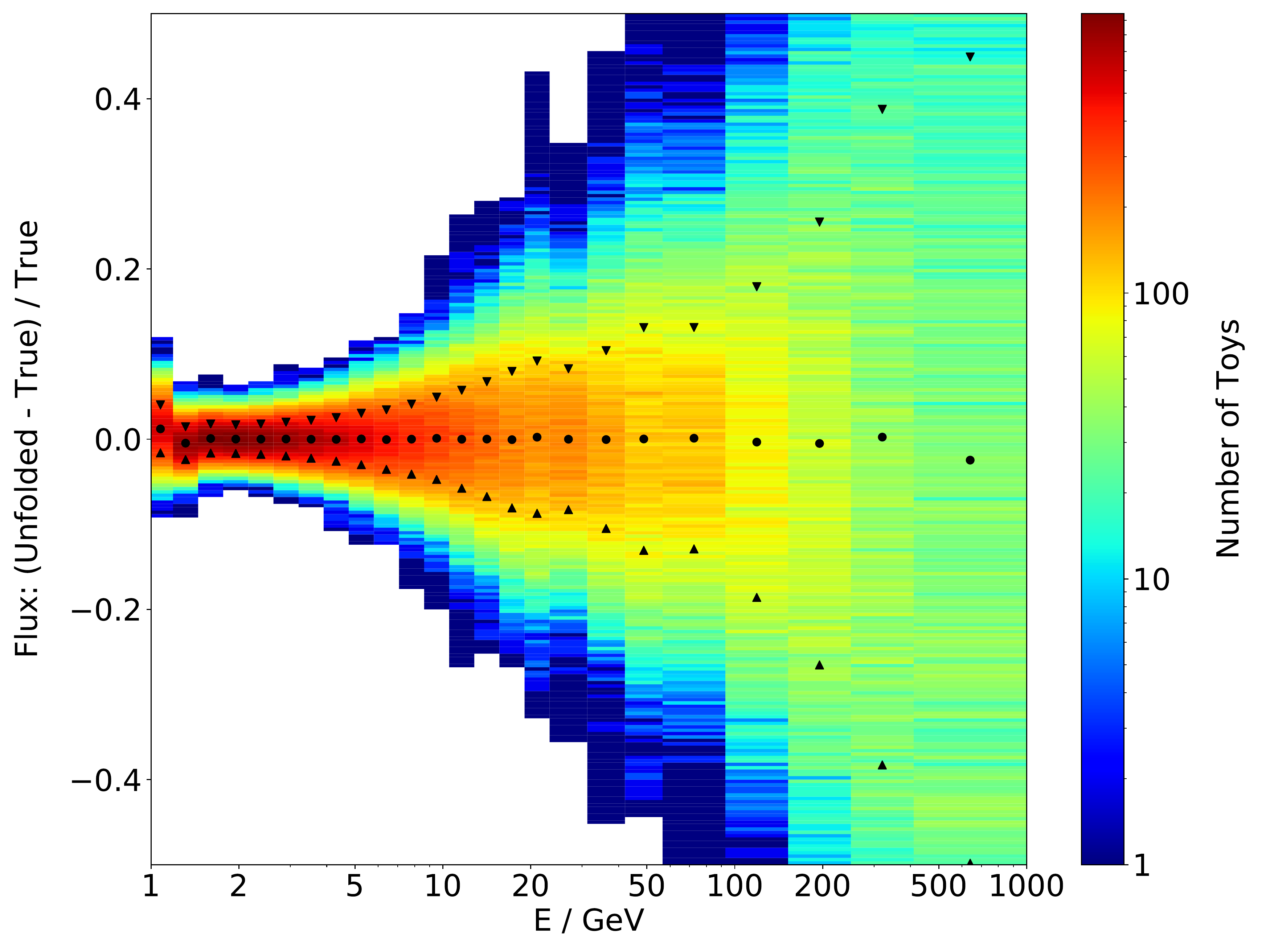}
  \caption{Distribution of the relative difference between the unfolded flux and the true flux for
    10000 toy experiments in the calorimeter analysis. The black markers correspond to the mean and
    RMS positions.}
  \label{fig:unfolding-correction-difference-unfolded-true-ecal}
\end{figure}

\begin{figure}[h]
  \centering
  \includegraphics[width=0.8\linewidth]{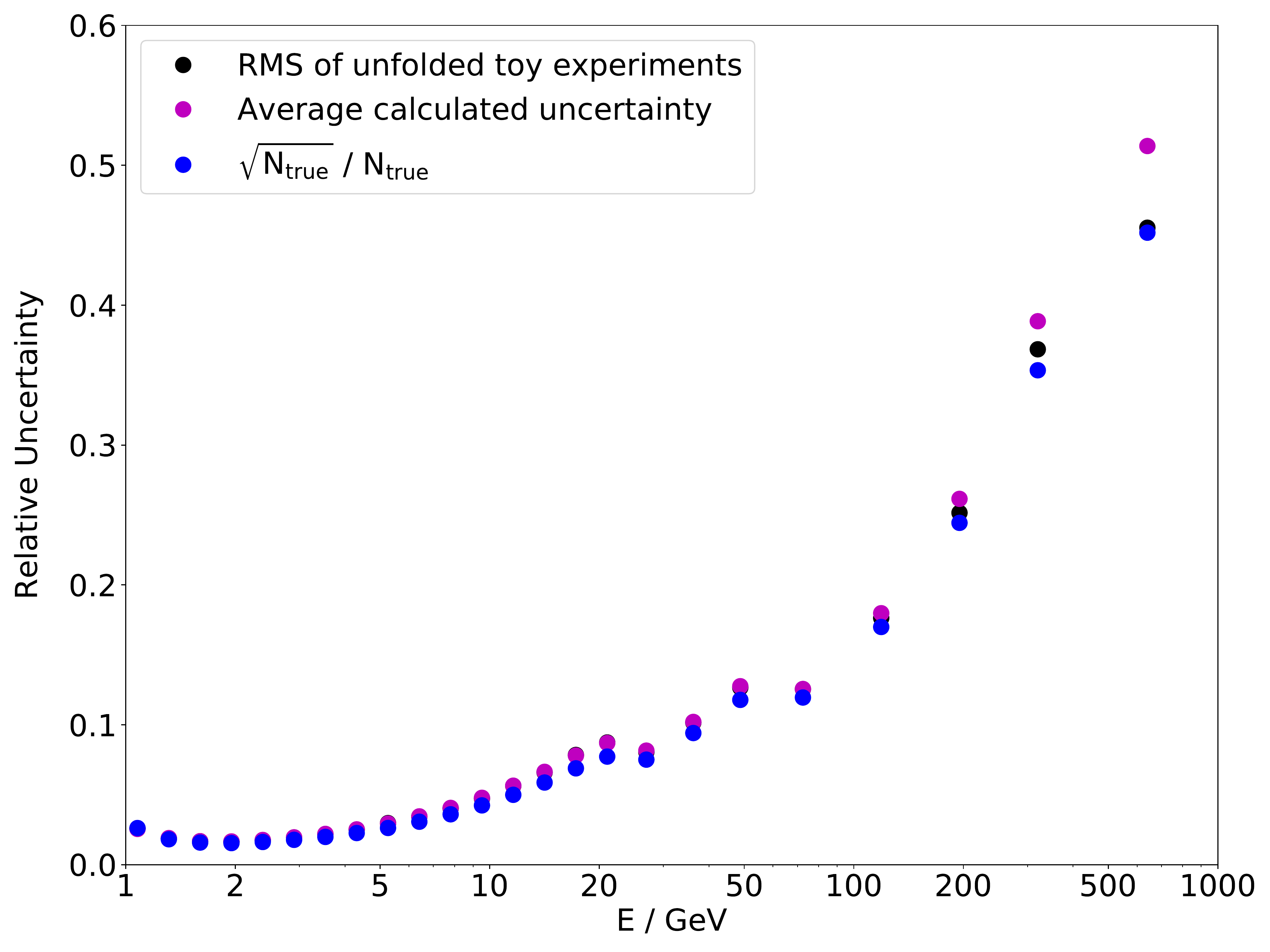}
  \caption{Relative uncertainty of the unfolded event counts compared with the inherent statistical
    uncertainty of the true distribution for the calorimeter unfolding toy.}
  \label{fig:unfolding-correction-uncertainty-ecal}
\end{figure}

\begin{figure}[h]
  \centering
  \includegraphics[width=0.95\linewidth]{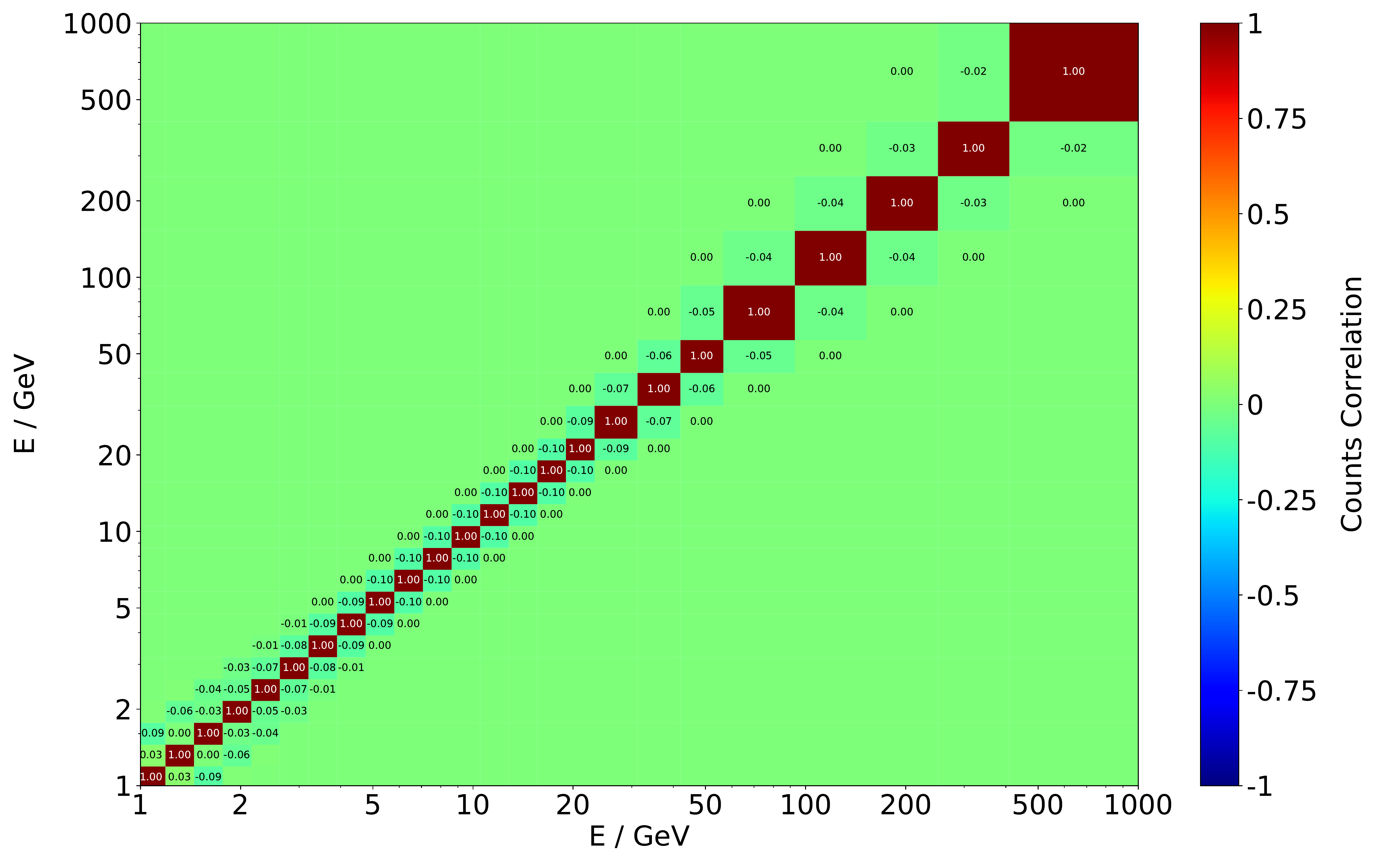}
  \caption{Correlation matrix of the unfolded counts in the calorimeter unfolding toy.}
  \label{fig:unfolding-correction-correlation-matrix-ecal}
\end{figure}

\KOMAoptions{open=any}
\chapter{Flux Measurements from Other Regions}
\label{sec:appendix-flux-other}

\begin{figure}[h]
  \centering
  \includegraphics[width=0.75\linewidth]{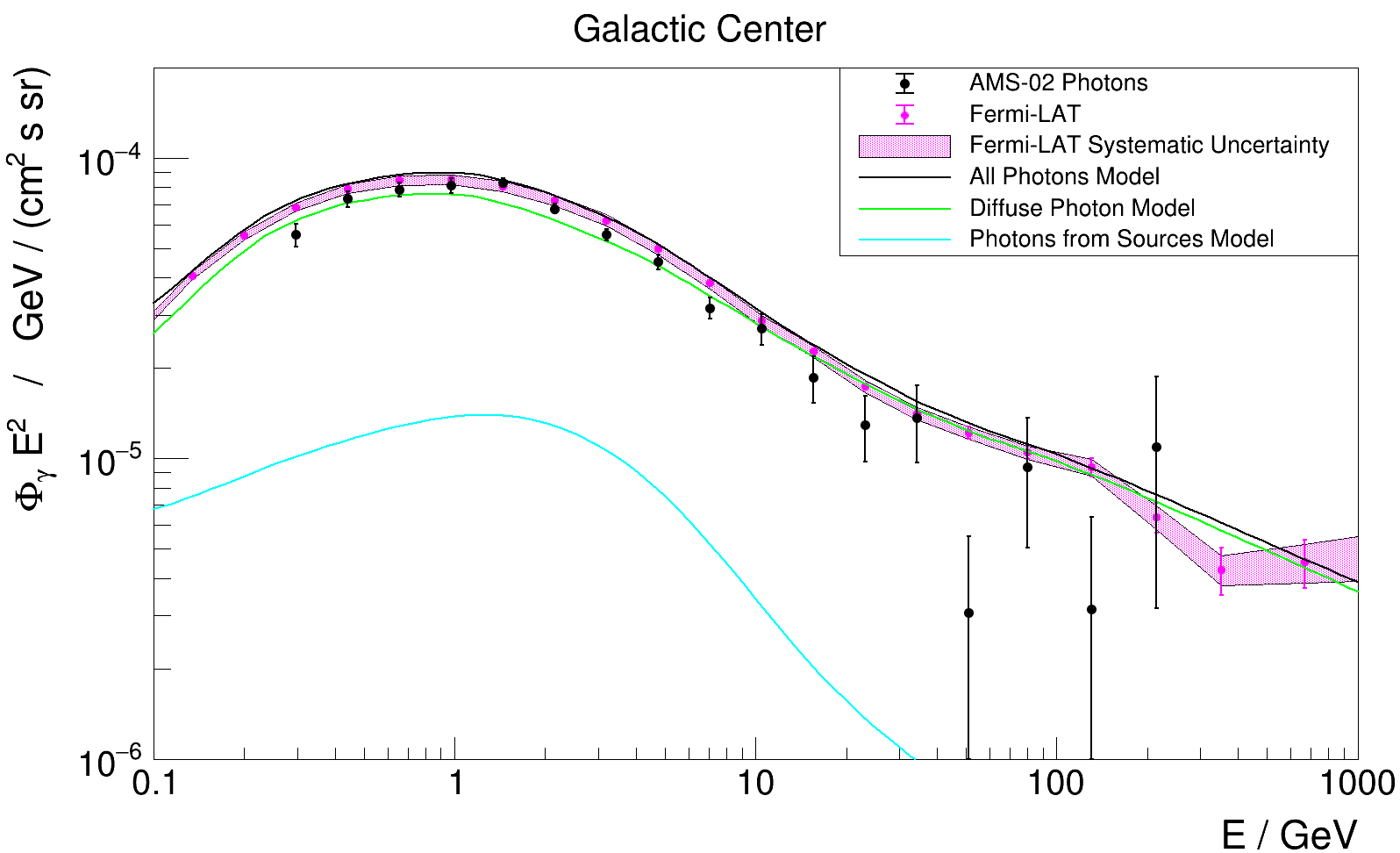}
  \caption{Average photon flux from the Galactic Center, multiplied by $E^2$ as
    a function of the photon energy. See figures~\ref{fig:flux-inner-galaxy}
    and~\ref{fig:flux-inner-galaxy-linear-average} for an explanation of the
    components.}
  \label{fig:flux-galactic-center-log-average}
\end{figure}

\begin{figure}[h]
  \centering
  \includegraphics[width=0.75\linewidth]{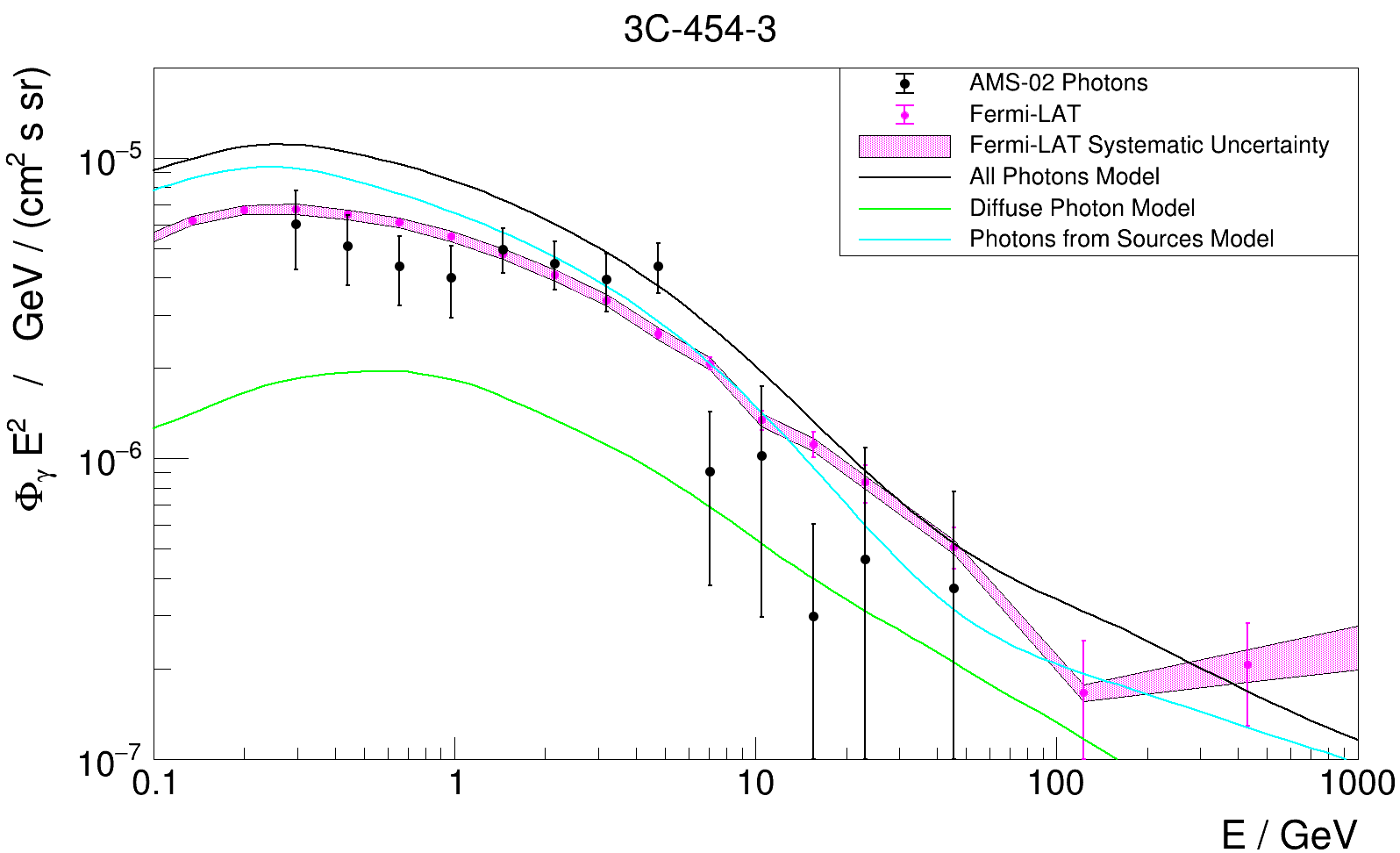}
  \caption{Average photon flux from the FSRQ quasar 3C 454.3, multiplied by
    $E^2$ as a function of the photon energy. See
    figures~\ref{fig:flux-inner-galaxy}
    and~\ref{fig:flux-inner-galaxy-linear-average} for an explanation of the
    components.}
  \label{fig:flux-3c-454-3-log-average}
\end{figure}

\begin{figure}[p]
  \centering
  \includegraphics[width=0.95\linewidth]{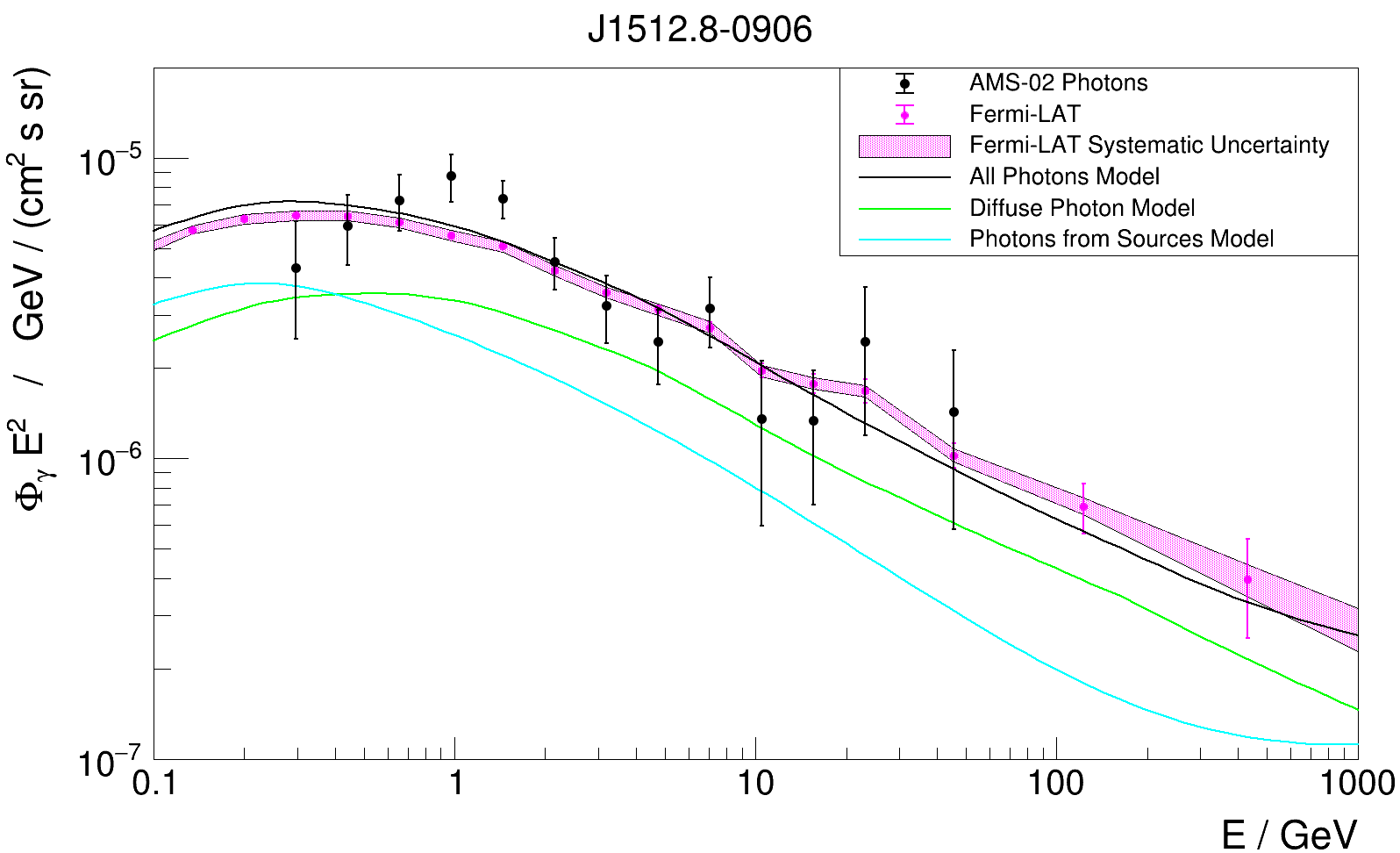}
  \caption{Average photon flux from the FSRQ quasar PKS 1510-089, multiplied by
    $E^2$ as a function of the photon energy. See
    figures~\ref{fig:flux-inner-galaxy}
    and~\ref{fig:flux-inner-galaxy-linear-average} for an explanation of the
    components.}
  \label{fig:flux-j1512-0906-log-average}

  \vspace*{3\floatsep}

  \centering
  \includegraphics[width=0.95\linewidth]{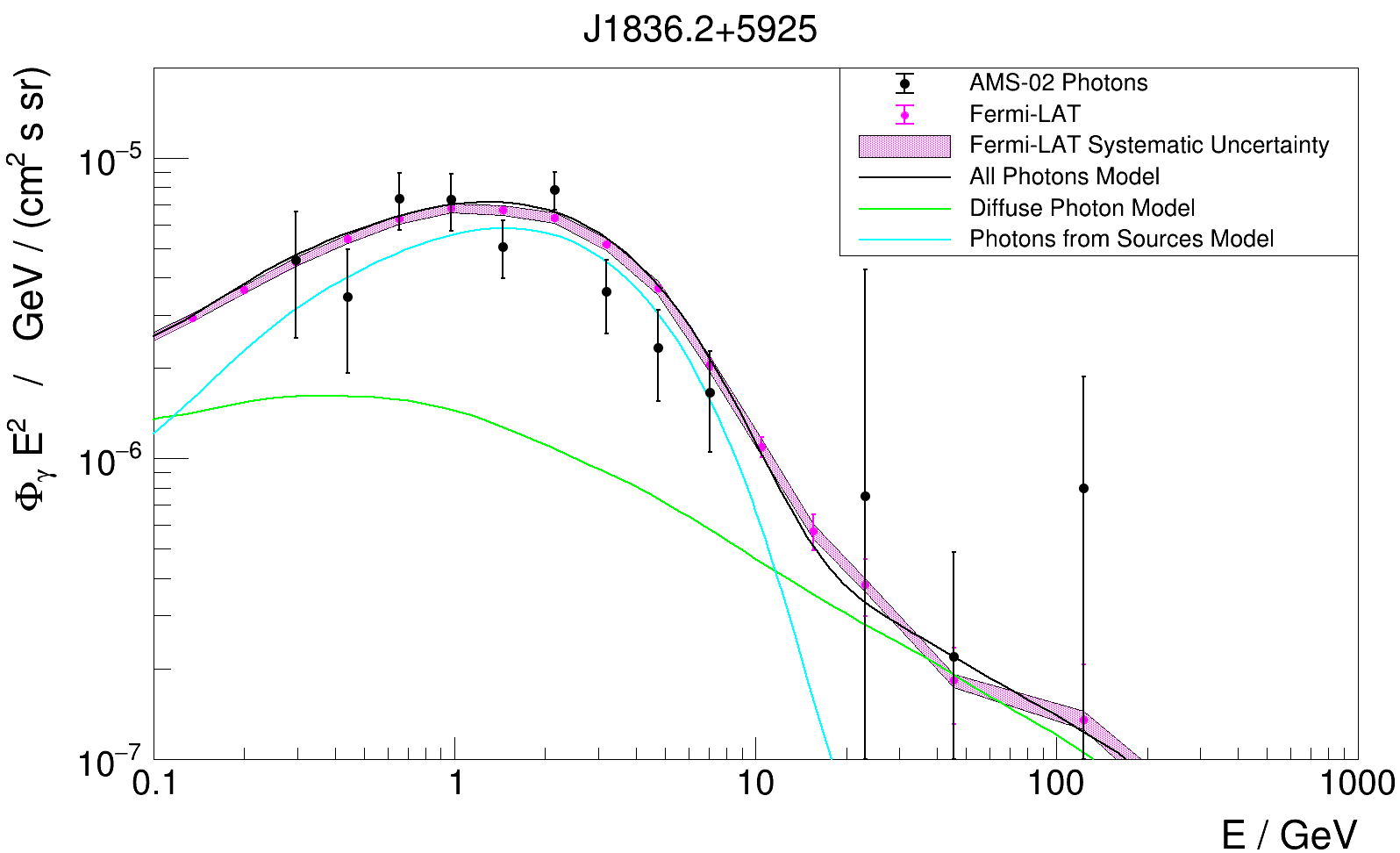}
  \caption{Average photon flux from the pulsar PSR J1836+5925, multiplied by
    $E^2$ as a function of the photon energy. See
    figures~\ref{fig:flux-inner-galaxy}
    and~\ref{fig:flux-inner-galaxy-linear-average} for an explanation of the
    components.}
  \label{fig:flux-j1836-5925-log-average}
\end{figure}

\end{appendices}

\addtocontents{toc}{\protect\setcounter{tocdepth}{0}}
\KOMAoptions{open=right}



%% file: acknowledgments.tex

\clearpage
\chapter*{Acknowledgments}
\addcontentsline{toc}{chapter}{Acknowledgments}

First and foremost I am deeply grateful to Prof. Dr. Stefan Schael who supervised this thesis and
provided me with the great opportunity to work in AMS. I have worked with him for a very long time
now and I still continue to learn new things from him to this day. I am very thankful for the
patience he has shown in working with me.

Prof. Dr. Christopher Wiebusch kindly agreed to serve as the second referee for this thesis, for
which I would like to thank him.

My time working in Aachen was made enjoyable by the great company of my colleagues and friends. I am
particularly thankful to Roman Greim, Hendrik Weber, Jens Wienkenhoever, Carsten Mai, Andreas
Bachlechner, Nikolas Zimmermann, Fabian Machate, Henning Gast, Sarah Beranek, Leila Ali Cavasonza
and Andreas G{\"u}th for all the fun we had, both in and outside the office.

In the first year of my time as a PhD student, before I joined the AMS group, I worked closely with
Roman Greim on the PERDaix experiment. It was a very pleasant and successful collaboration and I
would like to thank him for the great experience.

I have learned a lot about the operation of AMS, the TRD and TRD gas system from Thorsten
Siedenburg, Mike Capell, Joe Burger, Alessandro Bassili, Alexei Lebedev, Andreas Sabellek, Thomas
Kirn, Chan Hoon Chung and many others, both at CERN and in Aachen. I would like to thank all of them
for letting me profit from their enormous experience and knowledge.

My time as a TRD subdetector expert at CERN was very eventful, since AMS operations had just
begun. I would like to thank Thorsten Siedenburg, Thomas Kirn, Klaus L\"ubelsmeyer and Chan Hoon
Chung for their help in dealing with all the problems that occurred, be it during the day or at
night.

I had the greatest time in developing the ACsoft analysis software and the ACQt file format together
with Nikolas Zimmermann, from whom I learned a lot. I am also very grateful for the many fruitful
discussions with Henning Gast and Thorsten Siedenburg, who is the author of the original version of
the code. I also thoroughly enjoyed talking about all aspects of physics analysis with all of them.

The continuous availability of the AMS computing facilities was made possible by the enormous amount
of work put in by our colleagues in J{\"u}lich, in particular Alexander Schnurpfeil, Dorian Krause
and Philipp Th{\"o}rnig, with whom we had a great collaboration. The same is true for the staff of
the RWTH IT center, which is responsible for the High Performance Computing facilities. Christian
Terboven, Hans-J\"urgen Schnitzer, Marcus Wagner, Paul Kapinos and Sascha B\"ucken have helped us on
many occasions.

The administrative branch of the institute, in particular Georg Schwering, Natalie Driessen and
Tanja Bingler, have helped me with numerous problems over all the years. Thanks to all of you.

I would like to thank Sadakazu Haino and Kevin Flood, with whom I collaborated in the early stages
of the photon analysis. I'm also thankful to Andrei Kounine, Vitaly Choutko, Alberto Oliva, Qi Yan,
Paolo Zuccon, Marco Incaghli, Zhaoyi Qu and Weiwei Xu for helpful discussions about AMS analysis and
reconstruction of data.

My work on the production of AMS data and simulations was helped by the successful collaboration
with Vitaly Choutko, Baosong Shan and Alexandre Eline to whom I'm grateful for their continuous
support.

I admire Prof. Dr. Samuel Ting for his passion to move AMS forward. The things he accomplished for
AMS are simply astonishing.

Henning Gast and Sarah Beranek read large parts of this manuscript and gave very helpful comments
for which I am grateful.

My mother and father have always supported me wholeheartedly on my journey. I would not be where I
am today without their love and support. They have sparked my curiosity for physics and made me into
who I am today.

I am deeply grateful to Sarah for always being by my side. This thesis would not have been possible
without her love, patience and continuous support. Thank you!
